Doctoral Thesis in Electrical Engineering

# Optimal Security Response to Network Intrusions in IT Systems

KIM HAMMAR

KTH ROYAL INSTITUTE OF TECHNOLOGY

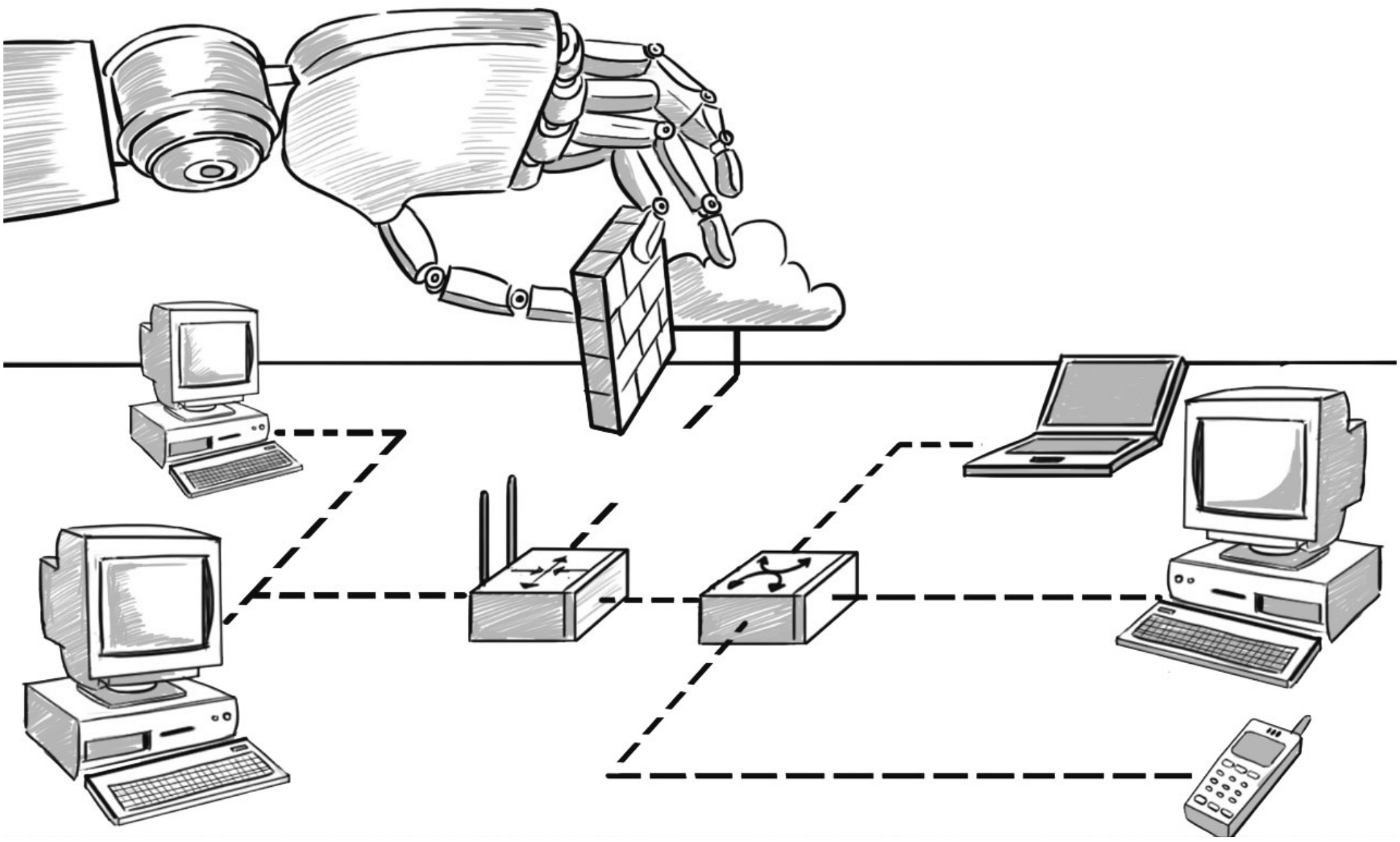

# Optimal Security Response to Network Intrusions in IT Systems

KIM HAMMAR

Academic Dissertation which, with due permission of the KTH Royal Institute of Technology, is submitted for public defence for the Degree of Doctor of Philosophy on Thursday the 5th December 2024, at 14:00 in F3, Lindstedtsvägen 26, Stockholm.

**Abstract**

Cybersecurity is one of the most pressing technological challenges of our time and requires measures from all sectors of society. A key measure is automated security response, which enables automated mitigation and recovery from cyber attacks. Significant strides toward such automation have been made due to the development of rule-based response systems. However, these systems have a critical drawback: they depend on domain experts to configure the rules, a process that is both error-prone and inefficient. Framing security response as an optimal control problem shows promise in addressing this limitation but introduces new challenges. Chief among them is bridging the gap between theoretical optimality and operational performance. Current response systems with theoretical optimality guarantees have only been validated analytically or in simulation, leaving their practical utility unproven.

This thesis tackles the aforementioned challenges by developing a practical methodology for optimal security response in IT infrastructures. It encompasses two systems. First, it includes an emulation system that replicates key components of the target infrastructure. We use this system to gather measurements and logs, based on which we identify a game-theoretic model. Second, it includes a simulation system where game-theoretic response strategies are optimized through stochastic approximation to meet a given objective, such as quickly mitigating potential attacks while maintaining operational services. These strategies are then evaluated and refined in the emulation system to close the gap between theoretical and operational performance.

We present CSLE, an open-source platform that implements our methodology. This platform allows us to experimentally validate the methodology on several instances of the security response problem, including intrusion prevention, intrusion response, intrusion tolerance, and defense against advanced persistent threats. We prove structural properties of optimal response strategies and derive efficient algorithms for computing them. This enables us to solve a previously unsolved problem: demonstrating optimal security response against network intrusions on an IT infrastructure.

**Sammanfattning**

Cybersäkerhet är en av vår tids mest angelägna teknologiska utmaningar och kräver åtgärder från alla samhällssektorer. En nyckelåtgärd är automatisering av säkerhetsrespons, vilket möjliggör automatisk avvärjning och återhämtning från cyberangrepp. Betydande framsteg mot sådan automatisering har gjorts genom utvecklingen av regelbaserade responssystem. Dessa system har dock en kritisk nackdel: de är beroende av domänexperter för att konfigurera reglerna, en process som är både felbenägen och ineffektiv. Modellering av säkerhetsrespons som ett reglertekniskt optimeringsproblem är ett lovande sätt att hantera denna begränsning men medför nya utmaningar. Främst bland dem är att överbrygga gapet mellan teoretisk optimalitet och operativ prestanda. Nuvarande responssystem med teoretiska optimalitetsgarantier har endast validerats i simulering, vilket lämnar deras praktiska nytta oprövad.

Den här avhandlingen behandlar ovannämnda utmaningar genom att utveckla en praktisk metodik för optimal säkerhetsrespons. Metodiken omfattar två system. För det första inkluderar den ett emuleringssystem som replikerar IT-infrastrukturer i en virtuell miljö. Från detta system samlar vi in mätvärden och loggar som vi sedan använder för att identifiera en spelteoretisk modell. För det andra innefattar metodiken ett simuleringssystem där spelteoretiska responsstrategier optimeras genom stokastisk approximation för att uppnå ett givet mål, exempelvis att minimera responssystemets driftkostnad samt maximera dess förmåga att automatiskt stävja potentiella cyberangrepp. De optimerade responsstrategierna utvärderas och förfinas sedan i emuleringssystemet för att minska klyftan mellan teoretisk och operativ prestanda.

Vi presenterar CSLE, en originell plattform med öppen källkod som implementerar vår metodik. Med hjälp av denna plattform utvärderar vi vår metodik experimentellt på flera användningsområden, inklusive förebyggande av intrång, intrångssvar, intrångstolerans och försvar mot avancerade bestående hot. Vi bevisar strukturella egenskaper hos optimala responsstrategier och härleder effektiva algoritmer för att beräkna dem. Detta gör det möjligt för oss att lösa ett tidigare olöst problem: att demonstrera optimal säkerhetsrespons mot nätverksintrång på en IT-infrastruktur.

# ACKNOWLEDGMENTS

It has been a pleasure to be a graduate student at KTH and starting this journey is one of the best decisions I have made. I want to thank my advisor, Prof. Rolf Stadler, for his guidance, helpful suggestions, and criticism throughout my doctoral studies. He has given me the freedom to pursue my ideas, and I owe much of my research style to him. Thanks are also due to Prof. Pontus Johnson, who provided useful suggestions and contacts. I am also thankful to Prof. Gunnar Karlsson, under whom I had the privilege of serving as a teaching assistant.

I sincerely thank my opponent, Prof. Tansu Alpcan, for his thoughtful comments. His expertise has contributed to deepening my understanding of the subject matter. I extend my gratitude to the members of the grading committee, Prof. Emil Lupu, Prof. Karl H. Johansson, Prof. Alina Oprea, and thesis reviewer Prof. Henrik Sandberg, for their support and thorough evaluation of this thesis. I am also grateful to Prof. Viktoria Fodor for chairing the thesis defense.

During my doctoral studies, I had the privilege to visit several universities and research labs, which broadened my academic perspective. I thank the people who hosted me at their institutions: Assoc. Prof. Quanyan Zhu at New York University (NYU), Prof. Tansu Alpcan at the University of Melbourne, Asst. Prof Maria Apostolaki at Princeton University, Dr. Una-May O'Reilly and Dr. Erik Hemberg at Massachusetts Institute of Technology, Dr. Neil Dhir at the Alan Turing Institute, and Dr. Enrico Lovat at Siemens Research. A special thanks goes to Tao Li at NYU for our excellent collaboration.

Thank you to all the students, faculty, and researchers at the network and systems engineering division at KTH. What a tremendous place it has been to learn and grow. Special thanks to Dr. Forough Shahab for helping me settle when I first arrived at the division and for being a great friend. I appreciate Xiaoxuan Wang, Jakob Nyberg, Yeongwoo Kim, and Huy Le for their attentive proofreading of this thesis. Thanks to Prof. György Dán for his dedication to organizing the division's reading group on security, which fostered a collaborative environment. I am grateful to Feridun Tütüncüoglu, Dr. Ezzeldin Shereen, Dr. Henrik Hellström, and Afsaneh Mahmoudi for the memorable activities outside of work, with particular attention to the trips to Turkey with Feridun and Serap.

Above all, I would like to thank my family for being very enthusiastic about my studies despite not fully understanding what I was doing (which is understandable, as there were times when I didn't either). Most importantly, I am grateful to my father, whose memory motivates me daily, and my dear mother, who is the reason for most of my accomplishments in life. I am also thankful to my siblings for providing wonderful moments away from the academic world. Tack!

*Stockholm, December 2024.*

# Contents

# List of Publications

This thesis is based on the following papers.

1. **K. Hammar** and R. Stadler (2022), "Intrusion Prevention through Optimal Stopping [179]." *IEEE Transactions on Network and Service Management (TNSM)*, vol. 19, no. 3, pp. 2333-2348.

2. **K. Hammar** and R. Stadler (2024), "Learning Near-Optimal Intrusion Responses Against Dynamic Attackers [183]." *IEEE Transactions on Network and Service Management (TNSM)*, vol. 21, no. 1, pp. 1158-1177.

3. **K. Hammar** and R. Stadler (2023), "Scalable Learning of Intrusion Response through Recursive Decomposition [185].". Springer Lecture Notes in Computer Science, vol 14167, pp. 172–192. *Decision and Game Theory for Security (GameSec)*, Avignon, France.

4. **K. Hammar** and R. Stadler (2024), "Intrusion Tolerance for Networked Systems through Two-Level Feedback Control [181]." *IEEE Dependable Systems and Networks Conference (DSN)*, Brisbane, Australia, pp. 338-352.

   A companion to this paper is published as **K. Hammar** and R. Stadler (2024) "Intrusion Tolerance as a Two-Level Game [180]" in Springer Lecture Notes in Computer Science, vol 14908, pp. 3-23. *Decision and Game Theory for Security (GameSec)*, New York, USA.

5. **K. Hammar**, T. Li, R. Stadler, and Q. Zhu (2024), "Automated Security Response through Online Learning with Adaptive Conjectures [174].". To appear in *IEEE Transactions on Information Forensics and Security (TIFS).*

6. **K. Hammar**, N. Dhir, and R. Stadler (2024), "Optimal Defender Strategies for CAGE-2 using Causal Modeling and Tree Search [173]". Submitted to *IEEE Transactions on Dependable and Secure Computing (TDSC)* (Under review).

***Contributions*** K. Hammar conceptualized the original idea, conducted the mathematical modeling, wrote the code, derived the theoretical results, performed the experiments, and led the writing of each paper. R. Stadler contributed to each paper's conceptualization, modeling, and writing. Q. Zhu and N. Dhir contributed to the writing of Paper 5 and Paper 6, respectively. T. Li contributed to the modeling and writing related to Paper 5; he wrote the proof of Thm. 5.4.

# List of Figures

# List of Tables

# List of Algorithms

# Code Listings

# Experimental Platform

All experimental results presented in this thesis were obtained using our custom-developed platform called the **C**yber **S**ecurity **L**earning **E**nvironment (CSLE). The source code is released under the CC-BY-SA 4.0 license and available at `https://github.com/Limmen/csle`, with documentation at `https://limmen.dev/csle/`.

# List of Acronyms

| | |
|---|---|
| **API** | Application Programming Interface. |
| **APT** | Advanced Persistent Threat. |
| **A.S** | Almost Surely. |
| **BNE** | Berk-Nash Equilibrium. |
| **BO** | Bayesian Optimization. |
| **BTR** | Bounded Time to Recovery. |
| **CAGE-2** | Cyber Autonomy Gym for Experimentation 2. |
| **CAP** | Consistency, Availability, and Partition tolerance. |
| **CDF** | Cumulative Distribution Function. |
| **CEM** | Cross-Entropy Method. |
| **CLI** | Command-Line Interface. |
| **CMDP** | Constrained Markov Decision Process. |
| **COL** | Conjectural Online Learning. |
| **C-POMCP** | Causal-Partially Observed Monte-Carlo Planning. |
| **CPU** | Central Processing Unit. |
| **CSLE** | Cyber Security Learning Environment. |
| **CUSUM** | Cumulative Sum. |
| **CVE** | Common Vulnerabilities and Exposures. |
| **CWE** | Common Weakness Enumeration. |
| **DAG** | Directed Acyclic Graph. |
| **DE** | Differential Evolution. |
| **DFP** | Decompositional Fictitious Play. |
| **DMZ** | Demilitarized Zone. |
| **DNC** | Does Not Converge. |
| **DNS** | Domain Name System. |
| **DP** | Dynamic Programming. |
| **DVWA** | Damn Vulnerable Web Application. |
| **EM** | Expectation Maximization. |
| **FIFO** | First In First Out. |
| **FLP** | Fischer, Lynch, and Paterson theorem. |
| **FP** | Fictitious Play. |
| **FTP** | File Transfer Protocol. |
| **GPU** | Graphics Processing Unit. |
| **GRPC** | Google Remote Procedure Call. |
| **GW** | Gateway. |
| **HDFS** | Hadoop Distributed File System. |
| **HSVI** | Heuristic Search Value Iteration. |
| **HTTP** | Hypertext Transfer Protocol. |
| **ICMP** | Internet Control Message Protocol. |
| **IDPS** | Intrusion Detection and Prevention System. |
| **IDS** | Intrusion Detection System. |
| **I.I.D** | Independent and Identically Distributed. |
| **IoT** | Internet of Things. |
| **IP** | Internet Protocol. |
| **IPM** | Integral Probability Metric. |
| **IPS** | Intrusion Prevention System. |
| **IRC** | Internet Relay Chat. |
| **IT** | Information Technology. |
| **KL** | Kullback-Leibler. |
| **LP** | Linear Programming. |
| **MAB** | Multi-Armed Bandit. |
| **MC** | Monte-Carlo. |
| **MCTS** | Monte-Carlo Tree Search. |
| **MDP** | Markov Decision Process. |
| **MDS** | Martingale Difference Sequence. |
| **MINBFT** | Minimal Byzantine Fault Tolerance. |
| **MLR** | Monotone Likelihood Ratio. |
| **MPE** | Markov Perfect Equilibrium. |
| **MTTF** | Mean Time To Failure. |
| **NE** | Nash Equilibrium. |
| **NEXP** | Nondeterministic Exponential. |
| **NFSP** | Neural Fictitious Self-Play. |
| **NP** | Nondeterministic Polynomial. |
| **NTP** | Network Time Protocol. |
| **ODE** | Ordinary Differential Equation. |
| **OS-POSG** | One-Sided Partially Observed Stochastic Game. |
| **OVS** | Open vSwitch. |
| **P** | Polynomial. |
| **PBE** | Perfect Bayesian Equilibrium. |
| **PBFT** | Practical Byzantine Fault Tolerance. |
| **PDF** | Probability Density Function. |
| **POMCP** | Partially Observed Monte-Carlo Planning. |
| **POMDP** | Partially Observed Markov Decision Process. |
| **POMIS** | Possibly Optimal Minimal Intervention Set. |
| **PO-POSG** | Public Observation-Partially Observed Stochastic Game. |
| **POSG** | Partially Observed Stochastic Game. |
| **PPAD** | Polynomial Parity Arguments on Directed Graphs. |

**PPO** Proximal Policy Optimization.
**PSPACE** Polynomial Space.
**QoS** Quality of Service.
**RAM** Random-Access Memory.
**RELU** Rectified Linear Unit.
**REST** REpresentational State Transfer.
**RQ** Research Question.
**RSA** Rivest Shamir Adleman.
**R&D** Research and Development.
**SCADA** Supervisory Control and Data Acquisition.
**SCM** Structural Causal Model.
**SDN** Software-Defined Network.
**SG** Stochastic Game.
**SIEM** Security Information and Event Management.
**SMTP** Simple Mail Transfer Protocol.
**SNMP** Simple Network Management Protocol.
**SOC** Security Operations Center.
**SPSA** Simultaneous Perturbation Stochastic Approximation.
**SQL** Structured Query Language.
**SSH** Secure Shell.
**TCP** Transmission Control Protocol.
**T-FP** Threshold-Fictitious Play.
**TOLERANCE** **T**w**o**-**le**vel **r**ecovery **an**d repli**c**ation control with f**e**edback.
**TP-2** Totally Positive of Order 2.
**T-SPSA** Threshold-Simultaneous Perturbation Stochastic Approximation.
**UCB** Upper Confidence Bound.
**UCT** Upper Confidence Bounds for Trees.
**UDP** User Datagram Protocol.
**VXLAN** Virtual eXtensible Local-Area Network.
**W.P** With Probability.

# Nomenclature

| Terminology | Description |
|---|---|
| Security response | Actions taken to contain, mitigate, and recover from cyber attacks. |
| Digital twin | Virtual replica of an operational infrastructure. |
| Emulation | Creation of a digital twin. |
| Simulation | Sampling from a mathematical model. |
| Attacker | Entity aiming to intrude on a system. |
| Defender | Organization responding to potential attacks. |
| Client | Process that consumes services of an infrastructure. |
| Observation | System measurements, e.g., log files and alerts. |
| State | Values describing the current condition of a system. |
| Action | Decision made by either the attacker or the defender. |
| Strategy | Function that maps observation sequences to actions. |
| Optimal strategy | Strategy that is most advantageous according to an objective. |
| Dynamic attacker | Attacker that adapts its strategy based on the defender strategy. |
| Data-driven | A methodology where decisions are informed by system measurements. |

# INTRODUCTION

*By 1985, machines will be capable of doing any work man can do.*

— Herbert Simon ***1965**, The shape of automation.*

Far from Herbert Simon's early prediction, an organization's *response* to a cyber attack—i.e., the actions taken to mitigate the attack—is still defined and implemented by humans. Although this approach can provide basic response capabilities for an organization, a mounting concern is the complexity of modern IT infrastructures, which has grown exponentially with the rise of cloud computing, distributed networks, and IoT services; see Fig. 1. Managing security responses for these systems alongside their service requirements and physical infrastructures is an arduous technical challenge. Automation addresses this challenge by ensuring that response measures can scale with the growing demands. However, achieving such automation remains an unsolved problem due to its inherent difficulties. Chief among them is that response actions must be executed in real-time amidst uncertainty about the scope and nature of a potential attack.

These research challenges have engaged security experts, control engineers, and game theorists for over two decades (Alpcan and Basar, 2003). Traditional approaches use static rules that map infrastructure statistics to automated responses (Stakhanova et al., 2007). While such approaches can automate responses to *known* attacks, they rely on humans for configuration. As a result, they cannot scale with the growing complexity and dynamicity of IT infrastructures (Fig. 1). A promising solution to this limitation is to frame security response as an optimal control problem[1], which enables the automatic computation of optimal responses based on system measurements. Such framing facilitates a rigorous treatment of security response where trade-offs between different security objectives can be studied through mathematical models. Prior research demonstrates the advantages of this approach in analytical and simulated settings (Nguyen and Reddi, 2023). However, its feasibility for operational use in IT infrastructures has yet to be proven.

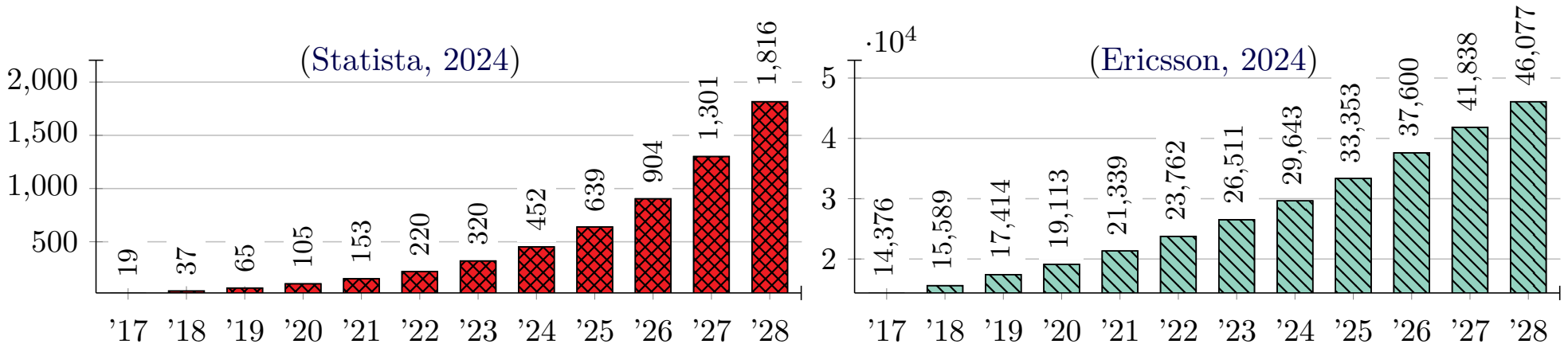

**(a)** Cost/y of cyber attacks in the U.S. (in billion $). **(b)** # connected devices worldwide (in millions).

***Figure 1:** Estimated trends in cyber attacks and network connectivity during 2017-2028.*

[1]Here "*control problem*" encompasses both control- and game-theoretic problem formulations.

Addressing the above limitations is the purpose of this thesis. We do so from three directions: via mathematical modeling, via experimental evaluation, and via systems engineering. Our main contribution is a practical methodology for *optimal security response* in IT infrastructures. This methodology is grounded in engineering principles for self-adaptive systems and has a rich mathematical foundation, which we systematically develop throughout this thesis. It draws on the theories of stochastic approximation, control, causality, and games. We prove theoretically and experimentally that our methodology is superior to present solutions on several instances of the security response problem, including intrusion prevention, intrusion response, intrusion tolerance, and defense against advanced persistent threats. Our key experimental finding is that the most important factor for scalable and optimal security response is to *exploit structure*, both structure in theoretical models (e.g., optimal substructure) and structure of the IT environment (e.g., network topology). The former enables efficient computation of optimal responses and the latter is key to managing the growing complexity of IT infrastructures (Fig. 1).

## ■ Background

Modern IT infrastructures involve a combination of application servers, data warehouses, and communication networks; see Fig. 2. Security operations within these infrastructures are managed through Security Operations Centers (SOCs), where human operators monitor and respond to incidents in real-time. Such centers are supported by Security and Event Management (SIEM) systems that aggregate data from various sources within the infrastructure. The operators use this data to assess the severity of incidents and decide how to respond. A common response, though drastic, is to shut down compromised systems. This response was key to mitigate the WANNACRY attack in 2017 (Morse, 2017). Another typical response is updating the network segmentation, which partitions the organization's communication network into separated segments. Such a segmentation was implemented in response to the attack against Sony in 2014 (Steinberg et al., 2021). When choosing responses, an operator must balance the need to mitigate the attack against the risk of disrupting services. While humans can manage such decisions given sufficient monitoring data, a growing concern is the escalating frequency of cyber attacks (Fig. 1), which drives a need for response automation and motivates this thesis.

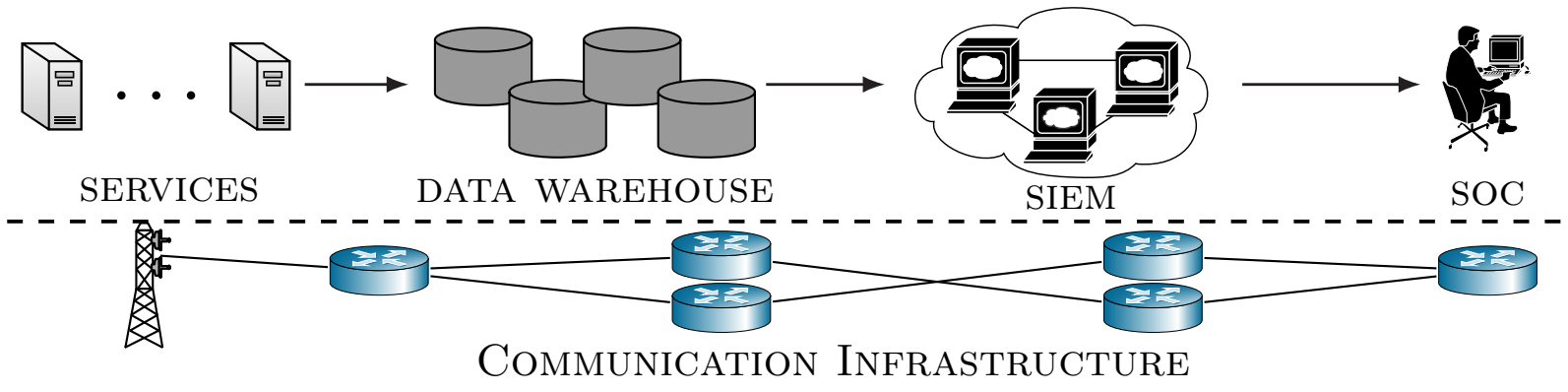

***Figure 2:*** *Security operations in modern IT infrastructures; Security and Event Management (SIEM) systems aggregate data for a Security Operations Center (SOC) where human operators respond to security incidents and potential attacks.*

# ■ Use Case

In our study of automated security response, we focus on a general response use case that involves the IT infrastructure of an organization. The operator of this infrastructure, which we call the *defender*, takes measures to protect it against an *attacker* while providing services to a client population. The infrastructure includes hardware, software, and networks, all of which work combined to provide operational services. These services range from web services and cloud storage to video and music streaming, which often have strict security requirements, such as guaranteed uptime and data integrity. Continuously meeting these requirements demands timely and effective security response.

***Infrastructure*** An example infrastructure is shown in Fig. 3. This infrastructure is segmented into zones with interconnected servers, which clients access through a public gateway. Though intended for service delivery, this gateway is also accessible to a possible attacker. Each zone has a security policy that regulates access control, allowing only authorized network traffic between zones. At the edge of each zone are intrusion detection systems (IDSs) that monitor traffic and log events in real-time. The defender accesses and analyzes these events to detect possible attacks.

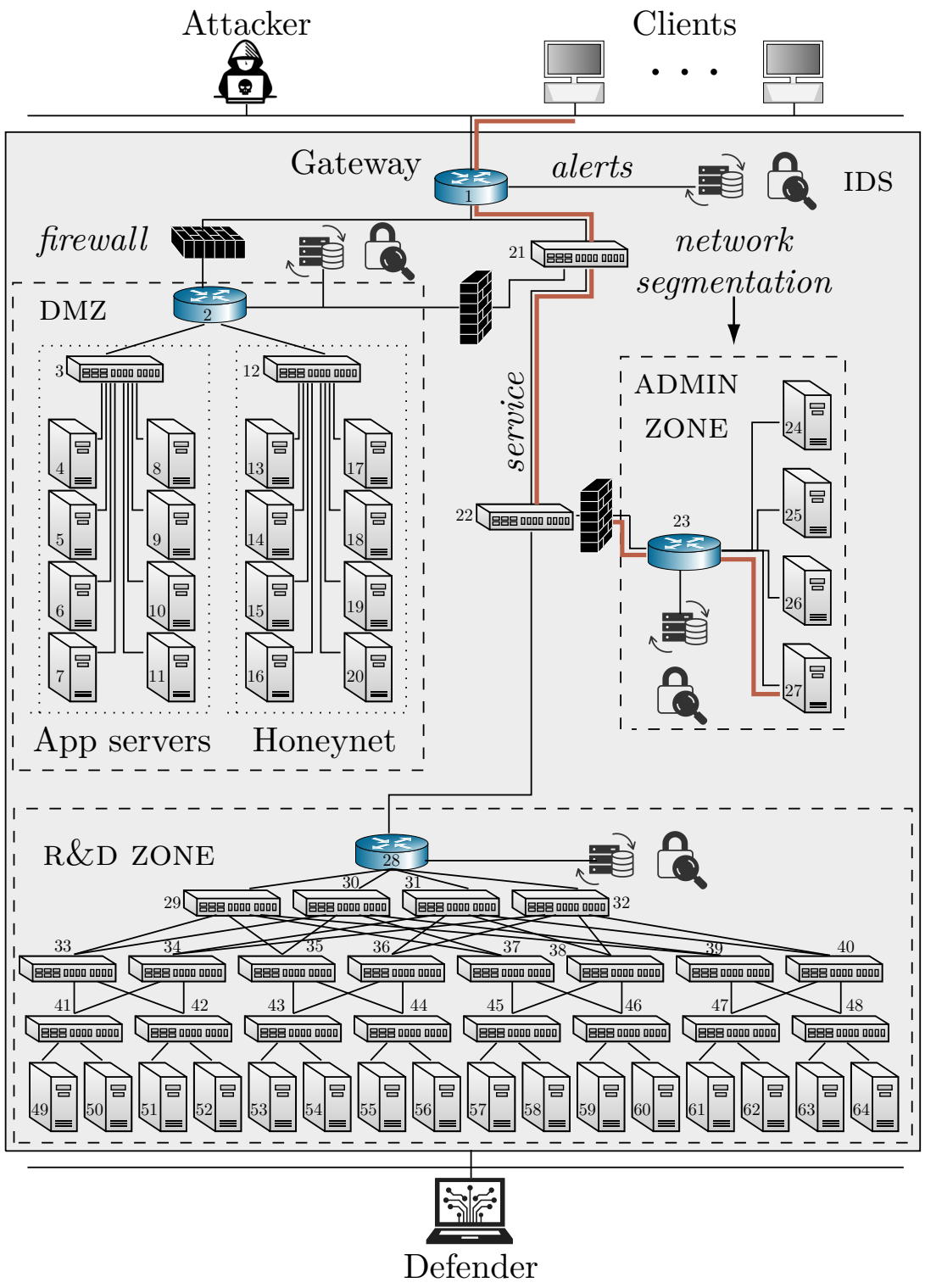

***Figure 3:*** *The actors and systems involved in the security response use case.*

***Services*** Services are deployed on the infrastructure[2]. They are built using architectural concepts such as client-server models, microservices, or hybrid cloud architectures. Each of them presents its own challenges in terms of security and operation. For example, microservices may require dynamic routing to reduce response times and avoid bottlenecks. Additionally, services are often part of larger service chains that include virtual network functions, such as firewalls, intrusion detection systems (IDS), or load balancers; see Fig. 4.

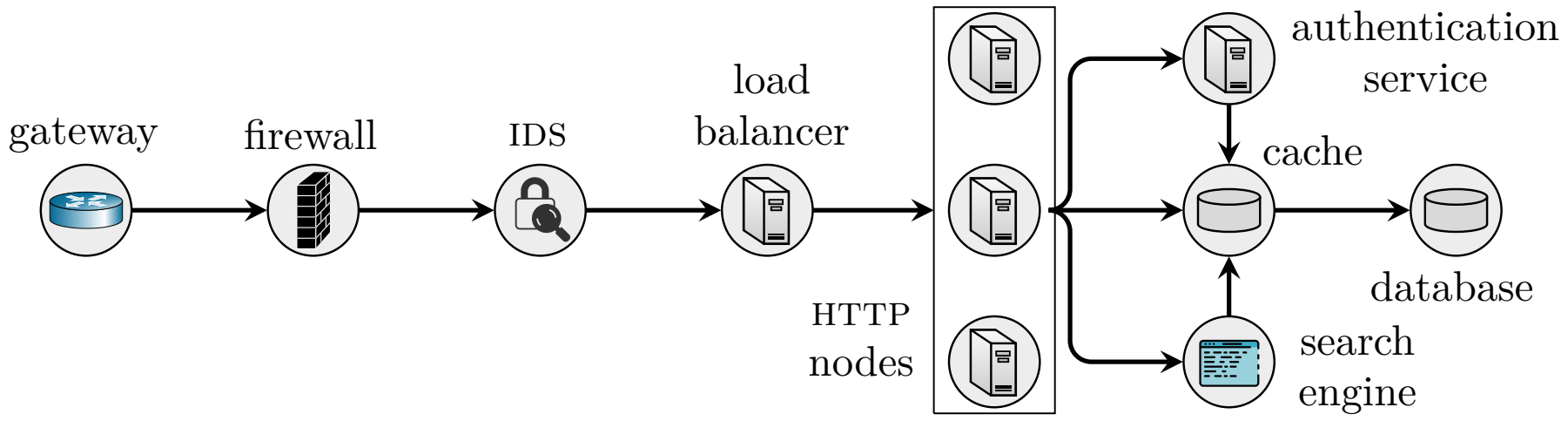

***Figure 4:*** *An example service chain with virtual network functions as microservices.*

***Infrastructure statistics*** During operation, services on the infrastructure produce real-time data, such as logs, performance metrics, and other statistics, which is critical for monitoring the security and performance of systems. Detecting anomalies within such data streams requires distinguishing between false and true security alerts. Figure 5 shows distributions of alerts on our testbed[3]. We observe many false alerts but also a clear difference between the distributions during normal operation (blue histograms) and during attacks (red histograms).

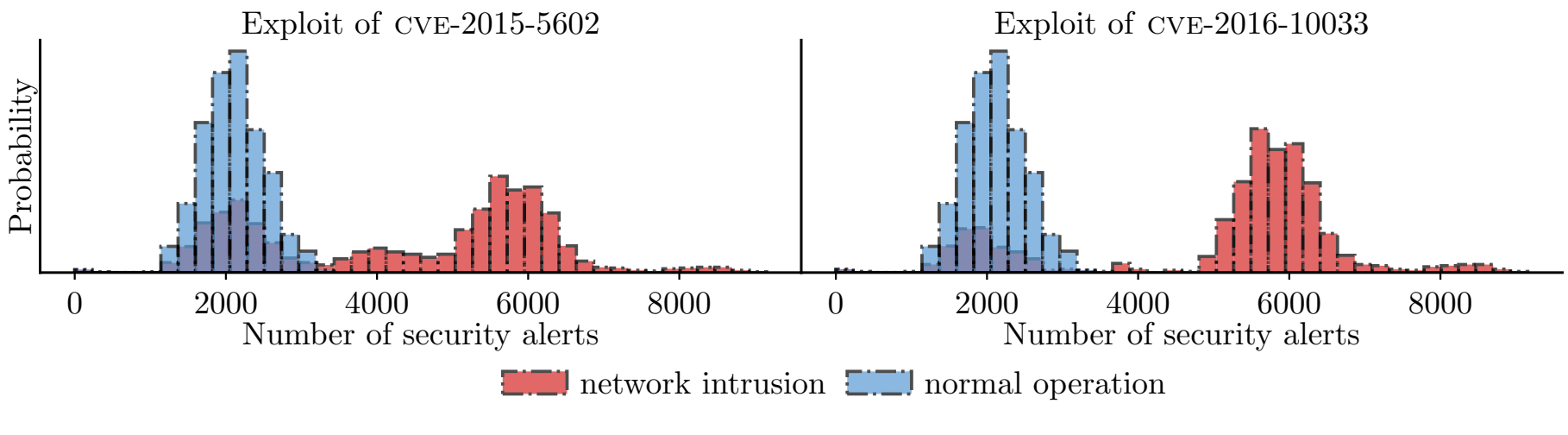

***Figure 5:*** *Distributions of security alerts on our testbed during two network intrusions.*

***Client population*** Clients influence the infrastructure statistics when interacting with services. For example, metrics like traffic rate and CPU utilization may spike during peak client load, mimicking patterns associated with attacks. Figure 6 shows alert distributions during an experiment where we loaded our testbed with client arrivals drawn from a Poisson process. We note a strong correlation between the clients' arrival rate $\lambda(t)$ and the number of alerts. ($\lambda(t)$ is defined in Paper 5.)

[2] The configuration of the infrastructure shown in Fig. 3 is given in Paper 3.
[3] See Paper 4 for details.

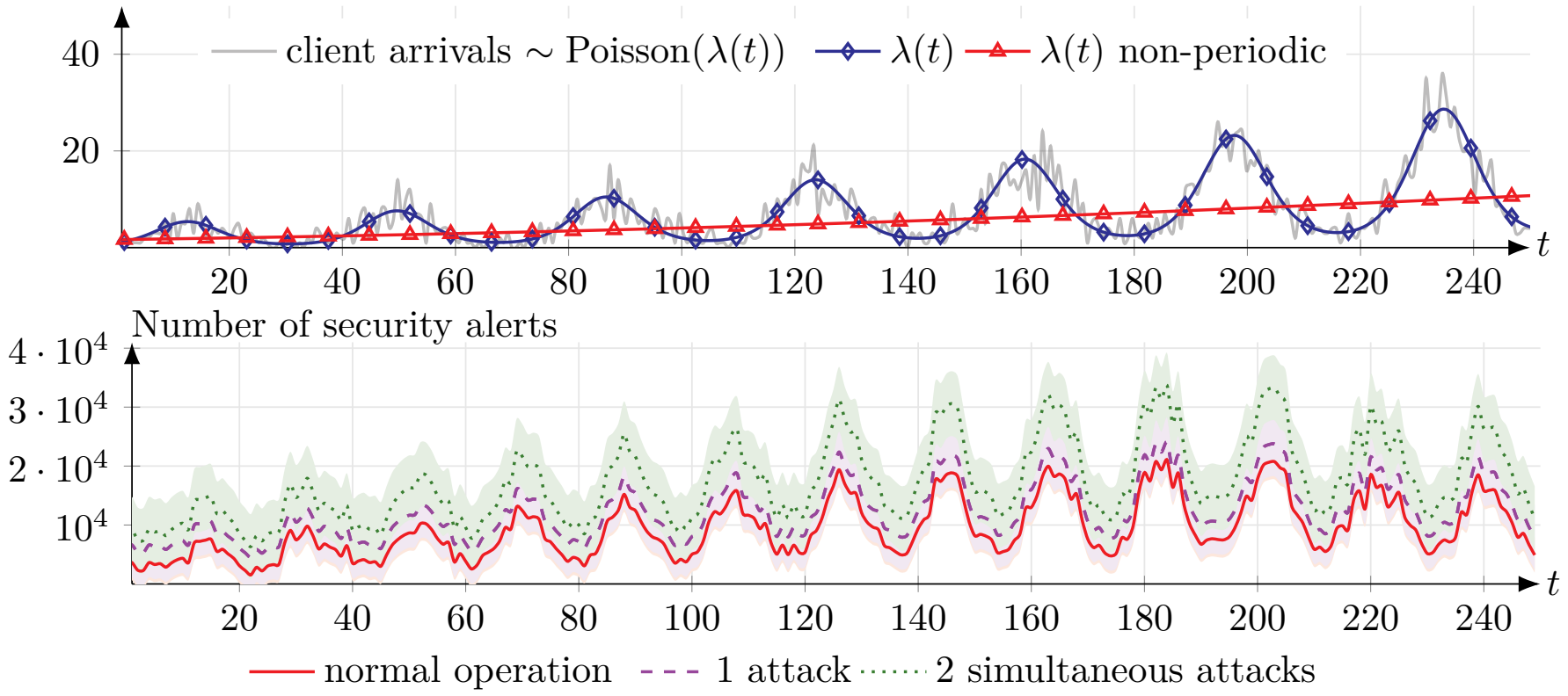

***Figure 6:*** *Time series of client arrival rates and distributions of alerts on our testbed.*

***Attacker*** The attacker, much like a client, accesses infrastructure services through the public gateway. However, instead of consuming these services as a client, the attacker aims to exploit them to compromise servers. To achieve this goal, it can perform different actions, e.g., reconnaissance, brute-force attacks, and exploits; see Fig. 7. The attacker's intent in compromising servers is irrelevant to our study and can vary from financial gain and espionage to hacktivism and geopolitical objectives (Anderson, 2001). Typically, the attacker starts with reconnaissance to gather information about the target system. This phase is followed by exploitation, where the attacker exploits vulnerabilities to compromise infrastructure components. For example, the attacker could exploit a buffer overflow to execute malicious code on the target system. After compromising a component, the attacker employs lateral movement techniques and may disrupt critical services.

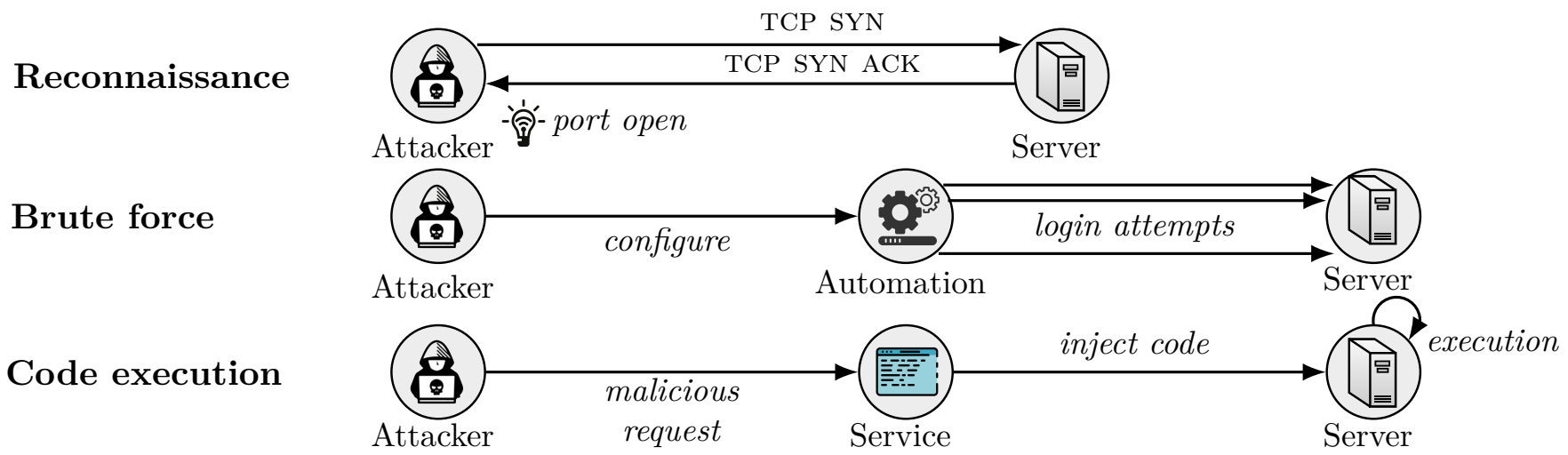

***Figure 7:*** *Attacks often involve reconnaissance, brute force, and malicious code execution.*

Today, it is widely recognized that IT infrastructures are unlikely to ever be capable of preventing all attacks (Maloof, 2005). They are simply too complex. Hence, IT infrastructures require the ability to quickly respond and mitigate attacks when they occur, which motivates the following problem formulation.

## ■ Problem

Given the above use case, we study the problem of developing *optimal defender strategies* that map infrastructure statistics to automated responses for optimizing a given objective, such as promptly mitigating potential attacks while minimizing service disruption. (By *mitigation*, we understand the ability to apply controls that restore operational services and steer the infrastructure to a secure state.) We consider responses on the physical layer, the network layer, the operating system layer, and the service layer. They include shutdown, recovery, access control, network resegmentation, replication control, malware removal, and cyber deception. These actions can be used to mitigate, prevent, and recover from cyber attacks. We provide three examples of response actions below.

**Example 1** (Flow control). Flow control is a common response to counter network intrusions. By redirecting suspicious traffic to a honeypot, for example, the defender can isolate and analyze malicious behavior without exposing critical parts of the network; see Fig. 8. Such responses are particularly effective in adaptable network infrastructures, e.g., software-defined networks.

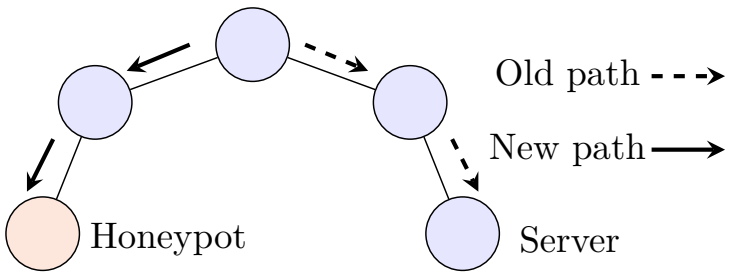

***Figure 8:*** *Flow control.*

**Example 2** (Access control). Access control is a traditional mechanism for responding to attacks. By adjusting resource permissions, defenders can prevent attackers from compromising critical assets; see Fig. 9. Such actions can also limit an attacker's ability to move laterally within the infrastructure (Bishop, 2004).

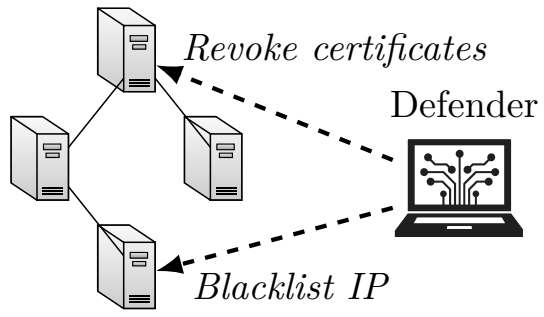

***Figure 9:*** *Access control.*

**Example 3** (Replication control). Replication control in distributed systems can be used to respond to network intrusions by ensuring that multiple replicas of critical services remain available even when some replicas are compromised; see Fig. 10. This control problem involves balancing the number of replicas and the timing of recovery actions to maintain service availability and limit operational cost (Castro and Liskov, 2002).

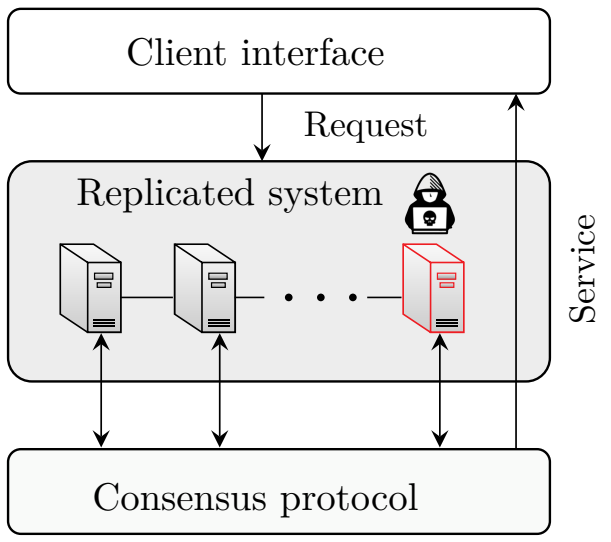

***Figure 10:*** *Replication control.*

***Challenges*** When choosing responses, the defender balances two *conflicting objectives*. On the one hand, it aims to maintain service to the clients. On the other hand, it must quickly respond to attacks and reduce their impact. In balancing this trade-off, the defender faces uncertainty about a potential attack and the causes of changes in the infrastructure statistics: load patterns shift, bandwidth fluctuates, components fail, etc. This uncertainty is both *epistemic*, stemming from incomplete knowledge of the attacker's actions, and *aleatory*, due to inherent randomness and variability in the infrastructure's operations. Responding to attacks amid such uncertainties involves two interdependent tasks: deciding *when* to take action and deciding *which* action to execute. The first task is linked to *attack detection*, which is essential for determining the optimal moment to intervene. The second task is related to *situational awareness* about the state of the infrastructure and the nature of the attack, which is key to selecting effective responses. This selection must also account for the possible adaptive behavior of the attacker, who may update its strategy to circumvent encountered defenses.

***Performance metrics*** Figure 11 illustrates the phases and performance metrics of the security response problem. The x-axis represents time, and the y-axis measures service quality and the level of security. The × on the horizontal axis indicates the time of an attack, which causes a drop in the level of security and service quality, leading to operational costs. Following the attack is a *response time* interval, during which response actions are deployed to mitigate and recover from the attack. Ideally, these actions should rapidly mitigate the attack while minimizing operational cost. The system's ability to continue functioning after the attack is reflected by its *tolerance*, also known as *resilience*. Similarly, the system's ability to prevent the attack in the first place is captured by its *survivability*.

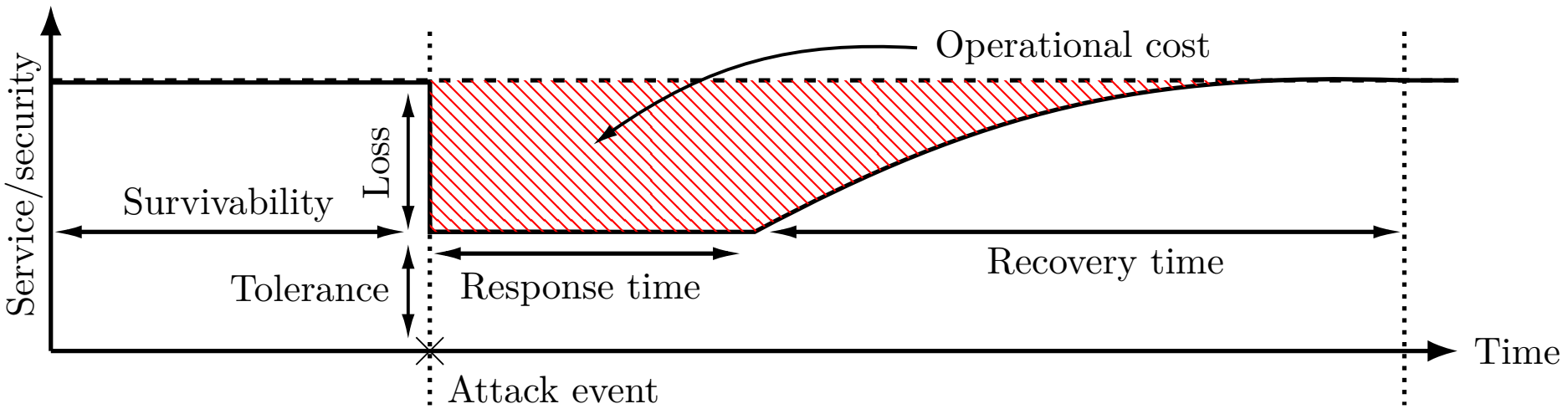

***Figure 11:*** *Phases and performance metrics of the security response problem.*

## ■ Approach

As evidenced by the problem description above, security response is, at its core, a decision-making problem. From choosing the right way to mitigate an attack to configuring access control policies or controlling network flows, each decision involves balancing security requirements and service availability amidst uncertainty about a potential attack. Relying on humans to make such decisions in real-time is unsustainable given the growing complexity of IT infrastructures (Fig. 1). For this reason, this thesis focuses on an alternative, *data-driven approach* to security response, which is rooted in the broader disciplines of autonomic computing (Kephart and Chess, 2003) and operations research (Kantorovich, 1960). This approach leverages mathematical models and real-time system measurements to automatically optimize responses through numerical and quantitative methods.

***Methodology*** We propose the data-driven methodology illustrated in Fig. 12 for automating security responses in IT infrastructures. It includes two systems. First, we use an *emulation system* for creating a virtual replica of the target infrastructure, i.e., a *digital twin.* This twin closely approximates the functionality and timing behavior of the target infrastructure, allowing us to run attack scenarios and defender responses. Such runs produce system measurements and logs, based on which we identify a game-theoretic model with two players: an attacker and a defender (von Neumann, 1928). Second, we use a *simulation system* where optimal response strategies are incrementally learned through stochastic approximation (Robbins and Monro, 1951). Learned strategies are extracted from the simulation system and evaluated on the digital twin. This process can be performed iteratively to provide progressively better response strategies[4], where strategy performance is measured by operational cost and ability to automatically mitigate cyber attacks.

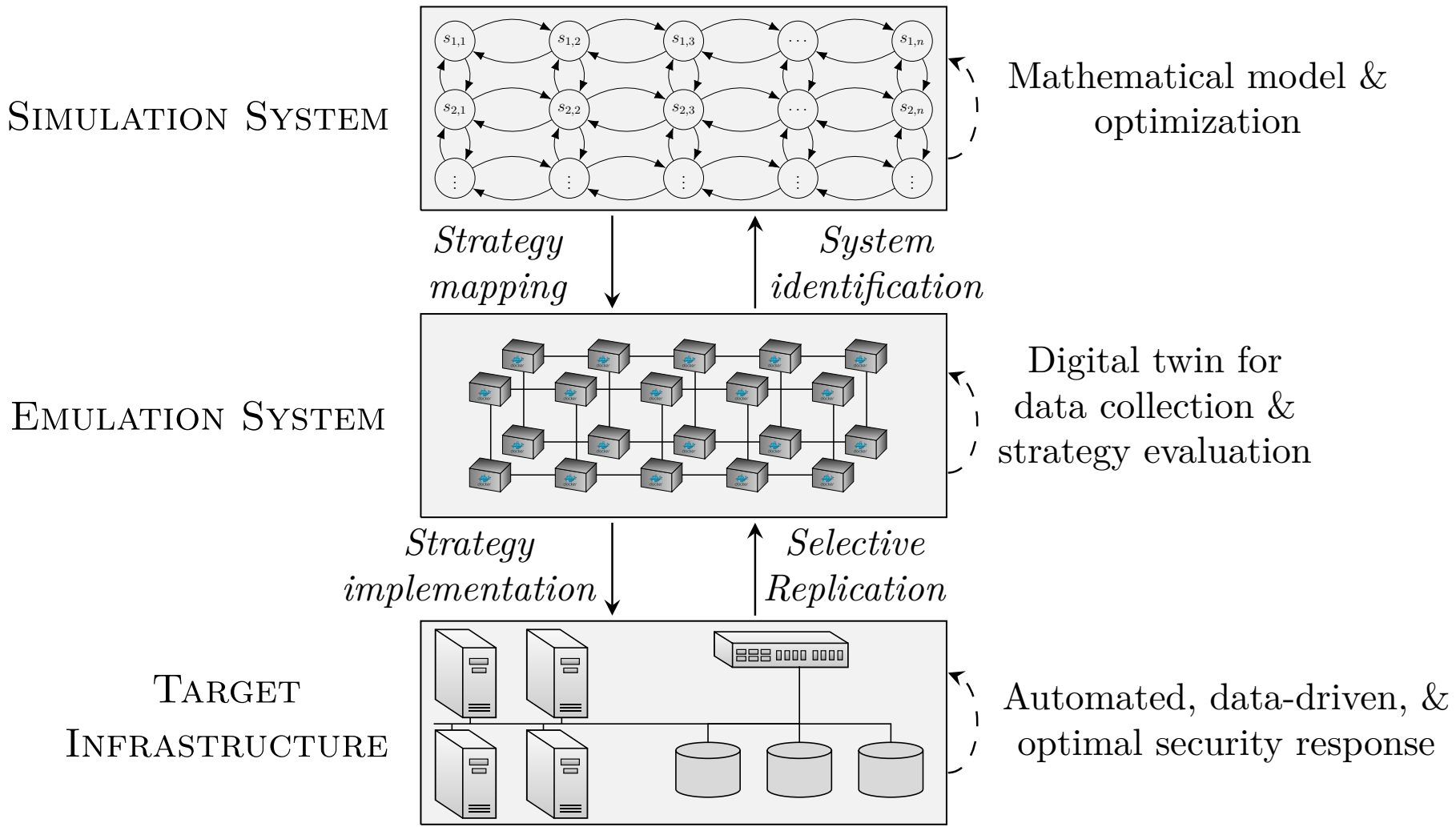

***Figure 12:*** *Our methodology for automated, data-driven, and optimal security response.*

***Emulation system*** The emulation system creates a digital twin of the target infrastructure consisting of virtual containers and networks. Such a twin allows us to test response strategies under different conditions, including varying attacks, workloads, and network latencies. Note that while offering this flexibility, a digital twin still closely approximates the target infrastructure by running the same software and controlling network delays. As a consequence, evaluating a response strategy on a digital twin reveals potential issues that simulations cannot feasibly replicate, such as unexpected resource constraints or software vulnerabilities.

[4] In control-theoretic terms, it is an *adaptive control* process (Åström and Wittenmark, 1995).

***Simulation system*** With a *simulation*, we mean an execution of a *discrete-time dynamical system* of the form

$$s_{t+1} \sim f(s_t, a_t^{(D)}, a_t^{(A)}), \qquad \text{(see Fig. 13)}$$

where $s_t$ is the system state at time $t$, $a_t^{(D)}$ is the defender action, $a_t^{(A)}$ is the attacker action, and $s \sim f$ means that $s$ is sampled from $f$. Such simulations run in milliseconds on modern computers and enable data-driven optimization[5]. Each simulation path $s_1, s_2, \ldots, s_t$ is associated with security consequences and costs; the aim is to find those defender actions that control the simulation in an optimal manner according to a stipulated objective. When performing such optimization, we face the complexity of the simulation model. The core issue is the level of abstraction at which the model is defined. The more detailed we construct the model, the closer it can capture the dynamics of the target infrastructure. At the same time, a detailed model incurs high computational costs and potentially a lack of generalization. To address this trade-off, we iteratively refine the model with the help of the digital twin; see Fig. 12. Specifically, after computing a theoretically optimal defender strategy, we validate it on the digital twin. If the evaluation is unsatisfactory, we update the simulation model accordingly. This procedure is repeated until satisfactory performance is obtained.

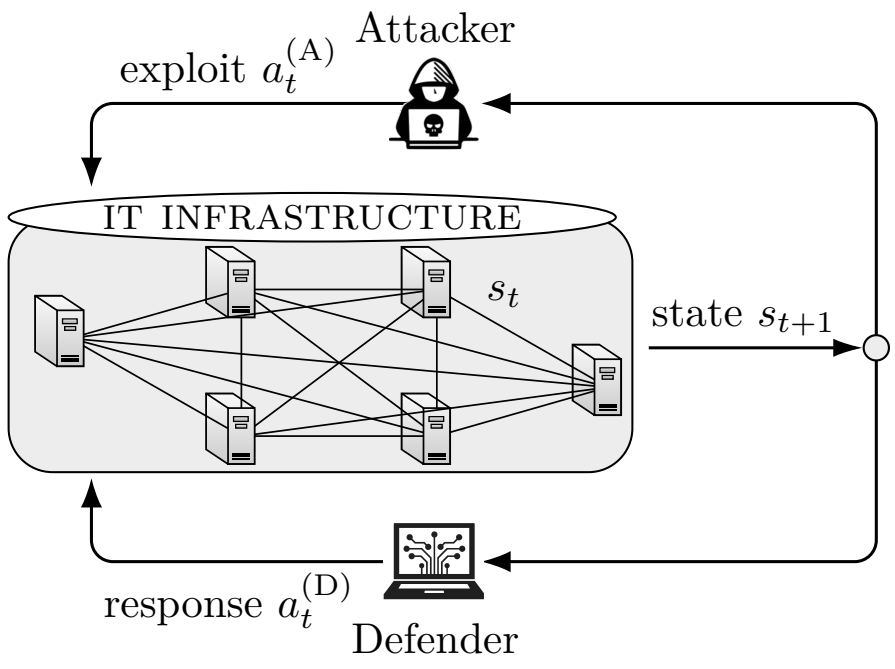

***Figure 13:*** *Security response as a discrete-time dynamical system.*

The above methodology is broadly applicable, extending beyond any specific response scenario, infrastructure configuration, optimization technique, or identification method. It can be instantiated with concepts from diverse fields, including control theory, game theory, causality, and stochastic approximation. These concepts will be developed in-depth throughout this thesis. For now, we turn to a review of the relevant literature to put our methodology in context.

## ■ Related Research

Since the early 1980s, there has been a broad interest in automating security functions, especially in intrusion detection and security response (Anderson, 1980). Traditional methods for intrusion detection rely on static rules that map infrastructure statistics to security alerts (Denning, 1987). For instance, an alert might be triggered if multiple failed login attempts occur within a short period. The main drawback of these methods is their dependence on domain experts to configure the rules, a process that is both labor-intensive and error-prone. Substantial effort has been devoted to addressing this limitation by developing statistical methods for detecting intrusions, e.g., anomaly detection (Dromard et al., 2017). As a result of this effort, modern intrusion detection systems now incorporate such techniques.

[5] Optimization on the digital twin is impractical due to time constraints of system commands.

In contrast to detection, *security response* remains a manual task performed by human operators. Attempts to automate this task include rule-based response systems (Wazuh Inc, 2022) and incident response playbooks (Applebaum et al., 2018). These systems respond to security incidents based on preconfigured rules. Although such systems can automate responses to *known* attacks, they rely on humans for configuration and cannot adapt to dynamic or evolving attacks. Current research—including this thesis—investigates data-driven methods to address these limitations; see Fig. 14. Three predominant approaches have emerged from this research: control-theoretic, reinforcement learning, and game-theoretic approaches. Below, we review their strengths and limitations, highlighting the research gaps.

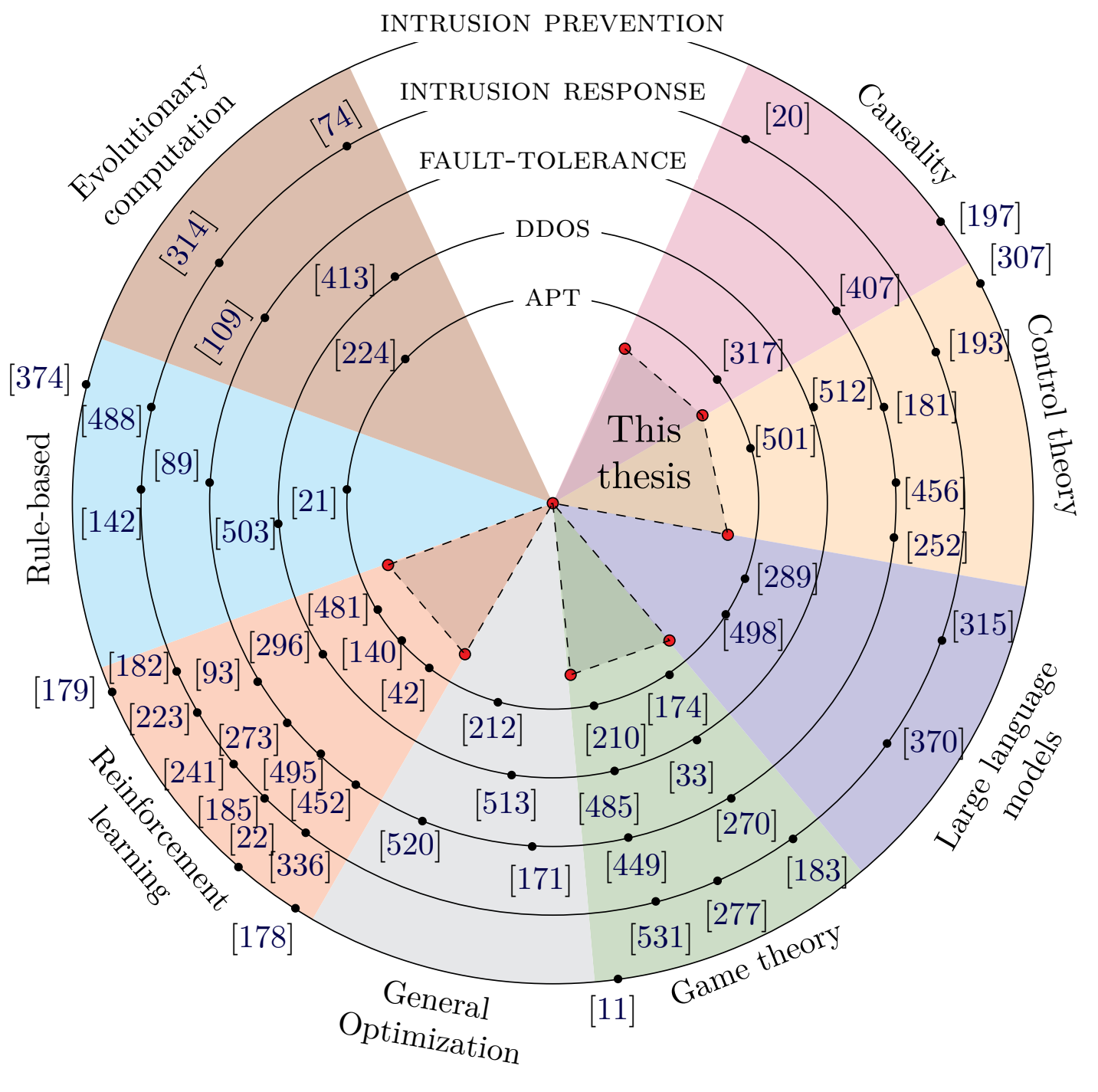

***Figure 14:*** *A categorization of the literature on automated security response and positioning of this thesis; we present a general methodology for automated security response; unlike other approaches, this methodology is not limited to a specific scenario or formal framework; throughout this thesis, we apply it to several instances of the security response problem, including intrusion prevention, intrusion response, intrusion tolerance, and defense against Advanced Persistent Threats (APTs); we also integrate it with several formal frameworks, including game theory, reinforcement learning, control theory, and causality.*

### Control theory for automated security response

Control theory provides a well-established mathematical framework for studying automatic systems. Therefore, it provides a foundational theory for automated security response. Classical control systems involve actuators in the physical world (e.g., electric power systems (Dán and Sandberg, 2010)), and many studies have focused on applying control theory to automate intrusion responses in cyber-physical systems; see survey (Teixeira et al., 2015). The control framework can also be applied to computing systems, and interest in control theory among researchers in IT security is growing (Miehling et al., 2019). Unlike traditional control theory, which often focuses on *continuous-time* systems, control-theoretic approaches for IT systems mainly use *discrete-time* models[6]. This focus is because measurements from IT systems are solicited on a sampled basis, which is best described by a discrete-time model (Hellerstein et al., 2004). Previous works that apply control theory to security response in IT systems include: [220, 252, 217, 306, 362, 363, 308][7], all of which model security response as the problem of controlling a discrete-time dynamical system and obtain optimal strategies through dynamic programming (Bellman, 1957). Their main limitation is that dynamic programming does not scale to problems of practical size due to the curse of dimensionality (Bertsekas and Tsitsiklis, 1996). Exceptions to this inefficiency are cases where specific problem structures allow the reduction of computational complexity (Miehling et al., 2018).

**Limitation 1: Scalability.**

Traditional control-theoretic approaches (e.g., dynamic programming) for security response do not scale to infrastructures of practical size.

### Reinforcement learning for automated security response

Reinforcement learning has emerged as a promising approach to approximate optimal control strategies in scenarios where dynamic programming is not applicable, and fundamental breakthroughs demonstrated by systems like ALPHAGO (Silver et al., 2016) and OPENAI FIVE (Berner et al., 2019) have inspired researchers to study reinforcement learning to automate security functions. Four seminal papers: (Georgia, 2000), (Xu and Xie, 2005), (Servin and Kudenko, 2008), and (Zhu and Basar, 2009) analyze security response and apply traditional reinforcement learning algorithms. They have catalyzed much follow-up research [328, 178, 182, 176, 130, 397, 510, 261, 66, 369, 525, 469, 152, 207, 150, 4, 286, 353, 175, 518, 125, 123, 508, 529, 528, 287, 391, 283, 292, 64, 213, 284, 519, 239, 486, 187, 493, 219, 218, 484, 179, 22, 521, 273, 223, 185, 278, 260, 141, 296, 336, 241, 368, 490, 495, 93, 452, 245,

[6] The theoretical framework for discrete-time systems is often referred to as *decision theory*.

[7] We use (author, year)-style referencing when we want to emphasize when the research was conducted; when this is not important, we use numeric references to save space.

215, 474, 500, 335, 47, 110]. These works show that *deep* reinforcement learning is a scalable technique for approximating optimal response strategies[8]. However, such methods lack convergence guarantees and often diverge when facing dynamic attackers, i.e., attackers that adapt their strategies to encountered defenses.

**Limitation 2: Lack of theoretical guarantees.**

Current deep reinforcement learning approaches for security response lack optimality guarantees and often diverge.

### Game theory for automated security response

Game theory stands out from control theory and reinforcement learning by focusing on decision-makers that *reason strategically* about the opponents' behavior. The formulation of security response as a game can be traced back to the early 2000s with works such as (Buttyán and Jean-Pierre, 2001), (Braynov and Sandholm, 2003), (Alpcan and Basar, 2004), (Theodorakopoulos and Baras, 2006), (Altman et al., 2007), (Grossklags et al., 2008), (Buchegger and Alpcan, 2008), (Kiekintveld et al., 2009), and (Saad et al., 2010). In addition to these early pioneers, numerous researchers have contributed to this line of research in the last two decades. Notable contributions include the textbooks (Buttyan and Hubaux, 2007), (Alpcan and Basar, 2010), (Tambe, 2011), (Kamhoua et al., 2021) and the doctoral dissertations (Alpcan, 2006), (Sallhammar, 2007), (Chen, 2008), (Grossklags, 2009), (Zonouz, 2011), (Zhu, 2013), (Malialis, 2014), (Teixeira, 2014), (Fang, 2016), (Bao, 2018), (Horák, 2019), (Umsonst, 2021), (Huang, 2022), (Guoxin, 2024), which have sparked significant follow up research [116, 210, 398, 36, 483, 517, 186, 312, 502, 446, 124, 202, 354, 476, 528, 393, 33, 471, 78, 144, 399, 327, 450, 451, 13, 399, 327, 450, 451, 13, 170, 364, 325, 178, 150, 268, 527, 205, 531, 284, 493, 484, 521, 16, 270, 346, 394, 149, 383, 275, 420, 462, 65]. These works study various aspects of security games, including the existence, uniqueness, and structure of equilibria, as well as computational methods. However, nearly all of them are simulation studies, and how they generalize to operational infrastructures is unproven. Few emulation studies exist, e.g., (Zonouz et al., 2009) and (Aydeger et al., 2021), but they do not present a general methodology[9].

**Limitation 3: Simulation-based evaluations.**

Most of the proposed game-theoretic approaches for security response have only been validated analytically or in simulation.

[8]By *scalable*, we mean that deep reinforcement learning methods can handle large state and action spaces, which traditional dynamic programming methods struggle with.

[9]Technical comparisons between the contributions of this thesis and the prior work can be found in the included papers; see §1.7, §2.6, §3.2, §4.8, §5.8, and §6.2.

**Research gap**

In summary, while previous research on data-driven methods for automated security response has made strides in various domains, critical problems remain unaddressed. Current control-theoretic approaches exhibit limitations in scalability when applied to IT infrastructures of practical size. In contrast, deep reinforcement learning approaches are scalable but lack theoretical guarantees, particularly when facing dynamic attackers. On the other hand, game-theoretic approaches allow modeling dynamic attackers but are often evaluated through simulation only. Hence, despite extensive prior research, the question of how to build automated response systems that are both theoretically sound and practically viable remains open. To address this challenge, we study the following research question.

> **RQ:** What methodology can be used to develop an automated security response system that guarantees scalability and optimality, and how can the system be rigorously validated on a testbed?

## ■ Contributions

Addressing the above question is the main focus of this thesis. We posit that the methodology illustrated in Fig. 12 provides the answer. This claim will be supported by applying the methodology to several instances of the security response problem, including intrusion prevention, intrusion response, intrusion tolerance, and defense against advanced persistent threats, each detailed in one of the papers presented in this thesis; see Papers 1–6. The advancements made through these papers encompass not only experimental findings but also theoretical insights and practical implementations. Our main contributions can be summarized as follows.

★ 1. **Scalable algorithms for computing optimal response strategies.**

To address Limitation 1, we design and implement seven scalable algorithms for computing optimal response strategies (Algs. 1.1–6.1), for which we prove convergence (Thms. 5.3, 5.4, and 6.4). They build on techniques from stochastic approximation, game theory, reinforcement learning, linear and dynamic programming, and causality. We demonstrate that these algorithms outperform state-of-the-art methods in the scenarios we study.

★ 2. **Mathematical formulations of optimal security response.**

We introduce six novel mathematical models of security response. With these models, we show that a) optimal stopping is a suitable framework for deriving the optimal times to take response actions; b) partially observed stochastic games effectively model the security response use case; and c) the Berk-Nash equilibrium allows capturing model misspecification in security games.

★ **3. Proving structural properties of optimal response strategies.**

To address Limitation 2, we develop fundamental mathematical tools for security response and prove structural properties of optimal response strategies, such as decomposability (Thm. 3.2) and threshold structure (Thms. 1.1, 2.1, 4.3, 4.5, and 5.1). These structural results enable scalable computation and efficient implementation of optimal strategies in operational systems.

★ **4. General methodology for optimal security response.**

We design a general methodology for optimal security response; see Fig. 12. Additionally, we present the **C**yber **S**ecurity **L**earning **E**nvironment (CSLE), an open-source platform that implements our methodology (Hammar, 2023). This platform allows us to experimentally validate the methodology on several instances of the security response problem, including intrusion prevention, intrusion response, intrusion tolerance, and defense against advanced persistent threats. Unlike previous simulation-based solutions, our methodology provides practical insights beyond a specific response scenario.

## ■ Organization

The remainder of this thesis is organized as follows. The first chapter covers theoretical background on decision and game theory. Readers already well-versed in these theories may skip this chapter without missing essential context. Following this background chapter are two chapters that cover the problem formulation and the methodology. These chapters set the stage for the remaining chapters, which contain Papers 1–6, each detailing an application of our methodology to a specific instance of the security response problem. Finally, we conclude and provide suggestions for future research.

# DECISION-THEORETIC FOUNDATIONS

*The purpose of models is not to fit the data but to sharpen the questions.*

— Samuel Karlin ***1983**, Fisher Memorial Lecture.*

MATHEMATICAL models allow us to rigorously formulate new ideas, provide a common basis for study, make assumptions explicit, and help categorize various solutions. It is with these reasons in mind that each paper presented in this thesis begins with the formulation of a mathematical model. In this chapter, we review the formalisms that underpin these models, namely Markov decision theory and game theory, laying the groundwork for the subsequent chapters. The exposition is brief but includes pointers to the relevant literature.

### Notation

Our usage of mathematical notation is fairly standard. For the reader's convenience, we summarize the most frequently used notations here. Boldface lowercase letters denote column vectors, e.g., $\mathbf{x} = (x_1, x_2, \ldots)$. Upper case calligraphy letters (e.g., $\mathcal{V}$) represent sets. $\mathbb{P}$ is a probability measure. The set of probability measures over $\mathcal{V}$ is written as $\Delta(\mathcal{V})$. A random variable is written in upper case (e.g., $X$), a random vector in boldface (e.g., $\mathbf{X}$). The range of a random variable $X$ is denoted as $\mathcal{R}_X$. The expectation of an expression $\phi$ with respect to $X$ is written as $\mathbb{E}_X[\phi]$. If $\phi$ includes many random variables that depend on $\pi$, we simply write $\mathbb{E}_\pi[\phi]$. $x \sim f$ means that $x$ is sampled from the distribution $f$. We use $\mathbb{P}[x \mid y]$ as a shorthand for $\mathbb{P}[X = x \mid Y = y]$. In a game with a set of players $\mathcal{N}$, we use $-k$ as a shorthand for $\mathcal{N} \setminus \{\mathrm{k}\}$. We use big $O$ notation, e.g., $O(g(x))$. The $i$-th standard basis vector is written as $\mathbf{e}_i$. We use "increasing" and "decreasing" to mean *strictly* increasing and *strictly* decreasing, respectively. Conversely, we use "non-decreasing" and "non-increasing" to mean *weakly* increasing and *weakly* decreasing, respectively. $\|\cdot\|_\infty$ is the supremum norm. Further notation is listed in Table 1 on the next page.

### Measurability

We predominantly study spaces endowed with the *discrete* topology. Consequently, the construction of the underlying probability space is standard and shall be omitted for brevity. Specifically, the probability space is defined as $(\Omega, \mathcal{F}, \mathbb{P})$, where $\Omega$ is a finite sample space, the $\sigma$-algebra of events is $\mathcal{F} = 2^\Omega$, and all sets $\mathcal{Y} \in \mathcal{F}$ are $\mathbb{P}$-measurable (Kolmogorov, 1933). A random variable $X : \Omega \to \mathbb{R}$ is a measurable function, i.e., for all $\mathcal{X} \in \mathcal{B}(\mathbb{R})$, we have that $X^{-1}(\mathcal{X}) \in \mathcal{F}$, where $\mathcal{B}(\mathbb{R})$ is the Borel $\sigma$-algebra on $\mathbb{R}$. A filtration $(\mathcal{F}_t)_{t \geq 0}$ is an increasing sequence of sub $\sigma$-algebras where each $\mathcal{F}_t \subset \mathcal{F}$ represents events observable up to time $t$. For example, $\mathcal{F}_t = \sigma(X_k \mid k \leq t)$ is the $\sigma$-algebra generated by the random variables $X_1, \ldots, X_t$.

| Notation | Description |
|---|---|
| $\mathbf{x}^T$ | The transpose of $\mathbf{x}$. |
| $\mathcal{X} \times \mathcal{Y}$ | The Cartesian product of the sets $\mathcal{X}$ and $\mathcal{Y}$. |
| $\bigtimes_{i=1}^{n} \mathcal{X}_i$ | $\mathcal{X}_1 \times \mathcal{X}_2 \times \dots \mathcal{X}_n$. |
| $\lvert\mathcal{X}\rvert$ | Cardinality of the set $\mathcal{X}$. |
| $f \triangleq \phi(x)$ | $f$ is defined as $\phi(x)$. |
| $\nabla_{\boldsymbol{\theta}} J(\boldsymbol{\theta})$ | Gradient of $J$ with respect to $\boldsymbol{\theta}$. |
| $(x_i)_{i=1,\dots,n}$ | Either a sequence or a vector $(x_1, x_2, \dots, x_n)$. |
| $\{x_i\}_{i=1,\dots,n}$ | The set $\{x_1, x_2, \dots, x_n\}$. |
| $\langle x_1, x_2, \dots, x_n \rangle$ | A tuple with $n$ components. |
| $\mathbb{1}_{\phi(x)}$ | The indicator function, which is 1 if $\phi(x)$ is true; 0 otherwise. |
| $\delta_i(\cdot)$ | The Dirac delta function centered at $i$. |
| $\mathbb{R}$ | The set of real numbers. |
| $\mathbb{R}_+$ | The set of non-negative real numbers. |
| $\mathbb{R}^{m \times n}$ | The set of $m \times n$ matrices with real entries. |
| $\mathbb{N}$ | The set of natural numbers $0, 1, 2, \dots$. |
| $\mathrm{pa}(X)_\mathcal{G}$ | Parents of node $X$ in graph $\mathcal{G}$. |
| $\mathrm{ch}(X)_\mathcal{G}$ | Children of node $X$ in graph $\mathcal{G}$. |
| $\mathrm{an}(X)_\mathcal{G}$ | Ancestors of node $X$ in graph $\mathcal{G}$. |
| $\mathrm{de}(X)_\mathcal{G}$ | Descendants of node $X$ in graph $\mathcal{G}$. |
| $\mathcal{G}[\mathcal{V}]$ | The subgraph obtained by restricting $\mathcal{G}$ to the nodes in $\mathcal{V}$. |
| $D_{\mathrm{KL}}(P \parallel Q)$ | The Kullback-Leibler (KL) divergence between $P$ and $Q$. |
| $\mathbf{0}_n$ | The $n$-dimensional zero vector $(0, \dots, 0) \in \mathbb{R}^n$. |
| $\mathbf{1}_n$ | The $n \times n$ identity matrix. |
| $2^\mathcal{S}$ | The power set of $\mathcal{S}$. |
| $\oplus$ | Vector concatenation operation. |
| $\Omega, \mathcal{F}, \mathbb{P}$ | Sample space, $\sigma$-algebra of events, probability measure. |
| $\mathcal{B}(\mathbb{R})$ | The Borel $\sigma$-algebra on $\mathbb{R}$. |
| $\mathcal{F}_t, \sigma(X_k \mid k \leq t)$ | Filtration at time $t$, $\sigma$-algebra generated by $X_1, \dots, X_t$. |

***Table 1:*** *Mathematical notation.*

## ■ The Markov Decision Process

A Markov Decision Process (MDP) models the control of a *discrete-time* dynamical system and is defined by the seven-tuple

$$\mathcal{M} \triangleq \langle \mathcal{S}, \mathcal{A}, f, r, \gamma, \mathbf{b}_1, T \rangle. \qquad \text{(Bellman, 1957)} \tag{1}$$

It evolves in time steps from $t = 1$ to $t = T$, which constitutes one *episode.* $\gamma \in [0, 1]$ is a discount factor, $\mathcal{S}$ is the set of states, and $\mathcal{A}$ is the set of actions. The initial state is drawn from $\mathbf{b}_1 \in \Delta(\mathcal{S})$ and $f(s_{t+1} \mid s_t, a_t)$ is the probability of transitioning from state $s_t$ to state $s_{t+1}$ when taking action $a_t$, which has the Markov property

$$f(s_{t+1} \mid s_t, a_t) = f(s_{t+1} \mid s_1, \dots, s_t, a_t). \tag{2}$$

Each state transition is associated with a reward $r(s_t, a_t) \in \mathbb{R}$. If $f$ (2) and $r$ are independent of the time step $t$, the MDP is *stationary.* Similarly, if $\mathcal{S}$ and $\mathcal{A}$ are finite, the MDP is *finite.*

**Assumption 1** (Finiteness and stationarity)**.** *Unless otherwise stated, all models considered in this thesis are assumed to be finite and stationary.*

**Assumption 2** (Bounded rewards)**.** $\exists M \in \mathbb{R}, |r(s,a)| \leq M < \infty, \forall (s,a) \in \mathcal{S} \times \mathcal{A}.$

Actions are decided by a *strategy* $\pi(\mathbf{h}_t)$, which is a function of the *history*

$$\mathbf{h}_t \triangleq (s_1, a_1, \ldots, s_{t-1}, a_{t-1}, s_t) = (\mathbf{h}_{t-1}, a_{t-1}, s_t) \in \mathcal{H}_t, \tag{3}$$

where $\mathcal{H}_t$ is the set of histories of length $t$.

If $\pi$ only depends on $\mathbf{h}_t$ through $s_t$, it is said to be *Markovian*. If $\pi$ maps each history to a unique action, it is *deterministic* (also called *pure*) and written as $a_t = \pi(\mathbf{h}_t)$. Otherwise, it is *stochastic* (also called *behavioral*) and written as $a_t \sim \pi(\cdot \mid \mathbf{h}_t)$. If $\pi$ is independent of the time step $t$, it is *stationary*. Otherwise, it is *non-stationary* and written as $\pi_t$.

An *optimal* strategy $\pi^\star$ maximizes the expected cumulative discounted reward

$$J(\pi^\star) = \sup_{\pi \in \Pi} J(\pi), \quad \text{where} \quad J(\pi) \triangleq \mathbb{E}_\pi \left[ \sum_{t=1}^{T} \gamma^{t-1} r(S_t, A_t) \right]. \tag{4}$$

Here $J : \Pi \to \mathbb{R}$ is an objective functional and $\mathbb{E}_\pi$ denotes the expectation of the random vector $\mathbf{H}_T \in \mathcal{H}_T$ (3) when following strategy $\pi$.

**Remark 1** (Measurability)**.** Assuming $f$ (2), $\mathbf{b}_1$, and $\pi$ are well-defined probability measures, then, by the extension theorem of (Ionescu Tulcea, 1949), there exists a unique probability measure $\mathbb{P}$ on $\mathcal{H}_t$ for all $t$. Consequently, the expectation in (4) is well-defined under general conditions (Prop. 7.28, Bertsekas and Shreve, 1978).

**Remark 2** (Constraints)**.** Two types of constraints can be added to (4): local and global. Local constraints restrict actions based on the state $s_t$ and can be handled by standard algorithms for MDPs. Global constraints, on the other hand, restrict the state trajectory $(s_t)_{t=1}^{T}$ and require a different formal framework, namely constrained Markov decision processes; we cover this framework in Paper 4.

If the strategy space $\Pi$ contains the class of Markovian strategies and the MDP is finite, then there exists a strategy that achieves the maximum in (4); see Thm. 1 below. However, if $\pi$ is parameterized, or if the MDP is not finite, the maximum might not be attained. Nevertheless, by the completeness of $\mathbb{R}$, the supremum in (4) always exists under the following assumption and Assumption 2.

**Assumption 3** (Finite objective (4))**.**

- *The time horizon $T$ is finite; or*
- *$T$ is a random variable and $\mathbb{E}_{\pi^\star}[T] < \infty$ for any optimal strategy $\pi^\star$; or*
- *$T$ is infinite and $\gamma < 1$.*

**Theorem 1** (Existence of optimal strategies in MDPs)**.** *Given a finite MDP and Assumptions 2–3, then an optimal, deterministic, Markovian strategy $\pi_t^\star$ exists. If the MDP is stationary and the horizon $T$ is infinite or random, then an optimal, deterministic, stationary, Markovian strategy $\pi^\star$ exists*[10]*.*

**Remark 3** (Dominance of Markovian strategies)**.** Throughout this thesis, whenever an optimal Markovian strategy exists, we shall implicitly restrict the strategy space $\Pi$ to the class of Markovian strategies, which simplifies notation and analysis.

The complexity of computing an optimal strategy for an MDP is polynomial in $|\mathcal{S}| + |\mathcal{A}| + B$, where $B$ is the maximum number of bits required to represent any component of $f$ (2) or $r$ (Littman et al., 1995). Such computation involves maximizing the objective in (4), which is *separable* in the sense that it is additive across time steps. This separability underlies the Bellman equation and enables efficient computation through dynamic programming, as described below.

With some abuse of notation, we use $J(\pi)$ to denote the value of a strategy $\pi$ (4), while also using $J^\pi(s)$ to refer to the value of a state $s$ under strategy $\pi$[11]:

Expected future reward.

$$J^\pi(s) \triangleq r(s, \pi(s)) + \gamma \sum_{s' \in \mathcal{S}} f(s' \mid s, \pi(s)) J^\pi(s') \qquad \forall s \in \mathcal{S}. \tag{5}$$

Immediate reward.

Likewise, $Q^\pi(s, a)$ is the value of taking action $a$ in state $s$:

$$Q^\pi(s, a) \triangleq r(s, a) + \gamma \sum_{s' \in \mathcal{S}} f(s' \mid s, a) J^\pi(s') \qquad \forall (s, a) \in \mathcal{S} \times \mathcal{A}. \tag{6}$$

Bellman's optimality equation relates $\pi^\star$ to the *optimal value function* $J^\star$:

$$\pi^\star(s) \in \arg\max_{a \in \mathcal{A}} \left[ r(s, a) + \gamma \sum_{s' \in \mathcal{S}} f(s' \mid s, a) J^\star(s') \right] \quad \text{(Eq. 1, Bellman, 1957)} \tag{7}$$

for all $s \in \mathcal{S}$. Since a strategy that maximizes (4) also satisfies (7), this equation effectively provides an alternative optimality condition (sufficient and necessary) to (4) (Thm. 4.3.3, Puterman, 1994). Note that $\pi^\star(s) \in \arg\max_{a \in \mathcal{A}} Q^\star(s, a)$.

**The principle of optimality (Bellman, 1957).**

The Bellman equation is grounded in the principle of optimality, which states that the optimal action in any state leads to a sequence of subsequent actions that are optimal starting from the next state.

[10] See (Thms. 4.4.2, 6.2.10, 8.4.5, Puterman, 1994) for the proofs, which are based on backward induction and Banach's fixed-point theorem (Thm. 6, p. 160, Banach, 1922).

[11] $J^\pi$ may depend on time if the MDP is non-stationary or if $T < \infty$, we then write it as $J_t^\pi$.

***Dynamic and linear programming*** Dynamic programming algorithms, e.g., value and policy iteration (§6.3-6.4, Puterman, 1994), use (7) to obtain an optimal strategy through successive approximations of $J^\star$. Specifically, value iteration implements the recursion $J_{k+1} = \mathscr{T} J_k$, where $k$ is the iteration index and $\mathscr{T}$ is a contraction mapping ($\|\mathscr{T}J - \mathscr{T}J'\|_\infty \leq \gamma \|J - J'\|_\infty$) defined as

$$(\mathscr{T}J)(s) \triangleq \max_{a \in \mathcal{A}} \left[ r(s,a) + \gamma \sum_{s' \in \mathcal{S}} f(s' \mid s,a) J(s') \right] \qquad \forall s \in \mathcal{S}. \tag{8}$$

Similarly, policy iteration is defined as

$$\pi_{k+1}(s) \in \arg\max_{a \in \mathcal{A}} \left[ r(s,a) + \gamma \sum_{s' \in \mathcal{S}} f(s' \mid s,a) J^{\pi_k}(s') \right] \qquad \forall s \in \mathcal{S}, \tag{9}$$

where $J^{\pi_k} = \mathscr{T}_\pi J^{\pi_k}$ and

$$(\mathscr{T}_\pi J)(s) \triangleq r(s,\pi(s)) + \gamma \sum_{s' \in \mathcal{S}} f(s' \mid s,\pi(s)) J(s') \qquad \forall s \in \mathcal{S}. \tag{10}$$

Note that $\mathscr{T}_{\pi_k} J^{\pi_{k-1}} = \mathscr{T} J^{\pi_{k-1}}$, where $\mathscr{T}_{\pi_k}$ is linear and $\mathscr{T}$ is non-linear. Hence, policy iteration effectively linearizes the Bellman operator around $J^{\pi_{k-1}}$ and then solves for the fixed point, analogous to Newton's method; see Fig. 15.

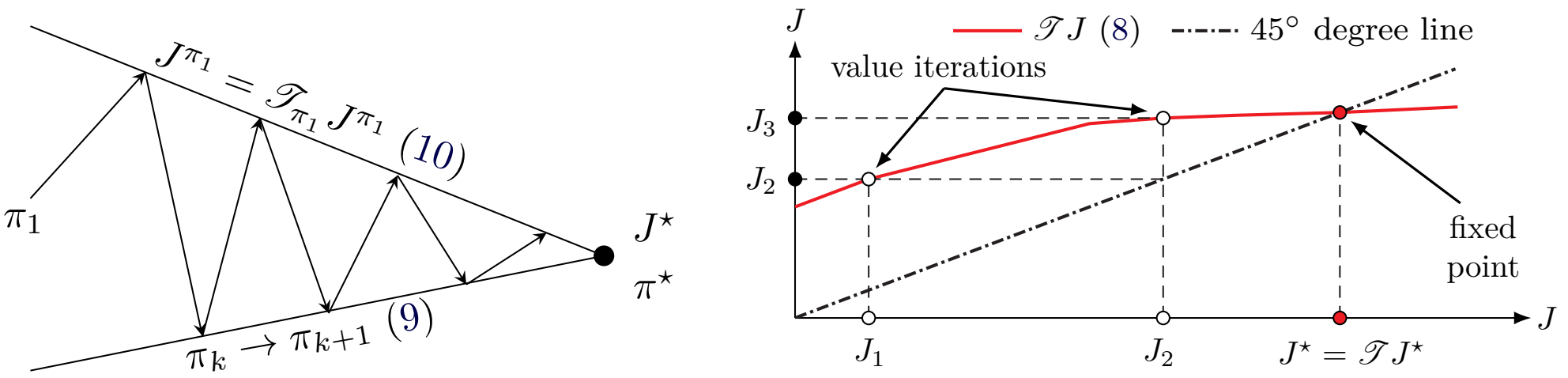

***(a)*** *Policy iteration [448].* ***(b)*** *Value iteration when* $|\mathcal{S}| = 1$ *[51].*

***Figure 15:*** *Dynamic programming; (a) policy iteration computes* $J^\pi = \mathscr{T}_\pi J^\pi$ *(10) and updates* $\pi$ *using (9); (b) value iteration implements the recursion* $J_{k+1} = \mathscr{T} J_k$ *(8).*

Under Assumptions 1–3, the recursion $J_{k+1} = \mathscr{T} J_k$ satisfies $\lim_{k \to \infty} J_k = J^\star$ (Thm. 6.3.1, Puterman, 1994). Likewise, the strategies produced by (9) satisfy $\lim_{k \to \infty} \pi_k = \pi^\star$ (Thm 6.4.2, Puterman, 1994). These convergence results are based on the observation that $J^\star = \mathscr{T} J^\star$, i.e., $J^\star$ is a fixed point of $\mathscr{T}$ (8). Therefore, $J^\star$ is the smallest vector $J \in \mathcal{J}$ that satisfies the *Bellman inequality*

$$J(s) \geq \max_{a \in \mathcal{A}} \left[ r(s,a) + \gamma \sum_{s' \in \mathcal{S}} f(s' \mid s,a) J(s') \right] \qquad \forall s \in \mathcal{S}, \tag{11}$$

where $\mathscr{J}$ is the Banach space of $|\mathcal{S}|$-dimensional vectors (Hordijk and Kallenberg, 1979). Consequently, $J^\star$ is the unique solution to the following linear program

$$\underset{J \in \mathscr{J}}{\text{minimize}} \left\{ \sum_{s \in \mathcal{S}} J(s)\beta(s) \mid J \text{ satisfies (11)} \right\} \quad \text{subject to } \beta(s) > 0\ \forall s \in \mathcal{S}, \tag{12}$$

where $\beta$ is a probability distribution over $\mathcal{S}$.

***Stochastic approximation*** Stochastic approximation provides an alternative method for solving (4) when dynamic programming cannot be applied, e.g., when only samples of $f$ (2) are available. Another key advantage of stochastic approximation is its ability to efficiently estimate near-optimal solutions to (4) in cases when dynamic programming is computationally prohibitive.

Let $\mathscr{H}$ be an operator on $\mathcal{S} \times \mathcal{A}$ defined as

$$(\mathscr{H}Q)(s,a) \triangleq r(s,a) + \gamma \sum_{s' \in \mathcal{S}} f(s' \mid s,a) \max_{a' \in \mathcal{A}} Q(s',a') - Q(s,a).$$

Note that $\mathscr{H}Q^\star = \mathbf{0}_{|\mathcal{S}|\times|\mathcal{A}|}$. Hence, solving (4) amounts to finding a root of $\mathscr{H}$. Suppose that we can obtain samples of the form $(\mathscr{H}Q)(s,a) + w$, where $w \in \mathbb{R}$ is a random noise. Using such samples, we can implement the stochastic approximation

$$\begin{aligned} Q_{n+1}(s,a) &= Q_n(s,a) + \alpha_n((\mathscr{H}Q_n)(s,a) + w_n) \\ &= Q_n(s,a) + \alpha_n \Bigg( (\mathscr{H}Q_n)(s,a) + \\ &\qquad \underbrace{\left( r(s,a) + \gamma \max_{a' \in \mathcal{A}} Q_n(s',a') - Q_n(s,a) \right) - (\mathscr{H}Q_n)(s,a)}_{\text{Noise term } w_n;\ \mathbb{E}[W_n]=0;\ s' \sim f(\cdot \mid s,a).} \Bigg) \\ &= Q_n(s,a) + \alpha_n \left( r(s,a) + \gamma \max_{a' \in \mathcal{A}} Q_n(s',a') - Q_n(s,a) \right), \end{aligned} \tag{13}$$

where $(\alpha_n)_{n=0}^{\infty}$ is a sequence of step sizes satisfying

$$\sum_{n=0}^{\infty} \alpha_n = \infty \quad \text{and} \quad \sum_{n=0}^{\infty} \alpha_n^2 < \infty. \qquad \text{(Robbins and Monro, 1951)}$$

This stochastic approximation algorithm is known as *Q-learning* (Watkins, 1989). Under Assumptions 1–3, $\sum_{n=0}^{\infty} \alpha_n w_n$ is almost surely convergent (martingale convergence theorem). Therefore, (13) asymptotically behaves as

$$Q_{n+1}(s,a) = Q_n(s,a) + \alpha_n(\mathscr{H}Q_n)(s,a). \tag{14}$$

Hence, $\lim_{n\to\infty} Q_n = Q^\star$ (Thm. 2, Jaakkola et al., 1994).

Note that (14) can be interpreted as an Euler scheme that approximately solves $\frac{dQ(s,a)}{dn} = (\mathscr{H}Q)(s,a)$. This interpretation enables the application of ODE theory to analyze the convergence properties of (13) (Borkar, 2008) (Meyn, 2022).

# ■ The Partially Observed Markov Decision Process

A Partially Observed Markov Decision Process (POMDP) is an extension of an MDP where the states are hidden; it is defined by the nine-tuple

$$\mathcal{M} \triangleq \langle \mathcal{S}, \mathcal{A}, f, r, \gamma, \mathbf{b}_1, T, \mathcal{O}, z \rangle. \qquad \text{(Åström, 1965)} \qquad (15)$$

The first seven elements define an MDP. $\mathcal{O}$ denotes the set of observations and $z(o_t \mid s_t)$ is the *observation function*, where $o_t \in \mathcal{O}$. If $\mathcal{O}$, $\mathcal{S}$, and $\mathcal{A}$ are finite, the POMDP is finite. If $f$, $r$, and $z$ are time-independent, the POMDP is *stationary*.

A control *strategy* $\pi(\mathbf{h}_t)$ is a function of the *history*

$$\mathbf{h}_t \triangleq (\mathbf{b}_1, a_1, o_2, a_2, \ldots, a_{t-1}, o_t) = (\mathbf{h}_{t-1}, a_{t-1}, o_t) \in \mathcal{H}_t. \qquad (16)$$

Based on this history, the controller computes the *belief state*

$$\begin{aligned}
\mathbf{b}_t(s) &\triangleq \mathbb{P}[S_t = s \mid \mathbf{h}_t] = \mathbb{P}[S_t = s \mid o_t, a_{t-1}, \mathbf{h}_{t-1}] \qquad (17)\\
&\overset{\text{(Bayes)}}{=} \frac{\mathbb{P}[o_t \mid s, a_{t-1}, \mathbf{h}_{t-1}]\mathbb{P}[s \mid a_{t-1}, \mathbf{h}_{t-1}]}{\mathbb{P}[o_t \mid a_{t-1}, \mathbf{h}_{t-1}]} \overset{\text{(Markov)}}{=} \frac{z(o_t \mid s)\mathbb{P}[s \mid a_{t-1}, \mathbf{h}_{t-1}]}{\mathbb{P}[o_t \mid a_{t-1}, \mathbf{h}_{t-1}]}\\
&\overset{\text{(Markov)}}{=} \frac{z(o_t \mid s)\sum_{s_{t-1} \in \mathcal{S}} \mathbf{b}_{t-1}(s_{t-1}) f(s \mid s_{t-1}, a_{t-1})}{\sum_{\hat{s} \in \mathcal{S}} \sum_{s' \in \mathcal{S}} z(o_t \mid s') f(s' \mid \hat{s}, a_{t-1}) \mathbf{b}_{t-1}(\hat{s})} \triangleq \mathbb{B}(\mathbf{b}_{t-1}, a_{t-1}, o_t)(s),
\end{aligned}$$

↑ Independent of $s$, i.e., a normalizing constant.

where $\mathbb{B}$ is a recursive *belief operator*. This operator performs $O(|\mathcal{S}|^2)$ scalar multiplications. (We assume that $\mathcal{S}$ is finite (Assumption 1); otherwise, the summations in (17) have to be replaced with integrals[12].)

Since $\mathbf{b}_t$ is a sufficient statistic for $s_t$ (Def. 4.2, Lem. 5.1, Thm. 7.1, Kumar and Varaiya, 1986), we can define a *Markovian strategy* as $\pi : \mathcal{B} \to \Delta(\mathcal{A})$, where $\mathcal{B} \triangleq \Delta(\mathcal{S})$ is the *belief space*, i.e., the unit $(|\mathcal{S}| - 1)$-simplex; see Fig. 16.

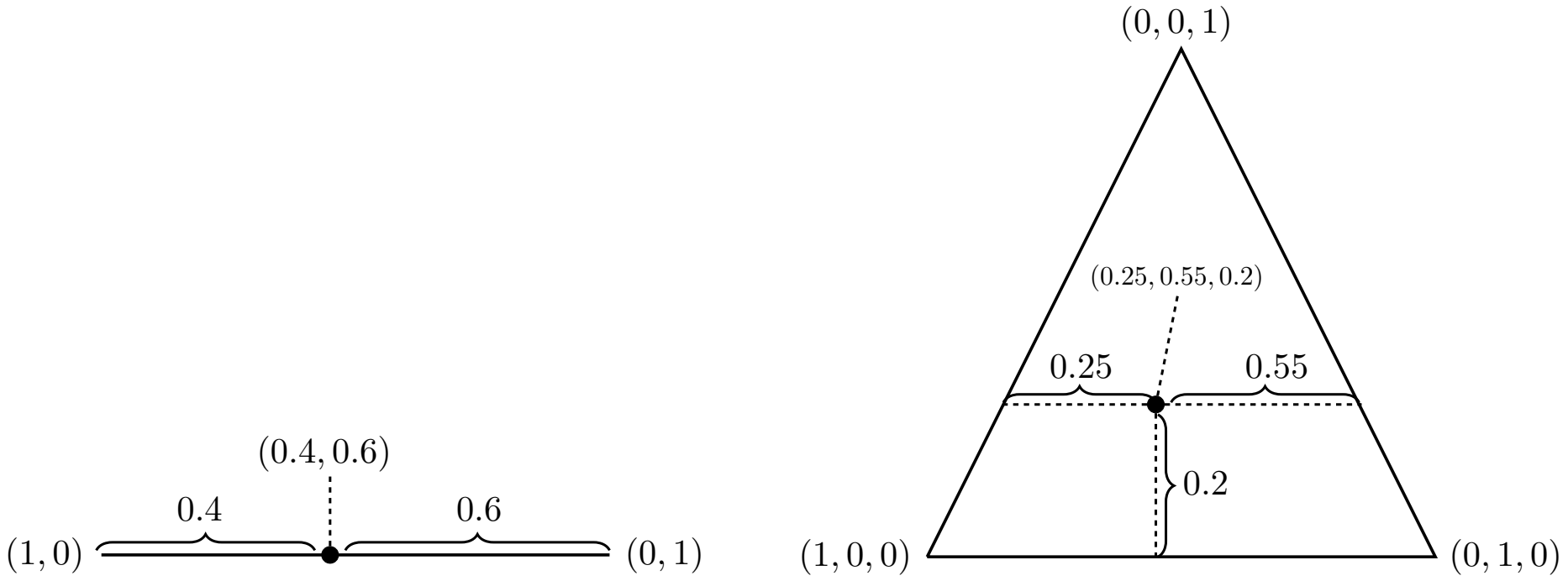

***(a)*** $\Delta(\{0,1\})$: *1-dimensional unit-simplex.* ***(b)*** $\Delta(\{0,1,2\})$: *2-dimensional unit-simplex.*

***Figure 16:*** *The belief state* $\mathbf{b} \in \Delta(\mathcal{S})$ *(17) is an* $|\mathcal{S}|$*-dimensional probability vector, which can be represented geometrically as a point in the unit simplex of dimension* $|\mathcal{S}| - 1$.

[12] When the state space is large, (17) generally has to be approximated; one common approximation technique is the *particle filter*, which we describe in Paper 6.

**Theorem 2** (Existence of optimal strategies in POMDPs).
*Given a finite POMDP and Assumptions 2–3, then*[13]

1. *There exists an optimal, deterministic, Markovian strategy $\pi_t^\star$ and an optimal value function $J_t^\star : \mathcal{B} \rightarrow \mathbb{R}$ that is piece-wise linear and convex*[14] *and satisfies*

$$J_t^\star(\mathbf{b}) = \max_{a \in \mathcal{A}} \left[ \mathbb{E}_S[r(S,a) \mid \mathbf{b}] + \gamma \sum_{o \in \mathcal{O}} J_t^\star(\mathbb{B}(\mathbf{b},a,o))\mathbb{P}[o \mid \mathbf{b},a] \right]. \quad (18)$$

2. *If the POMDP is stationary and $T$ is infinite or random, then there exists a stationary, deterministic, Markovian $\pi^\star$ and a stationary $J^\star$.*

Computing an optimal strategy for a POMDP is PSPACE-hard (Thm. 6, Papadimitriou and Tsitsiklis, 1987). To understand this complexity, note that a POMDP can be treated as a continuous-state MDP by defining $\mathbf{b}_t$ as the state and using $\mathbb{B}$ (17) as the transition function (Thm. 7.2.2, Krishnamurthy, 2016).

## ■ The One-Sided Partially Observed Stochastic Game

A zero-sum, *one-sided*, Partially Observed Stochastic Game (POSG) can be regarded as a generalized POMDP with two controllers (players)[15]. It is defined as

$$\Gamma \triangleq \langle \mathcal{N}, \mathcal{S}, (\mathcal{A}_\mathrm{k})_{\mathrm{k} \in \mathcal{N}}, f, r, \gamma, \mathbf{b}_1, T, z, \mathcal{O} \rangle. \quad \text{(Def. 3.1, Horák et al., 2023)} \quad (19)$$

$\mathcal{N} \triangleq \{1,2\}$ is the set of players, $\mathcal{S}$ is the set of states, and $\mathcal{A}_\mathrm{k}$ is the set of actions for player k. $f(s_{t+1} \mid s_t, \mathbf{a}_t)$ is the transition function and $r(s_t, \mathbf{a}_t)$ is the reward function, where $\mathbf{a}_t \triangleq (a_t^{(1)}, a_t^{(2)}) \in \mathcal{A}_1 \times \mathcal{A}_2$ is the *action profile* at time $t$. $\gamma$ is a discount factor, $\mathbf{b}_1$ is the initial state distribution, $T$ is the time horizon, and $z(o_t \mid s_t)$ is the observation function, where $o_t \in \mathcal{O}$. $\Gamma$ (19) is finite if $\mathcal{S}$, $\mathcal{A}_1 \times \mathcal{A}_2$, and $\mathcal{O}$ are finite. Similarly, $\Gamma$ is *stationary* if $f$, $r$, and $z$ are time-independent.

Player k follows a *behavior strategy* $\pi_\mathrm{k}$, where $a_t^{(\mathrm{k})} \sim \pi_\mathrm{k}(\mathbf{h}_t^{(\mathrm{k})})$ and

$$\mathbf{h}_t^{(\mathrm{k})} \triangleq (\mathbf{b}_1, a_1^{(\mathrm{k})}, \mathbf{i}_2^{(\mathrm{k})}, a_2^{(\mathrm{k})}, \ldots, a_{t-1}^{(\mathrm{k})}, \mathbf{i}_t^{(\mathrm{k})}) = (\mathbf{h}_{t-1}^{(\mathrm{k})}, a_{t-1}^{(\mathrm{k})}, \mathbf{i}_t^{(\mathrm{k})}) \in \mathcal{H}_t^{(\mathrm{k})}. \quad (20)$$

Here $\mathcal{H}_t^{(\mathrm{k})}$ is the history space and $\mathbf{i}_t^{(\mathrm{k})}$ is the *information feedback*:

$$\mathbf{i}_t^{(1)} \triangleq (o_t) \quad \text{and} \quad \mathbf{i}_t^{(2)} \triangleq (o_t, s_t, a_{t-1}^{(1)}). \quad (21)$$

Partial observability. Complete observability.

[13]See (Thms. 7.2.2, 7.4.1, 7.6.1, Krishnamurthy, 2016) for the proof.
[14]E.J. Sondik originally proved this property (Thm. 2, Sondik, 1978). A more accessible proof can be found in (Thms. 7.6.1–7.6.2, Krishnamurthy, 2016).
[15]A stochastic game (Shapley, 1953) is a specific type of dynamic (and extensive-form) game where state transitions can be stochastic (Def. 5.4, Basar and Olsder, 1999).

**Definition 1** (Perfect recall, (Def. 7, Kuhn, 1953))**.** *A stochastic game has perfect recall if each player* $\mathrm{k} \in \mathcal{N}$ *remembers* $\mathbf{h}_t^{(\mathrm{k})}$ *at each time step* $t$ *(20).*

**Assumption 4** (Perfect recall)**.** *Throughout this thesis, games have perfect recall.*

Based on the history (20), the *belief state* of player 1 is computed as

$$\mathbf{b}_t(s_t) \triangleq \mathbb{B}(\mathbf{b}_{t-1}, a_{t-1}^{(1)}, o_t, \pi_2)(s_t) = \mathbb{B}(\mathbf{h}_t^{(1)}, \pi_2)(s_t) \triangleq \mathbb{P}[S_t = s_t \mid \mathbf{h}_t^{(1)}, \pi_2] = \tag{22}$$
$$\frac{z(o_t \mid s_t) \sum_{s_{t-1}\in\mathcal{S}} \sum_{a_{t-1}^{(2)}\in\mathcal{A}_2} \pi_2(a_{t-1}^{(2)} \mid \mathbf{b}_{t-1}, s_{t-1})\mathbf{b}_{t-1}(s_{t-1}) f(s_t \mid s_{t-1}, a_{t-1}^{(1)}, a_{t-1}^{(2)})}{\sum_{a_{t-1}^{(2)}\in\mathcal{A}_2} \sum_{s',s\in\mathcal{S}} z(o_t \mid s')\pi_2(a_{t-1}^{(2)} \mid s, \mathbf{b}_{t-1}) f(s' \mid s, a_{t-1}^{(1)}, a_{t-1}^{(2)})\mathbf{b}_{t-1}(s)},$$

which requires $O(|\mathcal{A}_2||\mathcal{S}|^2)$ scalar multiplications. (If $\mathcal{S} \cup \mathcal{A}_2$ is not finite, the summations in (22) are replaced with integrals.) Note that this computation can be performed by *both* players, i.e., player 2 knows the true state and player 1's belief state (21). Consequently, the players' *behavior Markov strategies* can be defined as $\pi_1 : \mathcal{B} \rightarrow \Delta(\mathcal{A}_1)$ and $\pi_2 : \mathcal{B} \times \mathcal{S} \rightarrow \Delta(\mathcal{A}_2)$; see Fig. 17.

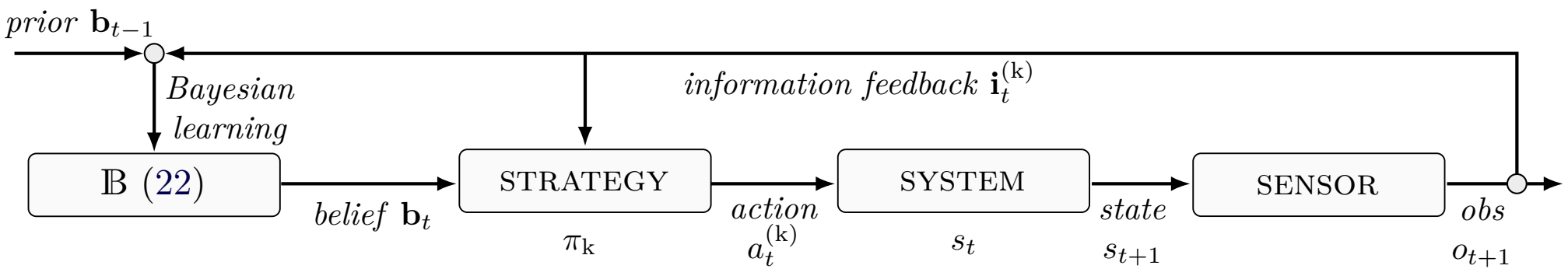

***Figure 17:*** *The control loop of player* k *in a partially observed stochastic game (19).*

$\pi_1$ is a *best response* against $\pi_2$ if it maximizes

$$J^{(\pi_1,\pi_2)}(\mathbf{b}_1) \triangleq \mathbb{E}_{(\pi_1,\pi_2)}\left[\sum_{t=1}^{T} \gamma^{t-1} r(S_t, \mathbf{A}_t) \mid \mathbf{b}_1\right]. \tag{23}$$

Similarly, $\pi_2$ is a best response against $\pi_1$ if it *minimizes* $J^{(\pi_1,\pi_2)}(\mathbf{b}_1)$. Hence, the *best response correspondences* are

$$\mathscr{B}_1(\pi_2) \triangleq \arg\max_{\pi_1\in\Pi_1} J^{(\pi_1,\pi_2)}(\mathbf{b}_1) \qquad \text{correspondence for player 1} \tag{24a}$$
$$\mathscr{B}_2(\pi_1) \triangleq \arg\min_{\pi_2\in\Pi_2} J^{(\pi_1,\pi_2)}(\mathbf{b}_1) \qquad \text{correspondence for player 2.} \tag{24b}$$

Computation of these correspondences amount to computing the optimal strategies in a POMDP and an MDP, respectively.

We refer to $\boldsymbol{\pi} \triangleq (\pi_1, \pi_2)$ as the *strategy profile*. When each player follows a best response, $\boldsymbol{\pi}$ forms a *Nash equilibrium* (NE), which is defined as[16]

$$\boldsymbol{\pi}^\star \triangleq (\pi_1^\star, \pi_2^\star) \in \mathscr{B}_1(\pi_2^\star) \times \mathscr{B}_2(\pi_1^\star). \qquad \text{(Eq. 1, Nash, 1951)} \tag{25}$$

[16]The complexity of computing an NE in a POSG is $\text{NEXP}^{\text{NP}}$ (Thm. 3.5, Goldsmith and Mundhenk, 2007). Recall that $\text{P} \subseteq \text{NP} \subseteq \text{PSPACE} \subseteq \text{NEXP} \subseteq \text{NEXP}^{\text{NP}}$.

This equilibrium can be refined as a *Perfect Bayesian Equilibrium* (PBE) by requiring the belief (22) to be consistent across subgames.

**Definition 2** (Subgame)**.** *An instance of* $\Gamma$ *starting from the history* $\mathbf{h}_t^{(2)}$ *is a subgame of* $\Gamma$ *and denoted as* $\Gamma|_{\mathbf{h}_t^{(2)}}$.

**Definition 3** (Reachable subgame)**.** *Given a strategy profile* $\boldsymbol{\pi}$ *and a game* $\Gamma$*, a subgame* $\Gamma|_{\mathbf{h}_t^{(2)}}$ *is reachable iff* $\mathbb{P}[\mathbf{h}_t^{(2)} \mid \boldsymbol{\pi}, \mathbf{b}_1] > 0$.

**Definition 4** (Perfect Bayesian Equilibrium (PBE))**.** $(\boldsymbol{\pi}^\star, \mathbb{B})$ *(22) is a PBE iff*

1. OPTIMALITY. $\boldsymbol{\pi}^\star = (\pi_1^\star, \pi_2^\star)$ *is a NE in* $\Gamma|_{\mathbf{h}_t^{(2)}}$ $\forall \mathbf{h}_t^{(2)} \in \mathcal{H}_t^{(2)}$.

2. BELIEF CONSISTENCY. *For any* $\mathbf{h}_t^{(1)} \in \mathcal{H}_t^{(1)}$ *with* $\mathbb{P}[\mathbf{h}_t^{(1)} \mid \boldsymbol{\pi}^\star, \mathbf{b}_1] > 0$*, then*

$$\mathbb{B}(\mathbf{h}_t^{(1)}, \pi_2^\star) = \mathbb{B}(\underbrace{\mathbb{B}(\mathbf{h}_{t-1}^{(1)}, \pi_2^\star)}_{\mathbf{b}_{t-1}.}, \underbrace{\pi_1^\star(\mathbb{B}(\mathbf{h}_{t-1}^{(1)}, \pi_2^\star))}_{a_{t-1}^{(1)}.}, o_t, \pi_2^\star).$$

**Theorem 3** (Existence of equilibria in one-sided POSGs)**.** *Under Assumptions 1–4, a zero-sum, one-sided POSG* $\Gamma$ *has a NE and a PBE in Markovian behavior strategies. Further, stationary versions of such equilibria exist if* $T$ *is random or* $T = \infty$.

*Proof.* Without loss of generality, we assume that $T$ is deterministic[17]. If $T < \infty$, then $\Gamma$ can be represented in extensive form[18]. Consequently, it has a NE in Markovian behavior strategies (Thm. 4.3, Thm. 4.6, Myerson, 1997). Denote the value of the game when restricting the horizon to be $T < \infty$ as $v_T$ and let $(\pi_{1,T}, \pi_{2,T})$ be the corresponding NE. Next, let $\pi_{1,T_\infty}$ be an infinite-horizon extension of $\pi_{1,T}$ where player 1 follows strategy $\pi_{1,T}$ for the first $T$ time steps and then follows an arbitrary strategy in the rest of the game. Define $\underline{r} \triangleq \min r(\cdot)$ and $\overline{r} \triangleq \max r(\cdot)$. It follows that the reward obtained by $\pi_{1,T_\infty}$ is at most $\overline{v}_T \triangleq v_T + \sum_{t=T+1}^{\infty} \gamma^{t-1}\overline{r} = v_T + \gamma^T \frac{\overline{r}}{1-\gamma}$ and at least $\underline{v}_T \triangleq v_T + \gamma^T \frac{\underline{r}}{1-\gamma}$. Since $\gamma^T \to 0$ as $T \to \infty$, the bounds $[\underline{v}_T, \overline{v}_T]$ converge to a single value, denoted $v_\infty$, which is achieved by stationary strategies (Thm. 2). Let $v^\star = \sup_{\pi_1} \inf_{\pi_2} [J^{(\pi_1, \pi_2)}(\mathbf{b})]$ (23). By definition, $\underline{v}_T \leq v^\star \leq \overline{v}_T$. Hence, $v_\infty = v^\star$ is the value of the game (squeeze theorem). Consequently, any *reachable* subgame has a stationary NE in Markovian behavior strategies. As this claim is independent of $\mathbf{b}_1$ and $s_1$, we obtain a PBE by combining the NEs of all reachable subgames with those of the unreachable subgames and $\mathbb{B}$ (22)[19]. □

**Corollary 1** (Convexity of the value function)**.** *Under Assumptions 1–4, the value function* $J_t^\star(\mathbf{b}) \triangleq J_t^{(\pi_1^\star, \pi_2^\star)}(\mathbf{b})$ *(23) of a zero-sum, two-player, one-sided POSG* $\Gamma$ *is piece-wise linear and convex*[20]*. Further,* $J^\star$ *is stationary if* $T$ *is random or* $T = \infty$.

[17] Otherwise $\Gamma$ can be reformulated with $T = \infty$; see Lemma 4.1 in Paper 4.
[18] See Lemma 4.2 of Paper 4 for a proof that a POSG can be represented in extensive form.
[19] Similar proofs are given in (Thm. 2.3, Horák, 2019) and (§3, Hespanha and Prandini, 2001).
[20] A proof is given in (Thm. 4.5, Horák, 2019).

**Remark 4** (Game-theoretic optimality)**.** In contrast to optimal control problems, e.g., MDPs and POMDPs, where optimality has an unambiguous meaning (4), in game-theoretic settings, optimality is not a well-defined concept. Throughout this thesis, we consider the PBE as a specific form of optimality, which, due to the zero-sum structure, aligns with *minimax optimality* (Blackwell and Girshick, 1979).

***Notation and terminology*** Our game- and decision-theoretic notation is summarized in Table 2 below.

| Notation(s) | Description |
|---|---|
| $\mathcal{M}, \Gamma, \Gamma\|_{\mathbf{h}^{(2)}}$ | A decision process (1) (15), a game (19), and a subgame. |
| $\mathcal{S}, \mathcal{O}, \mathcal{A}$ | Sets of states, observations, and actions in a decision process. |
| $\mathcal{A}_{\mathrm{k}}$ | Set of actions of player k in a game. |
| $T, \gamma$ | Time horizon and discount factor. |
| $f, r, z$ | Transition (2), reward, and observation functions. |
| $a_t, a_t^{(\mathrm{k})}$ | Action at time $t$ and action of player k in a game. |
| $\mathbf{a}_t$ | Action profile at time $t$ in a game. |
| $s_t, r_t$ | State at time $t$ and reward at time $t$. |
| $o_t$ | Observation at time $t$. |
| $\pi, \pi^\star$ | Strategy in a decision process and an optimal strategy. |
| $\pi_{\mathrm{k}}, \boldsymbol{\pi}$ | Strategy of player k in a game and strategy profile. |
| $\boldsymbol{\pi}^\star, \pi_{\mathrm{k}}^\star$ | Equilibrium strategy profile in a game and equilibrial strategy of player k. |
| $\tilde{\pi}_{\mathrm{k}}$ | Best response strategy of player k in a game. |
| $\Pi, \Pi_{\mathrm{k}}$ | Strategy space in a decision process and of player k in a game. |
| $J$ | Objective functional (4) or value function (5) (overloaded notation). |
| $J^\pi, J^{\boldsymbol{\pi}}$ | Value functions of strategy $\pi$ (7) and strategy profile $\boldsymbol{\pi}$ (23). |
| $Q^\pi, Q^{\boldsymbol{\pi}}$ | Q-functions of strategy $\pi$ and strategy profile $\boldsymbol{\pi}$ (6). |
| $J^\star, Q^\star$ | Optimal value function (7) (18) (23) and Q function (6). |
| $\mathscr{T}, \mathscr{T}_\pi$ | The Bellman operator (8) and the strategy evaluation operator (10). |
| $\mathscr{H}$ | The Q-learning operator (13). |
| $\mathbf{h}_t, \mathcal{H}_t$ | History at time $t$ in a decision process and set of histories of length $t$. |
| $\mathbf{h}_t^{(\mathrm{k})}, \mathcal{H}_t^{(\mathrm{k})}$ | History at time $t$ of player k in a game and set of histories of length $t$. |
| $\mathbf{i}_t^{(\mathrm{k})}$ | Information feedback of player k at time $t$ in a game. |
| $\mathbf{b}_t, \mathcal{B}, \mathbb{B}$ | Belief state at time $t$, set of beliefs, and belief operator (17) (22). |
| $\mathcal{N}, \mathscr{B}_{\mathrm{k}}$ | Set of players in a game, best response correspondence of player k (24). |

***Table 2:*** *Notation for our game- and decision-theoretic models.*

Our terminology for classifying different game and decision-theoretic models is summarized in Table 3 on the next page.

| Terminology | Meaning |
|---|---|
| Finite MDP | $\|\mathcal{S} \cup \mathcal{A}\| < \infty$. |
| Stationary MDP | $f$ and $r$ are independent of the time step $t$. |
| Markovian $\pi$ in an MDP | $\pi$ is a function of the state $s$. |
| Finite POMDP | $\|\mathcal{S} \cup \mathcal{A} \cup \mathcal{O}\| < \infty$. |
| Markovian $\pi$ in a POMDP | $\pi$ is a function of the belief state $\mathbf{b}$ (17). |
| Stationary POMDP | $f$, $r$, and $z$ are independent of the time step $t$. |
| Finite OS-POSG | $\|\mathcal{S} \cup \mathcal{A}_1 \cup \mathcal{A}_2 \cup \mathcal{O}\| < \infty$. |
| Stationary OS-POSG | $f$, $r$, and $z$ are independent of the time step $t$. |
| Markovian $\pi_1$ in a OS-POSG | $\pi_1$ is a function of the belief state $\mathbf{b}$ (22). |
| Markovian $\pi_2$ in a OS-POSG | $\pi_2$ is a function of the belief state $\mathbf{b}$ (22) and the state $s$. |
| Best response | Optimal strategy against a fixed opponent strategy in a OS-POSG. |
| Stationary $\pi$ | $\pi$ is conditionally independent of the time step $t$. |
| Pure $\pi$ | $\pi$ is a deterministic function. |
| Behavior $\pi$ | $\pi$ is a probability distribution over actions at each time step. |
| Episode | Execution of a decision process or a game for $T$ time steps. |

***Table 3:*** *Terminologies for our game- and decision-theoretic models.*

## ■ Summary

This chapter reviews two mathematical frameworks for modeling decision problems: Markov decision theory and (noncooperative) game theory. They are particularly well-suited in the context of security response due to their ability to model sequential decision-making in uncertain, complex, and dynamic environments. The next chapter formalizes the security response problem using these frameworks.

## ■ Bibliographic Remarks

The Markov process, which underpins Markov decision theory, was introduced by A.A Markov in 1906 [299, 300]. Markov's seminal papers laid the groundwork for Markov decision theory, which can be traced back to the mid-20th century with the contributions of R. Bellman [45], C.E. Shannon [404], A.L Samuel [388], S. Karlin [231], and L.S. Shapley [405]. In 1957, Bellman established the framework of MDPs in the landmark book "*Dynamic Programming*" [46].

The foundations of game theory were laid out in the classic book by J. von Neumann and O. Morgenstern, "*The Theory of Games and Economic Behavior*" in 1944 [480]. Precursors to this seminal work were E. Zermelo's work on chess in 1913 [511], a sequence of short papers by E. Borel in the 1920s [67, 68, 70, 69], and von Neumann's 1928 paper on finite zero-sum games [479]. The field blossomed in the 1950s with J.F. Nash's introduction of the Nash equilibrium, a concept that earned him the Nobel prize in 1994 [322]. Concurrently, L.S. Shapley made significant advances in dynamic games and introduced the stochastic game in 1953 [405]. Another significant advancement came from H.W. Kuhn, who introduced the concepts of perfect recall and behavior strategies in 1953 [257].

# FORMALIZING THE SECURITY RESPONSE PROBLEM

*A problem well stated is a problem half solved.*

— Charles Kettering ***1876-1958***.

THE security response problem is not intrinsically mathematical and can be stated with minimal formalism, as demonstrated in the introduction chapter. However, a mathematical formulation allows us to define concepts precisely, verify the consistency of ideas, and study the implications of assumptions. For these reasons, this chapter is devoted to formalizing the security response problem and articulating our assumptions in precise mathematical terms.

To accomplish this formalization, we need a vocabulary in which to talk about the systems and actors involved. Following the terminology of (Alpcan and Basar, 2010), we refer to the operator of the infrastructure as the *defender*, and we refer to the entity causing the attack as the *attacker*[21]; see Fig. 18. Both interact with the infrastructure by taking *actions*, which affect the infrastructure's *state*. When selecting these actions, the defender and the attacker consider measurements collected from the infrastructure, which we refer to as *observations*. A function that maps a sequence of observations to an action is called a *strategy*, and a strategy that is most advantageous according to some objective is *optimal*. We illustrate this terminology with an example on the next page.

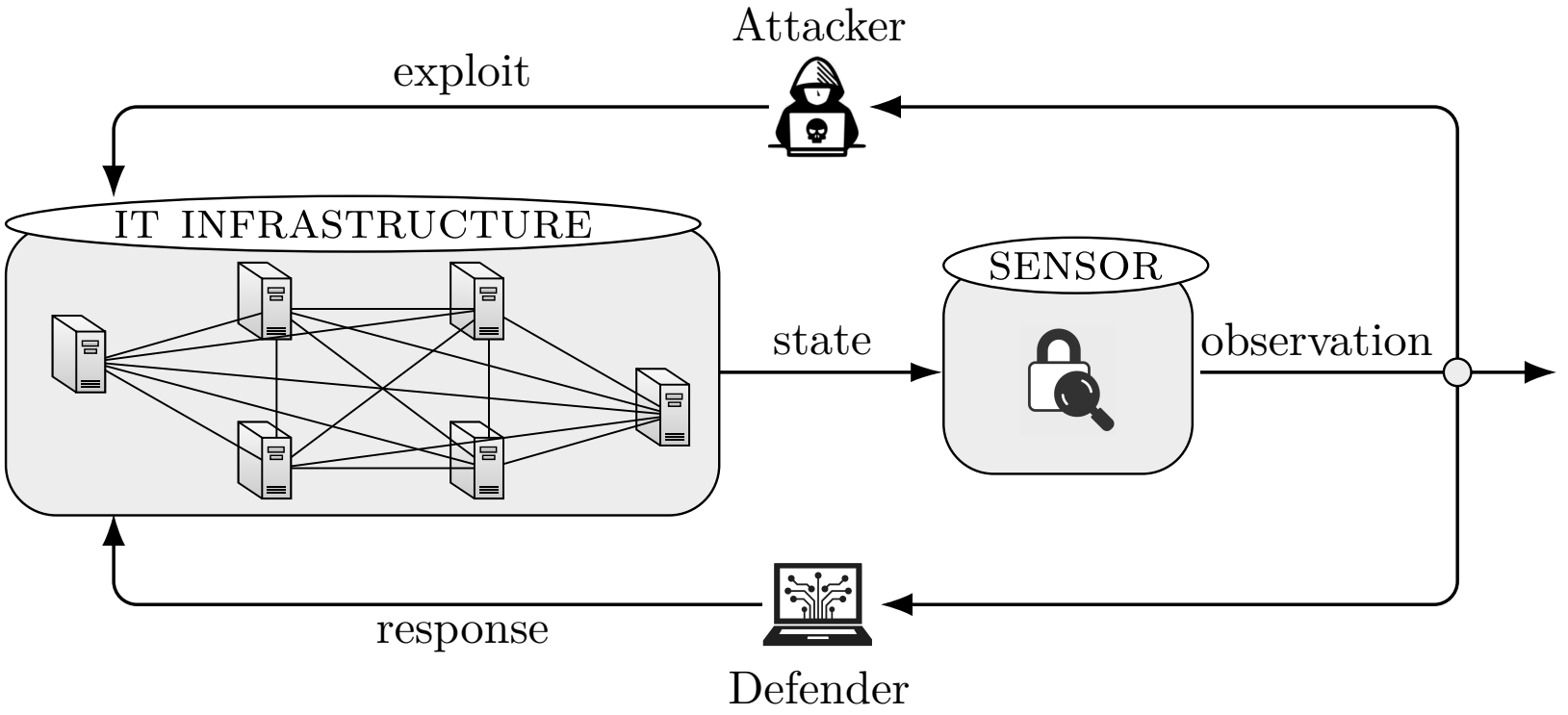

***Figure 18:*** *The security response problem; an attacker exploits vulnerabilities of an* IT *infrastructure; the operator of the infrastructure, which we refer to as the defender, monitors the network and executes responses.*

[21] Though several entities may be involved, defining a single "attacker" often simplifies the analysis without sacrificing generality.

**Example: Instance of the security response problem.**

A *defender* monitors an IT infrastructure by *observing* alerts from an intrusion detection system. The infrastructure includes a set of servers that provide service to *clients* through a public gateway, which is also open to an *attacker*. To prevent the attacker from intruding, the defender can block the gateway. In deciding when to take this *action*, the defender balances two *objectives*: maintain service and keep the attacker out. The *optimal strategy* can be to maintain service until the moment the attacker enters through the gateway, at which time the gateway must be blocked. (We prove in Paper 1 that this strategy has a threshold structure.) What makes the defender's task difficult is that it has to infer that an intrusion occurs from the sequence of alerts.

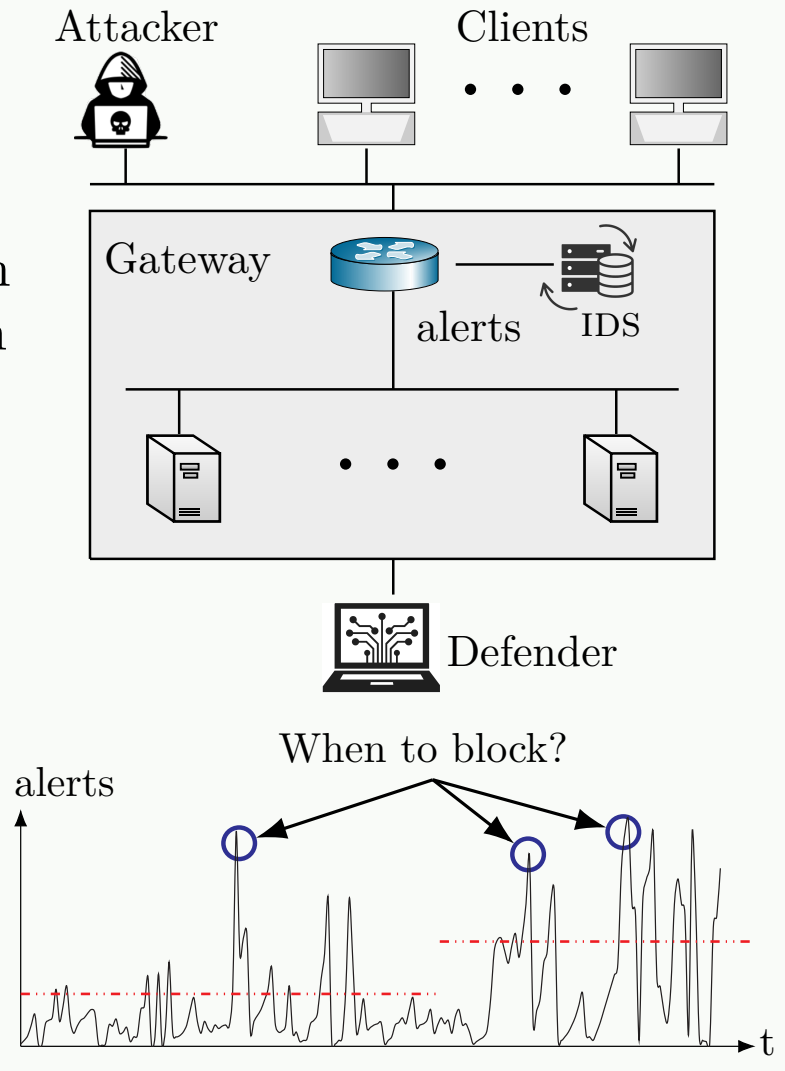

Mathematically, the response problem in its simplest form can be stated as

$$\sup_{a^{(\mathrm{D})} \in \mathcal{A}_{\mathrm{D}}} J(a^{(\mathrm{D})}), \qquad \text{(SECURITY RESPONSE AS A STATIC OPTIMIZATION)}$$

where $\mathcal{A}_{\mathrm{D}}$ is the set of defender actions and $J(a^{(\mathrm{D})})$ is an objective function that encodes the effect of taking action $a^{(\mathrm{D})}$ in terms of security and service utility.

While the above formulation models the response problem for a simple system, it only models the optimization of a *single* response action without addressing the state or dynamic aspects of the infrastructure. In practice, security response involves optimizing a sequence of interdependent actions. Such dynamic optimization can be captured by modeling the infrastructure as a *discrete-time dynamical system* whose evolution depends on the sequence of response actions. This approach allows us to frame security response as a Markov Decision Process (MDP)[22]:

$$\begin{aligned} &\sup_{\pi_{\mathrm{D}} \in \Pi_{\mathrm{D}}} && \mathbb{E}_{\pi_{\mathrm{D}}}[J(\pi_{\mathrm{D}}) \mid s_1] && \text{(SECURITY RESPONSE AS AN MDP)} \\ &\text{subject to} && s_{t+1} \sim f(\cdot \mid s_t, a_t^{(\mathrm{D})}) && \forall t \geq 1 \\ & && a_t^{(\mathrm{D})} \sim \pi_{\mathrm{D}}(\cdot \mid s_t) && \forall t \geq 1, \end{aligned} \tag{26}$$

[22] The components of an MDP are defined the background chapter; see (1).

where $t = 1, \dots, T$, $\pi_{\mathrm{D}}$ is the defender strategy, $\Pi_{\mathrm{D}}$ is the strategy space, $J(\pi_{\mathrm{D}})$ is an objective functional, and the attacker strategy is assumed to be static and implicitly modeled by $f$.

**Remark 5** (Axioms of rational choice)**.** The defender's objective to maximize the expected value of $J$ (26) is justified by the von Neumann-Morgenstern axioms (p. 26, von Neumann and Morgenstern, 1944). We show in Paper 4 that these axioms generally hold for the security response use case. If they do not hold, prospect theory can be used to model the objective (Kahneman and Tversky, 1979).

Though an MDP allows us to model the dynamic aspects of the response problem, it assumes full observability of the infrastructure's state – a condition rarely met in practice. Usually, the defender has partial observability of the state, e.g., through log files and alerts. This partial observability can be modeled with a Partially Observed Markov Decision Process (POMDP)[23]:

$$
\begin{aligned}
\sup_{\pi_{\mathrm{D}} \in \Pi_{\mathrm{D}}} \quad & \mathbb{E}_{\pi_{\mathrm{D}}}[J(\pi_{\mathrm{D}}) \mid \mathbf{b}_1] && \text{(SECURITY RESPONSE AS A POMDP)} \\
\text{subject to} \quad & s_1 \sim \mathbf{b}_1 && \\
& s_{t+1} \sim f(\cdot \mid s_t, a_t^{(\mathrm{D})}) && \forall t \geq 1 \\
& o_t \sim z(\cdot \mid s_t) && \forall t \geq 2 \\
& \mathbf{b}_{t+1} = \mathbb{B}(\mathbf{b}_t, a_t^{(\mathrm{D})}, o_{t+1}) && \forall t \geq 1 \\
& a_t^{(\mathrm{D})} \sim \pi_{\mathrm{D}}(\cdot \mid \mathbf{b}_t) && \forall t \geq 1,
\end{aligned}
$$

where $o_t$ is the defender's observation (15).

While the POMDP captures partial observability, it assumes a fixed attacker strategy that is implicitly modeled by the dynamics $f$. This assumption is unrealistic since attackers are dynamic and adapt their methods in response to the defender's actions (Anderson, 2001). A game-theoretic model is suitable to represent such adaptability. To define such a model, we must make assumptions about the attacker's behavior, resources, and objectives. Throughout this thesis, we make the following two assumptions.

**Assumption 5** (Zero-sum)**.** *The attacker's objective is inverse to the defender's.*

Assumption 5 expresses that the game is zero-sum. We can justify this assumption in three ways. First, since the defender is unaware of the attacker's goal, it is reasonable to assume that the attacker's objective is adversarial to the defender's. Second, it reduces the computational complexity of solving the game and implies that every equilibrium leads to the same value. Third, it captures a scenario where the attacker's goal is to inflict maximal harm to the defender, which was the case in the NOTPETYA attack in 2017 (U.S. Department of Justice, 2020).

[23] The components of a POMDP are defined the background chapter; see (15).

**Assumption 6** (Omniscent attacker)**.** *The attacker has complete observability.*

Assumption 6 expresses that the game between the attacker and the defender has *one-sided* partial observability. That is, while the defender has partial observability of the infrastructure's state, the attacker knows the state as well as the defender's observations. We can motivate this assumption in two ways. First, the assumption holds for insider attacks, such as the attack against Ukraine's power grid in 2015 (Case, 2016). Second, it reflects that it is generally not known what information is available to the attacker; therefore, a worst-case scenario is assumed.

Given the above assumptions, we formulate security response as a zero-sum, one-sided Partially Observed Stochastic Game (POSG)[24]:

$$\sup_{\pi_{\mathrm{D}} \in \Pi_{\mathrm{D}}} \inf_{\pi_{\mathrm{A}} \in \Pi_{\mathrm{A}}} \mathbb{E}_{\pi_{\mathrm{D}},\pi_{\mathrm{A}}}[J(\pi_{\mathrm{D}}, \pi_{\mathrm{A}}) \mid \mathbf{b}_1] \qquad \text{(SECURITY RESPONSE AS A POSG)} \tag{27a}$$

$$\begin{aligned}
\text{subject to} \quad & s_1 \sim \mathbf{b}_1 && && \text{(27b)}\\
& s_{t+1} \sim f(\cdot \mid s_t, a_t^{(\mathrm{D})}, a_t^{(\mathrm{A})}) && \forall t \geq 1 && \text{(27c)}\\
& o_t \sim z(\cdot \mid s_t) && \forall t \geq 2 && \text{(27d)}\\
& \mathbf{b}_{t+1} = \mathbb{B}(\mathbf{b}_t, a_t^{(\mathrm{D})}, o_{t+1}, \pi_{\mathrm{A}}) && \forall t \geq 1 && \text{(27e)}\\
& a_t^{(\mathrm{D})} \sim \pi_{\mathrm{D}}(\cdot \mid \mathbf{b}_t) && \forall t \geq 1 && \text{(27f)}\\
& a_t^{(\mathrm{A})} \sim \pi_{\mathrm{A}}(\cdot \mid \mathbf{b}_t, s_t) && \forall t \geq 1, && \text{(27g)}
\end{aligned}$$

where $a_t^{(\mathrm{A})}$ is the attacker action and $\pi_{\mathrm{A}}$ is the attacker strategy. Almost any security response scenario can be modeled as a game of this form; we provide several examples in the subsequent chapters.

**Remark 6** (Modeling clients)**.** The clients of the IT infrastructure are implicitly modeled by the observation distribution $z$ (27d), i.e., the clients' interactions with the infrastructure's services affect the defender's observations.

**Remark 7** (Maximin optimality)**.** The sup inf objective in (27) expresses that the defender should maximize the minimal value of $J$ (27a) across all possible attacker strategies. This objective is reasonable since a) the defender does not know the attacker's strategy; and b) the attacker likely uses the best possible strategy.

**Remark 8** (Bounded rationality)**.** In practice, the defender may not be able to maximize (27) due to computational or cognitive constraints, in which case we say that the defender has *bounded rationality* (Simon, 1955). However, assuming full rationality in the problem formulation remains valuable for theoretical development. We present an approach for modeling bounded rationality in Paper 5.

[24]The components of a POSG are defined the background chapter; see (19).

Solving (27) corresponds to finding a Nash equilibrium[25]. We know from Thm. 3 in the background chapter that such an equilibrium exists under general conditions. However, many difficulties are encountered when one attempts to compute it. Chief among them is that we must obtain the functions $f$ (27c) and $z$ (27d), and the functional $J$ (27a). Due to the complexity of IT infrastructures, these parameters need to be derived from a combination of system measurements and domain knowledge about the infrastructure's architecture. Hence, we have the following challenge.

**Challenge 1: System identification.**

The parameters of the security response problem must be determined from infrastructure measurements and architectural domain knowledge.

The next major difficulty is the complexity of computing the equilibrium, which is $\text{NEXP}^{\text{NP}}$ (Thm. 3.5, Goldsmith and Mundhenk, 2007)[26]. To make matters worse, this complexity often grows exponentially with the infrastructure size due to the curse of dimensionality (Bellman, 1957), as summarized in the following challenge.

**Challenge 2: Computational complexity.**

The complexity of computing an optimal defender strategy often grows exponentially with the size of the infrastructure configuration[a].

[a] See Paper 3 for a proof.

Lastly, once a (theoretically) optimal strategy has been obtained, it must be experimentally validated. Simulations are typically used for this purpose (Nguyen and Reddi, 2023). However, they do not adequately model many functional and timing details of an operational infrastructure. Therefore, a response strategy must be experimentally validated on a testbed, as stated in the following challenge.

**Challenge 3: Experimental strategy validation.**

Defender strategies must be experimentally validated on a testbed where attacks and response actions can be executed.

[25] Which also can form a stronger equilibrium, namely a Perfect Bayesian equilibrium (PBE); see the background chapter for details.

[26] Recall that $\text{P} \subseteq \text{NP} \subseteq \text{PSPACE} \subseteq \text{NEXP} \subseteq \text{NEXP}^{\text{NP}}$.

## ■ Summary

We model security response against a static attacker as a POMDP[27], and we model security response against a dynamic attacker as a one-sided POSG[28]. Given these models, we identify three primary challenges in devising an optimal response strategy: system identification (Challenge 1), computational complexity (Challenge 2), and strategy validation (Challenge 3). In the next chapter, we introduce our methodology for optimal security response, which tackles Challenge 1 and Challenge 3. Subsequently, we present six papers where we apply our methodology to different instances of the security response problem. In these papers, we demonstrate how to overcome Challenge 2 by deriving structural properties of optimal response strategies and leveraging stochastic approximation techniques.

[27] We use the POMDP model in Paper 1 and Paper 6.
[28] We use the POSG model in Paper 2, Paper 3, Paper 4, and Paper 5.

# PLATFORM AND METHODOLOGY FOR OPTIMAL SECURITY RESPONSE

*The best material model of a cat is another, or preferably the same, cat.*

— Norbert Wiener ***1945***, *The role of models in science.*

Our methodology for optimal security response in an IT infrastructure is centered around a *digital twin*, i.e., a virtual replica of the infrastructure; see Fig. 19. We use this twin to run attack scenarios and defender responses. Such runs produce system measurements and logs, from which we estimate infrastructure statistics. These statistics allow us to instantiate a mathematical model of the target infrastructure through *system identification* (Ljung, 1998). We then leverage this model to *optimize* response strategies, whose performance is assessed using the digital twin. This closed-loop process can be executed iteratively to provide progressively better response strategies.

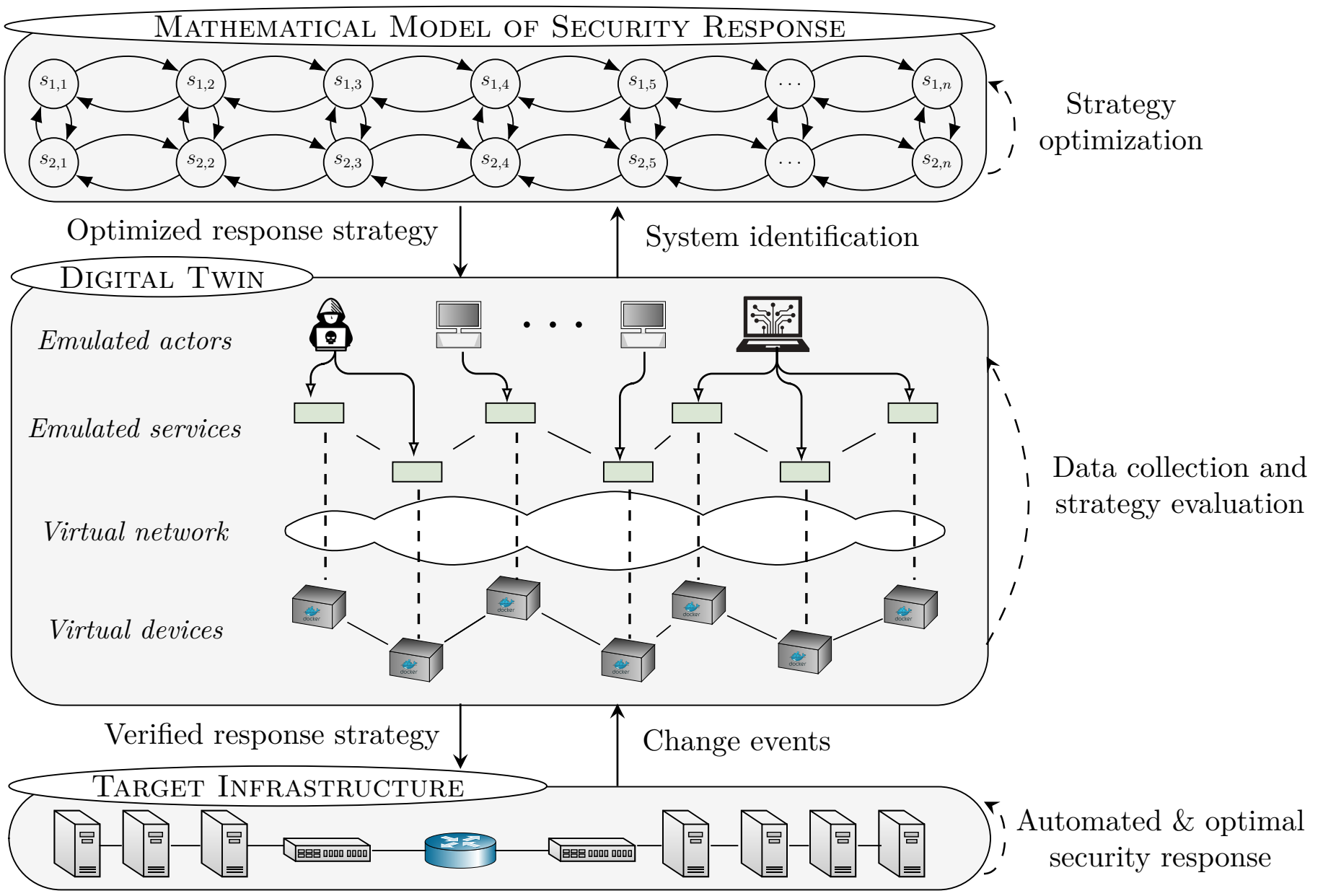

***Figure 19:*** *Our methodology for obtaining an optimal response strategy for a target infrastructure; the digital twin is a virtual replica of the target infrastructure; it is used to collect statistics and to evaluate strategies; the collected data enables us to identify a mathematical model of the infrastructure, which allows for strategy optimization.*

We have developed a platform based on the methodology captured in Fig. 19. This platform allows for creating digital twins, running simulations, performing system identification, and optimizing response strategies. It is deployed on commodity hardware, and the source code is released under the CC-BY-SA 4.0 license (Hammar, 2023)[29]. We refer to the platform as CSLE, which stands for the "***C**yber **S**ecurity **L**earning **E**nvironment*." The rest of this chapter delves into the technical details of CSLE, covering its architecture, implementation, and programming interface. Special attention is given to the digital twin, as it provides the main tool for the experimental results presented in the subsequent chapters.

## ■ Architecture

CSLE runs on a distributed system with $N \geq 1$ physical servers connected through an IP network. Each server runs a virtualization layer provided by DOCKER containers and virtual links (Merkel, 2014). CSLE is implemented in Python (Van Rossum and Drake Jr, 1995), JavaScript (Eich, 2005), and Bash (GNU, 2007). It can be accessed through Python libraries, a web interface, a command-line interface, and a GRPC interface (Google, 2022). It stores metadata in a distributed database referred to as the *metastore*, which is based on POSTGRES and CITUS (Cubukcu et al., 2021). This database consists of $N$ replicas, one per server; see Fig. 20. A quorum-based two-phase commit scheme is used to achieve consensus among replicas. One replica is a designated *leader* and is responsible for coordination. The others are *workers*. The leader is elected using the protocol described in (Alg. 1, Niazi et al., 2015). A new leader is elected by a quorum whenever the current leader fails or becomes unresponsive. CSLE thus tolerates up to $\frac{N}{2} - 1$ failing servers.

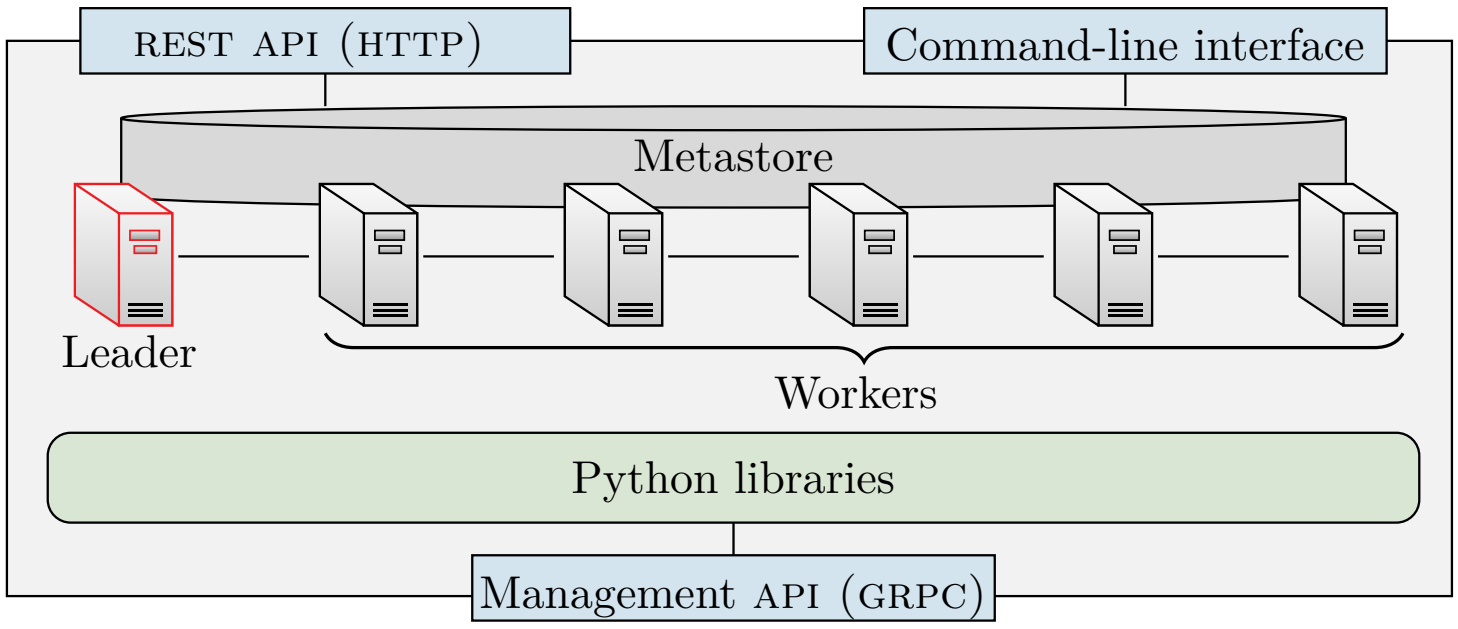

***Figure 20:*** *CSLE architecture; it is a distributed system with a database and four interfaces: a Python API, a GRPC API, a REST API, and a command-line interface.*

[29]The code repository also contains video demonstrations, usage examples, pre-built virtual containers, technical documentation, and a dataset of attack traces.

We run CSLE on our cluster of commodity servers at KTH; see Fig. 21. All experimental results presented in this thesis have been obtained through experimentation using this cluster. Server specifications can be found in Table 4. The servers are connected through an Ethernet switch. Deployment of CSLE on both on-premise and cloud infrastructures is automated using ANSIBLE (Red Hat, 2024).

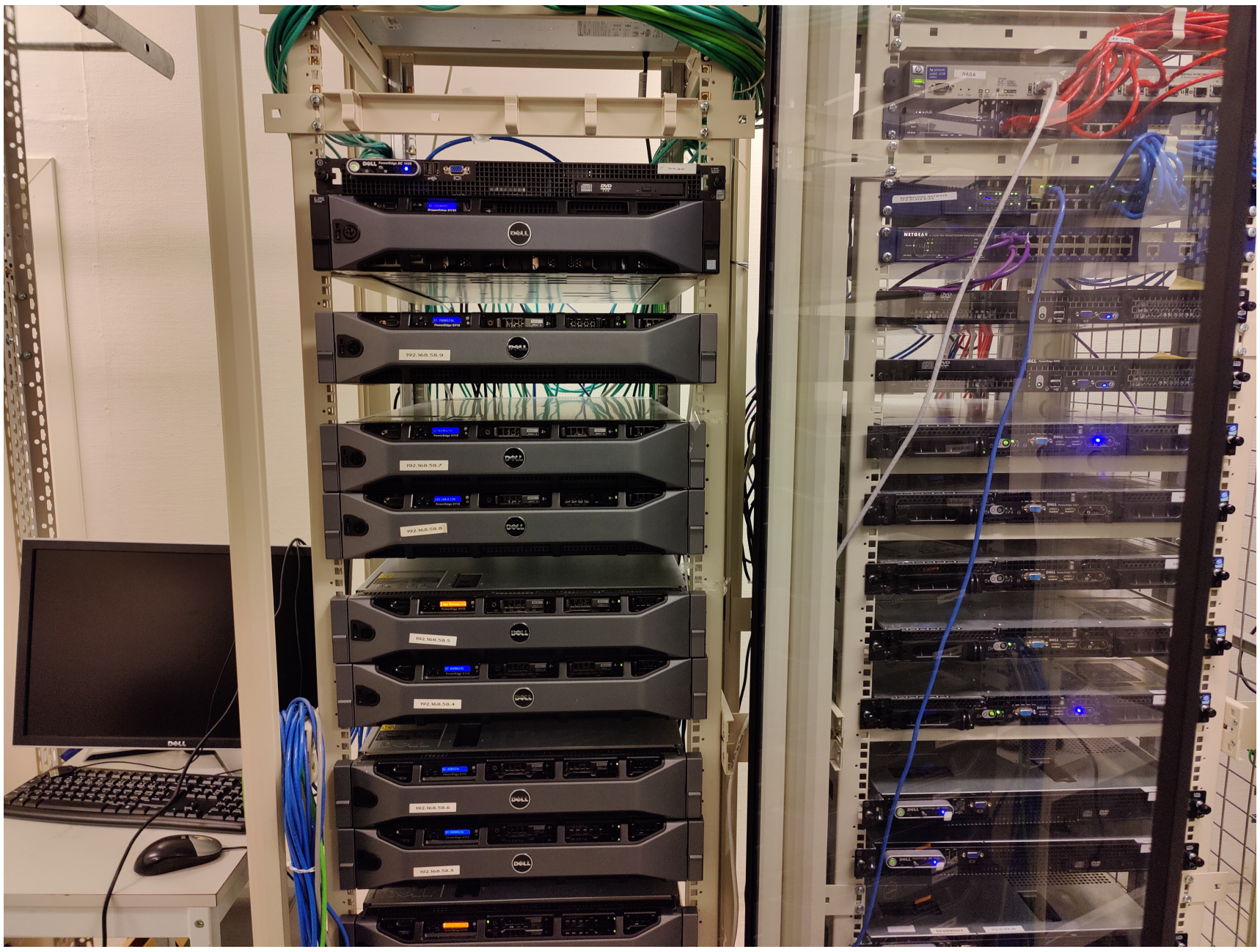

***Figure 21:*** *Our server rack at KTH where we run CSLE.*

| *Server* | *Processors* | *Network* | RAM *(GB)* |
|---|---|---|---|
| 1, R 715 2U | two 12-core AMD OPTERON | 12×GBE | 64. |
| 2, R 715 2U | two 12-core AMD OPTERON | 12×GBE | 64. |
| 3, R 715 2U | two 12-core AMD OPTERON | 12×GBE | 64. |
| 4, R 715 2U | two 12-core AMD OPTERON | 12×GBE | 64. |
| 5, R 715 2U | two 12-core AMD OPTERON | 12×GBE | 64. |
| 6, R 715 2U | two 12-core AMD OPTERON | 12×GBE | 64. |
| 7, R 715 2U | two 12-core AMD OPTERON | 12×GBE | 64. |
| 8, R 715 2U | two 12-core AMD OPTERON | 12×GBE | 64. |
| 9, R 715 2U | two 12-core AMD OPTERON | 12×GBE | 64. |
| 10, R 630 2U | two 12-core INTEL XEON E 5- 2680 | 12×GBE | 256. |
| 11, R 740 2U | 1 20-core INTEL XEON GOLD 5218R | 2 × 10GBE | 32. |
| 12, SUPERMICRO 7049 | 2 TESLA P 100, 1 16-core INTEL XEON | 100MBE | 126. |
| 13, SUPERMICRO 7049 | 4 RTX 8000, 1 24-core INTEL XEON | 10GBE | 768. |

***Table 4:*** *Specifications of the servers in our rack (Fig. 21).*

# Digital Twin

The concept of a *digital twin* emerged in the 1960s when NASA used virtual environments to evaluate failure scenarios for lunar landers (Allen B. Danette, 2021). In the context of our methodology, a digital twin is a virtual replica of an IT infrastructure that provides a controlled environment for virtual operations, the outcomes of which can be used to optimize operations in the physical infrastructure. It enables us to systematically test response strategies under different conditions, including varying attacks, workloads, and network latencies.

Creating a twin involves three main tasks: (*i*) replicating relevant parts of the physical infrastructure, such as processors, network interfaces, and network conditions; (*ii*) emulating actors, i.e., attackers, defenders, and clients; and (*iii*) instrumenting the twin with monitoring and management capabilities. Each of these three tasks is detailed below.

### Emulating hosts and switches

Hosts and switches are emulated with DOCKER containers (Merkel, 2014), i.e., lightweight executable packages that include runtime systems, code, libraries, and configurations. This virtualization lets us quickly instantiate large emulated infrastructures; see Fig. 22. Resource allocation to containers, e.g., CPU and memory, is enforced using CGROUPS. Containers that emulate switches run OVS (Pfaff et al., 2015) and connect to controllers through OPENFLOW (McKeown et al., 2008). Since the switches are programmed through flow tables, they can act as layer-two switches or as routers.

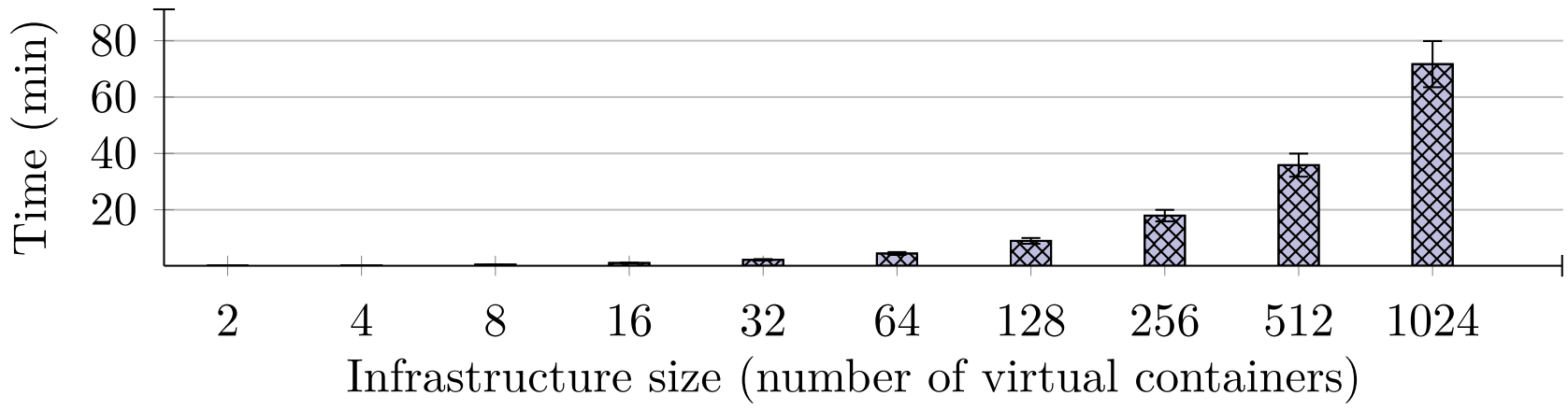

***Figure 22:*** *Time required to deploy a digital twin in function of the infrastructure size; deploying the twin involves creating containers and attaching them to virtual networks; the time measurements were performed for a digital twin with a single virtual network running on a server with a* 24*-core* INTEL XEON GOLD 2.10 *GHz* CPU *and* 768 GB RAM*.*

### Emulating network links

Network connectivity between containers in a digital twin is emulated with virtual links implemented by LINUX bridges and network namespaces, which create logical copies of the physical host's network stack. If an emulated network spans multiple physical servers, the traffic is tunneled over the physical network using VXLAN (Mahalingam et al., 2014). In other words, the physical network provides a substrate, on top of which the emulated networks are overlaid; see Fig. 23.

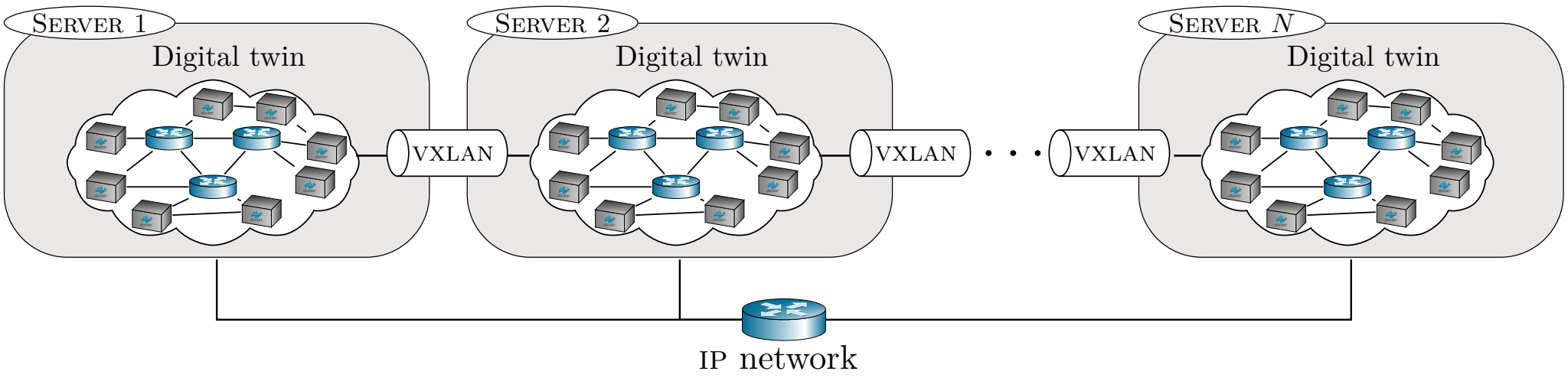

***Figure 23:*** *A distributed digital twin on CSLE; physical servers are connected through an IP network, over which virtual networks are created using VXLAN tunnels.*

Network conditions of virtual links are created using the NETEM module in the LINUX kernel (Hemminger, 2005). This module allows setting bit rates, packet delays, packet loss probabilities, and jitter. For the experimental results presented in this thesis, we emulate connections between servers in an IT infrastructure with full-duplex loss-less connections of 1 Gbit/s capacity in both directions. Similarly, we emulate connections between servers and external clients with full-duplex connections of 100 Mbit/s capacity and 0.1% packet loss with random bursts of 1% packet loss. These numbers are based on measurements on enterprise and wide-area networks (Kushida and Shibata, 2002)(Paxson, 1997)(Elliott, 1963).

### Management

Each emulated device in a digital twin runs a *management agent*, which exposes a GRPC API (Google, 2022). This API is invoked to perform control actions, e.g., restarting services and updating configurations. The communication channels to the agents are provided by a *management network*. The reason for using a separate network to carry management traffic is to avoid interference and simplify control of the digital twin (Clemm and Cisco Systems, 2007).

### Emulating actors

Client populations are emulated through processes that access services on emulated hosts. Client arrivals are controlled by a Poisson process with exponentially distributed service times. The sequence of service invocations is selected according to a Markov process. Similarly, attackers are emulated by programs that select actions from a pre-defined set that includes reconnaissance commands, brute-force attacks, and exploits. Examples of attacker actions are listed in Table 5 on the next page. Likewise, defender actions are emulated by executing system commands through the GRPC API described above. Examples of defender actions are listed in Table 6.

| *Type* | *Actions* | MITRE ATT&CK technique |
|---|---|---|
| Reconnaissance | TCP SYN scan, UDP scan | T1046 service scanning. |
| | TCP XMAS scan | T1046 service scanning. |
| | VULSCAN | T1595 active scanning. |
| | ping-scan | T1018 system discovery. |
| Brute-force | TELNET, SSH | T1110 brute force. |
| | FTP, CASSANDRA | T1110 brute force. |
| | IRC, MONGODB, MYSQL | T1110 brute force. |
| | SMTP, POSTGRES | T1110 brute force. |
| Exploit | CVE-2017-7494 | T1210 service exploitation. |
| | CVE-2015-3306 | T1210 service exploitation. |
| | CVE-2010-0426 | T1068 privilege escalation. |
| | CVE-2015-5602 | T1068 privilege escalation. |
| | CVE-2015-1427 | T1210 service exploitation. |
| | CVE-2014-6271 | T1210 service exploitation. |
| | CVE-2016-10033 | T1210 service exploitation. |
| | SQL injection (CWE-89) | T1210 service exploitation. |

***Table 5:*** *Examples of emulated attacker actions on CSLE; actions are identified by the vulnerability identifiers in the Common Vulnerabilities and Exposures (CVE) database (The MITRE Corporation, 2022) and the Common Weakness Enumeration (CWE) list (The MITRE Corporation, 2023); the actions are also linked to the corresponding attack techniques in the MITRE ATT&CK taxonomy (Strom et al., 2018).*

| *Index* | *Action* | MITRE D3FEND technique |
|---|---|---|
| 1 | Revoke user certificates | D3-CBAN certificate revocation. |
| 2 | Blacklist IPs | D3-NTF network traffic filtering. |
| $3-37$ | Drop traffic that generates alerts of priority $1-34$ | D3-NTF network traffic filtering. |
| 38 | Block gateway | D3-NI network isolation. |
| 39 | Migrate servers between different network zones | D3-NI network isolation. |
| 40 | Redirect traffic from one server to another | D3-NTF network traffic filtering. |
| 41 | Isolate a server | D3-NI network isolation. |
| 42 | Deploy new security functions (e.g., firewalls) | D3-NTPM network policy mapping. |
| 43 | Shutdown a server | D3-HS host shutdown. |
| 44 | Replicate a service | D3-SVCDM service mapping. |
| 45 | Start decoy services | D3-DE decoy environment. |

***Table 6:*** *Examples of emulated defender actions on CSLE; the actions are linked to the corresponding defense techniques in the MITRE D3FEND taxonomy (Kaloroumakis and Smith, 2021).*

### Monitoring

We use a monitoring system based on a publish-subscribe architecture; see Fig. 24. Following this architecture, each emulated device in a digital twin runs a *monitoring agent*, which reads local metrics of the host and pushes those metrics to an event bus implemented with KAFKA (Kreps, 2011). The data in this bus is consumed by data pipelines, which process the data and write it to storage systems. The number of metrics collected per time step is in the order of thousands and scales linearly with the number of emulated devices in the digital twin.

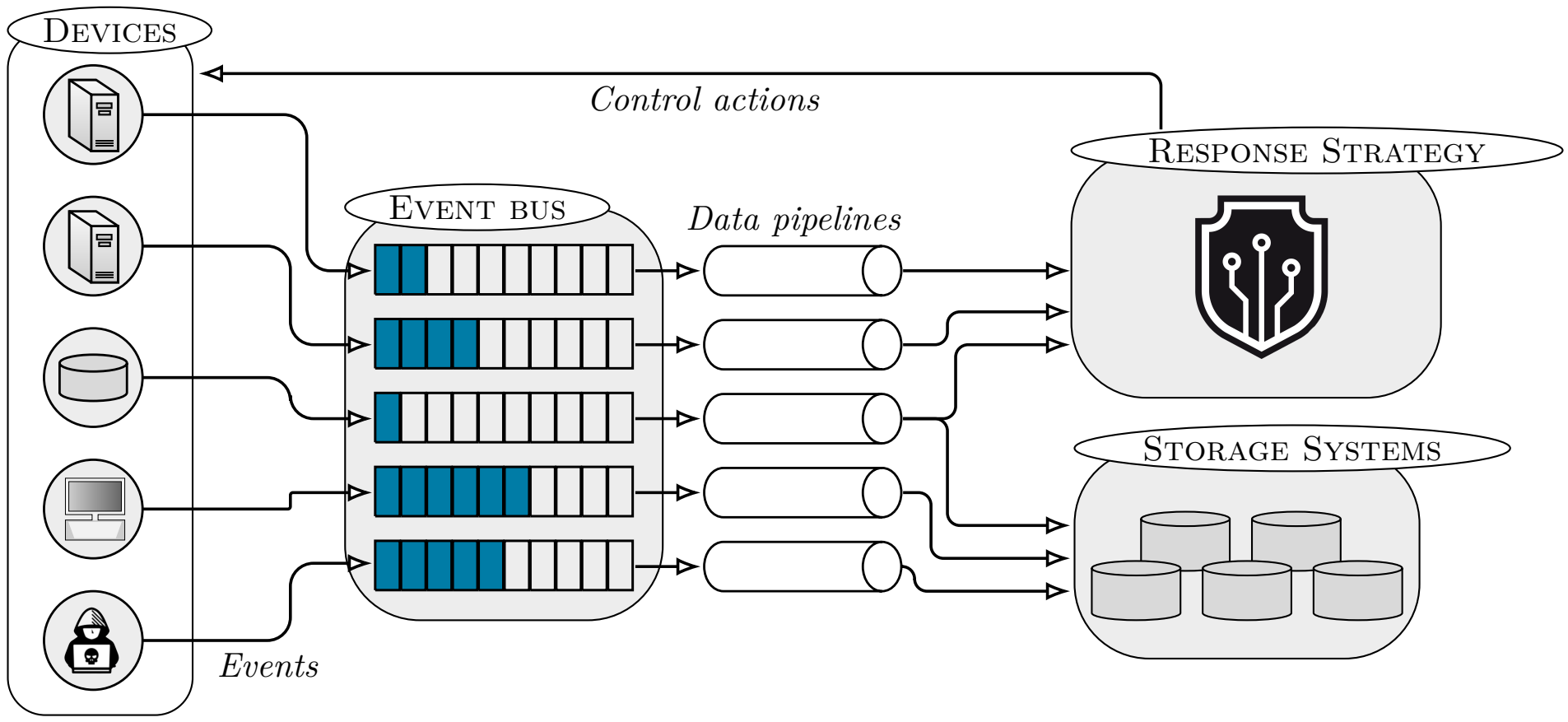

***Figure 24:*** *Monitoring system of a digital twin on CSLE; emulated devices run monitoring agents that periodically push metrics to an event bus; the data in this bus is consumed by data pipelines that process the data and write to storage systems; the processed data is also used by an automated response strategy to decide on control actions.*

## ■ System Identification

The emulation and monitoring capabilities described above enable *system identification*, which is a procedure to create mathematical models of dynamic systems based on measured data (Ljung, 1998). In our methodology, this procedure involves collecting data from a digital twin and then applying statistical techniques to fit a model that predicts the twin's behavior under various conditions. CSLE integrates several estimation algorithms to support this process, such as expectation-maximization (Dempster et al., 1977), Gaussian process regression (Rasmussen and Williams, 2006), and Markov Chain Monte Carlo (Robert and Casella, 2004). A code example of using CSLE for system identification is provided on the next page.

```python
1 import Metastore, Emulator, Monitor, EM from csle
2 dt = Metastore.get_digital_twin(..)
3 attacker_seq = ["nmap -p 22 --script ssh-brute 15.12.8.33", ..]
4 defender_seq = ["iptables -A INPUT -s 15.12.9.120 -j DROP", ..]
5 Emulator.run(dt, attacker_seq, defender_seq)
6 stats = Monitor.get_statistics(dt)
7 pomdp = Metastore.get_pomdp(..)
8 fitted_model = EM.fit(stats, pomdp)
```

***Listing 1:*** *Sample Python program that uses CSLE for system identification; line 2 fetches metadata of the digital twin; lines 3–4 define sequences of system commands to execute; line 5 runs the commands on the digital twin; the commands are executed at discrete time intervals of a specific length (e.g., 30s); line 6 extracts the collected data from the monitoring system (e.g., log files and infrastructure statistics); line 7 fetches the parameters of a POMDP from the metastore (see the background chapter for the definition of a POMDP); and line 8 fits the POMDP parameters (e.g., the observation distribution $z$ (15)) to the monitoring data using expectation-maximization (Dempster et al., 1977). (We define one time step in the POMDP as one monitoring interval on the digital twin, i.e., if we collect measurements at 30s intervals from the digital twin, then one time step in the POMDP corresponds to 30s on the digital twin.)*

## ■ Optimization and Evaluation

After completing the system identification, the next step in our methodology is to compute an optimal response strategy based on the identified model; see Fig. 19. CSLE has a suite of algorithms to perform this optimization. These algorithms are based on different algorithmic frameworks, including dynamic and linear programming, reinforcement learning, computational game theory, stochastic approximation, evolutionary computation, Bayesian optimization, and causality. We provide a code example below.

```python
1 import Metastore, CrossEntropyMethod from csle
2 dt = Metastore.get_digital_twin(..)
3 pomdp = Metastore.get_pomdp(..)
4 response_strategy = CrossEntropyMethod.train(pomdp)
5 results = dt.evaluate(response_strategy)
```

***Listing 2:*** *Sample Python program that uses CSLE for optimizing and evaluating a security response strategy; line 2 gets metadata of the digital twin from the metastore; line 3 extracts the parameters of a POMDP (see the background chapter for the definition of a POMDP); line 4 optimizes a response strategy for the POMDP using the cross-entropy method (Alg. 1, Moss, 2020); and line 5 evaluates the optimized response strategy on the digital twin.*

## ■ Related Platforms

Recently, major IT vendors, defense organizations, and academic institutions have started efforts to build similar platforms as ours (CSLE). They include CyberBattleSim by Microsoft (Blum, 2021), CYBORG by the Australian department of defense (Standen et al., 2021), NASIM by the University of Queensland (Schwartz et al., 2020), YAWNING TITAN by the UK defense science and technology laboratory (Andrew et al., 2022), CYGIL by Canada's department of defense (Li et al., 2021), NASIMEMU by the Czech Technical University in Prague (Janisch et al., 2023), ATMOS by the University of Waterloo (Akbari et al., 2020), GYM-FLIPIT by Northeastern University (Oakley and Oprea, 2019), GYM-IDSGAME by KTH Royal Institute of Technology (Hammar and Stadler, 2020), MAB-MALWARE by the University of California (Riverside) (Song et al., 2022), MALWARE-RL by the University of Virginia (Anderson et al., 2018), PENGYM by Japan's advanced institute of science and technology (Huynh Phuong Thanh et al., 2024), FARLAND by USA's national security agency (Molina-Markham et al., 2021), CYBERWHEEL by the Oak Ridge national laboratory (Oesch et al., 2024), CYBERSHIELD by the University of Malaga (Carrasco et al., 2024), and CYBORG++ by the Alan Turing Institute (Emerson et al., 2024). Like CSLE, all of them include capabilities for data-driven optimization of response strategies. The main differences are that CSLE is open-source and has been experimentally validated on several instances of the security response problem (as described in the next chapter of the thesis); see Table 7.

| Platform | Simulation | Emulation | Open-source | Library | Validated |
|---|---|---|---|---|---|
| CSLE **(ours)** | ✓ | ✓ | ✓ | ✓ | ✓ |
| CyberBattleSim | ✓ | ✗ | ✓ | ✓ | ✗ |
| CYBORG | ✓ | ✗ | ✓ | ✗ | ✗ |
| YAWNING TITAN | ✓ | ✗ | ✓ | ✗ | ✗ |
| NASIM | ✓ | ✗ | ✓ | ✗ | ✗ |
| ATMOS | ✗ | ✓ | ✓ | ✗ | ✓ |
| GYM-FLIPIT | ✓ | ✗ | ✓ | ✗ | ✗ |
| GYM-IDSGAME | ✓ | ✗ | ✓ | ✗ | ✗ |
| MAB-MALWARE | ✗ | ✓ | ✓ | ✗ | ✓ |
| MALWARE-RL | ✗ | ✓ | ✓ | ✗ | ✓ |
| PENGYM | ✓ | ✓ | ✓ | ✗ | ✗ |
| CYGIL | ✗ | ✓ | ✗ | ✗ | ✗ |
| NASIMEMU | ✓ | ✓ | ✓ | ✗ | ✗ |
| FARLAND | ✓ | ✓ | ✗ | ✗ | ✓ |
| CYBERWHEEL | ✓ | ✓ | ✓ | ✗ | ✓ |
| CYBERSHIELD | ✓ | ✗ | ✗ | ✗ | ✗ |
| CYBORG++ | ✓ | ✗ | ✓ | ✗ | ✗ |

***Table 7:*** *Comparison between platforms for data-driven security response based on key features: support for simulation-based optimization; support for emulation-based evaluation; open source code; whether the platform provides a library with algorithms to facilitate strategy optimization and system identification; and whether the platform has been experimentally validated on different instances of the security response problem.*

## ■ Summary

This chapter describes our methodology for optimal security response. The methodology encompasses system identification, strategy optimization, and evaluation on a digital twin, effectively addressing Challenge 1 and Challenge 3 from the previous chapter. The methodology is general in the sense that it is not limited to a specific response scenario, optimization technique, or identification method. Additionally, we present CSLE, an experimental platform that implements our methodology. In the following chapters, we leverage this platform to apply the methodology to six different instances of the security response problem, through which we demonstrate how to overcome Challenge 2 by deriving structural properties of optimal response strategies and applying stochastic approximation techniques.

## ■ Acknowledgments

The following people have contributed to the development of CSLE: Forough Shahab Samani, Antonio Frederico Nesti Lopes, Jakob Stymne, Arvid Lagerqvist, Nils Forsgren, Bength Pappila, Yu Hu, Yan Wang, Aws Jaber, and Xiaoxuan Wang.

# CASE STUDIES

*No isolated experiment, however significant in itself, can suffice for the experimental demonstration of any phenomenon.*

— Ronald Fisher ***1935***, *The Design of Experiments.*

We pursue the research question posed in the introduction chapter by applying our methodology to six instances of the security response problem, each detailed in one of the included papers; see Fig. 25. We summarize the contents of each paper below, together with references to publications[30]. The full papers are presented in the subsequent chapters.

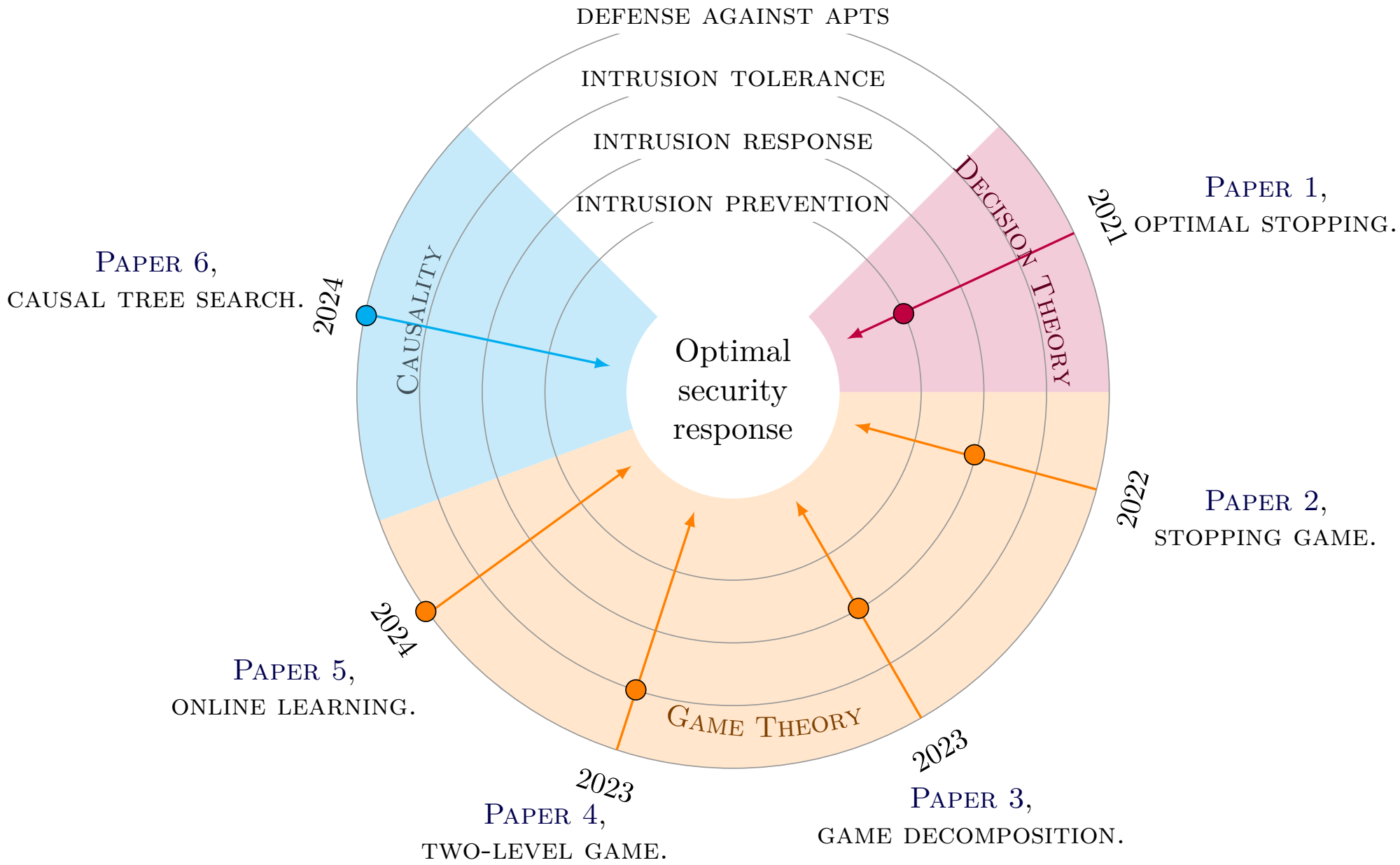

***Figure 25:*** *Each paper included in this thesis formalizes an aspect of the security response problem, presents analytical results, and evaluates the results through experimentation, all following the methodology from the previous chapter; Paper 1 studies intrusion prevention using optimal stopping; Paper 2 tackles intrusion response using a stopping game; Paper 3 addresses intrusion response through game decomposition; Paper 4 studies intrusion tolerance using a two-level game; Paper 5 focuses on online learning of response strategies against advanced persistent threats (APTs); and Paper 6 presents a causal tree search algorithm for defense against APTs.*

[30] All papers have been published (or accepted for publication) in refereed conference proceedings or journals except Paper 6, which is currently under review.

## Paper 1 – Intrusion Prevention through Optimal Stopping

The main question we answer in this paper answers is:

> At which points in time should a network operator take defensive actions given periodic but limited observational data from an IT infrastructure?

We propose a model based on optimal stopping theory to study this question; see Fig. 26. We prove that the optimal stopping times can be obtained through thresholds, which enables efficient computation of an optimal defender strategy. Based on this insight, we design T-SPSA, an efficient stochastic approximation algorithm for estimating the thresholds. We validate our methodology on a digital twin of an infrastructure with 31 servers. The validation results attest that our methodology outperforms state-of-the-art reinforcement learning and change detection methods.

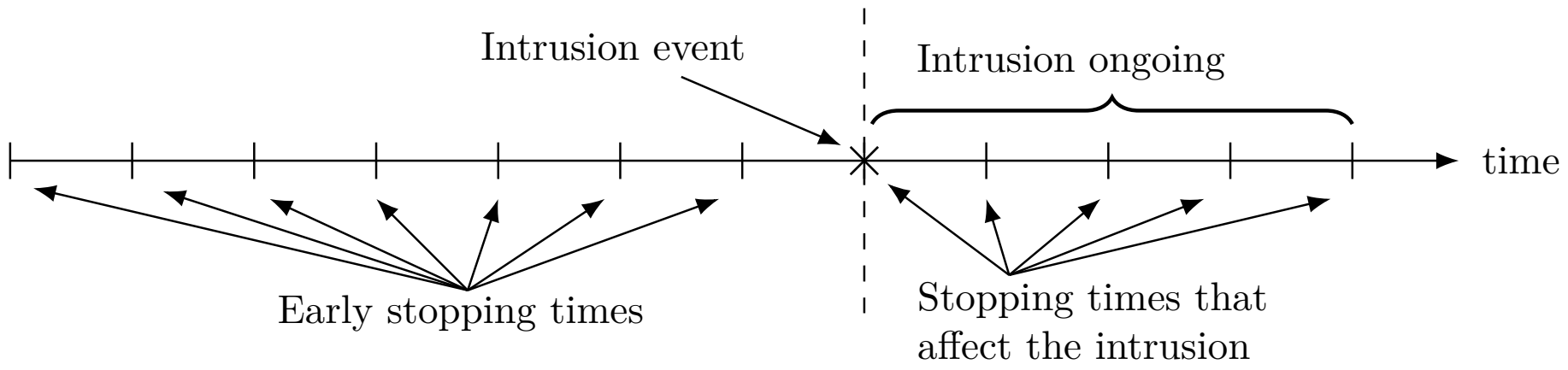

***Figure 26:*** *The optimal stopping formulation of intrusion prevention presented in Paper 1; the horizontal axis represents time; the dashed line shows the moment the intrusion starts, which is the optimal stopping time.*

The paper is published as

- K. Hammar and R. Stadler (2022), "Intrusion Prevention through Optimal Stopping [179]." *IEEE Transactions on Network and Service Management (TNSM)*, vol. 19, no. 3, pp. 2333-2348.

Related publications are

- K. Hammar and R. Stadler (2021), "Learning Intrusion Prevention Policies through Optimal Stopping [182].". *International Conference on Network and Service Management (CNSM)*, Izmir, Turkey, 2021, pp. 509-517.
- K. Hammar and R. Stadler (2022), "A System for Interactive Examination of Learned Security Policies [175]." *IEEE Network Operations and Management Symposium (NOMS)*, Budapest, Hungary, 2022, pp. 1-3.

## Paper 2 – Learning Near-Optimal Intrusion Responses Against Dynamic Attackers

This paper reports on a continuation of the work in Paper 1. It extends the optimal stopping problem to a game-theoretic formulation, which enables us to find optimal stopping strategies against a *dynamic* attacker, i.e., an attacker that adapts its strategy based on the defender strategy. In this stopping game, each player faces an optimal stopping problem. The problem for the defender is to decide when to take defensive actions and the problem for the attacker is to decide when to begin and end the intrusion. We prove the existence of equilibria and that the best responses have threshold properties. Leveraging these properties, we develop T-FP, a fictitious play algorithm that estimates equilibria through stochastic approximation; see Fig. 27. We show that T-FP outperforms a state-of-the-art fictitious play algorithm and that the obtained strategies are effective on a digital twin.

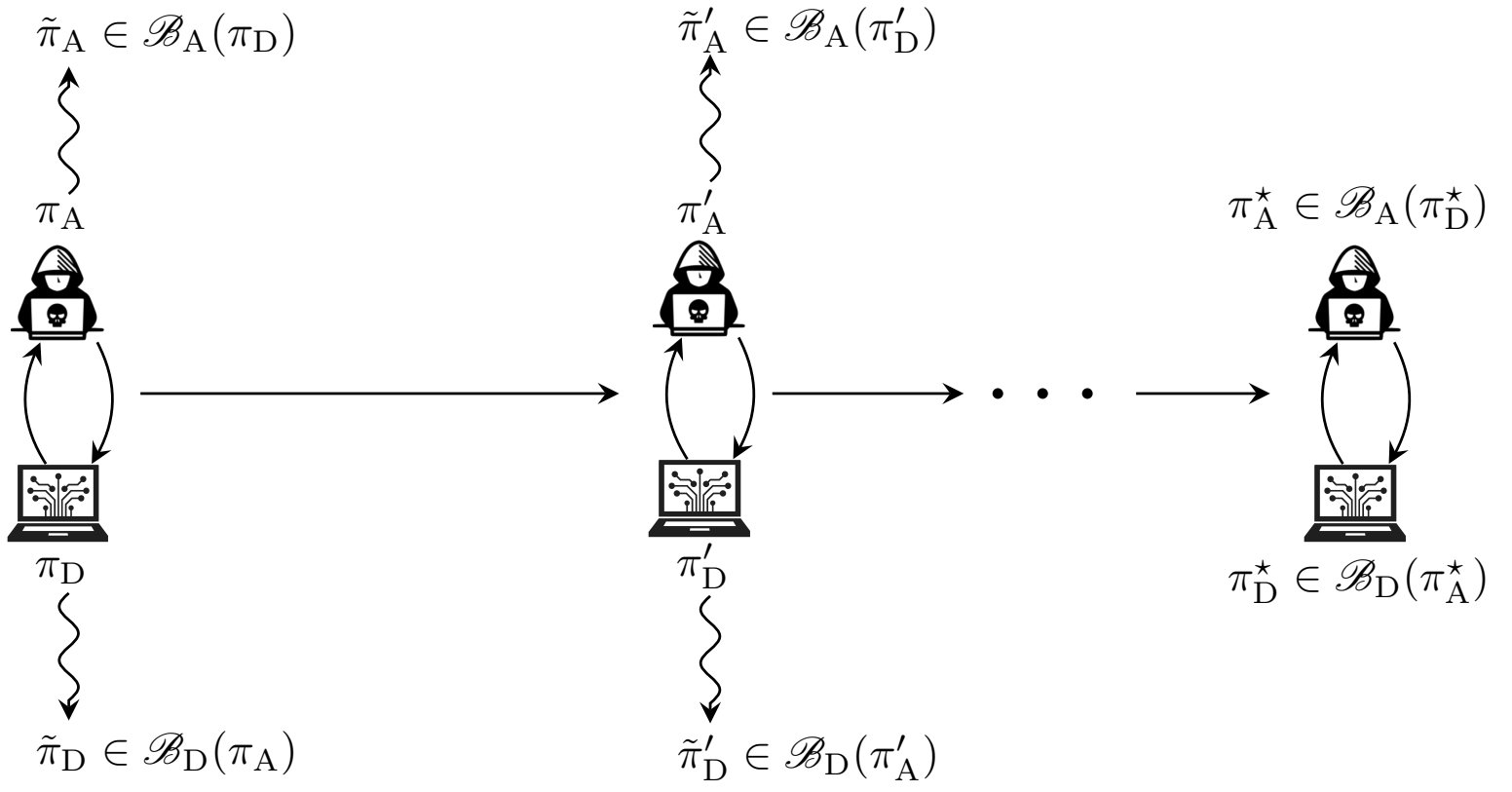

***Figure 27:*** *The fictitious play process between the attacker and the defender in Paper 2; horizontal arrows indicate iterations of fictitious play and vertical arrows indicate the learning of best responses; the process converges to an equilibrium* $(\pi^\star_{\mathrm{D}}, \pi^\star_{\mathrm{A}})$.

The paper is published as

- K. Hammar and R. Stadler (2024), "Learning Near-Optimal Intrusion Responses Against Dynamic Attackers [183]." *IEEE Transactions on Network and Service Management (TNSM)*, vol. 21, no. 1, pp. 1158-1177.

Related publications are

- K. Hammar and R. Stadler (2022), "Learning Security Strategies through Game Play and Optimal Stopping [184].". *Machine Learning for Cyber Security Workshop, International Conference on Machine Learning (ICML)*, Baltimore, USA. pp. 1-9.
- K. Hammar and R. Stadler (2022), "An Online Framework for Adapting Security Policies in Dynamic IT Environments [176]." *International Conference on Network and Service Management (CNSM)*, Thessaloniki, Greece, pp. 359-363.

### Paper 3 – Scalable Learning of Intrusion Response through Recursive Decomposition

In contrast to Paper 1 and Paper 2, this paper considers not only the problem of *when* defensive actions need to be taken but also the selection of *which* action to execute. We formulate this problem as a stochastic game where the attacker and the defender can execute different actions on components of the infrastructure. This detailed modeling means the game's complexity grows exponentially with the infrastructure's size due to the *curse of dimensionality* (Bellman, 1957); see Fig. 28. To manage this complexity, we recursively decompose the game into simpler subgames. We prove that this decomposition is optimal and that the best responses exhibit threshold structures. Building on this theoretical understanding, we develop DFP – an efficient algorithm for approximating equilibria. We validate our approach on a digital twin of an infrastructure with 64 servers. The results show that DFP outperforms a state-of-the-art reinforcement learning algorithm.

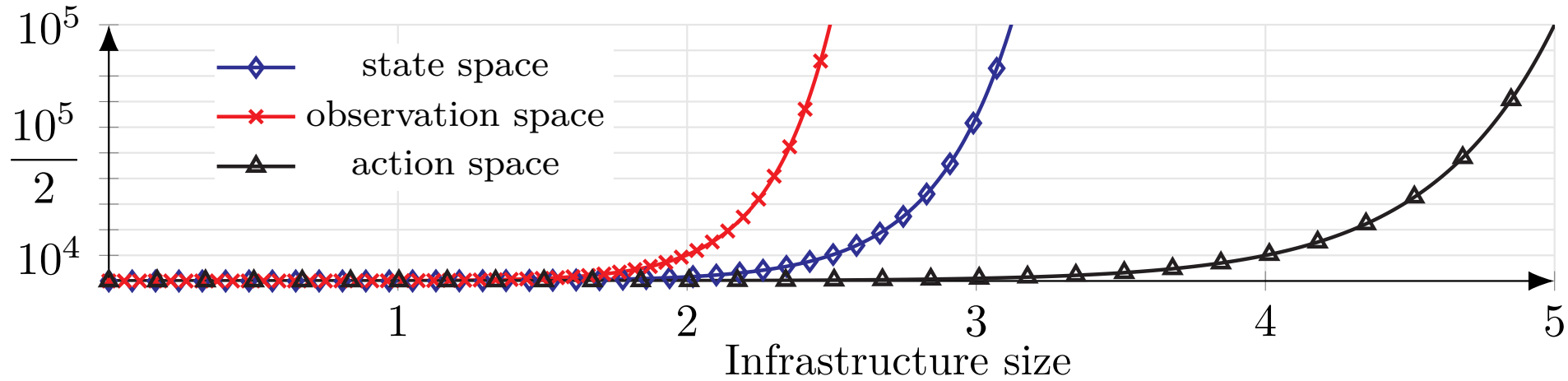

***Figure 28:*** *Size of the security response game considered in Paper 3 in function of the infrastructure size; the exponential growth exemplifies the curse of dimensionality: more state variables result in a combinatorial explosion of possible states (Bellman, 1957).*

---

The paper is published as

- K. Hammar and R. Stadler (2023), "Scalable Learning of Intrusion Response through Recursive Decomposition [185].". Springer Lecture Notes in Computer Science, vol 14167, pp. 172–192. *Decision and Game Theory for Security (GAMESEC)*, Avignon, France.

Related publications are

- K. Hammar and R. Stadler (2020), "Finding Effective Security Strategies through Reinforcement Learning and Self-Play [178].". *International Conference on Network and Service Management (CNSM)*, Izmir, Turkey, 2020, pp. 1-9.
- K. Hammar and R. Stadler (2023), "Digital Twins for Security Automation [177].". *IEEE Network Operations and Management Symposium (NOMS)*, Miami, USA, pp. 1-6.

## Paper 4 – Intrusion Tolerance for Networked Systems through Two-Level Feedback Control

Expanding on Papers 1–3, this paper demonstrates the generality of our methodology by applying it to *intrusion tolerance*. The main question we answer is:

> What are optimal recovery and replication strategies for a distributed system to maximize service availability and minimize operational cost in the presence of network intrusions?

We present a two-level game to study this question: a local game models intrusion recovery and a global game models replication control; see Fig. 29. For both games, we prove the existence of equilibria and show that the best responses have a threshold structure, which enables efficient computation of strategies. We argue that state-of-the-art intrusion-tolerant systems can be understood as instantiations of our game with heuristic control strategies. Our analysis shows the conditions under which such heuristics can be significantly improved through game-theoretic reasoning. Such reasoning allows us to derive optimal (equilibrial) control strategies and to evaluate them on a digital twin. The evaluation results demonstrate that our game-theoretic strategies can significantly improve service availability and reduce operational cost of state-of-the-art intrusion-tolerant systems. In addition, our game strategies can ensure any chosen level of service availability and time-to-recovery, bridging the gap between theoretical and operational performance.

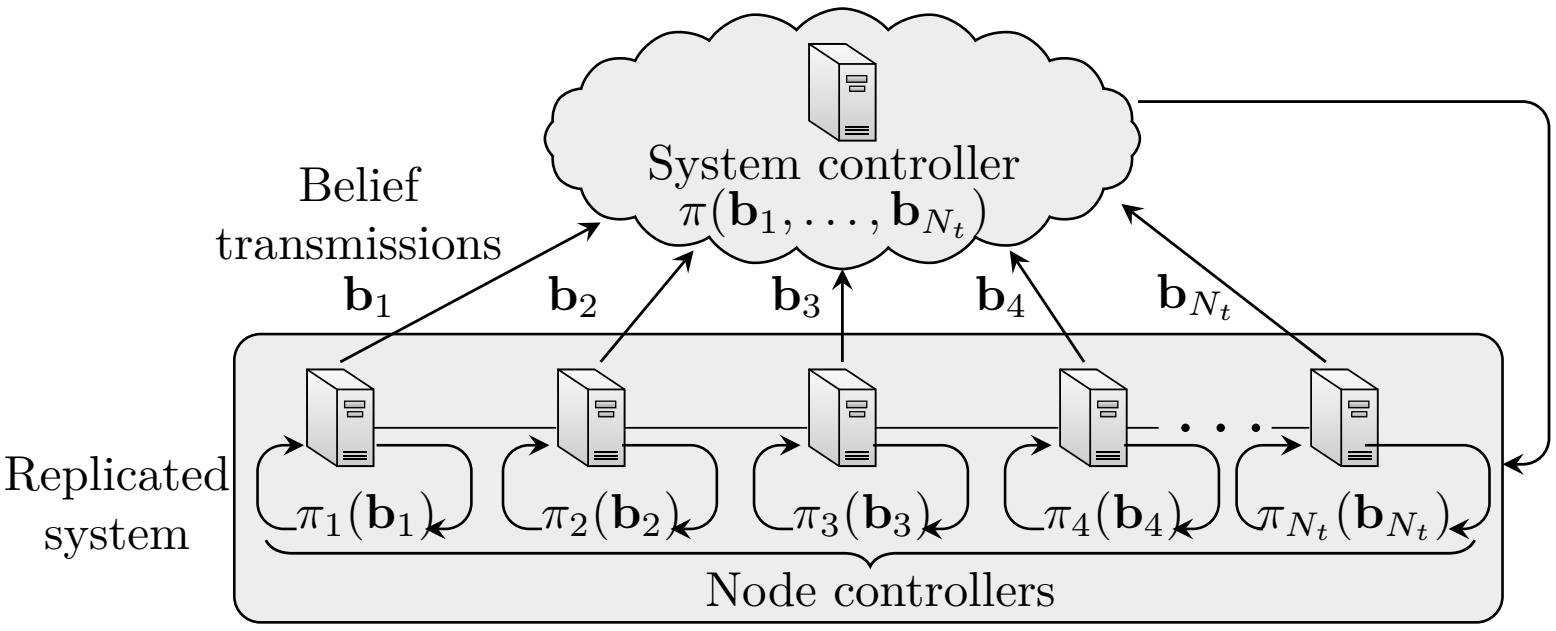

***Figure 29:*** *The intrusion-tolerant control architecture presented in Paper 4; node controllers with strategies $\pi_1, \ldots, \pi_{N_t}$ make local recovery decisions; a global system controller with strategy $\pi$ manages the replication factor $N_t$.*

The paper is published as

- K. Hammar and R. Stadler (2024), "Intrusion Tolerance for Networked Systems through Two-Level Feedback Control [181]." *IEEE Dependable Systems and Networks Conference (DSN)*, Brisbane, Australia, pp. 338-352.

A related publication is

- K. Hammar and R. Stadler (2024), "Intrusion Tolerance as a Two-Level Game [180].". Springer Lecture Notes in Computer Science, vol 14908, pp. 3-23. *Decision and Game Theory for Security (GameSec)*, New York, USA.

## Paper 5 – Automated Security Response through Online Learning with Adaptive Conjectures

This paper addresses a limitation of Papers 1–4, in which we assume that a perfect model of the underlying IT infrastructure can be obtained. To relax this assumption, we introduce **C**onjectural **O**nline **L**earning (COL), a security response algorithm which accounts for *model misspecification*. In COL, both the attacker and the defender have probabilistic *conjectures* about the model, which may be misspecified. These conjectures are iteratively adapted via Bayesian learning and used to update the strategies through rollout; see Fig. 30. We prove that the conjectures converge to best fits, and we provide a bound on the performance improvement that rollout enables with a conjectured model. To characterize the steady state when both the attacker and the defender run COL, we define a variant of the Berk-Nash equilibrium. We validate COL on a digital twin of an infrastructure with 64 servers. The evaluation results show that COL adapts to a changing environment, enables faster convergence than reinforcement learning, and outperforms the SNORT IDPS (Roesch, 1999) in several key metrics.

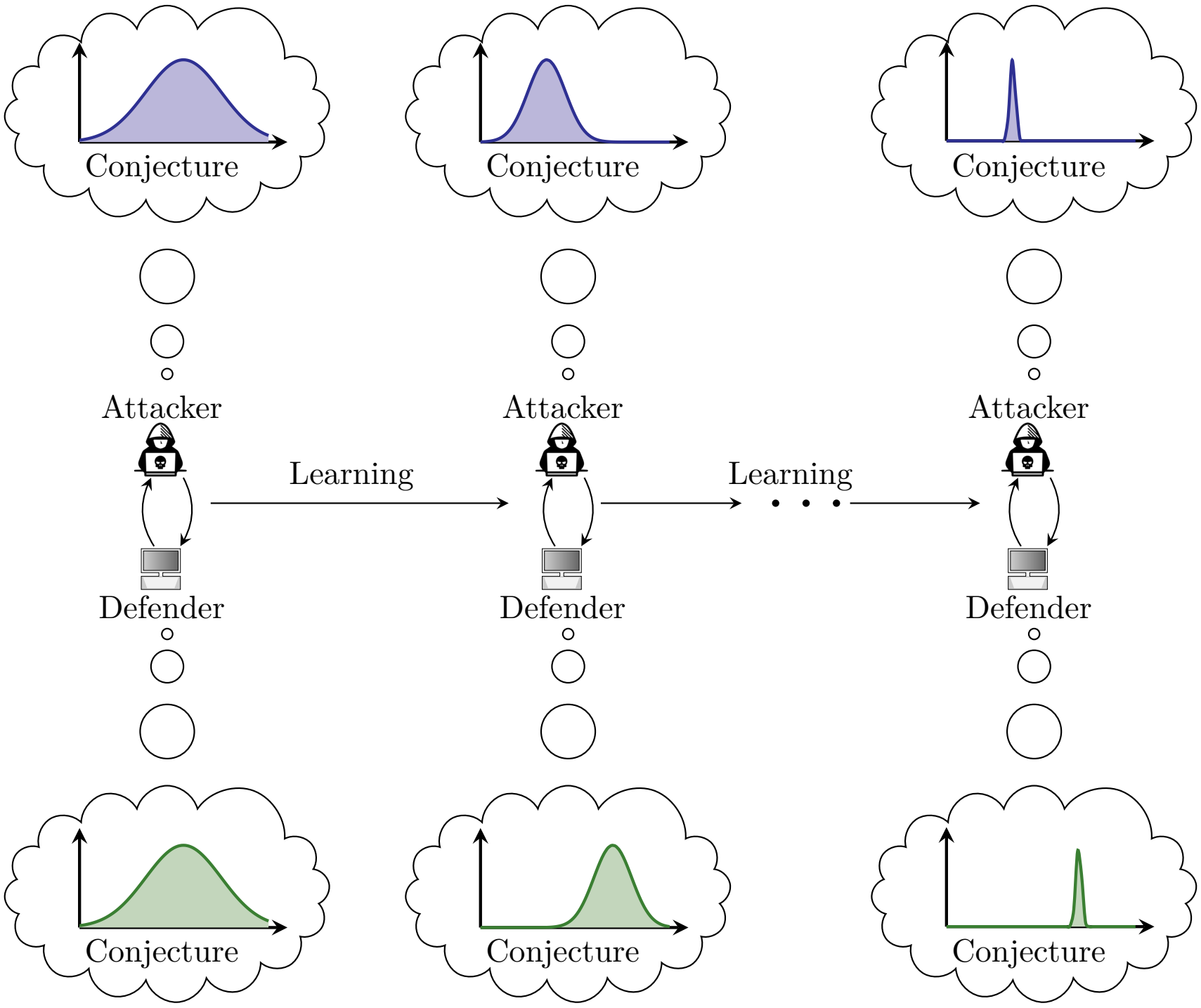

***Figure 30:*** ***C**onjectural **O**nline **L**earning (COL): Paper 5 formulates the interaction between an attacker and a defender as a game where each player has a probabilistic conjecture about the game model, which may be misspecified in the sense that the true model has probability* 0*; the conjectures are iteratively adapted through Bayesian learning.*

---

The paper is published as

- K. Hammar, T. Li, R. Stadler, and Q. Zhu (2024), "Automated Security Response through Online Learning with Adaptive Conjectures [174].". *To appear in IEEE Transactions on Information Forensics and Security (TIFS).*

Related publications are

- F.S Samani, K. Hammar, and R. Stadler (2024), "Online Policy Adaptation for Networked Systems using Rollout [387]". *IEEE Network Operations and Management Symposium (NOMS)*, Seoul, South Korea, pp. 1-9.

- T. Li, K. Hammar, R. Stadler, and Q. Zhu (2024), "Conjectural Online Learning with First-order Beliefs in Asymmetric Information Stochastic Games [277]". To appear in the proceedings of *IEEE Conference on Decision and Control (CDC)*, Milan, Italy.

## Paper 6 – Optimal Defender Strategies for CAGE-2 using Causal Modeling and Tree Search

This paper differs from the first five papers by using a different environment and scenario for evaluation, namely CAGE-2, which involves defending a networked system against an advanced persistent threat (CAGE-2, 2022). This departure from the evaluation on a digital twin facilitates direct comparisons with the existing literature. Our main contribution is a causal model of the CAGE-2 scenario, based on which we prove the existence of optimal defender strategies and design an iterative method that converges to such a strategy. The method, called C-POMCP, leverages the causal structure to prune, construct, and traverse a search tree; see Fig. 31. We show that C-POMCP achieves better performance and is two orders of magnitude more efficient than state-of-the-art methods.

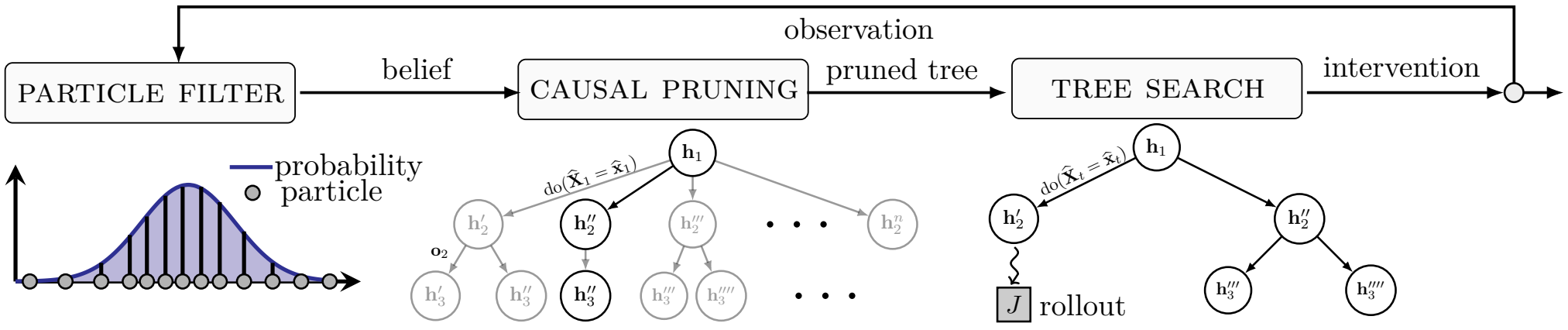

***Figure 31:*** *Paper 6 presents **C**ausal-**P**artially **O**bserved **M**onte-**C**arlo **P**lanning (C-POMCP), which is an online method for optimal security response; the figure illustrates one time step of C-POMCP during which (i) a particle filter is used to compute a belief state; (ii) a causal graph is used to prune a search tree; and (iii) tree search is used to find an optimal response intervention.*

The paper is submitted for publication as

- K. Hammar, N. Dhir, and R. Stadler (2024), "Optimal Defender Strategies for CAGE-2 using Causal Modeling and Tree Search [173]". *IEEE Transactions on Dependable and Secure Computing (TDSC).*

# Paper 1†

# INTRUSION PREVENTION THROUGH OPTIMAL STOPPING

Kim Hammar and Rolf Stadler

**Abstract**

We study automated intrusion prevention using stochastic approximation. Following a novel approach, we formulate the problem of intrusion prevention as an (optimal) multiple-stopping problem. This formulation gives us insight into the structure of optimal strategies, which we show to have threshold properties. For most practical cases, obtaining an optimal defender strategy using dynamic programming is not feasible. We therefore develop a stochastic approximation approach to estimate an optimal threshold strategy. We introduce T-SPSA, an efficient algorithm that learns threshold strategies through stochastic approximation. We show that T-SPSA outperforms state-of-the-art algorithms for our use case. Our methodology for learning and validating strategies includes a simulator where defender strategies are incrementally learned and a digital twin where statistics are produced that drive simulation runs and where learned strategies are evaluated. We show that this methodology can produce effective defender strategies for a practical IT infrastructure.

† The paper is published as

- K. Hammar and R. Stadler (2022), "Intrusion Prevention through Optimal Stopping [179]." *IEEE Transactions on Network and Service Management (TNSM)*, vol. 19, no. 3, pp. 2333-2348.

Related publications are

- K. Hammar and R. Stadler (2021), "Learning Intrusion Prevention Policies through Optimal Stopping [182].". *International Conference on Network and Service Management (CNSM)*, Izmir, Turkey, 2021, pp. 509-517.
- K. Hammar and R. Stadler (2022), "A System for Interactive Examination of Learned Security Policies [175]." *IEEE Network Operations and Management Symposium (NOMS)*, Budapest, Hungary, 2022, pp. 1-3.

*To stop or not to stop, that is the question.*

— William Shakespeare ***1601**, Hamlet (stop=be).*

## 1.1 Introduction

An organization's security strategy has traditionally been defined, implemented, and updated by domain experts (Fuchsberger, 2005). Although this approach can provide basic cybersecurity for an organization's communication and computing infrastructure, a growing concern is that infrastructure update cycles become shorter and attacks increase in sophistication (Zouave et al., 2020). Consequently, the security requirements become increasingly difficult to meet. This paper presents a novel approach to address this challenge by learning defender strategies automatically. We apply this approach to an *intrusion prevention* use case. Here, we use the term "intrusion prevention" as suggested in the literature, e.g., in (Fuchsberger, 2005). It means that a defender prevents an attacker from reaching its goal rather than preventing it from accessing any part of the infrastructure. The main question we answer is:

> At which points in time should a network operator take defensive actions given periodic but limited observational data from an IT infrastructure?

We study this question within the framework of discrete-time dynamical systems and formulate it as an *(optimal) multiple-stopping problem*; see Fig. 1.1. In this formulation, the defender can take a finite number of *stops*. Each stop is associated with a defensive action, and the objective is to decide the optimal time for stopping. This approach gives us insight into the structure of optimal defender strategies through the theory of optimal stopping (Wald, 1947). We prove that an optimal *multi-threshold strategy* exists that can be efficiently computed and implemented. Based on this insight, we design T-SPSA, an efficient stochastic approximation algorithm for estimating the thresholds. We evaluate our approach on a digital twin[2] of an infrastructure with 31 servers. The evaluation results attest that our approach outperforms state-of-the-art methods for our use case.

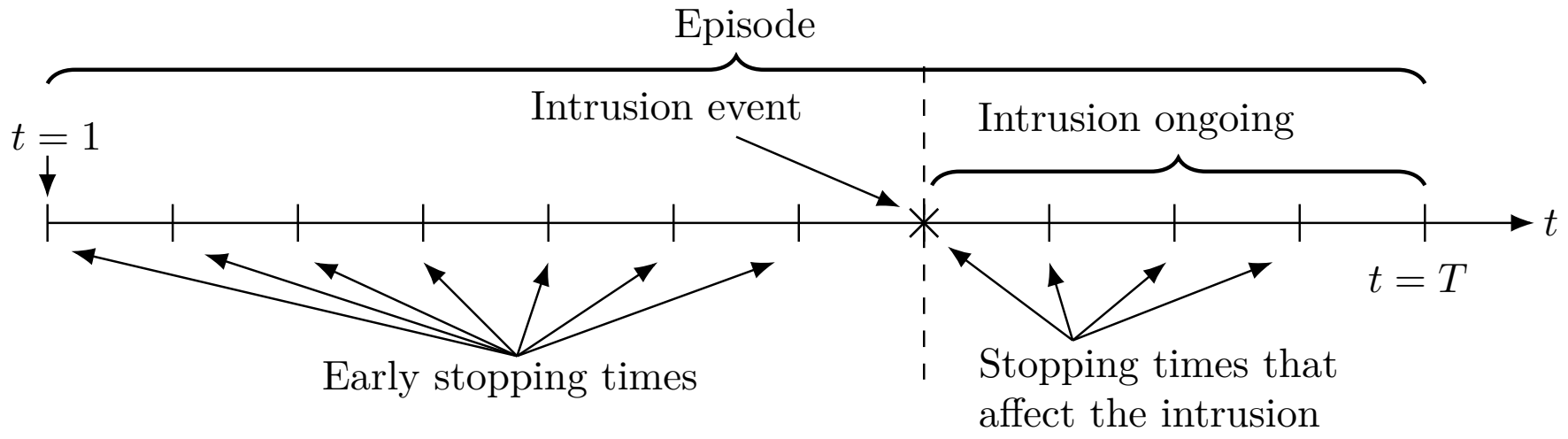

***Figure 1.1:*** *Optimal multiple-stopping formulation of intrusion prevention; the horizontal axis represents time; $T$ is the time horizon; the dashed line shows the intrusion start time; the optimal strategy is to prevent the attacker at the time of intrusion.*

[2]The digital twin is created using CSLE, as described in the methodology chapter.

## 1.2 The Intrusion Prevention Use Case

We consider an intrusion prevention use case that involves the IT infrastructure of an organization. The operator of this infrastructure, which we call the *defender*, takes measures to protect it against an *attacker* while providing services to a client population; see Fig. 1.2[3]. The infrastructure includes a set of servers that run the services and an Intrusion Detection System (IDS) that logs events in real-time. Clients access the services through a public gateway, which is also open to the attacker. We assume that the attacker intrudes into the infrastructure through the gateway, performs reconnaissance, and exploits vulnerabilities to compromise servers. We model the attacker as an agent that starts the intrusion at a random time and then takes a predefined sequence of actions, i.e., we consider an attacker with a *static* strategy.

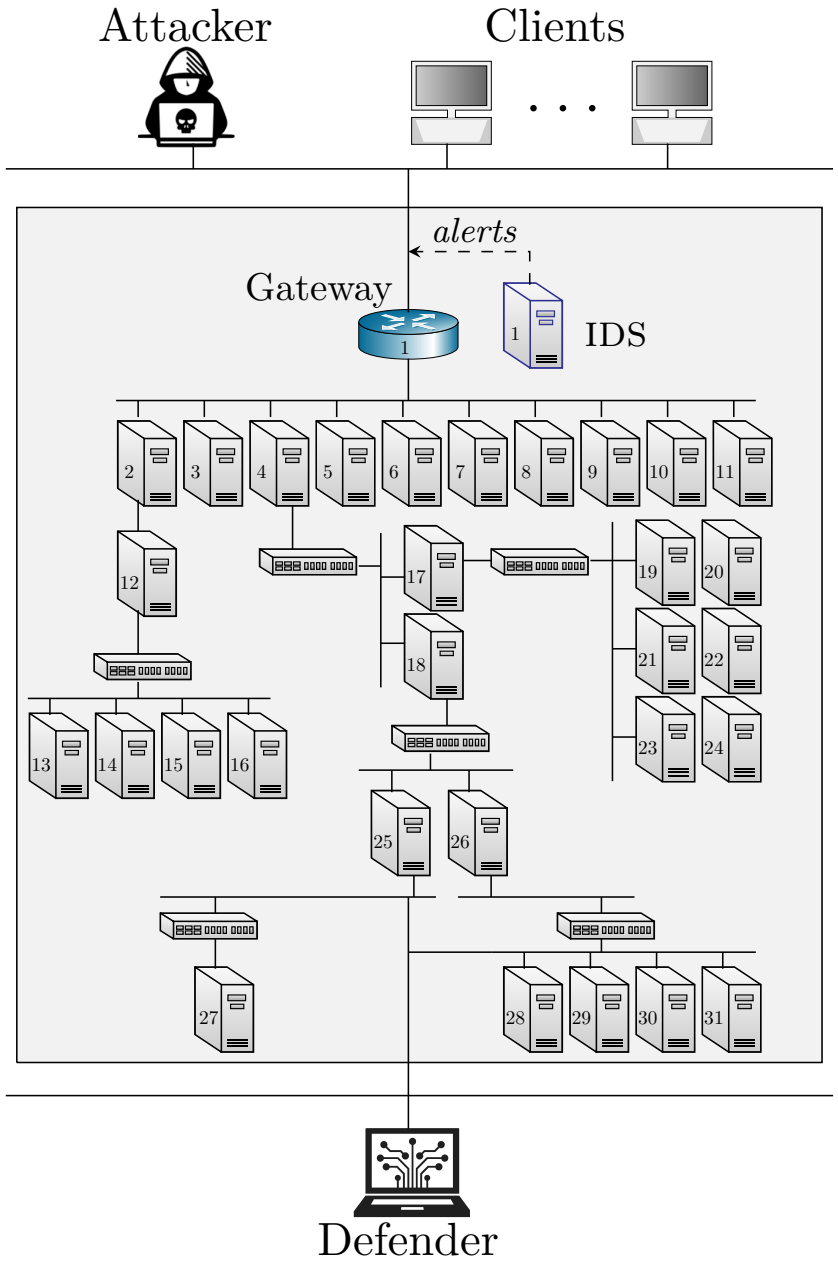

***Figure 1.2:*** *The IT infrastructure and the actors in the intrusion prevention use case.*

The defender monitors the infrastructure by accessing and analyzing IDS statistics and login attempts at the servers. It can take a fixed number of defensive actions to prevent the attacker. A defensive action is, for example, to revoke user certificates in the infrastructure to recover compromised accounts. We assume that the defender takes the defensive actions in a predetermined order. The final action the defender can take is blocking all external access to the gateway. As a consequence of this action, the service and any ongoing intrusion are disrupted. In deciding when to take action, the defender has two objectives: maintain service to its clients and keep a possible attacker out of the infrastructure. The optimal strategy is to monitor the infrastructure and maintain service until the attacker enters through the gateway. At this time, the attacker must be prevented by taking defensive actions. The defender's challenge is identifying when this moment occurs, which corresponds to an optimal stopping problem.

[3]The infrastructure configuration is listed in Appendix C.

## 1.3 The Markovian Optimal Stopping Problem

Optimal stopping is a classical problem domain with a well-developed theory [482, 408, 350, 94, 50, 377, 43, 358, 356]. Example use cases for this theory include: asset selling (Bertsekas, 2005), change detection (Tartakovsky et al., 2006), machine replacement (Krishnamurthy, 2016), hypothesis testing (Wald, 1947), gambling (Chow et al., 1971), selling decisions (Toit and Peskir, 2009), queue management (Roy et al., 2019), industrial control (Rabi and Johansson, 2008), and the secretary problem (Kleinberg, 2005). Many variants of the optimal stopping problem have been studied. For instance, discrete-time and continuous-time problems, finite horizon and infinite horizon problems, problems with fully observed and partially observed state spaces, problems with finite and infinite state spaces, Markovian and non-Markovian problems, and single-stop and multi-stop problems. Consequently, different solution approaches have been developed. The most common are the *martingale* (Snell, 1952) and the *Markovian* (Bather, 2000) approaches.

This paper investigates the multiple-stopping problem with $L$ stops, a finite time horizon $T$, discrete-time progression, a finite state space, and the Markov property. We use the Markovian solution approach and model the problem as a Partially Observed Markov Decision Process (POMDP), where the state evolves as a discrete-time Markov process $(S_t)_{t=1}^{T}$ that is partially observed. Two actions are available at each time step: $(\mathsf{S})$top and $(\mathsf{C})$ontinue. Action $a$ in state $s$ yields a reward $r(s, a)$. If $a = \mathsf{S}$ and only one of the $L$ stops remains, the decision process terminates. Otherwise, the process transitions to the next state.

The *stopping time* with $l$ stops remaining is a random variable

$$\mathcal{T}_l \triangleq \inf\{t \mid t > \mathcal{T}_{l+1}, a_t = \mathsf{S}\}, \text{ where } l \in \{1, .., L\} \text{ and } \mathcal{T}_{L+1} \triangleq 0.$$

This variable is measurable with respect to the filtration $\mathcal{F}_t = \sigma(\mathbf{H}_k \mid k \leq t)$, where $\mathbf{H}_t$ is the POMDP history (16)[4] and $\{\mathcal{T}_l = t\} \in \mathcal{F}_t$ for all $t$ (Peskir and Shiryaev, 2006)[5]. The objective is to find a stopping strategy $\pi_l^\star$ that maximizes

$$\sup_{\pi_l \in \Pi} \mathbb{E}_{\pi_l}\left[\sum_{t=1}^{\mathcal{T}_L - 1} r(S_t, \mathsf{C}) + \underbrace{r(S_{\mathcal{T}_L}, \mathsf{S})}_{\text{First stop.}} + \ldots + \sum_{t=\mathcal{T}_2+1}^{\mathcal{T}_1 - 1} r(S_t, \mathsf{C}) + \underbrace{r(S_{\mathcal{T}_1}, \mathsf{S})}_{\text{Final stop.}}\right].$$

[4]See the background chapter for the definitions of the history $\mathbf{H}_t$ and the POMDP.

[5]Note that we allow $\mathcal{T}_l$ to be infinite; stopping times that are not almost surely finite are sometimes called Markov times (Poor and Hadjiliadis, 2008).

## 1.4 Formalizing the Intrusion Prevention Use Case

We formulate the intrusion prevention use case as an instance of the multiple-stopping problem described above, where each stop is associated with a defensive action. We model this problem as a POMDP; see (15) in the background chapter for the definition of a POMDP. (A game-theoretic model is not needed since the attacker is static and follows a fixed strategy.) The requisite notation is given in Table 1.1 and the POMDP components $\langle \mathcal{S}, \mathcal{A}, f, r, \gamma, \mathbf{b}_1, T, \mathcal{O}, z \rangle$ are defined below.

| *Notation(s)* | *Description* |
|---|---|
| $L, l$ | Number of stop actions and stops remaining. |
| $\mathsf{S} = 1, \mathsf{C} = 0$ | Stop and continue actions. |
| $\mathscr{S}_l, \mathscr{C}_l$ | Stop and continue sets with $l$ stops remaining. |
| $\alpha_l^\star$ | Optimal stopping threshold with $l$ stops remaining. |
| $\pi_l, J(\pi_l), \Pi_l$ | Defender strategy, objective functional (1.4), and strategy space. |
| $\pi_{\boldsymbol{\theta},l}$ | Parameterized defender strategy. |
| $\mathcal{S}, \mathcal{A}, \mathcal{O}$ | Sets of states, actions, and observations. |
| $s_t, \mathbf{o}_t, r_t$ | State, observation, and reward at time $t$. |
| $S_t, \mathbf{O}_t, R_t$ | Random variables (vectors) with realizations $s_t, \mathbf{o}_t, r_t$. |
| $\Delta x_t, \Delta y_t, \Delta z_t$ | Severe/warning IDS alerts and login attempts during time step $t$. |
| $I, p$ | Intrusion start time, probability that an intrusion starts. |
| $f_l, z, r$ | Transition (1.1), observation (1.2), and reward (1.3) functions. |
| $\mathbf{B}_t, \mathbf{b}_t, \mathcal{B}$ | Belief state at time $t$, its realization, and the belief space. |
| $T_\emptyset$ | Time horizon. |
| $\boldsymbol{\theta}^{(i)}, \boldsymbol{\Delta}^{(i)}$ | Threshold and perturbation vectors at iteration $i$ of T-SPSA (Alg. 1.1). |
| $\mathcal{T}_l, \tau_l$ | Stopping time with $l$ stops remaining and its realization. |
| $\mathcal{T}_l^\star, \tau_l^\star$ | Optimal stopping time with $l$ stops remaining and its realization. |

***Table 1.1:*** *Variables and symbols used in the model.*

***Actions*** $\mathcal{A}$ The defender has two actions: (S)top and (C)ontinue. The action space is thus $\mathcal{A} \triangleq \{\mathsf{S}, \mathsf{C}\}$. We encode $\mathsf{S}$ with 1 and $\mathsf{C}$ with 0 to simplify the formal description below. The number of stops the defender must execute to prevent an intrusion is $L \geq 1$, which is a predefined parameter of our use case. The number of stop actions remaining is denoted by $l \in \{1, \dots, L\}$.

***States*** $\mathcal{S}$ The state $s_t$ is 0 if no intrusion occurs and 1 if an intrusion is ongoing. The terminal state $\emptyset$ is reached after the defender takes the final stop action. The state space is thus $\mathcal{S} \triangleq \{0, 1, \emptyset\}$. The initial state is $s_1 = 0$. Hence, $\mathbf{b}_1 \in \Delta(\mathcal{S})$ is the degenerate distribution $\mathbf{b}_1(0) = 1$.

***Observations*** $\mathcal{O}$ The defender has a partial view of the system and observes $\mathbf{o}_t \triangleq (\Delta x_t, \Delta y_t, \Delta z_t)$, where $\Delta x_t$, $\Delta y_t$, and $\Delta z_t$ are bounded counters that denote the number of severe IDS alerts, warning IDS alerts, and login attempts generated during time step $t$, respectively. Hence, the observation space is $\mathcal{O} \triangleq \{0, \dots, \Delta x_{\max}\} \times \{0, \dots, \Delta y_{\max}\} \times \{0, \dots, \Delta z_{\max}\}$.

***Belief space*** $\mathcal{B}$ Based on its history $\mathbf{h}_t$ (16)[6], the defender computes the *belief state* $\mathbf{b}_t(s_t) \triangleq \mathbb{P}[S_t = s_t \mid \mathbf{h}_t] \in \mathcal{B}$ through (17), as defined in the background chapter. Since $\emptyset$ is a terminal state, the only two reachable states during a POMDP episode are 0 and 1. Therefore, $\mathcal{B} = \Delta(\{0, 1\}) = [0, 1]$.

**Proposition 1.1** (Reachable beliefs)**.** *The number of reachable beliefs from the initial belief* $\mathbf{b}_1$ *in* $t$ *time steps is upper bounded by* $(|\mathcal{A}||\mathcal{O}|)^t$*.*

*Proof.* Since the denominator in (17) is a normalization constant (defined in the background chapter), it suffices to consider the numerator. The numerator depends on $\mathbf{o}_t$, which can take on $|\mathcal{O}|$ different values; and $a_{t-1}$, which can take on $|\mathcal{A}|$ different values. Therefore, the number of reachable beliefs from $\mathbf{b}_1$ in 1 step is upper bounded by $|\mathcal{A}||\mathcal{O}|$. Assume by induction that the number of reachable beliefs in $k - 1 > 1$ time steps is upper bounded by $(|\mathcal{A}||\mathcal{O}|)^{k-1}$. For each of those beliefs, we can reach a maximum of $|\mathcal{A}||\mathcal{O}|$ new beliefs at time $k$, which means that the total number of beliefs reachable in $k$ steps is upper bounded by $(|\mathcal{A}||\mathcal{O}|)^k$. □

***Transition function*** $f_l(s' \mid s, a)$ We model the start of an intrusion by a Bernoulli process $(Q_t)_{t=1}^T$, where $Q_t \sim \text{Ber}(p)$ is a Bernoulli random variable with $p > 0$. The first occurrence of $Q_t = 1$ defines the intrusion start time $I$, which thus is geometrically distributed; see Fig. 1.3.

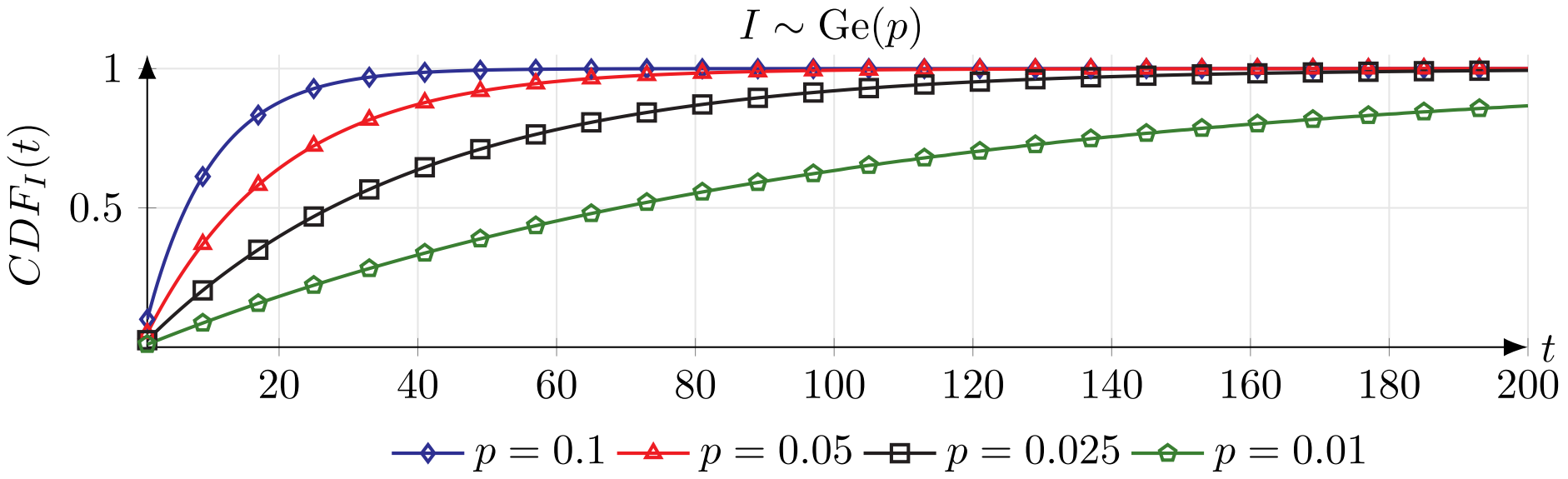

***Figure 1.3:*** *The Cumulative Distribution Function (CDF) of the intrusion start time I.*

Consequently, we can define the transition function as

$$f_1(\emptyset \mid \cdot, 1) \triangleq f_l(\emptyset \mid \emptyset, \cdot) \triangleq 1 \tag{1.1a}$$

$$f_l(0 \mid 0, a) \triangleq 1 - p \quad \text{if } l - a > 0 \tag{1.1b}$$

$$f_l(1 \mid 0, a) \triangleq p \quad \text{if } l - a > 0 \tag{1.1c}$$

$$f_l(1 \mid 1, a) \triangleq 1 \quad \text{if } l - a > 0, \tag{1.1d}$$

where $a \in \mathcal{A}$[7]. All other state transitions occur with probability 0. (1.1a) defines the transitions to the terminal state $\emptyset$, which is reached when the *final* stop action is taken (i.e., when $l = 1$ and $a = 1$). If (1.1a) is not applicable, i.e., if the system does not reach the terminal state, then the transitions are defined by (1.1b)-(1.1d). (1.1b) captures the case where no intrusion occurs; (1.1c) specifies the case when the intrusion starts; and (1.1d) describes the case where an intrusion is in progress. The state transition diagram is shown in Fig. 1.4. Note that the intrusion state $s = 1$ is absorbing until $L$ stop actions have been taken.

[6] The history is defined in the background chapter; see (16).

[7] Recall that we encode $(\mathsf{S}, \mathsf{C}) \triangleq (1, 0)$; hence $l_{t+1} = l_t - a_t$.

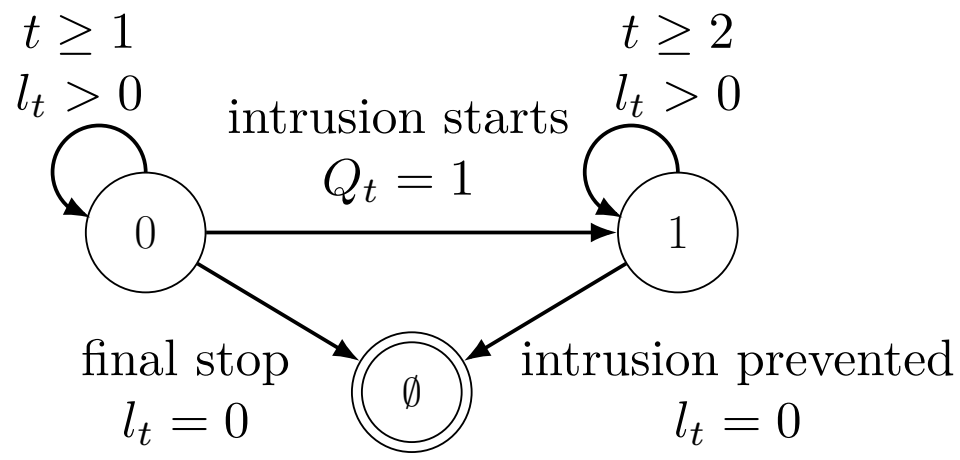

***Figure 1.4:*** *State transition diagram of the* POMDP*: each circle represents a state; an arrow represents a state transition; a label indicates the event that triggers the transition; an episode starts in state $s_1 = 0$ with $l_1 = L$.*

**Remark 1.1** (An intrusion will almost surely occur)**.**
The intrusion start time $I$ is almost surely finite since

$$\mathbb{P}[I < \infty] = 1 - \mathbb{P}[I = \infty] = 1 - \lim_{t \to \infty} (1-p)^t = 1.$$

***Observation function*** $z(\mathbf{o}_t \mid s_t)$ We assume that the number of IDS alerts and login attempts during a time step are discrete random variables $X, Y, Z$ that depend on the state. Consequently, the probability that $\Delta x_t$ severe alerts, $\Delta y_t$ warning alerts, and $\Delta z_t$ login attempts occur during time step $t$ can be expressed as

$$z(\Delta x_t, \Delta y_t, \Delta z_t \mid s_t) \triangleq z(\mathbf{o}_t \mid s_t) \triangleq \mathbb{P}[\mathbf{O}_t = \mathbf{o}_t \mid s_t], \tag{1.2}$$

where $\mathbf{o}_t \triangleq (\Delta x_t, \Delta y_t, \Delta z_t)$ realizes the random vector $\mathbf{O}_t \triangleq (X, Y, Z)$. Note that the distribution $z$ (1.2) depends on the attacker and the clients that interact with services of the infrastructure, i.e., they are implicitly modeled by $z$.

***Reward function*** $r(s, a)$ The objective of the intrusion prevention use case is to maintain service on the infrastructure while preventing a possible intrusion. Therefore, we define the reward function to give the maximal reward if the defender maintains service until the intrusion starts and then prevents the intrusion by taking $L$ stop actions. The reward per time step $r(s, a)$ is parameterized by the reward that the defender receives for stopping an intrusion ($R_{\text{st}} > 0$), the reward for maintaining service ($R_{\text{sla}} > 0$), and the loss of being intruded ($R_{\text{int}} < 0$):

$$r(\emptyset, \cdot) \triangleq 0 \tag{1.3a}$$

$$r(s,\mathsf{C}) \triangleq R_{\text{sla}} + \frac{sR_{\text{int}}}{L} \qquad s \in \{0,1\} \tag{1.3b}$$

$$r(s,\mathsf{S}) \triangleq \frac{sR_{\text{st}}}{L} \qquad s \in \{0,1\}. \tag{1.3c}$$

(1.3a) states that the reward in the terminal state is zero. (1.3b) states that the defender receives a positive reward ($\text{R}_{\text{sla}}$) for maintaining service and a loss ($\frac{R_{\text{int}}}{L}$) for each time step that it is under intrusion. Lastly, (1.3c) indicates that each stop incurs a cost by interrupting service (i.e., no $R_{\text{sla}}$) and possibly a reward ($\frac{R_{\text{st}}}{L}$) if it affects an ongoing intrusion. (The constants $R_{\text{st}}, R_{\text{sla}}$, and $R_{\text{int}}$ should be configured to satisfy Assumption 2 in the background chapter.)

**Assumption 1.1** (Intrusion cost exceeds service utility)**.** $\left(\frac{R_{\text{int}}}{L} + R_{\text{sla}}\right) < 0$.

***Time horizon*** $T_\emptyset$ The time horizon $T_\emptyset$ is a random variable that indicates the time $t > 1$ when the terminal state $\emptyset$ is reached. It follows from (1.1) that $\mathbb{E}_{\pi_l}[T_\emptyset] < \infty$ for any strategy $\pi_l$ that uses $L$ stops as $t \to \infty$.

***Strategy space*** $\Pi_l$ As the POMDP is stationary and the time horizon $T_\emptyset$ is not predetermined, it suffices to consider stationary deterministic strategies; see Thm. 2 in the background chapter. Despite this sufficiency, we consider the space of stochastic strategies $\pi_l \in \Pi_l \triangleq \{1,\ldots,L\} \times \mathcal{B} \to \Delta(\mathcal{A})$ to enable smooth optimization.

**Proposition 1.2.** *$\mathcal{T}_l = \inf_t\{t \mid a_t = \mathsf{S}, a_t \sim \pi_l(\mathbf{b}_t), l_t = l\}$ is a stopping time.*

*Proof.* Let $\mathcal{F}_t$ be the filtration generated by the history sequence $(\mathbf{H}_k)_{k=1}^t$ (16). Since $\pi_l(\mathbf{b}_t) = \pi_l(\mathbb{B}(\mathbf{h}_t))$ (17), $\{\mathcal{T}_l = t\} \in \mathcal{F}_t$ for all $t$. □

***Objective*** With some abuse of notation, we use $J(\pi)$ to denote the value of a strategy $\pi$ (4), while also using $J^\pi(\mathbf{b})$ to refer to the value of a belief state $\mathbf{b}$ under strategy $\pi$, as defined in the background chapter. An *optimal* strategy $\pi_l^\star \in \Pi_l$ maximizes the expected cumulative reward over the time horizon $T_\emptyset$, i.e.,

$$\pi_l^\star \in \operatorname*{arg\,max}_{\pi_l \in \Pi_l} J(\pi_l), \qquad \text{where} \qquad J(\pi_l) \triangleq \mathbb{E}_{\pi_l}\left[\sum_{t=1}^{T_\emptyset} r(S_t, A_t) \mid \mathbf{b}_1\right]. \tag{1.4}$$

**Proposition 1.3.** *$J(\pi_l^\star)$ (1.4) is finite for any optimal strategy $\pi_l^\star$.*

*Proof.* The objective in (1.4) is undiscounted, which means that it is unbounded for non-finite stopping times. Hence, we must show that the optimal stopping times $\mathcal{T}_1^\star, \ldots, \mathcal{T}_L^\star$ are almost surely finite. Note that

$$\begin{aligned}\mathbb{E}\left[\mathbf{B}_{t+1}(1) \mid \mathcal{F}_t\right] &= \mathbb{E}\left[\mathbb{P}[I \leq t+1 \mid \mathcal{F}_{t+1}] \mid \mathcal{F}_t\right] \overset{(a)}{=} \mathbb{E}\left[\mathbb{E}[\mathbb{1}_{I \leq t+1} \mid \mathcal{F}_{t+1}] \mid \mathcal{F}_t\right] \\ &= \mathbb{E}\left[\mathbb{1}_{I \leq t+1} \mid \mathcal{F}_t\right] \qquad \text{(Law of iterated expectations)}\end{aligned}$$

$$= \mathbb{P}\left[I \leq t+1 \mid \mathcal{F}_t\right] = \mathbb{P}\left[I \leq t \mid \mathcal{F}_t\right] + \mathbb{P}\left[I = t+1 \mid \mathcal{F}_t\right]$$
$$\geq \mathbb{P}\left[I \leq t \mid \mathcal{F}_t\right] = \mathbf{b}_t(1),$$

where (a) uses $\mathbb{E}[\mathbb{1}_{I \leq t+1}] = 1 \cdot \mathbb{P}[I \leq t+1] + 0 \cdot \mathbb{P}[I > t+1] = \mathbb{P}[I \leq t+1]$.

Moreover, $|\mathbf{B}_t(1)| \leq 1$. Therefore, $(\mathbf{B}_t(1))_{t\geq 1}$ is an $\mathcal{F}_t$-submartingale. It follows from the martingale convergence theorem that $\mathbf{B}_t(1)$ converges almost surely as $t \to \infty$ (Thm. 6.4.3, Ash, 1972). Thus, by the bounded convergence theorem,

$$\mathbb{E}\left[\lim_{t\to\infty} \mathbf{B}_t(1)\right] = \lim_{t\to\infty} \mathbb{E}[\mathbf{B}_t(1)] \qquad \text{(Ch. 4, Prop. 6, Royden, 1988)}$$
$$= \lim_{t\to\infty} \mathbb{E}[\mathbb{P}[I \leq t \mid \mathcal{F}_t]] = \lim_{t\to\infty} \mathbb{E}[\mathbb{E}[\mathbb{1}_{I\leq t} \mid \mathcal{F}_t]] \stackrel{(a)}{=} \lim_{t\to\infty} \mathbb{E}[\mathbb{1}_{I\leq t}]$$
$$= \lim_{t\to\infty} \mathbb{P}[I \leq t] = 1 - \lim_{t\to\infty} (1-p)^t = 1,$$

where (a) follows from the law of iterated expectations.

Hence, $\mathbf{b}_t(1)$ converges almost surely to 1 as $t \to \infty$. By definition, the optimal action when $\mathbf{b}_t(1) = 1$ is $\mathsf{S}$[8]. Thus, $\mathcal{T}_1^\star, \ldots, \mathcal{T}_L^\star$ are almost surely finite. □

Proposition 1.3 implies that the optimal value function $J^\star(\mathbf{b})$ (18)[9] is well-defined and finite for each $\mathbf{b} \in \mathcal{B}$. We provide an example of $J^\star$ below.

**Example: Instantiation of the POMDP.**

Consider the POMDP instantiated with $L \triangleq 1$, $p \triangleq 0.01$, $R_{\text{int}} \triangleq -1.1$, $R_{\text{sla}} \triangleq 1$, $R_{\text{st}} \triangleq 1$, $\mathcal{O} \triangleq \{1, 2, \ldots, 10\}$, $z(\cdot \mid 0) \triangleq \text{BetaBin}(n = 10, \alpha = 0.7, \beta = 3)$, and $z(\cdot \mid 1) \triangleq \text{BetaBin}(n = 10, \alpha = 1, \beta = 0.7)$.

The value $J^\star(\mathbf{b}) = \mathbb{E}_{\pi_l^\star}\left[\sum_{t=1}^{T_\emptyset} r(S_t, A_t) \mid \mathbf{b}\right]$ is shown in Fig. 1.5. We note that $J^\star$ is piece-wise linear and convex, as expected from Thm. 2 in the background chapter. Moreover, we observe that $J^\star$ has a threshold structure, as formally stated in Thm. 1.1 on the next page.

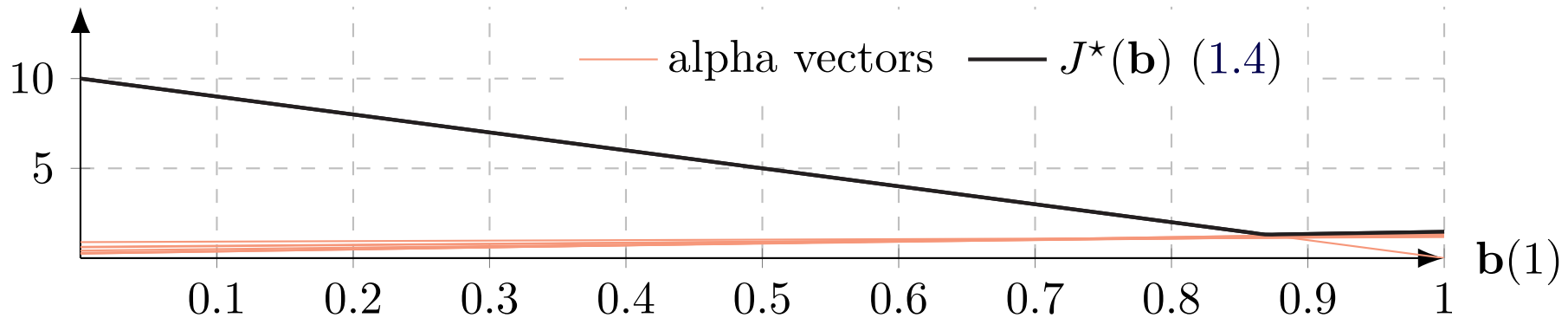

***Figure 1.5:*** *The value of an optimal defender strategy for the example; the dashed red lines indicate alpha-vectors $\boldsymbol{\alpha}^{(1)}, \boldsymbol{\alpha}^{(2)}, \ldots$, where $J^\star(\mathbf{b}) = \max_i [1 - \mathbf{b}(1), \mathbf{b}(1)]^T \boldsymbol{\alpha}^{(i)}$; see (Def. 1, Sondik, 1978) for the definition of the alpha vectors; we computed $J^\star$ using Heuristic Search Value Iteration (HSVI) (Alg. 1, Smith and Simmons, 2004).*

[8]i.e., it is optimal to stop when an intrusion occurs; see Appendix A for a proof of this fact.
[9]The value function is defined in the background chapter; see (18).

### Threshold properties of an optimal strategy

A strategy that solves the multiple-stopping problem is a solution to (1.4). Such a strategy satisfies the Bellman equation; see Thm. 2 in the background chapter. Based on this equation, we partition the belief space $\mathcal{B}$ into two sets

$$\mathscr{S}_l \triangleq \{\mathbf{b} \mid \mathbf{b} \in \mathcal{B}, \pi^\star_l(\mathbf{b}) = \mathsf{S}\} \qquad \text{the stopping set; and}$$

$$\mathscr{C}_l \triangleq \{\mathbf{b} \mid \mathbf{b} \in \mathcal{B}, \pi^\star_l(\mathbf{b}) = \mathsf{C}\} \qquad \text{the continuation set.}$$

Applying the stopping theory developed in (Nakai, 1985) and (Krishnamurthy et al., 2018), we obtain the following structural result.

**Theorem 1.1** (Threshold structure of an optimal strategy in the POMDP). *Given the definitions above and Assumption 1.1, the following holds.*

*(A)*

$$\mathscr{S}_{l-1} \subseteq \mathscr{S}_l \quad \forall l \in \{2, \ldots L\} \qquad \textit{(nested stopping sets).} \tag{1.5}$$

*(B) If $L = 1$, then there exists an optimal strategy $\pi^\star_L$ that satisfies*

$$\pi^\star_L(\mathbf{b}) = \mathsf{S} \iff \mathbf{b}(1) \geq \alpha^\star \qquad \textit{for some threshold } \alpha^\star \in [0,1]. \tag{1.6}$$

*(C) If $L \geq 1$ and $z$ (1.2) is totally positive of order 2 (i.e., TP-2 (Def. 10.2.1, Krishnamurthy, 2016)), then there exist $L$ values $\alpha^\star_1 \geq \alpha^\star_2 \geq \ldots \geq \alpha^\star_L$ and an optimal strategy $\pi^\star_l$ that satisfies*

$$\pi^\star_l(\mathbf{b}) = \mathsf{S} \iff \mathbf{b}(1) \geq \alpha^\star_l \qquad \forall l \in \{1, \ldots, L\}, \ \textit{where } \alpha^\star_l \in [0,1]. \tag{1.7}$$

Theorem 1.1.A states that the stopping sets have a nested structure. This structure means that if it is optimal to stop given $\mathbf{b}$ when $l-1$ stops remain, it is also optimal to stop when $l$ or more stops remain. Theorem 1.1.B–C state that there exists an optimal strategy with threshold properties; see Fig. 1.6. If $L \geq 1$, an additional condition applies: the stochastic matrix with the rows $z(\cdot \mid 0)$ and $z(\cdot \mid 1)$ (1.2) must be TP-2 (Karlin, 1964). This condition is satisfied, for example, if $z(\cdot \mid s)$ is stochastically monotone in $s$. We provide proof in Appendix A.

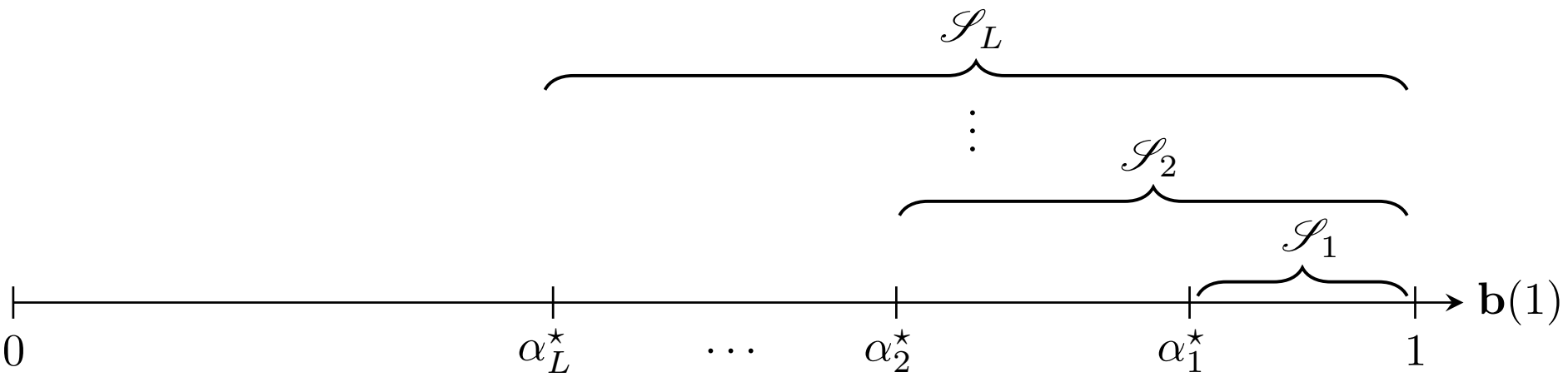

***Figure 1.6:*** *Illustration of Thm. 1.1: there exist $L$ thresholds $\alpha^\star_1 \geq \alpha^\star_2, \ldots, \geq \alpha^\star_L$ in the unit interval $[0,1]$ and an optimal threshold strategy $\pi^\star_l$ that satisfies (1.5)–(1.7).*

**Key insight.**

Theorem 1.1 implies that the optimal strategy for the defender is to monitor the infrastructure, update the belief $\mathbf{b}$, and when $\mathbf{b}(1)$ exceeds a threshold, trigger a defensive action.

Knowing that there exist optimal strategies with particular structures has two benefits. First, insight into the structure of optimal strategies often leads to a concise formulation and efficient implementation of the strategies (Puterman, 1994). This is obvious in the case of threshold strategies. Second, the complexity of computing an optimal strategy can be reduced by exploiting structural properties (Roy et al., 2019). The following section describes a stochastic approximation algorithm that exploits the structural result in Thm. 1.1.

### Our stochastic approximation algorithm: T-SPSA

Theorem 1.1 states that under given assumptions and given $L \geq 1$ stop actions, there exists an optimal strategy that uses $L$ thresholds $\alpha_1^\star \geq \alpha_2^\star, \ldots, \geq \alpha_L^\star$, where $\alpha_l^\star \in [0,1]$. We now present an algorithm, which we call **T**hreshold-**S**imultaneous **P**erturbation **S**tochastic **A**pproximation (T-SPSA), that computes these thresholds through stochastic approximation (Robbins and Monro, 1951).

We parameterize $\pi_{l,\boldsymbol{\theta}}$ with a vector $\boldsymbol{\theta} \in \mathbb{R}^L$. The component $\boldsymbol{\theta}_l$ relates to the threshold with $l \in \{1,\ldots L\}$ stops remaining. T-SPSA updates $\boldsymbol{\theta}$ through stochastic gradient ascent with the gradient $\nabla_{\boldsymbol{\theta}} J(\boldsymbol{\theta})$, where $J(\boldsymbol{\theta})$ is a shorthand for $J(\pi_{l,\boldsymbol{\theta}})$ (1.4). To ensure differentiability, we define $\pi_{\boldsymbol{\theta},l}$ to be a smooth stochastic strategy that approximates a threshold strategy:

$$\pi_{\boldsymbol{\theta},l}\left(\mathsf{S} \mid \mathbf{b}\right) \triangleq \left(1 + \left(\frac{\mathbf{b}(1)(1-\sigma(\boldsymbol{\theta}_l))}{\sigma(\boldsymbol{\theta}_l)(1-\mathbf{b}(1))}\right)^{-20}\right)^{-1}, \tag{1.8}$$

where $\sigma(\cdot)$ is the sigmoid function and $(\sigma(\boldsymbol{\theta}_l))_{l=1}^{L}$ are the $L$ thresholds[10]; see Fig. 1.7.

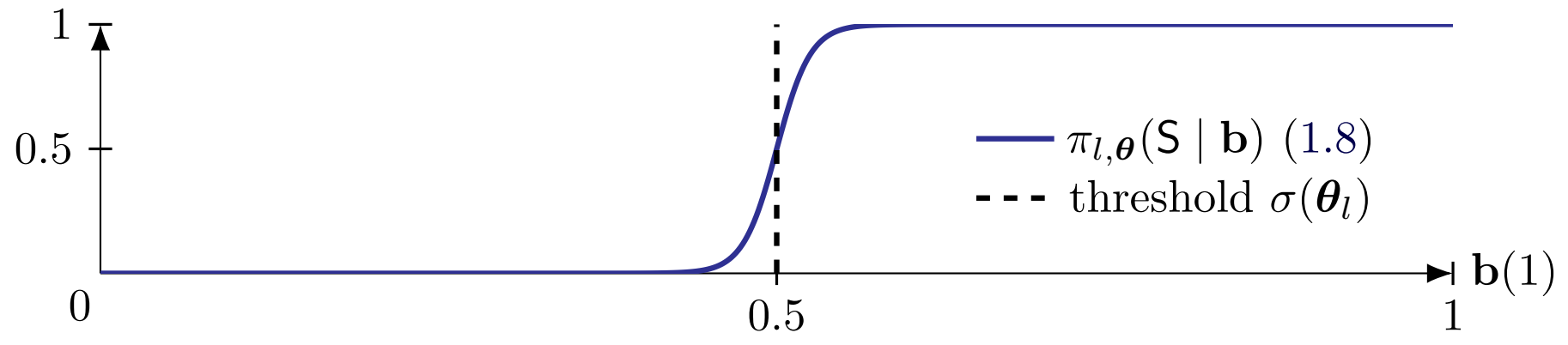

***Figure 1.7:*** *The stochastic threshold strategy in (1.8); $\sigma$ is the sigmoid function; $\sigma(\boldsymbol{\theta}_l)$ is the threshold (0.5 in this example); the x-axis indicates the belief state $\mathbf{b}(1) \in [0,1]$; and the y-axis indicates the probability prescribed by $\pi_{l,\boldsymbol{\theta}}$ to the stop action $\mathsf{S}$.*

[10] To avoid division by zero, we only use (1.8) when $\sigma(\boldsymbol{\theta}_l) \neq 0$ and $\mathbf{b}(1) \neq 1$; if $\sigma(\boldsymbol{\theta}_l) = 0$ or $\mathbf{b}(1) = 1$, then $\pi_{\boldsymbol{\theta},l}\left(\mathsf{S} \mid \mathbf{b}\right) = 1$.

We learn the threshold vector $\boldsymbol{\theta}$ through simulation of the POMDP as follows. First, we initialize $\boldsymbol{\theta}^{(1)} \in \mathbb{R}^L$ randomly. Second, for each iteration $n \in \{1, 2, \dots\}$ of T-SPSA, we perturb $\boldsymbol{\theta}^{(n)}$ to obtain $\boldsymbol{\theta}^{(n)} + c_n \boldsymbol{\Delta}^{(n)}$ and $\boldsymbol{\theta}^{(n)} - c_n \boldsymbol{\Delta}^{(n)}$, where $c_n = \frac{c}{n^\lambda} \in \mathbb{R}$ is the perturbation size, $(\lambda, c)$ are hyperparameters, and $\boldsymbol{\Delta}^{(n)} \in \mathbb{R}^L$ is a perturbation vector defined as

$$\boldsymbol{\Delta}_i^{(n)} \triangleq \begin{cases} +1 & \text{with probability } \frac{1}{2} \\ -1 & \text{with probability } \frac{1}{2} \end{cases} \qquad \forall i \in \{1, \dots, L\}.$$

Subsequently, we run two POMDP episodes where stop actions are prescribed according to the perturbed threshold vectors (1.8). We then use the obtained episode outcomes $\hat{J}(\boldsymbol{\theta}^{(n)} + c_n \boldsymbol{\Delta}^{(n)})$ and $\hat{J}(\boldsymbol{\theta}^{(n)} - c_n \boldsymbol{\Delta}^{(n)})$ to estimate the gradient $\nabla_{\boldsymbol{\theta}^{(n)}} J(\boldsymbol{\theta}^{(n)})$ using the **S**imultaneous **P**erturbation **S**tochastic **A**pproximation (SPSA) estimator

$$\left(\hat{\nabla}_{\boldsymbol{\theta}^{(n)}} J\left(\boldsymbol{\theta}^{(n)}\right)\right)_i \triangleq \frac{\hat{J}\left(\boldsymbol{\theta}^{(n)} + c_n \boldsymbol{\Delta}^{(n)}\right) - \hat{J}\left(\boldsymbol{\theta}^{(n)} - c_n \boldsymbol{\Delta}^{(n)}\right)}{2 c_n \boldsymbol{\Delta}_i^{(n)}}, \qquad \text{(Spall, 1992)}$$

where $i \in \{1, \dots, L\}$ is the component index of the gradient.

Next, we use the estimated gradient and the stochastic approximation algorithm to update the parameter vector

$$\boldsymbol{\theta}^{(n+1)} = \boldsymbol{\theta}^{(n)} + a_n \hat{\nabla}_{\boldsymbol{\theta}^{(n)}} J(\boldsymbol{\theta}^{(n)}),$$

where $a_n \triangleq \frac{a}{(n+A)^\epsilon}$ is the step size and $(A, \epsilon)$ are hyperparameters (Spall, 1998).

$\boldsymbol{\theta}^{(n)}$ converges almost surely to a *local* maximum of $J$ (1.4) as $n \to \infty$ if

$$\underbrace{\begin{array}{l} a_n, c_n > 0 \ \forall n; \quad a_n \to 0, c_n \to 0 \text{ as } n \to \infty; \\ \displaystyle\sum_{n=0}^{\infty} a_n = \infty; \quad \sum_{n=0}^{\infty} \left(\frac{a_n}{c_n}\right)^2 < \infty. \end{array}}_{\text{Step size conditions (Robbins and Monro, 1951)}} \qquad \text{(Prop. 1, Spall, 1992)}$$

We list the pseudocode of T-SPSA in Appendix D.

## 1.5 Creating a Digital Twin of the Target Infrastructure

To instantiate the POMDP defined above, we must know the observation distribution $z(\mathbf{o}_t \mid s_t)$ (1.2), i.e., the distribution of IDS alerts and login attempts conditioned on the state. We estimate this distribution using measurements from a digital twin of the target infrastructure. We create this digital twin using CSLE, as described in the methodology chapter (Hammar, 2023). The topology of the target infrastructure is shown in Fig. 1.2, and the configuration is listed in Appendix C.

The digital twin comprises virtual containers and networks that replicate the functionality and the timing behavior of the target infrastructure. These containers run the same software and processes as the physical infrastructure. For example, the container that emulates the gateway shown in Fig. 1.2 runs the SNORT IDS with community ruleset v2.9.17.1 (Roesch, 1999). This IDS produces real-time data, such as logs and alerts. We collect this data from the digital twin at 30-second intervals, which allows us to compute the defender observation $\mathbf{o}_t$ (1.2). (We define $30s$ in the digital twin to be 1 time step in the POMDP.) The distribution of the data produced by the digital twin depends on the actions of the defender, the clients, and the attacker, as defined below.

### Emulating the defender

We have implemented $L = 3$ stop actions on the digital twin to emulate the defender; they are listed in Table 1.2. The *first* stop recovers user accounts compromised by the attacker. The *second* and *third* stops update the firewall configuration of the gateway. Specifically, the *second* stop adds a rule to the firewall that drops incoming traffic from IP addresses that have been flagged by the IDS, and the *third* stop blocks *all* incoming traffic.

| *l* | *Action* | *Command on the digital twin* | MITRE D3FEND technique |
|---|---|---|---|
| 3 | Revoke certificates | `openssl ca -revoke <certificates>` | D3-CBAN certificate revocation. |
| 2 | Blacklist IPs | `iptables -A INPUT -s <ip> -j DROP` | D3-NTF network traffic filtering. |
| 1 | Block gateway | `iptables -A INPUT -i eth0 -j DROP` | D3-NI network isolation. |

***Table 1.2:*** *Defender stop commands on the digital twin; the commands are linked to the corresponding defense techniques in the MITRE D3FEND taxonomy (Kaloroumakis and Smith, 2021); $l$ denotes the number of remaining stops.*

### Emulating the clients

The *client population* is emulated by processes that interact with application servers through the gateway by performing a sequence of functions on a sequence of servers, both of which are selected uniformly at random from Table 1.3 on the next page. Client arrivals per time step are emulated using a stationary Poisson process with mean $\lambda = 20$ and exponentially distributed service times with mean $\mu = 4$.

### Emulating the attacker

We have implemented three *attackers* on the digital twin: NoviceAttacker, ExperiencedAttacker, and ExpertAttacker, which execute the sequences of actions listed in Table 1.4 (shown on the next page). The actions consist of reconnaissance commands and exploits. During each time step, one action is executed. The attackers differ in their reconnaissance command and the number of stops required to prevent the attack; see Table 1.5 on the next page.

| *Functions* | *Application servers* |
|---|---|
| HTTP, SSH, SNMP, ICMP | $N_2, N_3, N_{10}, N_{12}$. |
| IRC, POSTGRES, SNMP | $N_{31}, N_{13}, N_{14}, N_{15}, N_{16}$. |
| FTP, DNS, TELNET | $N_{10}, N_{22}, N_4$. |

***Table 1.3:*** *Emulated client population in the digital twin; each client invokes functions on application servers; shell commands for invoking the functions are listed in (Hammar, 2023); the server configurations $N_1, \dots, N_{31}$ are listed in Appendix C and the network topology is shown in Fig. 1.2. (Note that each component in Fig. 1.2 is labeled with an identifier $N_i$.)*

| *Time steps t* | NoviceAttacker | ExperiencedAttacker | ExpertAttacker |
|---|---|---|---|
| 1-$I \sim \text{Ge}(p)$ | (Intrusion not started) | (Intrusion not started) | (Intrusion not started) |
| $I+1$-$I+6$ | Recon$_1$, brute-force attacks (SSH,TELNET,FTP) $N_2, N_4, N_{10}$, login($N_2, N_4, N_{10}$), backdoor($N_2, N_4, N_{10}$) | Recon$_2$, CVE-2017-7494 $N_4$, brute-force attack (SSH) $N_2$, login($N_2, N_4$), backdoor($N_2, N_4$), Recon$_2$ | Recon$_3$, CVE-2017-7494 $N_4$, login($N_4$), backdoor($N_4$) Recon$_3$, SQL Injection $N_{18}$ |
| $I+7$-$I+10$ | Recon$_1$, CVE-2014-6271 $N_{17}$, login($N_{17}$), backdoor($N_{17}$) | CVE-2014-6271 $N_{17}$, login($N_{17}$) backdoor($N_{17}$), SSH brute-force attack $N_{12}$ | login($N_{18}$), backdoor($N_{18}$), Recon$_3$, CVE-2015-1427 $N_{25}$ |
| $I+11$-$I+14$ | SSH brute-force attack $N_{12}$, login($N_{12}$) CVE-2010-0426 $N_{12}$, Recon$_1$. | login($N_{12}$), CVE-2010-0426 $N_{12}$, Recon$_2$, SQL Injection $N_{18}$ | login($N_{25}$), backdoor($N_{25}$), Recon$_3$, CVE-2017-7494 $N_{27}$ |
| $I+15$-$I+16$ | | login($N_{18}$), backdoor($N_{18}$) | login($N_{27}$), backdoor($N_{27}$). |
| $I+17$-$I+19$ | | Recon$_2$, CVE-2015-1427 $N_{25}$, login($N_{25}$). | |

***Table 1.4:*** *Attacker actions on the digital twin; $I$ is the start time of the intrusion (see Fig. 1.3); shell commands and scripts for executing the actions are listed in (Hammar, 2023); the server configurations $N_1, \dots, N_{31}$ are listed in Appendix C and the network topology is shown in Fig. 1.2; if the attacker sequences complete before the defender performs the final stop, the sequences are restarted. (Note that each component in Fig. 1.2 is labeled with an identifier $N_i$.)*

| *Attacker* | $L$ | Reconnaissance |
|---|---|---|
| NoviceAttacker | 1 | TCP/UDP scan. |
| ExperiencedAttacker | 2 | ICMP ping scan. |
| ExpertAttacker | 3 | ICMP ping scan. |

***Table 1.5:*** *Number of stops ($L$) required to prevent the attacker and the reconnaissance commands of the attackers.*

### Estimating the distributions of alerts and login attempts

We estimate the distributions of IDS alerts and login attempts using data from the digital twin. At the end of every 30s interval on the twin, we collect the metrics $\Delta x$, $\Delta y$, $\Delta z$, which contain the alerts and login attempts that occurred during the interval. We use $M = 21,000$ i.i.d. samples to compute the empirical distribution $\widehat{z}(\cdot \mid s_t)$ as an estimate of $z(\cdot \mid s_t)$ (1.2), where $\widehat{z} \overset{\text{a.s.}}{\to} z$ as $M \to \infty$ (Glivenko and Cantelli, 1933). Figure 1.8 shows some of the estimated (smoothed) distributions (aggregated over all three attackers). The distributions during normal operation and intrusion mostly overlap. However, the distributions during intrusions tend to have more probability mass at larger values of $\Delta x, \Delta y$, and $\Delta z$. From these estimated distributions, we note that the TP-2 assumption in Thm. 1.1.C is reasonable.

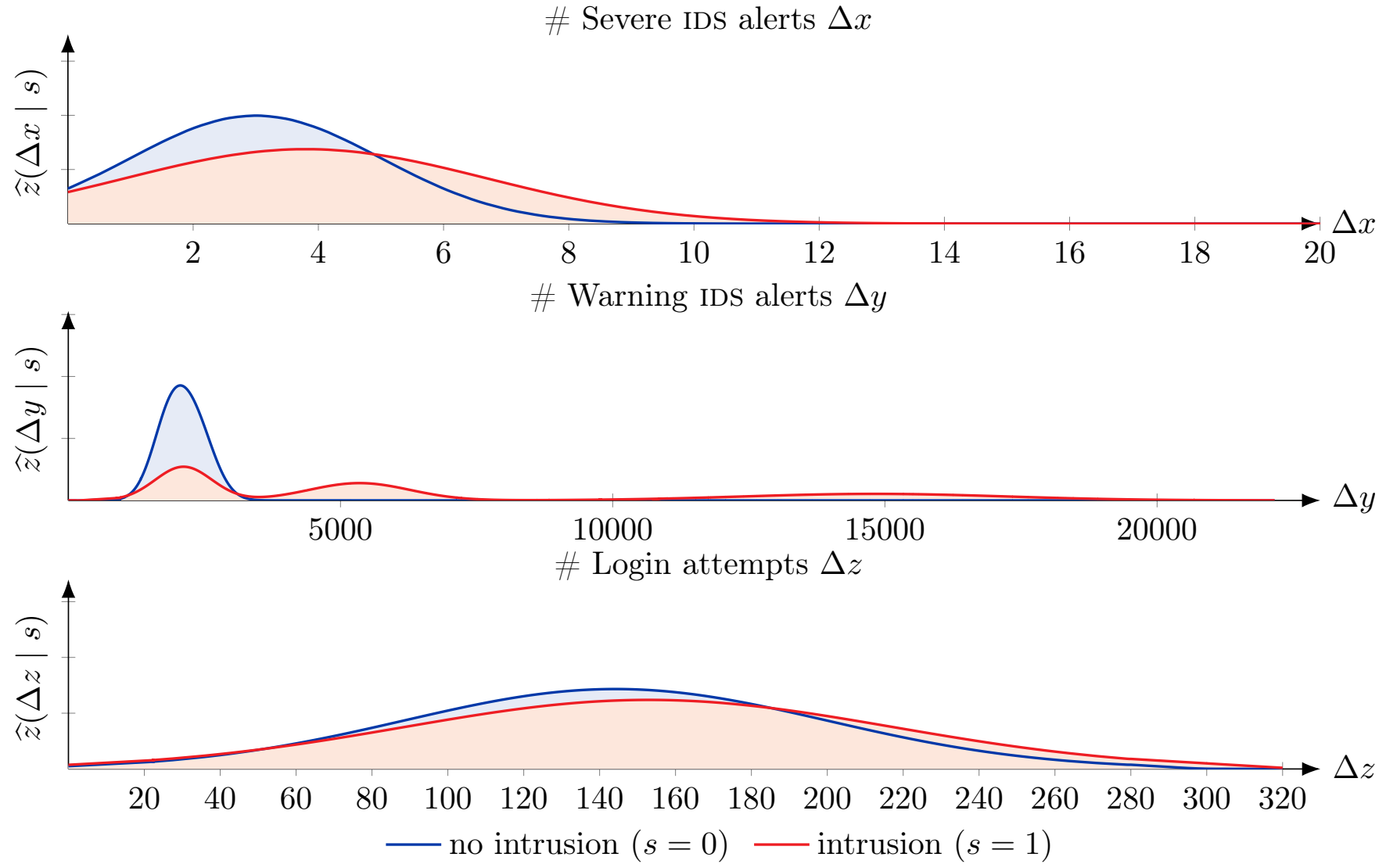

***Figure 1.8:*** *Estimated (smoothed) distributions of severe IDS alerts $\Delta x$ (top row), warning IDS alerts $\Delta y$ (middle row), and login attempts $\Delta z$ (bottom row) based on measurements from the digital twin.*

### Simulating an episode of the POMDP

A simulated episode evolves as follows. The episode starts in state $s_1 = 0$ with $l_1 = L$. During each time step, the action is sampled from the defender strategy as $a_t \sim \pi_{l,\boldsymbol{\theta}}(\cdot \mid \mathbf{b}_t)$. If the action is stop and $l_t = 1$, the episode ends. Otherwise, the remaining number of stop actions is updated: $l_{t+1} = l_t - a_t$[11]. The next state and observation are then sampled as $s_{t+1} \sim f_l(\cdot \mid s_t, a_t)$ (1.1) and $\mathbf{o}_{t+1} \sim \widehat{z}(\cdot \mid s_{t+1})$ (Fig. 1.8), respectively. Subsequently, the belief $\mathbf{b}_{t+1}$ is computed using (17)[12] and the reward $r_{t+1}$ is computed using (1.3). The sequence of time steps continues until the defender performs the final stop, after which the episode ends.

[11] Recall that we encode $(\mathsf{S}, \mathsf{C}) = (1, 0)$.

[12] As defined in the background chapter.

**Remark 1.2.** The clients' activities are not simulated but are captured by $\widehat{z}$.

### Emulating an episode of the POMDP on the digital twin

Like a simulated episode, an emulated episode on the digital twin starts with the same initial conditions, evolves in discrete time steps, and experiences an intrusion event at a random time. However, an episode on the digital twin differs from a simulated episode in the following ways. First, attacker and defender actions on the digital twin include computing and networking functions with side-effects; see Table 1.2 and Table 1.4. Second, the defender observations in the digital twin are not sampled but are obtained through reading log files and metrics of emulated servers in the digital twin. Third, the emulated client population in the digital twin performs requests to the emulated application servers just like on an operational infrastructure; see Table 1.3. Due to these differences, running an episode on the digital twin takes much longer time than simulating a similar episode (e.g., minutes vs milliseconds).

## 1.6 Experimental Evaluation

Our methodology for finding effective defender strategies includes simulation of POMDP episodes to learn defender strategies through T-SPSA, as well as evaluating them on the digital twin; see Fig. 12 in the introduction chapter. This section describes our evaluation results.

### Computing environment

The environment for training strategies and running simulations is a TESLA P100 GPU. The hyperparameters for the training algorithm are listed in Appendix B. The digital twin is deployed on a server with a 24-core Intel Xeon Gold 2.10GHz CPU and 768 GB RAM; see Fig. 21 in the methodology chapter.

### Evaluation process

We train three defender strategies against the NOVICE, EXPERIENCED, and EXPERT attackers until convergence (the attackers are defined in Table 1.4). We run 10′000 training episodes for each attacker to estimate an optimal defender strategy using T-SPSA. After each episode, we evaluate the current defender strategy by running evaluation episodes on the digital twin and computing various performance metrics. The 10′000 training episodes and the evaluation constitute one *training run*. We run five training runs with different random seeds. A single training run takes about 4 hours of processing time on a P100 GPU to perform the simulations and the strategy training, as well as around 12 hours for evaluating the strategies on the digital twin.

### Baseline strategies

We compare the strategies learned through T-SPSA with three baseline defender strategies. The first baseline prescribes the stop action whenever an IDS alert occurs, i.e., whenever $(\Delta x + \Delta y) \geq 1$. The second baseline is obtained by configuring the SNORT IDS as an Intrusion Detection and Prevention System (IDPS), which drops network traffic following its internal recommendation system; see Appendix C for the SNORT configuration. To calculate the reward, we define 100 dropped IP packets of the SNORT IDPS to be a stop action of the defender. Lastly, the third baseline is an ideal strategy that presumes knowledge of the intrusion time and performs all stop actions at that time.

### Baseline algorithms

We evaluate T-SPSA by comparing it with three baseline algorithms: **P**roximal **P**olicy **O**ptimization (PPO) (Alg. 1, Schulman et al., 2017)[13], **H**euristic **S**earch **V**alue **I**teration (HSVI) (Alg. 1, Smith and Simmons, 2004), and Shiryaev's algorithm (Shiryaev, 1963). PPO is a state-of-the-art deep reinforcement learning algorithm (see Fig. 1.9), HSVI is a state-of-the-art dynamic programming algorithm for POMDPs, and Shiryaev's algorithm is an optimal algorithm for change detection. The main difference between T-SPSA and the first two baselines (PPO and HSVI) is that T-SPSA exploits the threshold structure expressed in Thm. 1.1. The main difference between T-SPSA and Shiryaev's algorithm is that T-SPSA learns $L$ thresholds whereas Shiryaev's algorithm uses a single predefined threshold. We set this threshold to 0.75 based on a hyperparameter search; see Appendix B.

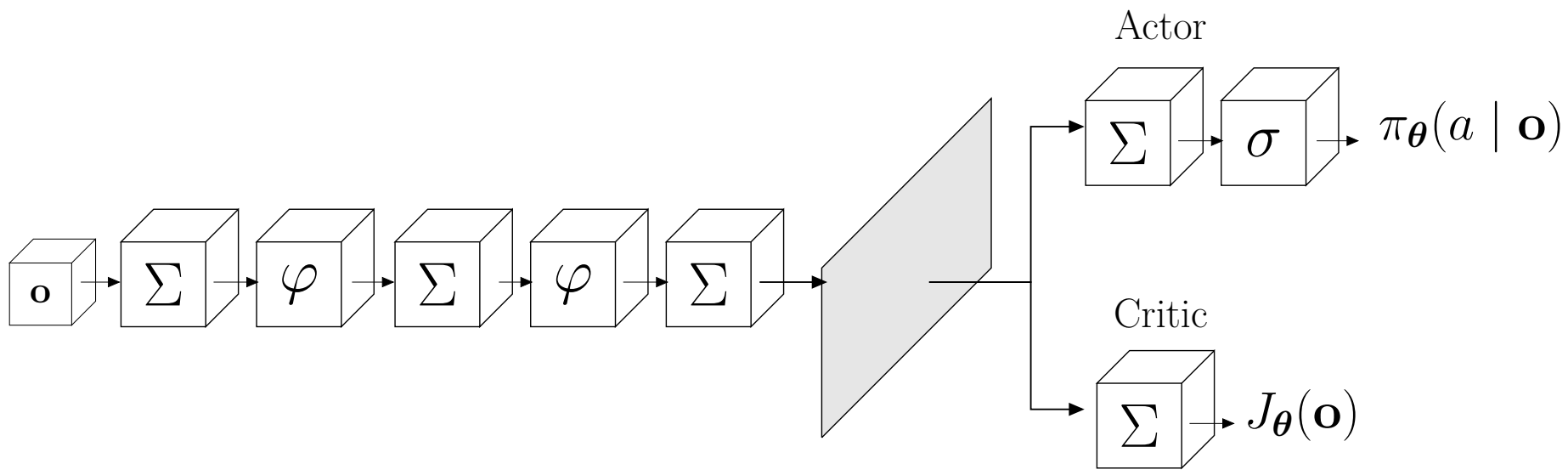

***Figure 1.9:*** *Architecture of the neural network used by PPO (Alg. 1, Schulman et al., 2017); the strategy $\pi_{\boldsymbol{\theta}}$ and the value function $J_{\boldsymbol{\theta}}$ are parameterized by a neural network with the actor-critic architecture (Sutton and Barto, 1998); $\Sigma$ represents a linear sum; $\sigma$ represents the softmax function; and $\varphi$ represents the ReLu function.*

[13] See Appendix D of Paper 3 for a derivation of the PPO algorithm.

### Learning intrusion prevention strategies

Figure 1.10 shows the performance of the learned strategies against the three attackers defined in Table 1.4. The red and blue curves represent the results from the simulator and the digital twin, respectively. The purple and orange curves give the performance of the SNORT IDPS and the baseline strategy that mandates a stop action whenever an IDS alert occurs, respectively. The dashed black curves give the performance of the baseline that assumes knowledge of the exact intrusion time.

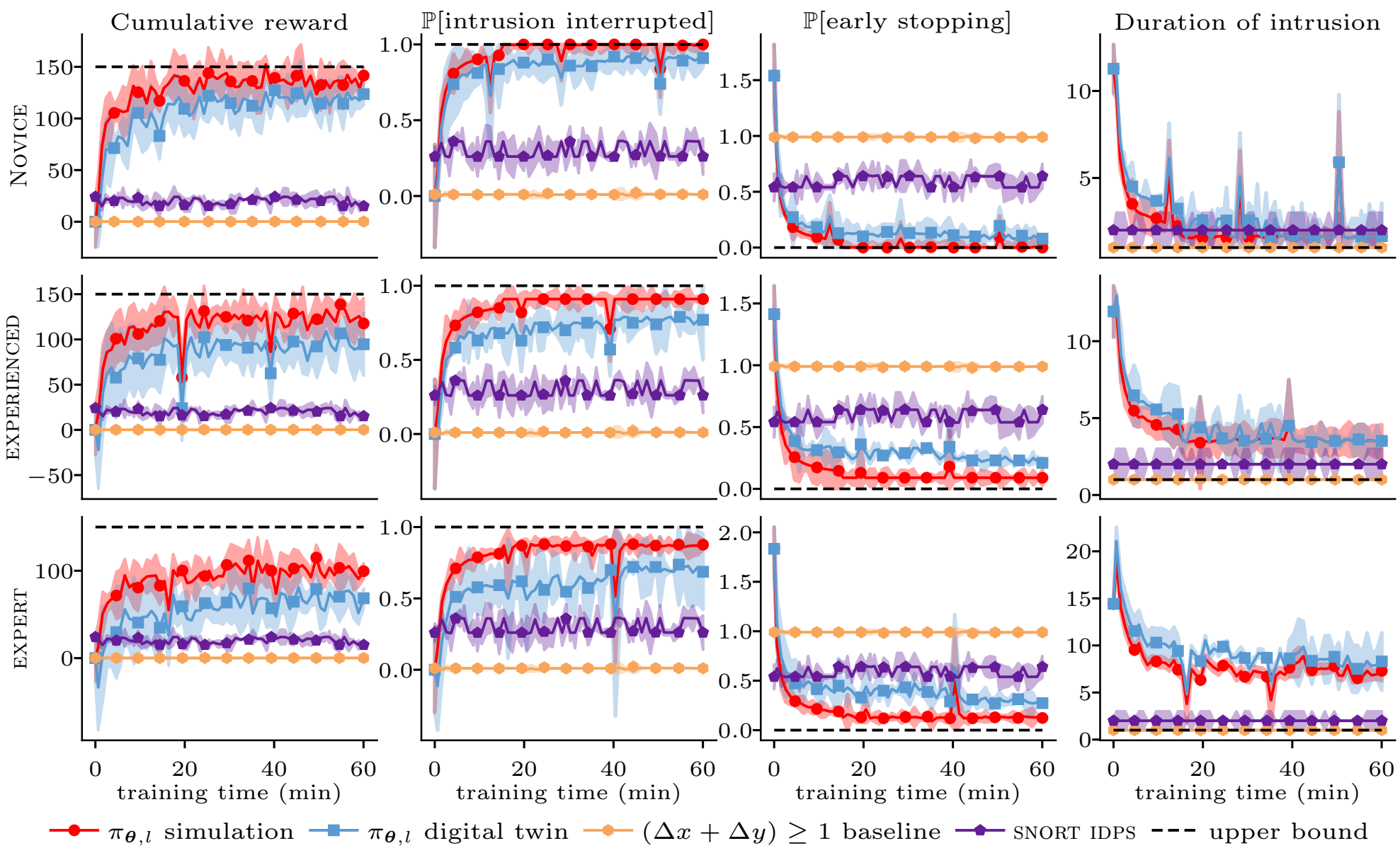

***Figure 1.10:*** *Learning curves obtained during training of T-SPSA; red curves show simulation results and blue curves show results from the digital twin; the purple, orange, and black curves relate to baseline strategies; the rows from top to bottom relate to: NOVICEATTACKER, EXPERIENCEDATTACKER, and EXPERTATTACKER; the columns from left to right show performance metrics: episodic reward, empirical prevention probability, empirical early stopping probability, and the duration of intrusion; the curves show the mean and 95% confidence interval for five training runs with different random seeds.*

An analysis of the graphs in Fig. 1.10 leads us to the following conclusions. The learning curves converge quickly to constant mean values for all attackers and across all investigated performance metrics. From this observation, we conclude that the learned strategies have also converged. Second, we observe that the converged values of the learning curves are close to the dashed black curves, which give an upper bound to an optimal strategy. In addition, we see that the empirical probability of preventing an intrusion is close to 1 (second leftmost column) and that the empirical probability of stopping before the intrusion starts is close to 0 (second rightmost column). This suggests that the learned strategies are close to optimal. We also observe that all learned strategies do significantly better than the SNORT IDPS baseline and the baseline that stops whenever an IDS alert occurs (leftmost column). Third, although the learned strategies, as expected, perform better on the simulator than on the digital twin, we are encouraged by the fact that the curves of the digital twin are close to those of the simulator (cf. the blue and red curves).

**Key result.**

The performance of the strategies learned on the simulator transfers to the digital twin; see Fig. 12 in the introduction chapter.

We note from Fig. 1.10 that the learned strategies do better against NOVICEATTACKER than against EXPERIENCEDATTACKER and EXPERTATTACKER. For instance, the learned strategies against EXPERIENCEDATTACKER and EXPERTATTACKER are more likely to stop before an intrusion has started (second rightmost column of Fig. 1.10). This indicates that NOVICEATTACKER is easier to detect for the defender as its actions create more IDS alerts than those of the other attackers.

Figure 1.11 shows a comparison between our stochastic approximation algorithm (T-SPSA) and the three baseline algorithms.

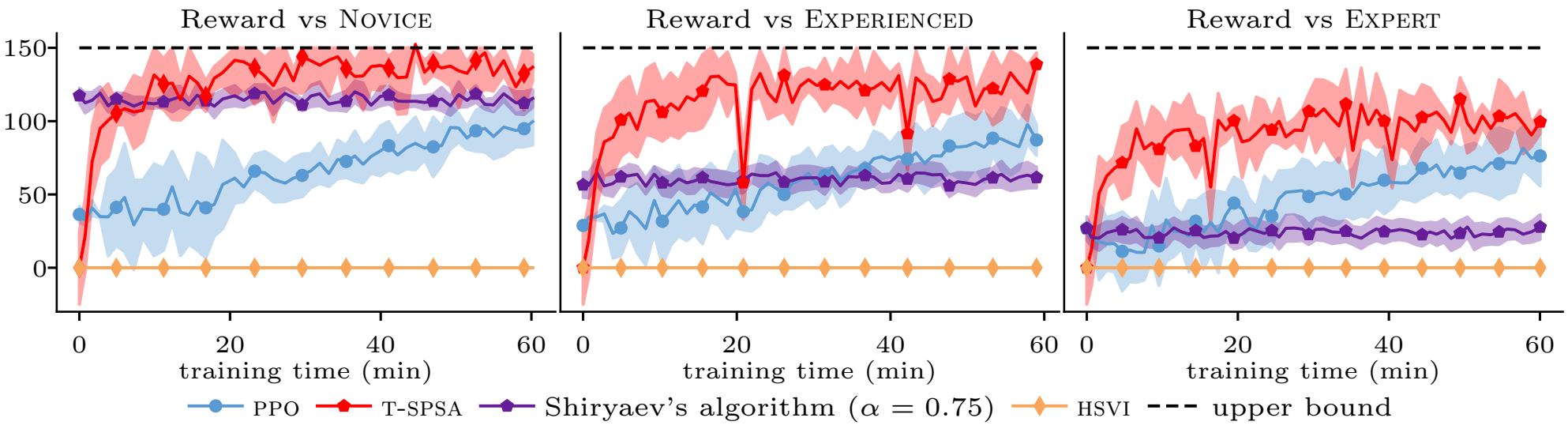

***Figure 1.11:*** *Comparison between T-SPSA and the baselines; all curves show simulation results; red curves relate to T-SPSA; blue curves relate to PPO; orange curves relate to HSVI; purple curves relate to Shiryaev's algorithm with threshold $\alpha = 0.75$; the columns from left to right relate to: NOVICEATTACKER, EXPERIENCEDATTACKER, and EXPERTATTACKER; all curves show the mean and 95% confidence interval for five training runs.*

We observe in Fig. 1.11 that both T-SPSA (red curves) and PPO (blue curves) converge to close approximations of an optimal strategy within an hour of training whereas HSVI (orange curves) does not converge within the measured time. The slow convergence of HSVI manifests the intractability of using dynamic pro-gramming to compute optimal strategies for POMDPs (Thm. 6, Papadimitriou and Tsitsiklis, 1987). We also see in Fig. 1.11 that T-SPSA converges significantly faster than PPO. This is expected since T-SPSA considers a smaller space of strategies than PPO. Finally, we note in Fig. 1.11 that T-SPSA outperforms Shiryaev's algorithm (purple curves), which demonstrates the benefit of using $L$ thresholds instead of a single threshold.

## 1.7 Related Work

Recent works that study security automation using stochastic approximation include [328, 178, 182, 176, 130, 397, 510, 261, 66, 369, 525, 469, 152, 207, 150, 4, 286, 353, 175, 518, 125, 123, 508, 529, 528, 287, 391, 283, 292, 64, 213, 284, 519, 239, 486, 187, 493, 219, 218, 484, 179, 22, 521, 273, 223, 185, 278, 260, 141, 296, 336, 241, 368, 490, 495, 93, 452, 245, 215, 474]. These works use a variety of algorithms, including Q-learning [130, 510, 369, 494, 391, 64, 486, 22], SARSA [261], PPO [178, 182, 123, 125, 529], hierarchical reinforcement learning [469], DQN [152, 508, 529, 528, 287, 284, 219], Thompson sampling [207], MUZERO [150], NFQ [4], DDQN [353, 493], NFSP [274, 497], A2C [292], A3C [187], and DDPG [286, 518].

The previous works differ from this paper in two main ways. First, we formulate the intrusion prevention problem as a multiple-stopping problem. The other works formulate the problem as a general MDP[14], POMDP, or stochastic game. The advantage of our approach is that we obtain structural properties of optimal strategies, which have practical benefits. Second, our methodology for experimental evaluation involves a digital twin in addition to simulations. The advantage of our method compared to the simulation-only approaches [178, 182, 130, 397, 510, 261, 66, 369, 525, 469, 152, 207] is that the parameters of our simulator are determined by measurements from the digital twin instead of being chosen by a human. Further, the learned strategies are evaluated on the digital twin, not the simulator. As a consequence, the evaluation results give a higher confidence in the obtained strategies' operational performance than what simulation results would provide.

Problem formulations based on stopping theory can be found in prior research on change detection [409, 344, 261, 453, 283, 182]. Compared to these papers, our approach is more general by allowing multiple stop actions within an episode. Another difference is that we model intrusion *prevention* rather than intrusion *detection*. Further, compared with traditional change detection algorithms, e.g., CUSUM (Page, 1954) and Shiryaev's algorithm (Shiryaev, 1963), our algorithm *learns* thresholds and does not assume them to be preconfigured.

[14]The components of an MDP are defined the background chapter; see (1).

## 1.8 Conclusion

In this paper, we propose a novel formulation of intrusion prevention based on the theory of optimal stopping. This formulation allows us to derive that a threshold strategy is optimal, which has practical benefits. To find and evaluate strategies, we use a methodology that includes simulation-based optimization and evaluation on a digital twin. In contrast to a simulation-only approach, our methodology produces strategies that can be executed in a target infrastructure with a practical configuration (Figs. 1.10-1.11). The evaluation results show that our stochastic approximation algorithm (T-SPSA), which takes advantage of the threshold structure (Thm. 1.1), outperforms state-of-the-art algorithms on our use case.

We make assumptions in this paper that limit the practical applicability of the results: the attacker follows a static strategy, and the defender learns only the times of taking defensive actions but not the types of actions. Therefore, the question arises whether our methodology can be extended so that (*i*) the attacker can pursue a wide range of realistic strategies and (*ii*) the defender learns optimal strategies that express not only when defensive actions need to be taken but also the specific measure to be executed. We address (*i*) in Paper 2 and we address (*ii*) in Paper 3.

# ■ Acknowledgments

The authors would like to thank Pontus Johnson for his useful input to this research and Vikram Krishnamurthy for helpful discussions. The authors are also grateful to Forough Shahab Samani and Xiaoxuan Wang for their constructive comments on a draft of this paper.

# ■ Appendix

## A Proof of Theorem 1.1

The main idea behind the proof of Thm. 1.1 is to show that the stopping sets $(\mathscr{S}_l)_{l=1,\dots,L}$ have the form $\mathscr{S}_l = [\alpha^\star_l, 1] \subseteq \mathcal{B}$ and that $\alpha^\star_l \geq \alpha^\star_{l+1}$. Toward this goal, we state the following five lemmas.

**Remark 1.3** (Notation)**.** *Since* $\mathbf{b}$ *is determined by* $\mathbf{b}(1)$*, i.e.,* $\mathbf{b}(0) = 1 - \mathbf{b}(1)$*, we use* $\mathbf{b}$ *as a shorthand for* $\mathbf{b}(1)$*.*

**Lemma 1.1.** $\mathscr{S}_1$ *is a convex subset of* $\mathcal{B} = [0, 1]$*.*

*Proof.* This result was originally proven in (Thm. 12.2.1, Krishnamurthy, 2016). For completeness, we give the proof here since it is very short. We need to show that $\mathbf{b}', \mathbf{b}'' \in \mathscr{S}_1 \implies (\lambda\mathbf{b}' + (1-\lambda)\mathbf{b}'') \in \mathscr{S}_1$ for any $\lambda \in [0, 1]$. We obtain

$$J^\star(\lambda\mathbf{b}' + (1-\lambda)\mathbf{b}'') \leq \lambda J^\star(\mathbf{b}') + (1-\lambda)J^\star(\mathbf{b}'') \qquad \text{(Convexity of } J^\star\text{; Thm. 2)}$$

$$
\begin{aligned}
&= \lambda Q^{\star}(\mathbf{b}', \mathsf{S}) + (1-\lambda)Q^{\star}(\mathbf{b}'', \mathsf{S}) && (\mathbf{b}', \mathbf{b}'' \in \mathscr{S}_1 \ (6))\\
&= (\lambda\mathbf{b}' + (1-\lambda)\mathbf{b}'')R_{\text{st}} && (L = 1\ (1.3))\\
&= Q^{\star}(\lambda\mathbf{b}' + (1-\lambda)\mathbf{b}'', \mathsf{S})\\
&\leq J^{\star}\big(\lambda\mathbf{b}' + (1-\lambda)\mathbf{b}''\big) && (J^{\star}(\mathbf{b}) \triangleq \max_{a \in \mathcal{A}} Q^{\star}(\mathbf{b}, a))\\
&\implies Q^{\star}(\lambda\mathbf{b}' + (1-\lambda)\mathbf{b}'', \mathsf{S}) = J^{\star}(\lambda\mathbf{b}' + (1-\lambda)\mathbf{b}'')\\
&\implies (\lambda\mathbf{b}' + (1-\lambda)\mathbf{b}'') \in \mathscr{S}_1 \qquad \forall \lambda \in [0,1].
\end{aligned}
$$

□

**Lemma 1.2.** *For each $a \in \mathcal{A}$ and $l \in \{1, \dots, L\}$, $f_l(\cdot \mid \cdot, a)$ (1.1) is* TP*-2 [254, Def. 10.2.1].*

*Proof.* Given any combination of $a \in \mathcal{A}$ and $l \in \{1, \dots, L\}$, $f_l(\cdot \mid \cdot, a)$ (1.1) can be represented by one of following two row-stochastic matrices:

$$
\begin{array}{c} \\ 0 \\ 1 \\ \emptyset \end{array}
\begin{array}{c} \begin{array}{ccc} 0 & 1 & \emptyset \end{array} \\ \begin{bmatrix} 1-p & p & 0 \\ 0 & 1 & 0 \\ 0 & 0 & 1 \end{bmatrix} \end{array},
\qquad
\begin{array}{c} \\ 0 \\ 1 \\ \emptyset \end{array}
\begin{array}{c} \begin{array}{ccc} 0 & 1 & \emptyset \end{array} \\ \begin{bmatrix} 0 & 0 & 1 \\ 0 & 0 & 1 \\ 0 & 0 & 1 \end{bmatrix} \end{array}.
$$

The left matrix corresponds to $f_l(\cdot \mid \cdot, \mathsf{C})$ and $f_{l>1}(\cdot \mid \cdot, \mathsf{S})$. The right matrix represents $f_1(\cdot \mid \cdot, \mathsf{S})$. To show that $f_l(\cdot \mid \cdot, a)$ (1.1) is TP-2 [254, Def. 10.2.1], it suffices to show that all $\binom{3}{2}^2$ second order minors of both matrices are non-negative. The second-order minors of the left matrix are

$$
\begin{aligned}
M_{1,2} = M_{1,3} = M_{2,3} = M_{3,1} = M_{3,2} &= 0\\
M_{1,1} &= 1\\
M_{2,1} &= p\\
M_{2,2} = M_{3,3} &= 1-p,
\end{aligned}
$$

where $M_{i,j}$ denotes the determinant of the submatrix formed by deleting the $i$th row and $j$th column. For the right matrix, all second-order minors are zero. □

**Lemma 1.3.** *$\mathbb{E}_S[r(S, \mathsf{S}) - r(S, \mathsf{C}) \mid \mathbf{b}]$ is increasing in $\mathbf{b}$ for $l \in \{1, \dots, L\}$.*

*Proof.* By definition of $r$ (1.3), we obtain

$$
\mathbb{E}_S[r(S, \mathsf{S}) - r(S, \mathsf{C}) \mid \mathbf{b}] = \frac{\mathbf{b}R_{\text{st}}}{L} - \left(R_{\text{sla}} + \frac{\mathbf{b}R_{\text{int}}}{L}\right) = \mathbf{b}\left(\frac{R_{\text{st}} - R_{\text{int}}}{L}\right) - R_{\text{sla}},
$$

which is increasing in $\mathbf{b}$ since $R_{\text{st}} > 0$ and $R_{\text{int}} < 0$. □

**Lemma 1.4** ((Thm. 10.3.1.1–3, p. 225, Krishnamurthy, 2016)[15].). *Given two beliefs* $\mathbf{b}' \geq \mathbf{b}$ *and two observations* $\mathbf{o} > \bar{\mathbf{o}}$[16], *if* $f_l(\cdot \mid \cdot, a)$ *(1.1) and* $z(\cdot \mid \cdot)$ *(1.2) are* TP*-2 for all* $a \in \mathcal{A}$ *and* $l \in \{1, \dots, L\}$ *[254, Def. 10.2.1], then the following holds for each* $a \in \mathcal{A}$, $\tilde{\mathbf{o}} \in \mathcal{O}$, *and* $l \in \{1, \dots, L\}$:

1. $\mathbb{B}(\mathbf{b}', a, \tilde{\mathbf{o}}) \geq \mathbb{B}(\mathbf{b}, a, \tilde{\mathbf{o}})$;
2. $\mathbb{P}[\mathbf{o} \geq \tilde{\mathbf{o}} \mid \mathbf{b}', a] \geq \mathbb{P}[\mathbf{o} \geq \tilde{\mathbf{o}} \mid \mathbf{b}, a]$[17]; *and*
3. $\mathbb{B}(\mathbf{b}, a, \mathbf{o}) \geq \mathbb{B}(\mathbf{b}, a, \bar{\mathbf{o}})$,

*where* $\mathbb{B}$ *is the belief operator (17), as defined in the background chapter.*

We now use Lemmas 1.1-1.4 to prove Thm. 1.1. Our proof is based on the same approach as in (Props. 4.5-4.8, pp. 437-441, Nakai, 1985) and (Thm. 1.C, Thm. 8, pp. 389-397, Krishnamurthy et al., 2018). The main idea is to show that, if $\mathbf{b}' \geq \mathbf{b}$, then $\mathbf{b} \in \mathscr{S}_l \implies \mathbf{b}' \in \mathscr{S}_l$ and $\mathbf{b} \in \mathscr{S}_{l-1} \implies \mathbf{b} \in \mathscr{S}_l$. To prove these properties, we use the value iteration algorithm to establish structural properties of $J_l^\star$ (18) (Smallwood and Sondik, 1973). Specifically, let $J_l^{(k)}$, $\mathscr{S}_l^{(k)}$, and $\mathscr{C}_l^{(k)}$, denote the value function, the stopping set, and the continuation set at iteration $k$ of the value iteration algorithm, respectively. Then, $\lim_{k\to\infty} J_l^{(k)} = J_l^\star, \lim_{k\to\infty} \mathscr{S}_l^{(k)} = \mathscr{S}_l$, and $\lim_{k\to\infty} \mathscr{C}_l^{(k)} = \mathscr{C}_l$ (Thms. 7.2.1, 7.6.1, Krishnamurthy, 2016). This convergence means that we can use mathematical induction on the value iterations to establish structural properties of $J_l^\star$. Let $J_l^{(0)}(\mathbf{b}) = 0\ \forall \mathbf{b} \in \mathcal{B}$. By showing that if the property holds for $J_l^{(k)}$ (induction hypothesis), it also holds for $J_l^{(k+1)}$, we establish (by induction) that the property holds for $J_l^\star$ (18).

For ease of notation, let $r(\mathbf{b}, a) \triangleq \mathbb{E}_S[r(S, a) \mid \mathbf{b}]$, $\mathbb{P}^{\mathbf{o}}_{a,\mathbf{b}} \triangleq \mathbb{P}[\mathbf{o} \mid \mathbf{b}, a]$, $\mathbb{P}^{\mathbf{o}}_{\mathbf{b}} \triangleq \mathbb{P}[\mathbf{o} \mid \mathbf{b}]$, and $\mathbf{b}^{\mathbf{o}}_a \triangleq \mathbb{B}(\mathbf{b}, a, \mathbf{o})$ (17). When $\mathbf{b}^{\mathbf{o}}_{\mathsf{S}} = \mathbf{b}^{\mathbf{o}}_{\mathsf{C}}$, we simply write $\mathbf{b}^{\mathbf{o}}$.

*Proof of Theorem 1.1.A.* It follows from the Bellman equation (18) that

$$\begin{aligned}
\mathbf{b} \in \mathscr{S}_{l-1} &\iff r(\mathbf{b}, \mathsf{S}) + \sum_{\mathbf{o}\in\mathcal{O}} \mathbb{P}^{\mathbf{o}}_{\mathsf{S},\mathbf{b}} J^\star_{l-2}(\mathbf{b}^{\mathbf{o}}_{\mathsf{S}}) \geq r(\mathbf{b}, \mathsf{C}) + \sum_{\mathbf{o}\in\mathcal{O}} \mathbb{P}^{\mathbf{o}}_{\mathsf{C},\mathbf{b}} J^\star_{l-1}(\mathbf{b}^{\mathbf{o}}_{\mathsf{C}}) \\
&\iff r(\mathbf{b}, \mathsf{S}) - r(\mathbf{b}, \mathsf{C}) + \sum_{\mathbf{o}\in\mathcal{O}} \mathbb{P}^{\mathbf{o}}_{\mathsf{S},\mathbf{b}} J^\star_{l-2}(\mathbf{b}^{\mathbf{o}}_{\mathsf{S}}) - \mathbb{P}^{\mathbf{o}}_{\mathsf{C},\mathbf{b}} J^\star_{l-1}(\mathbf{b}^{\mathbf{o}}_{\mathsf{C}}) \geq 0 \\
&\overset{(a)}{\iff} r(\mathbf{b}, \mathsf{S}) - r(\mathbf{b}, \mathsf{C}) + \sum_{\mathbf{o}\in\mathcal{O}} \mathbb{P}^{\mathbf{o}}_{\mathbf{b}} \underbrace{\left(J^\star_{l-2}(\mathbf{b}^{\mathbf{o}}) - J^\star_{l-1}(\mathbf{b}^{\mathbf{o}})\right)}_{\triangleq W_{l-1}(\mathbf{b}^{\mathbf{o}})} \geq 0,
\end{aligned} \tag{1.9}$$

[15] In the original proof, the monotone likelihood ratio (MLR) order among beliefs is considered [254, Def. 10.1.1]; Since $\mathcal{B} = [0, 1]$ in our case, the MLR order reduces to the natural order on $\mathbb{R}$. (To see this, plug in $\mathbf{b}'(0) = 1 - \mathbf{b}'(1)$ in the MLR definition $\mathbf{b}'(0)\mathbf{b}(1) \leq \mathbf{b}'(0)\mathbf{b}(1)$.)

[16] Here $\geq$ refers to the natural order on $\mathbb{R}$. To order the three-dimensional space $\mathcal{O}$, we map it to a one-dimensional space by associating each observation with a unique natural number.

[17] The left-hand side distribution stochastically dominates the distribution on the right-hand side in the first order (Def. 9.2.1, Krishnamurthy, 2016).

where (a) follows because $(i)$ $\mathbf{o}_t$ (1.2) is independent of $a_{t-1}$ when $l > 1$ since $f_{l>1}(\cdot \mid \cdot, \mathsf{S}) = f_{l>0}(\cdot \mid \cdot, \mathsf{C})$ (1.1); and $(iii)$ the belief operator (17) is independent of $a_{t-1}$ except for the case when $a_{t-1} = \mathsf{S}$ and $l = 1$, but in that case we obtain $J^{\star}_0(\mathbf{b}) = 0\ \forall \mathbf{b}$, which means that $J^{\star}_0(\mathbf{b}^{\mathbf{o}}_{\mathsf{S}}) = J^{\star}_0(\mathbf{b}^{\mathbf{o}}_{\mathsf{C}})$ for all $\mathbf{o} \in \mathcal{O}$.

We will use (1.9) to show that $\mathbf{b} \in \mathscr{S}_{l-1} \implies \mathbf{b} \in \mathscr{S}_l$ and thus $\mathscr{S}_{l-1} \subseteq \mathscr{S}_l$. Since $r(\mathbf{b}, \mathsf{S}) - r(\mathbf{b}, \mathsf{C})$ and $\mathbb{P}^{\mathbf{o}}_{\mathbf{b}}$ are independent of $l$, it suffices to show that $W_l(\mathbf{b}) - W_{l-1}(\mathbf{b}) \geq 0$ for all $\mathbf{b} \in \mathcal{B}$ and $l \in \{2, 3, \dots, L\}$. We show this statement by mathematical induction on $k = 0, 1, \dots$, where $k$ is the iteration of value iteration (Smallwood and Sondik, 1973).

For iteration $k = 0$, we have that $W^{(0)}_l(\mathbf{b}) = J^{(0)}_{l-1}(\mathbf{b}) - J^{(0)}_l(\mathbf{b}) = 0 = W^{(0)}_{l-1}(\mathbf{b})$ for all $l \in \{2, 3, \dots, L\}$ and $\mathbf{b} \in \mathcal{B}$. Hence, the inductive base case holds. Assume by induction that $W^{(k-1)}_l(\mathbf{b}) - W^{(k-1)}_{l-1}(\mathbf{b}) \geq 0$ for all $l \in \{2, 3, \dots, L\}$ and $\mathbf{b} \in \mathcal{B}$. We show below that this assumption implies that $W^{(k)}_l(\mathbf{b}) - W^{(k)}_{l-1}(\mathbf{b}) \geq 0$.

Expanding $W^{(k)}_l(\mathbf{b}) - W^{(k)}_{l-1}(\mathbf{b})$ gives

$$\begin{aligned}
W^{(k)}_l(\mathbf{b}) - W^{(k)}_{l-1}(\mathbf{b}) &= J^{(k)}_{l-1}(\mathbf{b}) - J^{(k)}_l(\mathbf{b}) - J^{(k)}_{l-2}(\mathbf{b}) + J^{(k)}_{l-1}(\mathbf{b}) \\
&= 2J^{(k)}_{l-1}(\mathbf{b}) - J^{(k)}_l(\mathbf{b}) - J^{(k)}_{l-2}(\mathbf{b}) \\
&= 2r(\mathbf{b}, a^{(k)}_{l-1,\mathbf{b}}) - r(\mathbf{b}, a^{(k)}_{l,\mathbf{b}}) - r(\mathbf{b}, a^{(k)}_{l-2,\mathbf{b}}) \\
&\quad + \sum_{\mathbf{o} \in \mathcal{O}} \mathbb{P}^{\mathbf{o}}_{\mathbf{b}} \left( 2J^{(k-1)}_{l-1-a^{(k)}_{l-1,\mathbf{b}}}(\mathbf{b}^{\mathbf{o}}) - J^{(k-1)}_{l-a^{(k)}_{l,\mathbf{b}}}(\mathbf{b}^{\mathbf{o}}) - J^{(k-1)}_{l-2-a^{(k)}_{l-2,\mathbf{b}}}(\mathbf{b}^{\mathbf{o}}) \right),
\end{aligned}$$

where $a^{(k)}_{l,\mathbf{b}}$ denotes the optimal action in belief state $\mathbf{b}$ with $l$ stops remaining at iteration $k$ of value iteration[18]. Hence, $W^{(k)}_l(\mathbf{b}) - W^{(k)}_{l-1}(\mathbf{b})$ depends on the actions $a^{(k)}_{l,\mathbf{b}}, a^{(k)}_{l-1,\mathbf{b}}$, and $a^{(k)}_{l-2,\mathbf{b}}$. Four combinations of these actions need to be considered:

1. If $\mathbf{b} \in \mathscr{S}^{(k)}_l \cap \mathscr{S}^{(k)}_{l-1} \cap \mathscr{S}^{(k)}_{l-2}$, then:

$$\begin{aligned}
&W^{(k)}_l(\mathbf{b}) - W^{(k)}_{l-1}(\mathbf{b}) \\
&= 2r(\mathbf{b}, \mathsf{S}) - 2r(\mathbf{b}, \mathsf{S}) + \sum_{\mathbf{o} \in \mathcal{O}} \mathbb{P}^{\mathbf{o}}_{\mathbf{b}} \left( 2J^{(k-1)}_{l-2}(\mathbf{b}^{\mathbf{o}}) - J^{(k-1)}_{l-1}(\mathbf{b}^{\mathbf{o}}) - J^{(k-1)}_{l-3}(\mathbf{b}^{\mathbf{o}}) \right) \\
&= \sum_{\mathbf{o} \in \mathcal{O}} \mathbb{P}^{\mathbf{o}}_{\mathbf{b}} \left( J^{(k-1)}_{l-2}(\mathbf{b}^{\mathbf{o}}) - J^{(k-1)}_{l-1}(\mathbf{b}^{\mathbf{o}}) - \left( J^{(k-1)}_{l-3}(\mathbf{b}^{\mathbf{o}}) - J^{(k-1)}_{l-2}(\mathbf{b}^{\mathbf{o}}) \right) \right) \\
&= \sum_{\mathbf{o} \in \mathcal{O}} \mathbb{P}^{\mathbf{o}}_{\mathbf{b}} \left( W^{(k-1)}_{l-1}(\mathbf{b}^{\mathbf{o}}) - W^{(k-1)}_{l-2}(\mathbf{b}^{\mathbf{o}}) \right),
\end{aligned}$$

which is non-negative by the induction assumption.

[18] Recall that we encode $\mathsf{S}$ and $\mathsf{C}$ with 1 and 0, respectively; see §1.4.

2. If $\mathbf{b} \in \mathscr{S}_l^{(k)} \cap \mathscr{S}_{l-1}^{(k)} \cap \mathscr{C}_{l-2}^{(k)}$, then:

$$\begin{aligned}
&W_l^k(\mathbf{b}) - W_{l-1}^{(k)}(\mathbf{b}) \\
&= 2r(\mathbf{b},\mathsf{S}) - r(\mathbf{b},\mathsf{S}) - r(\mathbf{b},\mathsf{C}) + \sum_{\mathbf{o}\in\mathcal{O}} \mathbb{P}_{\mathbf{b}}^{\mathbf{o}} \left(2J_{l-2}^{(k-1)}(\mathbf{b}^{\mathbf{o}}) - J_{l-1}^{(k-1)}(\mathbf{b}^{\mathbf{o}}) - J_{l-2}^{(k-1)}(\mathbf{b}^{\mathbf{o}})\right) \\
&= r(\mathbf{b},\mathsf{S}) - r(\mathbf{b},\mathsf{C}) + \sum_{\mathbf{o}\in\mathcal{O}} \mathbb{P}_{\mathbf{b}}^{\mathbf{o}} \left(J_{l-2}^{(k-1)}(\mathbf{b}^{\mathbf{o}}) - J_{l-1}^{(k-1)}(\mathbf{b}^{\mathbf{o}})\right),
\end{aligned}$$

which is non-negative because $\mathbf{b} \in \mathscr{S}_{l-1}^{(k)}$ (1.9).

3. If $\mathbf{b} \in \mathscr{S}_l^{(k)} \cap \mathscr{C}_{l-1}^{(k)} \cap \mathscr{C}_{l-2}^{(k)}$, then:

$$\begin{aligned}
&W_l^{(k)}(\mathbf{b}) - W_{l-1}^{(k)}(\mathbf{b}) \\
&= 2r(\mathbf{b},\mathsf{C}) - r(\mathbf{b},\mathsf{S}) - r(\mathbf{b},\mathsf{C}) + \sum_{\mathbf{o}\in\mathcal{O}} \mathbb{P}_{\mathbf{b}}^{\mathbf{o}} \left(2J_{l-1}^{(k-1)}(\mathbf{b}^{\mathbf{o}}) - J_{l-1}^{(k-1)}(\mathbf{b}^{\mathbf{o}}) - J_{l-2}^{(k-1)}(\mathbf{b}^{\mathbf{o}})\right) \\
&= r(\mathbf{b},\mathsf{C}) - r(\mathbf{b},\mathsf{S}) + \sum_{\mathbf{o}\in\mathcal{O}} \mathbb{P}_{\mathbf{b}}^{\mathbf{o}} \left(J_{l-1}^{(k-1)}(\mathbf{b}^{\mathbf{o}}) - J_{l-2}^{(k-1)}(\mathbf{b}^{\mathbf{o}})\right),
\end{aligned}$$

which is non-negative because $\mathbf{b} \in \mathscr{C}_{l-1}^{(k)}$[19].

4. If $\mathbf{b} \in \mathscr{C}_l^{(k)} \cap \mathscr{C}_{l-1}^{(k)} \cap \mathscr{C}_{l-2}^{(k)}$, then:

$$\begin{aligned}
&W_l^{(k)}(\mathbf{b}) - W_{l-1}^{(k)}(\mathbf{b}) \\
&= 2r(\mathbf{b},\mathsf{C}) - r(\mathbf{b},\mathsf{C}) - r(\mathbf{b},\mathsf{C}) + \sum_{\mathbf{o}\in\mathcal{O}} \mathbb{P}_{\mathbf{b}}^{\mathbf{o}} \left(2J_{l-1}^{(k-1)}(\mathbf{b}^{\mathbf{o}}) - J_{l}^{(k-1)}(\mathbf{b}^{\mathbf{o}}) - J_{l-2}^{(k-1)}(\mathbf{b}^{\mathbf{o}})\right) \\
&= \sum_{\mathbf{o}\in\mathcal{O}} \mathbb{P}_{\mathbf{b}}^{\mathbf{o}} \left(J_{l-1}^{(k-1)}(\mathbf{b}^{\mathbf{o}}) - J_{l}^{(k-1)}(\mathbf{b}^{\mathbf{o}}) - \left(J_{l-2}^{(k-1)}(\mathbf{b}^{\mathbf{o}}) - J_{l-1}^{(k-1)}(\mathbf{b}^{\mathbf{o}})\right)\right) \\
&= \sum_{\mathbf{o}\in\mathcal{O}} \mathbb{P}_{\mathbf{b}}^{\mathbf{o}} \left(W_{l}^{(k-1)}(\mathbf{b}^{\mathbf{o}}) - W_{l-1}^{(k-1)}(\mathbf{b}^{\mathbf{o}})\right),
\end{aligned}$$

which is non-negative by the induction assumption.

The other cases, e.g., $\mathbf{b} \in \mathscr{C}_l^{(k)} \cap \mathscr{C}_{l-1}^{(k)} \cap \mathscr{S}_{l-2}^{(k)}$, can be discarded due to the induction assumption. Therefore, $W_l(\mathbf{b}) - W_{l-1}(\mathbf{b}) \geq 0$ for all $l \in \{2,3,\dots,L\}$ and $\mathbf{b} \in \mathcal{B}$. □

*Proof of Theorem 1.1.B.* The proof follows the same chain of reasoning as used in (Corollary 12.2.2, p. 258, Krishnamurthy, 2016). We know from Lemma 1.1 that the stopping set $\mathscr{S}_1$ is a convex subset of $\mathcal{B} = [0,1]$. Thus, $\mathcal{B} = [\alpha^\star, \beta^\star]$ for some values $\alpha^\star$ and $\beta^\star$ where $0 \leq \alpha^\star \leq \beta^\star \leq 1$. Hence, it suffices to show that $\beta^\star = 1$.

[19] It follows by the same derivation as used in (1.9) except that the first inequality is reversed.

If $\mathbf{b} = L = 1$, the Bellman equation implies that

$$\begin{aligned}
\pi_1^\star(\mathbf{b}) &\in \underset{\{\mathsf{S},\mathsf{C}\}}{\arg\max} \Bigg[\underbrace{R_{\text{st}} + J_0^\star(\emptyset)}_{a=\mathsf{S}}, \underbrace{R_{\text{sla}} + R_{\text{int}} + \mathbb{E}_{\mathbf{B}'}[J_1^\star(\mathbf{B}')]}_{a=\mathsf{C}}\Bigg] \\
&= \underset{\{\mathsf{S},\mathsf{C}\}}{\arg\max} \Bigg[R_{\text{st}}, R_{\text{sla}} + R_{\text{int}} + \mathbb{E}_{\mathbf{B}'}[J_1^\star(\mathbf{B}')]\Bigg] && (J_0^\star(\emptyset) = 0) \\
&= \underset{\{\mathsf{S},\mathsf{C}\}}{\arg\max} \Bigg[R_{\text{st}}, R_{\text{sla}} + R_{\text{int}} + J_1^\star(\mathbf{b})\Bigg] && (\mathbf{b} \text{ is absorbing until the stop}) \\
&= \underset{\{\mathsf{S},\mathsf{C}\}}{\arg\max} \Bigg[R_{\text{st}}, (\tau - 1)(R_{\text{sla}} + R_{\text{int}}) + R_{\text{st}}\Bigg], && (\tau \text{ is the stopping time})
\end{aligned}$$

where

$$\begin{aligned}
&\tau > 1 \implies (\tau - 1)(R_{\text{sla}} + R_{\text{int}}) + R_{\text{st}} < R_{\text{st}} && \text{(Assumption 1.1)} \\
&\implies \underset{\{\mathsf{S},\mathsf{C}\}}{\arg\max} \Bigg[R_{\text{st}}, (\tau - 1)(R_{\text{sla}} + R_{\text{int}}) + R_{\text{st}}\Bigg] = \{\mathsf{S}\} \\
&\implies \pi_1^\star(\mathbf{b}) = \mathsf{S} \implies \mathbf{b} \in \mathscr{S}_1 \implies \beta^\star = 1 \implies \mathscr{S}_1 = [\alpha^\star, 1].
\end{aligned}$$

□

**Corollary 1.1.** *The stopping set $\mathscr{S}_l$ is connected for all $l \in \{1, \dots, L\}$.*

*Proof.* We use the same shorthand notations as in the proof of Theorem 1.1.A. Since $\mathcal{B} = [0, 1]$ (§1.4), the beliefs are totally ordered according to the standard order on $\mathbb{R}$. This fact together with the properties $\mathscr{S}_{l-1} \subseteq \mathscr{S}_l$ (Thm. 1.1.A) and $\mathscr{S}_1 = [\alpha_1^\star, 1]$ (Thm. 1.1.B) means that $\mathscr{S}_l$ is connected if $\mathbf{b} \in \mathscr{S}_l \implies \mathbf{b}' \in \mathscr{S}_l$ for all $\mathbf{b}' \geq \mathbf{b}$ and $l = L, L-1, \dots, 1$. We know from (1.9) that

$$\mathbf{b} \in \mathscr{S}_l \iff r(\mathbf{b}, \mathsf{S}) - r(\mathbf{b}, \mathsf{C}) + \sum_{\mathbf{o} \in \mathcal{O}} \mathbb{P}_{\mathbf{b}}^{\mathbf{o}} \underbrace{\left(J_{l-1}^\star(\mathbf{b}^{\mathbf{o}}) - J_l^\star(\mathbf{b}^{\mathbf{o}})\right)}_{=W_l(\mathbf{b}^{\mathbf{o}})} \geq 0.$$

Since a) $\mathbb{P}_{\mathbf{b}'}^{\mathbf{o}}$ stochastically dominates $\mathbb{P}_{\mathbf{b}}^{\mathbf{o}}$ in the first order for $\mathbf{b}' \geq \mathbf{b}$; and b) $\mathbb{B}(\mathbf{b}, a, \mathbf{o})$ is non-decreasing in both $\mathbf{b}$ and $\mathbf{o}$ (Lemma 1.4), it suffices to show that $r(\mathbf{b}, \mathsf{S}) - r(\mathbf{b}, \mathsf{C}) + W_l(\mathbf{b})$ is non-decreasing in $\mathbf{b}$ for all $l \in \{1, 2, \dots, L\}$. We show this property by mathematical induction on $k = 0, 1, \dots$, where $k$ is the iteration of value iteration (Smallwood and Sondik, 1973).

For iteration $k = 0$, we have $J_l^{(0)}(\mathbf{b}) = 0 \ \forall \mathbf{b} \in \mathcal{B}$ and $l \in \{1, \dots, L\}$. Thus,

$$\begin{aligned}
r(\mathbf{b}, \mathsf{S}) - r(\mathbf{b}, \mathsf{C}) + W_l^{(0)}(\mathbf{b}) &= r(\mathbf{b}, \mathsf{S}) - r(\mathbf{b}, \mathsf{C}) + J_{l-1}^{(0)}(\mathbf{b}) - J_l^{(0)}(\mathbf{b}) \\
&= r(\mathbf{b}, \mathsf{S}) - r(\mathbf{b}, \mathsf{C}) \qquad \forall l \in \{1, 2, \dots, L\}, \mathbf{b} \in \mathcal{B}.
\end{aligned}$$

Since $r(\mathbf{b}, \mathsf{S}) - r(\mathbf{b}, \mathsf{C})$ is increasing in $\mathbf{b}$ (Lemma 1.3), the inductive base case holds. Assume by induction that $r(\mathbf{b}, \mathsf{S}) - r(\mathbf{b}, \mathsf{C}) + W_l^{(k-1)}(\mathbf{b})$ is non-decreasing in $\mathbf{b}$ for all $l \in \{1, 2, \dots, L\}$. We will show that this assumption implies that $r(\mathbf{b}, \mathsf{S}) - r(\mathbf{b}, \mathsf{C}) + W_l^{(k)}(\mathbf{b})$ is non-decreasing in $\mathbf{b}$ for all $l \in \{1, 2, \dots, L\}$.

Expanding $r(\mathbf{b}, \mathsf{S}) - r(\mathbf{b}, \mathsf{C}) + W_l^{(k)}(\mathbf{b})$ gives

$$\begin{aligned}
r(\mathbf{b}, \mathsf{S}) - r(\mathbf{b}, \mathsf{C}) + W_l^{(k)}(\mathbf{b}) &= r(\mathbf{b}, \mathsf{S}) - r(\mathbf{b}, \mathsf{C}) + J_{l-1}^{(k)}(\mathbf{b}) - J_l^{(k)}(\mathbf{b}) \\
&= r(\mathbf{b}, \mathsf{S}) - r(\mathbf{b}, \mathsf{C}) + r(\mathbf{b}, a_{l-1,\mathbf{b}}^{(k)}) - r(\mathbf{b}, a_{l,\mathbf{b}}^{(k)}) \\
&\quad + \sum_{\mathbf{o} \in \mathcal{O}} \mathbb{P}_{\mathbf{b}}^{\mathbf{o}} \left( J_{l-1-a_{l-1,\mathbf{b}}^{(k)}}^{(k-1)}(\mathbf{b}^{\mathbf{o}}) - J_{l-a_{l,\mathbf{b}}^{(k)}}^{(k-1)}(\mathbf{b}^{\mathbf{o}}) \right),
\end{aligned}$$

where $a_{l,\mathbf{b}}^{(k)}$ denotes the optimal action in belief state $\mathbf{b}$ with $l$ stops remaining at iteration $k$ of value iteration[20]. Hence, $W_l^{(k)}(\mathbf{b})$ depends on the actions $a_{l,\mathbf{b}}^{(k)}$ and $a_{l-1,\mathbf{b}}^{(k)}$. Since $\mathscr{S}_{l-1}^{(k)} \subseteq \mathscr{S}_l^{(k)}$ (Thm. 1.1.A) and $\mathscr{S}_1^{(k)} = [\alpha_1^\star, 1]$ (Thm. 1.1.B), only three combinations of these actions need to be considered:

1. If $\mathbf{b} \in \mathscr{S}_l^{(k)} \cap \mathscr{S}_{l-1}^{(k)}$, then:

$$\begin{aligned}
&r(\mathbf{b}, \mathsf{S}) - r(\mathbf{b}, \mathsf{C}) + W_l^{(k)}(\mathbf{b}) \\
&= r(\mathbf{b}, \mathsf{S}) - r(\mathbf{b}, \mathsf{C}) + r(\mathbf{b}, \mathsf{S}) - r(\mathbf{b}, \mathsf{S}) + \sum_{\mathbf{o} \in \mathcal{O}} \mathbb{P}_{\mathbf{b}}^{\mathbf{o}} \left( J_{l-2}^{(k-1)}(\mathbf{b}^{\mathbf{o}}) - J_{l-1}^{(k-1)}(\mathbf{b}^{\mathbf{o}}) \right) \\
&= \sum_{\mathbf{o} \in \mathcal{O}} \mathbb{P}_{\mathbf{b}}^{\mathbf{o}} \left( r(\mathbf{b}, \mathsf{S}) - r(\mathbf{b}, \mathsf{C}) + W_{l-1}^{(k-1)}(\mathbf{b}^{\mathbf{o}}) \right),
\end{aligned}$$

   which is non-decreasing in $\mathbf{b}$ by the induction assumption and Lemma 1.4.

2. If $\mathbf{b} \in \mathscr{S}_l^{(k)} \cap \mathscr{C}_{l-1}^{(k)}$, then:

$$\begin{aligned}
&r(\mathbf{b}, \mathsf{S}) - r(\mathbf{b}, \mathsf{C}) + W_l^{(k)}(\mathbf{b}) \\
&= r(\mathbf{b}, \mathsf{S}) - r(\mathbf{b}, \mathsf{C}) + r(\mathbf{b}, \mathsf{C}) - r(\mathbf{b}, \mathsf{S}) + \sum_{\mathbf{o} \in \mathcal{O}} \mathbb{P}_{\mathbf{b}}^{\mathbf{o}} \left( J_{l-1}^{(k-1)}(\mathbf{b}^{\mathbf{o}}) - J_{l-1}^{(k-1)}(\mathbf{b}^{\mathbf{o}}) \right) \\
&= 0.
\end{aligned}$$

3. If $\mathbf{b} \in \mathscr{C}_l^{(k)} \cap \mathscr{C}_{l-1}^{(k)}$, then:

$$\begin{aligned}
&r(\mathbf{b}, \mathsf{S}) - r(\mathbf{b}, \mathsf{C}) + W_l^{(k)}(\mathbf{b}) \\
&= r(\mathbf{b}, \mathsf{S}) - r(\mathbf{b}, \mathsf{C}) + r(\mathbf{b}, \mathsf{C}) - r(\mathbf{b}, \mathsf{C}) + \sum_{\mathbf{o} \in \mathcal{O}} \mathbb{P}_{\mathbf{b}}^{\mathbf{o}} \left( J_{l-1}^{(k-1)}(\mathbf{b}^{\mathbf{o}}) - J_{l}^{(k-1)}(\mathbf{b}^{\mathbf{o}}) \right)
\end{aligned}$$

[20] Recall that we encode $\mathsf{S}$ and $\mathsf{C}$ with 1 and 0, respectively; see §1.4.

$$= \sum_{\mathbf{o} \in \mathcal{O}} \mathbb{P}^{\mathbf{o}}_{\mathbf{b}} \left( r(\mathbf{b}, \mathsf{S}) - r(\mathbf{b}, \mathsf{C}) + W_l^{(k-1)}(\mathbf{b}^{\mathbf{o}}) \right),$$

which is non-decreasing in $\mathbf{b}$ by the induction assumption and Lemma 1.4.

□

*Proof of Theorem 1.1.C.* Since a) $\mathcal{B} = [0, 1]$; b) $z$ (1.2) is TP-2 by assumption; and c) $f_l(\cdot \mid \cdot, a)$ (1.1) is TP-2 for all $a \in \mathcal{A}$ and $l \in \{1, \ldots, L\}$ by Lemma 1.2, it follows from Cor. 1.1 that $\mathscr{S}_l$ is a connected subset of $[0, 1]$ for $l \in \{1, \ldots, L\}$. Further, from Thm. 1.1.B we know that $\mathscr{S}_1 = [\alpha_1^\star, 1]$ and from Thm. 1.1.A we know that $\mathscr{S}_l \subseteq \mathscr{S}_{l+1}$ for $l \in \{1, \ldots, L-1\}$. Therefore, $1 \in \mathscr{S}_l$ for all $l \in \{1, \ldots, L\}$. We conclude that $\mathscr{S}_l = [\alpha_l^\star, 1]$ for $l \in \{1, \ldots, L\}$ and that $\alpha_l^\star \geq \alpha_{l+1}^\star$ for $l \in \{1, \ldots, L-1\}$. □

## B Hyperparameters

The hyperparameters used for the evaluation are listed in Table 1.6 and were obtained through grid search.

| *Parameters for the POMDP* | *Values* |
|---|---|
| $\Delta x_{max}, \Delta y_{max}, \Delta z_{max}$ | $6 \cdot 10^2$, $3 \cdot 10^2$, $10^2$ |
| $p, R_{\text{st}}, R_{\text{int}}, R_{\text{sla}}$ | 0.01, 50, −10, 1 |
| *Parameters for T-SPSA* | *Values* |
| $c, \epsilon, \lambda, A, a$ | 1, 0.101, 0.602, 100, 1 |
| *Parameters for PPO* (Alg. 1, Schulman et al., 2017) | *Values* |
| lr $\alpha$, batch, # layers, # neurons, clip $\epsilon$, GAE $\lambda$, ent-coef | $10^{-4}$, $4 \cdot 10^3 t$, 3, 32, 0.2, 0.95, $10^{-4}$ |
| activation function | RELU |
| *Parameters for HSVI* (Alg. 1, Smith and Simmons, 2004) | *Values* |
| $\epsilon$ | 0.01 |
| *Parameter for Shiryaev's algorithm* (Shiryaev, 1963) | *Value* |
| $\alpha$ | 0.75 |

***Table 1.6:*** *Hyperparameters of the POMDP and the algorithms used for evaluation.*

## C Configuration of the Infrastructure in Fig. 1.2

The configurations of the components $N_1, \ldots, N_{31}$ of the target infrastructure (Fig. 1.2) are available in Table 1.7 on the next page. (Note that each component in Fig. 1.2 is labeled with an identifier $N_i$.)

| *ID (s)* | *OS:Services:Exploitable Vulnerabilities* |
|---|---|
| $N_1$ | UBUNTU20:SNORT(community ruleset v2.9.17.1),SSH:-. |
| $N_2$ | UBUNTU20:SSH,HTTP Erl-Pengine,DNS:SSH-PW. |
| $N_4$ | UBUNTU20:HTTP,TELNET,SSH:TELNET-PW. |
| $N_{10}$ | UBUNTU20:FTP,MONGODB,SMTP,TOMCAT,TS3,SSH:FTP-PW. |
| $N_{12}$ | JESSIE:TS3,TOMCAT,SSH:CVE-2010-0426,SSH-PW. |
| $N_{17}$ | WHEEZY:APACHE2,SNMP,SSH:CVE-2014-6271. |
| $N_{18}$ | DEB9.2:IRC,APACHE2,SSH:SQL injection. |
| $N_{22}$ | JESSIE:PROFTPD,SSH,APACHE2,SNMP:CVE-2015-3306. |
| $N_{23}$ | JESSIE:APACHE2,SMTP,SSH:CVE-2016-10033. |
| $N_{24}$ | JESSIE:SSH:CVE-2015-5602,SSH-PW. |
| $N_{25}$ | JESSIE: ELASTICSEARCH,APACHE2,SSH,SNMP:CVE-2015-1427. |
| $N_{27}$ | JESSIE:SAMBA,NTP,SSH:CVE-2017-7494. |
| $N_3$,$N_{11}$,$N_5$-$N_9$ | UBUNTU20:SSH,SNMP,POSTGRESQL,NTP:-. |
| $N_{13-16}$,$N_{19-21}$,$N_{26}$,$N_{28-31}$ | UBUNTU20:NTP, IRC, SNMP, SSH, POSTGRESQL:-. |

***Table 1.7:*** *Configuration of the target infrastructure (Fig. 1.2).*

## D The T-SPSA Algorithm

Algorithm 1.1 contains the pseudocode of T-SPSA.

***Algorithm 1.1:*** ***T****hreshold-****S****imultaneous* ***P****erturbation* ***S****tochastic* ***A****pproximation.*

**Input:** $\mathcal{M}_{\mathcal{P}}, \boldsymbol{\theta}_{(1)} \in \mathbb{R}^L$: the POMDP, initial $L$ thresholds.
$N$: number of iterations, $a, c, \lambda, A, \epsilon$: scalar coefficients.
**Output:** $\boldsymbol{\theta}^{(N+1)}$: learned threshold vector.

**procedure** T-SPSA($\mathcal{M}_{\mathcal{P}}$, $\boldsymbol{\theta}_{(1)}$, $N$, $a$, $c$, $\lambda$, $A$, $\epsilon$)
  **for** $n \in \{1, \dots, N\}$ **do**
    $a_n \leftarrow \frac{a}{(n+A)^{\epsilon}}, c_n \leftarrow \frac{c}{n^{\lambda}}$.
    **for** $i \in \{1, \dots, L\}$ **do**
      $\boldsymbol{\Delta}_i^{(n)} \sim \mathcal{U}(\{-1, 1\})$. ▷ Discrete uniform distribution on $\{-1, 1\}$.
    $R_{\text{high}} \sim \hat{J}(\boldsymbol{\theta}^{(n)} + c_n \boldsymbol{\Delta}^{(n)})$, $R_{\text{low}} \sim \hat{J}(\boldsymbol{\theta}^{(n)} - c_n \boldsymbol{\Delta}^{(n)})$.
    **for** $i \in \{1, \dots, L\}$ **do**
      $\left(\hat{\nabla}_{\boldsymbol{\theta}^{(n)}} J(\boldsymbol{\theta}^{(n)})\right)_i \leftarrow \frac{R_{\text{high}} - R_{\text{low}}}{2c_n \boldsymbol{\Delta}_i^{(n)}}$. ▷ SPSA gradient estimator.
    $\boldsymbol{\theta}^{(n+1)} \leftarrow \boldsymbol{\theta}^{(n)} + a_n \hat{\nabla}_{\boldsymbol{\theta}^{(n)}} J(\boldsymbol{\theta}^{(n)})$. ▷ Stochastic approximation.
  **return** $\boldsymbol{\theta}^{(N+1)}$.

# Paper 2[†]

## LEARNING NEAR-OPTIMAL INTRUSION RESPONSE AGAINST DYNAMIC ATTACKERS

Kim Hammar and Rolf Stadler

**Abstract**

We study automated intrusion response and formulate the interaction between an attacker and a defender as an optimal stopping game where attack and defense strategies evolve through stochastic approximation and fictitious play. The game-theoretic modeling enables us to find effective defender strategies against a dynamic attacker, i.e., an attacker that adapts its strategy in response to the defender strategy. Further, the optimal stopping formulation allows us to prove that the best responses have threshold properties. To obtain near-optimal defender strategies, we develop **T**hreshold-**F**ictitious **P**lay (T-FP), an algorithm that learns equilibria through stochastic approximation. We show that T-FP outperforms a state-of-the-art algorithm for our use case. The experimental part of this investigation includes two systems: a simulator where defender strategies are incrementally learned and a digital twin where statistics are collected that drive simulation runs and where learned strategies are evaluated. We argue that this methodology can produce effective defender strategies for a practical IT infrastructure.

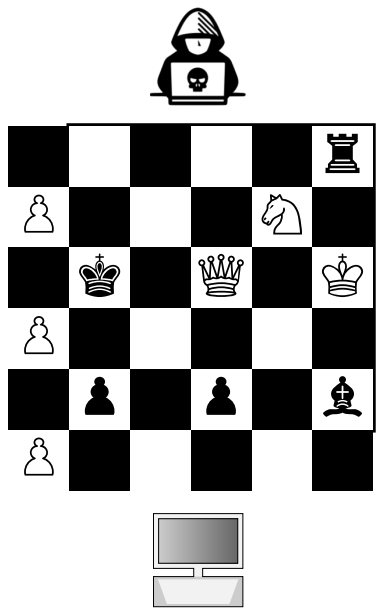

*In life, as in a game of chess, forethought wins.*

— Charles Buxton ***1873**, Notes of thought.*

## 2.1 Introduction

THIS paper extends the intrusion prevention use case in Paper 1 to a *response* use case that involves a *dynamic* attacker, i.e., an attacker that adapts its strategy based on the defender strategy. Like in Paper 1, we consider a defender that monitors the infrastructure in Fig. 1.2 and can take a fixed number of actions to respond to potential attacks. However, unlike the previous case, this paper considers a dynamic attacker that strategically decides when to begin and end its intrusion to avoid detection. This contrasts with Paper 1, where it is assumed that the attacker starts the intrusion at a random time and continues until it is stopped. We formulate the new use case as an *optimal stopping game*, namely a stochastic game where both players face an optimal stopping problem (Dynkin, 1969). The stopping problem for the defender is to decide when to take defensive actions and the stopping problem for the attacker is to decide when to begin and end its intrusion. This problem formulation enables us to gain insight into the structure of best responses, which we prove to have threshold properties. Based on these properties, we design **T**hreshold-**F**ictitious **P**lay (T-FP), an efficient algorithm that iteratively computes near-optimal defender strategies against a dynamic attacker. We show that T-FP outperforms a state-of-the-art algorithm for our use case and that the obtained strategies are effective on a digital twin[2].

## 2.2 Formalizing the Intrusion Response Use Case

We formulate the intrusion response use case described above as a zero-sum stochastic game with one-sided partial observability (a POSG)[3]

$$\Gamma \triangleq \langle \mathcal{N}, \mathcal{S}, (\mathcal{A}_{\mathrm{k}})_{\mathrm{k}\in\mathcal{N}}, f, r, \gamma, \mathbf{b}_1, T, z, \mathcal{O}\rangle. \quad \text{(Def. 3.1, Horák et al., 2023)} \quad (2.1)$$

[2]The digital twin is created using CSLE, as described in the methodology chapter.

[3]The components of a POSG are defined the background chapter; see (19).

The game has two players: the (D)efender and the (A)ttacker. In the following, we describe the components of the game, its evolution, and the players' objectives. The requisite notation is listed in Table 2.1.

| *Notation(s)* | *Description* |
|---|---|
| $\Gamma$ | The intrusion response POSG (2.1). |
| $\mathrm{D}, \mathrm{A}$ | The defender player and the attacker player. |
| $t, T, \gamma$ | Time step, time horizon, and discount factor. |
| $l_t$ | Defender stops remaining at time step $t$. |
| $L$ | Maximum number of defender stops. |
| $\pi_{\mathrm{D}}, \pi_{\mathrm{A}}$ | Defender and attacker strategies. |
| $\tilde{\pi}_{\mathrm{D}}, \tilde{\pi}_{\mathrm{A}}$ | Best response strategies (2.6a)–(2.6b). |
| $\pi^{\star}_{\mathrm{D}}, \pi^{\star}_{\mathrm{A}}$ | Optimal (equilibrium) strategies (2.7). |
| $\mathcal{N}, \mathcal{S}, \mathcal{O}$ | Sets of players, states, and observations. |
| $\mathcal{A}_{\mathrm{D}}, \mathcal{A}_{\mathrm{A}}$ | Sets of defender and attacker actions. |
| $f_l, r_l, z$ | Transition (2.2), reward (2.3), and observation (2.4) functions. |
| $s_t, o_t, \mathbf{a}_t = (a^{\mathrm{D}}_t, a^{\mathrm{A}}_t)$ | State, observation, and actions at time step $t$. |
| $\mathbf{b}_t \in \mathcal{B}, r_t$ | Defender belief and reward at time step $t$. |
| $\mathsf{S}, \mathsf{C} = 1, 0$ | Stop and continue actions. |
| $\tau_{\mathrm{k},i}$ | $i$th stopping time of player k (a realization of the r.v. $\mathcal{T}_{\mathrm{k},i}$). |
| $\mathscr{B}_{\mathrm{D}}, \mathscr{B}_{\mathrm{A}}$ | Best response correspondences (2.6a)–(2.6b). |
| $J_{\mathrm{D}}, J_{\mathrm{A}}$ | Defender and attacker objectives (2.5a)–(2.5b). |
| $\mathcal{M}^P, \mathcal{M}$ | Best response POMDP and MDP for players D and A. |
| $J^{\star}_{l,\pi_{\mathrm{A}}}, J^{\star}_{l,\pi_{\mathrm{D}}}$ | Value functions of $\mathcal{M}^P$ and cost-to-go function of $\mathcal{M}$. |
| $\mathscr{S}^{(\mathrm{k})}, \mathscr{C}^{(\mathrm{k})}$ | Stopping and continuation sets of player k. |
| $S_t, \mathbf{A}_t, O_t$ | Random variables with realizations $s_t, \mathbf{a}_t, o_t$. |
| $R_t, \mathbf{B}_t$ | Random variable with realization $r_t$, and random vector with realization $\mathbf{b}_t$. |

***Table 2.1:*** *Variables and symbols used in the model.*

### Time horizon $T$

The time horizon $T > 1$ is a random variable representing the time when the attacker stops its intrusion or is prevented, depending on which event occurs first.

### State space $\mathcal{S}$

The game has three states: $s_t = 0$ if no intrusion occurs, $s_t = 1$ if an intrusion is ongoing, and $s_T = \emptyset$ if the game has ended. Hence, $\mathcal{S} \triangleq \{0, 1, \emptyset\}$. The initial state is $s_1 = 0$. Therefore, the initial state distribution $\mathbf{b}_1 \in \Delta(\mathcal{S})$ is the degenerate distribution $\mathbf{b}_1(0) = 1$.

### Action spaces $\mathcal{A}_{\mathrm{k}}$

Each player $\mathrm{k} \in \mathcal{N}$ can invoke two actions: ($\mathsf{S}$)top and ($\mathsf{C}$)ontinue. The action spaces are thus $\mathcal{A}_{\mathrm{D}} \triangleq \mathcal{A}_{\mathrm{A}} \triangleq \{\mathsf{S}, \mathsf{C}\}$. Executing action $\mathsf{S}$ triggers a change in the game, while action $\mathsf{C}$ is a passive action. The attacker can invoke the stop action twice: the first to start the intrusion and the second to terminate it. The defender can invoke the stop action $L \geq 1$ times. Each invocation corresponds to a defensive action against a possible intrusion. The number of stop actions remaining to the defender at time $t$ is known to both players and is denoted by $l_t \in \{1, \dots, L\}$. Using the encoding $(\mathsf{S}, \mathsf{C}) \triangleq (1, 0)$, we can write $l_{t+1} = l_t - a_t^{(\mathrm{D})}$, where $a_t^{(\mathrm{D})}$ is the defender action at time $t$. At each time step, the attacker and the defender simultaneously choose their actions $\mathbf{a}_t \triangleq (a_t^{(\mathrm{D})}, a_t^{(\mathrm{A})})$, where $a_t^{(\mathrm{k})} \in \mathcal{A}_{\mathrm{k}}$.

### Observation space $\mathcal{O}$

The attacker has complete observability[4] and knows the game state, the defender's actions, and the defender's observations. In contrast, the defender has a limited set of observations $o_t \in \mathcal{O}$, where $\mathcal{O}$ is a finite set[5].

### Belief space $\mathcal{B}$

Based on its history $\mathbf{h}_t^{(\mathrm{D})}$ (20)[6], the defender computes the *belief state* $\mathbf{b}_t(s_t) \triangleq \mathbb{P}[S_t = s_t \mid \mathbf{h}_t^{(\mathrm{D})}] \in \mathcal{B}$ through (22), as defined in the background chapter. Since $\emptyset$ is a terminal state, the only two reachable states while $t < T$ are 0 and 1; hence, $\mathcal{B} = \Delta(\{0, 1\}) = [0, 1]$; see Fig. 16 in the background chapter.

**Remark 2.1.** Due to the complete observability of the attacker, it can compute the defender's belief $\mathbf{b}_t$ using the same equation as the defender (22).

### Transition function $f_l(s' \mid s, a^{(\mathrm{D})}, a^{(\mathrm{A})})$

At each time step $t$, a transition from $s_t$ to $s_{t+1}$ occurs with probability $f_l(s_{t+1} \mid s_t, a_t^{(\mathrm{D})}, a_t^{(\mathrm{A})})$, where $f_l$ is defined as

$$f_{l>1}(0 \mid 0, \mathsf{S}, \mathsf{C}) \triangleq f_l(0 \mid 0, \mathsf{C}, \mathsf{C}) = 1 \tag{2.2a}$$

$$f_{l>1}(1 \mid 1, \cdot, \mathsf{C}) \triangleq f_l(1 \mid 1, \mathsf{C}, \mathsf{C}) \triangleq 1 - \phi_l \tag{2.2b}$$

$$f_{l>1}(1 \mid 0, \cdot, \mathsf{S}) \triangleq f_l(1 \mid 0, \mathsf{C}, \mathsf{S}) \triangleq 1 \tag{2.2c}$$

$$f_{l>1}(\emptyset \mid 1, \cdot, \mathsf{C}) \triangleq f_l(\emptyset \mid 1, \mathsf{C}, \mathsf{C}) \triangleq \phi_l \tag{2.2d}$$

$$f_{l=1}(\emptyset \mid \cdot, \mathsf{S}, \cdot) \triangleq f_l(\emptyset \mid \emptyset, \cdot, \cdot) \triangleq f_l(\emptyset \mid 1, \cdot, \mathsf{S}) \triangleq 1. \tag{2.2e}$$

All other state transitions have probability 0. (2.2a)–(2.2b) define the probabilities of the recurrent transitions $0 \rightarrow 0$ and $1 \rightarrow 1$. The game stays in state 0 with probability 1 if the attacker selects action $\mathsf{C}$ and $l_t - a_t^{(\mathrm{D})} > 0$. Similarly, the game stays in state 1 with probability $1 - \phi_l$ if the attacker chooses action $\mathsf{C}$ and $l_t - a_t^{(\mathrm{D})} > 0$. Here, $\phi_l$ denotes the probability that the defender stops the intrusion, which is a parameter of the use case. The intrusion can be stopped at any time step either because the attacker terminates the intrusion or as a consequence of previous stop actions by the defender, i.e., the effect of a defensive action is non-immediate. We assume that $\phi_l$ increases with each stop action that the defender takes.

[4] See Assumption 6 in the problem chapter.

[5] In our use case, $o_t$ relates to the weighted sum of IDS alerts triggered during time step $t$. We focus on the IDS alert metric as it provides sufficient information for detecting the type of intrusions we consider; see Appendix C for a comparison between different metrics.

[6] The history is defined in the background chapter; see (20).

**Remark 2.2.** We model the effect of a defensive stop action as non-immediate to capture that the full effect of the defense may not be realized until the attacker reaches a certain stage in its intrusion.

(2.2c) captures the transition $0 \rightarrow 1$, which occurs when the attacker chooses action $\mathsf{S}$ and $l_t - a_t^{(\mathrm{D})} > 0$. (2.2d)–(2.2e) define the probabilities of the transitions to the terminal state $\emptyset$, which is reached in three cases: $(i)$ when $l_t = 1$ and the defender takes the final stop action $\mathsf{S}$ (i.e., when $l_t - a_t^{(\mathrm{D})} = 0$); $(ii)$ when the intrusion is stopped by the defender with probability $\phi_l$; and $(iii)$ when $s_t = 1$ and the attacker terminates the intrusion ($a_t^{(\mathrm{A})} = \mathsf{S} = 1$). The transition diagram is shown in Fig. 2.1.

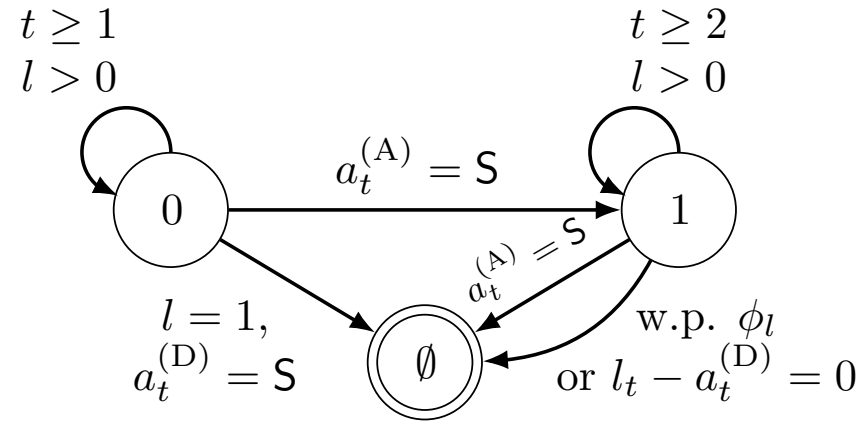

***Figure 2.1:*** *State transition diagram of a game episode: each disk represents a state; an arrow represents a state transition; a label indicates the conditions for the state transition; a game episode starts in state $s_1 = 0$ with $l = L$ and ends in state $s_T = \emptyset$.*

### Reward function $r_l(s, a^{(\mathrm{D})}, a^{(\mathrm{A})})$

At time step $t$, the defender receives the reward $r_t = r_l(s_t, a_t^{(\mathrm{D})}, a_t^{(\mathrm{A})})$ and the attacker receives the reward[7] $-r_t$. The reward function is parameterized by the defender's reward for stopping an intrusion ($\mathrm{R_{st}} > 0$), its cost of taking a defensive action ($\mathrm{R_{cost}} < 0$), and its cost while an intrusion occurs ($\mathrm{R_{int}} < 0$):

$$r_l(\emptyset, \cdot) \triangleq 0 \tag{2.3a}$$

$$r_l(1, \cdot, \mathsf{S}) \triangleq 0 \tag{2.3b}$$

$$r_l(0, \mathsf{C}, \cdot) \triangleq 0 \tag{2.3c}$$

$$r_l(0, \mathsf{S}, \cdot) \triangleq \frac{\mathrm{R_{cost}}}{l_t} \qquad l_t \in \{1, 2, \ldots, L\} \tag{2.3d}$$

[7]In other words, the game is zero-sum; see Assumption 5 in the problem chapter.

$$r_l(1, \mathsf{S}, \mathsf{C}) \triangleq \frac{\mathrm{R}_{\mathrm{st}}}{l_t} \qquad l_t \in \{1, 2, \ldots, L\} \tag{2.3e}$$

$$r_l(1, \mathsf{C}, \mathsf{C}) \triangleq \mathrm{R}_{\mathrm{int}}. \tag{2.3f}$$

(2.3a)–(2.3b) state that the reward is zero in the terminal state and when the attacker terminates an intrusion. (2.3c) states that the defender incurs no cost when no attack occurs and it does not take a defensive action. (2.3d) indicates that the defender incurs a cost when taking a defensive action if no intrusion is ongoing. (2.3e) states that the defender receives a reward when taking a stop action while an intrusion occurs. Lastly, (2.3f) indicates that the defender incurs a cost for each time step during which an intrusion occurs. (The constants $R_{\mathrm{st}}, R_{\mathrm{cost}}$, and $R_{\mathrm{int}}$ should be configured to satisfy Assumption 2 in the background chapter.)

### Observation function $z$

At time step $t$, $o \in \mathcal{O}$ is drawn from a random variable $O$ with distribution

$$z(o \mid s) \triangleq \mathbb{P}[O = o \mid S = s]. \tag{2.4}$$

Note that this distribution depends on the clients that consume services of the infrastructure, i.e., the clients are implicitly modeled by $z$ (2.4).

### Player strategies $\pi_{\mathrm{k}}$

A (behavior) defender strategy is a function $\pi_{\mathrm{D}} \in \Pi_{\mathrm{D}} \triangleq \{1, \ldots, L\} \times \mathcal{B} \rightarrow \Delta(\mathcal{A}_{\mathrm{D}})$. Likewise, a (behavior) attacker strategy is a function $\pi_{\mathrm{A}} \in \Pi_{\mathrm{A}} \triangleq \{1, \ldots, L\} \times \mathcal{B} \times \mathcal{S} \rightarrow \Delta(\mathcal{A}_{\mathrm{A}})$. The *strategy profile* is $\boldsymbol{\pi} \triangleq (\pi_{\mathrm{D}}, \pi_{\mathrm{A}})$. (Note that the strategies depend on $l$, which is a known parameter; for ease of notation, we use $\pi_{\mathrm{D}}(\mathbf{b})$ and $\pi_{\mathrm{A}}(\mathbf{b}, s)$ as shorthands for $\pi_{\mathrm{D}}(\mathbf{b}, l)$ and $\pi_{\mathrm{A}}(\mathbf{b}, s, l)$, respectively.)

### Objective functionals $J_{\mathrm{k}}$

The goal of the defender is to *maximize* the expected discounted cumulative reward over the time horizon $T$. Similarly, the attacker's goal is to *minimize* the same quantity[8]. Consequently, the objective functionals $J_{\mathrm{D}}$ and $J_{\mathrm{A}}$ are

$$J_{\mathrm{D}}(\pi_{\mathrm{D}}, \pi_{\mathrm{A}}) \triangleq \mathbb{E}_{(\pi_{\mathrm{D}}, \pi_{\mathrm{A}})}\left[\sum_{t=1}^{T} \gamma^{t-1} r_l(S_t, \mathbf{A}_t)\right] \tag{2.5a}$$

$$J_{\mathrm{A}}(\pi_{\mathrm{D}}, \pi_{\mathrm{A}}) \triangleq -J_{\mathrm{D}}(\pi_{\mathrm{D}}, \pi_{\mathrm{A}}), \tag{2.5b}$$

where $\gamma \in [0, 1)$ is the discount factor and $\mathbb{E}_{(\pi_{\mathrm{D}}, \pi_{\mathrm{A}})}$ denotes the expectation of the random vectors $(\mathbf{H}_t^{(\mathrm{D})}, \mathbf{H}_t^{(\mathrm{A})})_{t \in \{1, \ldots, T\}}$ (20) when the game is played according to the strategy profile $(\pi_{\mathrm{D}}, \pi_{\mathrm{A}})$.

[8] It follows from the zero-sum assumption; see Assumption 5 in the problem chapter.

### Best response strategies $\tilde{\pi}_{\text{k}}$

A defender strategy $\tilde{\pi}_{\text{D}}$ is a *best response* against $\pi_{\text{A}}$ if it *maximizes* $J_{\text{D}}$ (2.5a). Conversely, an attacker strategy $\tilde{\pi}_{\text{A}}$ is a best response against $\pi_{\text{D}}$ if it *minimizes* $J_{\text{D}}$ (2.5b). Hence, the best response correspondences are

$$\mathscr{B}_{\text{D}}(\pi_{\text{A}}) = \underset{\pi_{\text{D}} \in \Pi_{\text{D}}}{\arg\max}\, J_{\text{D}}(\pi_{\text{D}}, \pi_{\text{A}}) \tag{2.6a}$$

$$\mathscr{B}_{\text{A}}(\pi_{\text{D}}) = \underset{\pi_{\text{A}} \in \Pi_{\text{A}}}{\arg\min}\, J_{\text{D}}(\pi_{\text{D}}, \pi_{\text{A}}). \tag{2.6b}$$

**Remark 2.3.** Throughout this paper, we write max min instead of sup inf as the optimization problems that we consider have solutions; see Thm. 2.1 below.

**Remark 2.4.** Computation of (2.6a) and (2.6b) are equivalent to computation of optimal strategies in a POMDP[9] and an MDP[10], respectively.

### Optimal (equilibrium) strategies $\pi^{\star}_{\text{k}}$

An optimal defender strategy $\pi^{\star}_{\text{D}}$ is a best response against any attacker strategy that *minimizes* $J_{\text{D}}$. Similarly, an optimal attacker strategy $\pi^{\star}_{\text{A}}$ is a best response against any defender strategy that *maximizes* $J_{\text{D}}$. Hence, when both players follow optimal strategies, they play best responses against each other:

$$(\pi^{\star}_{\text{D}}, \pi^{\star}_{\text{A}}) \in \mathscr{B}_{\text{D}}(\pi^{\star}_{\text{A}}) \times \mathscr{B}_{\text{A}}(\pi^{\star}_{\text{D}}). \tag{2.7}$$

Since no player has an incentive to change its strategy, $(\pi^{\star}_{\text{D}}, \pi^{\star}_{\text{A}})$ is a Nash equilibrium (Eq. 1, Nash, 1951). Such a strategy pair can also form a stronger equilibrium, namely a Perfect Bayesian equilibrium (PBE); see Def. 4 in the background chapter.

**Remark 2.5** (Finite and stationary game)**.** $\Gamma$ satisfies assumptions 1–6 in the background chapter and the problem chapter.

**Remark 2.6** (Equilibrium uniformity)**.** Since $\Gamma$ is zero-sum, every equilibrium leads to the same value (Ch. 3, von Neumann and Morgenstern, 1944), regardless of the strategies employed at equilibrium. Consequently, we do not need to concern ourselves with equilibrium selection.

## 2.3 Game-Theoretic Analysis

Finding optimal strategies that satisfy (2.7) is equivalent to finding a perfect Bayesian equilibrium (PBE) for the POSG $\Gamma$. We know from Thm. 3 in the background chapter that $\Gamma$ has at least one PBE. In this section, we first analyze the structure of best responses in $\Gamma$ using optimal stopping theory, and then we describe an efficient fictitious play algorithm for approximating equilibria.

[9] The components of a POMDP are defined the background chapter; see (15).

[10] The components of an MDP are defined the background chapter; see (1).

### Analyzing best responses using optimal stopping theory

A best response for the defender is obtained by solving a POMDP $\mathcal{M}^P$ (15) and a best response for the attacker is obtained by solving an MDP $\mathcal{M}$ (1). The problem for the defender is to find a stopping strategy $\pi_{\mathrm{D}}^{\star}$ that maximizes $J_{\mathrm{D}}$ (2.5a) and prescribes the optimal stopping times $\tau_{\mathrm{D},1}^{\star}, \tau_{\mathrm{D},2}^{\star}, \ldots, \tau_{\mathrm{D},L}^{\star}$. Likewise, the problem for the attacker is to find a stopping strategy $\pi_{\mathrm{A}}^{\star}$ that minimizes $J_{\mathrm{D}}$ (2.5b) and prescribes the optimal stopping times $\tau_{\mathrm{A},1}^{\star}$ and $\tau_{\mathrm{A},2}^{\star}$; see Fig. 2.2.

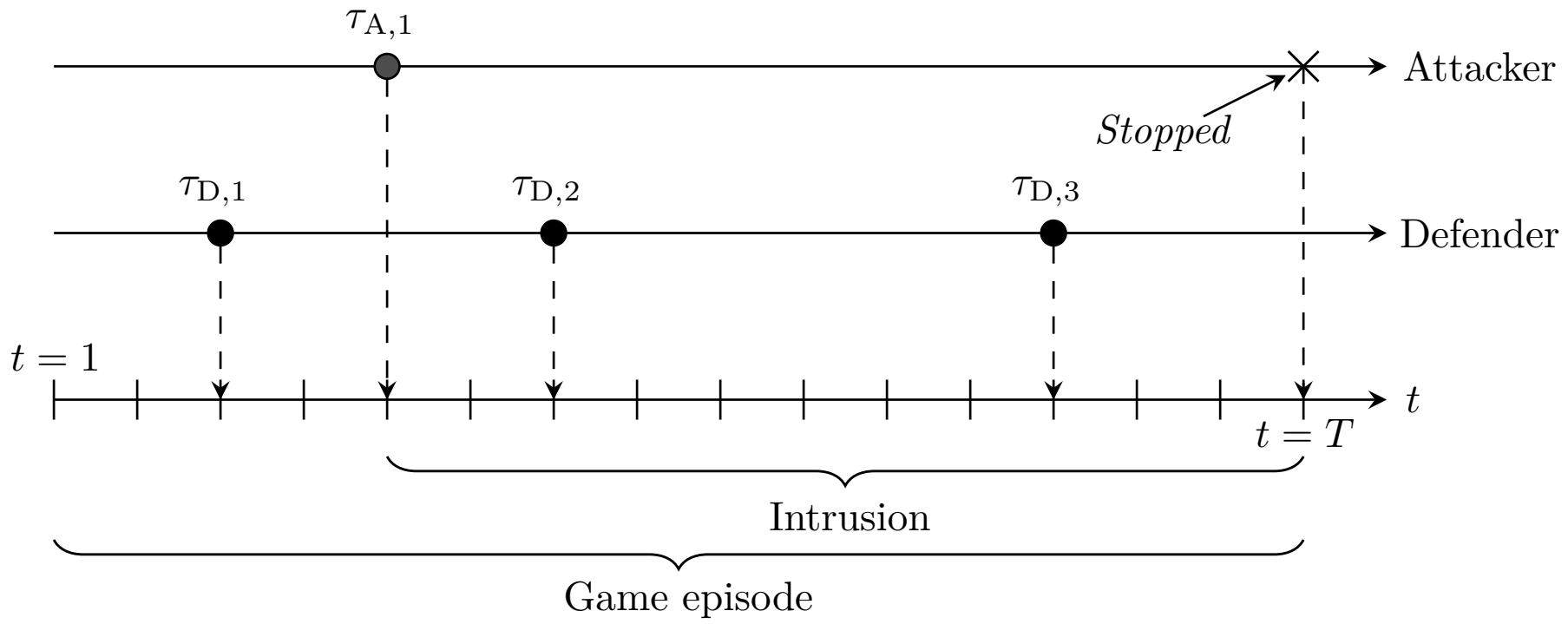

***Figure 2.2:*** *Stopping times of the defender and the attacker in a game episode; $\tau_{\mathrm{k},j}$ denotes the $j$th stopping time of player $\mathrm{k} \in \{\mathrm{D}, \mathrm{A}\}$; the cross shows the time the intrusion is stopped; an intrusion starts when the attacker takes the first stop action (at time $\tau_{\mathrm{A},1}$); an episode ends either when the attacker is stopped (as a consequence of defender actions) or when the attacker terminates its intrusion (at time $\tau_{\mathrm{A},2}$).*

It follows from Thms. 1–2 in the background chapter that for any strategy pair $(\pi_{\mathrm{D}}, \pi_{\mathrm{A}})$, a corresponding pair of *pure* (stationary) best responses $(\tilde{\pi}_{\mathrm{D}}, \tilde{\pi}_{\mathrm{A}}) \in \mathscr{B}_{\mathrm{D}}(\pi_{\mathrm{A}}) \times \mathscr{B}_{\mathrm{A}}(\pi_{\mathrm{D}})$ exists. Given a pair of stopping strategies $(\pi_{\mathrm{D}}, \pi_{\mathrm{A}})$ and their (pure) best responses $\tilde{\pi}_{\mathrm{D}} \in \mathscr{B}_{\mathrm{D}}(\pi_{\mathrm{A}})$ and $\tilde{\pi}_{\mathrm{A}} \in \mathscr{B}_{\mathrm{A}}(\pi_{\mathrm{D}})$, we define two subsets of the belief space $\mathcal{B}$: the *stopping sets* and the *continuation sets*. The stopping sets contain the belief states where $\mathsf{S}$ is a best response:

$$\mathscr{S}_{l,\pi_{\mathrm{A}}}^{(\mathrm{D})} \triangleq \{\mathbf{b} \mid \mathbf{b} \in \mathcal{B}, \tilde{\pi}_{\mathrm{D}}(\mathbf{b}) = \mathsf{S}\} \quad \text{and} \quad \mathscr{S}_{s,l,\pi_{\mathrm{D}}}^{(\mathrm{A})} \triangleq \{\mathbf{b} \mid \mathbf{b} \in \mathcal{B}, \tilde{\pi}_{\mathrm{A}}(\mathbf{b}, s) = \mathsf{S}\}.$$

Similarly, the continuation sets contain the belief states where $\mathsf{C}$ is a best response:

$$\mathscr{C}_{l,\pi_{\mathrm{A}}}^{(\mathrm{D})} \triangleq \{\mathbf{b} \mid \mathbf{b} \in \mathcal{B}, \tilde{\pi}_{\mathrm{D}}(\mathbf{b}) = \mathsf{C}\} \quad \text{and} \quad \mathscr{C}_{s,l,\pi_{\mathrm{D}}}^{(\mathrm{A})} \triangleq \{\mathbf{b} \mid \mathbf{b} \in \mathcal{B}, \tilde{\pi}_{\mathrm{A}}(\mathbf{b}, s) = \mathsf{C}\}.$$

Based on Thms. 2 and 3 in the background chapter, and Thm. 1.1 of Paper 1, we obtain Thm. 2.1 (shown on the next page), which contains an existence result for PBEs and a structural result for best responses in the game. We provide proof in Appendix A.

**Theorem 2.1** (Existence of equilibria and threshold structure of best responses)**.**

*(A)* $\Gamma$ *has a* PBE*. If* $s = 0 \iff \mathbf{b}(1) = 0$*, then it has a* PBE *in pure strategies.*

*(B) Assuming* $z$ *(2.4) is totally positive of order 2 (i.e.,* TP*-2 (Def. 10.2.1, p. 223, Krishnamurthy, 2016)). Then, given an attacker strategy* $\pi_{\mathrm{A}} \in \Pi_{\mathrm{A}}$*, there exist values* $\tilde{\alpha}_1 \geq \tilde{\alpha}_2 \geq \ldots \geq \tilde{\alpha}_L$ *and a best response* $\tilde{\pi}_{\mathrm{D}} \in \mathscr{B}_{\mathrm{D}}(\pi_{\mathrm{A}})$ *for the defender that satisfies*

$$\tilde{\pi}_{\mathrm{D}}(\mathbf{b}) = \mathsf{S} \iff \mathbf{b}(1) \geq \tilde{\alpha}_l \qquad \forall l \in \{1, \ldots, L\}, \text{ where } \tilde{\alpha}_l \in [0,1]. \tag{2.8}$$

*(C) Assuming* $\pi_{\mathrm{A}}(\mathbf{b}, 0) = \mathsf{S}$ *when* $\mathbf{b}(1) = 0$ *for all* $\pi_{\mathrm{A}} \in \Pi_{\mathrm{A}}$*. Then, given a defender strategy* $\pi_{\mathrm{D}} \in \Pi_{\mathrm{D}}$ *that satisfies (2.8), there exist values* $\tilde{\beta}_{0,1}$*,* $\tilde{\beta}_{1,1}$*, …,* $\tilde{\beta}_{0,L}$*,* $\tilde{\beta}_{1,L}$ *and a best response* $\tilde{\pi}_{\mathrm{A}} \in \mathscr{B}_{\mathrm{A}}(\pi_{\mathrm{D}})$ *for the attacker that satisfies*

$$\tilde{\pi}_{\mathrm{A}}(\mathbf{b}, 0) = \mathsf{C} \iff \mathbf{b}(1) \geq \tilde{\beta}_{0,l} \quad \forall l \in \{1, \ldots, L\}, \text{ where } \tilde{\beta}_{0,l} \in [0,1] \tag{2.9a}$$

$$\tilde{\pi}_{\mathrm{A}}(\mathbf{b}, 1) = \mathsf{S} \iff \mathbf{b}(1) \geq \tilde{\beta}_{1,l} \quad \forall l \in \{1, \ldots, L\}, \text{ where } \tilde{\beta}_{1,l} \in [0,1]. \tag{2.9b}$$

Theorem 2.1 tells us that $\Gamma$ has a PBE. Further, under assumptions generally met in practice, the best responses have threshold properties; see Fig. 2.3. In the following, we describe an algorithm that leverages these properties to efficiently approximate a PBE of $\Gamma$.

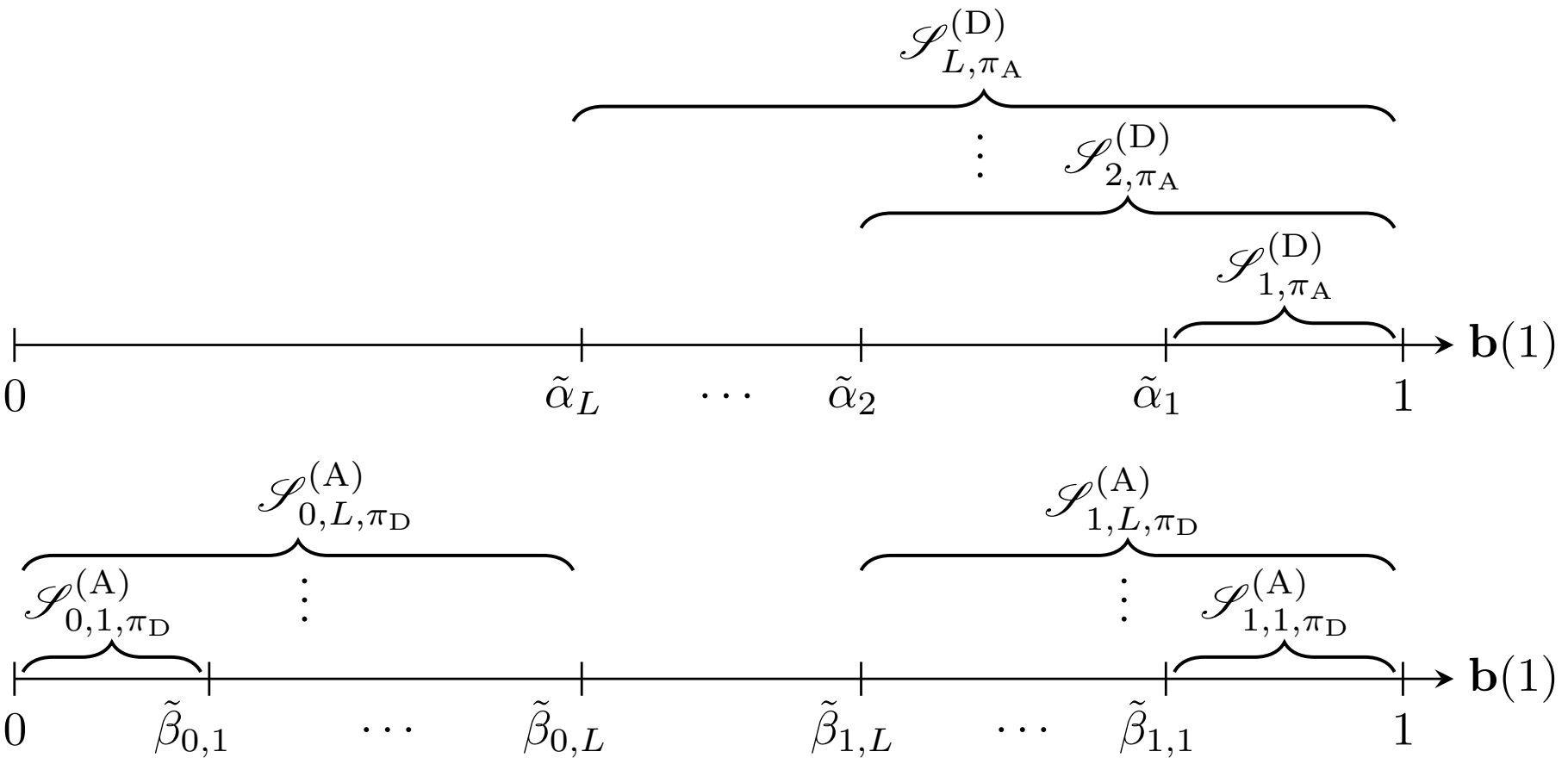

***Figure 2.3:*** *Illustration of Thm. 2.1; the upper part shows* $L$ *thresholds* $\tilde{\alpha}_1 \geq \tilde{\alpha}_2, \ldots, \geq \tilde{\alpha}_L$ *in the unit interval that define a best response* $\tilde{\pi}_{\mathrm{D}} \in \mathscr{B}_{\mathrm{D}}(\pi_{\mathrm{A}})$ *for the defender (2.8); the lower part shows* $2L$ *thresholds* $\tilde{\beta}_{0,1}, \tilde{\beta}_{1,1}, \ldots, \tilde{\beta}_{0,L}, \tilde{\beta}_{1,L}$ *in the unit interval that define a best response* $\tilde{\pi}_{\mathrm{A}} \in \mathscr{B}_{\mathrm{A}}(\pi_{\mathrm{D}})$ *for the attacker (2.9).*

### Finding equilibria through fictitious play

Computing Perfect Bayesian Equilibria (PBEs) for a POSG is generally a computationally intractable problem (Thm. 3.5, Goldsmith and Mundhenk, 2007). However, approximate solutions can be obtained through iterative methods. One such method is *fictitious play*, where players start from random strategies and iteratively update their strategies based on outcomes of played game episodes (Brown, 1951). This method evolves through a sequence of iteration steps, illustrated in Fig. 2.4. Each step includes two stages. First, both players learn best responses. Second, each player adopts a new strategy, determined by the empirical distribution over its past best responses. The sequence of iteration steps continues until the strategies of both players have sufficiently converged to a PBE (Thms. 7.2.4-7.2.5, Shoham and Leyton-Brown, 2009).

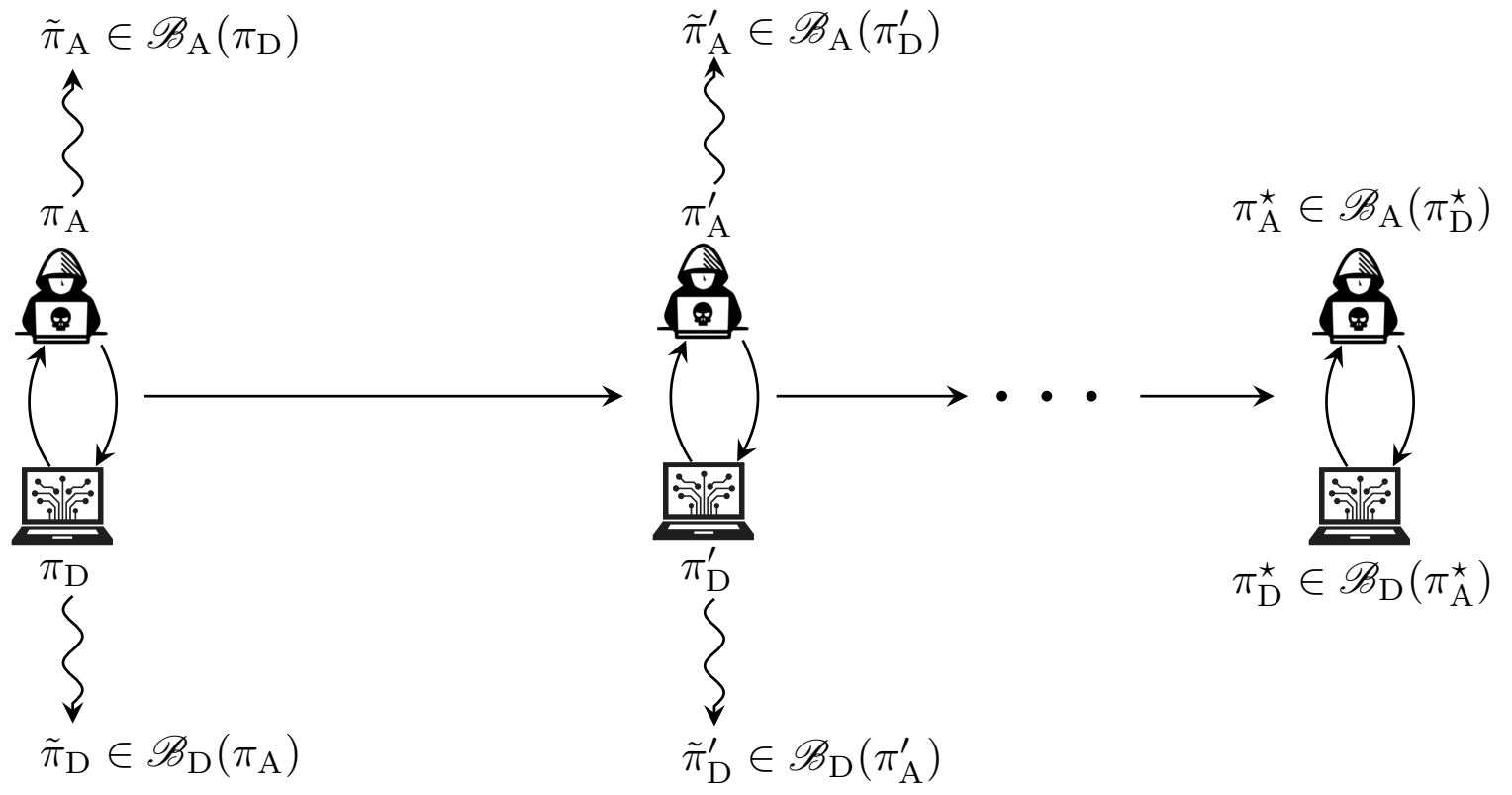

***Figure 2.4:*** *The fictitious play process; horizontal arrows indicate iterations of fictitious play and vertical arrows indicate the learning of best responses; the process converges to an equilibrium* $(\pi^\star_{\mathrm{D}}, \pi^\star_{\mathrm{A}})$ *(Thms. 7.2.4-7.2.5, Shoham and Leyton-Brown, 2009).*

### Our fictitious play algorithm: T-FP

We present a fictitious play algorithm called **T**hreshold-**F**ictitious **P**lay (T-FP), which implements the fictitious play process described above and generates a sequence of strategy profiles $(\pi_{\mathrm{D}}, \pi_{\mathrm{A}})$, $(\pi'_{\mathrm{D}}, \pi'_{\mathrm{A}})$, ... that converges to a PBE $(\pi^\star_{\mathrm{D}}, \pi^\star_{\mathrm{A}})$ (Thms. 7.2.4-7.2.5, Shoham and Leyton-Brown, 2009). During each step of this process, T-FP learns best responses against the players' current strategies and then updates the strategies of both players; see Fig. 2.4.

To learn the best responses, T-FP parameterizes them through threshold vectors according to Thm. 2.1. The defender's best response $\tilde{\pi}_{\mathrm{D}}$ is parameterized by the vector $\tilde{\boldsymbol{\theta}}^{(\mathrm{D})} \in \mathbb{R}^L$ as

$$\tilde{\pi}_{\mathrm{D},\tilde{\boldsymbol{\theta}}^{(\mathrm{D})}}(\mathsf{S} \mid \mathbf{b}) \triangleq \varphi\left(\tilde{\boldsymbol{\theta}}^{(\mathrm{D})}_l, \mathbf{b}\right), \qquad (2.10)$$

where[11]

$$\varphi(a, \mathbf{b}) \triangleq \left(1 + \left(\frac{\mathbf{b}(1)(1-\sigma(a))}{\sigma(a)(1-\mathbf{b}(1))}\right)^{-20}\right)^{-1}. \tag{2.11}$$

Here $\sigma(\cdot)$ is the sigmoid function and $a \in \mathbb{R}$. (Note that (2.10)–(2.11) is the same parameterization as used in Paper 1.) Correspondingly, the attacker's best response $\tilde{\pi}_{\mathrm{A}}$ is parameterized by $\tilde{\boldsymbol{\theta}}^{(\mathrm{A})} \in \mathbb{R}^{2L}$ as

$$\tilde{\pi}_{\mathrm{A},\tilde{\boldsymbol{\theta}}^{(\mathrm{A})}}(\mathsf{C} \mid \mathbf{b}, 0) \triangleq \varphi\left(\tilde{\boldsymbol{\theta}}^{(\mathrm{A})}_l, \mathbf{b}\right) \quad \text{and} \quad \tilde{\pi}_{\mathrm{A},\tilde{\boldsymbol{\theta}}^{(\mathrm{A})}}(\mathsf{S} \mid \mathbf{b}, 1) \triangleq \varphi\left(\tilde{\boldsymbol{\theta}}^{(\mathrm{A})}_{L+l}, \mathbf{b}\right). \tag{2.12}$$

The parameterized strategies are stochastic strategies that approximate threshold strategies; see Fig. 2.5. $\sigma(\tilde{\boldsymbol{\theta}}^{(\mathrm{D})}_1)$, $\sigma(\tilde{\boldsymbol{\theta}}^{(\mathrm{D})}_2)$, ..., $\sigma(\tilde{\boldsymbol{\theta}}^{(\mathrm{D})}_L)$ are the $L$ thresholds of the defender, where $\sigma(\tilde{\boldsymbol{\theta}}^{(\mathrm{D})}_l) \in [0,1]$; see Thm. 2.1.B. Likewise, $\sigma(\tilde{\boldsymbol{\theta}}^{(\mathrm{A})}_1)$, $\sigma(\tilde{\boldsymbol{\theta}}^{(\mathrm{A})}_2)$, ..., $\sigma(\tilde{\boldsymbol{\theta}}^{(\mathrm{A})}_{2L})$ are the $2L$ thresholds of the attacker, where $\sigma(\tilde{\boldsymbol{\theta}}^{(\mathrm{A})}_l) \in [0,1]$; see Thm. 2.1.C.

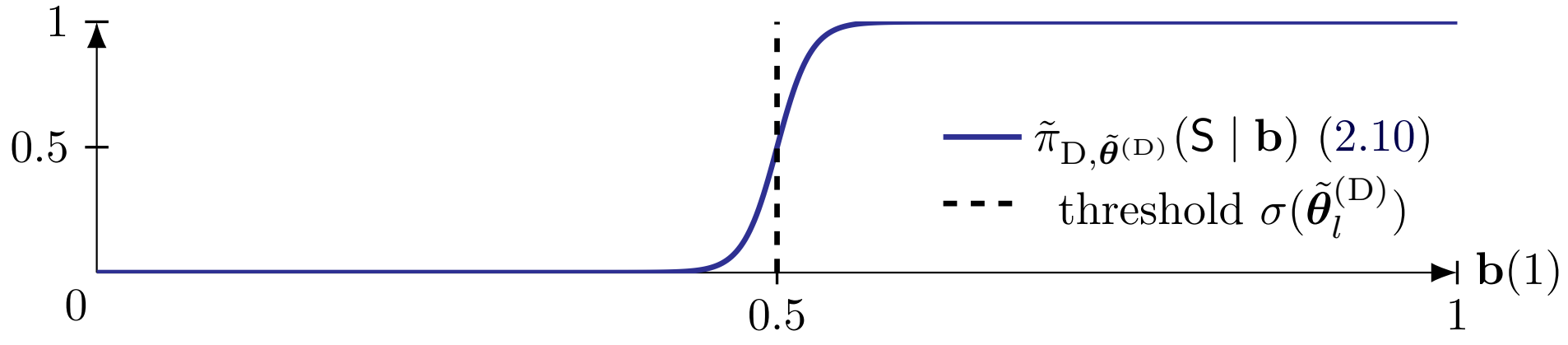

***Figure 2.5:*** *The stochastic threshold strategy in (2.10), where $\sigma$ is the sigmoid function and $\sigma(\tilde{\boldsymbol{\theta}}^{(\mathrm{D})}_l)$ is the threshold (0.5 in this example); the x-axis indicates the defender's belief state $\mathbf{b}(1) \in [0,1]$; and the y-axis indicates the probability prescribed by $\tilde{\pi}_{\mathrm{D},\tilde{\boldsymbol{\theta}}^{(\mathrm{D})}}$ to the stop action $\mathsf{S}$.*

Using the above parameterization, T-FP learns the best responses $\tilde{\pi}_{\mathrm{D},\tilde{\boldsymbol{\theta}}^{(\mathrm{D})}}$ and $\tilde{\pi}_{\mathrm{A},\tilde{\boldsymbol{\theta}}^{(\mathrm{A})}}$ by iteratively updating the threshold vectors $\tilde{\boldsymbol{\theta}}^{(\mathrm{D})}$ and $\tilde{\boldsymbol{\theta}}^{(\mathrm{A})}$ through the T-SPSA algorithm described in Paper 1[12]. After this process has finished, the threshold vectors $\tilde{\boldsymbol{\theta}}^{(\mathrm{D})}$ and $\tilde{\boldsymbol{\theta}}^{(\mathrm{A})}$ are added to buffers $\Theta^{(\mathrm{D})}$ and $\Theta^{(\mathrm{A})}$, which contain the vectors learned in previous iterations of T-FP. Subsequently, both players update their strategies based on the empirical distributions over the past vectors in the buffers. This process is repeated until the strategies have sufficiently converged to a PBE. The pseudocode of T-FP is listed in Algorithm 2.1 on the next page.

[11] To avoid division by zero, we only use (2.11) when $\sigma(a) \neq 0$ and $\mathbf{b}(1) \neq 1$; if $\sigma(\boldsymbol{\theta}_l) = 0$ or $\mathbf{b}(1) = 1$, then $\tilde{\pi}_{\mathrm{D},\tilde{\boldsymbol{\theta}}^{(\mathrm{D})}}(\mathsf{S} \mid \mathbf{b}) = 1$, $\tilde{\pi}_{\mathrm{A},\tilde{\boldsymbol{\theta}}^{(\mathrm{A})}}(\mathsf{S} \mid \mathbf{b}, 1) = 1$ and $\tilde{\pi}_{\mathrm{A},\tilde{\boldsymbol{\theta}}^{(\mathrm{A})}}(\mathsf{C} \mid \mathbf{b}, 0) = 1$.

[12] T-SPSA converges a.s. to a local optimum of (2.6); see (Prop. 1, Spall, 1992).

***Algorithm 2.1:*** T-FP: ***Threshold-Fictitious Play.***

**Input:** $\Gamma$, $\delta$: the POSG and the desired equilibrium precision.
$a, c, \lambda, A, \epsilon, N$: parameters for T-SPSA.
**Output:** $(\pi^\star_{\mathrm{D}}, \pi^\star_{\mathrm{A}})$: an approximate Perfect Bayesian Equilibrium (PBE).

**procedure** T-FP($\Gamma$, $N, a, c, \lambda, A, \epsilon, \delta$)
  $\Theta^{(\mathrm{D})} \leftarrow \emptyset, \quad \Theta^{(\mathrm{A})} \leftarrow \emptyset, \quad \hat{\delta} \leftarrow \infty.$
  **while** $\hat{\delta} \geq \delta$ **do**
    **for** $\mathrm{k} \in \{\mathrm{D}, \mathrm{A}\} \triangleq \{1, 2\}$ **do**
      $\boldsymbol{\theta}^{(\mathrm{k})}_{(1)} \sim \mathcal{U}_{\mathrm{k}L}(\{-1, 1\}).$ ▷ k$L$-dimensional uniform distribution.
      $\Theta^{(\mathrm{k})} \leftarrow \Theta^{(\mathrm{k})} \cup \{$T-SPSA$(\Gamma, \boldsymbol{\theta}^{(\mathrm{k})}_{(1)}, N, a, c, \lambda, A, \epsilon)\}.$
      $\pi_{\mathrm{k}} \leftarrow$ EmpiricalDistribution$(\Theta^{(\mathrm{k})})$.
    $\hat{\delta} \leftarrow$ Exploitability$(\pi_{\mathrm{D}}, \pi_{\mathrm{A}})$. ▷ See (2.13) on page 112.
  **return** $(\pi_{\mathrm{D}}, \pi_{\mathrm{A}})$.

## 2.4 Creating a Digital Twin of the Target Infrastructure

The T-FP algorithm described above approximates a PBE of $\Gamma$ by simulating game episodes and updating both players' strategies through stochastic approximation. To instantiate this process, we first need to obtain the observation function (2.4). We estimate this function using samples from a digital twin of the target infrastructure. We create this digital twin using CSLE, as described in the methodology chapter (Hammar, 2023). The network topology of the target infrastructure is shown in Fig. 1.2 of Paper 1; its configuration is given in Appendix C of Paper 1.

The digital twin comprises virtual containers and networks that replicate the functionality and the timing behavior of the target infrastructure. These containers run the same software and processes as the physical infrastructure. For example, the container that emulates the gateway in Fig. 1.2 of Paper 1 runs the SNORT IDS, which generates alerts in real-time. We collect these alerts from the digital twin at 30-second intervals, which allows us to compute the defender observation $o_t$ (2.4). (We define 30$s$ in the digital twin to be 1 time step in the game $\Gamma$.) We choose to define $o_t$ based on the IDS alert metric as it provides sufficient information for detecting the type of intrusions we consider in this paper; see Appendix C for a comparison between different metrics. The value of this metric depends on client behavior as well as the actions of the defender and the attacker, as described below.

### Emulating the clients

The *client population* is emulated by processes that interact with the virtual containers of the digital twin by performing a sequence of service invocations, which are selected uniformly at random from Table 1.3 in Paper 1. Client arrivals per time step are emulated using a stationary Poisson process with mean $\lambda = 20$ and exponentially distributed service times with mean $\mu = 4$. The duration of a time step is 30 seconds.

### Defender and attacker actions on the digital twin

The defender and the attacker take actions at time steps $t = 1, 2, \ldots, T-1$. The defender executes either a continue action or a stop action. A continue action is virtual in the sense that it does not trigger any function in the digital twin. A stop action, however, invokes specific functions in the digital twin. We have implemented $L = 7$ stop actions for the defender, listed in Table 2.2. The first stop action revokes user certificates and recovers user accounts expected to be compromised by the attacker. The second stop action updates the firewall configuration of the gateway to drop traffic from IP addresses flagged by the IDS[13]. Stop actions 3–6 trigger the dropping of traffic that generates IDS alerts of priorities 1–4. The final stop action blocks all incoming traffic.

| *Stop index* | *Action* | MITRE D3FEND technique |
|---|---|---|
| 1 | Revoke user certificates | D3-CBAN certificate revocation. |
| 2 | Blacklist IPs | D3-NTF network traffic filtering. |
| 3 | Drop traffic that generates IDS alerts of priority 1 | D3-NTF network traffic filtering. |
| 4 | Drop traffic that generates IDS alerts of priority 2 | D3-NTF network traffic filtering. |
| 5 | Drop traffic that generates IDS alerts of priority 3 | D3-NTF network traffic filtering. |
| 6 | Drop traffic that generates IDS alerts of priority 4 | D3-NTF network traffic filtering. |
| 7 | Block gateway | D3-NI network isolation. |

***Table 2.2:*** *Defender actions executed on the digital twin; shell commands for executing the actions are listed in (Hammar, 2023); the actions are linked to the corresponding defense techniques in the MITRE D3FEND taxonomy (Kaloroumakis and Smith, 2021).*

Like the defender, the attacker executes a stop or a continue action during each time step. The attacker can only take two stop actions during a game episode. The first determines when the intrusion starts and the second when it terminates; see §2.2. During an intrusion, the attacker executes a sequence of attack actions drawn randomly from the actions listed in Table 2.3 on the next page. An attack action is executed for each time step in the intrusion state $s = 1$ (2.2).

[13]The digital twin runs the SNORT IDS (Roesch, 1999); see Appendix C of Paper 1 for the SNORT configuration.

| *Type* | *Actions* | MITRE ATT&CK |
|---|---|---|
| Reconnaissance | TCP SYN scan, UDP port scan, TCP NULL scan, TCP XMAS scan, TCP FIN scan, ping scan, TCP connection scan, VULSCAN. | TA0043 reconnaissance. |
| Brute-force attack | TELNET, SSH, FTP, CASSANDRA, IRC, MONGODB, MYSQL, SMTP, POSTGRES. | T1110 brute force. |
| Exploit | CVE-2017-7494, CVE-2015-3306, CVE-2014-6271 CVE-2010-0426, CVE-2015-5602, CVE-2016-10033, CVE-2015-1427, exploit of the CWE-89 weakness on DVWA [454]. | T1210 service exploitation. |

***Table 2.3:*** *Attacker actions executed on the digital twin; actions that exploit vulnerabilities in specific software products are identified by the vulnerability identifiers in the Common Vulnerabilities and Exposures (CVE) database (The MITRE Corporation, 2022); actions that exploit vulnerabilities that are not described in the CVE database are categorized according to the types of the vulnerabilities they exploit based on the Common Weakness Enumeration (CWE) list (The MITRE Corporation, 2023); the actions are also linked to the corresponding attack techniques and tactics in the MITRE ATT&CK taxonomy (Strom et al., 2018); shell commands and scripts for executing the actions are listed in (Hammar, 2023); further details about the actions can be found in Appendix D.*

### Estimating the IDS alert distribution

At the end of every time step in the digital twin, i.e., at the end of each 30s interval, we collect the number of IDS alerts with priorities 1–4 that occurred during the time step, where priorities 1–4 refer to the SNORT priorities "very low", "low", "medium", and "high", respectively (Roesch, 1999)[14]. We do so for 23,000 time steps, which provides us with a dataset to estimate the distribution of IDS alerts. Using this dataset, we apply expectation-maximization (Dempster et al., 1977) to fit Gaussian mixture distributions $\widehat{z}(\cdot \mid 0)$ and $\widehat{z}(\cdot \mid 1)$ as estimates of $z(\cdot \mid 0)$ and $z(\cdot \mid 1)$ (2.4), which represent the true observation distributions in the target infrastructure.

Figure 2.6 on the next page shows the fitted models over the discrete observation space $\mathcal{O} \triangleq \{0, 1, \dots, 22000\}$. We note that $\widehat{z}(\cdot \mid 0)$ and $\widehat{z}(\cdot \mid 1)$ are (discretized) Gaussian mixtures with one and three components, respectively. Both mixtures have the most probability mass within 0–5000. $\widehat{z}(\cdot \mid 1)$ also has substantial probability mass at larger values. The stochastic matrix with the rows $\widehat{z}(\cdot \mid 0)$ and $\widehat{z}(\cdot \mid 1)$ has about $241 \times 10^6$ second-order minors, which are almost all non-negative. This suggests to us that the TP-2 assumption in Thm. 2.1 can be made (Def. 10.2.1, p. 223, Krishnamurthy, 2016).

[14] Note that according to SNORT's terminology (Roesch, 1999), 1 is the highest priority. We inverse the labeling in our framework for convenience.

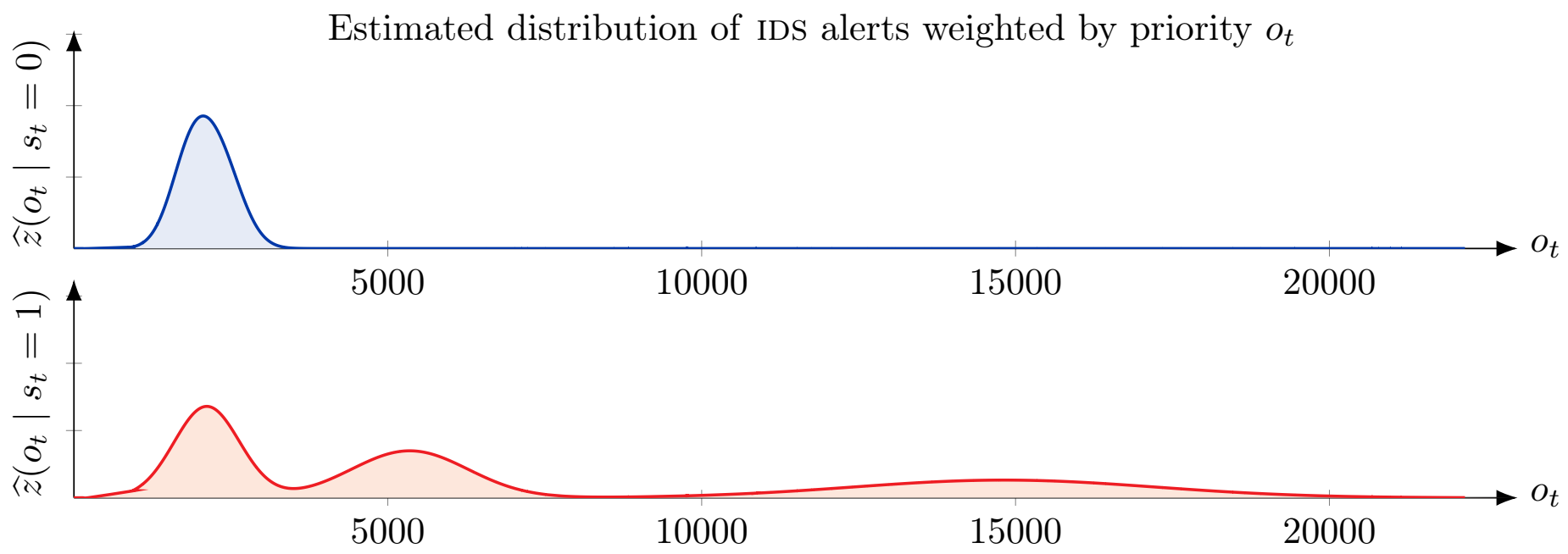

***Figure 2.6:*** *Fitted Gaussian mixture models of $z$ (2.4) when no intrusion occurs ($s_t = 0$) and during intrusion ($s_t = 1$).*

### Running a game episode

During a game episode, the state evolves according to the dynamics defined by (2.2), the defender's belief state evolves according to (22)[15], the players' rewards are calculated using the reward function $r$ (2.3), and the defender's observations are obtained from $\widehat{z}$; see Fig. 2.6. The actions of both players are determined by their respective strategies. If the game runs on the digital twin, the players' actions include executing networking and computing functions (see Tables 2.2–2.3), and the observations are obtained through reading metrics of the digital twin.

## 2.5 Experimental Evaluation

Our methodology for finding near-optimal defender strategies includes: (*i*) creating a digital twin of the target infrastructure to obtain statistics for instantiating the simulator; (*ii*) learning equilibrium strategies using the T-FP algorithm; and (*iii*) evaluating learned strategies on the digital twin; see Fig. 12 in the introduction chapter. This section describes the learning process and the evaluation results of the intrusion response use case.

### Learning equilibrium strategies through fictitious play

We run T-FP for 500 iteration steps to estimate a PBE, which is sufficient to meet the termination condition (line 3 in Algorithm 2.1). These iteration steps generate a sequence of strategy pairs $(\pi_{\mathrm{D}}, \pi_{\mathrm{A}})_1, (\pi_{\mathrm{D}}, \pi_{\mathrm{A}})_2, \ldots, (\pi_{\mathrm{D}}, \pi_{\mathrm{A}})_{500}$. At the end of each iteration step, we evaluate the current strategy pair $(\pi_{\mathrm{D}}, \pi_{\mathrm{A}})$ by simulating 500 evaluation episodes and executing 5 evaluation episodes on the digital twin. This process allows us to produce learning curves of the reward (2.3) and the exploitability (2.13); see Fig. 2.7 on page 113. The 500 training iterations and the associated evaluations constitute one *training run*. We run four training runs with different random seeds. A single training run takes about 5 hours of processing time on the simulator. In addition, it takes around 12 hours to evaluate the strategies on the digital twin. The hyperparameters of T-FP are listed in Appendix B.

[15] The equation to compute the belief state is defined in the background chapter.

### Computing environment

The environment for running simulations and training strategies is a TESLA P100 GPU. The emulated infrastructure is deployed on a server with a 24-core INTEL XEON GOLD 2.10 GHz CPU and 768 GB RAM; see Fig. 21 in the methodology chapter.

### Convergence metric for T-FP

To estimate the convergence of the sequence of strategy pairs generated by T-FP, we use the *approximate exploitability* metric

$$\hat{\delta} \triangleq J_{\text{D}}(\hat{\pi}_{\text{D}}, \pi_{\text{A}}) + J_{\text{A}}(\pi_{\text{D}}, \hat{\pi}_{\text{A}}), \qquad \text{(Eq. 3, Timbers et al., 2020)} \tag{2.13}$$

where $\hat{\pi}_{\text{k}}$ denotes an approximate best response for player k and $(J_{\text{D}}, J_{\text{A}})$ are defined in (2.5). The closer $\hat{\delta}$ becomes to 0, the closer $(\pi_{\text{D}}, \pi_{\text{A}})$ is to an equilibrium.

### Baseline algorithms

We compare the performance of T-FP with that of two widely-used algorithms from previous work [274, 497, 201, 200, 467]. The first algorithm is **N**eural **F**ictitious **S**elf-**P**lay (NFSP) (Alg. 1, Heinrich and Silver, 2016), which is a general fictitious play algorithm that does not exploit the threshold structures expressed in Thm. 2.1. The second algorithm is **H**euristic **S**earch **V**alue **I**teration (HSVI) (Alg. 1, Horák et al., 2017), a state-of-the-art dynamic programming algorithm for one-sided POSGs like $\Gamma$ (2.1), which is guaranteed to converge (Thm. 3, Horák et al., 2017).

### Baseline strategies

We compare the defender strategies learned through T-FP with three baseline strategies. The first baseline prescribes the stop action when an IDS alert occurs, i.e., when $o_t > 0$. The second baseline is derived from the SNORT IDPS (Roesch, 1999), which is a de-facto industry standard and can be considered state-of-the-art for our use case. This baseline uses the SNORT IDPS's recommendation system and takes a stop action when SNORT has dropped 100 IP packets (see Paper 1, Appendix C for the SNORT configuration). The third baseline assumes prior knowledge of the intrusion time and performs all $L$ stops during the $L$ subsequent time steps.

Although a growing body of work uses reinforcement learning and game theory to find intrusion response strategies (see §2.6 for a review of the related work), a direct comparison between the defender strategies learned through our methodology and those found in previous work is not feasible for two reasons. First, nearly all of the prior works have developed defender strategies for custom simulations [178, 182, 130, 397, 510, 261, 66, 369, 525, 469, 152, 207, 518, 123, 125, 494, 274, 528, 287, 217, 220, 306, 484, 327, 36, 483, 515, 205, 210, 398, 124, 202, 354, 393, 471, 325, 178, 268, 9, 527, 521, 16, 13, 457] and there is no obvious way to map their solutions to our digital twin. Second, the few prior works that study emulated infrastructures similar to our digital twin either consider static attackers in fully observed environments [150, 4, 286, 33, 353, 508, 529, 219, 486] or focus on different use cases [33, 531].

### Evaluating the learned strategies

Figure 2.7 shows the learning curves of the strategies obtained during the fictitious play process with T-FP and the baselines introduced above.

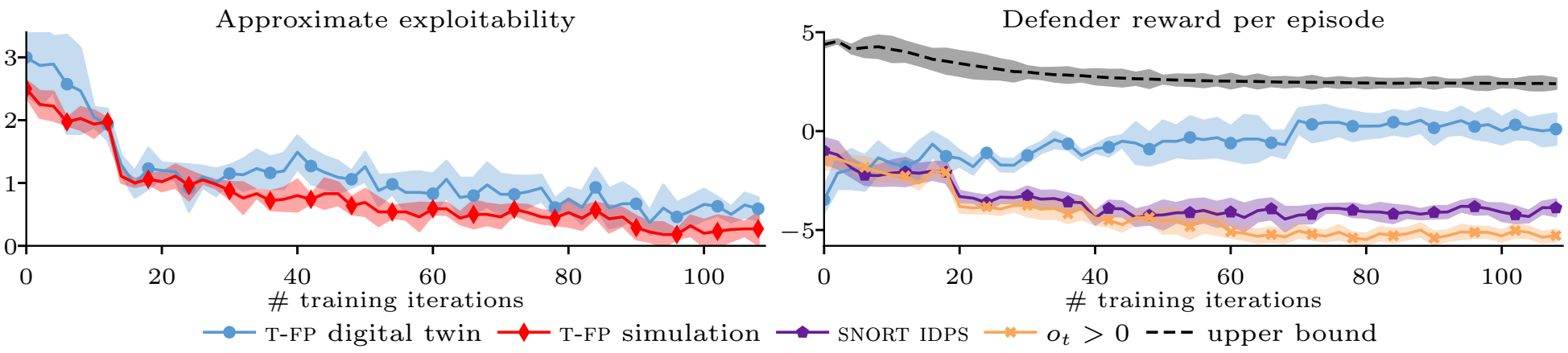

***Figure 2.7:*** *Learning curves from the fictitious play process with T-FP; the red curve shows simulation results and the blue curves show results from the digital twin; the purple, orange, and black curves relate to baseline strategies; the figures show two performance metrics: approximate exploitability (2.13) and reward (2.3); the curves indicate the mean and the 95% confidence interval over four training runs with different random seeds.*

We observe in Fig. 2.7 that the approximate exploitability (2.13) of the learned strategies converges to small values (left plot), which indicates that the learned strategies approximate a PBE both on the simulator and the digital twin. Further, we see in the right plot that both baseline strategies show decreasing performance as the attacker updates its strategy. In contrast, the defender strategy learned through T-FP improves its performance over time. This improvement shows the benefit of a game-theoretic approach where the defender strategy is optimized against a dynamic attacker, i.e., an attacker that adapts its strategy to the defender's strategy.

Figure 2.8 on the next page compares T-FP with the two baseline algorithms: NFSP and HSVI. NFSP implements fictitious play and can thus be compared with T-FP with respect to approximate exploitability (2.13). We observe in the left plot that T-FP converges much faster than NFSP. We explain the rapid convergence of T-FP by its design, which exploits the structural properties of the stopping game. The right plot shows that HSVI reaches an approximation error below 5 within an hour of processing time. Based on the recent literature, we anticipated a much longer processing time (Figs. 5-8, Horák et al., 2023). This result suggests to us that T-FP and HSVI have similar convergence properties. A more detailed comparison between T-FP and HSVI is hard to perform due to the different nature of the two algorithms.

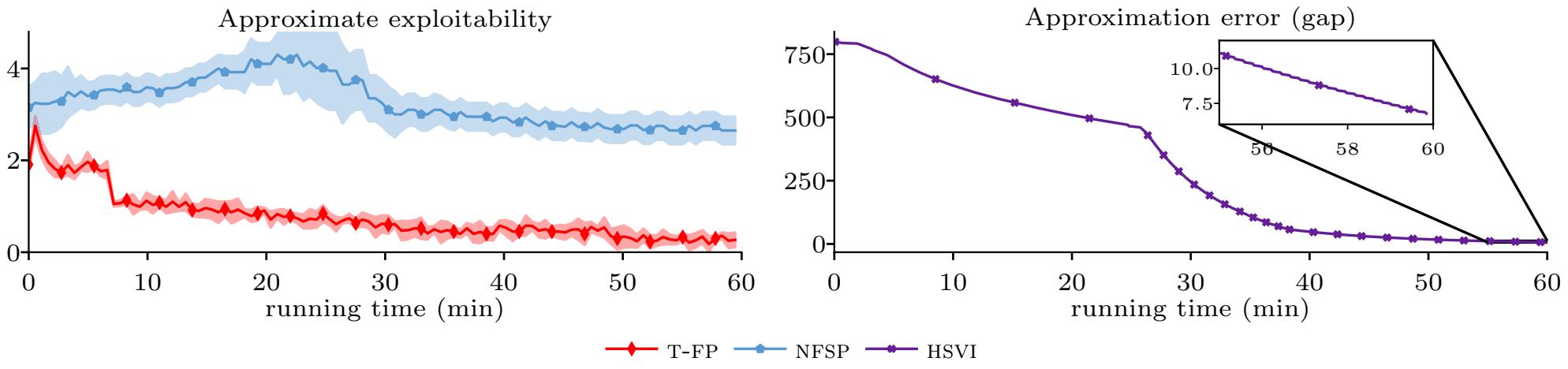

***Figure 2.8:*** *Comparison between T-FP and two baseline algorithms: NFSP and HSVI; all curves show simulation results; the red curve relates to T-FP; the blue curve to NFSP; the purple curve to HSVI; the left plot shows the approximate exploitability metric (2.13) and the right plot shows the HSVI approximation error; the curves depicting T-FP and NFSP show the mean and the 95% confidence interval over four random seeds.*

Figure 2.9 shows the estimated value function $\hat{J}_l^\star : \mathcal{B} \to \mathbb{R}$, where $\hat{J}_l^\star(\mathbf{b})$ is the expected cumulative discounted reward when the game starts in the belief state $\mathbf{b}$, the defender has $l$ stops remaining, and both players follow equilibrial strategies[16].

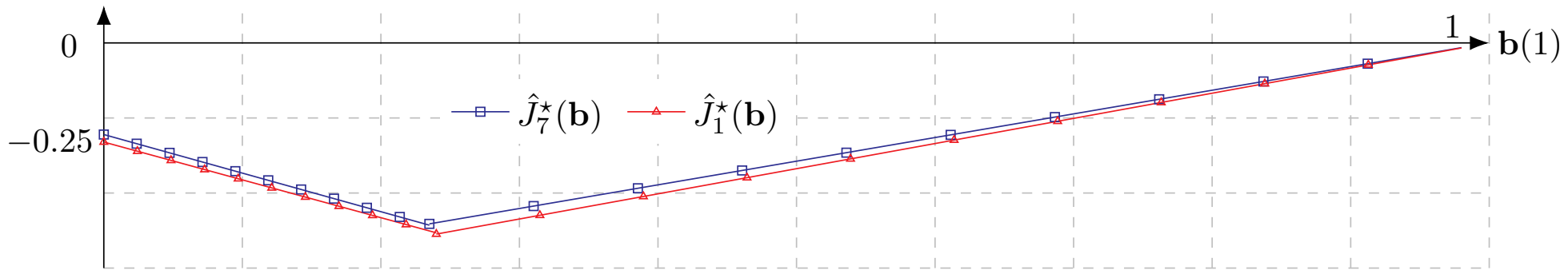

***Figure 2.9:*** *The value function $\hat{J}_l^\star(\mathbf{b})$ computed through the HSVI algorithm with approximation error 4; the blue and red curves relate to $l = 7$ and $l = 1$, respectively.*

We see in Fig. 2.9 that $\hat{J}_l^\star$ is piece-wise linear and convex, as expected from Cor. 1 in the background chapter. The figure indicates that $\hat{J}_l^\star(\mathbf{b}) \leq 0$ for all $\mathbf{b} \in \mathcal{B}$ and that $\hat{J}_l^\star(\mathbf{b}) = 0$ when $\mathbf{b}(1) = 1$. Further, we note that the value of $\hat{J}_l^\star$ is minimal when $\mathbf{b}(1)$ is around 0.25 and that the values for $l = 1$ and $l = 7$ are very close. That $\hat{J}_l^\star(\mathbf{b}) \leq 0$ for all $\mathbf{b} \in \mathcal{B}$ and all $l \in \{1, \ldots, L\}$ has an intuitive explanation. For any $\mathbf{b}$, the attacker has the option to never attack if $s = 0$ or to abort an attack if $s = 1$. Both options yield a cumulative reward less than or equal to 0 (2.3). As a consequence, $\hat{J}_l^\star(\mathbf{b}) \leq 0$ in any equilibrium[17]. The fact that $\hat{J}_l^\star(\mathbf{b}) = 0$ when $\mathbf{b}(1) = 1$ can be understood as follows. $\mathbf{b}(1) = 1$ means that the defender knows that an intrusion occurs and will take defensive actions; see Thm. 2.1.B. Hence, when $\mathbf{b}(1) = 1$, the only way for the attacker to avoid detection is to abort the intrusion, which causes the game to end and yields a reward of zero. We interpret the fact that $\arg\min_{\mathbf{b}(1)} \hat{J}_l^{\star}(\mathbf{b})$ is around 0.25 as follows. The value of $\mathbf{b}(1)$ that obtains the minimum corresponds to the belief state where the attacker achieves the lowest expected reward in the game. Negative rewards are obtained when the defender mistakes an intrusion for no intrusion and vice versa (2.3). Consequently, the attacker prefers belief states where the defender has a high uncertainty, e.g., $\mathbf{b}(1) = 0.5$. At the same time, the attacker does not want $\mathbf{b}(1)$ to be so large that the defender performs all its defensive actions before it gets a chance to attack, which can explain why we find the minimum to be around 0.25 rather than 0.5.

[16] See Cor. 1 in the background chapter.
[17] Recall that the attacker aims to minimize reward.

Lastly, Fig. 2.10 shows the percentage of blocked attacker and client traffic when running repeated game episodes on the digital twin with different defender strategies. The x-axis shows the running time, and the y-axis shows the percentage of blocked traffic per second.

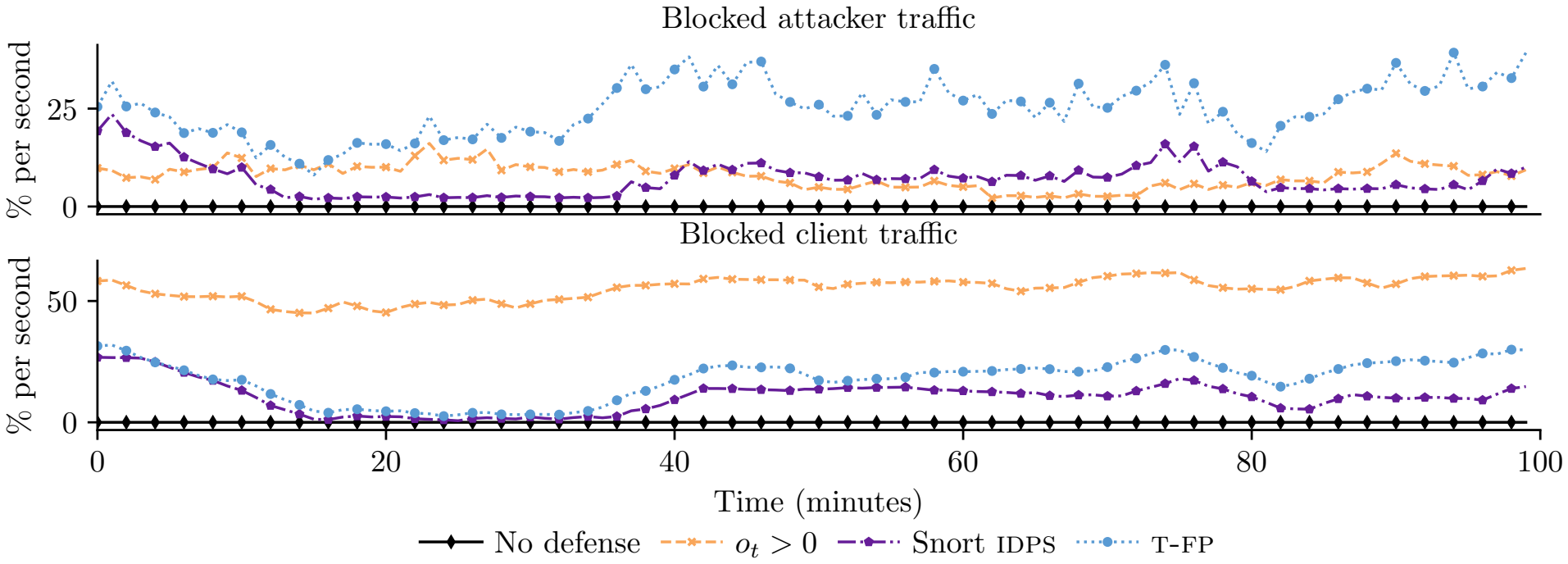

***Figure 2.10:*** *Percentage of blocked attacker and client traffic in the digital twin; the blue curves show results from the equilibrial strategy learned via* T-FP; *the purple, orange, and black curves relate to baseline strategies.*

We observe in the upper plot of Fig. 2.10 that all defender strategies block some client traffic, which is expected considering the false IDS alarms generated by the clients; see Fig. 2.6[18]. The $o_t > 0$ baseline strategy blocks the most *client* traffic (orange curves), and the SNORT IDPS baseline strategy blocks the least (purple curves), slightly less than the equilibrial strategy learned through T-FP (blue curves). We further observe in the lower plot that the equilibrial strategy blocks the most *attacker* traffic and that the $o_t > 0$ baseline strategy blocks the least. This observation suggests that the equilibrial strategy balances the trade-off between blocking clients and blocking the attacker. In comparison, the $o_t > 0$ baseline blocks nearly all traffic, and the SNORT IDPS baseline blocks too little traffic, failing to stop the intrusion.

[18]The defender actions that cause traffic to be dropped are listed in Table 2.2.

**Discussion of the evaluation results**

In this paper, we apply our methodology[19] to an intrusion response use case that involves a dynamic attacker, i.e., an attacker that adapts its strategy to the defender's strategy. The key findings can be summarized as follows:

- Our methodology can efficiently approximate optimal defender strategies for a practical IT infrastructure (Fig. 2.7). While we have not evaluated the learned strategies in the target infrastructure due to safety reasons, the fact that they achieve almost the same performance on the digital twin as on the simulator gives us confidence in their performance in the target infrastructure.

- The theory of optimal stopping provides insight about best responses for attackers and defenders, which enables efficient computation through stochastic approximation (Fig. 2.8). This finding can be explained by the threshold structures of the best response strategies, which drastically reduce the search space of possible strategies (Thm. 2.1 and Algorithm 2.1).

- The learned strategies can be efficiently implemented using the threshold properties. The computational complexity, which is dominated by the computation of the belief state, is upper bounded by $O(k|\mathcal{S}|^2|\mathcal{A}_{\mathrm{A}}|)$, where $k$ is a constant (22).

- Static defender strategies' performance deteriorates against a dynamic attacker, whereas defender strategies obtained through T-FP improve over time (Fig. 2.7). This finding is consistent with previous studies that use game-theoretic approaches (e.g., (Dijk et al., 2013)) and suggests limitations of static response systems, such as the SNORT IDPS (Roesch, 1999).

## 2.6 Related Work

The literature on game-theoretic modeling in cybersecurity is vast; a good introduction is offered in the textbook by (Alpcan and Basar, 2010). Security games are modeled in different ways depending on the use case. Examples include: APT games [116, 210, 398, 36, 483, 517, 186, 312, 502, 446], honeypot placement games [124, 202, 354], resource allocation games [476, 528], authentication games [393], distributed denial-of-service games [33, 471], situational awareness games [78, 144], moving target defense games [399, 327, 450, 451], jamming games [13], data corruption games [170], security provisioning games [364], and intrusion response games [325, 178, 150, 268, 9, 527, 205, 531, 284, 493, 484, 521, 16, 270]. These games are

[19] See the methodology chapter for details about the experimental methodology.

formulated using various game-theoretic models. For example: stochastic games [393, 398, 325, 527, 521, 13, 521], extensive-form games [9, 11], Blotto games [528], differential games [515, 16, 514, 502], hypergames [36, 483], partially observed stochastic games [471, 178, 150, 284, 493], Stackelberg games [471, 268, 531], graph-based games [476, 327], evolutionary games [515, 205, 450], continuous-kernel games [9], rivalry games [517], and Bayesian games [399]. This paper differs from the referenced works in two main ways. First, we model the intrusion response use case as an optimal stopping game. The benefit of our model is that it provides insight into the structure of best responses through the theory of optimal stopping. Second, we evaluate obtained strategies on a digital twin. This evaluation methodology contrasts with most of the game-theoretic approaches proposed in prior work, which evaluate strategies analytically or in simulation.

Game-theoretic models based on optimal stopping theory can be found in prior research on Dynkin games (Dynkin, 1969), (Alario-Nazaret et al., 1982), (Solan and Vieille, 2002), (Lempa and Matomäki, 2010), (Ekström et al., 2017). Compared to these models, our model is more general by (*i*) allowing each player to take multiple stop actions within an episode; and (*ii*) by not assuming a game of perfect information. Another difference is that the referenced papers either study purely mathematical problems or problems in mathematical finance. To the best of our knowledge, we are the first to apply the stopping game formulation to the use case of intrusion response. Our stopping game has similarities with the FLIPIT game (Dijk et al., 2013) and signaling games (Noe, 1988), both of which are commonplace in the security literature; see survey (Manshaei et al., 2013). Signaling games have the same information asymmetry as our game, and FLIPIT uses the same binary state space to model the state of an attack. The main differences are as follows. FLIPIT models the use case of advanced persistent threats and is a symmetric non-zero-sum game. In contrast, our game models an intrusion response use case and is asymmetric. Compared to signaling games, the main difference is that our game is a sequential and simultaneous-move game. Signaling games, in comparison, are typically two-stage games where one player moves in each stage.

Previous game-theoretic studies using digital twins like ours are (Aydeger et al., 2021) and (Zonouz et al., 2009). Specifically, in (Aydeger et al., 2021), a denial-of-service use case is formulated as a signaling game, for which a Nash equilibrium is derived. This equilibrium is then used to design a defense mechanism, which is evaluated in a software-defined network emulation based on MININET (Lantz et al., 2010). Compared to this paper, the main differences are that we focus on a different use case than (Aydeger et al., 2021) and that our solution method is based on stochastic approximation. Similar to this paper, the authors of (Zonouz et al., 2009) formulate an intrusion response use case as a POSG where the defender observes alerts from the SNORT IDS (Roesch, 1999). In contrast to our approach, however, the approach of (Zonouz et al., 2009) assumes access to attack-defense trees designed by human experts. Another difference between this paper and (Zonouz et al., 2009) is the POSG. The POSG in (Zonouz et al., 2009) has a larger state space than the POSG considered in this paper. Although this makes the POSG in (Zonouz et al., 2009) more expressive than ours, it also makes the computation of optimal defender strategies intractable. To estimate optimal defender strategies, the authors of (Zonouz et al., 2009) are forced to approximate their model with one that has a smaller state space and is fully observed. In comparison, we can efficiently approximate the equilibria of our game without relying on model simplifications and without assuming access to attack-defense trees designed by human experts.

## 2.7 Conclusion

In this work, we apply our methodology for automated security response to an intrusion response use case. We formulate the interaction between an attacker and a defender as an optimal stopping game. This formulation gives us insight into the structure of best responses, which we prove to have threshold properties. Based on this knowledge, we develop **T**hreshold-**F**ictitious **P**lay (T-FP), an efficient algorithm for learning equilibria. The results from running T-FP show that the learned strategies converge to an approximate Perfect Bayesian Equilibrium (PBE) and thus to near-optimal strategies (Fig. 2.7). The results also demonstrate that T-FP converges faster than a state-of-the-art fictitious play algorithm by taking advantage of threshold properties of best responses (Fig. 2.8). To assess the learned strategies in an operational environment, we evaluate them on digital twin of the target infrastructure (Fig. 1.2). The results attest that the strategies achieve almost the same performance on the digital twin as on the simulator. This result gives us confidence that the obtained strategies would perform as expected in the target infrastructure, which is not feasible to evaluate directly for safety reasons.

In the broader context of this thesis, this paper extends the optimal stopping formulation in Paper 1 to a game-theoretic formulation. The benefit of this approach is that it allows us to obtain stopping strategies that are optimal against a *dynamic* attacker, i.e., an attacker that updates its strategy to circumvent encountered defenses. In this paper, we have focused on the problem of learning the optimal *times* for taking defensive actions against a dynamic attacker. In the next chapter of the thesis (Paper 3), we extend the approach presented in this paper to not only compute when defensive actions need to be taken but also the specific measures to be executed.

## ■ Acknowledgments

The authors would like to thank Pontus Johnson for his useful input to this research and Forough Shahab Samani and Xiaoxuan Wang for their constructive comments on a draft of this paper. The authors are also grateful to Branislav Bosanský for sharing the code of the HSVI algorithm for one-sided POSGs and to Jakob Stymne for contributing to our implementation of NFSP.

# ■ Appendix

## A Proofs

### A.1 Proof of Theorem 2.1.A

The existence of Nash and Perfect Bayesian Equilibria (PBE) follows from Thm. 3 in the background chapter. We prove that a PBE in pure strategies exists when $s = 0 \iff \mathbf{b}(1) = 0$ using a proof by construction. It follows from (2.3) that if $s = 0 \iff \mathbf{b}(1) = 0$, then the defender strategy $\tilde{\pi}_{\mathrm{D}}(\mathbf{b}) = \mathsf{S} \iff \mathbf{b}(1) \neq 0$ is (weakly) dominating[20] (Def. 1.1, Fudenberg and Tirole, 1991). Given this defender strategy, it follows from (2.3) that the pure strategy

$$\tilde{\pi}_{\mathrm{A}}(\mathbf{b}, 0) = \mathsf{C} \quad \text{and} \quad \tilde{\pi}_{\mathrm{A}}(\mathbf{b}, 1) = \mathsf{S} \qquad \forall \mathbf{b} \in \mathcal{B}$$

is (weakly) dominating for the attacker. Hence, $(\tilde{\pi}_{\mathrm{D}}, \tilde{\pi}_{\mathrm{A}})$ is a pure PBE. □

### A.2 Proof of Theorem 2.1.B.

Given $\Gamma$ and a fixed attacker strategy $\pi_{\mathrm{A}}$, any best response for the defender $\tilde{\pi}_{\mathrm{D}} \in \mathscr{B}_{\mathrm{D}}(\pi_{\mathrm{A}})$ (2.6a) is an optimal strategy in a POMDP $\mathcal{M}^P$ (see §2.3). Hence, it suffices to show that there exists an optimal strategy $\pi^{\star}_{\mathrm{D}}$ in $\mathcal{M}^P$ that satisfies the threshold structure in (2.8). Conditions for (2.8) to hold and the existence proof are given in Thm. 1.1 of Paper 1. Since $z$ (2.4) is TP-2 by assumption and all of the remaining conditions hold by definition of $\Gamma$, the statement follows. □

### A.3 Proof of Theorem 2.1.C.

We start by proving (2.9b). Since $f(\emptyset \mid 1, a^{(\mathrm{D})}, \mathsf{S}) = 1$ for all $a^{(\mathrm{D})}$ (2.2e), the problem of selecting the best response action in state $s = 1$ for the attacker is an optimal stopping problem. Hence, it suffices to show that $\mathscr{S}^{(\mathrm{A})}_{1,l,\pi_{\mathrm{D}}} = [\tilde{\beta}_{1,l}, 1]$ for all $l \in \{1, \ldots, L\}$ and any $\pi_{\mathrm{D}} \in \Pi_{\mathrm{D}}$ that satisfies (2.8). To do this, we first establish a helpful lemma.

**Lemma 2.1.** *Let $J^{\star}_{\pi_{\mathrm{D}},l}$ denote the optimal cost-to-go function in the best response* MDP *for the attacker given a defender strategy $\pi_{\mathrm{D}}$*[21]*. Then, $J^{\star}_{\pi_{\mathrm{D}},l}(\mathbf{b}, s) \leq 0$ for all $s \in \mathcal{S}$, $\mathbf{b} \in \mathcal{B}$, and $l \in \{1, \ldots, L\}$.*

*Proof.* Let $\bar{\pi}_{\mathrm{A}}(\cdot, 0) = \mathsf{C}$ and $\bar{\pi}_{\mathrm{A}}(\cdot, 1) = \mathsf{S}$. Then it follows from (2.3) that $J^{\bar{\pi}_{\mathrm{A}}}_{\pi_{\mathrm{D}},l}(\mathbf{b}, s) \leq 0$ for any $\pi_{\mathrm{D}} \in \Pi_{\mathrm{D}}$, $s \in \mathcal{S}$, $\mathbf{b} \in \mathcal{B}$, and $l \in \{1, \ldots, L\}$. By optimality, $J^{\bar{\pi}_{\mathrm{A}}}_{\pi_{\mathrm{D}},l}(\mathbf{b}, s) \geq J^{\star}_{\pi_{\mathrm{D}},l}(\mathbf{b}, s)$. Hence, $J^{\star}_{\pi_{\mathrm{D}},l}(\mathbf{b}, s) \leq 0$. □

[20] A strategy is weakly dominant if it weakly dominates all other strategies; see (Def. 1.1, Fudenberg and Tirole, 1991) for the definition of weak and strict dominance.

[21] The cost is defined as the defender's reward (2.3), which the attacker seeks to minimize.

Returning to the proof of (2.9a). When $s = 1$ and $\mathbf{b}(1) < \tilde{\alpha}_l$ (2.8) we obtain[22] from the Bellman equation that the attacker's best response action $\tilde{a}^{(\mathrm{A})}$ satisfies

$$\tilde{a}^{(\mathrm{A})} \in \operatorname*{arg\,min}_{a^{(\mathrm{A})} \in \mathcal{A}_{\mathrm{A}}} \Bigg[ \underbrace{0}_{a^{(\mathrm{A})}=\mathsf{S}}, \underbrace{\mathbb{E}_{\mathbf{B}'}\Big[R_{\mathrm{int}} + \gamma \overbrace{J^{\star}_{l,\pi_{\mathrm{D}}}(\mathbf{B}',1)}^{\leq 0 \text{ (Lemma 2.1)}} \mid \mathbf{b}, a^{(\mathrm{A})} = \mathsf{C}\Big]}_{a^{(\mathrm{A})}=\mathsf{C}} \Bigg] \overset{(a)}{=} \{\mathsf{C}\},$$

where (a) follows from Lemma 2.1 and the fact that $R_{\mathrm{int}} < 0$. Therefore, $\mathscr{S}^{(\mathrm{A})}_l \subseteq [\tilde{\alpha}_l, 1]$. Now consider the case when $\mathbf{b}(1) = 1$, then the Bellman equation and (2.3) implies that

$$\tilde{a}^{(\mathrm{A})} \in \operatorname*{arg\,min}_{a^{(\mathrm{A})} \in \mathcal{A}_{\mathrm{A}}} \Bigg[ \overbrace{0}^{a^{(\mathrm{A})}=\mathsf{S}}, \overbrace{\frac{R_{\mathrm{st}}}{l} + \gamma J^{\star}_{l-1,\pi_{\mathrm{D}}}(\mathbf{b},1)}^{a^{(\mathrm{A})}=\mathsf{C}} \Bigg] \overset{(a)}{=} \operatorname*{arg\,min}_{a^{(\mathrm{A})} \in \mathcal{A}_{\mathrm{A}}} \left[0, \sum_{k=0}^{\min[l-1,\tau-1]} \gamma^k \frac{R_{\mathrm{st}}}{l-k}\right]$$
$$\overset{(b)}{=} \{\mathsf{S}\},$$

where $\tau \geq 1$ is the next stopping time of the attacker; (a) follows because $\mathbf{b}(1) = 1$ is an absorbing belief state until the game ends (2.2) and $\tilde{\alpha}_l \leq 1$ for all $l \in \{1, \dots, L\}$ (Thm. 2.1.B), which means that $a^{(\mathrm{D})} = \mathsf{S}$ until the game ends; and (b) follows because $R_{\mathrm{st}} > 0$. Hence, $\mathbf{b}(1) = 1 \implies \mathbf{b} \in \mathscr{S}^{(\mathrm{A})}_{1,l,\pi_{\mathrm{D}}}$. Since $\mathscr{S}^{(\mathrm{A})}_{1,l,\pi_{\mathrm{D}}}$ is convex (see Lemma 1.1 of Paper 1), it follows that $\mathscr{S}^{(\mathrm{A})}_{1,l,\pi_{\mathrm{D}}} = [\tilde{\beta}_{1,l}, 1]$ for some threshold $\tilde{\beta}_{1,l} \geq \tilde{\alpha}_l$. This proves (2.9b).

Now we turn our attention to (2.9a) and the case when $s = 0$. Define

$$\overline{r}_l(\mathbf{b}) \triangleq \mathbb{E}_{\pi_{\mathrm{D}},\pi_{\mathrm{A},\tilde{\beta}}}\left[\sum_{t=1}^{\infty} \gamma^{t-1} r_l(S_t, A^{(\mathrm{D})}, A^{(\mathrm{A})}) \mid s_1 = 1, \mathbf{b}_1 = \mathbf{b}\right],$$

where $\pi_{\mathrm{A},\tilde{\beta}}$ is the threshold strategy induced by (2.9b). Given this definition, we can rewrite the Bellman equation in state $s = 0$ as

$$J^{\star}_{\pi_{\mathrm{D}},l}(\mathbf{b},0) = \min_{a^{(\mathrm{A})} \in \mathcal{A}_{\mathrm{A}}} \Bigg[\mathbb{E}_{A^{(\mathrm{D})},\mathbf{B}'}\Big[r_l(0, A^{(\mathrm{D})}, \mathsf{S}) + \gamma J^{\star}_{\pi_{\mathrm{D}},l-A^{(\mathrm{D})}}(\mathbf{B}',1)\Big],$$
$$\mathbb{E}_{A^{(\mathrm{D})},\mathbf{B}'}\Big[r_l(0, A^{(\mathrm{D})}, \mathsf{C}) + \gamma J^{\star}_{\pi_{\mathrm{D}},l-A^{(\mathrm{D})}}(\mathbf{B}',0)\Big]\Bigg]$$
$$= \min_{\tau_{\mathrm{A},1}} \mathbb{E}_{\pi_{\mathrm{D}}}\Bigg[\sum_{t=1}^{\min[T-1,\tau_{\mathrm{A},1}-1]} \gamma^{t-1} r_{l_t}(0, A^{(\mathrm{D})}_t, \mathsf{C}) +$$
$$\mathbb{1}_{T > \tau_{\mathrm{A},1}}\left(\gamma^{\tau_{\mathrm{A},1}-1} r_{l_t}(0, A^{(\mathrm{D})}_t, \mathsf{S}) + \gamma^{\tau_{\mathrm{A},1}} \overline{r}_{l_{\tau_{\mathrm{A},1}+1}}(\mathbf{B}_{\tau_{\mathrm{A},1}+1})\right)\Bigg],$$

[22] Recall that $\tilde{\alpha}_l$ is the stopping threshold for the defender with $l$ stops remaining

where $\tau_{A,1}$ is the first stopping time of the attacker. Hence, the problem of selecting the best response action in state $s = 0$ for the attacker is an optimal stopping problem. Since $\mathscr{S}^{(A)}_{0,l,\pi_D} \subseteq [0,1]$ is convex (see Lemma 1.1 of Paper 1) and $0 \in \mathscr{S}^{(A)}_{0,l,\pi_D}$ by assumption, it follows that $\mathscr{S}^{(A)}_{0,l,\pi_D} = [0, \tilde{\beta}_{0,l}]$ for some threshold $\tilde{\beta}_{0,l} \in [0,1]$. □

## B Hyperparameters

The hyperparameters used for the evaluation in this paper are listed in Table 2.4 and were obtained through grid search.

| *Game parameters* | *Values* |
|---|---|
| $\mathrm{R_{st}}, \mathrm{R_{cost}}, \mathrm{R_{int}}, \gamma, \phi_l, L$ | 20, −2, −1, 0.99, 1/2$l$, 7 |
| T-FP *parameters* | *Values* |
| $c, \epsilon, \lambda, A, a, N, \delta$ | 10, 0.101, 0.602, 100, 1, 50, 0.2 |
| NFSP *parameters* [192, Alg. 1] | *Values* |
| lr RL, lr SL, batch, # layers, # neurons, $\mathcal{M}_{RL}$ | $10^{-2}$,$5 \cdot 10^{-3}$, 64, 2, 128, $2 \times 10^5$ |
| $\mathcal{M}_{SL}$, $\epsilon$, $\epsilon$-decay, $\eta$ | $2 \times 10^6$, 0.06, 0.001, 0.1 |
| HSVI *parameter* [201, Alg. 1] | *Value* |
| $\epsilon$ | 3 |

***Table 2.4:*** *Hyperparameters of the* POSG *and the algorithms used for evaluation.*

## C Distributions of Infrastructure Metrics

The digital twin collects hundreds of metrics every time step. To measure the information that a metric provides for detecting intrusions, we calculate the Kullback-Leibler (KL) divergence $D_{\mathrm{KL}}(\widehat{z}_{O|s=0} \parallel \widehat{z}_{O|s=1})$ between the distribution of the metric when no intrusion occurs $\widehat{z}_{O|s=0}$ and during an intrusion $\widehat{z}_{O|s=1}$ (Kullback and Leibler, 1951):

$$D_{\mathrm{KL}}(\widehat{z}_{O|s=0} \parallel \widehat{z}_{O|s=1}) = \sum_{o \in \mathcal{O}} \widehat{z}_{O|s=0}(o) \ln \left( \frac{\widehat{z}_{O|s=0}(o)}{\widehat{z}_{O|s=1}(o)} \right). \tag{2.14}$$

Here $o \in \mathcal{O}$ realizes the random variable $O$, which represents the metric's value. ($\mathcal{O}$ is the domain of $O$.)

Figure 2.11 on the next page shows empirical distributions of the collected metrics with the largest KL divergence. We see that the IDS alerts have the largest KL divergence and thus provide the most information for detecting intrusions.

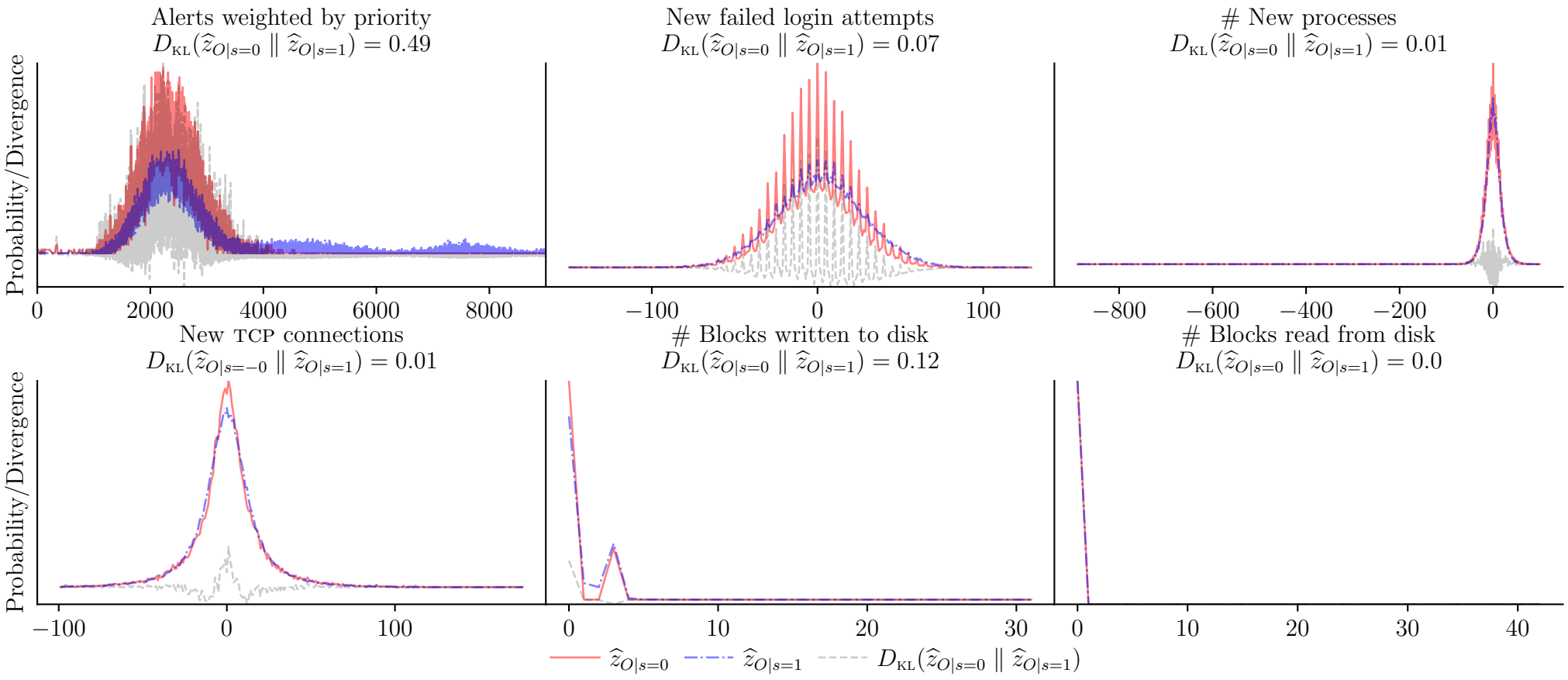

***Figure 2.11:*** *Empirical distributions of selected infrastructure metrics; the red and blue lines show the distributions when no intrusion occurs and during intrusion, respectively.*

## D Attacker Actions

The attacker actions and their descriptions are listed in Table 2.5.

| *Action* | *Description* |
|---|---|
| TCP scan | TCP port scan by using SYN packets using `nmap`, it allows detecting open TCP ports. |
| UDP port scan | UDP port scan by sending UDP packets using `nmap`, it allows detecting open UDP ports. |
| ping scan | IP scan with ICMP ping messages. |
| VULSCAN | vulnerability scan using `nmap`. |
| brute-force attack | dictionary attack against a login service using `nmap`. |
| CVE-2017-7494 exploit | remote code execution using the SAMBA. |
| CVE-2015-3306 exploit | uses `mod_copy` in `proftpd` for remote code execution. |
| CVE-2014-6271 exploit | uses a vulnerability in `bash` for remote code execution. |
| CVE-2016-10033 exploit | uses `phpmailer` for remote code execution. |
| CVE-2015-1427 exploit | uses `elasticsearch` for remote code execution. |
| exploit of CWE-89 weakness on DVWA [454] | injects SQL code for SQLITE3. |

***Table 2.5:*** *Descriptions of the attacker actions; shell commands and scripts for executing the actions are listed in (Hammar, 2023).*

# Paper 3[†]

## SCALABLE LEARNING OF INTRUSION RESPONSE THROUGH RECURSIVE DECOMPOSITION

Kim Hammar and Rolf Stadler

**Abstract**

We study automated intrusion response for an IT infrastructure and formulate the interaction between an attacker and a defender as a partially observed stochastic game. To solve the game, we follow an approach where attack and defense strategies co-evolve through fictitious play toward an equilibrium. Solutions proposed in previous work prove the feasibility of this approach for small infrastructures but do not scale to realistic scenarios due to the exponential growth in computational complexity with the infrastructure size. We address this problem by introducing a method that recursively decomposes the game into subgames with low computational complexity, which can be solved in parallel. Applying optimal stopping theory, we show that the best responses in these subgames exhibit threshold structures, which allows us to compute them efficiently. To solve the decomposed game, we introduce an algorithm called **D**ecompositional **F**ictitious **P**lay (DFP), which learns equilibria through stochastic approximation. We evaluate the learned strategies on a digital twin. The results demonstrate that the learned strategies approximate an equilibrium and that DFP significantly outperforms a state-of-the-art algorithm for a realistic infrastructure configuration.

*In the face of complexity, an in-principle reductionist may be at the same time a pragmatic holi.*

— Herbert Simon ***1962***, *The architecture of complexity.*

## 3.1 Introduction

In contrast to Paper 1 and Paper 2, this paper considers not only the problem of *when* defensive actions need to be taken but also the selection of *which* action to execute in order to effectively mitigate an attack. We formulate this problem as a stochastic game where the attacker and the defender have several possible actions per node in the infrastructure. This detailed modeling means the game's complexity grows exponentially with the number of nodes. To manage this complexity, we recursively decompose the game into simpler subgames, which allow detailed modeling while keeping computational complexity low.

The decomposition involves three steps. First, we partition the infrastructure according to workflows that are isolated from each other. This isolation allows us to decompose the game into *independent subgames* (one per workflow) that can be solved in parallel. Second, the graph structure of a workflow allows us to decompose the workflow games into node subgames. We prove that these subgames have *optimal substructure* (Ch. 15, Cormen et al., 2022), which means that a best response of the original game can be obtained from best responses of the node subgames. Third, we show that the problem of selecting *which* response action to apply on a node can be separated from that of deciding *when* to apply the action, which enables efficient learning of best responses through the application of *optimal stopping* theory (Wald, 1947). We use this insight to design an efficient algorithm called **D**ecompositional **F**ictitious **P**lay (DFP), which allows scalable approximation of perfect Bayesian equilibria (PBE).

We summarize the contributions in this paper as follows.

1. We formulate the intrusion response problem as a partially observed stochastic game and prove that, under assumptions often met in practice, the game decomposes into subgames that can be solved in parallel.

2. We show that the best responses in these subgames exhibit threshold structures, which allows us to compute them efficiently.

3. We design DFP, an efficient fictitious play algorithm for approximating a PBE of the decomposed game.

4. For a realistic use case, we evaluate the learned response strategies against network intrusions on a digital twin[2].

[2]The digital twin is created using CSLE, as described in the methodology chapter.

## 3.2 Related Work

Networked systems in engineering and science often exhibit a modular topological structure that can be exploited for designing control algorithms (Ouyang et al., 2017). Šiljak first suggested system decomposition for automatic control in 1978 (Šiljak, 1978) and approaches based on decomposition, such as divide and conquer, layering, and hierarchical structuring, are well established in the design of large-scale systems, a notable example being the Internet. Similar decomposition methods are frequently used in robotics and multi-agent systems, as exemplified by the subsumption architecture (Brooks, 1986). Within the fields of decision- and game-theory, decomposition is studied in the context of factored decision processes [402, 234, 422, 44], distributed decision processes [320], factored games [209, 340], and graph-structured games [235].

Decomposition to automate intrusion response has been studied first in (Huang et al., 2018), (Rasouli et al., 2018), (Zheng and Castañón, 2013), and (Zan et al., 2010). The work in (Huang et al., 2018) formulates the interaction between a defender and an attacker on a cyber-physical infrastructure as a factored Markov game. They introduce a decomposition based on linear programming. Following a similar approach, the work in (Zheng and Castañón, 2013) studies a Markov game formulation and shows that a multi-stage game can be decomposed into a sequence of one-stage games. In a separate line of work, (Rasouli et al., 2018) models intrusion response as a minimax control problem and develops a heuristic decomposition based on clustering and influence graphs. This approach resembles the work in (Zan et al., 2010), which studies a factored decision process and proposes a hierarchical decomposition.

In the above works, decomposition is key to obtaining effective strategies for large-scale systems. Compared to our work, some of them propose decomposition methods without optimal substructure (Rasouli et al., 2018). Others do not consider partial observability (Huang et al., 2018), (Zheng and Castañón, 2013), or dynamic attackers (Zan et al., 2010). Most importantly, the above works evaluate the obtained strategies in a simulation environment. They do not perform evaluation on a digital twin as we report in this paper, which gives higher confidence that the strategies are effective on the target infrastructure.

## 3.3 The Intrusion Response Use Case

We consider an intrusion response use case that involves an IT infrastructure that is segmented into *zones* with virtual *nodes* that run network services; see Fig. 3 in the introduction chapter. Services are realized by *workflows* that clients access through a gateway, which is also open to an attacker. The attacker aims to intrude on the infrastructure, compromise nodes, and disrupt workflows. To achieve this goal, it can take three types of actions: reconnaissance, brute-force attacks, and exploits; see Fig. 7 in the introduction chapter. To prevent the attacker, a defender continuously monitors the infrastructure by accessing and analyzing intrusion detection alerts and other statistics. It can take four types of defensive actions to respond to possible intrusions: migrate nodes between zones, redirect or block network flows, shut down nodes, and revoke node access; see Fig. 3.1 below. When deciding between these actions, the defender balances two conflicting objectives: maximize workflow utility towards clients and minimize the cost of intrusion.

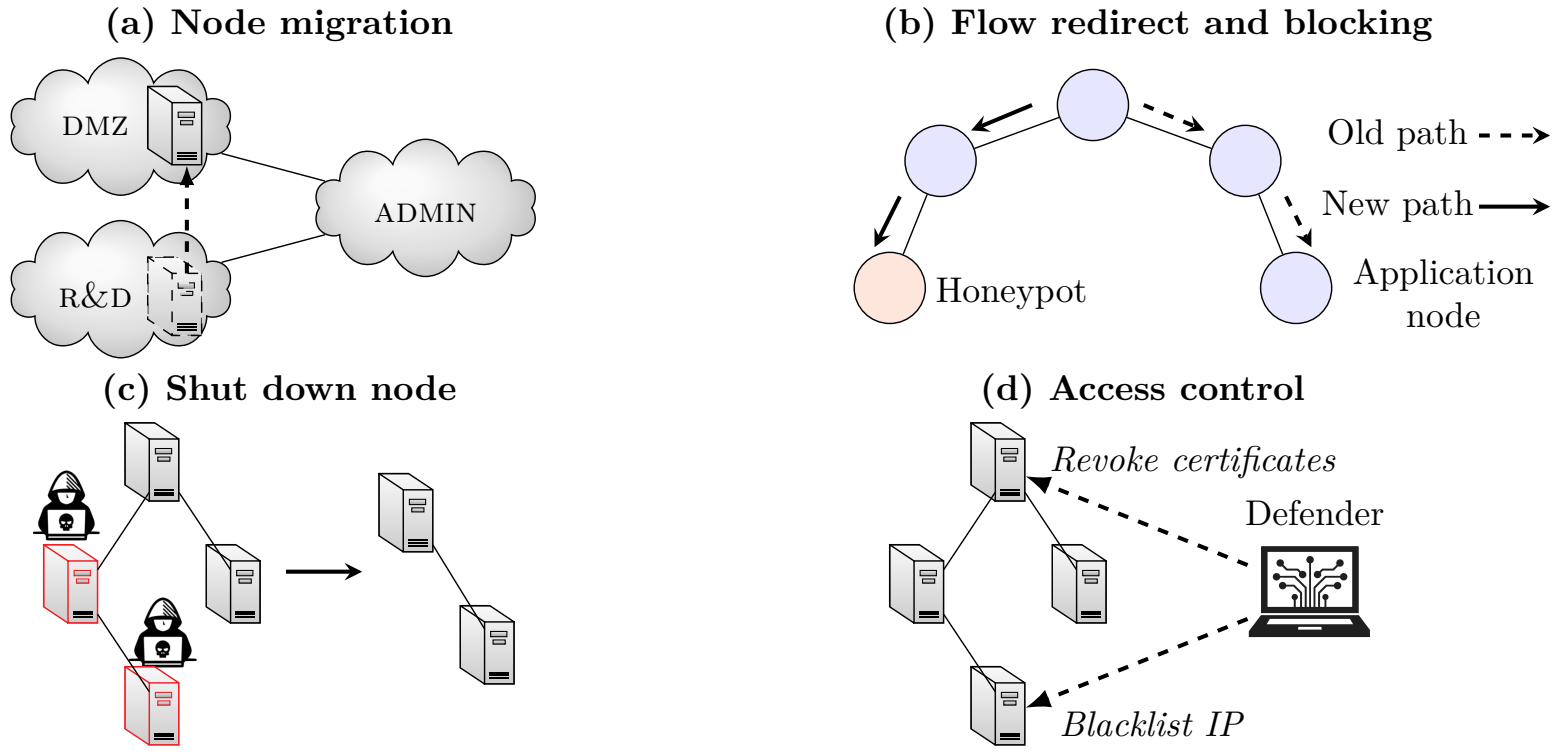

***Figure 3.1:*** *Defender actions: (a) migrate a node between two zones; (b) redirect or block traffic flows to a node; (c) shut down a node; and (d) revoke access to a node.*

## 3.4 Formalizing the Intrusion Response Problem

We formalize the above use case as an optimization problem where the goal is to select an optimal sequence of defender actions in response to a sequence of attacker actions. We assume a dynamic attacker, which leads to a game-theoretic formulation. The game is played on the IT infrastructure, which we model as a discrete-time dynamical system whose evolution depends on the actions of the attacker and the defender. Both actors have partial observability of the system state. Their observations depend on the traffic generated by clients requesting service, which we assume can be described by a stationary process.

### Modeling the infrastructure and services

Following the description in §3.3, we consider an IT infrastructure with application servers connected by a communication network that is segmented into zones; see Fig. 3 in the introduction chapter. Overlaid on this physical infrastructure is a virtual infrastructure with a tree structure that includes *nodes*, which collectively offer services to clients. A service is modeled as a *workflow*, which comprises a set of interdependent nodes. A dependency between two nodes reflects information exchange through service invocations. As an example of a virtual infrastructure, we can think of a microservice architecture where a workflow is defined as a tree of microservices; see Fig. 4 in the introduction chapter.

**Assumption 3.1.** *Each node belongs to exactly one workflow.*

Assumption 3.1 can always be satisfied in practice by splitting a node that belongs to multiple workflows into smaller virtual nodes.

***Infrastructure*** We model the virtual infrastructure as a (finite) directed graph $\mathcal{G} \triangleq \langle \{\mathrm{gw}\} \cup \mathcal{V}, \mathcal{E} \rangle$. The graph has a tree structure rooted at the gateway gw. Each node $i \in \mathcal{V}$ has three state variables. $v^{(\mathrm{R})}_{i,t}$ represents the reconnaissance state. $v^{(\mathrm{R})}_{i,t} = 1$ if the attacker has discovered the node, 0 otherwise. $v^{(\mathrm{I})}_{i,t}$ represents the intrusion state. $v^{(\mathrm{I})}_{i,t} = 1$ if the attacker has compromised the node, 0 otherwise. Lastly, $v^{(\mathrm{Z})}_{i,t}$ indicates the zone in which the node resides. We call a node *active* if it is functional as part of a workflow (denoted $\alpha_{i,t} = 1$). Due to a defender action (e.g., a shutdown), a node $i \in \mathcal{V}$ may become inactive ($\alpha_{i,t} = 0$). The active state is determined by $v^{(\mathrm{Z})}_{i,t}$, i.e, $\alpha_{i,t}$ is a function of $v^{(\mathrm{Z})}_{i,t}$.

***Workflows*** We model a workflow $\mathbf{w} \in \mathcal{W}$ as a subtree $\mathcal{G}_{\mathbf{w}} \triangleq \langle \{\mathrm{gw}\} \cup \mathcal{V}_{\mathbf{w}}, \mathcal{E}_{\mathbf{w}} \rangle$ of the infrastructure graph. Workflows do not overlap except for the gateway, which belongs to all workflows.

## Modeling actors

The intrusion response use case involves three types of actors: an attacker, a defender, and clients; see Fig. 3 in the introduction chapter.

***Attacker*** At each time $t$, the attacker takes an action $\mathbf{a}^{(\mathrm{A})}_t$, which is defined as the composition of the local actions on all nodes $\mathbf{a}^{(\mathrm{A})}_t \triangleq (\mathbf{a}^{(\mathrm{A})}_{1,t}, \ldots, \mathbf{a}^{(\mathrm{A})}_{|\mathcal{V}|,t}) \in \mathcal{A}_{\mathrm{A}}$, where $\mathcal{A}_{\mathrm{A}}$ is finite. A local action is either a null action (denoted with $\perp$) or an offensive action. An offensive action on a node $i$ may change the reconnaissance state $v^{(\mathrm{R})}_{i,t}$ or the intrusion state $v^{(\mathrm{I})}_{i,t}$. A node $i$ can only be compromised if it is discovered, i.e., if $v^{(\mathrm{R})}_{i,t} = 1$. We express this constraint as $\mathbf{a}^{(\mathrm{A})}_t \in \mathcal{A}_{\mathrm{A}}(\mathbf{s}^{(\mathrm{A})}_t)$.

The attacker state $\mathbf{s}^{(\mathrm{A})}_t \triangleq \big(v^{(\mathrm{R})}_{i,t}, v^{(\mathrm{I})}_{i,t}\big)_{i \in \mathcal{V}} \in \mathcal{S}_{\mathrm{A}}$ evolves as

$$\mathbf{s}^{(\mathrm{A})}_{t+1} \sim f_{\mathrm{A}}\big(\cdot \mid \mathbf{s}^{(\mathrm{A})}_t, \mathbf{a}^{(\mathrm{A})}_t, \mathbf{a}^{(\mathrm{D})}_t\big), \tag{3.1}$$

where $\mathbf{a}^{(\mathrm{D})}_t$ represents the defender action at time $t$, as defined below.

***Defender*** At each time $t$, the defender takes action $\mathbf{a}^{(\mathrm{D})}_t$, which is defined as the composition of the local actions on all nodes $\mathbf{a}^{(\mathrm{D})}_t \triangleq (\mathbf{a}^{(\mathrm{D})}_{1,t}, \ldots, \mathbf{a}^{(\mathrm{D})}_{|\mathcal{V}|,t}) \in \mathcal{A}_{\mathrm{D}}$, where $\mathcal{A}_{\mathrm{D}}$ is finite. A local action is either a defensive action or a null action $\perp$. Each defensive action $\mathbf{a}^{(\mathrm{D})}_{i,t} \neq \perp$ leads to $\mathbf{s}^{(\mathrm{A})}_{i,t+1} = (0,0)$ and may affect $v^{(\mathrm{Z})}_{i,t+1}$.

The defender state $\mathbf{s}^{(\mathrm{D})}_t \triangleq \big(v^{(\mathrm{Z})}_{i,t}\big)_{i \in \mathcal{V}} \in \mathcal{S}_{\mathrm{D}}$ evolves according to

$$\mathbf{s}^{(\mathrm{D})}_{t+1} \sim f_{\mathrm{D}}\big(\cdot \mid \mathbf{s}^{(\mathrm{D})}_t, \mathbf{a}^{(\mathrm{D})}_t\big). \tag{3.2}$$

**Remark 3.1.** *By introducing special zones that represent nodes that are shut down or have their traffic redirected, the effects of all defensive actions in Fig. 3.1 can be modeled by manipulating the zones* $v^{(\mathrm{Z})}_{1,t}, \ldots, v^{(\mathrm{Z})}_{|\mathcal{V}|,t}$.

***Clients*** Clients consume infrastructure services by accessing workflows. We model client behavior through stationary stochastic processes, which affect the observations available to the attacker and the defender. That is, the clients are implicitly modeled by the observation function $z$, as defined below.

### Observability and strategies

At each time $t$, the defender and the attacker both observe $\mathbf{o}_t \triangleq \big(\mathbf{o}_{1,t}, \ldots, \mathbf{o}_{|\mathcal{V}|,t}\big) \in \mathcal{O}$, where $\mathcal{O}$ is finite[3]. $\mathbf{o}_t$ is drawn from the random vector $\mathbf{O}_t \triangleq (\mathbf{O}_{1,t}, \ldots, \mathbf{O}_{|\mathcal{V}|,t})$ whose marginal distributions $z_{\mathbf{O}_1}, \ldots, z_{\mathbf{O}_{|\mathcal{V}|}}$ are stationary and conditionally independent given $\mathbf{s}_{i,t} \triangleq (\mathbf{s}^{(\mathrm{D})}_{i,t}, \mathbf{s}^{(\mathrm{A})}_{i,t})$. (Note that $z_{\mathbf{O}_i}$ depends on the traffic generated by clients.) As a consequence, the joint conditional distribution $z$ is given by

$$z(\mathbf{o} \mid \mathbf{s}) = \prod_{i=1}^{|\mathcal{V}|} z_{\mathbf{O}_i}(\mathbf{o}_i \mid \mathbf{s}_i) \qquad \forall \mathbf{o} \in \mathcal{O}, \mathbf{s} \in \mathcal{S}_{\mathrm{A}} \times \mathcal{S}_{\mathrm{D}}. \tag{3.3}$$

The sequence of observations and states at times $1, \ldots, t$ forms the histories

$$\mathbf{h}^{(\mathrm{D})}_t \triangleq (\mathbf{b}^{(\mathrm{D})}_1, \mathbf{s}^{(\mathrm{D})}_1, \mathbf{a}^{(\mathrm{D})}_1, \mathbf{o}_2, \ldots, \mathbf{a}^{(\mathrm{D})}_{t-1}, \mathbf{s}^{(\mathrm{D})}_t, \mathbf{o}_t) \in \mathcal{H}_{\mathrm{D}}$$
$$\mathbf{h}^{(\mathrm{A})}_t \triangleq (\mathbf{b}^{(\mathrm{A})}_1, \mathbf{s}^{(\mathrm{A})}_1, \mathbf{a}^{(\mathrm{A})}_1, \mathbf{o}_2, \ldots, \mathbf{a}^{(\mathrm{A})}_{t-1}, \mathbf{s}^{(\mathrm{A})}_t, \mathbf{o}_t) \in \mathcal{H}_{\mathrm{A}},$$

where $\mathbf{s}^{(\mathrm{A})} \sim \mathbf{b}^{(\mathrm{D})}_1$ and $\mathbf{s}^{(\mathrm{D})} \sim \mathbf{b}^{(\mathrm{A})}_1$ are the initial state distributions.

Based on their respective histories, the defender and the attacker select actions according to their strategies. The defender's behavior strategy is defined as $\pi_{\mathrm{D}} \in \Pi_{\mathrm{D}} \triangleq \mathcal{H}_{\mathrm{D}} \rightarrow \Delta(\mathcal{A}_{\mathrm{D}})$ and the attacker's behavior strategy is defined as $\pi_{\mathrm{A}} \in \Pi_{\mathrm{A}} \triangleq \mathcal{H}_{\mathrm{A}} \rightarrow \Delta(\mathcal{A}_{\mathrm{A}})$ (Def. 5, Kuhn, 1953).

### The intrusion response problem

When selecting the strategy $\pi_{\mathrm{D}}$, the defender must balance two conflicting objectives: maximize the workflow utility towards its clients and minimize the cost of intrusion. The weight $\eta \geq 0$ controls the trade-off between these two objectives, which results in the bi-objective

$$J \triangleq \sum_{t=1}^{\infty} \gamma^{t-1} \left( \sum_{\mathbf{w} \in \mathcal{W}} \sum_{i \in \mathcal{V}_{\mathbf{w}}} \underbrace{\eta u^{(\mathrm{W})}_{i,t}}_{\text{workflows utility}} - \underbrace{c^{(\mathrm{I})}_{i,t}}_{\text{intrusion cost}} \right), \tag{3.4}$$

[3]In our use case, $\mathbf{o}_{i,t}$ relates to the number of intrusion alerts associated with node $i$.

where $\gamma \in [0, 1)$ is a discount factor, $c^{(\mathrm{I})}_{i,t} \triangleq v^{(\mathrm{I})}_{i,t} + c^{(\mathrm{A})}(\mathbf{a}^{(\mathrm{D})}_{i,t})$ is the intrusion cost associated with node $i$ at time $t$[4], and $u^{(\mathrm{W})}_{i,t}$ expresses the workflow utility associated with node $i$ at time $t$.

**Assumption 3.2.** *$u^{(\mathrm{W})}_{i,t}$ is proportional to the number of active nodes in the subtree rooted at $i$.*

Given (3.4) and an attacker strategy $\pi_{\mathrm{A}}$, the intrusion response problem can be stated as

$$
\begin{aligned}
\underset{\pi_{\mathrm{D}} \in \Pi_{\mathrm{D}}}{\text{maximize}} \quad & \mathbb{E}_{(\pi_{\mathrm{D}}, \pi_{\mathrm{A}})}\left[J\right] && && \text{(3.5a)}\\
\text{subject to} \quad & \mathbf{s}^{(\mathrm{D})}_{t+1} \sim f_{\mathrm{D}}\big(\cdot \mid \mathbf{s}^{(\mathrm{D})}_{t}, \mathbf{a}^{(\mathrm{D})}_{t}\big) && \forall t \geq 1 && \text{(3.5b)}\\
& \mathbf{s}^{(\mathrm{A})}_{t+1} \sim f_{\mathrm{A}}\big(\cdot \mid \mathbf{s}^{(\mathrm{A})}_{t}, \mathbf{a}^{(\mathrm{A})}_{t}, \mathbf{a}^{(\mathrm{D})}_{t}\big) && \forall t \geq 1 && \text{(3.5c)}\\
& \mathbf{o}_{t} \sim z\big(\cdot \mid \mathbf{s}^{(\mathrm{D})}_{t}, \mathbf{s}^{(\mathrm{A})}_{t}\big) && \forall t \geq 2 && \text{(3.5d)}\\
& \mathbf{a}^{(\mathrm{A})}_{t} \sim \pi_{\mathrm{A}}\big(\cdot \mid \mathbf{h}^{(\mathrm{A})}_{t}\big), \quad \mathbf{a}^{(\mathrm{D})}_{t} \sim \pi_{\mathrm{D}}\big(\cdot \mid \mathbf{h}^{(\mathrm{D})}_{t}\big) && \forall t \geq 1 && \text{(3.5e)}\\
& \mathbf{s}^{(\mathrm{A})}_{1} \sim \mathbf{b}^{(\mathrm{A})}_{1}, \quad \mathbf{s}^{(\mathrm{D})}_{1} \sim \mathbf{b}^{(\mathrm{D})}_{1} && && \text{(3.5f)}\\
& \mathbf{a}^{(\mathrm{A})}_{t} \in \mathcal{A}_{\mathrm{A}}(\mathbf{s}^{(\mathrm{A})}_{t}), && && \text{(3.5g)}
\end{aligned}
$$

where $\mathbb{E}_{(\pi_{\mathrm{D}}, \pi_{\mathrm{A}})}$ denotes the expectation over the random vectors $(\mathbf{H}^{(\mathrm{D})}_{t}, \mathbf{H}^{(\mathrm{A})}_{t})_{t \in \{1,2,\ldots\}}$ when following the strategy profile $(\pi_{\mathrm{D}}, \pi_{\mathrm{A}})$; (3.5b)–(3.5c) are the dynamics constraints; (3.5d) describes the observations; (3.5e) captures the actions; (3.5f) defines the initial state distributions; and (3.5g) is the action constraint of the attacker.

**Remark 3.2.** As a maximizer of (3.5) exists (see Thm. 3.1 below), we write max instead of sup throughout this paper.

Solving (3.5) yields an optimal defender strategy against a *static* attacker with a fixed strategy. Note that this defender strategy is generally not optimal against a different attacker strategy. For this reason, we aim to find a defender strategy that maximizes the minimum value of $J$ (3.4) across all possible attacker strategies[5]. This objective can be formally expressed as

$$
\underset{\pi_{\mathrm{D}} \in \Pi_{\mathrm{D}}}{\text{maximize}}\ \underset{\pi_{\mathrm{A}} \in \Pi_{\mathrm{A}}}{\text{minimize}}\ \mathbb{E}_{(\pi_{\mathrm{D}}, \pi_{\mathrm{A}})}\left[J\right] \text{ subject to (3.5b)–(3.5g).} \tag{3.6}
$$

Solving (3.6) corresponds to finding a Nash equilibrium (NE) (Eq. 1, Nash, 1951) of a two-player game[6] and can thus be analyzed through game theory.

[4] $c^{(\mathrm{A})}$ is a non-negative function that represents the operational costs of defender actions.
[5] See Assumption 5 in the problem chapter.
[6] A solution to (3.6) can also form a stronger equilibrium, namely a Perfect Bayesian equilibrium (PBE), see Def. 4 in the background chapter.

## 3.5 The Intrusion Response Game

The maxmin problem in (3.6) defines a stationary, finite, and zero-sum Partially Observed Stochastic Game with Public Observations (a PO-POSG)

$$\Gamma \triangleq \langle \mathcal{N}, (\mathcal{S}_{\mathrm{k}}, \mathcal{A}_{\mathrm{k}}, f_{\mathrm{k}}, \mathbf{b}_1^{(\mathrm{k})})_{\mathrm{k}\in\mathcal{N}}, u, \gamma, \mathcal{O}, z\rangle. \quad \text{(Horák and Bošanský, 2019)} \tag{3.7}$$

$\Gamma$ has two players: $\mathcal{N} \triangleq \{\mathrm{D}, \mathrm{A}\}$, with D being the defender and A being the attacker. $(\mathcal{S}_{\mathrm{k}})_{\mathrm{k}\in\mathcal{N}}$ are the state spaces, $(\mathcal{A}_{\mathrm{k}})_{\mathrm{k}\in\mathcal{N}}$ are the action spaces, and $\mathcal{O}$ is observation space (as defined in §3.4). The transition functions $(f_{\mathrm{k}})_{\mathrm{k}\in\mathcal{N}}$ are defined by (3.5b)–(3.5c), the observation function $z$ is defined in (3.3), and the utility function $u(\mathbf{s}_t, \mathbf{a}_t^{(\mathrm{D})})$ is the expression within brackets in (3.4). $(\mathbf{b}_1^{(\mathrm{k})})_{\mathrm{k}\in\mathcal{N}}$ are the state distributions at $t = 1$ and $\gamma$ is the discount factor.

### Game play

When the game starts at $t = 1$, $\mathbf{s}_1^{(\mathrm{D})}$ and $\mathbf{s}_1^{(\mathrm{A})}$ are sampled from $\mathbf{b}_1^{(\mathrm{D})}$ and $\mathbf{b}_1^{(\mathrm{A})}$, respectively. A play of the game proceeds in time steps $t = 1, 2, \ldots$. At each time $t$, the defender observes $\mathbf{h}_t^{(\mathrm{D})}$ and the attacker observes $\mathbf{h}_t^{(\mathrm{A})}$. Based on these histories, both players select actions according to their respective strategies, i.e., $\mathbf{a}_t^{(\mathrm{D})} \sim \pi_{\mathrm{D}}(\cdot \mid \mathbf{h}_t^{(\mathrm{D})})$ and $\mathbf{a}_t^{(\mathrm{A})} \sim \pi_{\mathrm{A}}(\cdot \mid \mathbf{h}_t^{(\mathrm{A})})$. As a result of these actions, five events occur at time $t+1$: (*i*) $\mathbf{o}_{t+1}$ is sampled from $z$; (*ii*) $\mathbf{s}_{t+1}^{(\mathrm{D})}$ is sampled from $f_{\mathrm{D}}$; (*iii*) $\mathbf{s}_{t+1}^{(\mathrm{A})}$ is sampled from $f_{\mathrm{A}}$; (*iv*) the defender receives the utility $u(\mathbf{s}_t, \mathbf{a}_t^{(\mathrm{D})})$; and (*v*) the attacker receives the utility $-u(\mathbf{s}_t, \mathbf{a}_t^{(\mathrm{D})})$.

### Belief states

Based on their histories $\mathbf{h}_t^{(\mathrm{D})}$ and $\mathbf{h}_t^{(\mathrm{A})}$, both players form beliefs about the unobservable components of the state $\mathbf{s}_t$, which are expressed through the belief states $\mathbf{b}_t^{(\mathrm{D})}(\mathbf{s}_t^{(\mathrm{A})}) \triangleq \mathbb{P}[\mathbf{s}_t^{(\mathrm{A})} \mid \mathbf{h}_t^{(\mathrm{D})}]$ and $\mathbf{b}_t^{(\mathrm{A})}(\mathbf{s}_t^{(\mathrm{D})}) \triangleq \mathbb{P}[\mathbf{s}_t^{(\mathrm{D})} \mid \mathbf{h}_t^{(\mathrm{A})}]$. These beliefs are updated each time $t > 1$ as

$$\mathbf{b}_t^{(\mathrm{k})}(\mathbf{s}_t^{(-\mathrm{k})}) = C_{\mathrm{k}} \sum_{\mathbf{s}_{t-1}^{(-\mathrm{k})} \in \mathcal{S}_{-\mathrm{k}}} \sum_{\mathbf{a}_{t-1}^{(-\mathrm{k})} \in \mathcal{A}_{-\mathrm{k}}(\mathbf{s}_t)} \mathbf{b}_{t-1}^{(\mathrm{k})}(\mathbf{s}_{t-1}^{(-\mathrm{k})}) \pi_{-\mathrm{k}}^{(s)}(\mathbf{a}_{t-1}^{(-\mathrm{k})} \mid \mathbf{s}_{t-1}^{(-\mathrm{k})}) \cdot \tag{3.8}$$

$$z(\mathbf{o}_t \mid \mathbf{s}_t) f_{-\mathrm{k}}(\mathbf{s}_t^{(-\mathrm{k})} \mid \mathbf{s}_{t-1}^{(-\mathrm{k})}, \mathbf{a}_{t-1}), \qquad \text{(Eq. 1, Horák and Bošanský, 2019)}$$

where $\mathrm{k} \in \{\mathrm{D}, \mathrm{A}\}$ and $C_{\mathrm{k}} = 1/\mathbb{P}[\mathbf{o}_t \mid \mathbf{s}_t^{(\mathrm{k})}, \mathbf{a}_{t-1}^{(\mathrm{k})}, \pi_{-\mathrm{k}}, \mathbf{b}_{t-1}^{(\mathrm{k})}]$ is a normalizing factor that makes the components of $\mathbf{b}_t^{(\mathrm{k})}$ sum to 1. $\pi_{-\mathrm{k}}^{(s)} : \mathcal{S}_{-\mathrm{k}} \to \Delta(\mathcal{A}_{-\mathrm{k}})$ is the stage strategy for the opponent, which is assumed to be known. The initial beliefs at $t = 1$ are the degenerate distributions $\mathbf{b}_1^{(\mathrm{D})}(\mathbf{0}_{2|\mathcal{V}|}) = 1$ and $\mathbf{b}_1^{(\mathrm{A})}(\mathbf{s}_1^{(\mathrm{D})}) = 1$, where $\mathbf{0}_n$ is the n-dimensional zero-vector and $\mathbf{s}_1^{(\mathrm{D})}$ is given by the infrastructure configuration (see §3.4).

### Best response strategies

A defender strategy $\tilde{\pi}_{\mathrm{D}} \in \Pi_{\mathrm{D}}$ is a *best response* against $\pi_{\mathrm{A}} \in \Pi_{\mathrm{A}}$ if it *maximizes* $J$ (3.4). Similarly, an attacker strategy $\tilde{\pi}_{\mathrm{A}}$ is a best response against $\pi_{\mathrm{D}}$ if it *minimizes* $J$ (3.4). Hence, the best response correspondences are

$$\mathscr{B}_{\mathrm{D}}(\pi_{\mathrm{A}}) \triangleq \arg\max_{\pi_{\mathrm{D}} \in \Pi_{\mathrm{D}}} \mathbb{E}_{(\pi_{\mathrm{D}},\pi_{\mathrm{A}})}[J] \quad (3.9a)$$

$$\mathscr{B}_{\mathrm{A}}(\pi_{\mathrm{D}}) \triangleq \arg\min_{\pi_{\mathrm{A}} \in \Pi_{\mathrm{A}}} \mathbb{E}_{(\pi_{\mathrm{D}},\pi_{\mathrm{A}})}[J]. \quad (3.9b)$$

### Optimal (equilibrium) strategies

An optimal defender strategy $\pi^{\star}_{\mathrm{D}}$ is a best response against any attacker strategy that *minimizes* $J$. Similarly, an optimal attacker strategy $\pi^{\star}_{\mathrm{A}}$ is a best response against any defender strategy that *maximizes* $J$. Hence, when both players act optimally, their strategies form a Nash equilibrium (NE)

$$(\pi^{\star}_{\mathrm{D}}, \pi^{\star}_{\mathrm{A}}) \in \mathscr{B}_{\mathrm{D}}(\pi^{\star}_{\mathrm{A}}) \times \mathscr{B}_{\mathrm{A}}(\pi^{\star}_{\mathrm{D}}). \qquad \text{(Eq. 1, Nash, 1951)} \quad (3.10)$$

Since the players' beliefs are consistent, $(\pi^{\star}_{\mathrm{D}}, \pi^{\star}_{\mathrm{A}})$ can also form a PBE [7].

**Theorem 3.1** (Existence of equilibria and best responses)**.**

*(A) A game* $\Gamma$ *with instantiation described in §3.4 has a* PBE*.*

*(B) The best response correspondences (3.9) in* $\Gamma$ *with the instantiation described in §3.4 satisfy* $|\mathscr{B}_{\mathrm{D}}(\pi_{\mathrm{A}})| > 0$ *and* $|\mathscr{B}_{\mathrm{A}}(\pi_{\mathrm{D}})| > 0$ $\forall (\pi_{\mathrm{A}}, \pi_{\mathrm{D}}) \in \Pi_{\mathrm{A}} \times \Pi_{\mathrm{D}}$*.*

*Proof.* (A) follows from the following sufficient conditions: $(i)$ $\Gamma$ is stationary, finite, and zero-sum; $(ii)$ $\Gamma$ has public observations; and $(iii)$ $\gamma \in [0, 1)$. Due to these conditions, the existence proof in (Thm. 1, Horák and Bošanský, 2019) can be used, which shows that $\Gamma$ can be modeled as a finite game in extensive form, for which the proof of Thm. 3 in the background chapter applies. In the interest of space, we do not restate the proof. To prove (B), we note that obtaining a pair of best responses $(\tilde{\pi}_{\mathrm{D}}, \tilde{\pi}_{\mathrm{A}}) \in \mathscr{B}_{\mathrm{D}}(\pi_{\mathrm{A}}) \times \mathscr{B}_{\mathrm{A}}(\pi_{\mathrm{D}})$ for a given strategy pair $(\pi_{\mathrm{A}}, \pi_{\mathrm{D}}) \in \Pi_{\mathrm{A}} \times \Pi_{\mathrm{D}}$ amounts to solving two finite and stationary POMDPs[8] with discounted utilities. It then follows from Thm. 2 in the background chapter that a pair of pure best responses $(\tilde{\pi}_{\mathrm{D}}, \tilde{\pi}_{\mathrm{A}})$ exists. □

**Challenge: The curse of dimensionality (Bellman, 1957).**

The space complexity of $\Gamma$ increases exponentially with the number of nodes; see Fig. 3.2. This growth exemplifies the curse of dimensionality: more state variables result in a combinatorial explosion of possible states.

[7] While the definition of a PBE (Def. 4) in the background chapter is presented in terms of a one-sided POSG, the adaption of the definition to a PO-POSG is immediate.

[8] The components of a POMDP are defined the background chapter; see (15).

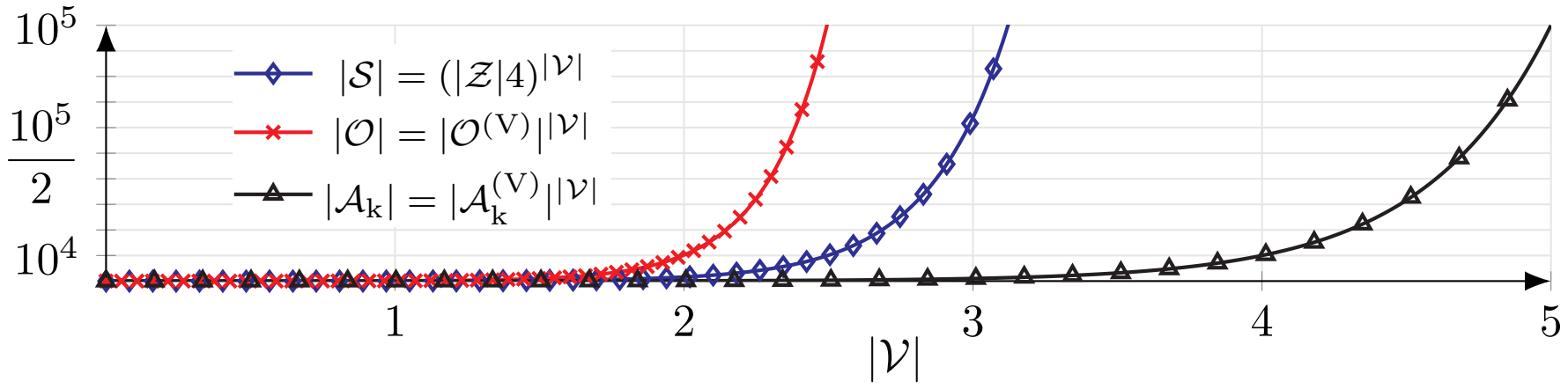

***Figure 3.2:*** *Growth of $|\mathcal{S}|$, $|\mathcal{O}|$, and $|\mathcal{A}_\mathrm{k}|$ in function of the number of nodes $|\mathcal{V}|$; the curves are computed using $|\mathcal{Z}| = 10$, $|\mathcal{O}^{(\mathrm{V})}| = 100$, and $|\mathcal{A}_\mathrm{D}^{(\mathrm{V})}| = |\mathcal{A}_\mathrm{A}^{(\mathrm{V})}| = 10$.*

## 3.6 Decomposing the Intrusion Response Game

In this section, we present the main contribution of the paper. We show how to address the curse of dimensionality by recursively decomposing the game $\Gamma$ into independent subgames. Due to their independence, the subgames have optimal substructure (Ch. 15, Cormen et al., 2022), which means that a best response of the original game can be obtained from best responses of the subgames. We further show that best responses of the subgames can be computed in parallel and that the space complexity of a subgame is independent of the number of nodes $|\mathcal{V}|$.

**Theorem 3.2** (Decomposition theorem)**.**
*Under assumptions 3.1–3.2, the following holds.*

*(A) A game $\Gamma$ with the instantiation described in §3.4 can be decomposed into independent workflow subgames $\Gamma^{(\mathbf{w}_1)}, \ldots, \Gamma^{(\mathbf{w}_{|\mathcal{W}|})}$.*

*(B) Each subgame $\Gamma^{(\mathbf{w})}$ can be further decomposed into independent node subgames $(\Gamma^{(i)})_{i \in \mathcal{V}_\mathbf{w}}$ with space complexities independent of $|\mathcal{V}|$.*

*(C) For each subgame $\Gamma^{(i)}$, a best response for the defender can be characterized by switching curves, under the assumption that the observation functions $z_{\mathbf{O}_1|\mathbf{s}^{(\mathrm{A})}}, \ldots, z_{\mathbf{O}_{|\mathcal{V}|}|\mathbf{s}^{(\mathrm{A})}}$ (3.3) are totally positive of order 2 (i.e., TP-2 (Def. 10.2.1, Krishnamurthy, 2016)).*

Statements A and B express that $\Gamma$ decomposes into simpler subgames, which consequently can be solved in parallel. This decomposition implies that the largest game tractable on a given compute platform scales linearly with the number of processors. Further, statement C says that a best response for the defender in each subgame can be characterized by switching curves, which can be estimated efficiently. In the following sections, we provide proofs of Thm. 3.2.A–C. The requisite notation is given in Table 3.1 on the next page.

| *Notation(s)* | *Description* |
|---|---|
| $\mathcal{G}, \mathcal{G}_{\mathbf{w}}$ | Infrastructure tree, subtree of $\mathbf{w}$. |
| $\mathcal{V}, \mathcal{E}$ | Sets of nodes and edges in $\mathcal{G}$. |
| $\mathcal{V}_{\mathbf{w}}, \mathcal{E}_{\mathbf{w}}$ | Sets of nodes and edges in $\mathcal{G}_{\mathbf{w}}$. |
| $\mathcal{Z}, \mathcal{W}$ | Sets of network zones and workflows. |
| $\mathcal{A}_{\mathrm{D}}, \mathcal{A}_{\mathrm{A}}(\mathbf{s}_t)$ | Defender and attacker action spaces at time $t$. |
| $\mathcal{A}_{\mathrm{D}}^{(\mathrm{V})}, \mathcal{A}_{\mathrm{A}}^{(\mathrm{V})}(\mathbf{s}_t)$ | Action spaces per node at time $t$, $\mathcal{A}_{\mathrm{k}} = (\mathcal{A}_{\mathrm{k}}^{(\mathrm{V})})^{\lvert\mathcal{V}\rvert}$. |
| $\mathcal{O}^{(\mathrm{V})}$ | Observation space per node at time $t$, $\mathcal{O} = (\mathcal{O}^{(\mathrm{V})})^{\lvert\mathcal{V}\rvert}$. |
| $v_{i,t}^{(\mathrm{I})}, v_{i,t}^{(\mathrm{Z})}, v_{i,t}^{(\mathrm{R})}$ | Intrusion state, zone, and reconnaissance state of $i \in \mathcal{V}$ at time $t$. |
| $V_{i,t}^{(\mathrm{I})}, V_{i,t}^{(\mathrm{Z})}, V_{i,t}^{(\mathrm{R})}$ | Random variables with realizations $v_{i,t}^{(\mathrm{I})}, v_{i,t}^{(\mathrm{Z})}, v_{i,t}^{(\mathrm{R})}$. |
| $\Gamma, \mathcal{N}$ | PO-POSG (3.7), set of players (§3.5). |
| $\mathcal{S}_{\mathrm{D}}, \mathcal{S}_{\mathrm{A}}$ | Defender and attacker state spaces (§3.5). |
| $\mathcal{S} \triangleq \mathcal{S}_{\mathrm{D}} \times \mathcal{S}_{\mathrm{A}}$ | State space (§3.5). |
| $u, \mathcal{O}$ | Utility function and observation space (§3.5). |
| $\mathbf{s}_t = (\mathbf{s}_t^{(\mathrm{D})}, \mathbf{s}_t^{(\mathrm{A})})$ | State at time $t$ (§3.5). |
| $\mathbf{a}_t = (\mathbf{a}_t^{(\mathrm{D})}, \mathbf{a}_t^{(\mathrm{A})})$ | Action at time $t$ (§3.5). |
| $\mathbf{o}_t, \mathbf{u}_t$ | Observation, utility at time $t$ (§3.5). |
| $\mathbf{a}_t^{(\mathrm{k})}, \mathbf{h}_t^{(\mathrm{k})}$ | Action and history of player k at time $t$ (§3.5). |
| $\mathcal{B}_{\mathrm{k}}, \mathbf{b}_t^{(\mathrm{k})}$ | Belief space and belief state of player $k$ (§3.5). |
| $\tilde{\pi}_{\mathrm{k}}, \tilde{\mathbf{a}}^{(\mathrm{k})}$ | Best response strategy and action of player $k$ (§3.5). |
| $\mathbf{S}_t, \mathbf{O}_t, \mathbf{A}_t$ | Random vectors with realizations $\mathbf{s}_t, \mathbf{o}_t, \mathbf{a}_t$ (§3.5). |
| $\mathbf{U}_t, \mathbf{B}_t^{(\mathrm{k})}, \mathbf{H}_t^{(\mathrm{k})}$ | Random vectors with realizations $\mathbf{u}_t, \mathbf{b}_t^{(\mathrm{k})}, \mathbf{h}_t^{(\mathrm{k})}$ (§3.5). |
| $\pi_{\mathrm{k}}, z$ | Strategy of player $k$, observation function (§3.5). |
| $u_{i,t}^{(\mathrm{w})}$ | Workflow utility of node $i$ at time $t$. |
| $\perp, \mathrm{an}(i)$ | Null action, set of $i$ and its ancestors in $\mathcal{G}$. |
| $\alpha_{i,t}$ | Active status of node $i$ at time $t$. |
| $f_{\mathrm{A}}, f_{\mathrm{D}}$ | Attacker and defender transition functions. |
| $\mathscr{B}_{\mathrm{k}}$ | Best response correspondence of player k (3.9). |
| $c_{i,t}^{(\mathrm{I})}$ | Intrusion cost associated with node $i$ at time $t$ (3.4). |
| $c^{(\mathrm{A})}$ | Action cost function. |

***Table 3.1:*** *Variables and symbols used in the model.*

## Proof of Theorem 3.2.A

Following the instantiation of $\Gamma$ described in §3.4, the state, observation, and action spaces factorize as

$$\mathcal{S} = (\mathcal{Z} \times \{0,1\}^2)^{\lvert\mathcal{V}\rvert}, \ \mathcal{O} = (\mathcal{O}^{(\mathrm{V})})^{\lvert\mathcal{V}\rvert}, \text{ and } \mathcal{A}_{\mathrm{k}} = (\mathcal{A}_{\mathrm{k}}^{(\mathrm{V})})^{\lvert\mathcal{V}\rvert} \quad \mathrm{k} \in \{\mathrm{D}, \mathrm{A}\}, \tag{3.11}$$

where $\mathcal{O}^{(\mathrm{V})}$, $\mathcal{A}_{\mathrm{D}}^{(\mathrm{V})}$, and $\mathcal{A}_{\mathrm{A}}^{(\mathrm{V})}$ denote the local observation and action spaces for each node. Assumption 3.1 together with (3.11) implies that $\Gamma$ can be decomposed into subgames $\Gamma^{(\mathbf{w}_1)}, \ldots, \Gamma^{(\mathbf{w}_{\lvert\mathcal{W}\rvert})}$. To show that the subgames are independent, it suffices to show that they are observation-independent, transition-independent, and utility-independent (Defs. 32, 33, 35, Seuken and Zilberstein, 2008).

From (3.3) we have

$$z\big(\mathbf{o}_{i,t} \mid \mathbf{s}_t^{(\mathrm{D})}, \mathbf{s}_t^{(\mathrm{A})}\big) = z\big(\mathbf{o}_{i,t} \mid \mathbf{s}_{i,t}^{(\mathrm{D})}, \mathbf{s}_{i,t}^{(\mathrm{A})}\big) \tag{3.12}$$

for all $\mathbf{o}_{i,t} \in \mathcal{O}^{(\mathrm{V})}$, $\mathbf{s}_t \in \mathcal{S}$, and $t \geq 1$. Therefore, $\Gamma^{(\mathbf{w}_1)}, \ldots, \Gamma^{(\mathbf{w}_{|\mathcal{W}|})}$ are observation-independent (Def. 33, Seuken and Zilberstein, 2008).

From the modeling in §3.4 and (3.1)–(3.2) we have

$$f_{\mathrm{D}}(\mathbf{s}^{(\mathrm{D})}_{i,t+1} \mid \mathbf{s}^{(\mathrm{D})}_{t}, \mathbf{a}^{(\mathrm{D})}_{t}) = f_{\mathrm{D}}(\mathbf{s}^{(\mathrm{D})}_{i,t+1} \mid \mathbf{s}^{(\mathrm{D})}_{i,t}, \mathbf{a}^{(\mathrm{D})}_{i,t}) \tag{3.13a}$$

$$f_{\mathrm{A}}(\mathbf{s}^{(\mathrm{A})}_{i,t+1} \mid \mathbf{s}^{(\mathrm{A})}_{t}, \mathbf{a}^{(\mathrm{A})}_{t}, \mathbf{a}^{(\mathrm{D})}_{t}) = f_{\mathrm{A}}(\mathbf{s}^{(\mathrm{A})}_{i,t+1} \mid \mathbf{s}^{(\mathrm{A})}_{i,t}, \mathbf{a}^{(\mathrm{A})}_{i,t}, \mathbf{a}^{(\mathrm{D})}_{i,t}) \tag{3.13b}$$

for all $\mathbf{s}_{t+1}, \mathbf{s}_t \in \mathcal{S}, \mathbf{a}_{i,t} \in \mathcal{A}^{(\mathrm{V})}, i \in \mathcal{V}$, and $t \geq 1$. Consequently, $\Gamma^{(\mathbf{w}_1)}, \ldots, \Gamma^{(\mathbf{w}_{|\mathcal{W}|})}$ are transition-independent (Def. 32, Seuken and Zilberstein, 2008).

Following (3.4) and the definition of $u^{(\mathrm{W})}_{i,t}$ we can rewrite $u(\mathbf{s}_t, \mathbf{a}^{(\mathrm{D})}_t)$ as

$$u(\mathbf{s}_t, \mathbf{a}^{(\mathrm{D})}_t) = \sum_{\mathbf{w}\in\mathcal{W}} \overbrace{\sum_{i\in\mathcal{V}_{\mathbf{w}}} \eta u^{(\mathrm{W})}_{i,t} - c^{(\mathrm{I})}_{i,t}(\mathbf{a}^{(\mathrm{D})}_{i,t}, v^{(\mathrm{I})}_{i,t})}^{\triangleq u_{\mathbf{w}}\left(\left(\mathbf{s}_{i,t}, \mathbf{a}^{(\mathrm{D})}_{i,t}\right)_{i\in\mathcal{V}_{\mathbf{w}}}\right)} = \sum_{\mathbf{w}\in\mathcal{W}} u_{\mathbf{w}}\left(\left(\mathbf{s}_{i,t}, \mathbf{a}^{(\mathrm{D})}_{i,t}\right)_{i\in\mathcal{V}_{\mathbf{w}}}\right). \tag{3.14}$$

The final expression in (3.14) is a sum of workflow utility functions, each of which depends only on the states and actions of one workflow. Hence, $\Gamma^{(\mathbf{w}_1)}, \ldots, \Gamma^{(\mathbf{w}_{|\mathcal{W}|})}$ are utility independent (Def. 35, Seuken and Zilberstein, 2008). □

Theorem 3.2.A is illustrated in Fig. 3.3. Arrows indicate inputs and outputs; $\oplus$ denotes vector concatenation; and $\mathrm{k} \in \{\mathrm{D}, \mathrm{A}\}$ denotes the player. The figure shows that the strategy $\pi_{\mathrm{k}}$ can be decomposed into $|\mathcal{W}|$ independent substrategies, one per workflow $\mathbf{w} \in \mathcal{W}$. These strategies are independent in the sense that each of them only depends on local information related to a specific workflow, i.e., the workflow history $\mathbf{h}^{(\mathrm{k})}_{\mathbf{w},t} \triangleq (\mathbf{h}^{(\mathrm{k})}_{j,t})_{j\in\mathcal{V}_{\mathbf{w}}}$. The output of each workflow strategy is a workflow action $\mathbf{a}^{(\mathrm{k})}_{\mathbf{w},t} \triangleq (\mathbf{a}^{(\mathrm{k})}_{j,t})_{j\in\mathcal{V}_{\mathbf{w}}}$, which defines the actions on each node that belongs to the workflow. These workflow actions are concatenated to obtain the action in $\Gamma$, i.e., $\mathbf{a}^{(\mathrm{k})}_t = \mathbf{a}^{(\mathrm{k})}_{\mathbf{w}_1,t} \oplus \mathbf{a}^{(\mathrm{k})}_{\mathbf{w}_2,t}, \oplus \ldots \oplus \mathbf{a}^{(\mathrm{k})}_{\mathbf{w}_{|\mathcal{W}|},t}$.

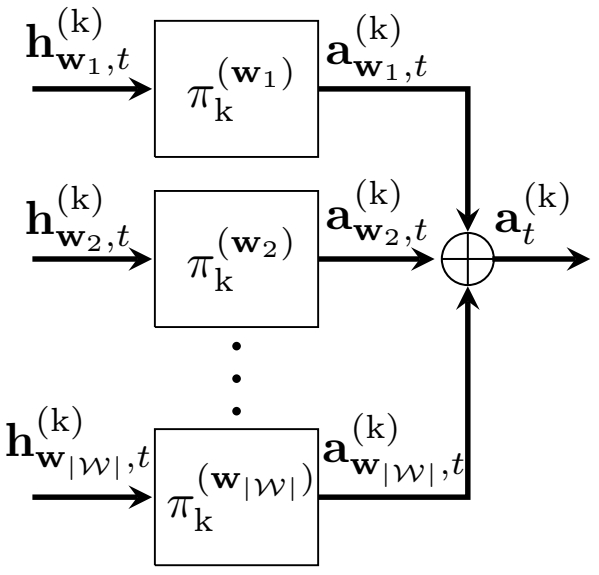

***Figure 3.3:*** *Theorem 3.2.A.*

## Proof of Theorem 3.2.B

Our goal is to show that a workflow subgame $\Gamma^{(\mathbf{w})}$ decomposes into independent node-level subgames $(\Gamma^{(i)})_{i\in\mathcal{V}_{\mathbf{w}}}$. We know from (3.11) that the space complexities of these subgames are independent of $|\mathcal{V}|$. Further, (3.12) and (3.13) imply that the subgames are observation-independent and transition-independent, respectively (Def. 32, 33, Seuken and Zilberstein, 2008). Hence, it only remains to show that they are utility independent (Def. 35, Seuken and Zilberstein, 2008). Using the decomposition in (3.14), we obtain

$$\begin{aligned}
u_{\mathbf{w}}\left(\left(\mathbf{s}_{i,t}, \mathbf{a}^{(\mathrm{D})}_{i,t}\right)_{i \in \mathcal{V}_{\mathbf{w}}}\right) &= \sum_{i \in \mathcal{V}_{\mathbf{w}}} \eta u^{(\mathrm{W})}_{i,t} - c^{(\mathrm{I})}_{i,t}(\mathbf{a}^{(\mathrm{D})}_{i,t}, v^{(\mathrm{I})}_{i,t}) \\
&= \sum_{i \in \mathcal{V}_{\mathbf{w}}} \eta u^{(\mathrm{W})}_{i,t} - v^{(\mathrm{I})}_{i,t} - c^{(\mathrm{A})}(\mathbf{a}^{(\mathrm{D})}_{i,t}) \\
&\overset{(a)}{=} \sum_{i \in \mathcal{V}_{\mathbf{w}}} \underbrace{\eta k |\mathrm{an}(i)| \alpha_{i,t} - v^{(\mathrm{I})}_{i,t} - c^{(\mathrm{A})}(\mathbf{a}^{(\mathrm{D})}_{i,t})}_{\triangleq u_i(\mathbf{s}_{i,t}, \mathbf{a}^{(\mathrm{D})}_{i,t})} \\
&= \sum_{i \in \mathcal{V}_{\mathbf{w}}} u_i(\mathbf{s}_{i,t}, \mathbf{a}^{(\mathrm{D})}_{i,t}),
\end{aligned} \tag{3.15}$$

where $\mathrm{an}(i)$ denotes the set of node $i$ and its ancestors in the infrastructure graph $\mathcal{G}$ and $k$ is a constant of proportionality. (a) follows from Assumption 3.2, which implies that the workflow utility of a node depends on its active state $\alpha_{i,t}$ and the number of ancestors in $\mathcal{G}$. Since $u_i$ in (3.15) is independent of the states and actions of the other nodes, Thm. 3.2.B follows. $\square$.

Theorem 3.2.B is illustrated in Fig. 3.4. Arrows indicate inputs and outputs; $\oplus$ denotes vector concatenation; and $\mathrm{k} \in \{\mathrm{D}, \mathrm{A}\}$ denotes the player. The figure shows that the workflow strategy $\pi^{(\mathbf{w})}_{\mathrm{k}}$ can be decomposed into $|\mathcal{V}_{\mathbf{w}}|$ independent substrategies, one per node $i \in \mathcal{V}_{\mathbf{w}}$. These strategies are independent in the sense that each of them only depends on local information related to a specific node, i.e., the node history $\mathbf{h}^{(\mathrm{k})}_{i,t} \triangleq (\mathbf{b}^{(\mathrm{k})}_{i,1}, \mathbf{s}^{(\mathrm{k})}_{i,1}, \mathbf{a}^{(\mathrm{k})}_{i,1}, \mathbf{o}_{i,2}, \mathbf{a}^{(\mathrm{k})}_{i,t-1}, \mathbf{s}^{(\mathrm{k})}_{i,t}, \mathbf{o}_{i,t})$. The output of each node strategy is a node action $\mathbf{a}^{(\mathrm{k})}_{i,t} \in \mathcal{A}^{(\mathrm{V})}_{\mathrm{k}}$. These node actions are concatenated to obtain the workflow action, i.e., $\mathbf{a}^{(\mathrm{k})}_{\mathbf{w},t} = \mathbf{a}^{(\mathrm{k})}_{1,t} \oplus \mathbf{a}^{(\mathrm{k})}_{2,t}, \oplus \ldots \oplus \mathbf{a}^{(\mathrm{k})}_{|\mathcal{V}_{\mathbf{w}}|,t}$.

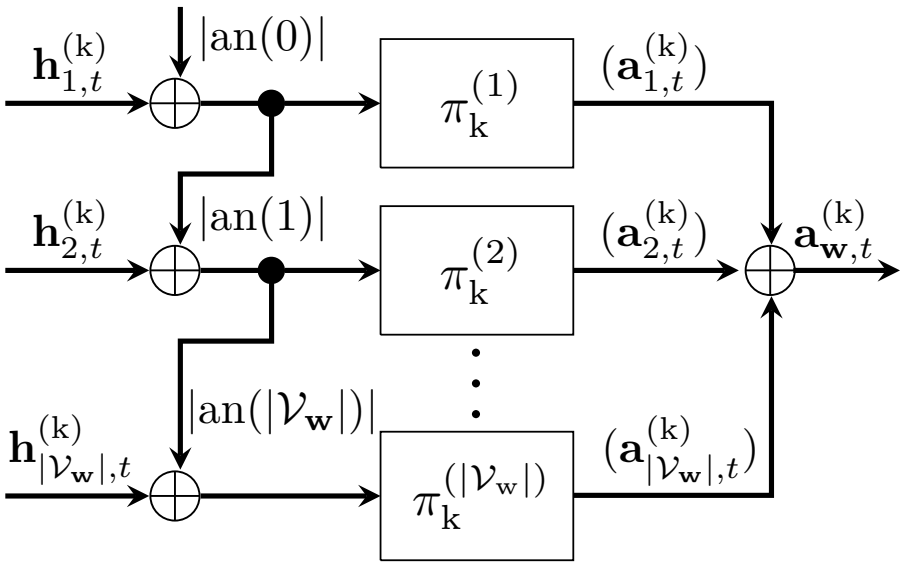

***Figure 3.4:*** *Theorem 3.2.B.*

**Proof sketch of Theorem 3.2.C**

The full proof of this result is technical and not needed elsewhere in the paper; we relegate it accordingly to the appendix; see Appendix E. The main idea of the proof can be outlined as follows. First, we note that a subgame $\Gamma^{(i)}$ (Thm. 3.2.B) can only be in three attack states: it can be unknown to the attacker, it can be known, or it can be compromised. Consequently, the defender's belief space $\mathcal{B}^{(i)}_{\mathrm{D}}$ is the unit 2-simplex. Second, we note that a defender strategy can be represented in an auto-regressive manner by appealing to the chain rule of probability. Such a representation allows to decompose the strategy into two (dependent) substrategies, one strategy for deciding *which* action to take and another strategy for deciding *when* to take it. Through this separation, we can analyze the latter problem using optimal stopping theory. In particular, we can apply (Thm. 12.3.4, Krishnamurthy, 2016), which states that there exists a switching curve $\Upsilon$ that partitions $\mathcal{B}_{\mathrm{D}}^{(i)}$ into two connected sets: a stopping set $\mathscr{S}_{\mathrm{D}}^{(i)}$ where it is a best response to take a defensive action and a continuation set $\mathscr{C}_{\mathrm{D}}^{(i)}$ where waiting is a best response.

The argument behind the existence of a switching curve is as follows. On any line segment $\mathcal{L}(\mathbf{e}_1, \widehat{\mathbf{b}}^{(\mathrm{D})})$ in $\mathcal{B}_{\mathrm{D}}^{(i)}$ that starts at $\mathbf{e}_1$ and ends at the subsimplex joining $\mathbf{e}_2$ and $\mathbf{e}_3$ (denoted with $\widehat{\mathbf{b}}^{(\mathrm{D})} \in \mathcal{B}_{\mathrm{D},\mathbf{e}_1}^{(i)}$), all belief states are totally ordered with respect to the Monotone Likelihood Ratio (MLR) order (Def. 10.1.1, Krishnamurthy, 2016). Furthermore, the utility function in the subgames is supermodular (3.4). As a consequence, Topkis's theorem (Thm. 6.3, Topkis, 1978) implies that the best response strategy on $\mathcal{L}(\mathbf{e}_1, \widehat{\mathbf{b}}^{(\mathrm{D})})$ is monotone with respect to the MLR order. Consequently, there exists a threshold belief state $\alpha_{\widehat{\mathbf{b}}^{(\mathrm{D})}}$ on $\mathcal{L}(\mathbf{e}_1, \widehat{\mathbf{b}}^{(\mathrm{D})})$ where the best response strategy switches from waiting to taking action. Since $\mathcal{B}_{\mathrm{D}}^{(i)}$ can be covered by the union of lines $\mathcal{L}(\mathbf{e}_1, \widehat{\mathbf{b}}^{(\mathrm{D})})$, the thresholds $\alpha_{\widehat{\mathbf{b}}_1^{(\mathrm{D})}}, \alpha_{\widehat{\mathbf{b}}_2^{(\mathrm{D})}}, \ldots$ yield a switching curve; see Fig. 3.5.

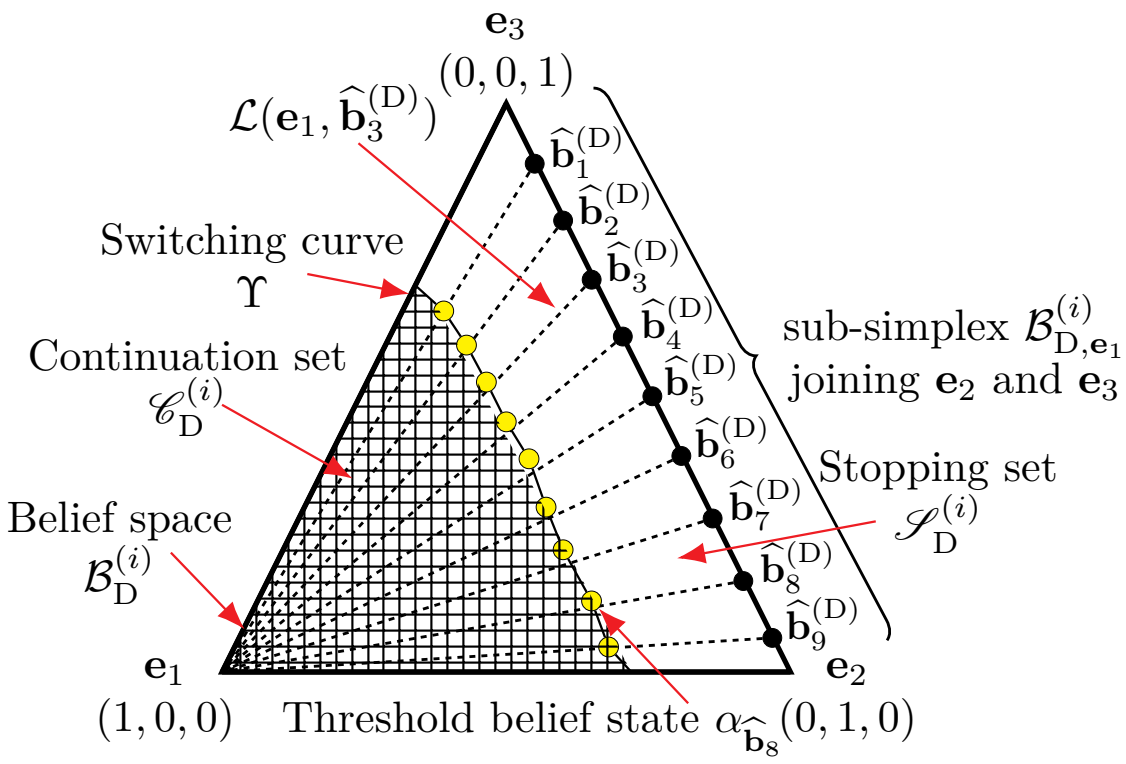

***Figure 3.5:*** *Theorem 3.2.C.*

## 3.7 Approximating Equilibria of the Decomposed Game

To approximate a perfect Bayesian equilibrium (PBE) of $\Gamma$ (3.7) we develop a *fictitious play* algorithm called **D**ecompositional **F**ictitious **P**lay (DFP), which estimates a PBE based on the decomposition presented above. The pseudocode is listed in Alg. 3.1[9] on the next page. DFP implements the fictitious play process described in (Brown, 1951) and generates a sequence of strategy profiles $(\pi_{\mathrm{D}}, \pi_{\mathrm{A}})$, $(\pi'_{\mathrm{D}}, \pi'_{\mathrm{A}})$, ... that converges to a PBE $(\pi^{\star}_{\mathrm{D}}, \pi^{\star}_{\mathrm{A}})$ (Thms. 7.2.4–7.2.5, Shoham and Leyton-Brown, 2009). During each step of this process, DFP learns best responses against the players' current strategies and then updates both players' strategies (lines 4–8 in Alg. 3.1). To obtain the best responses, it first finds best responses for the node subgames as constructed in the proof of Thm. 3.2.B (lines 10–14), and then it combines them using the method described in §3.6 (lines 15–20).

[9] In Alg. 3.1, $\oplus$ denotes vector concatenation.

***Algorithm 3.1:*** *DFP: Decompositional **F**ictitious **P**lay.*

**Input:** P-SOLVER: a POMDP solver, $\delta$: convergence criterion, $\Gamma$: the PO-POSG.
**Output:** An approximate PBE $(\pi_{\mathrm{D}}, \pi_{\mathrm{A}})$.

**procedure** DFP(P-SOLVER, $\delta$, $\Gamma$)
  Initialize $\pi_{\mathrm{D}}, \pi_{\mathrm{A}}, \widehat{\delta}$.
  **while** $\widehat{\delta} \geq \delta$ **do**
    **in parallel for** $\mathrm{k} \in \{\mathrm{D}, \mathrm{A}\}$ **then**
      $\boldsymbol{\pi}_{\mathrm{k}} \leftarrow$LOCAL-BEST-RESPONSES(P-SOLVER, $\Gamma, \mathrm{k}, \pi_{-\mathrm{k}}$).
      $\tilde{\pi}_{\mathrm{k}} \leftarrow$COMPOSITE-STRATEGY($\Gamma, \boldsymbol{\pi}_{\mathrm{k}}$).
      $\pi_{\mathrm{k}} \leftarrow$AVERAGE-STRATEGY($\pi_{\mathrm{k}}, \tilde{\pi}_{\mathrm{k}}$).
    $\widehat{\delta} \leftarrow$EXPLOITABILITY($\tilde{\pi}_{\mathrm{D}}$,$\tilde{\pi}_{\mathrm{A}}$) (3.16).
  **return** $(\pi_{\mathrm{D}}, \pi_{\mathrm{A}})$.
**procedure** LOCAL-BEST-RESPONSES(P-SOLVER, $\Gamma$, k, $\pi_{-\mathrm{k}}$)
  $\boldsymbol{\pi}_{\mathrm{k}} \leftarrow ()$.
  **in parallel for** $\mathbf{w} \in \mathcal{W}$, $i \in \mathcal{V}_{\mathbf{w}}$ **then**
    $\boldsymbol{\pi}_{\mathrm{k}} \leftarrow \boldsymbol{\pi}_{\mathrm{k}} \oplus$ P-SOLVER($\Gamma, \pi_{-\mathrm{k}}, \mathrm{k}, i$).
  **return** $\boldsymbol{\pi}_{\mathrm{k}}$.
**procedure** COMPOSITE-STRATEGY($\Gamma$, $\boldsymbol{\pi}_{\mathrm{k}}$)
  **return** $\pi_{\mathrm{k}} \leftarrow$ **Procedure** $\lambda(\mathbf{s}_t^{(\mathrm{k})}, \mathbf{b}_t^{(\mathrm{k})})$.
  $\mathbf{a}_t^{(\mathrm{k})} \leftarrow ()$.
  **for** $\mathbf{w} \in \mathcal{W}$, $i \in \mathcal{V}_{\mathbf{w}}$ **do**
    $\mathbf{a}_t^{(\mathrm{k})} \leftarrow \mathbf{a}_t^{(\mathrm{k})} \oplus (\boldsymbol{\pi}_{\mathrm{k}}^{(i)}(\mathbf{s}_{i,t}^{(\mathrm{k})}, \mathbf{b}_{i,t}^{(\mathrm{k})}))$.
  **return** $\mathbf{a}_t^{(\mathrm{k})}$.

***Computing best responses*** Each iteration of DFP involves computing a best response for the attacker and the defender. While Thm. 3.2 allows us to decompose this computation into $|\mathcal{V}|$ subproblems that can be solved in parallel, we are still left with the task of solving the subproblems. This task corresponds to computing best responses in the subgames $(\Gamma^{(i)})_{i \in \mathcal{V}}$ defined above. These computations amount to solving $2|\mathcal{V}|$ Partially Observed Markov Decision Processes (POMDPs)[10]. The principal method for solving a POMDP is dynamic programming. However, dynamic programming is intractable in our case, as demonstrated in Fig. 3.6 on the next page. To find the best responses, we instead resort to approximation algorithms. More specifically, we use the **P**roximal **P**olicy **O**ptimization (PPO) algorithm (Alg. 1, Schulman et al., 2017) to approximate a best response for the attacker[11], and we leverage the threshold structure of Thm. 3.2.C to approximate a best response for the defender through stochastic approximation[12].

[10] $|\mathcal{V}|$ POMDPs for the defender's best response and $|\mathcal{V}|$ POMDPs for the attacker's best response; the components of a POMDP are defined the background chapter; see (15).

[11] See Appendix D for details about PPO.

[12] See Appendix F for details about computing a local best response for the defender.

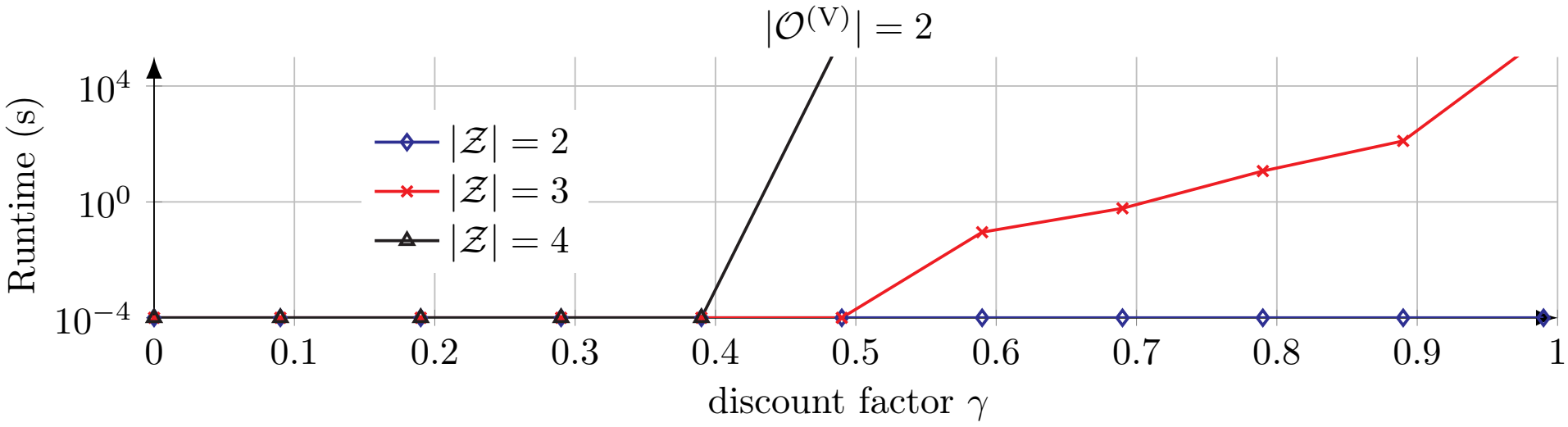

***Figure 3.6:*** *Runtimes of dynamic programming when computing a best response for the attacker in a subgame $\Gamma^{(i)}$ via Sondik's value iteration algorithm (Sondik, 1978); we note that even for a minimal observation space $\mathcal{O}^{(V)}$ the runtime increases exponentially with the discount factor $\gamma$ and the number of zones $|\mathcal{Z}|$.*

## 3.8 System Identification

The DFP algorithm described above approximates a PBE of $\Gamma$ (3.7) by simulating games and updating both players' strategies through PPO and stochastic approximation. We use a digital twin of the target infrastructure to identify the parameters required to instantiate these simulations and evaluate the learned strategies. We create this digital twin using CSLE, as described in the methodology chapter (Hammar, 2023). The topology of the target infrastructure is shown in Fig. 3 in the introduction chapter, and the configuration is listed in Appendix C.

The digital twin comprises virtual containers and networks that replicate the functionality and the timing behavior of the target infrastructure. These containers run the same software and processes as the physical infrastructure, including the SNORT intrusion detection system (IDS), which monitors network traffic and logs alerts in real-time. These alerts are tagged with IP addresses of the containers that caused them. We aggregate these alerts per container in the digital twin at 30-second intervals. (30$s$ in the digital twin corresponds to 1 time step in the game $\Gamma$.) This aggregation allows us to compute the observations $\mathbf{o}_{1,t}, \mathbf{o}_{2,t}, \ldots, \mathbf{o}_{|\mathcal{V}|,t}$, where $\mathbf{o}_{i,t}$ is defined as the number of intrusion alerts associated with node $i$ at time $t$, weighted by priority (3.3). These observations depend on client behavior as well as the actions of the attacker and the defender, as described below.

### Emulating the client population in the digital twin

The *client population* is emulated by processes in DOCKER containers. Clients interact with application nodes through the gateway by consuming workflows; see Fig. 3 in the introduction chapter. A client's workflow and sequence of service invocations are selected uniformly at random. Client arrivals per time step are emulated using a stationary Poisson process with rate $\lambda = 50$ and exponentially distributed service times with mean $\mu = 4$.

### Emulating the attacker in the digital twin

The attacker's actions are emulated by executing scripts that automate exploits; see Table 3.2.

| *Type* | *Actions* |
|---|---|
| Reconnaissance | TCP SYN scan, UDP port scan, TCP XMAS scan, VULSCAN, ping-scan. |
| Brute-force | TELNET, SSH, FTP, CASSANDRA,IRC, MONGODB, MYSQL, SMTP, POSTGRES. |
| Exploit | CVE-2017-7494, CVE-2015-3306, CVE-2010-0426, CVE-2015-5602 CVE-2014-6271, CVE-2016-10033, CVE-2015-1427, CWE-89 on DVWA [454]. |

***Table 3.2:*** *Attacker actions executed on the digital twin; further details about the actions can be found in Appendix D of Paper 2 and in the methodology chapter.*

### Emulating the defender in the digital twin

We implement four types of response actions to emulate the defender; see Fig. 3.1 on page 126. To emulate the *node migration* action, we remove all virtual network interfaces of the emulated node and add a new interface that connects it to the new zone. To emulate the *flow migration/blocking* action, we add rules to the flow tables of the emulated switches that match all flows towards the node and redirect them to a given destination. To emulate the *node shut down* action, we shut down the virtual container corresponding to the emulated node. Finally, to emulate the *access control* action, we reset all user accounts on the emulated node.

### Estimating the observation distributions

As our target infrastructure consists of 64 nodes (see Appendix C), there are 64 alert distributions $z_{\mathbf{O}_1}, \ldots, z_{\mathbf{O}_{64}}$ (3.3). We estimate these distributions using data from the digital twin. Specifically, at the end of every time step (i.e., every 30-second interval) in the digital twin, we collect the number of intrusion alerts during the time step and compute the vector $\mathbf{o}_t = (\mathbf{o}_{1,t}, \mathbf{o}_{2,t}, \ldots, \mathbf{o}_{|\mathcal{V}|,t})$, which contains the total number of intrusion alerts per node, weighted by priority.

For the evaluation in this paper, we collect $M = 10^4$ i.i.d. samples. Based on these samples, we compute the empirical distributions $\widehat{z}_{\mathbf{O}_1}, \ldots, \widehat{z}_{\mathbf{O}_{64}}$ as estimates of $z_{\mathbf{O}_1}, \ldots, z_{\mathbf{O}_{64}}$, where $\widehat{z}_{\mathbf{O}_i} \overset{\text{a.s.}}{\rightarrow} z_{\mathbf{O}_i}$ as $M \rightarrow \infty$[13]; see Fig. 3.7 on the next page. We observe in Fig. 3.7 that the distributions differ between nodes, which can be explained by the different services provided by the nodes; see Appendix C. We further observe that the distributions during intrusion have more probability mass at larger values than the distributions when no intrusion occurs.

[13]It follows by the Glivenko-Cantelli theorem; see (Glivenko and Cantelli, 1933).

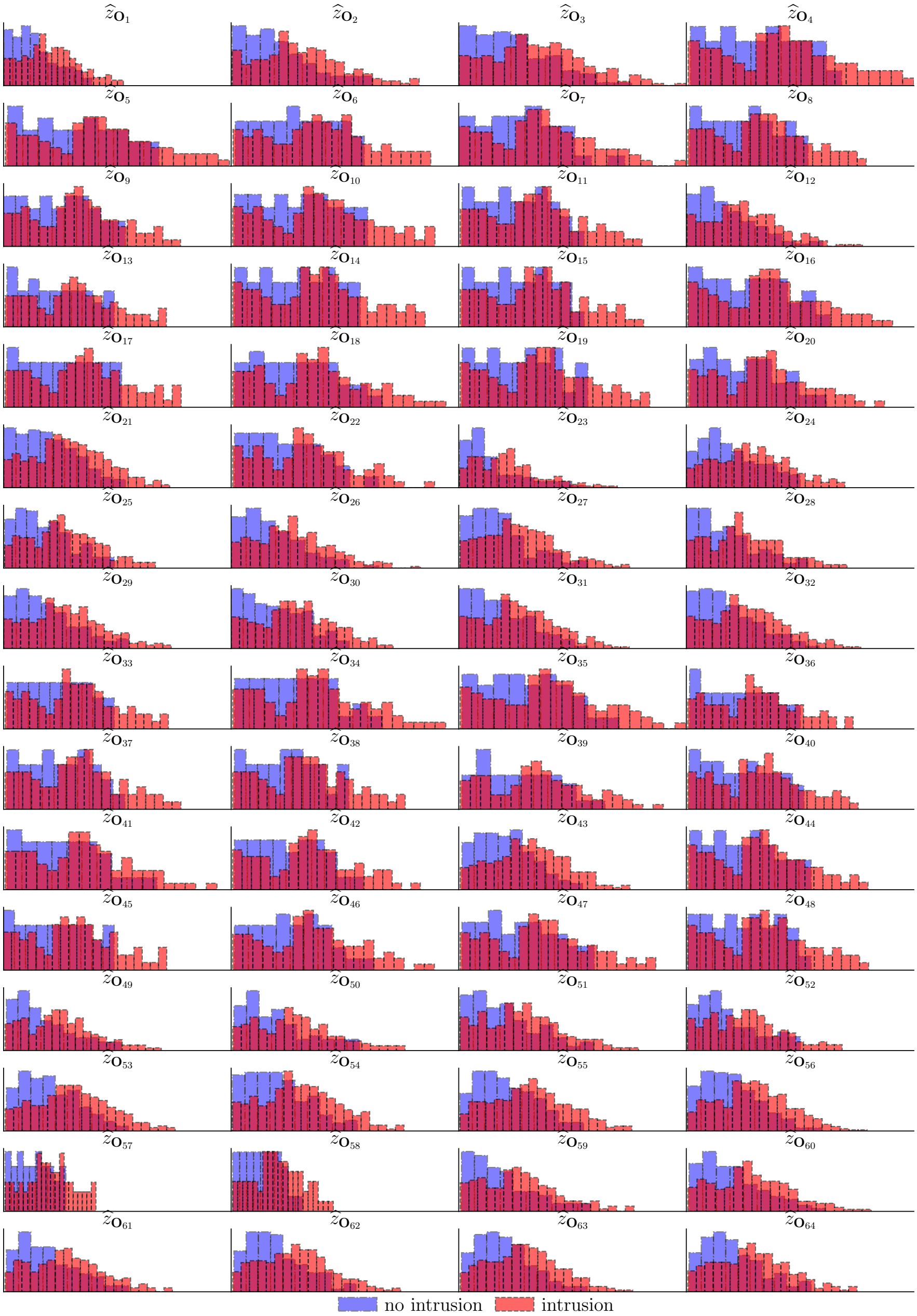

***Figure 3.7:*** *Empirical observation distributions $\widehat{z}_{\mathbf{O}_1}, \ldots, \widehat{z}_{\mathbf{O}_{|\mathcal{V}|}}$ as estimates of $z_{\mathbf{O}_1}, \ldots, z_{\mathbf{O}_{|\mathcal{V}|}}$ in the target infrastructure; $\mathbf{O}_i$ is a random variable representing the number of IDS alerts related to node $i \in \mathcal{V}$, weighted by priority; the x-axes show the node-local observation spaces $\mathcal{O}^{(\mathrm{V})}$; the y-axes show $\widehat{z}_{\mathbf{O}_i}(\mathbf{o}_i \mid \mathbf{s}_i)$ (3.3).*

**Remark 3.3.** The stochastic matrices with the rows $\widehat{z}_{\mathbf{O}_i|\mathbf{s}_i^{(\mathrm{A})}=(0,0)}$ and $\widehat{z}_{\mathbf{O}_i|\mathbf{s}_i^{(\mathrm{A})}\neq(0,0)}$ have $250 \times 10^9$ second-order minors, which are almost all non-negative. This suggests that the TP-2 assumption in Thm. 3.2.C can be made.

## 3.9 Experimental Evaluation

Our methodology for finding near-optimal defender strategies includes learning equilibrium strategies via the DFP algorithm and evaluating these strategies on the digital twin; see Fig. 12 in the introduction chapter. This section describes the evaluation results.

### Experiment setup

The instantiation of $\Gamma$ (3.7) and the hyperparameters are listed in Appendix A. The topology of the target infrastructure is depicted in Fig. 3 in the introduction chapter, and its configuration is available in Appendix C. The digital twin is deployed on a server with a 24-core INTEL XEON GOLD 2.10 GHz CPU and 768 GB RAM. Simulations of $\Gamma$ and executions of DFP run on a cluster with 2xTESLA P100 GPUS, 4xRTX8000 GPUS, and 3x16-core INTEL XEON 3.50 GHz CPUS; see Fig. 21 in the methodology chapter.

### Convergence metric

To estimate the convergence of the sequence of strategy pairs generated by DFP, we use the *approximate exploitability* metric

$$\widehat{\delta} = \mathbb{E}_{\widehat{\pi}_{\mathrm{D}},\pi_{\mathrm{A}}}[J] - \mathbb{E}_{\pi_{\mathrm{D}},\widehat{\pi}_{\mathrm{A}}}[J], \qquad \text{(Eq. 3, Timbers et al., 2020)} \qquad (3.16)$$

where $J$ is defined in (3.4) and $\widehat{\pi}_{\mathrm{k}}$ denotes an approximate best response for player k. The closer $\widehat{\delta}$ becomes to 0, the closer $(\pi_{\mathrm{D}}, \pi_{\mathrm{A}})$ is to an equilibrium.

### Baseline algorithms

We compare the performance of our approach ($\pi^{\text{decomposition}}$) with that of two baselines: $\pi^{\text{full}}$ and $\pi^{\text{workflow}}$. Baseline $\pi^{\text{full}}$ is the strategy obtained when attempting to solve the full game without decomposition, and $\pi^{\text{workflow}}$ is the strategy obtained when attempting to solve the game decomposed on the workflow level only (i.e., using the decomposition in Thm. 3.2.A but not the decomposition in Thm. 3.2.B). We compare the performance of DFP with that of **N**eural **F**ictitious **S**elf-**P**lay (NFSP) (Alg. 1, Heinrich and Silver, 2016) and PPO (Alg. 1, Schulman et al., 2017), which are the most popular algorithms among related works.

### Baseline strategies

We compare the defender strategies learned through DFP with three baselines. The first baseline selects actions uniformly at random. The second baseline assumes prior knowledge of the opponent's actions and acts optimally based on this information. The last baseline acts according to the following heuristic: shut down a node $i \in \mathcal{V}$ when an IDS alert occurs, i.e., when $\mathbf{o}_{i,t} > 0$.

### Learning best responses against static opponents

We first examine whether our method can discover effective strategies against a *static* opponent strategy, which in game-theoretic terms is the problem of finding best responses. The static strategies are defined in Appendix B. We compare the scalability of $\pi^{\text{decomposition}}$ with that of $\pi^{\text{workflow}}$ and $\pi^{\text{full}}$ on synthetic infrastructures with varying number of nodes $|\mathcal{V}|$ and workflows $|\mathcal{W}|$. Further, we compare the convergence rate when exploiting Theorem 3.2.C to compute a best response with that of PPO. Figure 3.8 shows the evaluation results.

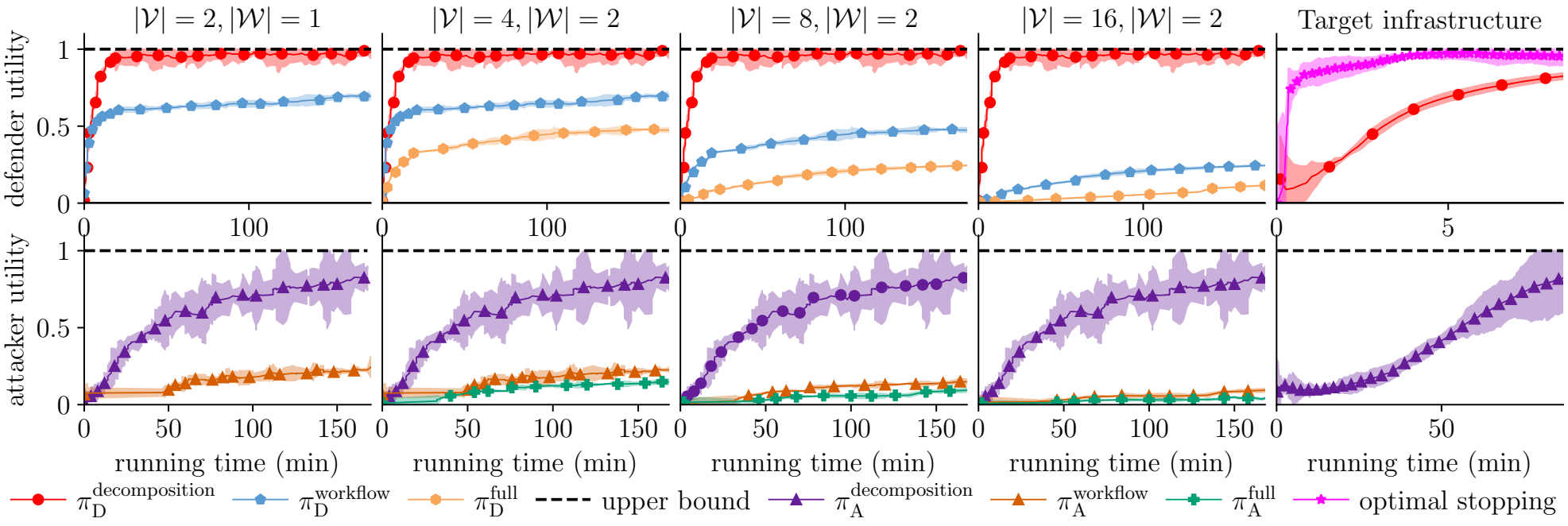

***(a)*** *Best response learning curves of normalized utility (3.4).*

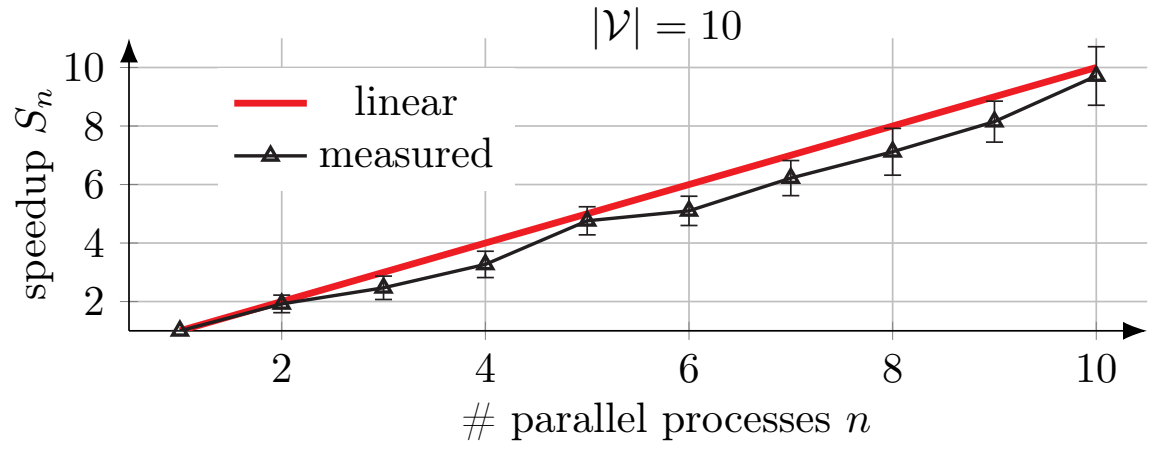

***(b)*** *Best response scalability.*

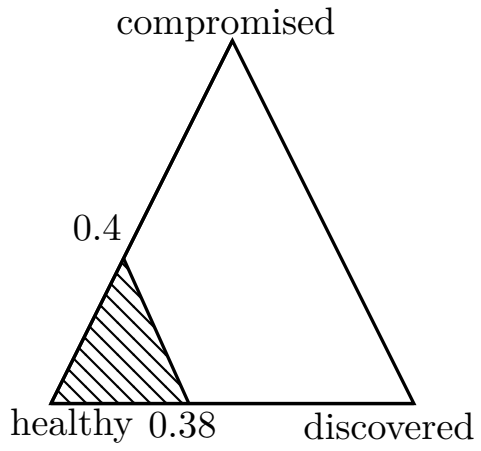

***(c)*** *Best response structure.*

***Figure 3.8:*** *Best response learning via decomposition; (a) shows learning curves in simulation; the curves show the mean and 95% confidence interval for five random seeds; (b) shows the speedup of our approach when computing best responses with different number of parallel processes; the speedup is calculated as $S_n = \frac{T_1}{T_n}$, where $T_n$ is the completion time with $n$ processes; and (c) shows an estimated switching curve (Thm. 3.2.C).*

The red, purple, and pink curves in Fig. 3.8.a represent the results obtained with $\pi^{\text{decomposition}}$; the blue and beige curves represent the results obtained with $\pi^{\text{workflow}}$; the orange and green curves represent the results obtained with $\pi^{\text{full}}$; and the dashed black lines relate to the baseline strategy that assumes prior knowledge of the opponent's strategy. We note that all the learning curves of $\pi^{\text{decomposition}}$ converge near the dashed black lines, which suggests that the learned strategies are close to best responses. In contrast, the learning curves of $\pi^{\text{workflow}}$ and $\pi^{\text{full}}$ do not converge near the dashed black lines within the measured time. This is expected as $\pi^{\text{workflow}}$ and $\pi^{\text{full}}$ cannot be parallelized like $\pi^{\text{decomposition}}$. The parallelization speedup is shown in Fig. 3.8.b. Lastly, we note in the rightmost plot of Fig. 3.8.a that the optimal stopping approach, which exploits the statement in Thm. 3.2.C, converges significantly faster than PPO. An example of a learned optimal stopping strategy based on the switching curve in Fig. 3.5 is shown in Fig. 3.8.c.

### Learning equilibrium strategies through fictitious play

Figure 3.9 shows the learning curves of the strategies obtained during the DFP execution and the baselines introduced above. The red curves represent the results from the simulator; the blue curves show the results from the digital twin; the green curve gives the performance of the random baseline; the orange curve relates to the $\mathbf{o}_{i,t} > 0$ baseline; and the dashed black line gives the performance of the baseline strategy that assumes prior knowledge of the attacker actions.

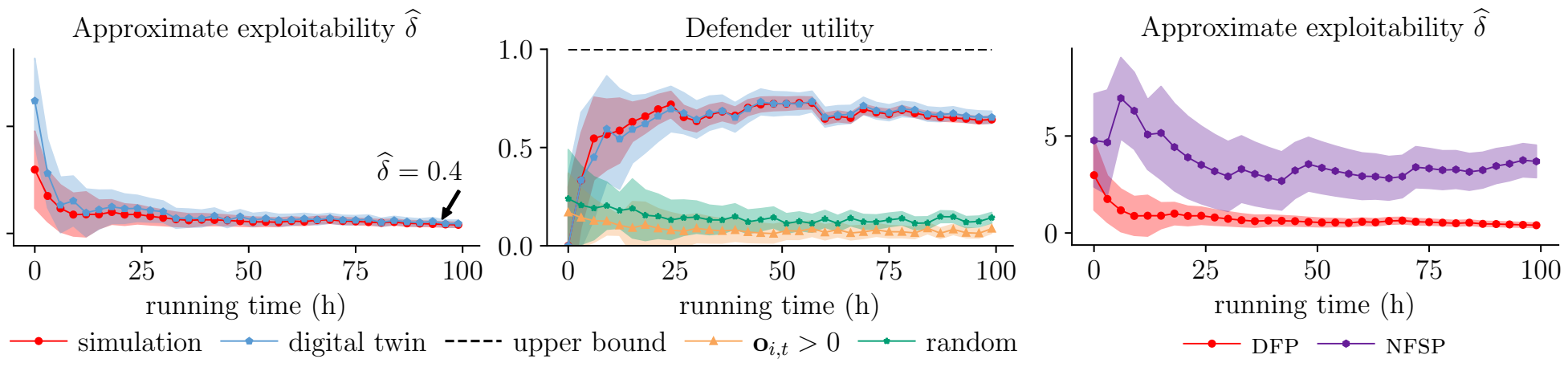

*(a) Equilibrium learning curves (utility is normalized).* *(b) Comparison DFP and NFSP.*

***Figure 3.9:*** *Equilibrium learning via DFP; the red curves show simulation results and the blue curves show results from the digital twin; the green, orange, purple, and black curves relate to baselines; the figures show approximate exploitability (3.16) and normalized utility; the curves indicate the mean and the shaded areas indicate the standard deviation over three random seeds.*

We observe in Fig. 3.9.a that the approximate exploitability (3.16) of the learned strategies converges to small values (left plot), which indicates that the learned strategies approximate an equilibrium both on the simulator and on the digital twin. Further, we see from the middle plot that both baseline strategies show decreasing performance as the attacker updates its strategy. In contrast, the defender strategy learned through DFP improves its performance over time.

Figure 3.9.b compares DFP with NFSP on the simulator. NFSP implements fictitious play and can thus be compared with DFP with respect to approximate exploitability (3.16). We observe that DFP converges significantly faster than NFSP. The fast convergence of DFP in comparison with NFSP is expected as DFP is parallelizable while NFSP is not.

### Discussion of the evaluation results

In this paper, we show how our methodology[14] for automated security response can scale to high-dimensional system models by leveraging recursive decomposition. The key findings can be summarized as follows.

- Our methodology approximates optimal defender strategies for a practical IT infrastructure (Fig. 3.9.a). While we have not evaluated the learned strategies on the target infrastructure for safety reasons, their performance on the digital twin gives us confidence in their expected performance on the target infrastructure.
- Decomposition provides a scalable approach to automate intrusion response for IT infrastructures (Fig. 3.8.a and Fig. 3.9.b). The intuition behind this finding is that decomposition allows the design of efficient divide-and-conquer algorithms that can be parallelized (Thm. 3.2.A–B and Alg. 3.1).
- The theory of optimal stopping provides insight into optimal defender strategies, which enables efficient computation of best responses (Fig. 3.9.a). This property can be explained by the threshold structures of the best responses, which drastically reduce the search space of possible strategies (Thm. 3.2.C).
- Static defender strategies' performance deteriorates against a dynamic attacker whereas defender strategies learned through DFP improve over time (Fig. 3.9.a). This finding is consistent with previous studies that use game-theoretic approaches and suggests fundamental limitations of static response systems, such as (Wazuh Inc, 2022).

## 3.10 Conclusion

We show how our methodology for automated security response can scale to high-dimensional system models by leveraging recursive decomposition. We present the decomposition through an intrusion response use case, which we model as a partially observed stochastic game. We prove a decomposition theorem stating that the game decomposes recursively into subgames that can be solved efficiently in parallel and that the best responses exhibit threshold structures. This decomposition provides us with a scalable approach to learn near-optimal defender strategies. We develop

[14] See the methodology chapter for details about the experimental methodology.

**D**ecompositional **F**ictitious **P**lay (DFP) – an algorithm for approximating equilibria. To assess the learned strategies for a target infrastructure, we evaluate them on a digital twin. The results demonstrate that DFP converges in reasonable time to near-optimal strategies, both in simulation and on the digital twin. At the same time, a state-of-the-art algorithm makes little progress toward an optimal strategy.

In the broader context of this thesis, this paper extends the problem formulations presented in Paper 1 and Paper 2. While those works focus on determining the optimal *timing* for defensive actions, this paper broadens the scope to also address the question of *which* actions should be executed. In the next chapter of the thesis, we show how our methodology can be generalized to a different use case, namely intrusion tolerance.

## ■ Acknowledgments

The authors would like to thank Quanyan Zhu for his useful input to this research. The authors are also grateful to Forough Shahab Samani and Xiaoxuan Wang for their constructive comments on a draft of this paper.

## ■ Appendix

### A Hyperparameters and Game Instantiation

We instantiate $\Gamma$ (3.7) for the experimental evaluation as follows. Client arrivals are sampled from a stationary Poisson process $\text{Po}(\lambda = 50)$, and service times are exponentially distributed with mean $\mu = 4$. In addition to migrating a node, the defender can shut it down or redirect its traffic to a honeynet, which we model with the zones $\mathfrak{S}, \mathfrak{R} \in \mathcal{Z}$. A node $i \in \mathcal{V}$ is shutdown if $v^{(\mathrm{Z})}_{i,t} = \mathfrak{S}$ and have its traffic redirected if $v^{(\mathrm{Z})}_{i,t} = \mathfrak{R}$. The set of local attacker actions is $\mathcal{A}^{(\mathrm{V})}_{\mathrm{A}} = \{\perp, \text{reconnaissance}, \text{brute-force}, \text{exploit}\}$, which we encode as $\{0, 1, 2, 3\}$. These actions have the following effects on the state $\mathbf{s}_t$: $\mathbf{a}^{(\mathrm{A})}_{i,t} = 1 \implies v^{(\mathrm{R})}_{i,t} = 1$, $\mathbf{a}^{(\mathrm{A})}_{i,t} = 2 \implies v^{(\mathrm{I})}_{i,t} = 1$ with probability 0.3, and $\mathbf{a}^{(\mathrm{A})}_{i,t} = 3 \implies v^{(\mathrm{I})}_{i,t}$ with probability 0.4. We enforce a tree structure on the target infrastructure by disregarding the redundant edges in the R&D zone; see Fig. 3 in the introduction chapter. The remaining parameters are listed in Table 3.3 on the next page.

| *Game parameters* | *Values* |
|---|---|
| $u_{\mathbf{w},t}$, $\mathcal{A}_{\mathrm{D}}^{(\mathrm{V})}$ | $\sum_{i\in\mathcal{V}_{\mathbf{w}}} \mathbb{1}_{\mathrm{gw}\rightarrow_t i}$, $\mathcal{Z}\cup\{\text{access control}, \perp\}$ |
| $\lvert\mathcal{O}^{(\mathrm{V})}\rvert$, $\gamma$, $\eta$, $\lvert\mathcal{Z}\rvert$, $\lvert\mathcal{W}\rvert$, $\lvert\mathcal{V}\rvert$ | $10^3$, 0.9, 0.4, 6, 10, 64 |
| $u_i^{(\mathrm{w})}(\perp,l)$, $u_i^{(\mathrm{w})}(\mathfrak{S},l)$,$u_i^{(\mathrm{w})}(\mathfrak{R},l)$, $u_i^{(\mathrm{w})}(2,l)$ | 0, $10+l$, $15+l$, $0.1+l$ |
| $u_i^{(\mathrm{w})}(3,l)$, $u_i^{(\mathrm{w})}(4,l)$, $u_i^{(\mathrm{w})}(5,l)$, $u_i^{(\mathrm{w})}(0.8,l)$ | $0.5+l$, $1+l$, $1.5+l$, $2+l$ |
| topology $\mathcal{G}$ and $\mathbf{s}_1^{(\mathrm{D})}$ | see Fig. 3 |
| $\lvert\mathcal{V}_{\mathbf{w}_1}\rvert$,$\lvert\mathcal{V}_{\mathbf{w}_2}\rvert$,$\lvert\mathcal{V}_{\mathbf{w}_3}\rvert$,$\lvert\mathcal{V}_{\mathbf{w}_4}\rvert$,$\lvert\mathcal{V}_{\mathbf{w}_5}\rvert$,$\lvert\mathcal{V}_{\mathbf{w}_6}\rvert$ | 16, 16, 16, 16, 6, 4 |
| $\lvert\mathcal{V}_{\mathbf{w}_7}\rvert$,$\lvert\mathcal{V}_{\mathbf{w}_8}\rvert$,$\lvert\mathcal{V}_{\mathbf{w}_9}\rvert$,$\lvert\mathcal{V}_{\mathbf{w}_{10}}\rvert$ | 6, 4, 6, 6 |
| *PPO parameters* [396, Alg. 1] | |
| lr $\alpha$, batch, # layers, # neurons, clip $\epsilon$ | $10^{-5}$, $4\cdot 10^3 t$, 4, 64, 0.2 |
| GAE $\lambda$, ent-coef, activation | 0.95, $10^{-4}$, ReLU |
| *NFSP parameters* [192, Alg. 1] | |
| lr RL, lr SL, batch, # layers,# neurons, $\mathcal{M}_{RL}$ | $10^{-2}$, $5\cdot 10^{-3}$, 64, 2,128, $2\times 10^5$ |
| $\mathcal{M}_{SL}$,$\epsilon$, $\epsilon$-decay, $\eta$ | $2\times 10^6$, 0.06, 0.001, 0.1 |
| *Stochastic approximation parameters* | |
| $c, \epsilon, \lambda, A, a, N, \delta$ | 10, 0.101, 0.602, 100, 1, 50, 0.2 |

***Table 3.3:*** *Hyperparameters.*

# B Static Defender and Attacker Strategies

The static defender and attacker strategies for the evaluation described in §3.9 are defined in (3.17)–(3.18). (w.p is short for "with probability".)

$$\pi_{\mathrm{D}}(\mathbf{h}_t^{(\mathrm{D})})_i = \begin{cases} \perp & \text{w.p } 0.95 \\ j\in\mathcal{Z} & \text{w.p } \dfrac{0.05}{\lvert\mathcal{Z}\rvert+1} \end{cases} \tag{3.17}$$

$$\pi_{\mathrm{A}}(\mathbf{h}_t^{(\mathrm{A})})_i = \begin{cases} \perp & \text{if } v_{i,t}^{(\mathrm{I})}=1 \\ \perp & \text{w.p } 0.8 \text{ if } v_{i,t}^{(\mathrm{R})}=0 \\ \perp & \text{w.p } 0.7 \text{ if } v_{i,t}^{(\mathrm{R})}=1, v_{i,t}^{(\mathrm{I})}=0 \\ \text{recon} & \text{w.p } 0.2 \text{ if } v_{i,t}^{(\mathrm{R})}=0 \\ \text{brute} & \text{w.p } 0.15 \text{ if } v_{i,t}^{(\mathrm{R})}=1, v_{i,t}^{(\mathrm{I})}=0 \\ \text{exploit} & \text{w.p } 0.15 \text{ if } v_{i,t}^{(\mathrm{R})}=1, v_{i,t}^{(\mathrm{I})}=0. \end{cases} \tag{3.18}$$

# C Configuration of the Infrastructure in Figure 3

The configuration of the target infrastructure is available in Tables 3.4 and 3.5 on the subsequent pages; the network topology is shown in Fig. 3 in the introduction chapter.

| *ID(s)* | *Type* | *Operating system* | *Zone* | *Services* | *Vulnerabilities* |
|---|---|---|---|---|---|
| 1 | Gateway | UBUNTU 20 | - | SNORT (ruleset v2.9.17.1), SSH, OPENFLOW v1.3, RYU SDN controller | - |
| 2 | Gateway | UBUNTU 20 | DMZ | SNORT (ruleset v2.9.17.1), SSH, OVS v2.16, OPENFLOW v1.3 | - |
| 28 | Gateway | UBUNTU 20 | R&D | SNORT (ruleset v2.9.17.1), SSH, OVS v2.16, OPENFLOW v1.3 | - |
| 3,12 | Switch | UBUNTU 22 | DMZ | SSH, OPENFLOW v1.3 , OVS v2.16 | - |
| 21, 22 | Switch | UBUNTU 22 | - | SSH, OPENFLOW v1.3, OVS v2.16 | - |
| 23 | Switch | UBUNTU 22 | ADMIN | SSH, OPENFLOW v1.3, OVS v2.16 | - |
| 29-48 | Switch | UBUNTU 22 | R&D | SSH, OPENFLOW v1.3, OVS v2.16 | - |
| 13-16 | Honeypot | UBUNTU 20 | DMZ | SSH, SNMP, POSTGRES, NTP | - |
| 17-20 | Honeypot | UBUNTU 20 | DMZ | SSH, IRC, SNMP, SSH, POSTGRES | - |
| 4 | App node | UBUNTU 20 | DMZ | HTTP, DNS, SSH | CWE-1391 |
| 5, 6 | App node | UBUNTU 20 | DMZ | SSH, SNMP, POSTGRES, NTP | - |
| 7 | App node | UBUNTU 20 | DMZ | HTTP, TELNET, SSH | CWE-1391 |
| 8 | App node | DEBIAN JESSIE | DMZ | FTP, SSH, APACHE 2,SNMP | CVE-2015-3306 |
| 9,10 | App node | UBUNTU 20 | DMZ | NTP, IRC, SNMP, SSH, POSTGRES | - |
| 11 | App node | DEBIAN JESSIE | DMZ | APACHE 2, SMTP, SSH | CVE-2016-10033 |
| 24 | Admin system | UBUNTU 20 | ADMIN | HTTP, DNS, SSH | CWE-1391 |
| 25 | Admin system | UBUNTU 20 | ADMIN | FTP, MONGODB, SMTP, TOMCAT, TS 3, SSH | - |
| 26 | Admin system | UBUNTU 20 | ADMIN | SSH, SNMP, POSTGRES, NTP | - |
| 27 | Admin system | UBUNTU 20 | ADMIN | FTP, MONGODB, SMTP, TOMCAT, TS 3, SSH | CWE-1391 |
| 49-59 | Compute node | UBUNTU 20 | R&D | SPARK, HDFS | - |
| 60 | Compute node | DEBIAN WHEEZY | R&D | SPARK, HDFS, APACHE 2,SNMP, SSH | CVE-2014-6271 |
| 61 | Compute node | DEBIAN 9.2 | R&D | IRC, APACHE 2, SSH | CWE-89 |
| 62 | Compute node | DEBIAN JESSIE | R&D | SPARK, HDFS, TS 3, TOMCAT, SSH | CVE-2010-0426 |
| 63 | Compute node | DEBIAN JESSIE | R&D | SSH, SPARK, HDFS | CVE-2015-5602 |
| 64 | Compute node | DEBIAN JESSIE | R&D | SAMBA, NTP, SSH, SPARK, HDFS | CVE-2017-7494 |

***Table 3.4:*** *Configuration of the target infrastructure shown in Fig. 3 of the introduction chapter; each row contains the configuration of one or more nodes; vulnerabilities in specific software products are identified by the vulnerability identifiers in the Common Vulnerabilities and Exposures (CVE) database (The MITRE Corporation, 2022); vulnerabilities that are not described in the CVE database are categorized according to the types of the vulnerabilities they exploit based on the Common Weakness Enumeration (CWE) list (The MITRE Corporation, 2023). (Note that each component in Fig. 3 is labeled with an identifier i.)*

| *ID* | *Name* | *Zone* | *Nodes* |
|---|---|---|---|
| 1 | SPARK 1 | R&D | 1, 21, 22, 28, (29 – 32), (33 – 34), (41 – 42), (49 – 52). |
| 2 | SPARK 2 | R&D | 1, 21, 22, 28, (29 – 32), (35 – 36), (43 – 44), (53 – 56). |
| 3 | SPARK 3 | R&D | 1, 21, 22, 28, (29 – 32), (37 – 38), (45 – 46), (57 – 60). |
| 4 | SPARK 4 | R&D | 1, 21, 22, 28, (29 – 32), (39 – 40), (47 – 48), (61 – 65). |
| 5 | Web 1 | DMZ | 1, 2, 3, 4, 5, 6. |
| 6 | Web 2 | DMZ | 1, 2, 3, 7. |
| 7 | Storage 1 | DMZ | 1, 2, 3, 8, 9, 10. |
| 8 | Mail 1 | DMZ | 1, 2, 3, 11. |
| 9 | Admin 1 | ADMIN | 1, 21, 22, 23, 24, 25. |
| 10 | Admin 2 | ADMIN | 1, 21, 22, 23, 25, 26. |

***Table 3.5:*** *Workflows of the target infrastructure (Fig. 3 in the introduction chapter).*

## D Proximal Policy Optimization

We use Proximal Policy Optimization (PPO) (Alg. 1, Schulman et al., 2017) to approximate a best response for the attacker; see §3.7. This procedure involves parameterizing the attacker strategy as $\pi_{\mathrm{A},\boldsymbol{\theta}}(\mathbf{a}_t^{(\mathrm{A})} \mid \mathbf{o}_t, \mathbf{s}_t^{(\mathrm{A})})$, where $\boldsymbol{\theta}$ is the parameter vector of a neural network with the actor-critic architecture; see Fig. 3.10 (Sutton and Barto, 1998). PPO implements the *policy gradient method* (Sutton et al., 1999) and builds on the classical REINFORCE policy gradient algorithm (Williams, 1987). In the following, we start by describing REINFORCE, and then we explain how PPO modifies the general algorithm.

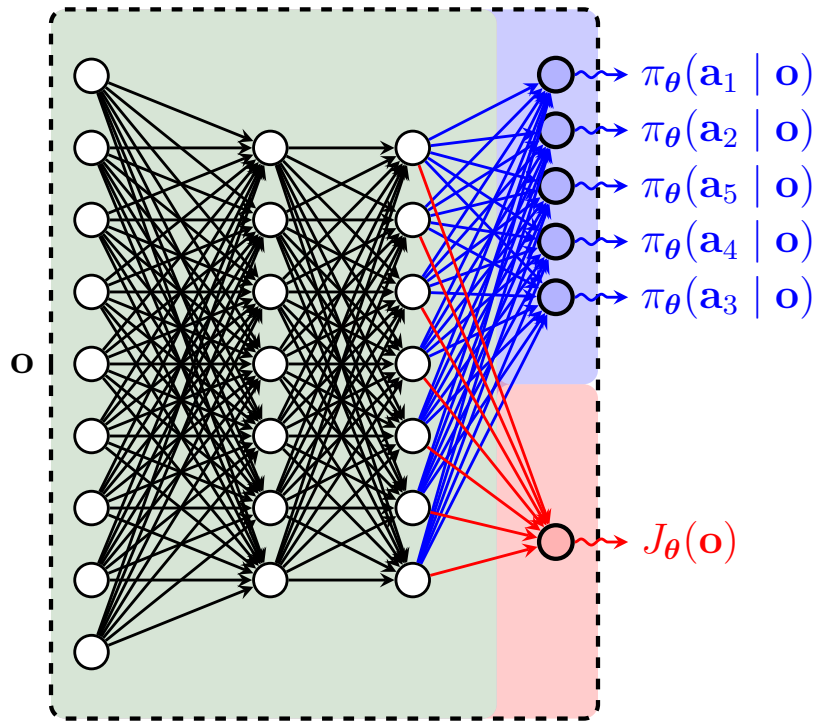

***Figure 3.10:*** *The actor-critic architecture in PPO; a neural network represents a control strategy $\pi_{\boldsymbol{\theta}}(\mathbf{a} \mid \mathbf{o})$ and a value function $J_{\boldsymbol{\theta}}(\mathbf{o})$.*

For ease of notation, we drop the player subscripts and superscripts, i.e., we write $\pi$ and $\mathbf{s}$ instead of $\pi_{\mathrm{A}}$ and $\mathbf{s}^{(\mathrm{A})}$. Moreover, instead of writing $\pi_{\boldsymbol{\theta}}(\mathbf{a}_t \mid \mathbf{o}_t, \mathbf{s}_t)$, we simply write $\pi_{\boldsymbol{\theta}}(\mathbf{a}_t \mid \mathbf{o}_t)$ as $\mathbf{s}_t$ does not affect the main steps of the derivation apart from making the expressions more verbose. The learning process of REINFORCE is iterative and produces a sequence of strategies $\pi_{\boldsymbol{\theta}^{(1)}}, \pi_{\boldsymbol{\theta}^{(2)}}, \ldots$. Each iteration consists of two steps. First, the current strategy $\pi_{\boldsymbol{\theta}^{(i)}}$ is executed on the simulator to collect *training histories* $(\mathbf{h}_T^{(k)})_{k=1}^M$. Second, the strategy is updated through stochastic gradient ascent: $\boldsymbol{\theta}^{(i+1)} \leftarrow \boldsymbol{\theta}^{(i)} + \eta \nabla_{\boldsymbol{\theta}^{(i)}} \mathbb{E}_{\pi_{\boldsymbol{\theta}^{(i)}}}[J]$ (3.4), where $\eta$ is the step size and $\nabla_{\boldsymbol{\theta}^{(i)}} \mathbb{E}_{\pi_{\boldsymbol{\theta}^{(i)}}}[J]$ is the (likelihood ratio) policy gradient, which can be derived as

$$\nabla_{\boldsymbol{\theta}^{(i)}} \mathbb{E}_{\pi_{\boldsymbol{\theta}^{(i)}}}[J] = \nabla_{\boldsymbol{\theta}^{(i)}} \left( \sum_{\mathbf{h}_T \in \mathcal{H}_T} \mathbb{P}\left[\mathbf{h}_T \mid \pi_{\boldsymbol{\theta}^{(i)}}\right] \overbrace{\sum_{(\mathbf{s}_t, \mathbf{a}_t) \in \mathbf{h}_T} u(\mathbf{s}_t, \mathbf{a}_t)}^{\triangleq u(\mathbf{h}_T)} \right)$$

$$
\begin{aligned}
&\overset{(a)}{=} \sum_{\mathbf{h}_T \in \mathcal{H}_T} \frac{\mathbb{P}\left[\mathbf{h}_T \mid \pi_{\boldsymbol{\theta}^{(i)}}\right]}{\mathbb{P}\left[\mathbf{h}_T \mid \pi_{\boldsymbol{\theta}^{(i)}}\right]} \nabla_{\boldsymbol{\theta}^{(i)}} \mathbb{P}\left[\mathbf{h}_T \mid \pi_{\boldsymbol{\theta}^{(i)}}\right] u(\mathbf{h}_T) \\
&= \mathbb{E}_{\mathbf{H}_T}\left[\nabla_{\boldsymbol{\theta}^{(i)}} \ln\left(\mathbb{P}\left[\mathbf{H}_T \mid \pi_{\boldsymbol{\theta}^{(i)}}\right]\right) u(\mathbf{H}_T)\right] \\
&= \mathbb{E}_{\mathbf{H}_T}\left[\nabla_{\boldsymbol{\theta}^{(i)}} \ln\left(\mathbf{b}_1(\mathbf{S}_1) \prod_{t=2}^{T} z(\mathbf{O}_t \mid \mathbf{S}_t) f(\mathbf{S}_t \mid \mathbf{S}_{t-1}, \mathbf{A}_{t-1}) \pi_{\boldsymbol{\theta}^{(i)}}(\mathbf{A}_{t-1} \mid \mathbf{O}_{t-1})\right) u(\mathbf{H}_T)\right] \\
&= \mathbb{E}_{\mathbf{H}_T}\left[\left(\sum_{t=1}^{T} \nabla_{\boldsymbol{\theta}^{(i)}} \ln \pi_{\boldsymbol{\theta}^{(i)}}(\mathbf{A}_t \mid \mathbf{O}_t)\right) u(\mathbf{H}_T)\right] \qquad (3.19) \\
&\overset{(b)}{\approx} \frac{1}{M} \sum_{k=1}^{M} \left(\sum_{t=1}^{T} \nabla_{\boldsymbol{\theta}^{(i)}} \ln \pi_{\boldsymbol{\theta}^{(i)}}(\mathbf{a}_t^{(k)} \mid \mathbf{o}_t^{(k)})\right) u(\mathbf{h}_T^{(k)}),
\end{aligned}
$$

where $\mathbf{o}_1 = (\mathbf{s}_1, \mathbf{b}_1)$; (a) uses the log-gradient trick; and (b) approximates the expectation $\mathbb{E}_{\mathbf{H}_T}$ using i.i.d. sample histories $\mathbf{h}_T^{(1)}, \ldots, \mathbf{h}_T^{(M)}$.

PPO extends REINFORCE with techniques from *trust-region optimization*. The key insight is that large changes to the strategy make the learning process unstable. To remedy this issue, PPO constrains the strategy updates to be within a *trust region*, where the changes to the strategy are small. This can be achieved by bounding the divergence $D_{\text{KL}}(\pi_{\boldsymbol{\theta}^{(i)}} \parallel \pi_{\boldsymbol{\theta}^{(i+1)}})$. However, enforcing this constraint is impractical due to the large computational complexity. Therefore, PPO adds a soft constraint to the policy gradient objective. In REINFORCE, this objective is

$$
\max_{\pi_{\boldsymbol{\theta}^{(i)}}} \mathbb{E}_{\mathbf{H}_T}\left[\left(\sum_{t=1}^{T} \pi_{\boldsymbol{\theta}^{(i)}}(\mathbf{A}_t \mid \mathbf{O}_t)\right) u(\mathbf{H}_T)\right].
$$

In PPO, this objective is changed to maximize

$$
\mathbb{E}_{\mathbf{H}_T}\left[\min\left[\sum_{t=1}^{T} \frac{\pi_{\boldsymbol{\theta}^{(i)}}(\mathbf{A}_t \mid \mathbf{O}_t)}{\pi_{\boldsymbol{\theta}^{(i-1)}}(\mathbf{A}_t \mid \mathbf{O}_t)} A(\mathbf{O}_t, \mathbf{A}_t), \sum_{t=1}^{T} \text{clip}\left(\frac{\pi_{\boldsymbol{\theta}^{(i)}}(\mathbf{A}_t \mid \mathbf{O}_t)}{\pi_{\boldsymbol{\theta}^{(i-1)}}(\mathbf{A}_t \mid \mathbf{O}_t)}, \epsilon\right) A(\mathbf{O}_t, \mathbf{A}_t)\right]\right],
$$

where $A(\mathbf{o}_t, \mathbf{a}_t) \triangleq u_t + \mathbb{E}[J_{\boldsymbol{\theta}}(\mathbf{o}_{t+1})] - J_{\boldsymbol{\theta}}(\mathbf{o}_t)$ is the advantage function and $\text{clip}(x, \epsilon)$ constrains the value of $x$ to the range $[1-\epsilon, 1+\epsilon]$, where $\epsilon$ is a hyperparameter[15]. If $x$ is within this range, it remains unchanged; otherwise, it is clipped to the nearest boundary value: $1-\epsilon$ or $1+\epsilon$. By taking the minimum of the regular objective and the clipped objective, the optimization process is incentivized to make small updates to $\pi_{\boldsymbol{\theta}^{(i)}}$ that improves the advantage $A(\mathbf{O}_t, \mathbf{A}_t)$.

**Remark 3.4** (Discount factor for the policy gradient)**.** *In our derivation of the policy gradient (3.19), we omit the discount factor $\gamma$ to obtain the most mathematically accurate gradient expression (Nota and Thomas, 2020). However, in practice, we use the discount factor to enhance convergence properties and focus the optimization on near-term utilities.*

[15] $J_{\boldsymbol{\theta}}$ is the critic output of the neural network; see Fig. 3.10.

## E Proof of Theorem 3.2.C

The idea behind this proof is that the problem of selecting *which* defensive action to apply in a subgame $\Gamma^{(i)}$ (Thm. 3.2.B) against a given attacker strategy can be separated from the problem of deciding *when* to apply it. Through this separation, we can analyze the latter problem using optimal stopping theory. We perform this separation by decomposing $\mathbf{a}^{(\mathrm{D})}_{i,t}$ into two subactions: $\mathbf{a}^{(\mathrm{D},1)}_{i,t}$ and $\mathbf{a}^{(\mathrm{D},2)}_{i,t}$. The first subaction $\mathbf{a}^{(\mathrm{D},1)}_{i,t} \neq \bot$ determines the defensive action and the second subaction $\mathbf{a}^{(\mathrm{D},2)}_{i,t} \in \{\mathsf{S}, \mathsf{C}\}$ determines when to take it. Specifically, if $\mathbf{a}^{(\mathrm{D},2)}_{i,t} = \mathsf{C}$, then $\mathbf{a}^{(\mathrm{D})}_{i,t} = \bot$; otherwise $\mathbf{a}^{(\mathrm{D})}_{i,t} = \mathbf{a}^{(\mathrm{D},1)}_{i,t}$. Using this action decomposition, at each time $t$, a strategy $\pi^{(i)}_{\mathrm{D}}$ in $\Gamma^{(i)}$ is a joint distribution over $\mathbf{A}^{(\mathrm{D},1)}_{i,t}$ and $\mathbf{A}^{(\mathrm{D},2)}_{i,t}$, which means that it can be represented in an auto-regressive manner as

$$
\begin{aligned}
&\pi^{(i)}_{\mathrm{D}}(\mathbf{a}^{(\mathrm{D},1)}_{i,t}, \mathbf{a}^{(\mathrm{D},2)}_{i,t} \mid \mathbf{h}^{(\mathrm{D})}_{i,t}) && (3.20)\\
&\overset{(a)}{=} \pi^{(i)}_{\mathrm{D}}(\mathbf{a}^{(\mathrm{D},1)}_{i,t} \mid \mathbf{h}^{(\mathrm{D})}_{i,t})\pi^{(i)}_{\mathrm{D}}(\mathbf{a}^{(\mathrm{D},2)}_{i,t} \mid \mathbf{h}^{(\mathrm{D})}_{i,t}, \mathbf{a}^{(\mathrm{D},1)}_{i,t})\\
&\overset{(b)}{=} \pi^{(i)}_{\mathrm{D}}(\mathbf{a}^{(\mathrm{D},1)}_{i,t} \mid \mathbf{b}^{(\mathrm{D})}_{i,t}, \mathbf{s}^{(\mathrm{D})}_{i,t})\pi^{(i)}_{\mathrm{D}}(\mathbf{a}^{(\mathrm{D},2)}_{i,t} \mid \mathbf{b}^{(\mathrm{D})}_{i,t}, \mathbf{s}^{(\mathrm{D})}_{i,t}, \mathbf{a}^{(\mathrm{D},1)}_{i,t})\\
&\overset{(c)}{=} \pi^{(i)}_{\mathrm{D}}(\mathbf{a}^{(\mathrm{D},1)}_{i,t} \mid \mathbf{s}^{(\mathrm{D})}_{i,t})\pi^{(i)}_{\mathrm{D}}(\mathbf{a}^{(\mathrm{D},2)}_{i,t} \mid \mathbf{b}^{(\mathrm{D})}_{i,t}, \mathbf{s}^{(\mathrm{D})}_{i,t}, \mathbf{a}^{(\mathrm{D},1)}_{i,t}),
\end{aligned}
$$

where (a) follows from the chain rule of probability; (b) holds because $(\mathbf{s}^{(\mathrm{D})}_{i,t}, \mathbf{b}^{(\mathrm{D})}_{i,t})$ is a sufficient statistic for $\mathbf{s}^{(\mathrm{A})}_{i,t}$ (Def. 4.2, Lem. 5.1, Thm. 7.1, Kumar and Varaiya, 1986); and (c) follows because $\mathbf{a}^{(\mathrm{D},1)}_{i,t} \neq \bot$ resets the belief state to $\mathbf{b}^{(\mathrm{D})}_{i,t+1}((0,0)) = 1$. Consequently, the action is independent of $\mathbf{b}^{(\mathrm{D})}_{i,t}$.

The strategy decomposition in (3.20) means that we can obtain a best response in $\Gamma^{(i)}$ by jointly optimizing two substrategies: $\pi^{(i,1)}_{\mathrm{D}}$ and $\pi^{(i,2)}_{\mathrm{D}}$. The former corresponds to solving an MDP[16] $\mathscr{M}^{(\mathrm{D},1)}$ with state space $\mathbf{s}^{(\mathrm{D})}_{i} \in \mathcal{Z}$ and the latter corresponds to solving a set of optimal stopping POMDPs $(\mathscr{M}^{(\mathrm{D},2)}_{i,s^{(\mathrm{D})},a^{(\mathrm{D})}})_{s^{(\mathrm{D})}\in\mathcal{Z},a^{(\mathrm{D})}\in\mathcal{A}^{(\mathrm{V})}_{\mathrm{D}}}$ with state space[17] $\mathbf{s}^{(\mathrm{A})}_{i} \in \{(0,0),(1,0),(1,1)\} \triangleq \{0,1,2\}$[18]. The transition matrices for each stopping problem are of the form:

$$
\begin{bmatrix} 1-p & p & 0 \\ 0 & 1-q & q \\ 0 & 0 & 1 \end{bmatrix} \quad \text{and} \quad \begin{bmatrix} 1 & 0 & 0 \\ 1 & 0 & 0 \\ 1 & 0 & 0 \end{bmatrix}, \tag{3.21}
$$

[16]The components of an MDP are defined the background chapter; see (1).

[17]The state of the stopping POMDP for node $i$ corresponds to $\mathbf{s}^{(\mathrm{A})}_{i,t}$. This state is defined as $\mathbf{s}^{(\mathrm{A})}_{i,t} = (v^{(\mathrm{R})}_{i,t}, v^{(\mathrm{I})}_{i,t})$, where $v^{(\mathrm{R})}_{i,t}$ and $v^{(\mathrm{I})}_{i,t}$ are binary variables that represent the reconnaissance state and the intrusion state, respectively.

[18]Since $\mathbf{a}^{(\mathrm{D},1)}_{i,t} \neq \bot$ effectively resets the stopping problem, each stopping problem can be defined with a *single* stop action rather than multiple stop actions; see Lemma 4.4 of Paper 4 for a proof.

where $p$ is the probability that the attacker performs reconnaissance and $q$ is the probability that the attacker compromises the node. The left matrix in (3.21) relates to $\mathbf{a}_{i,t}^{(\mathrm{D},2)} = \mathsf{C}$ and the right matrix relates to $\mathbf{a}_{i,t}^{(\mathrm{D},2)} = \mathsf{S}$. The *non-zero* second order minors of the matrices are $(1-p)(1-q)$, $pq$, $1-q$, $1-p$, $p$, and $(1-p)q$, which implies that the matrices are TP-2 (Def. 10.2.1, Krishnamurthy, 2016). Further, (3.4) and (3.15) implies that the utility function $\mathbf{u}_{i,2}$ of the stopping problems satisfies

$$\mathbf{u}_{i,2}(0,a) \geq \mathbf{u}_{i,2}(1,a) \geq \mathbf{u}_{i,2}(2,a) \qquad a \in \{\mathsf{S},\mathsf{C}\} \tag{3.22a}$$

$$\mathbf{u}_{i,2}(s,\mathsf{S}) - \mathbf{u}_{i,2}(s,\mathsf{C}) = \mathbf{u}_{i,2}(s+1,\mathsf{S}) - \mathbf{u}_{i,2}(s+1,\mathsf{C}) \qquad s \in \{0,1\}. \tag{3.22b}$$

(3.22a) follows because the defender's utility associated with a node $i$ reduces when the attacker compromises the node (3.15). (Recall that state 0 represents the healthy state, 1 represents the reconnaissance state, and 2 represents the intrusion state.) The structure in (3.22b) implies that the utility function is supermodular given the action encoding $(\mathsf{C},\mathsf{S}) = (0,1)$. This modularity is a direct consequence of (3.15), which implies that the immediate utility of taking a defensive action is independent of the state $s$.

Since the distributions $z_{\mathbf{O}_1|\mathbf{s}^{(\mathrm{A})}}, \ldots, z_{\mathbf{O}_{|\mathcal{V}|}|\mathbf{s}^{(\mathrm{A})}}$ are TP-2 by assumption, it follows from (Thm. 12.3.4, Krishnamurthy, 2016) that there exists a switching curve $\Upsilon$ that partitions $\mathcal{B}_{\mathrm{D}}^{(i)}$ into two connected sets: a stopping set $\mathscr{S}_{\mathrm{D}}^{(i)}$ where $\mathbf{a}_{i,t}^{(\mathrm{D},2)} = \mathsf{S}$ is a best response and a continuation set $\mathscr{C}_{\mathrm{D}}^{(i)}$ where $\mathbf{a}_{i,t}^{(\mathrm{D},2)} = \mathsf{C}$ is a best response. □

## F Computing a Local Best Response for the Defender

We use a combination of dynamic programming and stochastic approximation to find a local best response for the defender based on Thm. 3.2.C. We first solve the MDP defined in Appendix E via the value iteration[19] algorithm (Eq. 6.21, Krishnamurthy, 2016), which can be done efficiently due to full observability. After solving the MDP, we approximate the optimal switching curves defined in Appendix E with the following linear approximation (Eq. 12.18, Krishnamurthy, 2016).

$$\pi_{\mathrm{D}}(\mathbf{b}^{(\mathrm{D})}) = \begin{cases} \mathsf{S} & \text{if } \begin{bmatrix} 0 & 1 & \boldsymbol{\theta} \end{bmatrix} \begin{bmatrix} \mathbf{b}^{(\mathrm{D})} \\ -1 \end{bmatrix} > 0 \\ \mathsf{C} & \text{otherwise} \end{cases} \tag{3.23}$$
$$\text{subject to } \boldsymbol{\theta} \in \mathbb{R}^2,\ \boldsymbol{\theta}_2 > 0,\ \text{and } \boldsymbol{\theta}_1 \geq 1,$$

where $\mathbf{b}^{(\mathrm{D})} \in \Delta(\{0,1,2\})$ is a 2-dimensional probability vector in the unit 2-simplex.

The coefficients $\boldsymbol{\theta}$ in (3.23) are estimated through the T-SPSA algorithm described in Paper 1.

[19]Value iteration is defined in the background chapter; see (8).

# Paper 4†

# INTRUSION TOLERANCE FOR NETWORKED SYSTEMS THROUGH TWO-LEVEL FEEDBACK CONTROL

Kim Hammar and Rolf Stadler

**Abstract**

We formulate intrusion tolerance for a system with service replicas as a two-level game: a local game models intrusion recovery and a global game models replication control. For both games, we prove the existence of equilibria and show that the best responses have a threshold structure, which enables efficient computation of strategies. The local and global control problems can be formulated as classical problems in operations research, namely, the machine replacement problem and the inventory replenishment problem. Based on this formulation, we design TOLERANCE, a novel control architecture for intrusion-tolerant systems. State-of-the-art intrusion-tolerant systems can be understood as instantiations of our architecture with heuristic control strategies. Our analysis shows the conditions under which such heuristics can be significantly improved through game-theoretic reasoning. This reasoning allows us to derive the optimal (equilibrium) control strategies and evaluate them against 10 types of network intrusions on a testbed. The results demonstrate that our game-theoretic control strategies can significantly improve service availability and reduce the operational cost of state-of-the-art intrusion-tolerant systems. In addition, our game strategies can ensure any chosen level of service availability and time-to-recovery, bridging the gap between theoretical and operational performance.

*A distributed system is one in which the failure of a computer you didn't even know existed can render your own computer unusable.*

— Leslie Lamport ***1987**, email correspondence.*

## 4.1 Introduction

As our reliance on online services grows, there is increasing demand for reliable systems that provide service without disruption. Traditionally, the main causes of disruption in networked systems have been hardware failure and power outages (Avižienis, 1976). While tolerance against these types of failures is important, a growing source of disruptions is network intrusion, impacting an estimated 75% of organizations each year (Netwrix, 2024).

Intrusions fundamentally differ from hardware failures as an attacker can behave arbitrarily, i.e., Byzantine, which leads to unanticipated failure behavior. Due to the high costs of such failures and the fact that it is impractical to prevent all intrusions, the ability to *tolerate* intrusions becomes necessary (Ganin et al., 2016). This ability is particularly important for safety-critical systems, e.g., aircrafts (Wensley et al., 1978), real-time control systems (Sims, 1997)(Lala and Harper, 1994), power systems (Babay et al., 2019), nuclear plants (Lala, 1986), control-planes for software defined networks (Sakic et al., 2018), SCADA systems (Nogueira et al., 2018), (Kirsch et al., 2014), and e-commerce applications (Soikkeli et al., 2023).

The common approach to building an intrusion-tolerant system is to replicate the system across a set of *nodes*, which allows *compromised* and *crashed* nodes to be substituted by *healthy* nodes. This approach to intrusion tolerance includes three main building blocks: (*i*) a protocol for service replication that tolerates a subset of compromised and crashed nodes; (*ii*) a replication strategy that adjusts the replication factor; and (*iii*) a recovery strategy that determines when to recover potentially compromised nodes (Deswarte et al., 1991). Replication protocols that satisfy the condition in (*i*) are called *Byzantine fault-tolerant* (BFT) and have been studied extensively; see survey (Distler, 2021). Few prior works have studied (*ii*) and (*iii*). Current intrusion-tolerant systems typically use a fixed replication factor (Babay et al., 2018) and rely on inefficient recovery strategies, such as periodic recovery (Castro and Liskov, 2002), heuristic rule-based recovery (Veríssimo et al., 2003), or manual recovery by system administrators (Reiter, 1995).

In this paper, we address the above limitations and present a game-theoretic model that allows us to characterize optimal recovery and replication strategies for intrusion-tolerant systems. We formulate intrusion tolerance for a system with service replicas as a game with two levels: local and global. The local game involves *node controllers* that independently perform intrusion recovery, and the global game involves a *system controller* that manages the replication factor; see Fig. 4.1. Both games are modeled as stochastic zero-sum games and incorporate safety constraints. We prove the existence of constrained perfect Bayesian and Markov equilibria in the local and global games, respectively. We also derive a threshold structure of the best responses, which enables efficient computation of strategies. Based on these insights, we design a control architecture for intrusion-tolerant systems, which we call TOLERANCE: **T**w**o**-**le**vel **r**ecovery **an**d replication **c**ontrol with f**e**edback. To assess the performance of TOLERANCE, we implement it in an emulation environment where we run 10 types of network intrusions. The results show that TOLERANCE can achieve higher service availability and lower operational cost than state-of-the-art intrusion-tolerant systems.

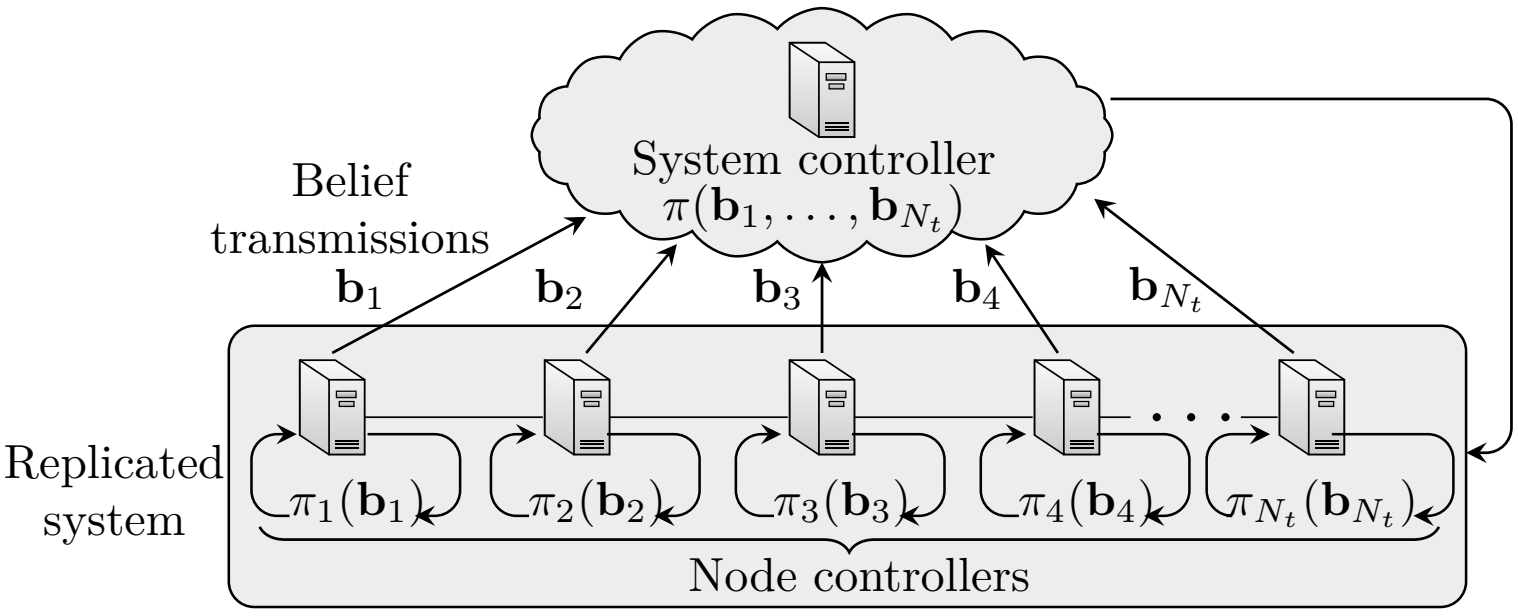

***Figure 4.1:*** *Two-level feedback control for intrusion tolerance; node controllers with strategies $\pi_1, \ldots, \pi_{N_t}$ compute belief states $\mathbf{b}_1, \ldots, \mathbf{b}_{N_t}$ and make local recovery decisions; a global system controller with strategy $\pi$ receives belief states and manages the replication factor $N_t$.*

In the context of this thesis, this paper demonstrates the generality of our methodology[2] by applying it to a different type of response use case than those studied in papers 1–3, namely intrusion tolerance. Our contributions can be summarized as follows:

1. We present a novel formulation of intrusion tolerance as a two-level game. The local game models intrusion recovery, and the global game models replication control. Leveraging this model, we derive optimal (equilibrium) control strategies against a dynamic attacker, for which we provide theoretical guarantees.

2. We prove the existence of equilibria and that the best responses have a threshold structure. Based on these insights, we design efficient algorithms for computing the best responses.

3. We present and evaluate TOLERANCE, a novel control architecture for intrusion-tolerant systems that uses two levels of control to decide when to perform recovery and when to increase the replication factor.

4. We implement TOLERANCE in an emulation environment and evaluate its performance against 10 types of network intrusions. The results show that TOLERANCE can improve service availability and reduce the operational cost of state-of-the-art intrusion-tolerant systems.

[2]See the methodology chapter for details about our experimental methodology.

## 4.2 The Intrusion Tolerance Use Case

We consider a set of nodes collectively offering a service to a client population; see Fig. 4.2. Each node is segmented into two domains: an application domain, which runs a service replica, and a privileged domain, which runs security and control functions. The replicas are coordinated through a replication protocol that relies on digital signatures and guarantees correct service if no more than $f$ nodes are compromised or crashed simultaneously.

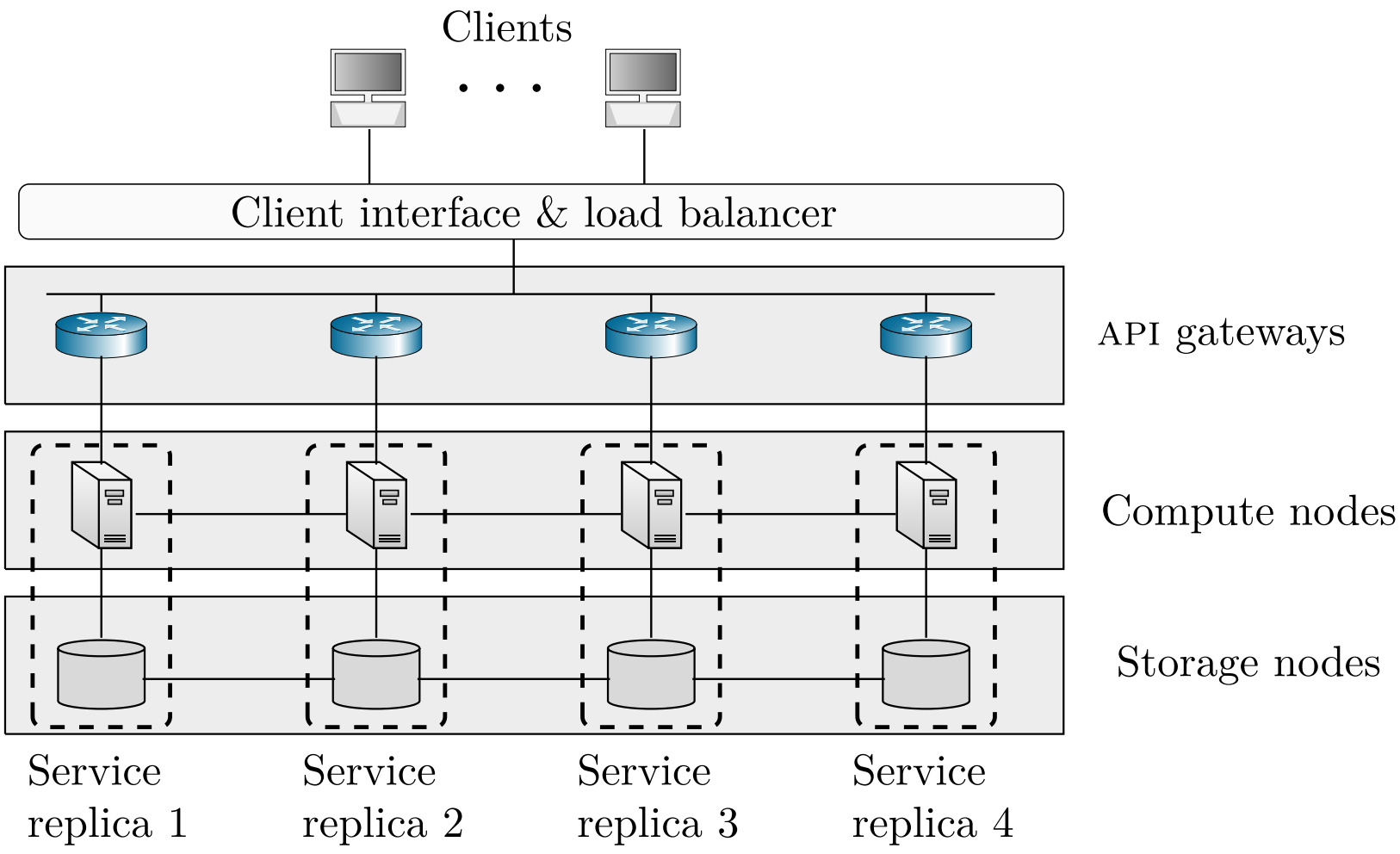

***Figure 4.2:*** *The intrusion tolerance use case: a replicated system offers a service to a client population; the system should maintain correct service to the clients even if an attacker compromises a subset of replicas.*

Clients access the service through gateways, which also are accessible to an attacker. The attacker aims to intrude on the system and compromise replicas while avoiding detection. We assume that the attacker a) does not have physical access to nodes; b) cannot forge digital signatures; and c) can only access the service replicas, not the privileged domains (i.e., we consider the *hybrid failure model* (Correia et al., 2007)). Apart from these restrictions, the attacker can control a compromised replica in arbitrary, i.e., Byzantine, ways. It can shut it down, delay service responses, communicate with other replicas, etc.

To prevent the number of compromised and crashed nodes from exceeding $f$, we consider three types of response actions: (*i*) recover compromised nodes; (*ii*) evict crashed nodes from the system; and (*iii*) add new nodes. Each action incurs a cost that must be weighed against the security benefit.

## 4.3 Background on Intrusion-Tolerant Systems

Research on fault-tolerant systems has almost a century-long history, with the seminal work made by von Neumann [478] and Shannon [311] in 1956. The early work focused on tolerance against hardware failures. Since then the field has broadened to include tolerance against software bugs, operator mistakes, and malicious attacks [30, 31, 32, 103, 266]. The common approach to build a fault-tolerant service is *redundancy*, whereby the service is provided by a set of replicas. Through such redundancy, compromised and crashed replicas can be substituted by healthy replicas as long as they can coordinate their service responses. This coordination problem is known as the *consensus* problem.

### Consensus

Consensus is the problem of reaching agreement among distributed nodes subject to failures (Cachin et al., 2011). This problem can be solved under synchrony and failure assumptions. The main synchrony options are (*i*) the *synchronous model*, which mandates an upper bound on the communication delay between nodes; (*ii*) the *partially synchronous model*, which warrants an upper bound but allows for periods of instability where the bound is violated; and (*iii*) the *asynchronous model*, where no bound exists. Similarly, the main failure options are (*i*) the *crash-stop failure model*, where nodes fail by crashing; (*ii*) the *Byzantine failure model*, where nodes fail arbitrarily; and (*iii*) the *hybrid failure model*, where nodes fail arbitrarily but are equipped with trusted components that fail by crashing (Dwork et al., 1988), (Dolev et al., 1987).

**Theorem 4.1** (Solvability of the consensus problem)**.**

1. *Consensus is not solvable in the asynchronous model.*
2. *Consensus is solvable in the partially synchronous model with $N$ nodes and at most $\frac{N-1}{2}$ crash-stop failures, $\frac{N-1}{3}$ Byzantine failures, and $\frac{N-1}{2}$ hybrid failures.*
3. *Consensus is solvable in the synchronous model with $N$ nodes and at most $N-1$ crash-stop failures, $\frac{N-1}{2}$ Byzantine failures, and $\frac{N-1}{2}$ hybrid failures.*

Theorem 4.1 summarizes several decades of research; hence, the proofs are scattered across the literature. The proof of Theorem 4.1.1 is available in (Thm. 1, Fischer et al., 1985). The proof of Theorem 4.1.2 is available in (Thm. 1, Bracha and Toueg, 1985), (Thms. 5.8, 5.11, Attiya and Welch, 2004), (Thm. 2, Santos Veronese, 2010), and (Thm. 1, Yandamuri et al., 2021). The proof of Theorem 4.1.3 is available in (Thm. 5.2, Attiya and Welch, 2004), (Cor. 14, Katz and Koo, 2006), and (Thm. 1, Abraham et al., 2019). Since several of these proofs span multiple pages, we omit them here for brevity. The main proof technique is to model the system using I/O automata (Lynch, 1996).

**Remark 4.1** (FLP)**.** While Thm. 4.1.1 states that deterministic consensus cannot be guaranteed in the asynchronous model (Thm. 1, Fischer et al., 1985), it does not rule out probabilistic consensus systems. In such systems, consensus can be reached with probability 1, but it is still possible (though with probability measure 0) that some execution of the system does not reach consensus.

### Intrusion-tolerant systems

Intrusion-tolerant systems extend fault-tolerant systems with intrusion detection, recovery, and response (Goseva-Popstojanova et al., 2001). We call a system *intrusion-tolerant* if it remains secure and operational while intrusions occur (Deswarte et al., 1991). Theorem 4.1 provides the basis for designing an intrusion-tolerant system and indicates the number of nodes required to tolerate $f$ compromised nodes. However, the theorem does not provide guidance on the likelihood that the threshold $f$ will be exceeded. Quantifying this likelihood is the objective of *reliability theory.*

### Reliability theory

The reliability of a system is defined as the probability that the system performs its task under the operating conditions encountered (Barlow and Proschan, 1965). If $T^{(\mathrm{F})}$ is a random variable representing the time to failure (e.g., compromise), then the reliability function can be defined as $R(t) \triangleq \mathbb{P}[T^{(\mathrm{F})} > t]$ and the mean time to failure (MTTF) is $\mathbb{E}[T^{(\mathrm{F})}]$. In the context of intrusion tolerance, we also consider the metrics *average time-to-recovery* $T^{(\mathrm{R})}$, *average availability* $T^{(\mathrm{A})}$, and *frequency of recovery* $F^{(\mathrm{R})}$. When measuring service availability for a replicated system, we assume the *primary-partition* model (Birman, 1997) to circumvent the CAP theorem (Thm. 2, Gilbert and Lynch, 2002). Specifically, in the case of a network partition, only one partition is permitted to remain operational (to maintain consistency) and we consider the system to be available as long as one partition is operational.

## 4.4 The TOLERANCE Control Architecture

In this section, we describe TOLERANCE: a two-level control architecture for intrusion-tolerant systems; see Fig. 4.3 on the next page. It is a distributed system with $N_t \geq 2f+1+k$ *nodes* connected through an authenticated network (Lamport et al., 2010), where $k$ is the number of nodes that can be in recovery simultaneously. Each node runs a *service replica.* The replicas are coordinated through a *reconfigurable* consensus protocol that guarantees *correct service* if no more than $f$ nodes are compromised or crashed simultaneously; e.g., reconfigurable MINBFT (§4.2, Santos Veronese, 2010)[3]. TOLERANCE uses two levels of control: local and global. On the local level, each node runs a *node controller* that monitors the service replica through alerts from an Intrusion Detection System (IDS). Based on these alerts, the controller estimates the replica's state, i.e., whether it is compromised or not, and decides when it should be recovered. As each recovery incurs a cost, the challenge for the controller is to balance the recovery costs against the security benefits. To guarantee correct service, at most $k$ nodes can recover simultaneously.

[3] A reconfigurable consensus protocol allows dynamic addition and removal of nodes.

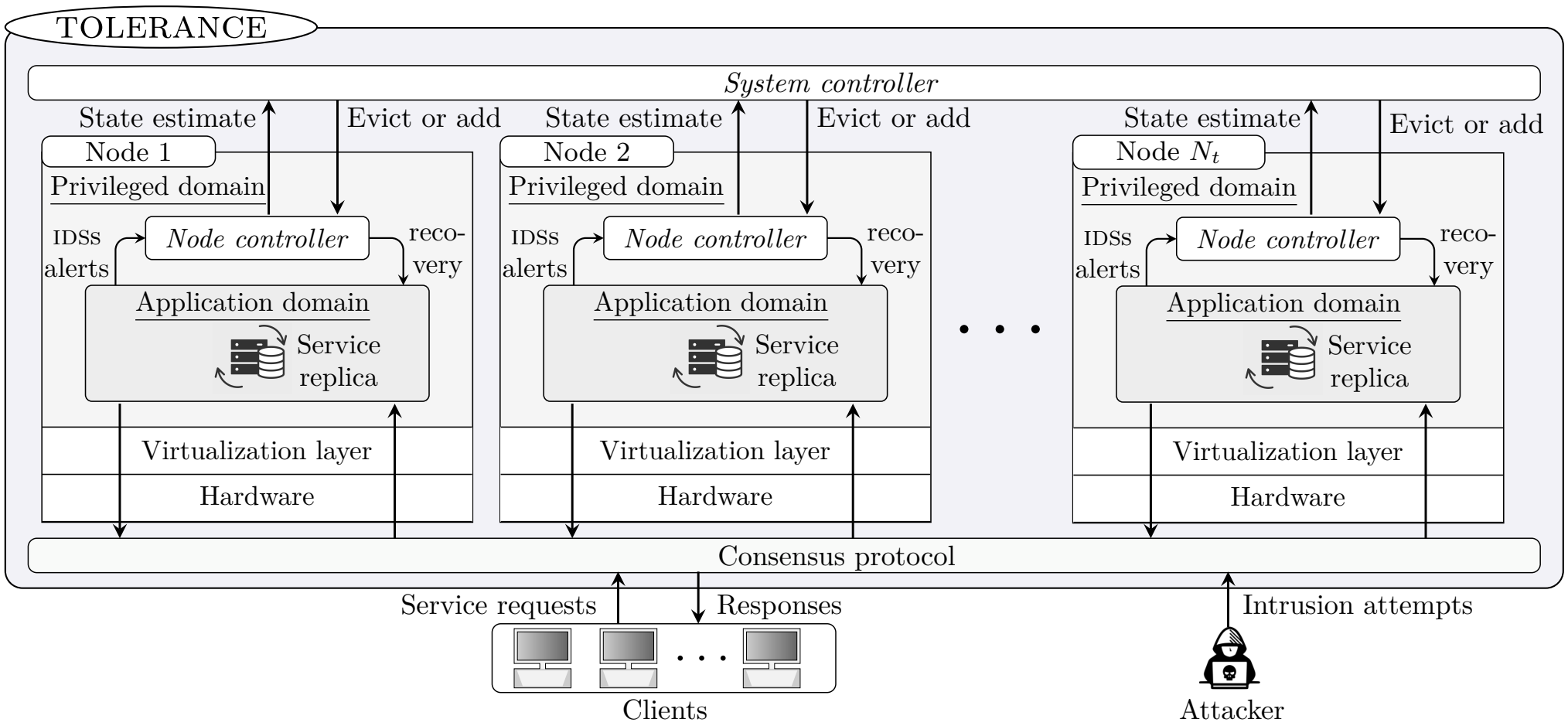

***Figure 4.3:*** *The TOLERANCE architecture: **T**w**o**-**le**vel **r**ecovery **an**d replication **c**ontrol with f**e**edback; $N_t$ nodes provide a replicated service to clients; service responses are coordinated through an intrusion-tolerant consensus protocol; local node controllers decide when to recover and a global system controller manages the replication factor $N_t$.*

The global level includes a *system controller* that collects state estimates from the nodes and adjusts the replication factor $N_t$. When deciding if $N_t$ should be increased, the controller faces a classical dilemma in reliability theory (Barlow and Proschan, 1965). On the one hand, it aims to achieve high redundancy to maximize service availability. On the other hand, it does not want an excessively large and costly system. Since the only task of the system controller is to execute control actions and communicate with the node controllers, it can be deployed on a standard crash-tolerant system, e.g., a RAFT-based system (Ongaro and Ousterhout, 2014). For this reason, we consider the probability that the system controller crashes negligible. (See Assumption 4.1 on page 161 for details.)

Similar to the VM-FIT and the WORM-IT architectures (Distler et al., 2011), (Reiser and Kapitza, 2007), (Correia et al., 2007), each node in TOLERANCE is segmented into two domains: a *privileged domain*, which can only fail by crashing, and an *application domain*, which may be compromised by an attacker. The controllers and the IDSs execute in the privileged domain, whereas the service replicas execute in the application domain. The separation between the two domains can be realized in several ways. One option is to use a secure coprocessor to execute the privileged domain (e.g., IBM 4758) (Castro and Liskov, 2002)(Yandamuri et al., 2021). Another option is implementing the privileged domain using dedicated hardware modules, such as a smart card or an FPGA (Distler, 2021). A third op-tion, which does not require special hardware, is to use a security kernel to run the privileged domain, as in the WORM-IT architecture (Correia et al., 2007). A fourth option, used in the VM-FIT architecture (Distler et al., 2011), is to separate the application domain from the privileged domain using a secure virtualization layer that can be formally verified (Dam et al., 2013). TOLERANCE implements the last option for the following reasons: (*i*) virtualization enables efficient recovery of a compromised replica by replacing its virtual container (Reiser and Kapitza, 2007); and (*ii*) virtualization simplifies implementation of software diversification, which reduces the correlation between compromise events across nodes (Garcia et al., 2011).

### Correctness

**Definition 4.1** (Correct service)**.** *TOLERANCE provides correct service if the healthy replicas satisfy the following properties:*

*Each request is eventually executed.* (Liveness)

*Each executed request was sent by a client.* (Validity)

*Each replica executes the same request sequence.* (Safety)

**Proposition 4.1** (Correctness)**.** *TOLERANCE provides correct service if*

*An attacker cannot forge digital signatures.* (P1.A)

*An attacker cannot access the privileged domains.* (P1.B)

*Network links are authenticated and reliable [84, p. 42]*[4]. (P1.C)

*At most $k$ nodes recover simultaneously.* (P1.D)

*At most $f$ nodes are compromised or crashed simultaneously.* (P1.E)

$N_t \geq 2f + 1 + k$. (P1.F)

*The system is partially synchronous [84, §2.5.3].* (P1.G)

*Proof.* (P1.A)–(P1.C) imply the hybrid failure model. (P1.D)-(P1.F) state that at least $f+1+k$ nodes are healthy. These properties together with the tolerance threshold $f = \frac{N_t-1-k}{2}$ of the consensus protocol (e.g., MINBFT (Santos Veronese, 2010), §4.2) imply (Safety) (Thm. 4.1, (Thms. 1–2, Santos Veronese, 2010)). Next, it follows from (P1.G) that the healthy nodes will eventually agree on the response to any service request, which allows to circumvent FLP (Thm. 1, Fischer et al., 1985) and achieve (Liveness). Finally, (Validity) is ensured by the consensus protocol (e.g., MINBFT (§4.2, Santos Veronese, 2010)). □

[4] In an authenticated network, nodes can verify each other's digital signatures. In a reliable network, the following properties are satisfied: a) RELIABLE DELIVERY: if a healthy node $p$ sends a message $m$ to a healthy node $q$, then $q$ eventually delivers $m$ (i.e., network partitions eventually heal); b) NO DUPLICATION: no message is delivered by a node more than once; c) NO CREATION: if some node $q$ delivers a message $m$ with sender $p$, then $m$ was previously sent to $q$ by node $p$ [84, p. 42].

Assumptions (P1.A), (P1.C), (P1.D), (P1.G) imply that the system uses standard cryptographic mechanisms and network equipment. Similarly, (P1.E)–(P1.F) can always be met by tuning $f$ and $N_t$. The strongest assumption is (P1.B), which implies that the controllers are securely separated from the service replicas. As explained on page 159, this separation can be realized in several ways.

Proposition 4.1 implies that, to guarantee correct service (Def. 4.1), the controllers must ensure (in expectation) that: a) the number of compromised and crashed nodes is at most $f$, which is achieved by recovery; and b) the number of nodes satisfies $N_t \geq 2f + 1 + k$, which is achieved by replacing crashed nodes. In the following section, we model the problem of meeting these two constraints while minimizing operational cost as a game with a local and a global level. On the local level, node controllers minimize cost while meeting a), and on the global level, the system controller minimizes cost while meeting b). At the same time, an attacker aims to maximize the cost of the system.

**Remark 4.2** (Extension of Prop. 4.1). TOLERANCE can be extended in two ways to provide confidentiality in addition to (Safety), (Liveness), and (Validity). One approach is forcing all replicas to send messages through a firewall that filters faulty messages containing confidential data (Yin et al., 2002). Another option is cryptographic techniques such as verifiable secret sharing (Padilha and Pedone, 2011).

## 4.5 Modeling Intrusion Tolerance as a Two-Level Game

Our game-theoretic model is based on the following assumptions.

**Assumption 4.1.** *The probability that the system controller crashes is negligible.*

**Assumption 4.2.** *Compromise and crash events are statistically independent across nodes.*

**Assumption 4.3.** *(P1.D) is enforced by the system implementation.*

**Assumption 4.4.** *The attacker has access to the controllers' observations.*

Since the system controller is not accessible by an attacker, Assumption 4.1 can be satisfied by deploying the system controller on a crash-tolerant system, e.g., a RAFT-based system. In such a system, we can make the failure probability negligible by using a large number of replicas, ensuring that the likelihood of a majority crashing at the same time is vanishingly small. Assumption 4.2 can be satisfied by a) distributing the nodes geographically, which reduces the likelihood of simultaneous power outages; and b) employing software diversification, which reduces the likelihood that different nodes have the same vulnerabilities (Garcia et al., 2011). Assumption 4.3 can be met through proper implementation design (Sousa et al., 2007). Lastly, Assumption 4.4 holds for insider attacks and reflects that it is generally not known what information is available to the attacker.

Our model is presented in the following; notation is listed in Table 4.1.

| *Notation(s)* | *Description* |
|---|---|
| $\mathcal{N}_t, N_t, f$ | Set of nodes, number of nodes, tolerance threshold (Prop. 4.1). |
| $T^{(\mathrm{R})}$ | Average time-to-recovery when an intrusion occurs. |
| $F^{(\mathrm{R})}, T^{(\mathrm{A})}$ | Frequency of recovery, average service availability (4.4). |
| $T^{(\mathrm{F})}, k$ | Time to system failure (Fig. 4.10), # parallel recoveries (Prop. 1). |
| $R(t), R_i(t)$ | Reliability function for the system and for a node (Fig. 4.10). |
| $J_i, J$ | Objectives of node controller $i$ (4.4) and the system controller (4.7). |
| $\pi^{(\mathrm{C})}_{i,t}, \pi^{(\mathrm{A})}_{i,t}$ | Control and attack strategy for node $i$ (4.5). |
| $\tilde{\pi}^{(\mathrm{C})}_{i,t}, \tilde{\pi}^{(\mathrm{A})}_{i,t}$ | Best response control and attack strategy for node $i$ (4.5). |
| $\pi^{(\mathrm{C}),\star}_{i,t}, \alpha^\star_i$ | Equilibrium recovery strategy and best response threshold for node $i$ (4.5). |
| $\boldsymbol{\pi}^\star_{i,t}, \pi^{(\mathrm{A}),\star}_{i,t}$ | Equilibrium strategy profile of Game 4.1 for node $i$, attacker equilibrium strategy. |
| $\mathbf{i}^{(\mathrm{C})}_{i,t}, \mathbf{i}^{(\mathrm{A})}_{i,t}$ | Information feedback for the controller and the attacker at time $t$ in Game 4.1 (4.2). |
| $\mathbf{I}^{(\mathrm{C})}_{i,t}, \mathbf{I}^{(\mathrm{A})}_{i,t}$ | Random vectors with realizations $\mathbf{i}^{(\mathrm{C})}_{i,t}, \mathbf{i}^{(\mathrm{A})}_{i,t}$ (4.2). |
| $s_{i,t}, o_{i,t}$ | State (4.1) and observation (4.2) of node $i$ at time $t$. |
| $\mathbf{b}_{i,t}$ | Belief state of node $i$ at time $t$ (4.3). |
| $S_{i,t}, O_{i,t}, \mathbf{B}_{i,t}$ | Random variables with realizations $s_{i,t}$ (4.1), $o_{i,t}$ (4.2), $\mathbf{b}_{i,t}$ (4.3). |
| $a^{(\mathrm{C})}_{i,t}, c_{i,t}$ | Action and cost of the node controller $i$ at time $t$ (4.5). |
| $A^{(\mathrm{C})}_{i,t}, C_{i,t}$ | Random variables with realizations $a^{(\mathrm{C})}_{i,t}$ and $c_{i,t}$ (4.5). |
| $a^{(\mathrm{A})}_{i,t}, A^{(\mathrm{A})}_{i,t}$ | Action of the attacker on node $i$ at time $t$ (4.5). |
| $\mathcal{S}_{\mathrm{N}}, \mathcal{O}_{\mathrm{N}}$ | State and observation spaces of nodes. |
| $\mathsf{W}, \mathsf{R} = 0, 1$ | The (W)ait and (R)ecovery actions (Fig. 4.6.a). |
| $\mathbb{H}, \mathbb{C} = 0, 1$ | The ($\mathbb{H}$)ealthy and ($\mathbb{C}$)ompromised node states (Fig. 4.6.a). |
| $\emptyset$ | The crashed node state (Fig. 4.6.a). |
| $f_{\mathrm{N},i}, z_i$ | Transition (4.1) and observation (4.2) functions for node $i$. |
| $c_{\mathrm{N}}(s_{i,t}, a_{i,t})$ | Cost function for a node (4.4). |
| $p_{\mathrm{A},i}$ | Probability that an attack against node $i$ is successful (4.1). |
| $p_{\mathrm{C},i}$ | Probability that node $i$ crashes in the healthy state (4.1). |
| $\Delta_{\mathrm{R}}$ | Maximum allowed time between node recoveries (4.5). |
| $S_t, s_t$ | State of the system controller at time $t$, $s_t$ realizes $S_t$. |
| $a^{(\mathrm{C})}_t, c_t$ | Action and cost of the system controller at time $t$. |
| $A^{(\mathrm{C})}_t, C_t$ | Random variables with realizations $a^{(\mathrm{C})}_t, c_t$ (4.8). |
| $\mathbf{a}^{(\mathrm{A})}_t, \mathbf{A}^{(\mathrm{A})}_t$ | Action of the attacker in Game 4.2 (4.8). |
| $f_{\mathrm{S}}$ | Transition function of Game 4.2. |
| $\mathcal{S}_{\mathrm{S}}$ | State space of Game 4.2. |
| $s_{\max}$ | Maximum number of nodes in Game 4.2. |
| $\pi^{(\mathrm{C})}, \pi^{(\mathrm{A})}$ | Control and attack strategy in Game 4.2 (4.8). |
| $\tilde{\pi}^{(\mathrm{C})}, \tilde{\pi}^{(\mathrm{A})}$ | Best response control and attack strategy in Game 4.2 (4.8). |
| $\pi^{(\mathrm{C}),\star}, \pi^{(\mathrm{A}),\star}$ | Equilibrium control and attack strategies in Game 4.2 (4.8). |
| $\boldsymbol{\pi}, \boldsymbol{\pi}^\star$ | Strategy profile and equilibrium strategy profile in Game 4.2 (4.8). |
| $\epsilon_{\mathrm{A}}$ | Lower bound on the average service availability (4.8). |

***Table 4.1:*** *Variables and symbols used in the model.*

### The local intrusion recovery game

The local game involves two players: a node controller that aims to minimize operational cost by performing intrusion recovery and an attacker that aims to maximize that cost. The attacker can perform two actions to achieve its goal:

($i$) compromise the node's service replica; and ($ii$) trigger excess recoveries by the deliberate generation of false intrusion alarms.

Let $\mathcal{N}_t \triangleq \{1, 2, \ldots, N_t\}$ be the set of nodes and $\pi^{(\mathrm{C})}_{i,t}$ the corresponding *behavior control strategy* at time $t$ (Def. 5, Kuhn, 1953). Controller $i$ takes one of two actions $a^{(\mathrm{C})}_{i,t}$: (R)ecover or (W)ait. Similarly, the attacker follows a *behavior strategy* $\pi^{(\mathrm{A})}_{i,t}$ and takes one of two actions $a^{(\mathrm{A})}_{i,t}$: (A)ttack or (F)alse alarm.

Node $i$ has state $s_{i,t} \in \mathcal{S}_{\mathrm{N}}$ with three values: $\emptyset$ if it is crashed, $\mathbb{C}$ if it is compromised, and $\mathbb{H}$ if it is healthy; see Fig. 4.4. The evolution of $s_{i,t}$ can be written as $s_{i,t+1} \sim f_{\mathrm{N},i}(\cdot \mid s_{i,t}, a^{(\mathrm{C})}_{i,t}, a^{(\mathrm{A})}_{i,t})$, where $f_{\mathrm{N},i}$ is defined as

$$f_{\mathrm{N},i}(\emptyset \mid \emptyset, \cdot, \cdot) \triangleq 1 \tag{4.1a}$$

$$f_{\mathrm{N},i}(\emptyset \mid \mathbb{H}, \cdot, \cdot) \triangleq f_{\mathrm{N},i}(\emptyset \mid \mathbb{C}, \cdot, \cdot) \triangleq p_{\mathrm{C},i} \tag{4.1b}$$

$$f_{\mathrm{N},i}(\mathbb{H} \mid \mathbb{H}, \mathsf{W}, \mathsf{A}) \triangleq (1 - p_{\mathrm{A},i})(1 - p_{\mathrm{C},i}) \tag{4.1c}$$

$$f_{\mathrm{N},i}(\mathbb{H} \mid \mathbb{H}, \cdot, \mathsf{F}) \triangleq f_{\mathrm{N},i}(\mathbb{H} \mid \mathbb{H} \text{ or } \mathbb{C}, \mathsf{R}, \cdot) \triangleq f_{\mathrm{N},i}(\mathbb{C} \mid \mathbb{C}, \mathsf{W}, \cdot) \triangleq (1 - p_{\mathrm{C},i}) \tag{4.1d}$$

$$f_{\mathrm{N},i}(\mathbb{C} \mid \mathbb{H}, \mathsf{W}, \mathsf{A}) \triangleq (1 - p_{\mathrm{C},i})p_{\mathrm{A},i}. \tag{4.1e}$$

$p_{\mathrm{A},i} \in (0,1)$ is the probability that an attack on node $i$ is successful and $p_{\mathrm{C},i} \in (0,1)$ is the probability that the node crashes during the time interval $[t, t+1]$. These parameters can be set based on domain knowledge or be obtained through system measurements. Companies such as Google, Meta, and IBM have documented procedures for estimating such parameters, e.g., (Ford et al., 2010), (Meza et al., 2018).

(4.1a)–(4.1b) capture the transitions to the crashed state $\emptyset$, which is absorbing[5]. Next, (4.1c)–(4.1d) define the transitions to the healthy state $\mathbb{H}$ after the controller takes action R. Lastly, (4.1e) captures the transition to the compromised state $\mathbb{C}$ when an intrusion occurs. All other transitions have probability 0. It follows from (4.1) that the number of time steps until a node fails (crash or compromise) is geometrically distributed; see Fig. 4.4.

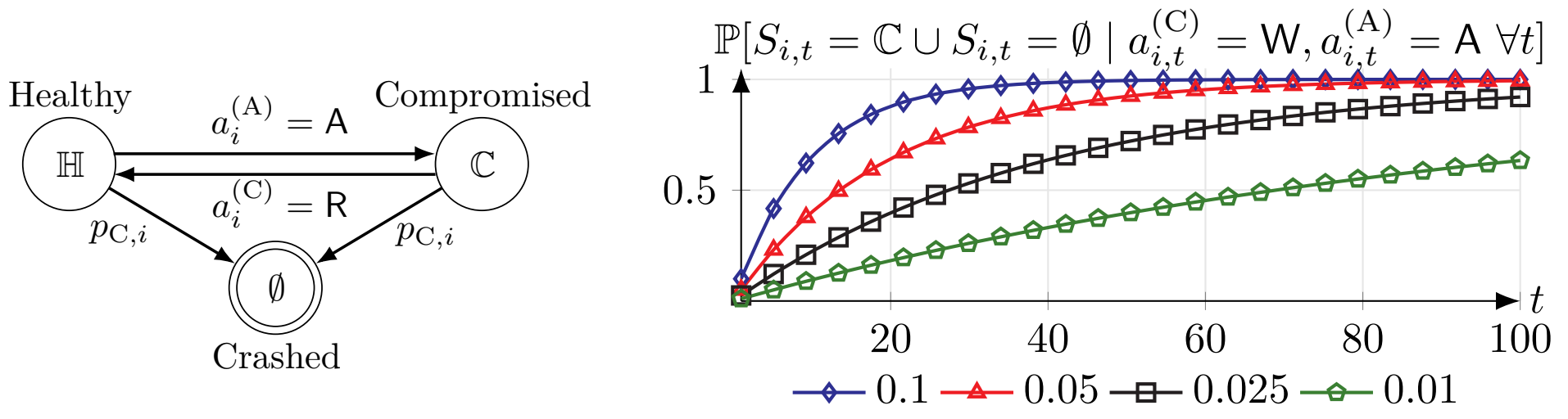

***(a)** State transition diagram of node $i$ (4.1).* ***(b)** Failure (crash or compromise) probability.*

***Figure 4.4:** a) disks represent states, arrows represent transitions, labels indicate conditions for transition, self-transitions are not shown; b) the probability that a node is compromised ($\mathbb{C}$) or crashed ($\emptyset$) if no recoveries occur; curves relate to* $\min[p_{\mathrm{A},i} + p_{\mathrm{C},i}, 1]$.

[5] A crashed node can be restarted and appears as a new node in our model.

***Observability*** The attacker has complete observability in the sense that it knows the state $s_{i,t}$, the controller's action $a^{(\mathrm{C})}_{i,t}$, and the controller's observation. In contrast, the controller has a restricted view. It only has access to an observation $o_{i,t} \in \mathcal{O}$, which is based on the number of IDS alerts received during the time interval $[t-1,t]$ ($\mathcal{O}$ is finite). Consequently, the *information feedback* for the controller and the attacker at time $t$ are

$$\mathbf{i}^{(\mathrm{C})}_{i,t} \triangleq (o_{i,t}) \quad \text{and} \quad \mathbf{i}^{(\mathrm{A})}_{i,t} \triangleq (s_{i,t}, a^{(\mathrm{C})}_{i,t-1}, o_{i,t}), \quad \text{where } o_{i,t} \sim z_i(\cdot \mid a^{(\mathrm{A})}_{i,t-1}). \tag{4.2}$$

**Remark 4.3** (General observation spaces)**.** While we focus on the IDS alert metric in this paper, alternative sources of metrics can be used. A comparison between different metrics is available in (Paper 2, Appendix C).

**Remark 4.4** (Modeling clients)**.** The clients are implicitly modeled by $z_i$ (4.2).

Both the controller and the attacker have *perfect recall*, which means that they remember their respective history $\mathbf{h}^{(\mathrm{j})}_{i,t} \triangleq (\mathbf{b}_{i,1}, (a^{(\mathrm{j})}_{i,l-1}, \mathbf{i}^{(\mathrm{j})}_{i,l})_{l=2,\ldots,t})$, where $\mathrm{j} \in \{\mathrm{C}, \mathrm{A}\}$ (Def. 7, Kuhn, 1953)[6]. Based on this history, the controller uses the *belief operator*[7] $\mathbb{B}$ (22) to compute the *belief state*

$$\mathbf{b}_{i,t}(s_i) \triangleq \mathbb{P}[S_{i,t} = s_i \mid \mathbf{h}^{(\mathrm{C})}_{i,t}], \tag{4.3}$$

as defined in the background chapter.

Since $\mathbf{b}_{i,t}(\mathbb{C})$ (4.3) is a sufficient statistic for $s_{i,t}$ (Def. 4.2, Lem. 5.1, Thm. 7.1, Kumar and Varaiya, 1986), we can define $\pi^{(\mathrm{C})}_{i,t}$ as a function $[0,1] \rightarrow \Delta(\{\mathsf{W}, \mathsf{R}\})$. Similarly, since the attacker has complete observability, it can also compute $\mathbf{b}_{i,t}(\mathbb{C})$, and hence we can define $\pi^{(\mathrm{A})}_{i,t}$ as a function $\mathcal{S}_{\mathrm{N}} \times [0,1] \rightarrow \Delta(\{\mathsf{A}, \mathsf{F}\})$. (Strategies can be time-dependent, as the subscript $t$ indicates.)

**Proposition 4.2.** *Let $X_{i,t}$ represent the number of recoveries of node $i$ that occurred by time $t$ and define $T_{i,\emptyset} \triangleq \inf_t[s_{i,t} = \emptyset]$. If (i) $(\pi^{(\mathrm{C})}_i, \pi^{(\mathrm{A})}_i)$ are stationary; and (ii) $\pi^{(\mathrm{C})}_i(\mathsf{R} \mid \mathbb{B}(\mathbf{h}^{(\mathrm{C})}_{i,t})(\mathbb{C})) > 0$ for some $\mathbf{h}^{(\mathrm{C})}_{i,t}$ where $\mathbb{P}[\mathbf{h}^{(\mathrm{C})}_{i,t} \mid (\pi^{(\mathrm{C})}_i, \pi^{(\mathrm{A})}_i)] > 0$ and $\mathbb{B}$ is the belief operator (22), then $(X_{i,t})_{t=1}^{T_{i,\emptyset}}$ is a renewal process.*

*Proof.* To establish that $(X_{i,t})_{t=1}^{T_{i,\emptyset}}$ is a renewal process, we need to show that a) the times between recoveries are independent and identically distributed (i.i.d.); and b) $(X_{i,t})_{t=1}^{T_{i,\emptyset}}$ are not all zero with probability 1 (Ch. 3.2, Barlow and Proschan, 1965). a) follows from (*i*) and the Markov properties of $(\pi^{(\mathrm{C})}_i, \pi^{(\mathrm{A})}_i)$ and $f_{\mathrm{N},i}$ (4.1). b) follows from (*ii*). □

[6] See Assumption 4 in the background chapter.

[7] The belief operator (22) defined in the background chapter assumes an observation function of the form $z_i(\cdot \mid s)$ whereas this paper assumes an observation function of the form $z_i(\cdot \mid a^{(\mathrm{A})}_{i,t-1})$; however, this change does not affect the derivation.

***Controller objective*** When selecting the strategy $\pi^{(\mathrm{C})}_{i,t}$, the controller balances two conflicting goals: minimize the average time-to-recovery $T^{(\mathrm{R})}_i$ and minimize the frequency of recovery $F^{(\mathrm{R})}_i$. The weight $\eta > 1$ controls the trade-off between these two objectives, which leads to the cost

$$J_i \triangleq T_{i,\emptyset}(\eta \underbrace{T^{(\mathrm{R})}_i}_{\text{Time-to-recovery.}} + \underbrace{F^{(\mathrm{R})}_i}_{\text{Frequency of recovery.}}) = \sum_{t=1}^{T_{i,\emptyset}} \eta s_{i,t}(1-a^{(\mathrm{C})}_{i,t}) + a^{(\mathrm{C})}_{i,t} = \sum_{t=1}^{T_{i,\emptyset}} c_{\mathrm{N}}(s_{i,t}, a^{(\mathrm{C})}_{i,t}), \quad (4.4)$$

where $T_{i,\emptyset} \triangleq \inf_t[s_{i,t} = \emptyset]$, $c_{\mathrm{N}}$ is the cost function, and $(\mathbb{H}, \mathbb{C}, \mathsf{W}, \mathsf{R}) \triangleq (0,1,0,1)$.

The objective in (4.4) corresponds to the cumulative cost optimality criterion. The following lemma establishes a relationship between (4.4) and the discounted optimality criterion. It is key for our subsequent analysis.

**Lemma 4.1** (Connection between discounted and cumulative optimality)**.**

$$\mathbb{E}_{\mathbf{H}^{(\mathrm{A})}_{i,T_{i,\emptyset}}, T_{i,\emptyset}}[J_i] = \mathbb{E}_{\mathbf{H}^{(\mathrm{A})}_{i,T_{i,\emptyset}}}\left[\sum_{t=1}^{\infty} \gamma^{t-1} c_{\mathrm{N}}(S_{i,t}, A^{(\mathrm{C})}_{i,t})\right], \quad \textit{where } \gamma \triangleq (1-p_{\mathrm{C},i}).$$

*Proof.* For ease of notation, let $C_t \triangleq c_{\mathrm{N}}(S_{i,t}, A^{(\mathrm{C})}_{i,t})$. Then

$$\begin{aligned}
\mathbb{E}_{\mathbf{H}^{(\mathrm{A})}_{i,T_{i,\emptyset}}, T_{i,\emptyset}}[J_i] &= \mathbb{E}_{\mathbf{H}^{(\mathrm{A})}_{i,T_{i,\emptyset}}, T_{i,\emptyset}}\left[\sum_{t=1}^{T_{i,\emptyset}} C_t\right] = \mathbb{E}_{\mathbf{H}^{(\mathrm{A})}_{i,T_{i,\emptyset}}}\left[\sum_{T_{i,\emptyset}=1}^{\infty}\sum_{t=1}^{T_{i,\emptyset}} \mathbb{P}[T_{i,\emptyset}]C_t\right] \\
&\overset{(a)}{=} \mathbb{E}_{\mathbf{H}^{(\mathrm{A})}_{i,T_{i,\emptyset}}}\left[\sum_{t=1}^{\infty}\sum_{T_{i,\emptyset}=t}^{\infty} \mathbb{P}[T_{i,\emptyset}]C_t\right] \overset{(b)}{=} \mathbb{E}_{\mathbf{H}^{(\mathrm{A})}_{i,T_{i,\emptyset}}}\left[\sum_{t=1}^{\infty}\sum_{T_{i,\emptyset}=t}^{\infty} p_{\mathrm{C},i}(1-p_{\mathrm{C},i})^{T_{i,\emptyset}-1}C_t\right] \\
&= \mathbb{E}_{\mathbf{H}^{(\mathrm{A})}_{i,T_{i,\emptyset}}}\left[\sum_{t=1}^{\infty} C_t p_{\mathrm{C},i} \sum_{T_{i,\emptyset}=t}^{\infty} \gamma^{T_{i,\emptyset}-1}\right] \overset{(c)}{=} \mathbb{E}_{\mathbf{H}^{(\mathrm{A})}_{i,T_{i,\emptyset}}}\left[\sum_{t=1}^{\infty} C_t(1-\gamma) \sum_{T_{i,\emptyset}=t}^{\infty} \gamma^{T_{i,\emptyset}-1}\right] \\
&= \mathbb{E}_{\mathbf{H}^{(\mathrm{A})}_{i,T_{i,\emptyset}}}\left[\sum_{t=1}^{\infty} C_t(1-\gamma)\gamma^{t-1} \sum_{T_{i,\emptyset}=1}^{\infty} \gamma^{T_{i,\emptyset}-1}\right] \overset{(d)}{=} \mathbb{E}_{\mathbf{H}^{(\mathrm{A})}_{i,T_{i,\emptyset}}}\left[\sum_{t=1}^{\infty} \gamma^{t-1} C_t\right].
\end{aligned}$$

In (a), we use the fact that $\sum_{T_{i,\emptyset}=1}^{\infty}\sum_{t=1}^{T_{i,\emptyset}} \varphi(t, T_{i,\emptyset})$ is an infinite series with constraints $1 \leq t \leq T_{i,\emptyset} \leq \infty$, which is equivalent to $\sum_{t=1}^{\infty}\sum_{T_{i,\emptyset}=t}^{\infty} \varphi(t, T_{i,\emptyset})$. (b) follows because $T_{i,\emptyset} \sim \mathrm{Ge}(p_{\mathrm{C},i})$ (4.1). (c) uses $1-\gamma = 1-(1-p_{\mathrm{C},i}) = p_{\mathrm{C},i}$. In (d), we use that $\sum_{T_{i,\emptyset}=1}^{\infty} \gamma^{T_{i,\emptyset}} = (1-\gamma)^{-1}$ is a convergent geometric series. □

Based on Lemma 4.1, we model intrusion recovery as a *zero-sum game* where the controller and the attacker aim to minimize and maximize $J_i$ (4.4), respectively.

**Game 4.1** (Local intrusion recovery game.).

$$\underset{\pi^{(\mathrm{C})}_{i,t}}{\text{minimize}}\ \underset{\pi^{(\mathrm{A})}_{i,t}}{\text{maximize}} \quad \mathbb{E}_{(\pi^{(\mathrm{C})}_{i,t},\pi^{(\mathrm{A})}_{i,t})}\left[J_i \mid \mathbf{b}_{i,1}(\mathbb{C}) = 0\right] \tag{4.5a}$$

$$\text{subject to} \quad \tau_{i,k} - \tau_{i,k-1} \leq \Delta_{\mathrm{R}}, \tau_{i,k} \triangleq \inf_{t>\tau_{i,k-1}} [a^{(\mathrm{C})}_{i,t} = \mathsf{R}] \quad \forall k \geq 1 \tag{4.5b}$$

$$s_{i,t+1} \sim f_{\mathrm{N},i}(\cdot \mid s_{i,t}, a^{(\mathrm{C})}_{i,t}, a^{(A)}_{i,t}) \quad \forall t \geq 1 \tag{4.5c}$$

$$o_{i,t+1} \sim z_i(\cdot \mid a^{(\mathrm{A})}_{i,t}) \quad \forall t \geq 1 \tag{4.5d}$$

$$a^{(\mathrm{C})}_{i,t} \sim \pi^{(\mathrm{C})}_{i,t}(\cdot \mid \mathbf{b}_{i,t}(\mathbb{C})) \quad \forall t \geq 1 \tag{4.5e}$$

$$a^{(\mathrm{A})}_{i,t} \sim \pi^{(\mathrm{A})}_{i,t}(\cdot \mid \mathbf{b}_{i,t}(\mathbb{C}), s_{i,t}) \quad \forall t \geq 1, \tag{4.5f}$$

where $t = 1, 2, \ldots$; $\tau_{i,0} \triangleq 0$; $\mathbf{b}_{i,1}(\mathbb{C})$ defines the initial state distribution; (4.5b) is a bounded-time-to-recovery (BTR) constraint; (4.5c) is the dynamics constraint; (4.5d) captures the observations; and (4.5e)–(4.5f) capture the actions. Game 4.1 is a zero-sum partially observed stochastic game[8] with one-sided partial observability that satisfies assumptions 1–4 in the background chapter.

**Remark 4.5.** Throughout this paper, we write min max (4.5a) instead of inf sup as the optimization problems we consider have solutions (see Thms. 4.2–4.4 below).

**Remark 4.6** (Expected cost)**.** We choose to minimize the expected cost (4.5a) to model the controllers' preferences. This approach is justified by the fact that the preference relations of the controllers satisfy the von Neumann-Morgenstern axioms (p. 26, von Neumann and Morgenstern, 1944), as we show in Appendix E.

**Remark 4.7** (Bounded time-to-recovery)**.** The BTR constraint (4.5b) with $\Delta_{\mathrm{R}} < \infty$ ensures that undetectable intrusions are eventually recovered. It also implies that the optimal recovery strategy may be time-dependent.

***Equilibrium analysis*** A control strategy $\tilde{\pi}^{(\mathrm{C})}_{i,t}$ in Game 4.1 is a *best response* against an attacker strategy $\pi^{(\mathrm{A})}_{i,t}$ if it minimizes (4.5). Likewise, an attacker strategy $\pi^{(\mathrm{A})}_{i,t}$ is a best response against $\pi^{(\mathrm{C})}_{i,t}$ if it maximizes (4.5). When both players follow best responses, their strategy pair is a *Nash equilibrium* (NE) $\boldsymbol{\pi}^{\star}_i = (\pi^{(\mathrm{C}),\star}_{i,t}, \pi^{(\mathrm{A}),\star}_{i,t})$. Such an equilibrium, together with the belief operator in (4.3), can also form a stronger equilibrium, namely a Perfect Bayesian equilibrium (PBE). Before showing that such equilibria exist, we prove the following lemma.

**Lemma 4.2.** *Game 4.1 can be represented in extensive form.*

*Proof.* The proof is illustrated in Fig. 4.5. □

[8]The components of a POSG are defined the background chapter; see (19).

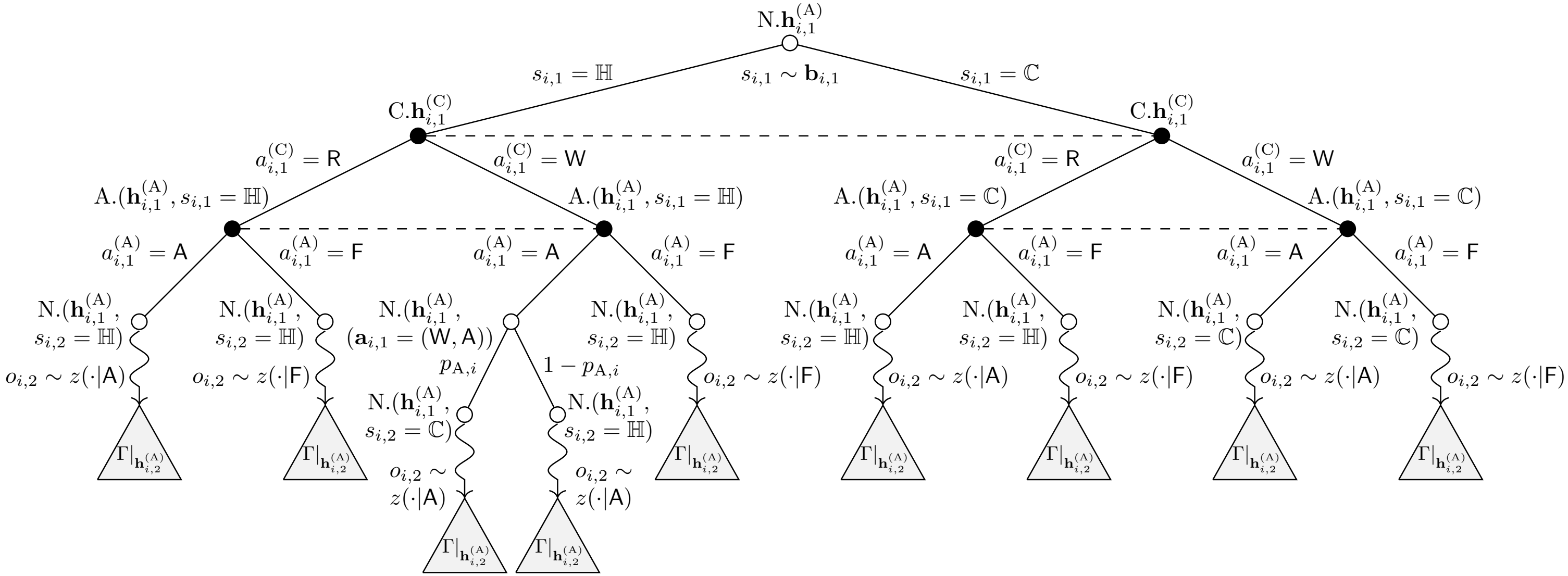

***Figure 4.5:*** *Game 4.1 in extensive form; we use the extensive form notation described in (Myerson, 1997); a filled circle denotes a decision node; an unfilled circle denotes a chance node; the first label next to each node indicates the player who makes the decision;* N *denotes nature; the second label next to each node and the dashed lines indicate the information sets; e.g., the label* $\mathrm{C}.\mathbf{h}_{i,1}^{(\mathrm{C})}$ *indicates a decision node for the* (C)*ontroller in the information set determined by* $\mathbf{h}_{i,1}^{(\mathrm{C})}$*; a zig-zag branch is a short-hand for many branches; a triangle indicates a subgame; see Def. 2 in the background chapter for the definition of a subgame; branches to the terminal state* $\emptyset$ *are omitted to improve readability (at each time step, the game transitions to* $\emptyset$ *with probability* $p_{\mathrm{C},i}$ *(4.1)).*

**Theorem 4.2** (Existence of equilibria and best responses in Game 4.1)**.**

*(A) For each strategy pair* $\boldsymbol{\pi}_i$ *in Game 4.1, there exists a pair of best responses.*

*(B) Game 4.1 has a* PBE*.*

*(C) If* $s_{i,t} = \mathbb{H} \iff \mathbf{b}_{i,t}(\mathbb{C}) = 0$*, then Game 4.1 has a pure* PBE*.*

*(D) The average equilibrium cost of Game 4.1 is not larger than* 1*.*

*Proof.* By definition, the best response problems in Game 4.1 correspond to finite Partially Observable Markov Decision Processes (POMDPs)[9]. Lemma 4.1 implies that these POMDPs can be formulated with the discounted cost optimality criterion. Claim (A) thus follows from Thm. 2 in the background chapter. (B) follows from Lemma 4.1 and Thm. 3 in the background chapter. (C) follows from Thm. 2.1.1 of Paper 2. Lastly, to see why (D) holds, consider the control strategy that always recovers, i.e., $\pi^{(\mathrm{C})}_{i,t}(\mathsf{R} \mid \cdot) = 1$. It follows from (4.4) that the average cost incurred by this strategy is 1. □

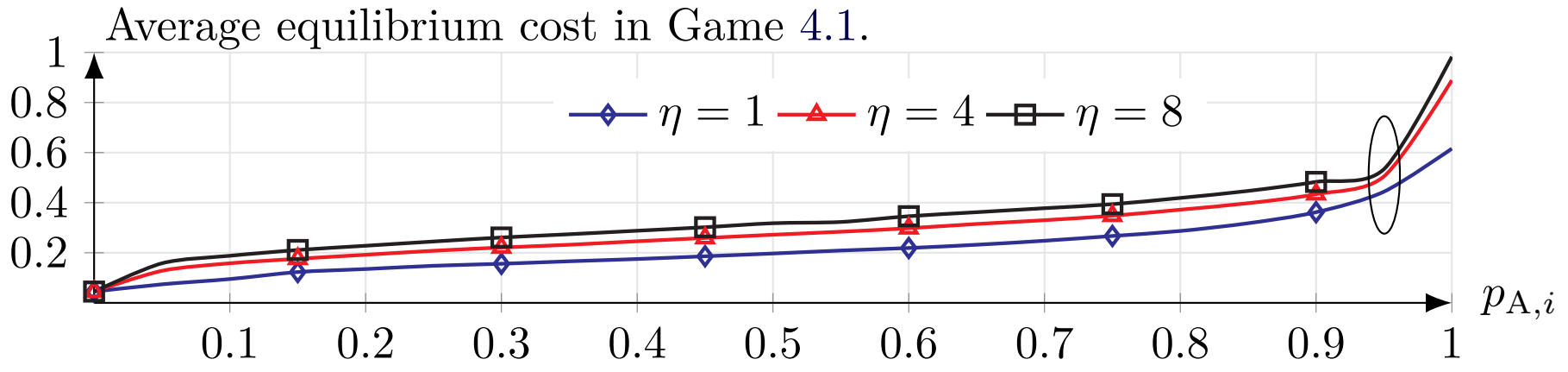

***(a)*** *Average equilibrium cost in Game 4.1 as a function of* $p_{\mathrm{A},i}$ *and* $\eta$ *(4.5).*

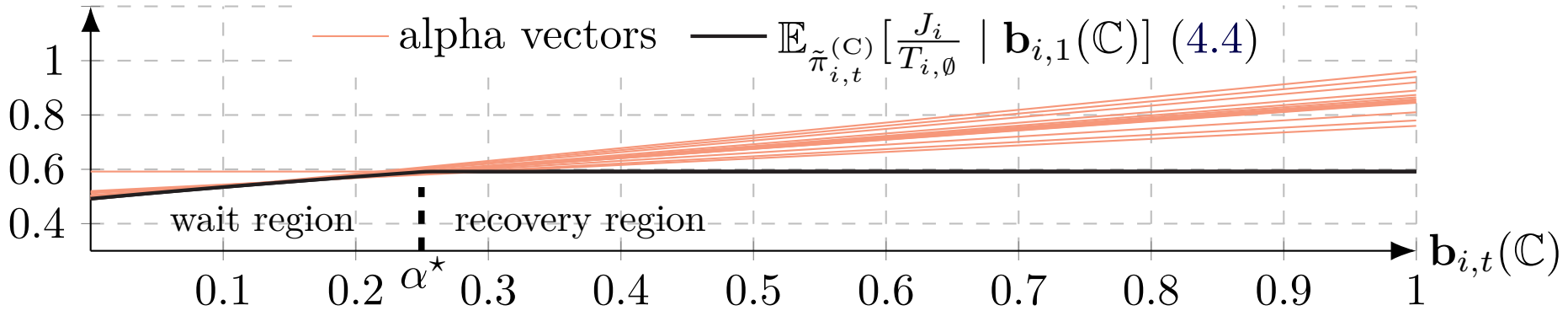

***(b)*** *The controller's average best response value (4.4).*

***Figure 4.6:*** *a) the ellipse indicates the place where the equilibrial strategy for the defender is to almost always recover; b) the dashed red lines indicate alpha-vectors* $\boldsymbol{\alpha}^{(1)}, \boldsymbol{\alpha}^{(2)}, \ldots$*, where* $\mathbb{E}_{\tilde{\pi}^{(\mathrm{C})}_{i,t}}[J_i \mid \mathbf{b}_{i,1}(\mathbb{C})] = \max_i [1 - \mathbf{b}_{i,1}(\mathbb{C}), \mathbf{b}_{i,1}(\mathbb{C})]^T \boldsymbol{\alpha}^{(i)}$ *(Def. 1, Sondik, 1978); hyperparameters are listed in Appendix G.*

Theorem 4.2 guarantees the existence of a strategy pair $\boldsymbol{\pi}^\star_i$ that solves (4.5). Such a pair can be computed using the HSVI algorithm (Alg. 3, Horák et al., 2023), see Fig. 4.6.a. The theorem also establishes that when one player's strategy is fixed, a best response for the opponent exists. Such a strategy can be computed using standard solution algorithms for POMDPs. Figure 4.6.b shows the expected cost for a best response $\tilde{\pi}^{(\mathrm{C})}_{i,t}$. We note that $\tilde{\pi}^{(\mathrm{C})}_{i,t}$ has a threshold structure, as stated below.

[9]The components of a POMDP are defined the background chapter; see (15).

**Theorem 4.3** (Threshold structure of best responses in Game 4.1)**.**
*For any* $\pi^{(\mathrm{A})}_{i,t}$ *in Game 4.1, there exists a best response* $\tilde{\pi}^{(\mathrm{C})}_{i,t}$ *that satisfies*

$$\tilde{\pi}^{(\mathrm{C})}_{i,t}(\mathbf{b}_{i,t}(\mathbb{C})) = \mathsf{R} \iff \mathbf{b}_{i,t}(\mathbb{C}) \geq \alpha^{\star}_{i,t}\ \forall t,\ \text{where } \alpha^{\star}_{i,t} \in [0,1] \text{ is a threshold.} \tag{4.6}$$

**Corollary 4.1** (Stationary threshold)**.** *As* $\Delta_{\mathrm{R}} \to \infty$*, all thresholds converge to* $\alpha^{\star}_{i}$*.*

Theorem 4.3 states that there exists a best response for the controller that performs recovery when the belief (4.3) exceeds a threshold (4.6). Further, Cor. 4.1 states that when there are no periodic recoveries (i.e., when $\Delta_{\mathrm{R}} = \infty$), the threshold is independent of time. We prove Thm. 4.3 and Cor. 4.1 by showing that the region of the belief space where recovery is a best response is a connected interval $[\alpha^{\star}_{i}, 1]$. To show this property, we leverage the optimal stopping theory developed in Paper 1 and the concavity of $\mathbb{E}_{\tilde{\pi}^{(\mathrm{C})}_{i,t}}[J_i \mid \mathbf{b}_{i,1}(\mathbb{C})]$ (4.4); see Thm. 2 in the background chapter. We provide detailed proof in Appendix A.

***Numerical evaluation*** Computing a best response in Game 4.1 is equivalent to solving a POMDP, which generally is PSPACE-hard (Thm. 6, Papadimitriou and Tsitsiklis, 1987). However, Thm. 4.3 and Cor. 4.1 imply that we can parameterize $\tilde{\pi}^{(\mathrm{C})}_{i,t}$ with a finite number of thresholds. Given such parametrization, we formulate the best response problem as a parametric optimization problem, which can be solved efficiently with standard optimization algorithms. Algorithm 4.1 contains the pseudocode of our solution.

***Algorithm 4.1:*** *Parametric optimization to obtain a best response in Game 4.1.*

**Input:** Game 4.1, an attacker strategy $\pi^{(\mathrm{A})}_{i,t}$, and a parametric optimizer PO.
**Output:** An approximate best response control strategy $\hat{\pi}_{i,\boldsymbol{\theta},t}$.

**procedure** THRESHOLD OPTIMIZATION(Game 4.1, $\pi^{(\mathrm{A})}_{i,t}$, PO)
  if $\Delta_{\mathrm{R}} < \infty$, $\mathrm{d} \leftarrow \Delta_{\mathrm{R}} - 1$, else $\mathrm{d} \leftarrow 1$.
  $\Theta \leftarrow [0,1]^{\mathrm{d}}$. ▷ Parameter space.
  $\pi_{i,\boldsymbol{\theta},t}(\mathbf{b}_{i,t}(\mathbb{C})) \triangleq \begin{cases} \mathsf{R} & \text{if } \mathbf{b}_{i,t}(\mathbb{C}) \geq \boldsymbol{\theta}_t. \\ \mathsf{W} & \text{otherwise.} \end{cases}$ ▷ Threshold strategy.
  $J_{i,\boldsymbol{\theta}} \leftarrow \mathbb{E}_{\pi_{i,\boldsymbol{\theta},t},\pi^{(\mathrm{A})}_{i,t}}[J_i]$ (4.4). ▷ Black-box objective.
  $\hat{\pi}_{i,\boldsymbol{\theta},t} \leftarrow \mathrm{PO}(\Theta, J_{i,\boldsymbol{\theta}})$. ▷ Simulation-based optimization.
  **return** $\hat{\pi}_{i,\boldsymbol{\theta},t}$.

**Remark 4.8.** Algorithm 4.1 is a generalization of the T-SPSA algorithm described in Paper 1. T-SPSA can be seen as an instantiation of Alg. 4.1 with SPSA (Fig. 1, Spall, 1998) as the parametric optimizer. However, Alg. 4.1 can be instantiated with many other types of parametric optimizers as well, as demonstrated below.

We evaluate Alg. 4.1 on instantiations of Game 4.1 with different values of $\Delta_{\mathrm{R}}$ for the BTR constraint (4.5b). Hyperparameters are listed in Appendix G. The computing environment for the evaluation is a server with a 24-core INTEL XEON GOLD 2.10 GHz CPU and 768 GB RAM; see Fig. 21 in the methodology chapter.

***Baselines*** For each instantiation of Game 4.1, we run Alg. 4.1 with four optimization algorithms: **S**imultaneous **P**erturbation **S**tochastic **A**pproximation (SPSA) (Fig. 1, Spall, 1998), **B**ayesian **O**ptimization (BO) (Alg. 1, Shahriari et al., 2016), **C**ross **E**ntropy **M**ethod (CEM) (Alg. 1, Moss, 2020), and **D**ifferential **E**volution (DE) (Fig. 3, Storn and Price, 1997). We compare the results with two baselines: **I**ncremental **P**runing (IP) (Fig. 4, Cassandra et al., 1997), which is a dynamic programming algorithm, and **P**roximal **P**olicy **O**ptimization (PPO) (Alg. 1, Schulman et al., 2017), a reinforcement learning algorithm[10]. The results are shown in Fig. 4.7 below and Table 4.2 and Fig. 4.8 on the subsequent page.

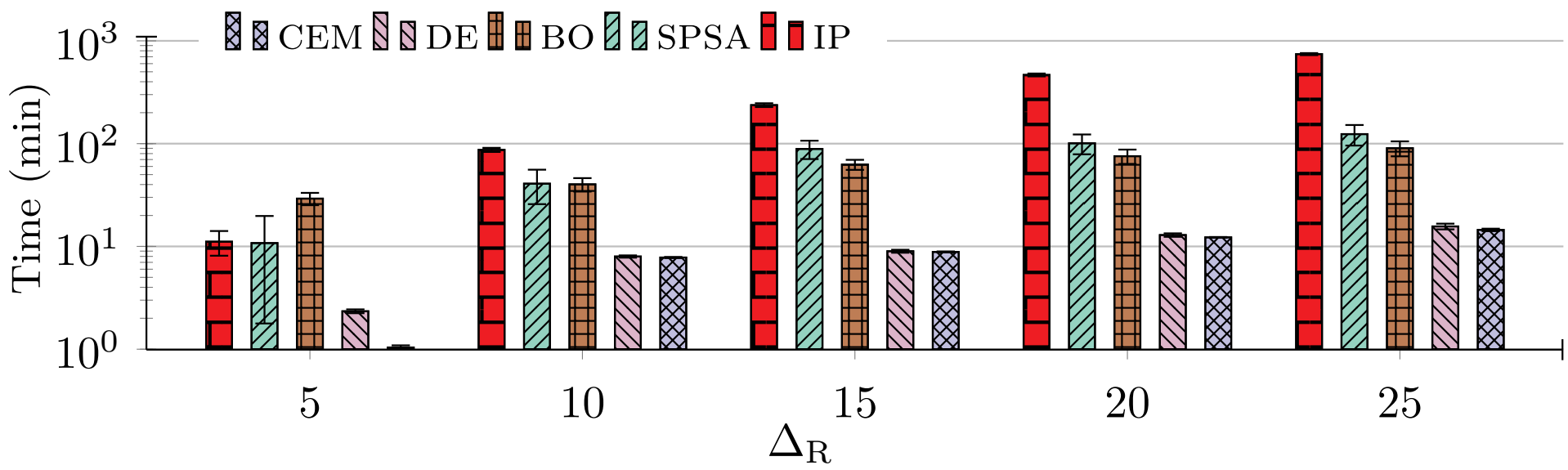

***Figure 4.7:*** *Mean compute time of Alg. 4.1 for different values of* $\Delta_{\mathrm{R}}$ *and different parametric optimizers:* SPSA*,* BO*,* CEM*, and* DE*, as well as a dynamic programming baseline:* IP*; the error bars indicate the* 95% *confidence interval based on* 20 *measurements; hyperparameters are available in Appendix G.*

[10] See Appendix D of Paper 3 for a derivation of the PPO algorithm.

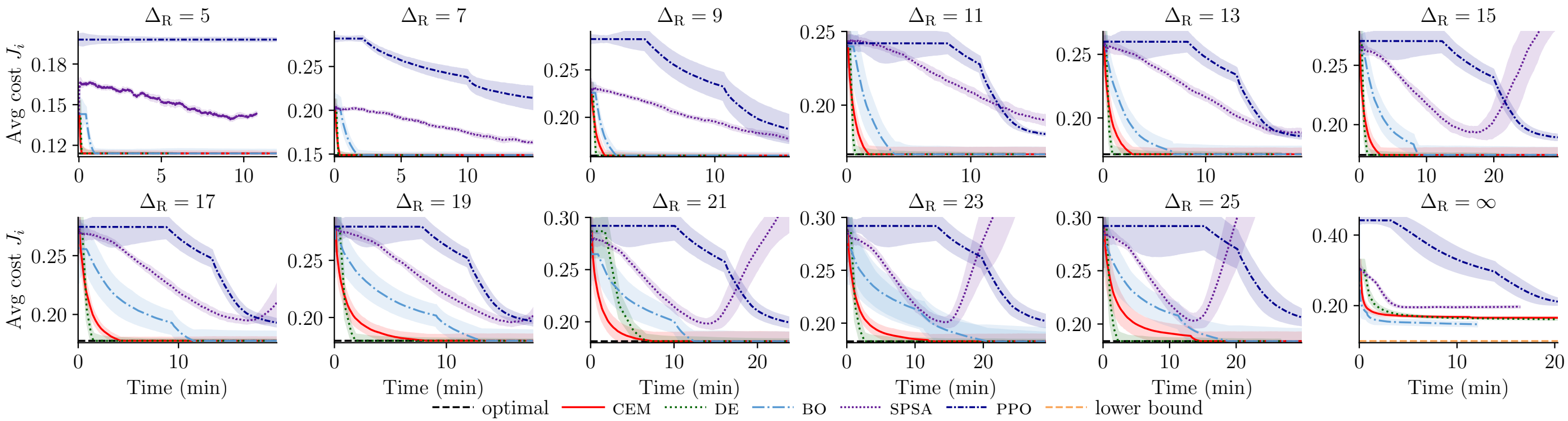

***Figure 4.8:*** *Convergence curves of Alg. 4.1 for computing a best response control strategy in Game 4.1 (the local intrusion recovery game); the curves relate to different parametric optimizers: SPSA, BO, CEM, DE, and PPO; the curves show the mean value from evaluations with* 20 *random seeds and the shaded areas indicate the* 95% *confidence interval divided by* 10.

| Method | $\Delta_R = 5$ Time (min) | $\Delta_R = 5$ $\frac{J_i}{T_{i,\emptyset}}$ (4.4) | $\Delta_R = 15$ Time (min) | $\Delta_R = 15$ $\frac{J_i}{T_{i,\emptyset}}$ (4.4) | $\Delta_R = 25$ Time (min) | $\Delta_R = 25$ $\frac{J_i}{T_{i,\emptyset}}$ (4.4) | $\Delta_R = \infty$ Time (min) | $\Delta_R = \infty$ $\frac{J_i}{T_{i,\emptyset}}$ (4.4) |
|---|---|---|---|---|---|---|---|---|
| CEM [316, Alg. 1] | **1.04** | $\mathbf{0.12 \pm 0.01}$ | **8.84** | $\mathbf{0.17 \pm 0.06}$ | **14.48** | $0.19 \pm 0.08$ | 11.81 | $0.16 \pm 0.01$ |
| DE [443, Fig. 3] | 2.35 | $\mathbf{0.12 \pm 0.03}$ | 8.98 | $\mathbf{0.17 \pm 0.01}$ | 15.45 | $\mathbf{0.18 \pm 0.02}$ | 22.68 | $0.16 \pm 0.01$ |
| BO [403, Alg. 1] | 29.18 | $\mathbf{0.12 \pm 0.02}$ | 62.57 | $\mathbf{0.17 \pm 0.05}$ | 90.26 | $\mathbf{0.18 \pm 0.12}$ | 9.07 | $\mathbf{0.15 \pm 0.06}$ |
| SPSA [436, Fig. 1] | 10.78 | $0.18 \pm 0.01$ | 88.35 | $0.58 \pm 0.40$ | 123.85 | $0.77 \pm 0.48$ | **4.20** | $0.20 \pm 0.02$ |
| PPO [396, Alg. 1] | 28.20 | $0.18 \pm 0.01$ | 30.01 | $0.19 \pm 0.02$ | 30.33 | $0.21 \pm 0.07$ | 28.95 | $0.21 + \pm 0.09$ |
| IP [88, Fig. 4] | 11.11 | **0.12** | 237.06 | **0.17** | 743.73 | **0.18** | > 10000 | not converged |

***Table 4.2:*** *Computing a best response for the controller in Game 4.1 using Alg. 4.1 (upper rows) and baselines (lower rows); columns represent* $\Delta_R$*; subcolumns indicate the computational time (left) and the average cost (right); numbers indicate the mean and the* 95% *confidence interval based on* 20 *random seeds.*

We observe in the first three rows of Table 4.2 that most of the algorithms that utilize Thm. 4.3 find near-optimal recovery strategies for all $\Delta_R$. By contrast, IP becomes computationally intractable as $\Delta_R \to \infty$ (bottom row of Table 4.2). The convergence times are shown in Fig. 4.7.b and Fig. 4.8. We observe that CEM, BO, DE, and PPO find near-optimal strategies within an hour of computation, whereas SPSA does not converge. The divergence of SPSA is probably due to a poor selection of hyperparameters.

Figure 4.9 shows a comparison between the operational cost (4.4) incurred by the equilibrium strategy in Game 4.1 and the periodic recovery strategy used in many state-of-the-art intrusion-tolerant systems (Distler, 2021).

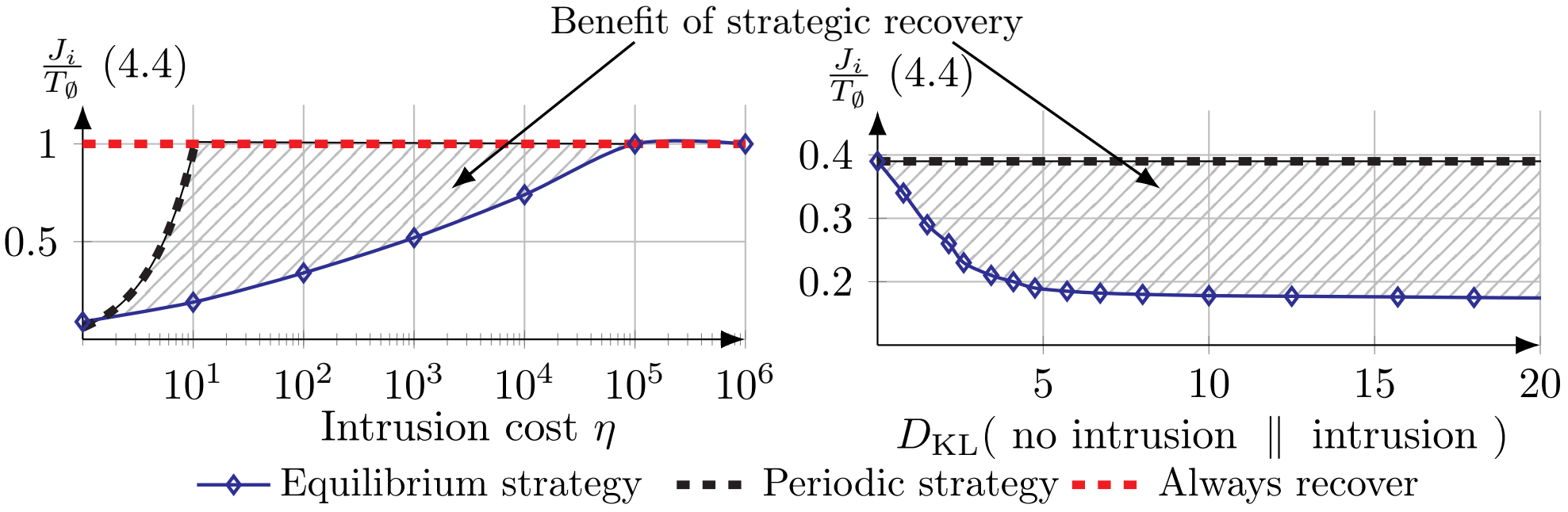

***Figure 4.9:*** *Comparison between the operational cost incurred by equilibrium and periodic strategies in Game 4.1 as functions of $\eta$ (4.4) and the KL divergence between $z_i(\cdot \mid \mathsf{F})$ and $z_i(\cdot \mid \mathsf{A})$ (4.2); hyperparameters are listed in Appendix G; the values were computed using the HSVI algorithm (Alg. 3, Horák et al., 2023); when computing the value of the periodic strategy, we restricted the strategy space of the controller to the class of periodic strategies.*

We note in Fig. 4.9 that the cost of the equilibrium strategy remains consistently lower than the cost of the periodic strategy. However, we also observe that the benefit of the equilibrium strategy reduces when a) the Kullback-Leibler (KL) divergence between $z_i(\cdot \mid \mathsf{F})$ and $z_i(\cdot \mid \mathsf{A})$ (4.2) decreases (right plot), i.e., when the intrusion detection accuracy decreases; and b) when the intrusion cost $\eta$ (4.4) becomes very large (left plot), in which case it is optimal to always perform recovery.

**Key insight.**

*Game-theoretic recovery strategies* can significantly reduce the operational cost of state-of-the-art intrusion-tolerant systems if an accurate intrusion detection model is available and the cost of recovery is significant. Otherwise, *periodic recovery strategies* can be optimal.

### The global replication game

The global game involves two players: a *system controller* that adjusts the replication factor $N_t$ to maintain service availability and minimize operational cost, and an attacker aiming to maximize that cost. At each time $t$, the system controller receives the belief states $\mathbf{b}_{1,t}, \ldots, \mathbf{b}_{N_t,t}$ from the nodes and decides whether or not $N_t$ should be increased; see Fig. 4.1 on page 155. Similarly, at each time $t$, the attacker selects a subset of nodes to attack. A node that fails to send $\mathbf{b}_{i,t}$ at time $t$ is considered crashed by the controller[11], which evicts the node and decrements $N_t$ by 1. We define the state of the game to represent the number of healthy nodes as estimated by the controller. The state space thus is $\mathcal{S}_{\mathrm{S}} \triangleq \{0, 1, \ldots, s_{\max}\}$ with initial state $s_1 = N_1$ Here $s_{\max}$ is the maximum number of nodes, which is defined by the available hardware. The state evolves as $s_{t+1} \sim f_{\mathrm{S}}(\cdot \mid s_t, a_t^{(\mathrm{C})}, \mathbf{a}_t^{(\mathrm{A})})$, where $a_t^{(\mathrm{C})} \in \{0, 1\}$ is the number of nodes added by the controller at time $t$, $\mathbf{a}_t^{(\mathrm{A})} \in \{\mathsf{F}, \mathsf{A}\}^{N_t}$ is the attacker action, and $f_{\mathrm{S}}$ is the transition function, which depends on the local control strategies in Game 4.1.

***System reliability analysis*** Proposition 4.1 implies that correct service is guaranteed if $s_t > f$, where $f$ is the tolerance threshold. The mean time to failure (MTTF) $\mathbb{E}[T^{(\mathrm{F})}]$ thus equals the mean hitting time of a state where $s_t \leq f$:

$$\mathbb{E}[T^{(\mathrm{F})} \mid S_1 = s_1] = \mathbb{E}_{(S_t)_{t \geq 1}}\Big[\inf\{t \mid t \in \mathbb{N}, t \geq 1, S_t \leq f\} \mid S_1 = s_1\Big].$$

Consider the case where the system controller and the node controllers are passive, i.e., when there are never recoveries or additions of nodes. In this case, the set of states where service is disrupted is absorbing. Therefore,

$$\mathbb{E}[T^{(\mathrm{F})} \mid S_1 = s_1] = \begin{cases} 0 & \text{if } s_1 \leq f \\ 1 + \displaystyle\sum_{s' \in \mathcal{S}_{\mathrm{S}}} \mathbf{P}_{s_1, s'} \mathbb{E}[T^{(\mathrm{F})} \mid S_1 = s'] & \text{if } s_1 > f, \end{cases}$$

which defines a system of $|\mathcal{S}_{\mathrm{S}}|$ linear equations, one for each state $s \in \mathcal{S}_{\mathrm{S}}$. ($\mathbf{P} \in [0,1]^{|\mathcal{S}_{\mathrm{S}}|^2}$ is the transition probability matrix.) The reliability function of the system is $R(t) = \mathbb{P}[T^{(\mathrm{F})} > t] = \mathbb{P}[S_t > f]$. Applying the Chapman-Kolmogorov equation, (Eq. 2.12, Krishnamurthy, 2016), we have that $R(t) = \sum_{s \in \mathcal{S}'_{\mathrm{S}}} \left(\mathbf{e}_{s_1}^T \mathbf{P}^t\right)_s$, where $\mathbf{e}_{s_1}$ is the $s_1$-basis vector and $\mathcal{S}'_{\mathrm{S}} \triangleq \{s \mid s > f, s \in \mathcal{S}_{\mathrm{S}}\}$; see Fig. 4.10.b on the next page.

[11] Note that this automatic eviction means the controller never has to explicitly remove a node; hence we can define the controller's action space as $\{0, 1\}$ rather than $\{-1, 0, 1\}$.

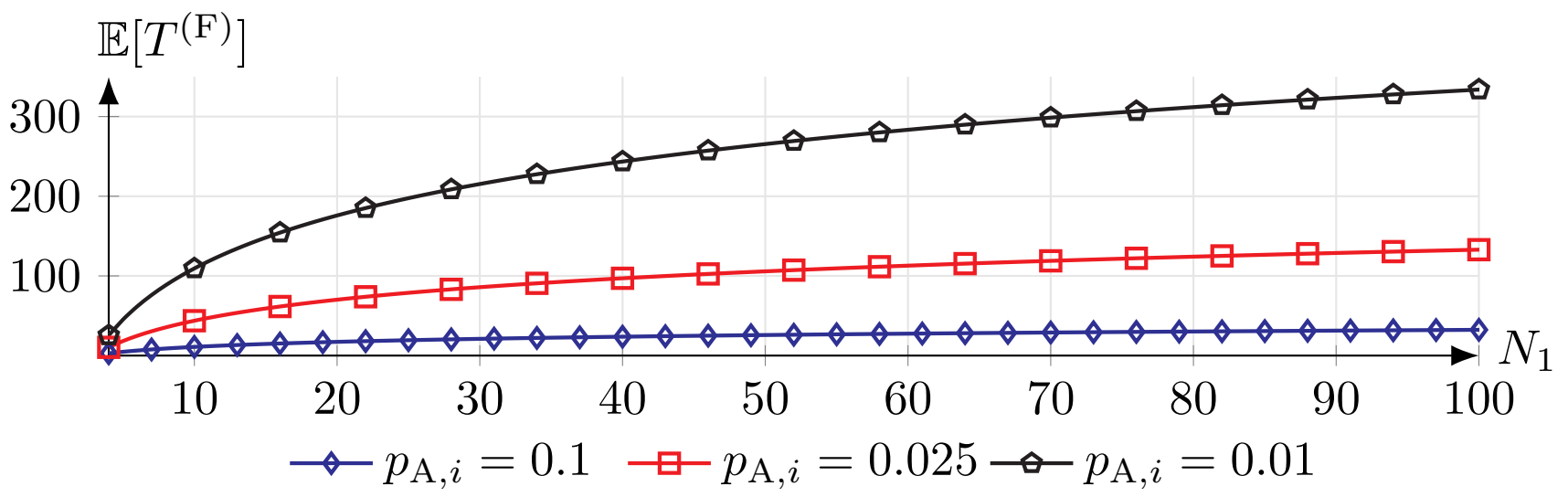

***(a)** Mean time to failure (MTTF).*

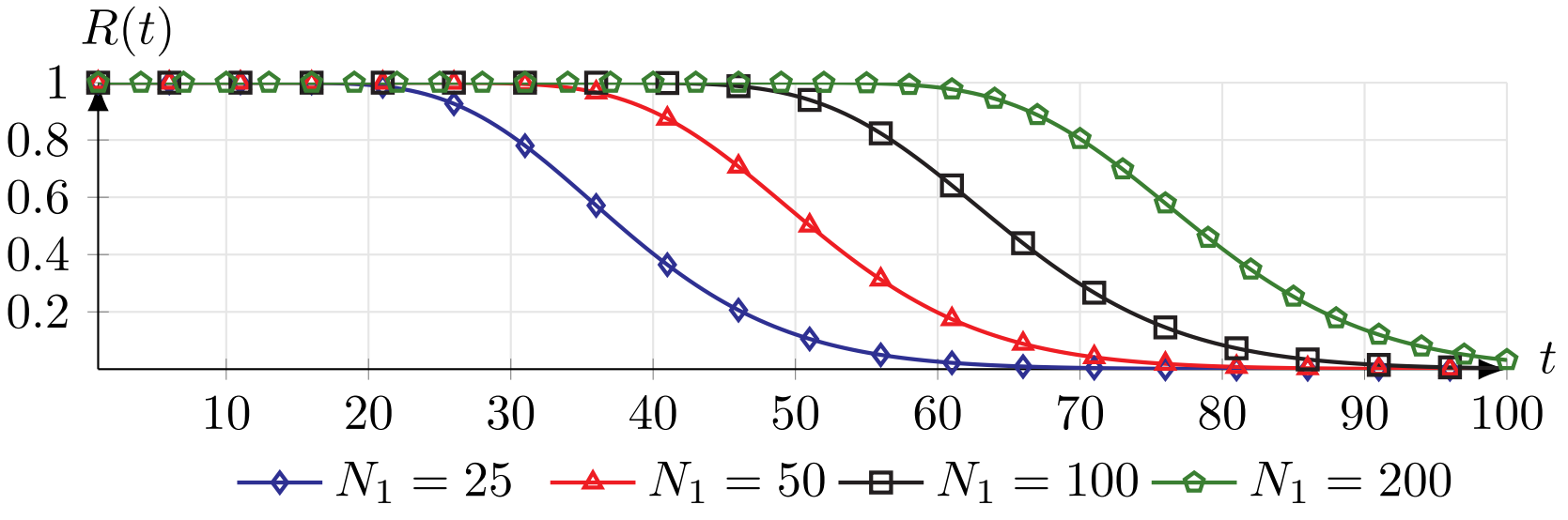

***(b)** The reliability function $R(t) \triangleq \mathbb{P}[T^{(f)} > t]$.*

***Figure 4.10:** The MTTF and the reliability function in Game 4.2 when all controllers are passive; $T^{(\mathrm{F})}$ is a random variable representing the time when $N_t < 2f+k+1$ with $f=3$ and $k=1$ (Prop. 4.1); $N_1$ is the initial number of nodes in the system; hyperparameters are listed in Appendix G.*

***Controller objective*** Increasing the replication factor $N_t$ improves service availability $T^{(\mathrm{A})}$ but increases cost; see Fig. 4.10.a. ($T^{(\mathrm{A})}$ is the fraction of time steps where service is available.) The goal of the controller is thus to find the optimal cost-redundancy trade-off, i.e., to minimize

$$J \triangleq \lim_{T\to\infty}\left[\sum_{t=1}^{T}\frac{a_t^{(\mathrm{C})}}{T}\right] \qquad \text{subject to } T^{(\mathrm{A})} \geq \epsilon_{\mathrm{A}}, \tag{4.7}$$

where $\epsilon_{\mathrm{A}}$ is the chosen lower bound on service availability. For instance, if $\epsilon_{\mathrm{A}} = 0.999$, then at most 8.4 hours of service disruption per year is allowed; see Table 4.3 on the next page. Note that the availability constraint can be written in terms of the state $s_t$ as

$$\lim_{T\to\infty}\left[\sum_{t=1}^{T}\frac{\mathbb{1}_{s_t \geq f+1}}{T}\right] \geq \epsilon_{\mathrm{A}}.$$

**Remark 4.9** (Nodes vs. throughput)**.** (4.7) indicates that the number of nodes should be minimized. This objective reflects the fact that the more nodes there are in the system, the lower the throughput of the consensus protocol. This trade-off is evident in our experimental evaluation of TOLERANCE (see Fig. 4.13 on page 180).

| $\epsilon_{\mathrm{A}}$ (4.7) | *Allowed service downtime per year* |
|---|---|
| 0.7 | 108 days. |
| 0.8 | 72 days. |
| 0.9 | 36 days. |
| 0.95 | 18 days. |
| 0.99 | 3 days. |
| 0.999 | 8 hours. |
| 0.9999 | 52 minutes. |
| 0.99999 | 5 minutes. |
| 0.999999 | 31 seconds. |
| 1 | 0 seconds. |

***Table 4.3:*** *Allowed service downtime for different values of $\epsilon_{\mathrm{A}}$ (4.7).*

Given (4.7) and the Markov property of $s_t$, we define the controller and the attacker strategies as $\pi^{(\mathrm{C})} : \mathcal{S}_{\mathrm{S}} \rightarrow \Delta(\{0,1\})$ and $\pi^{(\mathrm{A})} : \mathcal{S}_{\mathrm{S}} \rightarrow \Delta(\{\mathsf{F},\mathsf{A}\}^{N_t})$, respectively. (We restrict the strategies to be time-independent, as stationary best responses and equilibria exist; see Thm. 4.4 below.) Based on these definitions, we model replication control as a *constrained, zero-sum, stochastic game.*

**Game 4.2** (Global replication game.)**.**

$$\underset{\pi^{(\mathrm{C})}}{\text{minimize}}\ \underset{\pi^{(\mathrm{A})}}{\text{maximize}} \quad \mathbb{E}_{(\pi^{(\mathrm{C})},\pi^{(\mathrm{A})})}\left[J \mid s_1 = N_1\right] \tag{4.8a}$$

$$\text{subject to} \quad \mathbb{E}_{(\pi^{(\mathrm{C})},\pi^{(\mathrm{A})})}\left[T^{(\mathrm{A})}\right] \geq \epsilon_{\mathrm{A}} \tag{4.8b}$$

$$s_{t+1} \sim f_{\mathrm{S}}(\cdot \mid s_t, a_t^{(\mathrm{C})}, \mathbf{a}_t^{(\mathrm{A})}) \qquad \forall t \geq 1 \tag{4.8c}$$

$$a_t^{(\mathrm{C})} \sim \pi^{(\mathrm{C})}(\cdot \mid s_t), \quad a_t^{(\mathrm{C})} = 1 \text{ if } s_t \leq f \qquad \forall t \geq 1 \tag{4.8d}$$

$$\mathbf{a}_t^{(\mathrm{A})} \sim \pi^{(\mathrm{A})}(\cdot \mid s_t) \qquad \forall t \geq 1, \tag{4.8e}$$

where (4.8b) is the availability constraint; (4.8c) is the dynamics constraint; and (4.8d)–(4.8e) capture the actions. Note that Game 4.2 satisfies Assumption 1, Assumption 2, and Assumption 4 in the background chapter.

**Remark 4.10** (Partition tolerance)**.** To circumvent the CAP theorem (Thm. 2, Gilbert and Lynch, 2002) and satisfy (4.8b) in the presence of network partitions, we use the *primary-partition* model (Birman, 1997): in the case of a network partition, only one partition is permitted to remain operational (to maintain consistency) and we consider the system to be available as long as one partition is operational.

***Equilibrium analysis*** When both the controller and the attacker play best response, their strategy pair is a *Nash equilibrium* (NE) $\boldsymbol{\pi}^{\star}$. Due to the Markov property of the strategies, $\boldsymbol{\pi}^{\star}$ can also form a stronger equilibrium, namely a *Markov perfect equilibrium* (MPE).

**Definition 4.2** (Markov perfect equilibrium (MPE) (Def. 5, Deng et al., 2022))**.** *A strategy pair $\boldsymbol{\pi}^{\star}$ is a Markov perfect equilibrium if each player follows a Markovian behavior strategy and $\boldsymbol{\pi}^{\star}$ is a NE regardless of the initial state.*

Since Game 4.2 is a finite constrained stochastic game, we obtain the following theorem by combining (Thm. 4.3, Altman, 1999) and (Thm. 2.1, Altman and Shwartz, 2000).

**Theorem 4.4** (Existence of equilibria and best responses in Game 4.2)**.** *Assuming*

$$f_{\mathrm{S}}(s_{t+1} \mid s_t, a_t^{(\mathrm{C})}, \mathbf{a}_t^{(\mathrm{A})}) > 0 \qquad \forall s_{t+1}, s_t, a_t^{(\mathrm{C})}, \mathbf{a}_t^{(\mathrm{A})}. \tag{T4.1}$$

$$\exists \pi^{(\mathrm{C})} \text{ such that } \mathbb{E}_{(\pi^{(\mathrm{C})}, \pi^{(\mathrm{A})})}\left[T^{(\mathrm{A})}\right] \geq \epsilon_{\mathrm{A}} \qquad \forall \pi^{(\mathrm{A})}. \tag{T4.2}$$

*Then, the following holds.*

*(A) For each strategy pair $\boldsymbol{\pi}$ in Game 4.2, a pair of stationary best responses can be obtained through linear programming.*

*(B) Game 4.2 has a constrained, stationary MPE.*

**Definition 4.3** (Unichain)**.** *A stochastic game is unichain if, for any stationary, Markovian strategy profile $\boldsymbol{\pi}$, the state process $(S_t)_{t=1}^{T}$ has a single recurrent class.*

Assumption (T4.1) implies that Game 4.2 is unichain, i.e., the Markov chain induced by any strategy profile $\boldsymbol{\pi}$ is irreducible. (T4.2) implies that the constraint in (4.8b) is feasible, i.e., a Slater condition (Slater, 1959)(Altman, 1999). Under these assumptions, Thm. 4.4 guarantees the existence of an MPE for Game 4.2. The theorem also establishes that, when the strategy of one player is fixed, there always exists a best response for the opponent that can be computed in polynomial time using linear programming (Thm. 4.3, Altman, 1999). This linear program[12] is listed in Alg. 4.2 on page 178. By contrast, computing an MPE generally means solving a PPAD-complete problem (Thm. 1, Deng et al., 2022). Fortunately, upon examination of (4.8), we find that Game 4.2 has a special structure that allows efficient computation of equilibria.

**Theorem 4.5** (Threshold structure of best responses in Game 4.2)**.**
*Given (T4.1), (T4.2), any attacker strategy $\pi^{(\mathrm{A})}$, and assuming*

$$\mathbb{E}_{\boldsymbol{\pi}}[S_{t+1} \mid S_t = s+1] = \mathbb{E}_{\boldsymbol{\pi}}[S_{t+1} \mid S_t = s] + 1 \qquad \forall \boldsymbol{\pi}, t \geq 1, \tag{T5.1}$$

*then there exist two strategies $\pi_{\lambda_1}^{(\mathrm{C})}$ and $\pi_{\lambda_2}^{(\mathrm{C})}$ that satisfy*

$$\pi_{\lambda_1}^{(\mathrm{C})}(s) = 1 \iff s \leq \beta_1 \quad \text{and} \quad \pi_{\lambda_2}^{(\mathrm{C})}(s) = 1 \iff s \leq \beta_2 \qquad \forall s \in \mathcal{S}_{\mathrm{S}}, \tag{4.9}$$

*and a (stochastic) best response control strategy $\tilde{\pi}^{(\mathrm{C})}$ that satisfies*

$$\tilde{\pi}^{(\mathrm{C})}(s) = \kappa \pi_{\lambda_1}(s) + (1-\kappa)\pi_{\lambda_2}(s) \qquad \forall s \in \mathcal{S}_{\mathrm{S}}, \tag{4.10}$$

*where $\kappa \in [0,1]$, $\lambda_1, \lambda_2 \geq 0$ are Lagrange multipliers, and $\beta_1, \beta_2 \geq f$ are thresholds.*

[12] This linear program is the dual of (12) in the background chapter (Altman, 1999).

**Remark 4.11.** The randomization in (4.10) is required to ensure that the service availability constraint is satisfied in expectation (Thm. 4.4, Altman, 1999).

Assumption (T5.1) says that an additional healthy node at time $t$ increases the expected number of healthy nodes at time $t+1$ by 1. Under this assumption, Thm. 4.5 states that there exists a best response for the controller that can be written as a mixture of two threshold strategies; see Fig. 4.11. Such a strategy is (weakly) *decreasing* in the sense that the fewer healthy nodes there are, the more likely it is that the controller will add a node, which is intuitive. This structure means that a (weakly) dominating strategy for the attacker is to minimize the expected number of healthy nodes $\mathbb{E}[S]$ (Def. 1.1, Fudenberg and Tirole, 1991).

**Corollary 4.2** (Dominating attacker strategy)**.** *Given (T5.1) and assuming each $\pi^{(C)}$ satisfies (4.10), then an attacker strategy that minimizes $\mathbb{E}[S]$ is (weakly) dominating*[13] *(Def. 1.1, Fudenberg and Tirole, 1991).*

Corollary 4.2 means that, under assumption (T5.1), we can obtain an MPE of Game 4.2 by computing a best response of each player independently. Due to the independence, this computation can be done in polynomial time using the linear program of Thm. 4.4; see Alg. 4.2 on the next page.

**Remark 4.12** (Equilibrium uniformity)**.** Since Game 4.1 and Game 4.2 are zero-sum, every equilibrium leads to the same value (Ch. 3, von Neumann and Morgenstern, 1944), regardless of the strategies employed at equilibrium. Consequently, we do not need to concern ourselves with equilibrium selection.

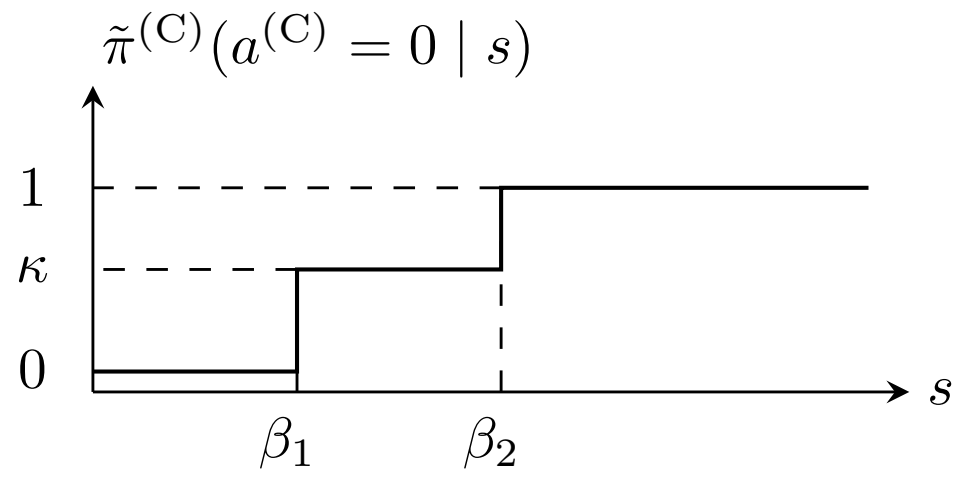

***Figure 4.11:*** *Illustration of Thm. 4.5.*

The proofs of Thm. 4.5 and Cor. 4.2 involve a combination of techniques from CMDP theory and lattice programming. We defer the proofs to Appendix C. However, for the coherence of our argument, we outline the main steps here. To prove Thm. 4.5, we formulate the best response CMDP for the controller as a discounted MDP through Lagrangian relaxation (Thm. 3.7, Altman, 1999). Then, leveraging Topkis' theorem, we show that there exists a best response threshold strategy for any non-negative Lagrange multiplier, discount factor in $[0, 1)$, and attacker strategy (Thm. 6.1, Topkis, 1978). Next, we use the vanishing discount method to establish that the threshold structure applies under the average cost optimality criterion (4.7). Then the proof of Thm. 4.5 follows from standard results in CMDP theory (Thm. 12.7, Altman, 1999). Finally, the corollary is obtained by analyzing the Bellman equation that is induced by Thm. 4.5.

[13] A strategy is weakly dominant if it weakly dominates all other strategies.

***Algorithm 4.2:*** *Linear program for computing a best response in Game 4.2.*

**Input:** Game 4.2, an attacker strategy $\pi^{(\mathrm{A})}$, a linear programming solver LPS.
**Output:** A best response control strategy $\tilde{\pi}^{(\mathrm{C})}$ in Game 4.2.

1: **procedure** LINEAR BEST RESPONSE PROGRAMMING(Game 4.2, $\pi^{(\mathrm{A})}$, LPS)
2: Let $f_{\mathrm{S}}(s' \mid s, a^{(\mathrm{C})}, \pi^{(\mathrm{A})})$ denote the transition function induced by $\pi^{(\mathrm{A})}$.
3: Solve the following linear program using LPS.

$$\underset{\boldsymbol{\rho}}{\text{minimize}} \sum_{s \in \mathcal{S}_{\mathrm{S}}} \sum_{a^{(\mathrm{C})} \in \{0,1\}} a^{(\mathrm{C})} \boldsymbol{\rho}(s, a^{(\mathrm{C})}) \qquad (a^{(\mathrm{C})} \text{ is the cost.}) \tag{4.11}$$

subject to

$$\boldsymbol{\rho}(s, a^{(\mathrm{C})}) \geq 0 \quad \forall s \in \mathcal{S}_{\mathrm{S}}, a^{(\mathrm{C})} \in \{0,1\} \qquad (\boldsymbol{\rho} \text{ is an occupancy measure.})$$

$$\sum_{s \in \mathcal{S}_{\mathrm{S}}} \sum_{a^{(\mathrm{C})} \in \{0,1\}} \boldsymbol{\rho}(s, a^{(\mathrm{C})}) = 1$$

$$\sum_{a^{(\mathrm{C})} \in \{0,1\}} \boldsymbol{\rho}(s', a^{(\mathrm{C})}) = \sum_{s \in \mathcal{S}_{\mathrm{S}}} \sum_{a^{(\mathrm{C})} \in \{0,1\}} \boldsymbol{\rho}(s, a^{(\mathrm{C})}) f_{\mathrm{S}}(s' \mid s, a^{(\mathrm{C})}, \pi^{(\mathrm{A})}) \quad \forall s' \in \mathcal{S}_{\mathrm{S}}$$

$$\sum_{s \in \mathcal{S}_{\mathrm{S}}} \sum_{a^{(\mathrm{C})} \in \{0,1\}} \boldsymbol{\rho}(s, a^{(\mathrm{C})}) \mathbb{1}_{s \geq f+1} \geq \epsilon_{\mathrm{A}}.$$

4: Let $\boldsymbol{\rho}^{\star}$ denote the solution to (4.11).
5: Define

$$\tilde{\pi}^{(\mathrm{C})}(a^{(\mathrm{C})} \mid s) \triangleq \begin{cases} \dfrac{\boldsymbol{\rho}^{\star}(s, a^{(\mathrm{C})})}{\sum_{\hat{a}^{(\mathrm{C})} \in \{0,1\}} \boldsymbol{\rho}^{\star}(s, \hat{a}^{(\mathrm{C})})} & \text{if } \displaystyle\sum_{\hat{a}^{(\mathrm{C})} \in \{0,1\}} \boldsymbol{\rho}^{\star}(s, \hat{a}^{(\mathrm{C})}) \geq 0 \\ \text{arbitrary} & \text{otherwise,} \end{cases}$$

6: **return** $\tilde{\pi}^{(\mathrm{C})}$.

***Numerical evaluation*** Figure 4.12.a on the next page shows the compute time to obtain a best response in Game 4.2 using Alg. 4.2. We note that Alg. 4.2 enables us to compute a best response control strategy within a couple of minutes, even for systems with up to 1000 nodes. Figure 4.12.b shows a comparison between the service availability achieved by the equilibrium strategy and the fixed-replication strategy that is used in many state-of-the-art intrusion-tolerant systems (Distler, 2021). As depicted in the figure, the equilibrium strategy guarantees a high service availability for the system's lifetime. In contrast, the availability of the fixed replication strategy degrades over time.

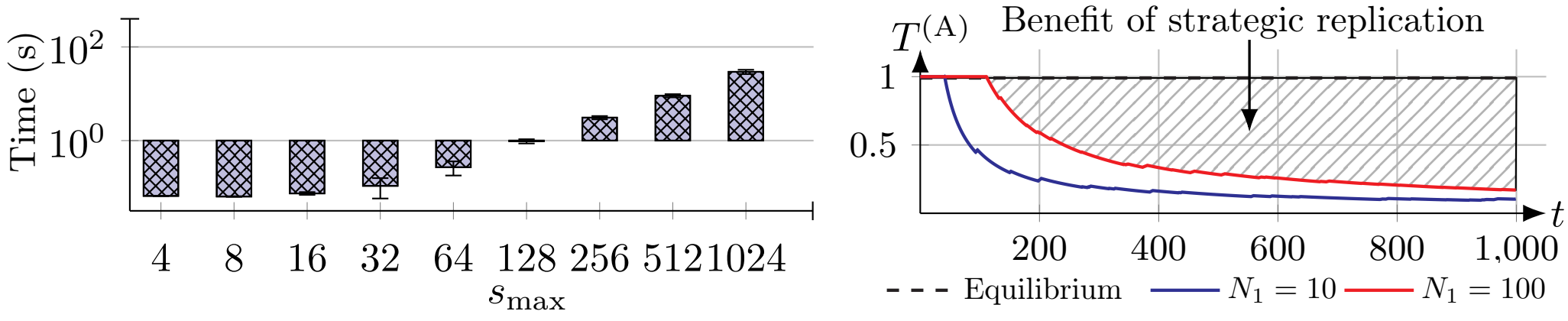

***(a)*** *Mean best response compute time.* ***(b)*** *Service availability* $T^{(\mathrm{A})}$.

***Figure 4.12:*** *Plot a) shows the compute time to obtain a best response in Game 4.2 via the linear program in Alg. 4.2; the error bars indicate the* 95% *confidence interval based on* 20 *runs;* $s_{\max}$ *is the maximum number of nodes; plot b) shows the availability of the equilibrium strategy and two fixed replication strategies with different number of nodes* $N_1$*; hyperparameters are listed in Appendix G.*

> **Key insight.**
>
> *Game-theoretic replication strategies* can guarantee a high service availability. By contrast, many state-of-the-art intrusion-tolerant systems are based on *fixed replication strategies*, for which no such guarantee has been given.

## 4.6 Testbed Implementation of TOLERANCE

We implement TOLERANCE as a proof-of-concept on our testbed. The implementation includes three layers.

### The physical layer

The physical layer contains a cluster with 13 nodes connected through an Ethernet network; see Fig. 4 in the methodology chapter. Specifications of the nodes can be found in Table 4.4 on the next page. Nodes communicate via message passing over authenticated channels[14]. Each node runs (*i*) a service replica in a DOCKER container (Merkel, 2014); (*ii*) a node controller (§4.4); and (*iii*) the SNORT IDS with ruleset v2.9.17.1 (Roesch, 1999).

[14]In an authenticated network, nodes can verify each other's digital signatures.

| *Server* | *Processors* | *Network* | *RAM (GB)* |
|---|---|---|---|
| 1, R 715 2U | two 12-core AMD OPTERON | 12×GbE | 64. |
| 2, R 715 2U | two 12-core AMD OPTERON | 12×GbE | 64. |
| 3, R 715 2U | two 12-core AMD OPTERON | 12×GbE | 64. |
| 4, R 715 2U | two 12-core AMD OPTERON | 12×GbE | 64. |
| 5, R 715 2U | two 12-core AMD OPTERON | 12×GbE | 64. |
| 6, R 715 2U | two 12-core AMD OPTERON | 12×GbE | 64. |
| 7, R 715 2U | two 12-core AMD OPTERON | 12×GbE | 64. |
| 8, R 715 2U | two 12-core AMD OPTERON | 12×GbE | 64. |
| 9, R 715 2U | two 12-core AMD OPTERON | 12×GbE | 64. |
| 10, R 630 2U | two 12-core INTEL XEON E 5- 2680 | 12×GbE | 256. |
| 11, R 740 2U | 1 20-core INTEL XEON GOLD 5218R | 2 × 10GbE | 32. |
| 12, SUPERMICRO 7049 | 2 TESLA P 100, 1 16-core INTEL XEON | 100MbE | 126. |
| 13, SUPERMICRO 7049 | 4 RTX 8000, 1 24-core INTEL XEON | 10GbE | 768. |

***Table 4.4:*** *Specifications of the physical nodes.*

### The virtualization layer

Each service replica is a state machine and runs a web service (Schneider, 1990). This service offers two deterministic operations: (*i*) a *read operation*, which returns the current state of the service; and (*ii*) a *write operation*, which updates the state. Replicas run reconfigurable MINBFT (§4.2, Santos Veronese, 2010) to coordinate these operations. The throughput of our implementation of MINBFT is shown in Fig. 4.13. A description of MINBFT is available in Appendix F.

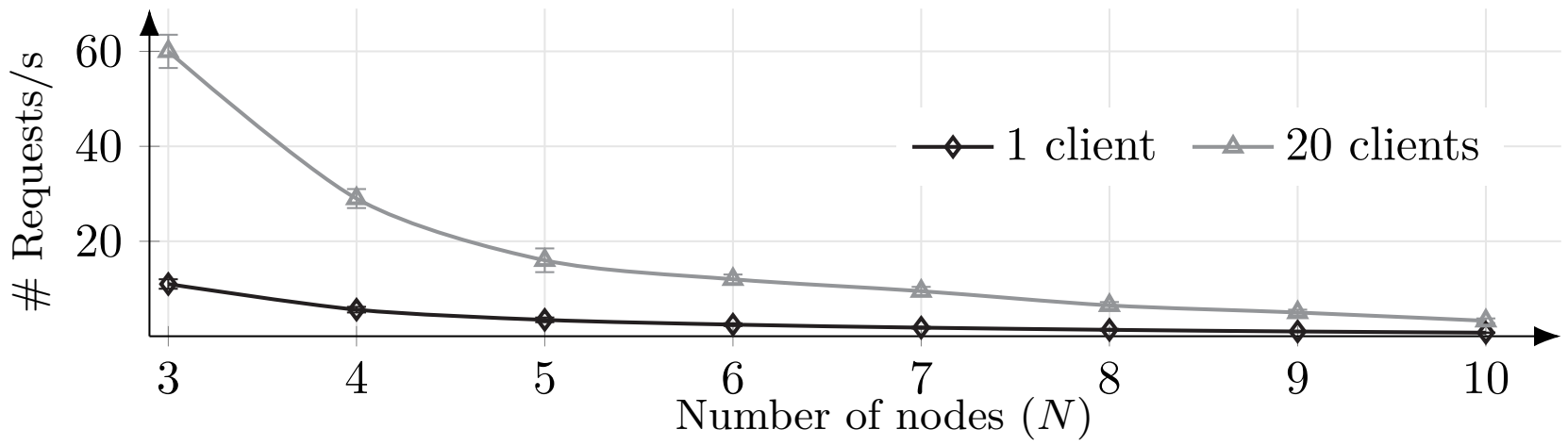

***Figure 4.13:*** *Average throughput of our implementation of MINBFT; error bars indicate the* 95% *confidence interval based on* 1000 *samples.*

**Remark 4.13.** Figure 4.13 motivates the cost function in (4.7), i.e., the number of nodes should be kept small to maximize throughput.

Clients access the service by issuing requests that are sent to all replicas. Each request has a unique identifier that is digitally signed. After sending a request, the client waits for a quorum of $f+1$ identical replies with valid signatures.

**Remark 4.14** (Pigeonhole principle)**.** A quorum is necessary to guarantee a correct response since the client does not know which replicas are compromised (Prop. 4.1).

### The control layer

Node controllers collect IDS alerts and decide when to recover service replicas. When a replica is recovered, it starts with a new container, and its state is initialized with the (identical) state from $f + 1$ other replicas (Garcia et al., 2019)(Reiser and Kapitza, 2007). The system controller is implemented by a crash-tolerant system that runs the RAFT protocol (Ongaro and Ousterhout, 2014). When it decides to evict or add a node, it triggers a view change in MINBFT (§4.2, Santos Veronese, 2010).

## 4.7 Experimental Evaluation of TOLERANCE

In this section, we evaluate our implementation of TOLERANCE and compare it with state-of-the-art intrusion-tolerant systems. We follow the experimental methodology described in the methodology chapter.

### Evaluation setup

An evaluation run evolves in time steps of 60 seconds and lasts for 120 time steps. It starts with $N_1$ nodes from Table 4.4, each of which runs a service replica; see Table 4.5. At each time step, one or more replicas may be recovered by the node controllers, and a new node may be added by the system controller. When a replica is recovered, its container is replaced with a container selected randomly from the list in Table 4.5. Similarly, when a new node is added, a node from the list in Table 4.4 is started.

| *Replica ID* | *Operating system* | *Vulnerabilities* |
|---|---|---|
| 1 | UBUNTU 14 | FTP weak password. |
| 2 | UBUNTU 20 | SSH weak password. |
| 3 | UBUNTU 20 | TELNET weak password. |
| 4 | DEBIAN 10.2 | CVE-2017-7494. |
| 5 | UBUNTU 20 | CVE-2014-6271. |
| 6 | DEBIAN 10.2 | CWE-89 on DVWA [454]. |
| 7 | DEBIAN 10.2 | CVE-2015-3306. |
| 8 | DEBIAN 10.2 | CVE-2016-10033. |
| 9 | DEBIAN 10.2 | CVE-2010-0426, SSH weak password. |
| 10 | DEBIAN 10.2 | CVE-2015-5602, SSH weak password. |

***Table 4.5:*** *Containers running the service replicas.*

Each replica has one or more vulnerabilities that can be exploited by the attacker using the steps listed in Table 4.6 on the next page. After compromising a replica, the attacker randomly chooses between a) participating in the consensus protocol; b) not participating; and c) participating with randomly selected messages. Replicas are interconnected through Gbit/s connections with 0.05% packet loss (emulated with NETEM (Hemminger, 2005)). They receive a stream of service requests sent by a client over 100 Mbit/s connections with 0.1% packet loss.

| *Replica ID* | *Intrusion steps* |
|---|---|
| 1 | TCP SYN scan, FTP brute force. |
| 2 | TCP SYN scan, SSH brute force. |
| 3 | TCP SYN scan, TELNET brute force. |
| 4 | ICMP scan, exploit of CVE-2017-7494. |
| 5 | ICMP scan, exploit of CVE-2014-6271. |
| 6 | ICMP scan, exploit of CWE-89 on on DVWA [454]. |
| 7 | ICMP scan, exploit of CVE-2015-3306. |
| 8 | ICMP scan, exploit of CVE-2016-10033. |
| 9 | ICMP scan, SSH brute force, exploit of CVE-2010-0426. |
| 10 | ICMP scan, SSH brute force, exploit of CVE-2015-5602. |

***Table 4.6:*** *Steps to compromise service replicas.*

Each replica runs a set of background services in addition to the replicated service to emulate IDS events for a realistic system; see Table 4.7.

| *Background services* | *Replica ID(s)* |
|---|---|
| FTP, SSH, MONGODB, HTTP, TEAMSPEAK | 1. |
| SSH, DNS, HTTP | 2. |
| SSH, TELNET, HTTP | 3. |
| SSH, SAMBA, NTP | 4. |
| SSH | $5, 7, 8, 10.$ |
| DVWA, IRC, SSH | 6. |
| TEAMSPEAK, HTTP, SSH | 9. |

***Table 4.7:*** *Background services of the service replicas.*

These background services are consumed by a population of background clients, who arrive with a Poisson rate $\lambda = 20$ and have exponentially distributed service times with mean $\mu = 4$ time steps. All parameters for the evaluation are listed in Appendix G except $z_i$ (4.2), which we estimate with the empirical distribution $\widehat{z}_i$; see Fig. 4.14 on the next page. We compute $\widehat{z}_i$ based on $M = 25,000$ i.i.d. samples from our testbed (Hammar, 2023), knowing that $\widehat{z}_i \xrightarrow{\text{a.s.}} z_i$ as $M \to \infty$[15]. In practice, $\widehat{z}_i$ may be implemented using any statistical intrusion detection method (e.g., anomaly detection (Fuchsberger, 2005)). Similarly, the model parameters (e.g., the probability that a node crashes) can be defined based on domain knowledge or system measurements (Ford et al., 2010).

**Remark 4.15** (System identification)**.** *The collection of samples and the estimation of $z_i$ (4.2) corresponds to the system identification step in our experimental methodology (see the methodology chapter for details).*

[15]It follows by the Glivenko-Cantelli theorem; see (Glivenko and Cantelli, 1933).

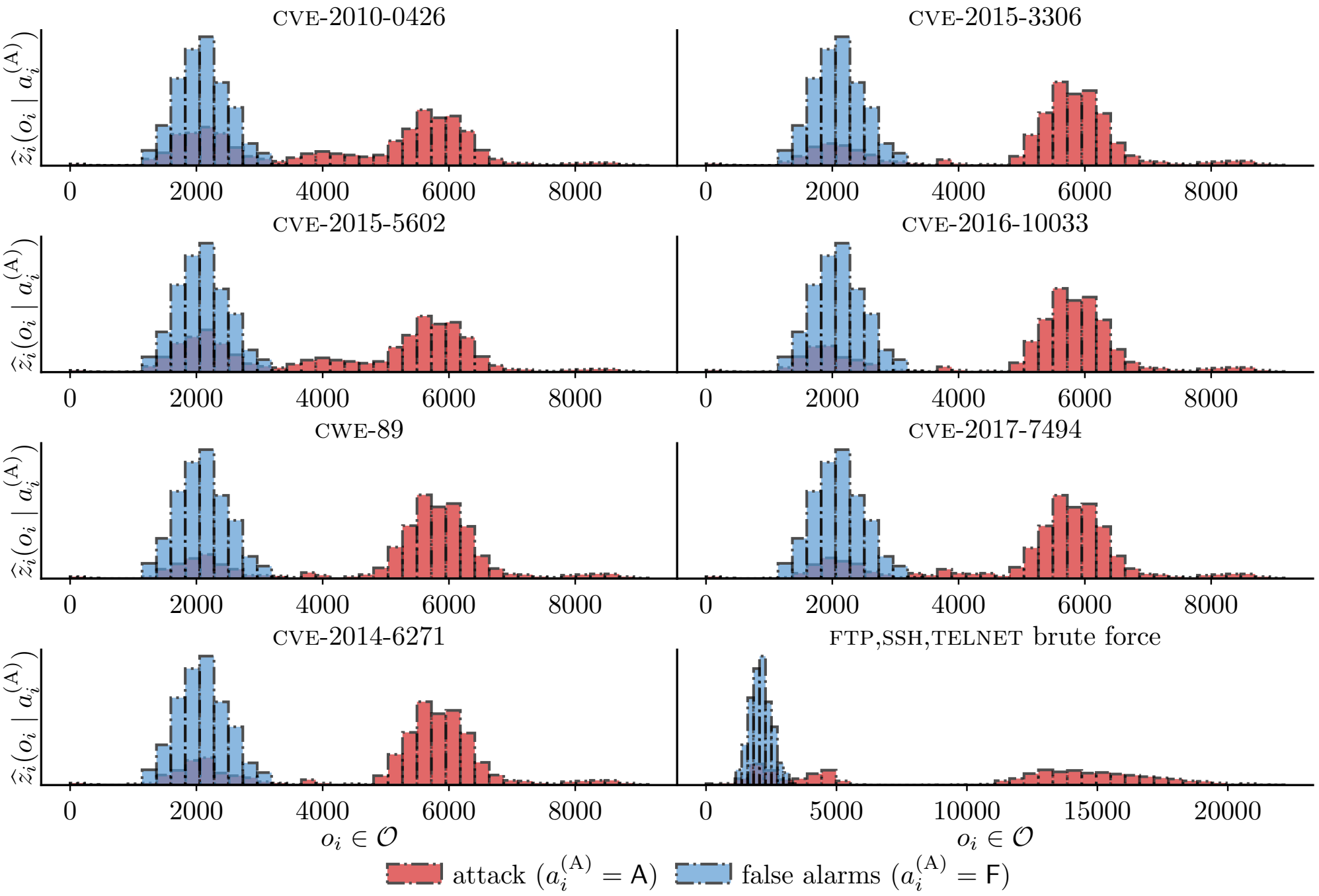

***Figure 4.14:*** *Empirical distributions $\widehat{z}_1(\cdot \mid s_i), \dots, \widehat{z}_{10}(\cdot \mid s_i)$ as estimates of $z_1, \dots, z_{10}$ (4.2) for the containers in Table 4.5.*

## Baseline control strategies

We compare the control strategies of TOLERANCE with those used in current intrusion-tolerant systems, whereby we choose three baseline strategies: NO-RECOVERY, PERIODIC and PERIODIC-ADAPTIVE. The first baseline, NO-RECOVERY, does not recover or add any nodes, which corresponds to the strategy used in traditional intrusion-tolerant systems, such as RAMPART [367] and SECURE-RING [242]. The second baseline, PERIODIC, recovers nodes every $\Delta_R$ time steps but does not add any new nodes. This is the strategy used in most of the intrusion-tolerant systems proposed in prior work, including PBFT [89], VM-FIT [119, 366], WORM-IT [100], PRRW [434, 433], MAFTIA [445], RECOVER [435], SCIT [214, 326], COCA [524], SPIRE [35], ITCIS-PRR [432], CRUTIAL [57], SBFT [161], BFT-SMART [56], UPRIGHT [97], and SKYNET [145]. The third baseline, PERIODIC-ADAPTIVE, recovers nodes every $\Delta_R$ time steps and adds a node when $o_{i,t} \geq 2\mathbb{E}[O_{i,t}]$ (4.2), which approximates the heuristic strategies used in [384], SITAR [77], ITSI [337], and ITUA [345, 421].

## Evaluation results

The results are summarized in Fig. 4.15 and Table 4.8. The brown bars relate to the equilibrial strategies of Games 4.1 and 4.2. The red bars relate to the case where the controllers follow best response strategies against a (different) static attacker. The blue, green, and pink bars relate to the baselines.

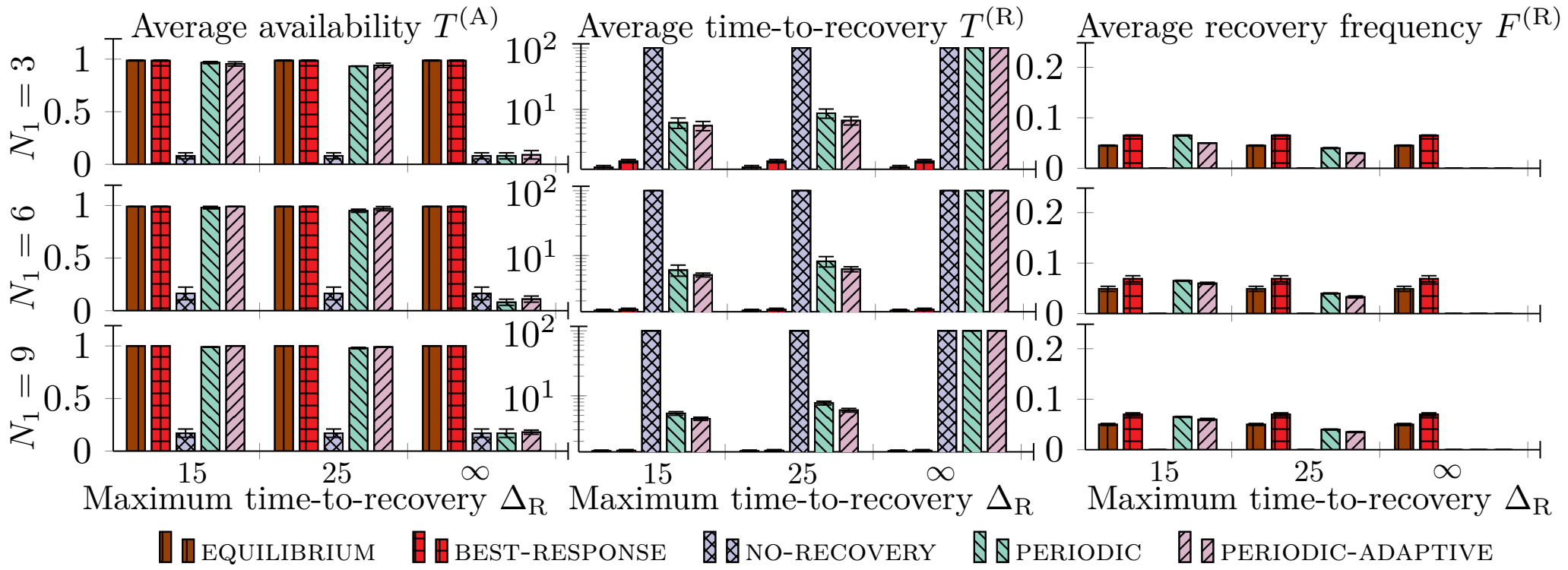

***Figure 4.15:*** *Comparison between our game-theoretic strategies and the baselines; columns represent performance metrics; x-axes indicate values of* $\Delta_R$*; rows relate to the number of initial nodes* $N_1$*; error bars indicate the* 95% *confidence interval from evaluations with* 20 *random seeds.*

The leftmost column in Fig. 4.15 shows the average availability for different values of $\Delta_R$. We observe that the game-theoretic strategies achieve close to 100% service availability in all cases we studied. By contrast, NO-RECOVERY leads to around 0% availability. The availability achieved by PERIODIC and PERIODIC-ADAPTIVE is in-between; they achieve a high availability when $\Delta_R$ is small (i.e., when recoveries are frequent) and a low availability when $\Delta_R \to \infty$. We note in Table 4.8 (on the next page) that increasing $N_1$ from 3 to 9 doubles the availability achieved by NO-RECOVERY but has a negligible impact on the performance of the other strategies.

The middle column in Fig. 4.15 shows the average time-to-recovery $T^{(R)}$. We observe that $T^{(R)}$ of the game-theoretic strategies is an order of magnitude smaller than that of PERIODIC and PERIODIC-ADAPTIVE and two orders of magnitude smaller than that of NO-RECOVERY. This result illustrates the benefit of feedback control, which allows the system to react promptly to intrusions.

Finally, the rightmost column in Fig. 4.15 shows the average frequency of recovery $F^{(R)}$. We note that $F^{(R)}$ of the equilibrium strategy is about the same as PERIODIC and PERIODIC-ADAPTIVE. As expected, $F^{(R)}$ is higher for the best response strategy than the equilibrium strategy. This result demonstrates the exploitability of the best response, allowing the attacker to trigger excess recoveries.

| Control strategy | $\Delta_{\mathrm{R}} = 15$ | | | $\Delta_{\mathrm{R}} = 25$ | | | $\Delta_{\mathrm{R}} = \infty$ | | |
|---|---|---|---|---|---|---|---|---|---|
| | $T^{(\mathrm{A})}$ | $T^{(\mathrm{R})}$ | $F^{(\mathrm{R})}$ | $T^{(\mathrm{A})}$ | $T^{(\mathrm{R})}$ | $F^{(\mathrm{R})}$ | $T^{(\mathrm{A})}$ | $T^{(\mathrm{R})}$ | $F^{(\mathrm{R})}$ |
| | | | | $N_1 = 3$ | | | | | |
| BEST RESPONSE | $\mathbf{0.99 \pm 0.01}$ | $1.43 \pm 0.09$ | $0.09 \pm 0.01$ | $\mathbf{0.99 \pm 0.01}$ | $\mathbf{1.43 \pm 0.09}$ | $0.09 \pm 0.01$ | $\mathbf{0.99 \pm 0.01}$ | $1.43 \pm 0.09$ | $0.09 \pm 0.01$ |
| EQUILIBRIUM | $\mathbf{0.99 \pm 0.01}$ | $\mathbf{1.138 \pm 0.08}$ | $0.06 \pm 0.01$ | $\mathbf{0.99 \pm 0.01}$ | $\mathbf{1.138 \pm 0.0}$ | $0.06 \pm 0.01$ | $\mathbf{0.99 \pm 0.01}$ | $\mathbf{1.138 \pm 0.08}$ | $0.06 \pm 0.01$ |
| NO-RECOVERY | $0.08 \pm 0.06$ | $10^3 \pm 0.00$ | $\mathbf{0.00 \pm 0.00}$ | $0.08 \pm 0.06$ | $10^3 \pm 0.00$ | $\mathbf{0.00 \pm 0.00}$ | $0.08 \pm 0.06$ | $10^3 \pm 0.00$ | $\mathbf{0.00 \pm 0.00}$ |
| PERIODIC | $0.97 \pm 0.01$ | $6.06 \pm 1.16$ | $0.065 \pm 0.01$ | $0.93 \pm 0.01$ | $8.64 \pm 1.48$ | $0.04 \pm 0.01$ | $0.08 \pm 0.06$ | $10^3 \pm 0.00$ | $\mathbf{0.00 \pm 0.00}$ |
| PERIODIC-ADAPTIVE | $0.95 \pm 0.02$ | $5.42 \pm 0.93$ | $0.05 \pm 0.01$ | $0.94 \pm 0.02$ | $6.57 \pm 1.01$ | $0.03 \pm 0.01$ | $0.09 \pm 0.04$ | $10^3 \pm 0.00$ | $\mathbf{0.00 \pm 0.00}$ |
| | | | | $N_1 = 6$ | | | | | |
| BEST RESPONSE | $\mathbf{0.99 \pm 0.01}$ | $1.47 \pm 0.07$ | $0.07 \pm 0.01$ | $\mathbf{0.99 \pm 0.01}$ | $1.47 \pm 0.07$ | $0.07 \pm 0.01$ | $\mathbf{0.99 \pm 0.01}$ | $1.47 \pm 0.07$ | $0.07 \pm 0.01$ |
| EQUILIBRIUM | $\mathbf{0.99 \pm 0.01}$ | $\mathbf{1.42 \pm 0.06}$ | $0.05 \pm 0.01$ | $\mathbf{0.99 \pm 0.01}$ | $\mathbf{1.42 \pm 0.07}$ | $0.05 \pm 0.01$ | $\mathbf{0.99 \pm 0.01}$ | $\mathbf{1.42 \pm 0.06}$ | $0.05 \pm 0.01$ |
| NO-RECOVERY | $0.16 \pm 0.06$ | $10^3 \pm 0.00$ | $\mathbf{0.00 \pm 0.00}$ | $0.16 \pm 0.06$ | $10^3 \pm 0.00$ | $\mathbf{0.00 \pm 0.00}$ | $0.16 \pm 0.06$ | $10^3 \pm 0.00$ | $\mathbf{0.00 \pm 0.00}$ |
| PERIODIC | $0.98 \pm 0.01$ | $5.96 \pm 1.16$ | $0.065 \pm 0.01$ | $0.95 \pm 0.02$ | $8.13 \pm 1.48$ | $0.04 \pm 0.01$ | $0.16 \pm 0.03$ | $10^3 \pm 0.00$ | $\mathbf{0.00 \pm 0.00}$ |
| PERIODIC-ADAPTIVE | $\mathbf{0.99 \pm 0.01}$ | $5.02 \pm 0.34$ | $0.06 \pm 0.01$ | $0.97 \pm 0.02$ | $6.16 \pm 0.54$ | $0.03 \pm 0.01$ | $0.17 \pm 0.03$ | $10^3 \pm 0.00$ | $\mathbf{0.00 \pm 0.00}$ |
| | | | | $N_1 = 9$ | | | | | |
| BEST RESPONSE | $\mathbf{1.00 \pm 0.00}$ | $1.44 \pm 0.05$ | $0.07 \pm 0.01$ | $\mathbf{1.00 \pm 0.00}$ | $1.44 \pm 0.05$ | $0.07 \pm 0.01$ | $\mathbf{1.00 \pm 0.00}$ | $1.44 \pm 0.05$ | $0.07 \pm 0.01$ |
| EQUILIBRIUM | $\mathbf{1.00 \pm 0.00}$ | $\mathbf{1.42 \pm 0.04}$ | $0.05 \pm 0.01$ | $\mathbf{1.00 \pm 0.00}$ | $\mathbf{1.42 \pm 0.04}$ | $0.05 \pm 0.01$ | $\mathbf{1.00 \pm 0.00}$ | $\mathbf{1.42 \pm 0.04}$ | $0.05 \pm 0.01$ |
| NO-RECOVERY | $0.17 \pm 0.04$ | $10^3 \pm 0.00$ | $\mathbf{0.00 \pm 0.00}$ | $0.17 \pm 0.04$ | $10^3 \pm 0.00$ | $\mathbf{0.00 \pm 0.00}$ | $0.17 \pm 0.04$ | $10^3 \pm 0.00$ | $\mathbf{0.00 \pm 0.00}$ |
| PERIODIC | $0.99 \pm 0.00$ | $5.37 \pm 0.34$ | $0.06 \pm 0.01$ | $0.98 \pm 0.01$ | $7.74 \pm 0.51$ | $0.04 \pm 0.01$ | $0.17 \pm 0.04$ | $10^3 \pm 0.00$ | $\mathbf{0.00 \pm 0.00}$ |
| PERIODIC-ADAPTIVE | $\mathbf{1.00 \pm 0.00}$ | $4.44 \pm 0.25$ | $0.06 \pm 0.01$ | $0.99 \pm 0.01$ | $6.01 \pm 0.39$ | $0.04 \pm 0.01$ | $0.18 \pm 0.02$ | $10^3 \pm 0.00$ | $\mathbf{0.00 \pm 0.00}$ |

***Table 4.8:*** *Comparison between our game-theoretic control strategies and the baselines; columns indicate values of* $\Delta_{\mathrm{R}}$*; subcolumns represent performance metrics; row groups relate to the number of initial nodes* $N_1$*; numbers indicate the mean and the* 95% *confidence interval from evaluations with* 20 *random seeds.*

**Discussion of the testbed results and the theoretical analysis**

The key findings from the evaluation and the analysis are summarized below.

- The game-theoretic strategies can achieve a lower time-to-recovery and a higher service availability than the fixed periodic strategies used in state-of-the-art intrusion-tolerant systems (Fig. 4.15). The BTR constraint and the availability constraint provide theoretical guarantees. No such guarantees have been given for the baselines.
- The performance of the game-theoretic strategies depends on the accuracy of the intrusion detection model $z_i$ (4.2); see Fig. 4.9.
- The best response strategies in both Game 4.1 and Game 4.2 have threshold properties (Thms. 4.3–4.5, Cor. 4.1), which allow us to compute them efficiently (Fig. 4.7, Fig. 4.12).
- The benefit of using an adaptive replication strategy as opposed to a fixed strategy is mainly prominent when node crashes are frequent; see Fig. 4.12.b and cf. the results of PERIODIC and PERIODIC-ADAPTIVE in Fig. 4.15.
- A non-equilibrium strategy is exploitable in the sense that it allows a strategic attacker to trigger excess recoveries; cf. the results of EQUILIBRIUM and BEST-RESPONSE in Fig. 4.15. However, the increase in operational cost caused by the excess recoveries is relatively small in the scenarios we studied.

While the results demonstrate clear benefits of TOLERANCE compared to current intrusion-tolerant systems, the performance of TOLERANCE depends on the accuracy of the intrusion detection model $\widehat{z}_i$; see Fig. 4.14 and Fig. 4.9. This dependence means that practical deployments of TOLERANCE require a statistical intrusion detection model for estimating the probability of intrusion (4.3). This model can be realized in many ways. It can, for example, be based on anomaly detection methods or machine learning techniques. Further, the detection model can use different data sources, e.g., log files, IDS alerts, etc. Our proof-of-concept implementation of TOLERANCE uses the SNORT IDS (Roesch, 1999) as the data source and obtains the distribution of IDS alerts using maximum likelihood estimation.

## 4.8 Related Work

Intrusion tolerance is studied in several broad areas of research, including: Byzantine fault tolerance [117], dependability [165, 81], reliability [60], survivability [252], and cyber resilience [153, 168, 250, 249, 132, 215]. This research effort has led to many mechanisms for implementing intrusion-tolerant systems, such as: intrusion-tolerant consensus protocols [117, 165, 81, 60, 118, 505, 161, 15, 97, 56], software diversification schemes [155], geo-replication schemes [238], cryptographic mechanisms [504, 343], and defenses against denial of service [237]. These mechanisms provide the foundation for TOLERANCE, which adds automated recovery and replication control. While TOLERANCE builds on all of the above works, we limit the following discussion to explain how TOLERANCE differs from current intrusion-tolerant systems and how it relates to prior work that uses feedback control.

### Intrusion-tolerant systems

Intrusion-tolerant systems developed in prior work include PBFT [89], ZYZZYVA [248], HQ [101], HOTSTUFF [505], VM-FIT [119, 366], WORM-IT [100], PRRW [434], RECOVER [435], SCIT [214, 326], COCA [524], [384], SPIRE [35], ITCIS-PRR [432], CRUTIAL [57], UPRIGHT [97], BFT-SMART [56], SBFT [161], SITAR [77], ITUA [345, 421], MAFTIA [445], ITSI [337], and SKYNET [145]. All of them are based on intrusion-tolerant consensus protocols and support recovery, either directly or indirectly through external recovery services, like PHOENIX [470]. TOLERANCE differs from these systems in two main ways.

First, TOLERANCE uses feedback control to decide when to perform intrusion recovery. This contrasts with all referenced systems, which use periodic or heuristic recovery schemes. (PRRW, RECOVER, CRUTIAL, SCIT, SITAR, ITSI, and ITUA can be implemented with feedback-based recovery, but they do not specify how to implement such recovery strategies.) Second, TOLERANCE uses an adaptive replication strategy. In comparison, all referenced systems use static replication strategies except SITAR, [384], ITUA, and ITSI. They implement adaptive replication based on time-outs and static rules instead of feedback control. The benefit of feedback control is that it allows the system to adapt promptly to intrusions, not having to wait for a time-out.

### Intrusion response through feedback control

Intrusion response through feedback control is an active area of research that uses concepts and methods from various emergent and traditional fields. Most notably from reinforcement learning (see surveys [215, 474, 328]), control theory (see examples [193, 251, 23, 252, 108]), and game theory (see textbooks [11, 149]). While these works have obtained promising results, none of them considers the integration with intrusion-tolerant systems as we do in this paper. Further, a drawback of the existing solutions is that many are inefficient (compared to our threshold-based solutions) and lack safety guarantees. Finally, and most importantly, nearly all of the previous works are limited to simulation environments for evaluation, and it is unclear how their results generalize to operational systems. In contrast, our game-theoretic strategies are useful in practice: they can be integrated with existing intrusion-tolerant systems, they satisfy safety constraints, and they are computationally efficient.

## 4.9 Conclusion

This paper presents TOLERANCE: a novel control architecture for intrusion-tolerant systems. TOLERANCE is based on a formulation of intrusion tolerance for a system with service replicas as a two-level game: a local game models intrusion recovery and a global game models replication control. We prove the existence of equilibria in both games and derive a threshold structure of the best responses, which enables efficient computation of control strategies. We implement and evaluate the game-theoretic control strategies on a testbed and assess their performance against 10 types of network intrusions. The testbed results demonstrate that our game-theoretic strategies can significantly improve service availability and reduce the operational cost of state-of-the-art intrusion-tolerant systems. In addition, our game strategies can meet any chosen service availability and time-to-recovery, bridging the gap between theoretical and operational performance.

In the broader context of this thesis, this paper demonstrates the generality of our methodology[16] by applying it to a different type of response use case than those studied in papers 1–3, namely intrusion tolerance. In this paper, we assume that both the attacker and the controller have correctly specified models. In the next chapter of the thesis, we relax this assumption and show how our methodology can handle cases where the attacker and the controller have *misspecified models.*

## Acknowledgments

The authors would like to thank Emil Lupu for his useful input to this research. The authors are also grateful to Forough Shahab Samani, Xiaoxuan Wang, and Duc Huy Le for their constructive comments on a draft of this paper.

## Appendix

### A Proof of Theorem 4.3 and Corollary 4.1

Given an attacker strategy $\pi^{(\mathrm{A})}_{i,t}$, a best response strategy $\tilde{\pi}^{(\mathrm{C})}_{i,t}$ is an optimal strategy in a POMDP $\mathcal{M}$ (Thm. 4.2). Hence, it suffices to show that there exists an optimal strategy in $\mathcal{M}$ that satisfies the threshold structure in (4.6). Towards this end, we state and prove the following four lemmas.

**Remark 4.16** (Notation). *For ease of notation, let $J_i(\mathbf{b}_{i,1}(\mathbb{C}))$ denote the cost-to-go function in the best response POMDP $\mathcal{M}$, i.e., $J_i(\mathbf{b}_{i,1}(\mathbb{C})) = \mathbb{E}_{\tilde{\pi}^{(\mathrm{C})}_{i,t}}[J_i \mid \mathbf{b}_{i,1}(\mathbb{C})]$ (4.4).*

**Lemma 4.3.** *The expectation in (4.5) is well-defined for each strategy pair $\boldsymbol{\pi}_i$.*

[16] See the methodology chapter.

*Proof.* Since the sample space of the random vectors $(\mathbf{I}_t^{(\mathrm{C})}, \mathbf{I}_t^{(\mathrm{A})})$ is finite (and measurable) for each $t$, the space of realizable histories $\mathbf{h}_t^{(\mathrm{C})} \times \mathbf{h}_t^{(\mathrm{A})} \in \mathcal{H}_t$ is countable. By the Ionescu Tulcea extension theorem, it thus follows that there exists a well-defined probability measure $\mathbb{P}$ over $\mathcal{H}_t$ for $t = 1, 2, \ldots$ (Ionescu Tulcea, 1949). Further, Lemma 4.1 implies that the expectation is finite. □

**Lemma 4.4.** *The controller's best response* POMDP $\mathcal{M}$ *defined in (4.5) with the* BTR *constraint can be converted into a sequence of unconstrained* POMDPs $(\mathcal{M}_k)_{k\geq 1}$.

*Proof.* Using Lemma 4.1 and Lemma 4.3 we obtain

$$\begin{aligned}
&\arg\min_{\pi_{i,t}^{(\mathrm{C})}} \mathbb{E}_{\pi_{i,t}^{(\mathrm{C})}}\left[\sum_{t=1}^{\infty} \gamma^{t-1} C_{i,t} \mid \mathbf{b}_{i,1}(\mathbb{C}) = 0\right] \stackrel{(a)}{=} \arg\min_{\pi_{i,t}^{(\mathrm{C})}} \Bigg[\mathbb{E}_{\pi_{i,t}^{(\mathrm{C})}}\Bigg[ \qquad (4.12)\\
&\sum_{t=1}^{\tau_1} \gamma^{t-1} C_{i,t} \mid \mathbf{b}_{i,t}(\mathbb{C}) = 0\Bigg] + \mathbb{E}_{\pi_{i,t}^{(\mathrm{C})}}\left[\sum_{t=\tau_1+1}^{\tau_2} \gamma^{t-1} C_{i,t} \mid \mathbf{b}_{i,\tau_1+1}(\mathbb{C}) = 0\right] + \ldots\Bigg]\\
&= \arg\min_{\pi_{i,t}^{(\mathrm{C})}} \mathbb{E}_{\pi_{i,t}^{(\mathrm{C})}}\left[\gamma^{\tau_1} J_i(0) + \sum_{t=1}^{\tau_1} \gamma^{t-1} C_{i,t} \mid \mathbf{b}_{i,1}(\mathbb{C}) = 0\right],
\end{aligned}$$

where $\tau_i \triangleq \inf\{t \mid t > \tau_{i-1}, a_t^{(\mathrm{C})} = \mathsf{R}\}$ is the $i$th recovery time (stopping time); $\tau_0 = 0$; $C_{i,t}$ is a random variable representing the cost of node $i$ at time $t$; and (a) follows from linearity of $\mathbb{E}$. Since $J_i(0)$ is completely determined by the recovery strategy, it can be seen as a fixed recovery cost. As a consequence, the final expression in (4.12) defines an optimal stopping problem. (Recall that due to Lemma 4.1, the crash probability $p_{\mathrm{C},i}$ is captured by $\gamma$.) □

## A.1 Proof of Theorem 4.3

Lemma 1.1 of Paper 1 implies that the recovery set has the form $\mathscr{S}_{i,t} = [\alpha^\star_{i,t}, \kappa_{i,t}]$, where $0 \leq \alpha^\star_{i,t} \leq \kappa_{i,t} \leq 1$. Thus, it suffices to show that $\kappa_{i,t} = 1$. Let $c(\mathsf{R}) \triangleq c_{\mathrm{N}}(s_{i,t}, \mathsf{R}) + \gamma J_i(0)$ be the stopping cost in (4.12). It follows from Bellman's optimality equation that

$$\begin{aligned}
\pi_{i,t}^{(\mathrm{C})}(1) &\in \arg\min_{a_i^{(\mathrm{C})} \in \{\mathsf{R},\mathsf{W}\}} \Bigg[\overbrace{c(\mathsf{R})}^{a_i^{(\mathrm{C})}=\mathsf{R}}, \overbrace{c_{\mathrm{N}}(\mathbb{C}, \mathsf{W}) + \gamma \mathbb{E}_{\pi_{i,t}^{(\mathrm{C})}}[J^\star_{i,t+1}(1)]}^{a_i^{(\mathrm{C})}=\mathsf{W}}\Bigg]\\
&\stackrel{(a)}{=} \arg\min_{a_i^{(\mathrm{C})} \in \{\mathsf{R},\mathsf{W}\}} \left[c(\mathsf{R}), \gamma^{\tau-1} c(\mathsf{R}) + \sum_{t=1}^{\tau-1} \gamma^{t-1} c_{\mathrm{N}}(\mathbb{C}, \mathsf{W})\right]\\
&= \arg\min_{a_i^{(\mathrm{C})} \in \{\mathsf{R},\mathsf{W}\}} \left[c(\mathsf{R}), \gamma^{\tau-1} c(\mathsf{R}) + \sum_{t=1}^{\tau-1} \gamma^{t-1} \eta\right]
\end{aligned}$$

$$= \operatorname*{arg\,min}_{a_i^{(\mathrm{C})} \in \{\mathsf{R},\mathsf{W}\}} \left[ c(\mathsf{R}), \gamma^{\tau-1} c(\mathsf{R}) + \frac{\eta(1-\gamma^{\tau-2})}{1-\gamma} \right] \overset{(b)}{=} \{\mathsf{R}\},$$

where (a) follows because $s_i = \mathbb{C}$ is an absorbing state until a recovery occurs (using the discounted formulation in Lemma 4.1); and (b) follows from (4.4) and the fact that $\eta > 1$, which means that the cost per time step is upper bounded by $\eta$ (4.4). Consequently, $\tilde{\pi}_{i,t}^{(\mathrm{C})}(1) = \mathsf{R} \implies 1 \in \mathscr{S}_{i,t} \implies \mathscr{S}_{i,t} = [\alpha^{\star}_{i,t}, 1]$. □

## A.2 Proof of Corollary 4.1

Theorem 2 in the background chapter states that the cost-to-go function $J_i^{\star}(\mathbf{b}(\mathbb{C}))$ becomes stationary when $\Delta_{\mathrm{R}} \to \infty$. Therefore, it follows from Thm. 4.3 and Lemma 1.1 of Paper 1 that there exists a stationary best response strategy. Such a strategy induces a stationary recovery set $\mathscr{S}_i$, which means that $\alpha_i^{\star}$ (4.6) is time-independent. □

# B Proof of Theorem 4.5

Computing a best response for the controller in Game 4.2 amounts to computing an optimal strategy in a CMDP with the average-cost optimality criterion. Hence, it suffices to show that there exists an optimal strategy in the CMDP that satisfies the threshold structure in (4.9).

By introducing a Lagrange multiplier $\lambda \geq 0$ and defining the immediate cost to be $c_{\lambda}(s_t, a_t^{(\mathrm{C})}) \triangleq a_t^{(\mathrm{C})} + \lambda \mathbb{1}_{s_t < f+1}$ we can reformulate the average-cost CMDP as a discounted Lagrangian MDP through Lagrangian relaxation (Thm. 3.7, Altman, 1999) and the vanishing discount method (Sennott, 1989). The best response strategy in this MDP satisfies

Lagrangian cost function.

$$\tilde{\pi}_{\lambda}^{(\mathrm{C})}(s_t) \in \operatorname*{arg\,min}_{a_t^{(\mathrm{C})} \in \{0,1\}} \left[ c_{\lambda}(s_t, a_t^{(\mathrm{C})}) + \gamma \mathbb{E}_{S_{t+1}} \left[ \tilde{J}_{\lambda}(S_{t+1}) \mid s_t, a_t^{(\mathrm{C})} \right] \right], \tag{4.13}$$

Best response cost-to-go function.

where $\gamma \in [0,1)$ is a discount factor and $\tilde{J}_{\lambda}$ is the best response cost-to-go function induced by a best response strategy $\tilde{\pi}_{\lambda}^{(\mathrm{C})}$ (Thm. 3.6, Altman, 1999).

Our approach to proving Thm. 4.5 follows the approach described in (Ngo and Krishnamurthy, 2010). The main steps of the proof are as follows. We first show that there exists an optimal threshold strategy in the discounted Lagrangian MDP for any non-negative Lagrange multiplier and discount factor $\gamma \in [0,1)$; see Lemma 4.8 on page 192. Next, we use the vanishing discount method to establish that the threshold structure applies under the average cost optimality criterion (Sennott, 1989); see Lemma 4.11 on page 193. We then prove that the optimal cost in the CMDP corresponds to $\inf_{\pi^{(\mathrm{C})}} \sup_{\lambda} J_{\lambda}$, where $J_{\lambda}$ is the cost-to-go function in the Lagrangian MDP with the average cost optimality criterion; see Lemma 4.12 on

page 194. As the sup and inf can be interchanged, an optimal strategy for the CMDP can be obtained by first computing an optimal strategy in the Lagrangian MDP and then maximizing with respect to the Lagrange multiplier $\lambda$. Then we note that since the CMDP only has a single constraint (4.7), an optimal strategy in the CMDP requires at most one randomization (Thm. 4.4, Altman, 1999). Consequently, there exists an optimal strategy in the CMDP that is a randomized mixture of two optimal deterministic strategies in the Lagrangian MDP for different Lagrange multipliers (Thm. 1, Ma et al., 1986). Since there exist optimal threshold strategies in the Lagrangian MDP for any non-negative Lagrange multiplier $\lambda$, the theorem statement follows. We provide proof below, starting with some lemmas.

**Lemma 4.5.** *The expectation in (4.8a) is well-defined for each strategy pair* $\boldsymbol{\pi}$.

*Proof.* Since the sample space of $(S_t, A_t^{(\mathrm{C})}, \mathbf{A}_t^{(\mathrm{A})})$ is finite (and measurable) for each $t$, the statement follows from Lemma 4.3. □

**Lemma 4.6.** *Given (T5.1), $\gamma \in [0, 1)$, $\lambda \geq 0$, and $s > f$, then $\tilde{J}_\lambda$ has non-decreasing differences, i.e., $\tilde{J}_\lambda(s+1) - \tilde{J}_\lambda(s)$ is non-decreasing in $s$ for all $s \in \mathcal{S}_\mathrm{S} \setminus \{s_{\max}\}$.*

*Proof.* We establish the non-decreasing differences property using the value iteration algorithm (Eq. 6.3.2–6.3.4, Puterman, 1994). Let $\tilde{J}_{\lambda,k}$ denote the optimal cost-to-go function at iteration $k$ of value iteration. Then, $\lim_{k\to\infty} \tilde{J}_{\lambda,k} = \tilde{J}_\lambda$ (Thm. 6.3.1, Puterman, 1994)(Thm. 6, p. 160, Banach, 1922). Let $\tilde{J}_{\lambda,1}(s) = 0\ \forall s$. Proceed by mathematical induction.

BASE CASE: $\tilde{J}_{\lambda,1}(s+1) - \tilde{J}_{\lambda,1}(s) = 0\ \forall s \in \mathcal{S}_\mathrm{S} \setminus \{s_{\max}\}$.

INDUCTIVE CASE: Assume that $\tilde{J}_{\lambda,l}(s+1) - \tilde{J}_{\lambda,l}(s)$ is non-decreasing in $s$ for $l = k-1, k-2, \ldots, 2$ and $s_{\max} > s > f$. We show that these assumptions also imply that the statement holds for $l = k$. There are four cases:

1. If $\tilde{\pi}_\lambda^{(\mathrm{C})}(s+1) = \tilde{\pi}_\lambda^{(\mathrm{C})}(s) = 1$, then it follows from (T5.1) that

$$\begin{aligned}
&\tilde{J}_{\lambda,k}(s+1) - \tilde{J}_{\lambda,k}(s)\\
&= \gamma\mathbb{E}_{S'}[\tilde{J}_{\lambda,k-1}(S'+1) \mid S = s+1] - \gamma\mathbb{E}_{S'}[\tilde{J}_{\lambda,k-1}(S'+1) \mid S = s]\\
&= \mathbb{E}_{S'}[\tilde{J}_{\lambda,k-1}(S'+2) - \tilde{J}_{\lambda,k-1}(S'+1) \mid S = s],
\end{aligned}$$

which is non-decreasing in $s$ by the induction hypothesis. ($S'$ is a random variable that represents the number of healthy nodes at the subsequent time step, given that the current state is $S = s$; this variable depends on the attacker's actions and the local control strategies.)

2. If $\tilde{\pi}_\lambda^{(\mathrm{C})}(s+1) = \tilde{\pi}_\lambda^{(\mathrm{C})}(s) = 0$, then it follows from (T5.1) that

$$\tilde{J}_{\lambda,k}(s+1) - \tilde{J}_{\lambda,k}(s)$$

$$
\begin{aligned}
&= \gamma \mathbb{E}_{S'}[\tilde{J}_{\lambda,k-1}(S') \mid S = s+1] - \gamma \mathbb{E}_{S'}[\tilde{J}_{\lambda,k-1}(S') \mid S = s] \\
&= \mathbb{E}_{S'}[\tilde{J}_{\lambda,k-1}(S'+1) - \tilde{J}_{\lambda,k-1}(S') \mid S = s],
\end{aligned}
$$

which is non-decreasing in $s$ by the induction hypothesis.

3. If $\tilde{\pi}^{(\mathrm{C})}_{\lambda}(s+1) = 0$ and $\tilde{\pi}^{(\mathrm{C})}_{\lambda}(s) = 1$, then (T5.1) implies that

$$
\begin{aligned}
\tilde{J}_{\lambda,k}(s+1) - \tilde{J}_{\lambda,k}(s) &= \gamma \mathbb{E}_{S'}[\tilde{J}_{\lambda,k-1}(S'+1) - \tilde{J}_{\lambda,k-1}(S'+1) \mid S = s] - 1 \\
&= -1.
\end{aligned}
$$

4. If $\tilde{\pi}^{(\mathrm{C})}_{\lambda}(s+1) = 1$ and $\tilde{\pi}^{(\mathrm{C})}_{\lambda}(s) = 0$, then (T5.1) implies that

$$
\begin{aligned}
\tilde{J}_{\lambda,k}(s+1) - \tilde{J}_{\lambda,k}(s) &= 1 + \mathbb{E}_{S'}[\tilde{J}_{\lambda,k-1}(S'+2) - \tilde{J}_{\lambda,k-1}(S') \mid S = s] \\
&= 1 + \mathbb{E}_{S'}[\tilde{J}_{\lambda,k-1}(S'+2) - \tilde{J}_{\lambda,k-1}(S'+1) \\
&\quad + \tilde{J}_{\lambda,k-1}(S'+1) - \tilde{J}_{\lambda,k-1}(S') \mid S = s],
\end{aligned}
$$

which is non-decreasing in $s$ by the induction hypothesis.

□

Let $\tilde{Q}_{\lambda}(s, a^{(\mathrm{C})})$ be the best response Q-function (6) associated with (4.13).

**Lemma 4.7.** *Given (T5.1), $\gamma \in [0,1)$, $\lambda \geq 0$, and $s > f$, then $\tilde{Q}_{\lambda}(s, a^{(\mathrm{C})})$ is supermodular.*

*Proof.* $\tilde{Q}_{\lambda}$ is supermodular if the following inequality holds:

$$
\begin{aligned}
&\tilde{Q}_{\lambda}(s,1) - \tilde{Q}_{\lambda}(s,0) \leq \tilde{Q}_{\lambda}(s+1,1) - \tilde{Q}_{\lambda}(s+1,0) \\
\overset{\text{(T5.1)}}{\Longleftrightarrow}\ &\mathbb{E}_{S'}[\tilde{J}_{\lambda}(S'+1) - \tilde{J}_{\lambda}(S') \mid S = s] \leq \mathbb{E}_{S'}[\tilde{J}_{\lambda}(S'+2) - \tilde{J}_{\lambda}(S'+1) \mid S = s],
\end{aligned}
$$

which follows from Lemma 4.6. □

**Lemma 4.8** (Threshold structure in the discounted Lagrangian MDP)**.** *Given (T5.1), $\gamma \in [0,1)$ and $\lambda \geq 0$, then there exists an optimal strategy in the discounted Lagrangian MDP that satisfies (4.13) and has the following form.*

$$
\tilde{\pi}^{(\mathrm{C})}_{\lambda}(s) = 1 \iff s \leq \beta \qquad \forall s \in \mathcal{S}_{\mathrm{S}}, \tag{4.14}
$$

*where $\beta \geq f$ is a threshold.*

*Proof.* We can rewrite (4.13) as

$$
\tilde{\pi}^{(\mathrm{C})}_{\lambda}(s_t) \in \underset{a^{(\mathrm{C})}_t \in \{0,1\}}{\arg\min} \left[ c_{\lambda}(s_t, a^{(\mathrm{C})}_t) + \gamma \mathbb{E}_{S_{t+1}} \left[ \tilde{J}_{\lambda}(S_{t+1}) \mid s_t, a^{(\mathrm{C})}_t \right] \right]
$$

$$= \underset{a_t^{(\mathrm{C})} \in \{0,1\}}{\arg\min}\ \tilde{Q}_\lambda(s_t, a_t^{(\mathrm{C})}). \tag{4.15}$$

Since $\tilde{Q}_\lambda$ is supermodular (Lemma 4.7), it follows from Topkis' theorem (Thm. 6.1, Topkis, 1978) that the minimizer of (4.15) is (weakly) decreasing in $s$. As a consequence, there exists a best response strategy that satisfies (4.14). □

**Lemma 4.9.** *For some $\lambda \geq 0$, let $(\gamma_n)_{n\geq 1}$ be any sequence of discount factors $\gamma_n \in [0,1]$ converging to 1. Let $(\pi^{(\mathrm{C})}_{\lambda,\gamma_n})_{n\geq 1}$ be an associated sequence of stationary, deterministic, best response strategies for the discounted Lagrangian MDP, each of which satisfies (4.14). Then there exists a subsequence $(\upsilon_n)_{n\geq 1}$ of $(\gamma_n)_{n\geq 1}$ and a stationary deterministic strategy $\tilde{\pi}^{(\mathrm{C})}_\lambda$ satisfying (4.14) that is a limit point of $(\pi^{(\mathrm{C})}_{\lambda,\upsilon_n})_{n\geq 1}$. That is, for each state $s \in \mathcal{S}_\mathrm{S}$, there exists an integer $N(s)$ such that $\tilde{\pi}^{(\mathrm{C})}_\lambda(s) = \pi^{(\mathrm{C})}_{\lambda,\upsilon_n}(s)$ for all $n \geq N(s)$.*

*Proof.* This statement is proven in (Hordijk, 1971) and also used in (Derman, 1970). A more accessible version of the proof is given in (Lemma 1, Sennott, 1989). For completeness, we give the proof here since it is very short. Let the controller's action space $\{0,1\}$ be endowed with the discrete topology. By Tychonoff's theorem (p. 383, Ash, 1972), the product of compact spaces is compact, and hence $\{0,1\}^{|\mathcal{S}_\mathrm{S}|}$ is compact and metrizable for any value of $s_{\max}$. (Recall that $|\mathcal{S}_\mathrm{S}| = s_{\max} + 1$.) Every stationary deterministic control strategy can be regarded as a point in this space. Hence, $(\pi^{(\mathrm{C})}_{\lambda,\gamma_n})_{n\geq 1}$ is a bounded sequence in this space. Therefore, by the Bolzano–Weierstrass theorem, there exists a subsequence $(\pi^{(\mathrm{C})}_{\lambda,\upsilon_n})_{n\geq 1}$ converging to a deterministic stationary strategy $\tilde{\pi}^{(\mathrm{C})}_\lambda$ that satisfies (4.14). □

**Lemma 4.10.** *Any stationary strategy $\tilde{\pi}^{(\mathrm{C})}_\lambda$ obtained by Lemma 4.9 is a best response in the Lagrangian MDP obtained by replacing the discounted optimality criterion in (4.13) with the average cost optimality criterion.*

The proof of this lemma is given in (Thm. 1, Sennott, 1989) and holds if Assumptions 1-3′ in (Thm. 1, Sennott, 1989) are satisfied. Assumption 1 asserts that $J^\star_\lambda(s)$ is finite for every state $s$, Lagrange multiplier $\lambda \geq 0$, and discount factor $\gamma \in [0,1)$. Assumptions 2–3′ require that the average cost is bounded. These assumptions follow directly from the definition of Game 4.2 and (T4.1)–(T4.2). In the interest of space, we do not repeat the full proof here. The proof primarily relies on Tychonoff's theorem (p. 383, Ash, 1972), which allows to establish that the average cost is the limit of a sequence in the space. The statement of the lemma is then demonstrated by applying Fatou's lemma (p. 257, Ash, 1972) and the dominated convergence theorem (p. 50, Ash, 1972).

**Lemma 4.11** (Threshold structure in the undiscounted Lagrangian MDP)**.** *The threshold structure of the discounted Lagrangian MDP defined in (4.14) is inherited by any best response strategy in the undiscounted Lagrangian MDP obtained by Lemma 4.10.*

The proof of this lemma follows from Lemmas 4.5–4.10; see (Thm. 2.2, Remark 1, Ross, 1983) for details.

**Lemma 4.12.** *Let $\tilde{J}$ (4.7) denote the optimal cost-to-go in the best response CMDP for the controller against an attacker strategy $\pi^{(\mathrm{A})}$ and let $J_\lambda$ be the cost-to-go function in the undiscounted Lagrangian MDP. Then*

$$\tilde{J} = \inf_{\pi^{(\mathrm{C})}} \sup_{\lambda \geq 0} \mathbb{E}_{\pi^{(\mathrm{C})}}[J_\lambda(S_1)] = \sup_{\lambda \geq 0} \inf_{\pi^{(\mathrm{C})}} \mathbb{E}_{\pi^{(\mathrm{C})}}[J_\lambda(S_1)].$$

*Proof.* That the sup and inf can be interchanged follows from standard minimax theorems, see (Thm. 3.6, Altman, 1999). Hence, it only remains to prove that the best response cost $\tilde{J}$ is equal the optimal cost $\inf_{\pi^{(\mathrm{C})}} \mathbb{E}_{\pi^{(\mathrm{C})}}[J_\lambda(S_1)]$ in the undiscounted Lagrangian MDP when taking the supremum with respect to the Lagrange multiplier $\lambda$. If the constraint (4.8b) is not feasible, then $\sup_{\lambda \geq 0} \mathbb{E}_{\pi^{(\mathrm{C})}}[J_\lambda(S_1)] = \infty$ for any $\pi^{(\mathrm{C})}$ and thus $\tilde{J} = \inf_{\pi^{(\mathrm{C})}} \sup_{\lambda \geq 0} \mathbb{E}_{\pi^{(\mathrm{C})}}[J_\lambda(S_1)]$. If (4.8b) is feasible, then the supremum is obtained by choosing $\lambda = 0$. Consequently, $\mathbb{E}_{\pi^{(\mathrm{C})}}[J_\lambda(S_1)] = J$ (4.7). Hence $\inf_{\pi^{(\mathrm{C})}} \sup_{\lambda \geq 0} \mathbb{E}_{\pi^{(\mathrm{C})}}[J_\lambda(S_1)] = \inf_{\pi^{(\mathrm{C})}} \mathbb{E}_{\pi^{(\mathrm{C})}}[J] = \tilde{J}$. □

## B.1 Proof of Theorem 4.5 (Threshold Structure in the CMDP)

A well-known result in Lagrangian dynamic programming theory is that, given a finite CMDP with a single constraint, there exists an optimal strategy that is a randomized mixture of two optimal, deterministic, stationary strategies of the induced Lagrangian MDP (4.13) with different Lagrange multipliers $\lambda_1$ and $\lambda_2$ (Thm. 1, Ma et al., 1986)(Thm. 6.6.2, Krishnamurthy, 2016)(Thm. 12.7, Altman, 1999). Lemma 4.12 implies that the Lagrange multipliers can be obtained by first computing the optimal strategies in the Lagrangian MDP and then taking the supremum over the set of non-negative Lagrange multipliers. Lemma 4.11 implies that for each non-negative Lagrange multiplier, there exists a stationary, deterministic strategy that has the threshold structure in (4.10). Combined, these three properties imply the statement of Thm. 4.5. □

# C Proof of Corollary 4.2

Consider the MDP faced by the attacker when fixing the control strategy $\pi^{(\mathrm{C})}$ and assume that $\pi^{(\mathrm{C})}$ satisfies (4.10). The reward function in this MDP is the cost function of the controller, i.e., $r(a^{(\mathrm{C})}) = a^{(\mathrm{C})}$ (4.7). Since this MDP is unichain and the objective is the average reward criterion, the optimality equation reads

$$0 = \max_{\mathbf{a}^{(A)}} \left[ \pi^{(C)}(s) - g + \sum_{s' \in \mathcal{S}_S} f_S(s' \mid s, \pi^{(C)}, \mathbf{a}^{(A)}) J(s') - J(s) \right],$$

where $f_S$ is the transition function and $g$ is the best response average reward:

$$g \triangleq \lim_{T \to \infty} \frac{1}{T} \mathbb{E}_{(S_t)_{t \geq 1}, \tilde{\pi}^{(A)}} \left[ \sum_{t=1}^{T} \pi^{(C)}(S_t) \mid s_1 \right]. \tag{4.16}$$

Here $J(s)$ is the (differential) value function, which measures the transient reward effects of being in state $s$ as opposed to the average reward $g$ (Eq. 8.4.2, Thm. 8.4.3, Puterman, 1994)(Thm. 6.5.2, Krishnamurthy, 2016). It follows from Thm. 4.4.A that the above limit exists (Thm. 8.4.5, Puterman, 1994). Therefore, any best response strategy $\tilde{\pi}^{(A)}$ for the attacker satisfies

$$\tilde{\pi}^{(A)}(s) \in \arg\max_{\mathbf{a}^{(A)}} \left[ \sum_{s' \in \mathcal{S}_S} f_S(s' \mid s, \pi^{(C)}, \mathbf{a}^{(A)}) J(s') \right].$$

We will show that this maximization is equivalent to minimizing $\mathbb{E}[S']$. Towards this end, we prove the following two lemmas.

**Lemma 4.13.** *(T5.1) implies that $f_S(s' \mid s, \boldsymbol{\pi})$ is stochastically monotone in $s$.*

*Proof.* (T5.1) states that

$$\begin{aligned}
&\mathbb{E}_{\boldsymbol{\pi}}[S_{t+1} \mid S_t = s+1] = \mathbb{E}_{\boldsymbol{\pi}}[S_{t+1} \mid S_t = s] + 1 \\
\implies &\sum_{s' \in \mathcal{S}_S} s' f_S(s' \mid S = s+1, \boldsymbol{\pi}) = 1 + \sum_{s' \in \mathcal{S}_S} s' f_S(s' \mid S = s, \boldsymbol{\pi}) \\
\implies &\sum_{s'' \in \{x \mid x \in \mathcal{S}_S, x \geq s'\}} f_S(s'' \mid S = s+1, \boldsymbol{\pi}) \geq \sum_{s'' \in \{x \mid x \in \mathcal{S}_S, x \geq s'\}} f_S(s'' \mid S = s, \boldsymbol{\pi}),
\end{aligned}$$

for all $s' \in \mathcal{S}_S$. □

**Lemma 4.14.** *Given (T5.1) and assuming each $\pi^{(C)}$ satisfies (4.10), then $J(s)$ (4.16) is decreasing in $s$ for each strategy profile $\boldsymbol{\pi}$.*

*Proof.* Applying the same argument as in Lemma 4.6–Lemma 4.11, we can use mathematical induction on the iterates of the value iteration algorithm $J_0, J_1, \ldots$ and use the vanishing discount method (Eq. 6.3.2–6.3.4, Puterman, 1994). Let $J_0(s) = 0 \ \forall s \in \mathcal{S}_S$. The inductive base case holds trivially. Assume by induction that $J_{k-1}(s)$ is decreasing in $s$. By definition,

$$J_k(s) = \mathbb{E}[A^{(C)} \mid \pi^{(C)}, s] + \gamma \sum_{s' \in \mathcal{S}_S} f_S(s' \mid s, \boldsymbol{\pi}) J_{k-1}(s').$$

(4.10) implies that

$$\mathbb{E}[A^{(\mathrm{C})} \mid \pi^{(\mathrm{C})}, s] = \begin{cases} 1 & \text{if } s \leq \min[\beta_1, \beta_2] \\ 1-\kappa & \text{if } \beta_1 < s \leq \beta_2 \\ \kappa & \text{if } \beta_2 < s \leq \beta_1 \\ 0 & \text{if } s > \max[\beta_1, \beta_2]. \end{cases} \tag{4.17}$$

Hence, $\mathbb{E}[A^{(\mathrm{C})} \mid \pi^{(\mathrm{C})}, s]$ is decreasing in $s$. Thus, it only remains to show that $\sum_{s' \in \mathcal{S}_{\mathrm{S}}} f_{\mathrm{S}}(s' \mid s, \boldsymbol{\pi}) J_{k-1}(s')$ is decreasing in $s$. (T5.1) implies that $f_{\mathrm{S}}(s' \mid s, \boldsymbol{\pi})$ is stochastically monotone in $s$ for each $\boldsymbol{\pi}$ (Lemma 4.13) and the induction assumption implies that $J_{k-1}(s)$ is decreasing in $s$. Consequently, given any two states $s_1 < s_2$, we have

$$\sum_{s' \in \mathcal{S}_{\mathrm{S}}} f_{\mathrm{S}}(s' \mid s_2, \boldsymbol{\pi}) J_{k-1}(s') \leq \sum_{s' \in \mathcal{S}_{\mathrm{S}}} f_{\mathrm{S}}(s' \mid s_1, \boldsymbol{\pi}) J_{k-1}(s').$$

Hence, $\sum_{s' \in \mathcal{S}_{\mathrm{S}}} f_{\mathrm{S}}(s' \mid s, \boldsymbol{\pi}) J_{k-1}(s')$ is decreasing in $s$. □

Finally, we are ready to the prove Corollary 4.2. Lemma 4.14 implies that

$$\tilde{\pi}^{(\mathrm{A})}(s) \in \arg\max_{\mathbf{a}^{(\mathrm{A})}} \mathbb{E}_{S'}\left[J(S') \mid s, \mathbf{a}^{(\mathrm{A})}, \pi^{(\mathrm{C})}\right] = \arg\min_{\mathbf{a}^{(\mathrm{A})}} \mathbb{E}_{S'}\left[S' \mid s, \mathbf{a}^{(\mathrm{A})}, \pi^{(\mathrm{C})}\right]. \quad \square$$

## D Preference Relations

It follows from (4.4) that the outcome of Game 4.1 is determined by the number of recoveries and the number of time steps in the compromised state $\mathbb{C}$. Let $\alpha$ and $\beta$ denote the average number of recoveries and time steps in state $\mathbb{C}$, respectively. The game's outcome can then be defined as $o \triangleq \eta\tilde{\alpha} + \tilde{\beta}$. Given this definition, the controller's preference relation $\succ_i$ induced by $J_i$ (4.4) can be expressed as

$$o \succ_i \tilde{o} \iff o < \tilde{o} \quad o \approx_i \tilde{o} \iff o = \tilde{o}. \tag{4.18}$$

**Theorem 4.6.** *The preference relation $\succeq_i$ (4.18) of Game 1 satisfies the von Neumann-Morgenstern axioms (p. 26, von Neumann and Morgenstern, 1944).*

*Proof.* By definition (4.18), $\succeq_i$ is reflexive, transitive and complete. Hence, it only remains to prove the continuity, monotonicity, and independence axioms. We start with the continuity axiom. Consider three outcomes $o_1 \preceq_i o_2 \preceq_i o_3$. We need to show that there exists a probability $\theta_i \in [0, 1]$ such that

$$o_2 = \theta_i o_3 + (1 - \theta_i) o_1.$$

Solving for $\theta_i$, we get:

$$\theta_i = \frac{o_2 - o_1}{o_3 - o_1}.$$

It follows from (4.18) that $o_2 \geq o_3$. Hence, $\theta_i \in [0,1]$.

Next, consider the monotonicity axiom. Suppose $o_2 \succeq_i o_1$ and let $a, b$ be values in $[0,1]$. We need to show that $ao_2 + (1-a)o_1 \leq bo_2 + (1-b)o_1 \implies a \geq b$. We have

$$ao_2 + (1-a)o_1 \leq bo_2 + (1-b)o_1 \implies o_2(a-b) \leq o_1(a-b).$$

Since $o_2 \leq o_1$, the above inequality implies that $a \geq b$.

Now consider the independence axiom. Given two outcomes $o_1 \preceq_i o_2$, any $z \in \mathbb{R}^+$, and any $p \in (0,1)$, the axiom asserts that we must have $po_1 + (1-p)z \leq po_2 + (1-p)z$. This can be verified as follows.

$$po_1 + (1-p)z \leq po_2 + (1-p)z \implies po_1 \leq po_2 \implies o_1 \leq o_2, \tag{4.19}$$

which holds by definition of $\succeq_i$ (4.18). □

**Corollary 4.3.** *The preference relation $\succeq$ of Game 2 satisfies the von Neumann-Morgenstern axioms (p. 26, von Neumann and Morgenstern, 1944).*

*Proof.* It follows from (4.7) that the outcome of the game (given that the constraint is feasible) is determined by the number of added nodes. Let $\alpha$ denote the average number of added nodes. Then, the game's outcome can be defined as $o \triangleq \alpha$, and the preference relation is equivalent to (4.18). □

## E Intrusion-Tolerant Consensus Protocol

The TOLERANCE architecture (Fig. 4.3 on page 159) is based on a reconfigurable consensus protocol for the partially synchronous system model with hybrid failures, a reliable network, and authenticated communication links (see Prop. 4.1)[17]. Examples of such protocols include MINBFT [392, §4.2], MINZYZZYVA [392, §4.3], REMINBFT [118, §5], and CHEAPBFT [230, §3]. Our implementation uses MINBFT. Correctness of MINBFT is proven in (Santos Veronese, 2010). MINBFT is based on PBFT (Castro and Liskov, 2002) with one crucial difference. While PBFT assumes Byzantine failures and tolerates $f = \frac{N-1}{3}$ failures, MINBFT assumes hybrid failures (Correia et al., 2007) and tolerates $f = \frac{N-1}{2}$ failures. The improved resilience of MINBFT is achieved by leveraging a trusted component that provides certain functions for the protocol. In particular, MINBFT relies on a tamperproof service at each node to assert whether a given sequence number was assigned to a message. This service allows MINBFT to prevent *equivocation*[18] (Chun et al., 2007) and imposes a first-in-first-out (FIFO) order on client requests. In TOLERANCE, the tamperproof service is provided by the virtualization layer; see Fig. 4.3 on page 159.

[17]In an authenticated network, nodes can verify each other's digital signatures.

[18]A compromised node is said to *equivocate* if it sends inconsistent information to different nodes in the system.

We extend minbft (§4.2, Santos Veronese, 2010) to be *reconfigurable* (Lamport et al., 2010), where the reconfiguration procedure is based on the method described in (§IV.B,, Hao et al., 2018). (A reconfigurable consensus protocol allows dynamic addition and removal of nodes from the system.) The different stages of the protocol are illustrated in Fig. 4.16 on the next page, and the throughput of our implementation is shown in Fig. 4.13 on page 180. The hyperparameters are listed in Appendix G.

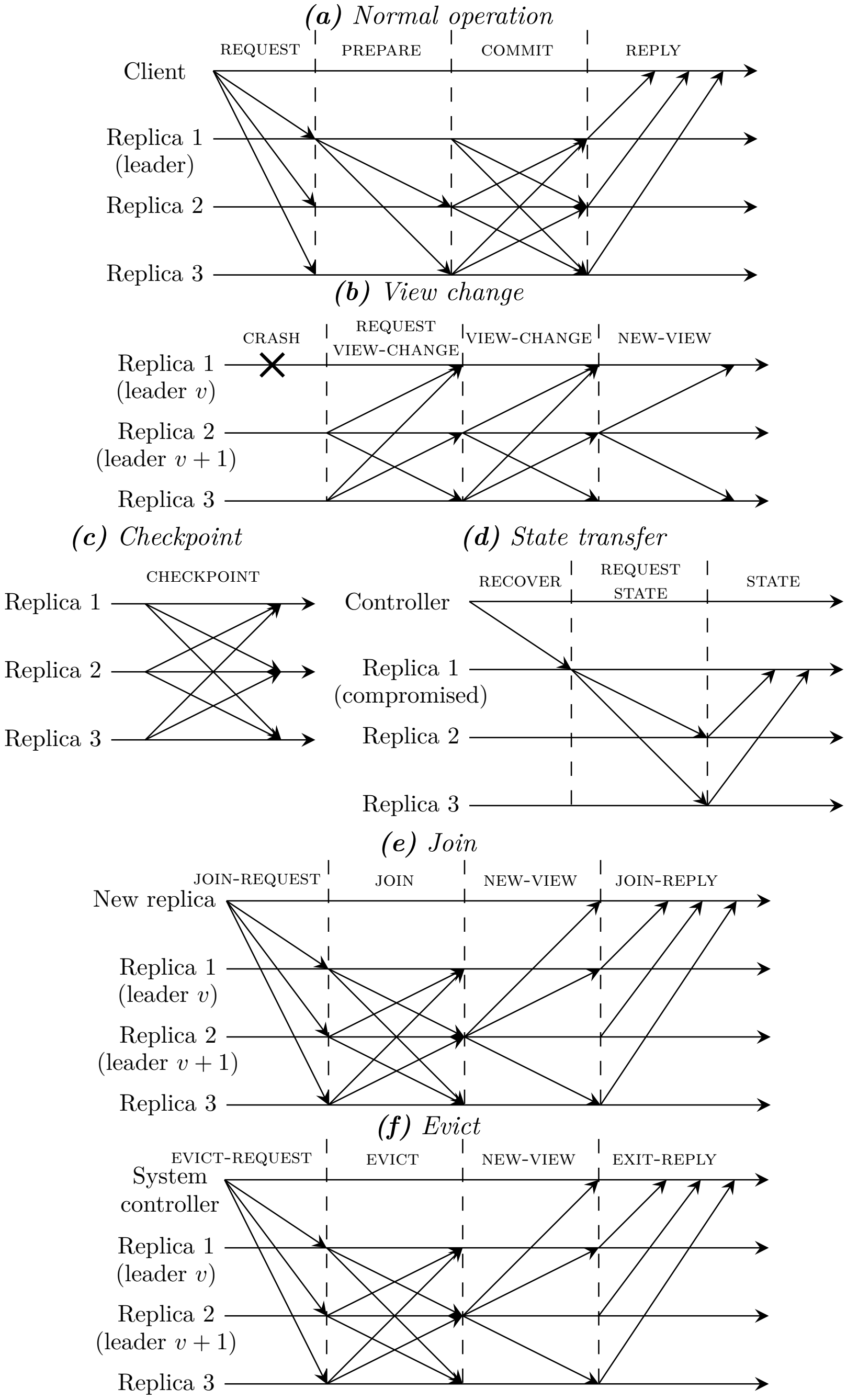

***Figure 4.16:*** *Time-space diagrams illustrating the message patterns of the* MINBFT *intrusion-tolerant consensus protocol (§4.2, Santos Veronese, 2010).*

# F Hyperparameters

Hyperparameters for the experimental results are listed in Table 4.9. Confidence levels for all figures are computed based on the Student-t distribution.

| *Figures* | *Values* |
|---|---|
| Figure 4.4.b | $p_{\mathrm{C},i} = 0$ |
| Figure 4.10 | $p_{\mathrm{C},i} = 10^{-5}$ |
| Figure 4.6.a | $\eta = 2$, $p_{\mathrm{A},i} = 0.05$, $p_{\mathrm{C},i} = 0.01$, $\Delta_{\mathrm{R}} = 100$, $\lvert\mathcal{O}\rvert = 100$ |
| | $z_i(\cdot \mid 0) = \mathrm{BetaBin}(n = 100, \alpha = 0.7, \beta = 3)$ |
| | $z_i(\cdot \mid 1) = \mathrm{BetaBin}(n = 100, \alpha = 1, \beta = 0.7$ |
| | $\pi^{(\mathrm{A})}_{i,t}(s, b) = \mathsf{A}$ |
| Figure 4.6.b | $p_{\mathrm{C},i} = 0.01$, $\Delta_{\mathrm{R}} = \infty$, $\lvert\mathcal{O}\rvert = 100$ |
| | $z_i(\cdot \mid 0) = \mathrm{BetaBin}(n = 100, \alpha = 0.7, \beta = 3)$ |
| | $z_i(\cdot \mid 1) = \mathrm{BetaBin}(n = 100, \alpha = 1, \beta = 0.7$ |
| Figure 4.15 | $p_{\mathrm{C},i} = 0.01$, $p_{\mathrm{A},i} = 0.5$, $\Delta_{\mathrm{R}} = \infty$, $\lvert\mathcal{O}\rvert = 100$ |
| | $z_i(\cdot \mid 0) = \mathrm{BetaBin}(n = 100, \alpha = 0.7, \beta = 3)$ |
| | $z_i(\cdot \mid 1) = \mathrm{BetaBin}(n = 100, \alpha = 1, \beta = 0.7$ |
| | $\Delta_{\mathrm{R}}$ for PERIODIC was selected using BO [403, Alg. 1]. |
| Figure 4.9 | $p_{\mathrm{C},i} = 0.05$, $p_{\mathrm{A},i} = 0.3$, $\Delta_{\mathrm{R}} = \infty$, $\lvert\mathcal{O}\rvert = 100$ |
| | $z_i(\cdot \mid 0) = \mathrm{BetaBin}(n = 100, \alpha = 0.7, \beta = 3)$ |
| | $z_i(\cdot \mid 1) = \mathrm{BetaBin}(n = 100, \alpha = 1, \beta = 0.7$ |
| Figure 4.12 | $p_{\mathrm{A},i} = 10^{-1}$, $p_{\mathrm{C},i} = 10^{-5}$, $\epsilon_{\mathrm{A}} = 0.9$ |
| | $s_{\max} = 13$, $\eta = 2$, $f = \min[\frac{N_1-1}{2}, 2]$ |
| | $f_{\mathrm{S}}$ estimated from simulations of Game 1 |
| | $z_i$ estimated from testbed measurements, see Fig. 4.14 |
| Figure 4.15 | $p_{\mathrm{A},i} = 10^{-1}$, $p_{\mathrm{C},i} = 10^{-5}$, $\epsilon_{\mathrm{A}} = 0.9$, |
| | $s_{\max} = 13$, $\eta = 2$, $f = \min[\frac{N_1-1}{2}, 2]$ |
| | $f_{\mathrm{S}}$ estimated from simulations of Game 4.2, |
| | $z_i$ estimated from testbed measurements, see Fig. 4.14 |
| *Incremental pruning [88, Fig. 4]* | |
| Variation, $\epsilon$ | normal, 0 |
| *SPSA [436, Fig. 1]* | |
| $c, \epsilon, \lambda, A, a, N, \delta$ | 10, 0.101, 0.602, 100, 1, 50, 0.2 |
| $M$ number of samples for each evaluation | 50 |
| *Cross-entropy method [380][316, Alg. 1]* | |
| $\lambda$ (fraction of samples to keep) | $0.15, 100$ |
| $K$ population size | 100 |
| $M$ number of samples for each evaluation | 50 |
| *Differential evolution [443, Fig. 3]* | |
| Population size $K$, mutate step | $10, 0.2$ |
| Recombination rate | 0.7 |
| $M$ number of samples for each evaluation | 50 |
| *Bayesian optimization [403, Alg. 1]* | |
| Acquisition function | lower confidence bound [438, Alg. 1] |
| $\beta$, Kernel | 2.5, Matern(2.5) |
| $M$ number of samples for each evaluation | 50 |
| *HSVI [200, Alg. 3]* | |
| $\epsilon$, PDELTA, PLIMIT | 0.001, 0.005, 2000 |
| *MINBFT [392, §4.2]* | |
| USIG implementation | RSA with key lengths 1024 bits [371] |
| $T_{\mathrm{exec}}$, $T_{\mathrm{vc}}$, cp, $L$ | 30 seconds, 280 seconds, $10^2$, $10^3$ |

***Table 4.9:*** *Hyperparameters.*

# Paper 5†

## AUTOMATED SECURITY RESPONSE THROUGH ONLINE LEARNING WITH ADAPTIVE CONJECTURES

Kim Hammar, Tao Li, Rolf Stadler, and Quanyan Zhu

**Abstract**

We study automated security response for an IT infrastructure and formulate the interaction between an attacker and a defender as a partially observed, non-stationary game. We relax the standard assumption that the game model is correctly specified and consider that each player has a probabilistic conjecture about the model, which may be misspecified in the sense that the true model has probability 0. This formulation allows us to capture uncertainty and misconceptions about the infrastructure and the opponent. To learn effective game strategies online, we design **C**onjectural **O**nline **L**earning (COL), a novel method where a player iteratively adapts its conjecture using Bayesian learning and updates its strategy through rollout. We prove that the conjectures converge to best fits, and we provide a bound on the performance improvement that rollout enables with a conjectured model. To characterize the steady state of the game, we propose a variant of the Berk-Nash equilibrium. We present COL through an advanced persistent threat use case. Testbed evaluations show that COL produces effective security strategies that adapt to a changing environment. We also find that COL enables faster convergence than current reinforcement learning techniques.

*All models are wrong, but some are useful.*

— George E. P. Box ***1976**, Science and statistics.*

## 5.1 Introduction

THIS paper addresses a limitation of papers 1–4, which assume that a perfect model of the underlying IT infrastructure can be obtained. This assumption is unrealistic because attackers and defenders often have incorrect prior knowledge about the infrastructure and the opponent, which means that they generally have *misspecified* models.

**Motivating example: The NOTPETYA attack.**

NOTPETYA is a malware that was used by the SANDWORM Advanced Persistent Threat (APT) in a worldwide attack in 2017 (U.S. Department of Justice, 2020). Security researchers initially conjectured that NOTPETYA was a version of the PETYA ransomware (hence the name) (The MITRE Corporation, 2024). As a result, many organizations focused on traditional ransomware response strategies. However, it later became evident that the malware was not financially motivated but designed for destruction. This *misspecification* delayed effective responses.

In this paper, we address the above misspecification and present **C**onjectural **O**nline **L**earning (COL), a game-theoretic method for *online learning* of security strategies that applies to dynamic IT environments where attackers and defenders have misconceptions about the environment and the opponent's strategy. Using this method, we formulate the interaction between an attacker and a defender as a non-stationary, partially observed game. We relax the standard assumption that the game model is correctly specified and consider the case where each player has a probabilistic *conjecture* about the model, i.e., a probability distribution over possible models, which may be *misspecified* in the sense that the true model has probability 0. Both players iteratively adapt their conjecture using *Bayesian learning* (see Fig. 5.1 on the next page) and update their strategies using *rollout*, which is a form of approximate dynamic programming (Bertsekas, 2021). We prove that the conjectures in COL converge to best fits, and we provide a bound on the performance improvement that rollout enables with a conjectured model. To characterize the steady state of the game, we define a variant of the *Berk-Nash Equilibrium* (BNE) (Def. 1, Esponda and Pouzo, 2016), which represents a fixed point where players act optimally given their conjectures.

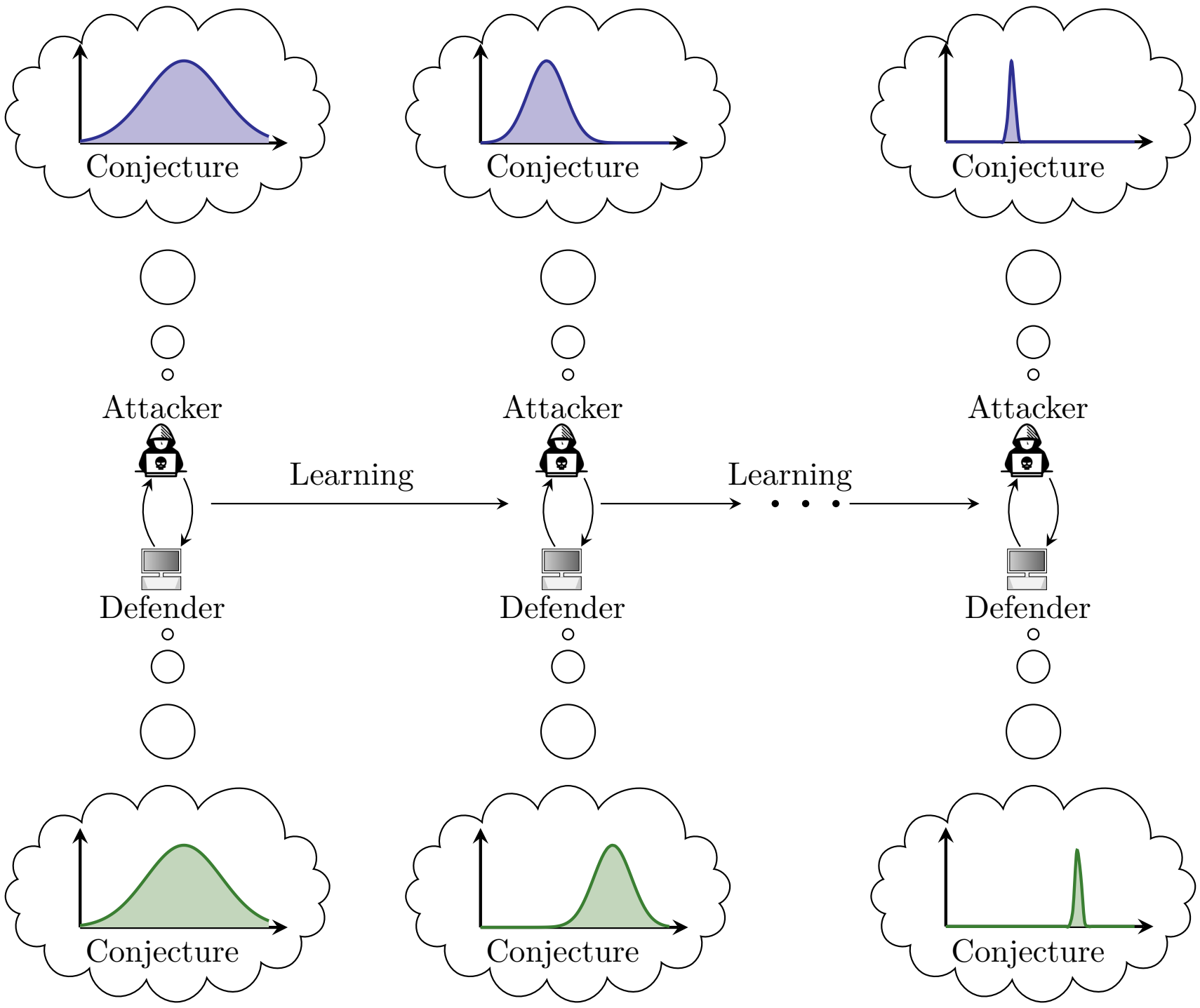

***Figure 5.1:*** *Conjectural **O**nline **L**earning (COL): we formulate the interaction between an attacker and a defender as a non-stationary game; each player has a probabilistic conjecture about the game model, which may be misspecified in the sense that the true model has probability* 0*; the conjectures are iteratively adapted through Bayesian learning.*

While the study of learning with misspecified models has attracted long-standing interest in economics (Arrow and Green, 1973), engineering (Kagel and Levin, 1986), and psychology (Rabin, 2002), it remains unexplored in the security context. Related research in the security literature include (*i*) game-theoretic approaches based on bounded rationality [418, 389, 376, 90, 390, 151, 91, 423, 483, 36, 1, 301]; (*ii*) game-theoretic approaches based on imperfect and incomplete information [471, 178, 183, 33, 11]; and (*iii*) model-free learning techniques [147, 507, 206, 281, 156, 279, 178, 179, 183, 185, 3]. (A review of related work can be found in §5.8.) To our knowledge, we provide the first study of learning with misspecified models in a security context. The benefit of this approach is threefold. First, it provides a new methodology to capture uncertainty and misspecification in security games. Second, as we show in this paper, it applies to dynamic, non-stationary, and partially observed games. Third, the model conjectures produced by our method are guaranteed to converge under reasonable conditions, and the worst-case performance of the learned strategies is bounded.

We present COL through a use case that involves an Advanced Persistent Threat (APT) on an IT infrastructure; see Fig. 3 in the introduction chapter. We emulate this infrastructure with a digital twin, which we create using CSLE, as described in the methodology chapter (Hammar, 2023). We use the twin to run APT actions and defender responses. During such runs, we collect measurements and logs, from which we estimate infrastructure statistics. This data is then used to instantiate simulations of the use case, based on which we evaluate the performance of COL. We find that COL produces effective security strategies that adapt to a changing environment. The simulations also show that COL enables faster convergence than current reinforcement learning techniques. In addition to the simulation studies, we evaluate COL on the digital twin and compare it against the SNORT Intrusion Detection and Prevention System (IDPS) (Roesch, 1999). The results attest that COL adapts to changes in the distribution of network traffic and outperforms SNORT in several key metrics. Specifically, it blocks a higher percentage of attack attempts and a lower percentage of client traffic.

***Contributions***

1. We introduce a novel game-theoretic formulation for the problem of automated security response where each player (i.e., attacker or defender) has a probabilistic conjecture about the game model. This formulation allows us to capture model misspecification and uncertainty.

2. We present COL, a new method for online learning of game strategies where a player iteratively adapts its conjecture using Bayesian learning and updates its strategy through rollout. This method allows us to automatically adapt security strategies to changes in the environment.

3. We prove that, when using COL, the conjectures of both players converge, and we characterize the steady state as a variant of the Berk-Nash equilibrium (Def. 1, Esponda and Pouzo, 2016). We also provide a bound on the performance improvement that rollout enables with a conjectured model.

4. We evaluate COL using simulation and emulation studies based on a digital twin running 64 virtualized servers and 10 different types of APTs (Hammar, 2023). This evaluation provides insights into how COL performs under different conditions and shows that it converges faster than current reinforcement learning techniques. It also shows that COL outperforms the SNORT IDPS in several key metrics (Roesch, 1999).

## 5.2 Use Case: Advanced Persistent Threat (APT)

We consider the problem of defending an organization's IT infrastructure against an APT caused by an *attacker* (Moothedath et al., 2020). The operator of the infrastructure, which we call the *defender*, takes measures to protect it against the attacker while providing services to a client population; see Fig. 3 in the introduction chapter. The infrastructure includes a set of servers and an Intrusion Detection System (IDS) that logs events in real-time. Clients access the services through a public gateway, which is also open to the attacker.

The attacker aims to intrude on the infrastructure over an extended period. It begins with reconnaissance to identify vulnerabilities, after which it attempts to compromise servers through exploits. Once inside the infrastructure, the attacker employs lateral movement techniques, escalates privileges, and uses advanced evasion tactics to avoid detection.

The defender monitors the infrastructure by observing IDS alerts. It can recover potentially compromised servers (e.g., by upgrading their software), which temporarily disrupts service for clients. When deciding to take this response action, the defender balances two conflicting objectives: (*i*) maintain services to its clients; and (*ii*) recover compromised servers.

## 5.3 Game Model of the APT Use Case

We formulate the above use case as a zero-sum stochastic game with one-sided partial observability (a POSG)[2]

$$\Gamma \triangleq \langle \mathcal{N}, \mathcal{S}, (\mathcal{A}_{\mathrm{k}})_{\mathrm{k}\in\mathcal{N}}, f, c, \gamma, \mathbf{b}_1, z, \mathcal{O} \rangle. \tag{5.1}$$

The game has two players: the (D)efender and the (A)ttacker. In the following subsections, we define the components of the game, its evolution, and the players' objectives. The requisite notation is listed in Table 5.1 on the next page.

### Actions

Both players can invoke two actions: (S)top and (C)ontinue. The action spaces are thus $\mathcal{A}_{\mathrm{D}} \triangleq \mathcal{A}_{\mathrm{A}} \triangleq \{\mathsf{S}, \mathsf{C}\}$. $\mathsf{S}$ triggers a change in the game state while $\mathsf{C}$ is a passive action that does not change the state. Specifically, $a_t^{(\mathrm{A})} = \mathsf{S}$ is the attacker's compromise action, and $a_t^{(\mathrm{D})} = \mathsf{S}$ is the defender's recovery action (as defined in the use case §5.2).

### Dynamics

The state $s_t \in \mathcal{S} \triangleq \{0, 1, \dots, N\}$ represents the number of compromised servers at time $t$, where $s_1 = 0$. The transition $s_t \rightarrow s_{t+1}$ occurs with probability $f\big(s_{t+1} \mid s_t, a_t^{(\mathrm{D})}, a_t^{(\mathrm{A})}\big)$:

$$f(S_{t+1} = 0 \mid s_t, \mathsf{S}, a_t^{(\mathrm{A})}) \triangleq 1 \tag{5.2a}$$

$$f(S_{t+1} = s_t \mid s_t, \mathsf{C}, \mathsf{C}) \triangleq f(S_{t+1} = N \mid N, \mathsf{C}, \mathsf{S}) \triangleq 1 \tag{5.2b}$$

[2]The components of a POSG are defined the background chapter; see (19).

$$f(S_{t+1} = s_t \mid s_t, \mathsf{C}, \mathsf{S}) \triangleq 1 - p_{\mathrm{A}} \qquad s_t < N \tag{5.2c}$$

$$f(S_{t+1} = s_t + 1 \mid s_t, \mathsf{C}, \mathsf{S}) \triangleq p_{\mathrm{A}} \qquad s_t < N, \tag{5.2d}$$

where $p_{\mathrm{A}}$ is the probability of a successful attack. All other transitions have probability 0; see Fig. 5.2 on the next page. (5.2a) defines the transition $s_t \to 0$, which occurs when the defender takes action $\mathsf{S}$. (5.2b)–(5.2c) define the recurrent transition $s_{t+1} = s_t$, which occurs when both players take action $\mathsf{C}$ or when the attacker is unsuccessful in compromising a server, which happens with probability $1 - p_{\mathrm{A}}$. Lastly, (5.2d) defines the transition $s_t \to s_t + 1$, which occurs with probability $p_{\mathrm{A}}$ when the attacker takes action $\mathsf{S}$, $s_t < N$, and the defender takes action $\mathsf{C}$.

| *Notation(s)* | *Description* |
|---|---|
| $\Gamma, c, N$ | The game (5.1), cost function (5.6), and # servers (5.2). |
| $\mathrm{D}, \mathrm{A}$ | The defender player and the attacker player (5.1). |
| $\mathcal{N}, \mathcal{S}, \mathcal{O}$ | Sets of players, states, and observations (5.1). |
| $\mathcal{A}_{\mathrm{D}}, \mathcal{A}_{\mathrm{A}}$ | Sets of defender and attacker actions (5.1). |
| $t, \gamma$ | Time step and discount factor (5.7). |
| $\pi_{\mathrm{D}}, \pi_{\mathrm{A}}$ | Defender and attacker strategies. |
| $\Pi = \Pi_{\mathrm{D}} \times \Pi_{\mathrm{A}}$ | Defender and attacker strategy spaces. |
| $\tilde{\pi}_{\mathrm{D}}, \tilde{\pi}_{\mathrm{A}}$ | Best response strategies (5.8). |
| $\boldsymbol{\pi}^\star = (\pi^\star_{\mathrm{D}}, \pi^\star_{\mathrm{A}})$ | Nash equilibrium strategies (5.10). |
| $\mathscr{B}_{\mathrm{D}}, \mathscr{B}_{\mathrm{A}}$ | Best response correspondences (5.8). |
| $J_{\mathrm{D}}, J_{\mathrm{A}}$ | Defender and attacker objectives (5.7). |
| $f, z$ | Transition function (5.2) and observation function (5.4). |
| $s_t, o_t$ | State (5.2) and observation (5.4) at time $t$. |
| $\mathbf{a}_t = (a_t^{(\mathrm{D})}, a_t^{(\mathrm{A})})$ | Actions at time $t$. |
| $S_t, O_t, \mathbf{A}_t$ | Random variables (vector) with realizations $s_t$ (5.2), $o_t$ (5.4), and $\mathbf{a}_t$. |
| $\mathbf{b}_t, \mathbf{B}_t$ | Defender belief ($\mathbf{b}_t$ realizes the random vector $\mathbf{B}_t$) (5.5). |
| $\mathcal{B}, \mathbb{B}$ | Belief space and belief operator of the defender (22). |
| $\mathbf{h}_t^{(\mathrm{k})}, \mathbf{h}_t$ | History of player k and joint history. |
| $\mathbf{H}_t^{(\mathrm{k})}, \mathbf{H}_t$ | Random vectors with realizations $\mathbf{h}_t^{(\mathrm{k})}$ and $\mathbf{h}_t$. |
| $\mathcal{H}_t = \mathcal{H}_t^{(\mathrm{D})} \times \mathcal{H}_t^{(\mathrm{A})}$ | History spaces. |
| $\mathsf{S}, \mathsf{C}$ | Stop and continue actions. |
| $\overline{\pi}_{-\mathrm{k},t}$ | Player k's conjecture of player $-$k's strategy (Alg. 5.1). |
| $\ell_{\mathrm{D}}, \ell_{\mathrm{A}}, \overline{\ell}_{-\mathrm{k},t}$ | Lookahead horizons (Alg. 5.1) and player k's conjecture. |
| $\boldsymbol{\theta}_t, \overline{\boldsymbol{\theta}}_t^{(\mathrm{k})}$ | Parameter vector of $\Gamma$ and player k's conjecture of $\boldsymbol{\theta}_t$ at time $t$ (5.14a). |
| $\mathcal{L}, \Theta_{\mathrm{k}}$ | Player k's sets of possible conjectures of $\ell_{\mathrm{A}}$ and $\boldsymbol{\theta}$ (5.14). |
| $\mathcal{L}^\star, \Theta^\star_{\mathrm{k}}$ | Sets of consistent conjectures (5.16). |
| $\mathbf{i}_t^{(\mathrm{k})}, \mathbf{I}_t^{(\mathrm{k})}$ | Information feedback of player k at time $t$ (5.3). |
| $\mu_t, \rho_t^{(\mathrm{k})}$ | Posteriors $\mathbb{P}[\ell_{\mathrm{A}} \mid \mathbf{h}_t^{(\mathrm{D})}]$ (5.14b) and $\mathbb{P}[\boldsymbol{\theta}_t^{(\mathrm{k})} \mid \mathbf{h}_t^{(\mathrm{k})}]$ (5.14a). |
| $\nu, K(\overline{\alpha}, \nu)$ | Occupancy measure and discrepancy of conjecture $\overline{\alpha}$ (5.15). |
| $\pi_{1,\mathrm{k}}, \pi_{t,\mathrm{k}}$ | Base and rollout strategy of player k at time $t$ (5.12). |
| $\boldsymbol{\pi}_{\mathbf{h}_t}$ | Strategy profile induced by Alg. 5.1 at time $t$. |
| $\mathscr{R}, \mathbb{P}^{\mathscr{R}}$ | Rollout operator (5.12), distribution over $\bigtimes_{t \geq 1}(\mathcal{H}_t^{(\mathrm{D})} \times \mathcal{H}_t^{(\mathrm{A})})$ (Thm. 5.4). |
| $K^\star_{\mathcal{L}}, K^\star_{\Theta_{\mathrm{k}}}$ | Minimal discrepancy values for $\mathcal{L}$ and $\Theta_{\mathrm{k}}$ (5.16). |

***Table 5.1:*** *Variables and symbols used in the model.*

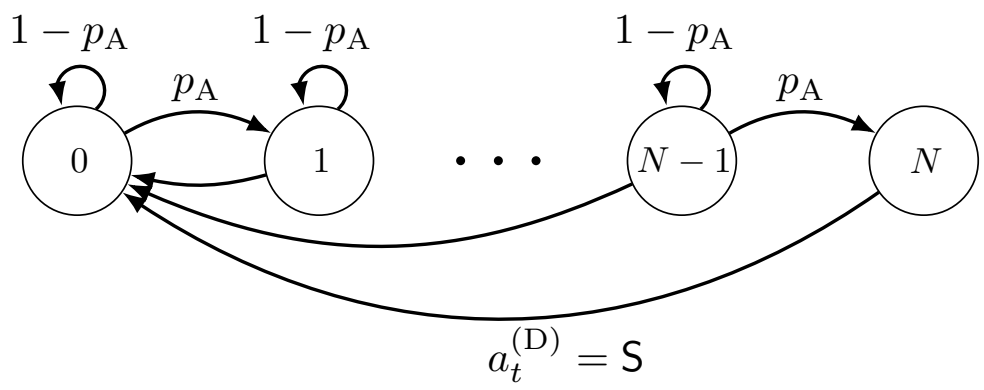

***Figure 5.2:*** *State transition diagram of the game* $\Gamma$*: disks represent states; arrows represent transitions; labels indicate conditions for transition; the initial state is* $s_1 = 0$.

### Observability

The attacker has complete observability. It knows the game state, the defender's actions, and the defender's observations[3]. In contrast, the defender has a finite set of observations $o_t \in \mathcal{O}$. Consequently, the *information feedbacks* for the attacker and the defender at time $t$ are

$$\mathbf{i}_t^{(\mathrm{A})} \triangleq (o_t, s_t, a_{t-1}^{(\mathrm{D})}) \quad \text{and} \quad \mathbf{i}_t^{(\mathrm{D})} \triangleq (o_t), \tag{5.3}$$

respectively, where $o_t$ is drawn from a random variable $O_t$ whose distribution depends on the state, i.e.,

$$o_t \sim z(\cdot \mid s_t). \tag{5.4}$$

**Remark 5.1** (Modeling clients)**.** The clients are implicitly modeled by $z$ (5.4).

Each player k has *perfect recall* (Def. 7, Kuhn, 1953), which means that it remember the history $\mathbf{h}_t^{(\mathrm{k})} \triangleq (\mathbf{b}_1, (a_l^{(\mathrm{k})}, \mathbf{i}_l^{(\mathrm{k})})_{l=1,2,\ldots}) \in \mathcal{H}_t^{(\mathrm{k})}$. Based on this history, the defender uses $\mathbb{B}$ (22)[4] to compute the *belief state*

$$\mathbf{b}_t(s_t) \triangleq \mathbb{P}[S_t = s_t \mid \mathbf{h}_t^{(\mathrm{D})}] \in \mathcal{B}, \tag{5.5}$$

as defined in the background chapter. ($\mathbf{b}_t$ can be computed by the attacker also.)

### Strategies and objectives

Since $\mathbf{b}_t$ is a sufficient statistic for $s_t$ (5.2) (Def. 4.2, Lem. 5.1, Thm. 7.1, Kumar and Varaiya, 1986), we can define the players' *behavior Markov strategies* as $\pi_\mathrm{D} \in \Pi_\mathrm{D} \triangleq \mathcal{B} \rightarrow \Delta(\mathcal{A}_\mathrm{D})$ and $\pi_\mathrm{A} \in \Pi_\mathrm{A} \triangleq \mathcal{B} \times \mathcal{S} \rightarrow \Delta(\mathcal{A}_\mathrm{A})$ (Def. 5, Kuhn, 1953). Their performances are quantified using the cost function

$$c(s_t, a_t^{(\mathrm{D})}) \triangleq \overbrace{s_t^p \mathbb{1}_{a_t^{(\mathrm{D})} \neq \mathsf{S}}}^{\text{intrusion cost}} + \overbrace{\mathbb{1}_{a_t^{(\mathrm{D})} = \mathsf{S}}(q - r\mathbb{1}_{s_t > 0})}^{\text{response action cost}}, \tag{5.6}$$

[3] See Assumption 6 in the problem chapter.
[4] The equation to compute the belief state is defined in the background chapter; see (22).

where $p \geq 1$, $q > 0$, and $r > 0$ are scalar constants satisfying $1 > q - r$; see Fig. 5.3. The first term in (5.6) encodes the intrusion cost $s_t^p$, which increases with the number of compromised servers $s_t$. The second term in (5.6) encodes the cost of the defender's stop action, which is $q - r$ if an intrusion occurs and $q$ otherwise.

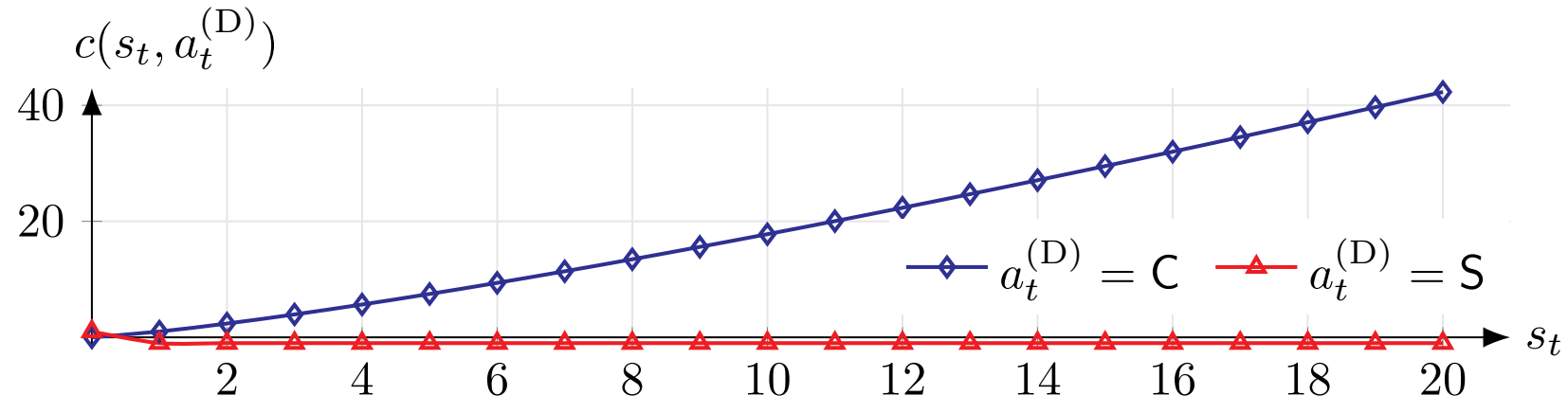

***Figure 5.3:*** *Example cost function $c(s_t, a_t^{(\mathrm{D})})$ (5.6); see Appendix F for hyperparameters.*

The goal of the defender is to *minimize* the expected cumulative (discounted) cost, and the goal of the attacker is to *maximize* the same quantity. Therefore, the objective functions are

$$J_{\mathrm{D}}^{(\pi_{\mathrm{D}},\pi_{\mathrm{A}})}(\mathbf{b}_1) \triangleq \mathbb{E}_{(\pi_{\mathrm{D}},\pi_{\mathrm{A}})}\left[\sum_{t=1}^{\infty} \gamma^{t-1} c(S_t, A_t^{(\mathrm{D})}) \mid \mathbf{b}_1\right] \tag{5.7a}$$

$$J_{\mathrm{A}}^{(\pi_{\mathrm{D}},\pi_{\mathrm{A}})}(\mathbf{b}_1) \triangleq -J_{\mathrm{D}}^{(\pi_{\mathrm{D}},\pi_{\mathrm{A}})}(\mathbf{b}_1), \tag{5.7b}$$

where $\gamma \in [0, 1)$ is a discount factor and $\mathbb{E}_{(\pi_{\mathrm{D}},\pi_{\mathrm{A}})}$ is the expectation over the random vectors $(\mathbf{H}_t^{(\mathrm{D})}, \mathbf{H}_t^{(\mathrm{A})})_{t\in\{1,2,\ldots\}}$ when the game is played according to $(\pi_{\mathrm{D}}, \pi_{\mathrm{A}})$.

A defender strategy $\tilde{\pi}_{\mathrm{D}} \in \Pi_{\mathrm{D}}$ is a *best response* against $\pi_{\mathrm{A}} \in \Pi_{\mathrm{A}}$ if it *minimizes* $J_{\mathrm{D}}^{(\pi_{\mathrm{D}},\pi_{\mathrm{A}})}$ (5.7a). Conversely, an attacker strategy $\tilde{\pi}_{\mathrm{A}}$ is a best response against $\pi_{\mathrm{D}}$ if it *maximizes* $J_{\mathrm{D}}^{(\pi_{\mathrm{D}},\pi_{\mathrm{A}})}$ (5.7b). Hence, the best response correspondences are

$$\mathscr{B}_{\mathrm{D}}(\pi_{\mathrm{A}}) \triangleq \operatorname*{arg\,min}_{\pi_{\mathrm{D}}\in\Pi_{\mathrm{D}}} J_{\mathrm{D}}^{(\pi_{\mathrm{D}},\pi_{\mathrm{A}})}(\mathbf{b}_1) \tag{5.8a}$$

$$\mathscr{B}_{\mathrm{A}}(\pi_{\mathrm{D}}) \triangleq \operatorname*{arg\,max}_{\pi_{\mathrm{A}}\in\Pi_{\mathrm{A}}} J_{\mathrm{D}}^{(\pi_{\mathrm{D}},\pi_{\mathrm{A}})}(\mathbf{b}_1). \tag{5.8b}$$

When the infrastructure contains a single server ($N = 1$), there exist best responses with *threshold structure*, as stated below.

**Theorem 5.1** (Threshold structure of best responses when $N = 1$).
*If $N = 1$, then*

*(A) For any $\pi_{\mathrm{A}} \in \Pi_{\mathrm{A}}$, there exists a value $\alpha^\star \in [0, 1]$ and a best response $\tilde{\pi}_{\mathrm{D}} \in \mathscr{B}_{\mathrm{D}}(\pi_{\mathrm{A}})$ (5.8a) that satisfies*

$$\tilde{\pi}_{\mathrm{D}}(\mathbf{b}) = \mathsf{S} \iff \mathbf{b}(1) \geq \alpha^\star. \tag{5.9a}$$

*(B) Assuming* $\pi_{\mathrm{A}}(\mathbf{e}_1, 0) = \mathsf{S}$ *for all* $\pi_{\mathrm{A}} \in \Pi_{\mathrm{A}}$*, where* $\mathbf{e}_1 = (1, 0)$*. Then, for any* $\pi_{\mathrm{D}} \in \Pi_{\mathrm{D}}$ *that satisfies (5.9a), there exists a value* $\beta^\star \in [0, 1]$ *and a best response* $\tilde{\pi}_{\mathrm{A}} \in \mathscr{B}_{\mathrm{A}}(\pi_{\mathrm{D}})$ *(5.8b) that satisfies*

$$\tilde{\pi}_{\mathrm{A}}(\mathbf{b}, s) = \mathsf{S} \iff s = 0,\ \mathbf{b}(1) \leq \beta^\star. \tag{5.9b}$$

The above theorem implies that when $N = 1$, the best responses can be parameterized by thresholds, which allows formulating (5.8a)–(5.8b) as parametric optimization problems that can be solved using Alg. 4.1 of Paper 4. Figure 5.4 shows the convergence curves when performing these optimizations with the **C**ross-**E**ntropy **M**ethod (cem) (Rubinstein, 1999). We provide proof in Appendix A.

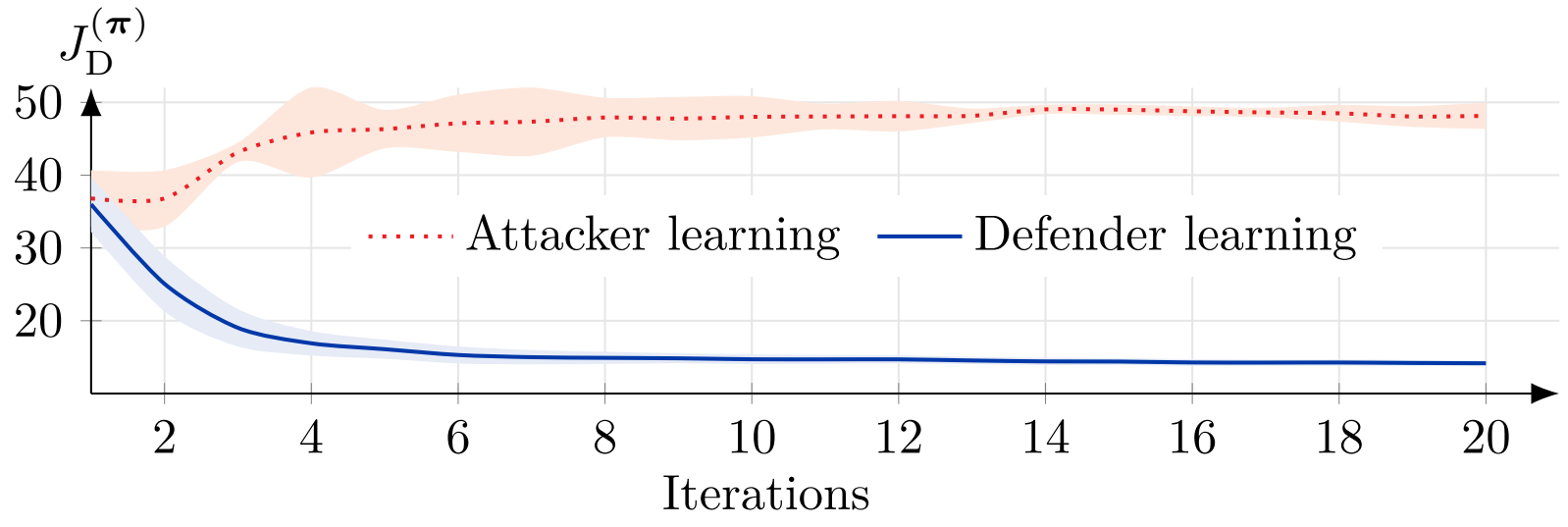

***Figure 5.4:*** *Best response learning when* $N = 1$ *using* cem *(Rubinstein, 1999) and the structural result in Thm. 5.1; the curves show the mean and the* 95% *confidence interval from evaluations with* 20 *random seeds; hyperparameters are listed in Appendix F.*

**Remark 5.2** (Finite and stationary zero-sum game)**.** $\Gamma$ is a finite, stationary, and zero-sum posg which satisfies assumptions 1–3 in the background chapter.

### Equilibria

When the attacker and the defender play best responses, their strategy pair is a *Nash equilibrium* (ne) and can be written as

$$\boldsymbol{\pi}^\star \triangleq (\pi_{\mathrm{D}}^\star, \pi_{\mathrm{A}}^\star) \in \mathscr{B}_{\mathrm{D}}(\pi_{\mathrm{A}}^\star) \times \mathscr{B}_{\mathrm{A}}(\pi_{\mathrm{D}}^\star). \qquad \text{(Eq. 1, Nash, 1951)} \tag{5.10}$$

This equilibrium solves the following minimax problem (von Neumann, 1928).

$$\underset{\pi_{\mathrm{D}} \in \Pi_{\mathrm{D}}}{\text{minimize}}\ \underset{\pi_{\mathrm{A}} \in \Pi_{\mathrm{A}}}{\text{maximize}}\ J_{\mathrm{D}}^{(\pi_{\mathrm{D}}, \pi_{\mathrm{A}})}(\mathbf{b}_1) \tag{5.11a}$$
$$\text{subject to}\ s_{t+1} \sim f\big(\cdot \mid s_t, \mathbf{a}_t\big) \qquad \forall t \geq 1 \tag{5.11b}$$
$$o_t \sim z\big(\cdot \mid s_t\big) \qquad \forall t \geq 2 \tag{5.11c}$$
$$a_t^{(\mathrm{A})} \sim \pi_{\mathrm{A}}\big(\cdot \mid \mathbf{b}_t, s_t\big) \qquad \forall t \geq 1 \tag{5.11d}$$
$$a_t^{(\mathrm{D})} \sim \pi_{\mathrm{D}}\big(\cdot \mid \mathbf{b}_t\big) \qquad \forall t \geq 1 \tag{5.11e}$$
$$s_1 \sim \mathbf{b}_1, \tag{5.11f}$$

where (5.11b) is the dynamics constraint; (5.11c) describes the observations; (5.11d)–(5.11e) capture the actions; and (5.11f) defines the initial state distribution.

**Remark 5.3.** We write min max instead of inf sup since (5.11) has a solution.

While any strategy pair $\boldsymbol{\pi}^\star$ that satisfies (5.10) is a NE, (5.11) implies that $\boldsymbol{\pi}^\star$ together with $\mathbb{B}$ (22) can form a stronger equilibrium, namely a Perfect Bayesian equilibrium (PBE), see Def. 4 in the background chapter. It follows from Thm. 3 in the background chapter that such an equilibrium exists, as formally stated below.

**Theorem 5.2** (Existence of equilibria and best responses).
*Given the instantiation of* $\Gamma$ *described in §5.3, the following holds.*

*(A)* $|\mathscr{B}_{\mathrm{D}}(\pi_{\mathrm{A}})| > 0$ *and* $|\mathscr{B}_{\mathrm{A}}(\pi_{\mathrm{D}})| > 0$ $\forall(\pi_{\mathrm{A}}, \pi_{\mathrm{D}}) \in \Pi_{\mathrm{A}} \times \Pi_{\mathrm{D}}$.

*(B)* $\Gamma$ *has a* PBE.

Figure 5.5 shows the value of a PBE when $N = 1$. Interestingly, the expected cost is the highest when $\mathbf{b}(1)$ (belief of compromise) is 0.35 rather than[5] 0.5.

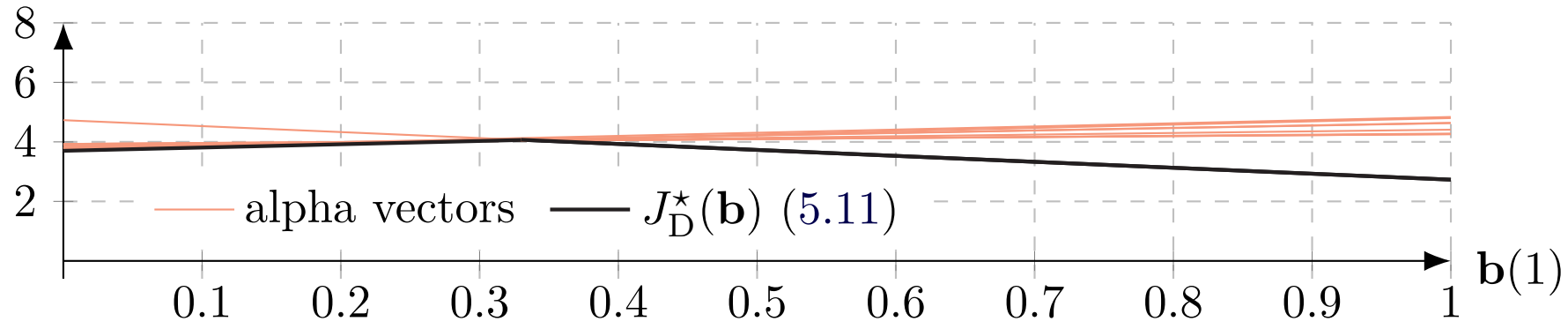

***Figure 5.5:*** *The equilibrium value (5.11) of* $\Gamma$ *when* $N = 1$ *(computed with the HSVI algorithm (Alg. 1, Horák et al., 2017)), see Fig. 5.6;* $J^\star_{\mathrm{D}}(\mathbf{b}) = \min_i [1-\mathbf{b}(1), \mathbf{b}(1)]^T \boldsymbol{\alpha}^{(i)}$*, where* $\boldsymbol{\alpha}^{(i)}$ *is an alpha vector (Def. 1, Sondik, 1978); see Appendix F for the hyperparameters.*

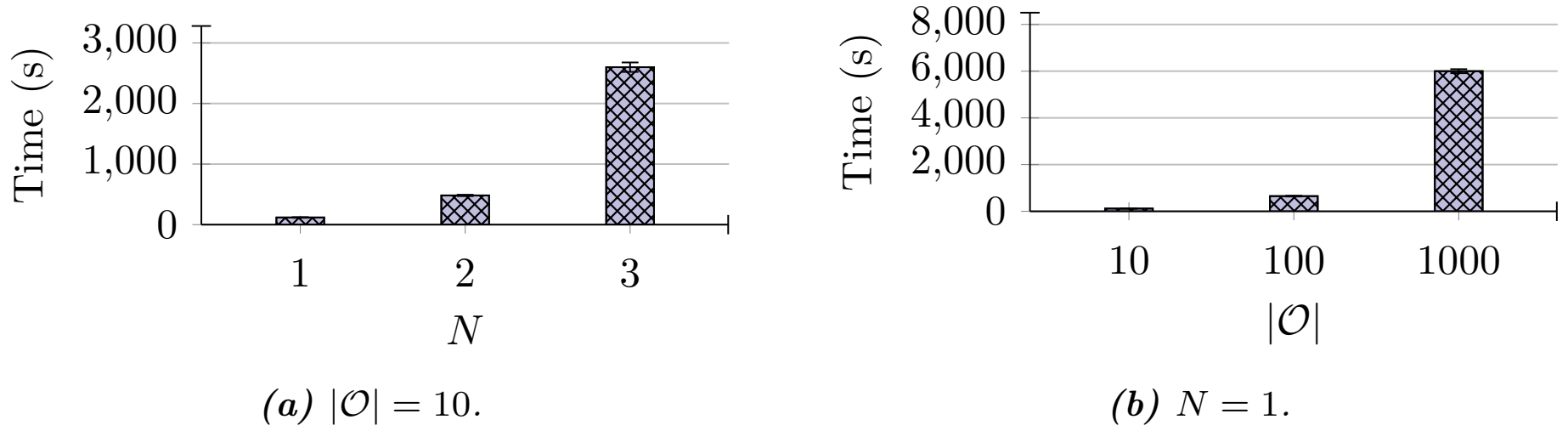

***Figure 5.6:*** *Time required to compute a Perfect Bayesian equilibrium (PBE) of* $\Gamma$ *(5.1) with HSVI (Alg. 1, Horák et al., 2017) for different values of* $N$ *(Fig. 5.6.a) and* $|\mathcal{O}|$ *(Fig. 5.6.b); error bars indicate the* 95% *confidence interval based on* 20 *measurements; hyperparameters are listed in Appendix F.*

[5] This finding is similar to the value function shown in Fig. 2.9 of Paper 2.

## 5.4 Problem Statement

While the equilibrium defined above describes how the game *ought to be played* by rational players, computing it is generally intractable, as illustrated in Fig. 5.6. Further, the equilibrium assumes a stationary game where both players have correctly specified models, which is unrealistic. To address these limitations, we relax the standard assumptions and consider a setting where the game is *non-stationary* and players have *misspecified* models. In particular, we consider the following problem.

**Problem 5.1** (Non-stationary game with misspecification)**.** We consider a game $\Gamma_{\boldsymbol{\theta}_t}$ based on (5.1) that is parameterized by a time-dependent vector $\boldsymbol{\theta}_t$, which is hidden from the players. This vector can represent the transition function (5.2), the observation function (5.4), etc. The game parameters not included in $\boldsymbol{\theta}_t$ are defined in §5.3 and known to both players. Player k has a *conjecture* of $\boldsymbol{\theta}_t$, denoted by $\overline{\boldsymbol{\theta}}_t^{(\mathrm{k})} \in \Theta_{\mathrm{k}}$, which is *misspecified* if $\boldsymbol{\theta}_t \notin \Theta_{\mathrm{k}}$. As $\boldsymbol{\theta}_t$ evolves, player k adapts its conjecture based on feedback $\mathbf{i}_t^{(\mathrm{k})}$ (5.3) and uses the conjecture to update its strategy $\pi_{\mathrm{k},t}$, starting from a *base strategy* $\pi_{\mathrm{k},1}$[6]. Strategy updates are parameterized by a *lookahead horizon* $\ell_{\mathrm{k}}$, which can be understood as a computational constraint. The defender conjectures $\ell_{\mathrm{A}}$ as $\overline{\ell}_{\mathrm{A}} \in \mathcal{L}$, which captures the defender's uncertainty about the attacker's computational capacity. Following Assumption 6 in the problem chapter, we assume the attacker knows $\ell_{\mathrm{D}}$ and the defender's conjectures.

We illustrate Prob. 5.1 using the following examples.

**Example: Backdoor.**

A defender detects an intrusion on the infrastructure and decides to block *incoming* connections at the gateway, which stops the intrusion with certainty according to its conjectured transition model $\overline{\boldsymbol{\theta}}_t^{(\mathrm{D})} \in \Theta_{\mathrm{D}}$ (5.2). However, it fails to realize that the attacker has installed a backdoor that allows access through *outgoing* connections. Hence, the defender is misspecified and $\boldsymbol{\theta}_t \notin \Theta_{\mathrm{D}}$.

**Example: Moving target defense.**

An attacker has performed reconnaissance and is *certain* about the configuration of the infrastructure, i.e., $\Theta_{\mathrm{A}} = \{\boldsymbol{\theta}_1\}$, where $\boldsymbol{\theta}_1$ represents the complete game model (5.1). However, it fails to realize that the defender employs moving target defense and regularly changes the configuration, leading to a misspecified model when $\boldsymbol{\theta}_t$ changes.

[6] Note that we do not make any assumption about the time evolution of $\boldsymbol{\theta}_t$.

Solving the game described in Prob. 5.1 leads to the following questions.

- ❓ What is an effective method for a player to update its conjecture and its strategy?
- ❓ Do the sequences of conjectures converge?
- ❓ Once the parameters $\boldsymbol{\theta}_t$ remain constant, how can the steady state of $\Gamma_{\boldsymbol{\theta}_t}$ be characterized?

## 5.5 Online Learning with Adaptive Conjectures

We address the above questions and develop **C**onjectural **O**nline **L**earning (COL), a game-theoretic method for *online learning* in $\Gamma_{\boldsymbol{\theta}_t}$ (Prob. 5.1). Using COL, each player iteratively adapts its conjecture through *Bayesian learning* and updates its strategy through *rollout*, which is a form of approximate dynamic programming that resembles model predictive control; see Fig. 5.7 (Bertsekas, 2021). The pseudocode of COL is listed in Alg. 5.1 on page 215. The main steps of COL are described below.

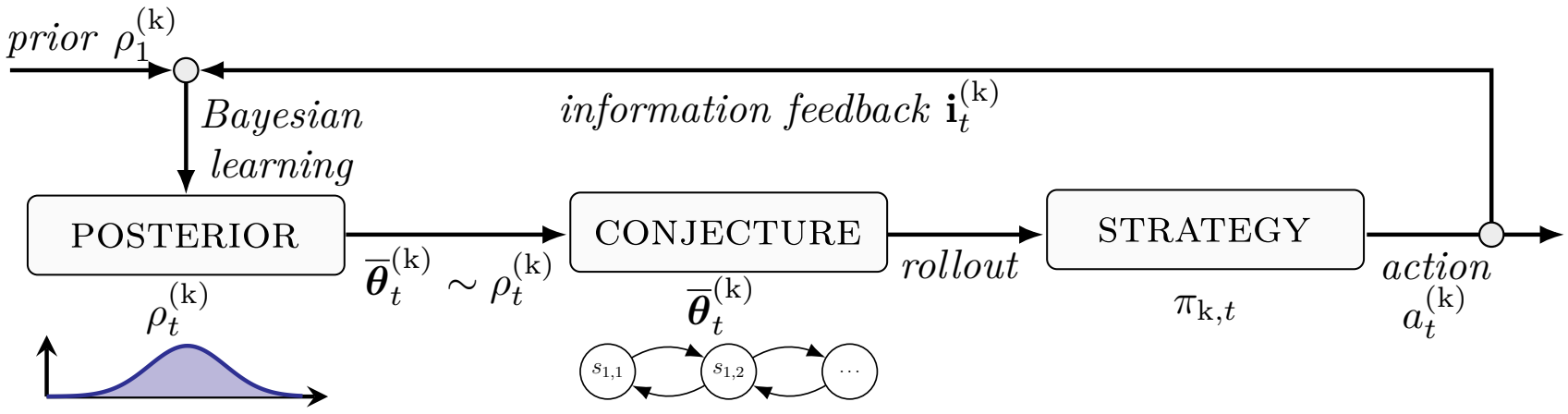

***Figure 5.7:*** *Conjectural Online Learning (COL); the figure illustrates a time step during which player* k *updates its conjecture* $\overline{\boldsymbol{\theta}}_t^{(\mathrm{k})}$ *and its strategy* $\pi_{\mathrm{k},t}$.

At time $t$, player k computes its action as follows[7].

$$(a_t^{(\mathrm{k})} \sim \pi_{\mathrm{k},t}) \in \mathscr{R}(\mathrm{k}, \overline{\boldsymbol{\theta}}_t^{(\mathrm{k})}, \mathbf{b}_t, \overline{J}_{\mathrm{k}}^{(\boldsymbol{\pi}_t)}, \ell_{\mathrm{k}}, \mathbf{h}_t^{(\mathrm{k})}) \triangleq \underset{a_t^{(\mathrm{k})}, a_{t+1}^{(\mathrm{k})}, \ldots, a_{t+\ell_{\mathrm{k}}-1}^{(\mathrm{k})}}{\arg\min} \tag{5.12}$$

$$\mathbb{E}_{\boldsymbol{\pi}_t}\left[\sum_{j=t}^{t+\ell_{\mathrm{k}}-1} \underbrace{\gamma^{j-t} c_{\mathrm{k}}(S_j, A_j^{(\mathrm{D})})}_{\text{Lookahead.}} + \underbrace{\gamma^{\ell_{\mathrm{k}}} \overline{J}_{\mathrm{k}}^{(\boldsymbol{\pi}_t)}(\mathbf{B}_{t+\ell_{\mathrm{k}}})}_{\text{Cost-to-go.}} \mid \mathbf{b}_t, \mathbf{h}_t^{(\mathrm{k})}\right],$$

where $\pi_{\mathrm{k},t}$ is the *rollout strategy*, $\ell_{\mathrm{k}}$ is the lookahead horizon, $\mathscr{R}$ is the rollout operator, $c_{\mathrm{D}} \triangleq c$ (5.6), $c_{\mathrm{A}} \triangleq -c$, $\boldsymbol{\pi}_t = (\pi_{\mathrm{k},1}, \overline{\pi}_{-\mathrm{k},t})$, $\overline{J}_{\mathrm{k}}^{(\boldsymbol{\pi}_t)}$ is the cost function induced by $\overline{\boldsymbol{\theta}}_t^{(\mathrm{k})}$, and $\overline{\pi}_{-\mathrm{k},t}$ is the conjectured strategy of the opponent. Since the base strategies $(\pi_{\mathrm{A},1}, \pi_{\mathrm{D},1})$ are known, this conjecture can be obtained from (5.12).

[7] $J_{\mathrm{A}}$ depends on $s$; however, to streamline the notation for the operations performed by the attacker and the defender, we do not explicitly denote this dependence in (5.12).

The Bellman equation in (5.12) corresponds to one step of policy iteration[8] with the base strategy as the starting point (Eqs. 6.4.1-22, Puterman, 1994). In fact, (5.12) can be interpreted as a Newton step (Bertsekas, 2022). The effect of $\ell_{\mathrm{k}} > 1$ is that the starting point of this Newton step is moved closer to the best response strategy through $\ell_{\mathrm{k}} - 1$ value iterations (Eq. 6.3.2-4, Puterman, 1994). Hence, the computational complexity of (5.12) grows exponentially with $\ell_{\mathrm{k}}$, as shown in Fig. 5.8. To manage this complexity for large instantiations of $\Gamma_{\boldsymbol{\theta}_t}$, we estimate $\mathbb{E}$ in (5.12) using Monte-Carlo samples. Note that (5.12) computes the next action as if the conjectures were true, i.e., the action is computed based on *(enforced) certainty equivalence* (p. 185, Bertsekas, 2021)(p. 232, Kumar and Varaiya, 1986).

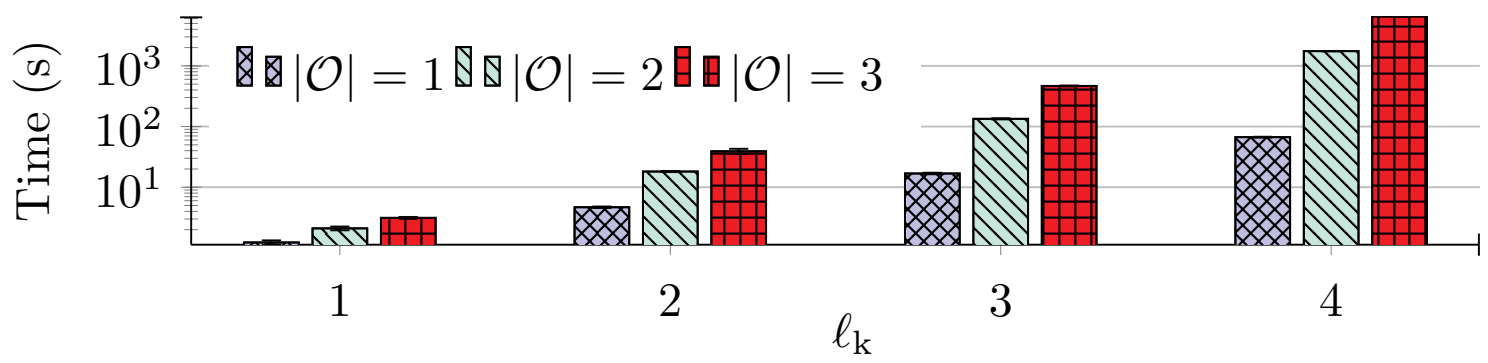

***(a)*** *Exact rollout.*

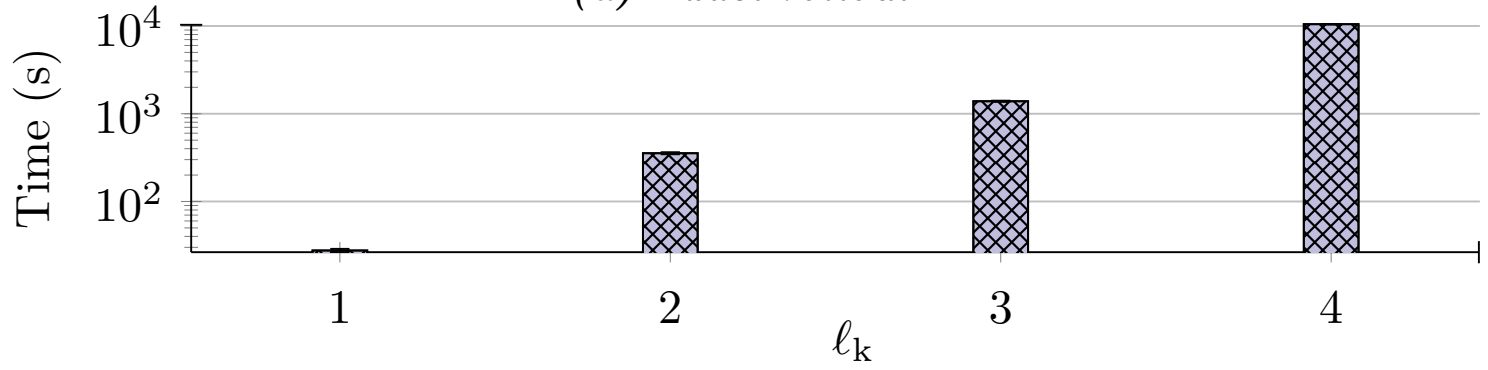

***(b)*** *Monte-Carlo rollout with* $|\mathcal{O}| = 26178$.

***Figure 5.8:*** *Compute time of the rollout operator (5.12) for varying lookahead horizons* $\ell_{\mathrm{k}}$ *and observation space sizes* $|\mathcal{O}|$*; hyperparameters are listed in Appendix F.*

We know from dynamic programming that $\pi_{\mathrm{k},t}$ (5.12) improves on the base strategy $\pi_{\mathrm{k},1}$ (Prop. 1, Bhattacharya et al., 2020). The extent of this improvement depends on the lookahead horizon and the accuracy of the conjectures as follows.

**Theorem 5.3** (Rollout performance bound)**.**
*The conjectured cost of player* k*'s rollout strategy* $\pi_{\mathrm{k},t}$ *satisfies*

$$\overline{J}_{\mathrm{k}}^{(\pi_{\mathrm{k},t},\overline{\pi}_{-\mathrm{k},t})}(\mathbf{b}) \leq \overline{J}_{\mathrm{k}}^{(\pi_{\mathrm{k},1},\overline{\pi}_{-\mathrm{k},t})}(\mathbf{b}) \qquad \forall \mathbf{b} \in \mathcal{B}. \tag{5.13a}$$

*Assuming* $(\overline{\boldsymbol{\theta}}_t^{(\mathrm{k})}, \overline{\ell}_{-\mathrm{k}})$ *predicts the game* $\ell_{\mathrm{k}}$ *steps ahead, then*

$$\|\overline{J}_{\mathrm{k}}^{(\pi_{\mathrm{k},t},\overline{\pi}_{-\mathrm{k},t})} - J_{\mathrm{k}}^{\star}\| \leq \frac{2\gamma^{\ell_{\mathrm{k}}}}{1-\gamma} \|\overline{J}_{\mathrm{k}}^{(\pi_{\mathrm{k},1},\overline{\pi}_{-\mathrm{k},t})} - J_{\mathrm{k}}^{\star}\|, \tag{5.13b}$$

Goes quickly to zero as $\ell_{\mathrm{k}} \to \infty$.

*where* $J_{\mathrm{k}}^{\star}$ *is the optimal cost-to-go when facing* $\pi_{-\mathrm{k},t}$ *and* $\|J\| \triangleq \max_x |J(x)|$.

[8] Policy iteration is defined in the background chapter; see (9)–(10).

Theorem 5.3 states that the performance bound improves superlinearly when the lookahead horizon $\ell_{\mathrm{k}}$ increases or when the conjectured cost function $\overline{J}_{\mathrm{k}}$ moves closer to $J^{\star}_{\mathrm{k}}$ (5.8), as shown in Fig. 5.9. In particular, (5.13b) suggests that $\ell_{\mathrm{k}}$ controls the trade-off between computational cost and rollout performance. We provide proof in Appendix B.

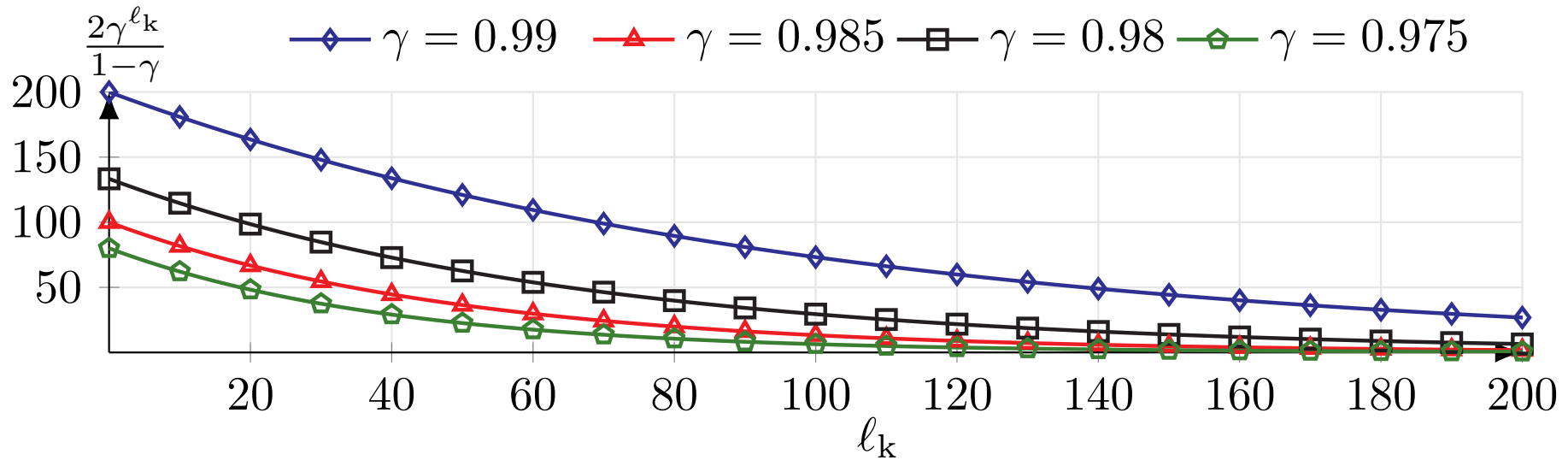

***Figure 5.9:*** *Illustration of Thm. 5.3; the x-axis indicates the lookahead horizon $\ell_{\mathrm{k}}$; the y-axis indicates the factor of the performance bound in (5.13b); the curves relate to different discount factors $\gamma$.*

After computing (5.12) and executing the corresponding action, player k receives the feedback $\mathbf{i}^{(\mathrm{k})}_t$ (5.3) and updates its conjectures as $\overline{\boldsymbol{\theta}}^{(\mathrm{k})}_t \sim \rho^{(\mathrm{k})}_t$ and $\overline{\ell}_{\mathrm{A},t} \sim \mu_t$, where $\rho^{(\mathrm{k})}_t$ and $\mu_t$ are adapted through *Bayesian learning* as

$$\rho^{(\mathrm{k})}_t(\overline{\boldsymbol{\theta}}^{(\mathrm{k})}_t) \triangleq \frac{\mathbb{P}[\mathbf{i}^{(\mathrm{k})}_t \mid \overline{\boldsymbol{\theta}}^{(\mathrm{k})}_t, \mathbf{b}_{t-1}]\rho^{(\mathrm{k})}(\overline{\boldsymbol{\theta}}^{(\mathrm{k})}_{t-1})}{\int_{\Theta_{\mathrm{k}}} \mathbb{P}[\mathbf{i}^{(\mathrm{k})}_t \mid \overline{\boldsymbol{\theta}}^{(\mathrm{k})}_t, \mathbf{b}_{t-1}]\rho^{(\mathrm{k})}_{t-1}(\mathrm{d}\overline{\boldsymbol{\theta}}^{(\mathrm{k})}_t)} \quad \forall \overline{\boldsymbol{\theta}}^{(\mathrm{k})}_t \in \Theta_{\mathrm{k}}, \mathrm{k} \in \mathcal{N} \tag{5.14a}$$

$$\mu_t(\overline{\ell}_{\mathrm{A},t}) \triangleq \frac{\mathbb{P}[\mathbf{i}^{(\mathrm{D})}_t \mid \overline{\ell}_{\mathrm{A},t}, \mathbf{b}_{t-1}]\mu_{t-1}(\overline{\ell}_{\mathrm{A},t})}{\sum_{\tilde{\ell}_{\mathrm{A}} \in \mathcal{L}} \mathbb{P}[\mathbf{i}^{(\mathrm{D})}_t \mid \tilde{\ell}_{\mathrm{A}}, \mathbf{b}_{t-1}]\mu_{t-1}(\tilde{\ell}_{\mathrm{A}})} \quad \forall \overline{\ell}_{\mathrm{A},t} \in \mathcal{L}. \tag{5.14b}$$

These Bayesian updates are well-defined using the following assumption.

**Assumption 5.1.** *(i) $\mathcal{L}$ is finite and $\Theta_{\mathrm{k}}$ is a compact subset of an Euclidean space; (ii) $\rho^{(\mathrm{k})}_1$ and $\mu_1$ have full support; and (iii) for all feasible $(\mathbf{i}^{(\mathrm{k})}, \mathbf{b})$, there exists $\overline{\boldsymbol{\theta}} \in \Theta_{\mathrm{k}}$ and $\overline{\ell}_{\mathrm{A}} \in \mathcal{L}$ for which the feedback $\mathbf{i}^{(\mathrm{k})}$ has positive probability in $\mathbf{b}$.*

The steps outlined above are executed online. Specifically, at each time step during online play, both players update their conjectures through Bayesian learning (5.14) and adjust their strategies using rollout (5.12). The pseudocode of COL, which implements these steps, is provided in Alg. 5.1 on the following page.

***Algorithm 5.1:** Conjectural **O**nline **L**earning (COL).*

**Input:** Initial belief $\mathbf{b}_1$, game model $\Gamma_{\boldsymbol{\theta}_1}$, base strategies $\boldsymbol{\pi}_1 \triangleq (\pi_{\mathrm{D},1}, \pi_{\mathrm{A},1})$, priors $(\mu_1, \rho_1^{(\mathrm{D})}, \rho_1^{(\mathrm{A})})$, discount factor $\gamma$, lookahead horizons $\ell_{\mathrm{D}}, \ell_{\mathrm{A}}$.
**Output:** A sequence of action profiles $\mathbf{a}_1, \mathbf{a}_2, \ldots$.

**procedure** COL($\mathbf{b}_1$, $\Gamma_{\boldsymbol{\theta}_1}$, $\boldsymbol{\pi}_1 \triangleq (\pi_{\mathrm{D},1}, \pi_{\mathrm{A},1})$, $(\mu_1, \rho_1^{(\mathrm{D})}, \rho_1^{(\mathrm{A})})$, $\gamma$, $\ell_{\mathrm{D}}, \ell_{\mathrm{A}}$)
  $\mathbf{h}_1^{(\mathrm{D})} \leftarrow (\mathbf{b}_1), \mathbf{h}_1^{(\mathrm{A})} \leftarrow (\mathbf{b}_1)$, $s_1 \sim \mathbf{b}_1$. ▷ Initialization
  $\overline{\pi}_{\mathrm{A},1} \leftarrow \pi_{\mathrm{A},1}, \overline{\pi}_{\mathrm{D},1} \leftarrow \pi_{\mathrm{D},1}$.
  $a_1^{(\mathrm{D})} \sim \pi_{\mathrm{D},1}(\mathbf{b}_1), a_1^{(\mathrm{A})} \sim \pi_{\mathrm{A},1}(\mathbf{b}_1, s_1)$.
  $s_2 \sim f(\cdot \mid s_1, a_1^{(\mathrm{D})}, a_1^{(\mathrm{A})})$.
  **for** $t = 2, 3, \ldots$ **do**
    $o_t \sim z(\cdot \mid s_t), \mathbf{i}_t^{(\mathrm{D})} \leftarrow (o_t), \mathbf{h}_t^{(\mathrm{D})} \leftarrow (\mathbf{h}_{t-1}^{(\mathrm{D})}, \mathbf{i}_t^{(\mathrm{D})}, a_{t-1}^{(\mathrm{D})})$. ▷ Defender learning
    Update $\rho_t^{(\mathrm{D})}$ and $\mu_t$ (5.14) and set $\overline{\boldsymbol{\theta}}_t^{(\mathrm{D})} \sim \rho_t^{(\mathrm{D})}$ and $\overline{\ell}_{\mathrm{A},t} \sim \mu_t$.
    $\mathbf{b}_t \leftarrow \mathbb{B}(\mathbf{h}_t^{(\mathrm{D})}, \overline{\pi}_{\mathrm{A},t-1})$ (22).
    Estimate $\overline{J}_{\mathrm{A}}^{(\pi_{\mathrm{D},t-1}, \pi_{\mathrm{A},1})}$ using $\Gamma_{\overline{\boldsymbol{\theta}}_t^{(\mathrm{D})}}$.
    Define $\overline{\pi}_{\mathrm{A},t}(\mathbf{b}_t)$ based on $\mathscr{R}(\mathrm{A}, \overline{\boldsymbol{\theta}}_t^{(\mathrm{D})}, \mathbf{b}_t, \overline{J}_{\mathrm{A}}^{(\pi_{\mathrm{D},t-1}, \pi_{\mathrm{A},1})}, \overline{\ell}_{\mathrm{A},t}, \mathbf{h}_t^{(\mathrm{D})})$.
    Estimate $\overline{J}_{\mathrm{D}}^{(\pi_{\mathrm{D},1}, \overline{\pi}_{\mathrm{A},t})}$ using $\Gamma_{\overline{\boldsymbol{\theta}}_t^{(\mathrm{D})}}$.
    $a_t^{(\mathrm{D})} \in \mathscr{R}(\mathrm{D}, \overline{\boldsymbol{\theta}}_t^{(\mathrm{D})}, \mathbf{b}_t, \overline{J}_{\mathrm{D}}^{(\pi_{\mathrm{D},1}, \overline{\pi}_{\mathrm{A},t})}, \ell_{\mathrm{D}}, \mathbf{h}_t^{(\mathrm{D})})$.
    $\mathbf{i}_t^{(\mathrm{A})} \leftarrow (o_t, s_t, a_{t-1}^{(\mathrm{D})}), \quad \mathbf{h}_t^{(\mathrm{A})} \leftarrow (\mathbf{h}_{t-1}^{(\mathrm{A})}, \mathbf{i}_t^{(\mathrm{A})}, a_{t-1}^{(\mathrm{A})})$. ▷ Attacker learning
    Update $\rho_t^{(\mathrm{A})}$ using (5.14a) and set $\overline{\boldsymbol{\theta}}_t^{(\mathrm{A})} \sim \rho_t^{(\mathrm{A})}$.
    Estimate $\overline{J}_{\mathrm{A}}^{(\pi_{\mathrm{A},1}, \pi_{\mathrm{D},t})}$ using $\Gamma_{\overline{\boldsymbol{\theta}}_t^{(\mathrm{A})}}$.
    $a_t^{(\mathrm{A})} \in \mathscr{R}(\mathrm{A}, \overline{\boldsymbol{\theta}}_t^{(\mathrm{A})}, \mathbf{b}_t, \overline{J}_{\mathrm{A}}^{(\pi_{\mathrm{A},1}, \pi_{\mathrm{D},t})}, \ell_{\mathrm{A}}, \mathbf{h}_t^{(\mathrm{A})})$.
    $s_{t+1} \sim f(\cdot \mid s_t, a_t^{(\mathrm{D})}, a_t^{(\mathrm{A})})$.

### Convergence and equilibrium analysis

When player k updates its conjectures through (5.14), the goal is to minimize the *discrepancy* between the feedback distributions induced by the conjectures and the observed feedback (5.3). We define this discrepancy as

$$K(\overline{\alpha}, \nu) \triangleq \mathbb{E}_{\mathbf{b} \sim \nu} \mathbb{E}_{\mathbf{I}^{(\mathrm{k})}} \left[ \ln \left( \frac{\mathbb{P}[\mathbf{I}^{(\mathrm{k})} \mid \alpha, \mathbf{b}]}{\mathbb{P}[\mathbf{I}^{(\mathrm{k})} \mid \overline{\alpha}, \mathbf{b}]} \right) \mid \alpha, \mathbf{b} \right], \tag{5.15}$$

where $\alpha \in \{\boldsymbol{\theta}, \ell_{\mathrm{A}}\}$ and $\nu \in \Delta(\mathcal{B})$ is an occupancy measure.

**Key insight.**

Minimizing (5.15) offers an alternative to Bayesian consistency by focusing the posterior on conjectures that produce consistent observations (5.4), thus shifting focus from the parameter to the observation process.

We say that conjectures that minimize (5.15) are *consistent* with $\nu$ (Kullback and Leibler, 1951)[9]. Hence, the sets of consistent conjectures at time $t$ are

$$\overline{\boldsymbol{\theta}}_t^{(\mathrm{k})} \in \Theta^\star_{\mathrm{k}}(\nu_t) \triangleq \arg\min_{\overline{\boldsymbol{\theta}}_t^{(\mathrm{k})} \in \Theta_{\mathrm{k}}} K(\overline{\boldsymbol{\theta}}_t^{(\mathrm{k})}, \nu_t) \qquad \forall \mathrm{k} \in \mathcal{N} \tag{5.16a}$$

$$\overline{\ell}_{\mathrm{A},t} \in \mathcal{L}^\star(\nu_t) \triangleq \arg\min_{\overline{\ell}_{\mathrm{A},t} \in \mathcal{L}} K(\overline{\ell}_{\mathrm{A},t}, \nu_t), \tag{5.16b}$$

where $\nu_t(\mathbf{b}) \triangleq \frac{1}{t}\sum_{\tau=1}^{t} \mathbb{1}_{\mathbf{b}=\mathbf{b}_\tau}$ is the empirical occupancy measure and $\boldsymbol{\pi}_{\mathbf{h}_t}$ is the empirical strategy profile induced by COL at time $t$. Intuitively, $\Theta^\star_{\mathrm{k}}$ and $\mathcal{L}^\star$ contain the conjectures that player k considers possible after observing feedback generated by $\nu_t$ and $\boldsymbol{\pi}_{\mathbf{h}_t}$ (Berk, 1966). A desirable property of the conjecture distributions (5.14) is, therefore, that they concentrate on $\Theta^\star_{\mathrm{k}}$ and $\mathcal{L}^\star$ (5.16). This property is guaranteed asymptotically under the following conditions.

**Assumption 5.2** (Regularity conditions). *For fixed values of* $\mathbf{i}^{(\mathrm{k})}$ *and* $\boldsymbol{\theta}$*,*

1. *The mapping* $\mathbf{b} \mapsto \ln \mathbb{P}[\mathbf{i}^{(\mathrm{k})} \mid \boldsymbol{\theta}, \mathbf{b}]$ *is Lipschitz w.r.t. the Wasserstein-*1 *distance, and the Lipschitz constant is independent of* $\mathbf{i}^{(\mathrm{k})}$ *and* $\boldsymbol{\theta}$*.*
2. *The mapping* $\boldsymbol{\theta} \mapsto \ln \mathbb{P}[\mathbf{i}^{(\mathrm{k})} \mid \boldsymbol{\theta}, \mathbf{b}]$ *is continuous and there exists an integrable function* $g_{\mathbf{b}}(\mathbf{i}^{(\mathrm{k})})$ *for all* $\mathbf{b} \in \mathcal{B}$ *such that* $|\ln \frac{\mathbb{P}[\mathbf{i}^{(\mathrm{k})}|\boldsymbol{\theta},\mathbf{b}]}{\mathbb{P}[\mathbf{i}^{(\mathrm{k})}|\overline{\boldsymbol{\theta}},\mathbf{b}]}| \leq g_{\mathbf{b}}(\mathbf{i}^{(\mathrm{k})})$ *for all* $\overline{\boldsymbol{\theta}} \in \Theta_{\mathrm{k}}$*.*

**Theorem 5.4** (Asymptotic consistency). *Given Assumptions 5.1–5.2, the following holds for any sequence* $(\boldsymbol{\pi}_{\mathbf{h}_t}, \nu_t)_{t\geq 1}$ *generated by* COL *(Alg. 5.1).*

$$\lim_{t\to\infty} \sum_{\overline{\ell}_{\mathrm{A}} \in \mathcal{L}} \left(K(\overline{\ell}_{\mathrm{A}}, \nu_t) - K^\star_{\mathcal{L}}(\nu_t)\right) \mu_{t+1}(\overline{\ell}_{\mathrm{A}}) = 0 \ a.s.\text{-}\mathbb{P}^{\mathscr{R}}, \tag{A}$$

*and provided that* $\boldsymbol{\theta}_t = \boldsymbol{\theta}_1$ *for all* $t$*,*

$$\lim_{t\to\infty} \int_{\Theta_{\mathrm{k}}} \left(K(\overline{\boldsymbol{\theta}}, \nu_t) - K^\star_{\Theta_{\mathrm{k}}}(\nu_t)\right) \rho^{(\mathrm{k})}_{t+1}(\mathrm{d}\overline{\boldsymbol{\theta}}) = 0 \ a.s.\text{-}\mathbb{P}^{\mathscr{R}}, \tag{B}$$

*where* $(K^\star_{\mathcal{L}}, K^\star_{\Theta_{\mathrm{k}}})$ *denote the minimal values of (5.16) and* $\mathbb{P}^{\mathscr{R}}$ *is a probability measure over the set of realizable histories* $\bigtimes_{t\geq 1}(\mathcal{H}_t^{(\mathrm{D})} \times \mathcal{H}_t^{(\mathrm{A})})$ *that is induced by* $(\boldsymbol{\pi}_{\mathbf{h}_t})_{t\geq 1}$*.*

Theorem 5.4 states that the conjectures produced by COL are asymptotically consistent (5.16); see Appendix C and Appendix D for the proof. This consistency means that if the sequence $(\boldsymbol{\pi}_{\mathbf{h}_t}, \nu_t)_{t\geq 1}$ generated by COL converges, then it converges to a Berk-Nash Equilibrium (BNE), defined on the next page.

[9] We use the standard convention that $-\ln 0 = \infty$ and $0 \ln 0 = 0$.

**Definition 5.1** (Berk-Nash Equilibrium (BNE), adapted from (Esponda and Pouzo, 2021))**.** *$(\boldsymbol{\pi}, \nu) \in \Pi \times \Delta(\mathcal{B})$ is a BNE of $\Gamma_{\boldsymbol{\theta}_t}$ (Prob. 5.1, COL) iff there exist a $\rho^{(\mathrm{k})} \in \Delta(\Theta_{\mathrm{k}})$ for each player $\mathrm{k} \in \{\mathrm{D}, \mathrm{A}\}$ such that*

*(i) BOUNDED RATIONALITY. $\pi_{\mathrm{k}}$ is a best response against $\pi_{-\mathrm{k}}$ for any $\mathbf{b}$ given the occupancy measure $\nu$, the conjectures $(\rho^{(\mathrm{k})}, \rho^{(-\mathrm{k})})$, the lookahead horizons $(\ell_{\mathrm{k}}, \ell_{-\mathrm{k}})$, and the base strategies $\boldsymbol{\pi}_1 = (\pi_{\mathrm{D},1}, \pi_{\mathrm{A},1})$.*

*(ii) CONSISTENCY. $\rho^{(\mathrm{k})} \in \Delta(\Theta^{\star}_{\mathrm{k}}(\nu))$.*

*(iii) STATIONARITY. $(\boldsymbol{\pi}, \nu)$ is a limit point of some sequence $(\boldsymbol{\pi}_{\mathbf{h}_t}, \nu_t)_{t \geq 1}$ generated by COL, satisfying*

$$\nu(\mathbf{b}') = \int_{\mathcal{B}} \mathbb{E}_{A^{(\mathrm{D})}, O, \Gamma_{\hat{\boldsymbol{\theta}}}} \left[\delta_{\mathbf{b}'}\left(\mathbb{B}(\mathbf{b}, A^{(\mathrm{D})}, O, \pi_{\mathrm{A}})\right)\right] \mathrm{d}\nu(\mathbf{b}),$$

*where $\mathbb{B}$ is the belief operator defined in (22), and $\Gamma_{\hat{\boldsymbol{\theta}}}$ is parameterized by $\hat{\boldsymbol{\theta}} \triangleq \int_{\Theta_{\mathrm{D}}} \overline{\boldsymbol{\theta}} \mathrm{d}\rho^{(\mathrm{D})}(\overline{\boldsymbol{\theta}})$, and $\delta_i(\cdot)$ is the Dirac delta function centered at $i$.*

**Origin of the term “Berk-Nash Equilibrium”.**

The term "Berk-Nash Equilibrium" (BNE) is a concept in game theory and economics that combines the ideas of R.H. Berk [48] and J.F. Nash [322]. The term was coined by I. Esponda and D. Pouzo in 2016 [135] and reflects that a) the rationality (optimality) condition of a BNE resembles the best response condition defined by Nash [322] in 1951; and b) the consistency condition of a BNE resembles Berk's definition of “asymptotic carrier” in 1966 [48].

**Corollary 5.1** (Connection between the BNE and the PBE)**.**

*(A) If the sequence $(\boldsymbol{\pi}_{\mathbf{h}_t}, \nu_t)_{t \geq 1}$ generated by COL converges to $(\boldsymbol{\pi}, \nu)$, then $(\boldsymbol{\pi}, \nu)$ is a BNE.*

*(B) Given a BNE $(\boldsymbol{\pi}, \nu)$ and assuming*

$$\overline{\boldsymbol{\theta}}^{(\mathrm{k})} \in \Theta^{\star}_{\mathrm{k}}(\nu) \implies \mathbb{P}[\mathbf{I}^{(\mathrm{k})} \mid \overline{\boldsymbol{\theta}}^{(\mathrm{k})}, \mathbf{b}] = \mathbb{P}[\mathbf{I}^{(\mathrm{k})} \mid \boldsymbol{\theta}, \mathbf{b}] \quad \textit{(IDENTIFIABLE)} \quad (5.17\text{a})$$

$$\ell_{\mathrm{D}} = \ell_{\mathrm{A}} = \infty, \quad \textit{(RATIONALIZABLE)} \quad (5.17\text{b})$$

*then $(\boldsymbol{\pi}, \mathbb{B})$ is a PBE.*

*Proof.* Condition $(i)$ in Def. 5.1 is ensured by rollout (5.12); condition $(ii)$ is ensured asymptotically by Thm. 5.4; and condition $(iii)$ is a consequence of convergence. Hence, statement (A) holds. Now consider statement (B). Condition 2) in Def. 4 is satisfied by definition of $\mathbb{B}$ (22). To see why condition 1) also must hold, note that assumption (5.17a) together with condition (*ii*) of Def. 5.1 implies that $\mathbb{P}[\mathbf{I}^{(\mathrm{k})} \mid \boldsymbol{\theta}, \mathbf{b}] = \mathbb{P}[\mathbf{I}^{(\mathrm{k})} \mid \overline{\boldsymbol{\theta}}^{(\mathrm{k})}, \mathbf{b}]$ for any $\overline{\boldsymbol{\theta}}^{(\mathrm{k})} \sim \rho^{(\mathrm{k})}$. Consequently, it follows from assumption (5.17b) that $\pi_{\mathrm{A}} \in \mathscr{B}_{\mathrm{A}}(\pi_{\mathrm{D}})$ and $\pi_{\mathrm{D}} \in \mathscr{B}_{\mathrm{D}}(\pi_{\mathrm{A}})$ for any $\mathbf{b}_1$ (5.8). □

Corollary 5.1 states that if the sequence $(\boldsymbol{\pi}_{\mathbf{h}_t}, \nu_t)_{t\geq 1}$ generated by COL converges, then it must converge to a BNE. Further, under the conditions defined in (5.17), this equilibrium is also a PBE. The converse is not necessarily true, however, since the perfect Bayesian equilibrium does not enforce condition (*iii*) of the BNE (Def. 5.1). This condition requires that $\nu_t$ converges to a stationary distribution, which is not guaranteed to exist. We provide an example in Appendix E. Whether a stationary distribution exists or not depends primarily on the parameter vector $\boldsymbol{\theta}_t$ and the observation function $z$ (5.4).

## 5.6 System Identification

To implement and evaluate the method described above for the APT use case (§5.2), we estimate the parameters of $\Gamma$ (5.1) using a digital twin of the target infrastructure. We create this digital twin using CSLE, as described in the methodology chapter (Hammar, 2023). The configuration of the target infrastructure is listed in Table 3.4 of Paper 3. The topology is shown in Fig. 3 in the introduction chapter. It consists of $N = 64$ servers, some of which are vulnerable to APTs. The digital twin comprises virtual containers and networks that replicate the functionality and the timing behavior of the target infrastructure. These containers run the same software and processes as the physical infrastructure.

Similar to papers 1–4, we define the observation $o_t$ (5.4) to be the priority-weighted sum of the number of IDS alerts at time $t$. We measure the value of $o_t$ in the digital twin at 30-second intervals. (30$s$ in the digital twin corresponds to 1 time step in the game $\Gamma$.) For the evaluation reported in this paper, we collect about $M = 10^5$ i.i.d. samples[10]. Using these samples, we estimate the observation distribution $z$ (5.4) with the empirical distribution $\widehat{z}$, where $\widehat{z} \overset{\text{a.s.}}{\rightarrow} z$ as $M \rightarrow \infty$[11].

## 5.7 Experimental Evaluation of COL

We implement COL (Fig. 5.7, Alg. 5.1) in Python and apply it to the APT use case using our methodology, which combines simulation-based optimization with evaluation on the digital twin[12]. The simulations and the digital twin run on a server with a 24-core INTEL XEON GOLD 2.10 GHz CPU and 768 GB RAM; see Fig. 21 in the methodology chapter. Hyperparameters for COL are listed in Appendix F.

[10] The measurements are available at (Hammar, 2023), where $|\mathcal{O}| = 26178$.
[11] It follows by the Glivenko-Cantelli theorem; see (Glivenko and Cantelli, 1933).
[12] The digital twin is created using CSLE, as described in the methodology chapter.

### Evaluation scenarios

We define five scenarios to evaluate the performance properties of COL. The first four scenarios are evaluated in simulation, and the fifth is evaluated on the digital twin. Our aim in evaluating these scenarios is to assess a) the computational requirements of rollout (5.12); b) the convergence rate of COL for different instantiations of $\Gamma_{\boldsymbol{\theta}_t}$ (Prob. 5.1); and c) the benefit of COL compared to existing intrusion response systems. Each evaluation scenario is based on an instantiation of Prob. 5.1 for the APT use case (§5.2) with the model described in §5.3.

In such an instantiation, the defender and the attacker take actions at time steps $t = 1, 2, \ldots$. During each step, they perform one action each: either a (passive) *continue* action or a *stop action*. The defender's stop action corresponds to hypervisor-based recovery of servers (Scenarios 5.1–5.4) or blocking of IP addresses (Scenario 5.5) (Reiser and Kapitza, 2007). The attacker's stop action is drawn randomly from Table 5.2. After executing the actions, the observations are either sampled from the estimated observation distribution (Scenario 5.1–Scenario 5.4) or measured directly from the digital twin (Scenario 5.5). The main difference between the evaluation scenarios is how the attacker and the defender models are misspecified, as explained below.

| *Type* | *Actions* | MITRE ATT&CK technique |
|---|---|---|
| Reconnaissance | TCP SYN scan, UDP scan | T1046 service scanning. |
| | TCP XMAS scan | T1046 service scanning. |
| | VULSCAN | T1595 active scanning. |
| | ping-scan | T1018 system discovery. |
| Brute-force | TELNET, SSH | T1110 brute force. |
| | FTP, CASSANDRA | T1110 brute force. |
| | IRC, MONGODB, MYSQL | T1110 brute force. |
| | SMTP, POSTGRES | T1110 brute force. |
| Exploit | CVE-2017-7494 | T1210 service exploitation. |
| | CVE-2015-3306 | T1210 service exploitation. |
| | CVE-2010-0426 | T1068 privilege escalation. |
| | CVE-2015-5602 | T1068 privilege escalation. |
| | CVE-2015-1427 | T1210 service exploitation. |
| | CVE-2014-6271 | T1210 service exploitation. |
| | CVE-2016-10033 | T1210 service exploitation. |
| | SQL injection (CWE-89) | T1210 service exploitation. |

***Table 5.2:*** *Attacker actions on the digital twin; see the methodology chapter for details.*

**Scenario 5.1** (Defender is uncertain about $\ell_{\mathrm{A}}$)**.** In this scenario, the game is stationary (i.e., $\boldsymbol{\theta}_t = \boldsymbol{\theta}_1$ for all $t$). $\boldsymbol{\theta}_1$ represents the complete game model (5.1) and is known to both players (i.e., $\rho_1^{(\mathrm{k})}(\boldsymbol{\theta}_1) = 1$), but the defender is uncertain about the attacker's computational capacity $\ell_{\mathrm{A}}$, i.e., $\mu_1(\ell_{\mathrm{A}}) < 1$.

**Scenario 5.2** (Non-stationary $\boldsymbol{\theta}_t$)**.** In this scenario, $\ell_{\mathrm{A}}$ is known to the defender, but the game is non-stationary, and $z$ (5.4) is parameterized by $\boldsymbol{\theta}_t$, which represents the number of clients. Hence, $\boldsymbol{\theta}_t$ changes whenever a client arrives or departs. Clients have exponential service times and arrive following a Poisson process with the following rate function

$$\lambda(t) = \exp\Bigg(\underbrace{\sum_{i=1}^{\dim(\boldsymbol{\psi})} \boldsymbol{\psi}_i t^i}_{\text{trend}} + \underbrace{\sum_{k=1}^{\dim(\boldsymbol{\chi})} \boldsymbol{\chi}_k \sin(\boldsymbol{\omega}_k t + \boldsymbol{\phi}_k)}_{\text{periodic}}\Bigg). \tag{5.18}$$

The function is illustrated in Figs. 5.10–5.11 (Kuhl et al., 1995); see Appendix F for the parameter values.

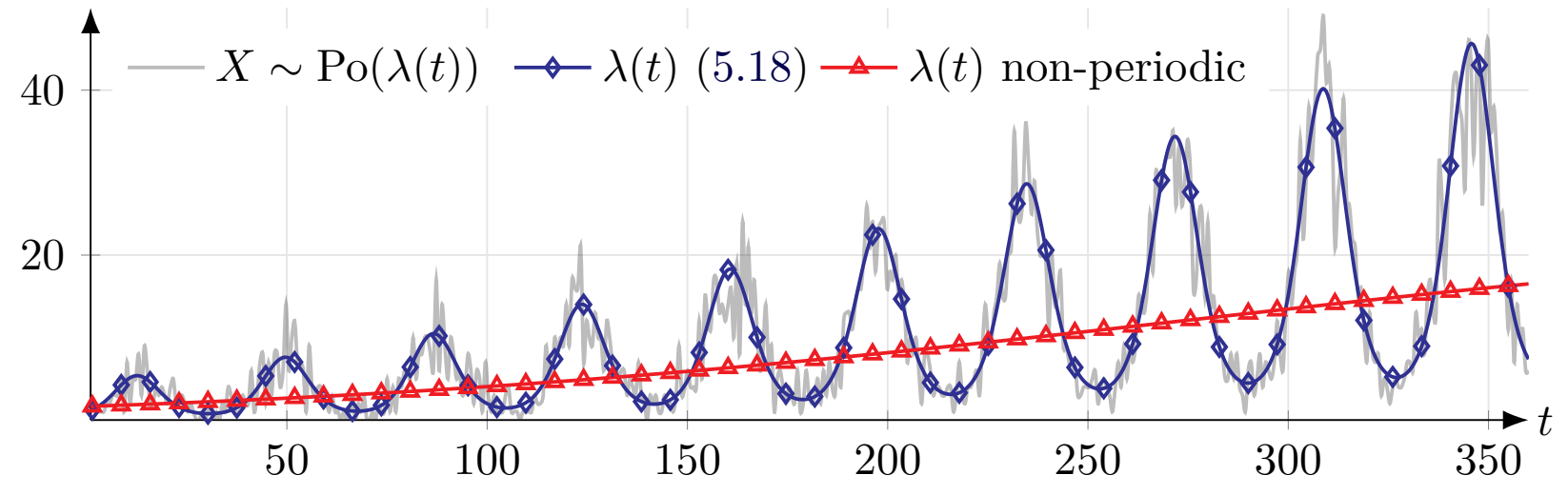

***Figure 5.10:*** *The arrival rate function (5.18) used in Scenario 5.2; the blue curve shows the arrival rate $\lambda(t)$; the red curve shows the trend of $\lambda(t)$ without the periodic effects; and the shaded black curve shows the number of arrivals.*

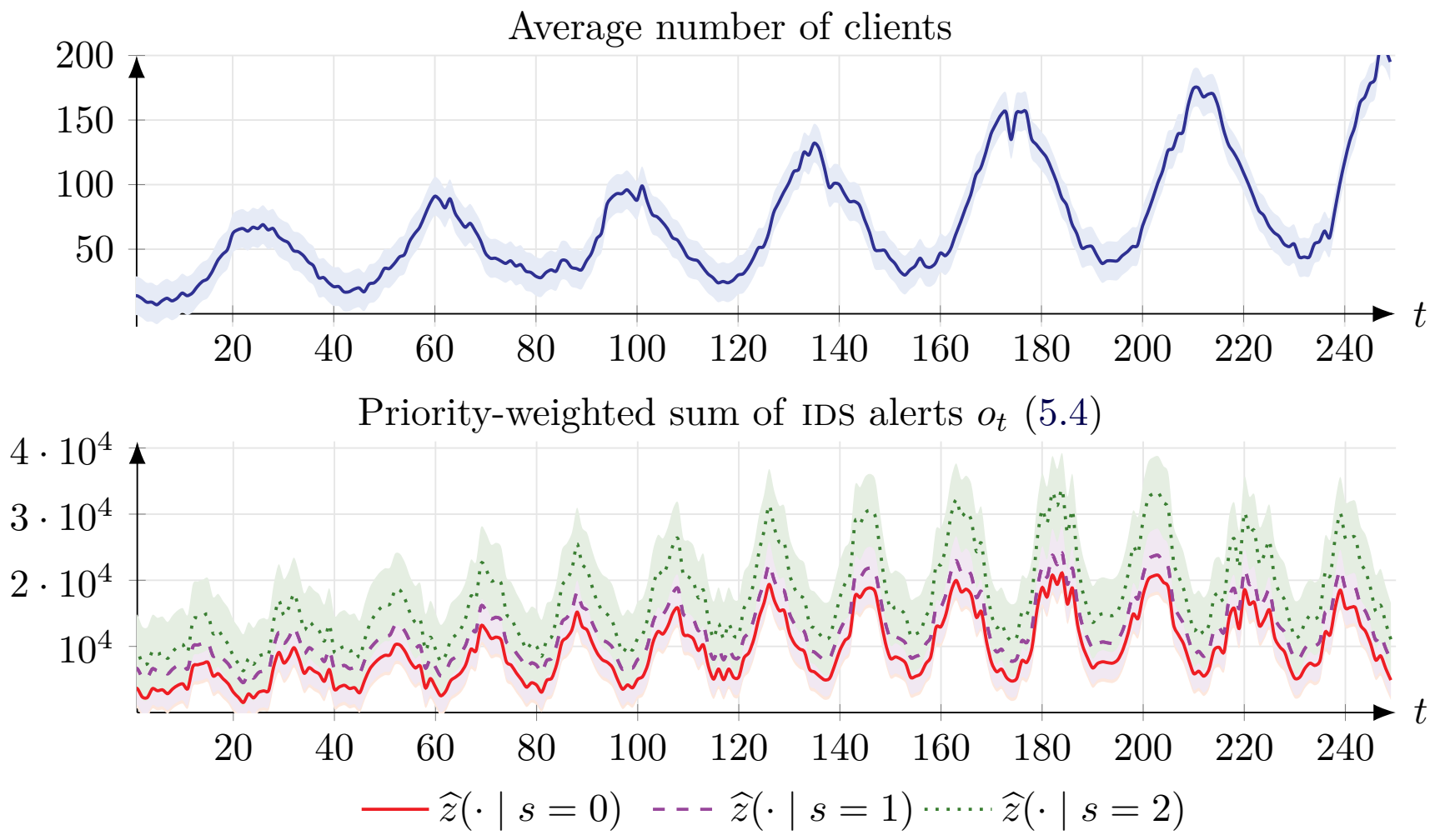

***Figure 5.11:*** *Estimated distributions of the number of clients and the priority-weighted sum of IDS alerts $o_t$ (5.4) during different arrival rates $\lambda(t)$ (5.18) based on the APT actions listed in Table 5.2 and the rate function shown in Fig. 5.10; the curves indicate mean values and the shaded areas indicate standard deviations from 3 measurements; the measurements are available at (Hammar, 2023).*

**Scenario 5.3** (Misspecified model conjectures $\rho_t^{(D)}, \rho_t^{(A)}$)**.** In this scenario, the game is stationary (i.e., $\boldsymbol{\theta}_t = \boldsymbol{\theta}_1$ for all $t$) and $\boldsymbol{\theta}_1$ represents the compromise probability $p_A$ (5.2). Further, the attacker's computational capacity $\ell_A$ is known to the defender, but both players are uncertain about $\boldsymbol{\theta}_1$ and have misspecified conjectures, i.e., $\boldsymbol{\theta}_1 \notin \Theta_A \cup \Theta_D$.

**Scenario 5.4** (Defender is uncertain about $\ell_A$ and $\boldsymbol{\theta}_t$)**.** This scenario is the same as Scenario 5.3, except that the defender is uncertain about $\ell_A$, i.e., $\mu_1(\ell_A) < 1$.

**Scenario 5.5** (Comparison with the snort idps (Roesch, 1999))**.** In this scenario, we compare col with the snort idps (ruleset v2.9.17.1). We use two baselines: snort-high and snort-medium, which block ip traffic that generates alerts with high and medium priority, respectively. The attacker follows the fixed strategy $\pi_A(\mathsf{S} \mid \cdot) = 1$ and spoofs its ip address. The observation $o_t$ (5.4) represents a snort alert, where $o_t = 0$ means no alert. The defender's stop action corresponds to blocking the ip address that generated the alert. $\boldsymbol{\theta}_t$ parameterizes $z$ (5.4) and represents the distributions of alert priorities generated by the clients and the attacker. Specifically, $\Theta_D = \{\boldsymbol{\theta}', \boldsymbol{\theta}''\}$ and

$$\boldsymbol{\theta}' \triangleq \begin{array}{c} \\ \text{N} \\ \text{M} \\ \text{H} \end{array} \!\!\begin{array}{c} \begin{array}{cc} \text{C} & \text{A} \end{array} \\ \begin{bmatrix} 0.85 & 0.4 \\ 0.1 & 0.3 \\ 0.05 & 0.3 \end{bmatrix} \end{array} \quad \boldsymbol{\theta}'' \triangleq \begin{array}{c} \\ \text{N} \\ \text{M} \\ \text{H} \end{array} \!\!\begin{array}{c} \begin{array}{cc} \text{C} & \text{A} \end{array} \\ \begin{bmatrix} 0.4 & 0.1 \\ 0.3 & 0.1 \\ 0.3 & 0.8 \end{bmatrix} \end{array}, \tag{5.19}$$

where $\boldsymbol{\theta}_t = \boldsymbol{\theta}'$ for $t < 50$ and $\boldsymbol{\theta}_t = \boldsymbol{\theta}''$ for $t \geq 50$. Here, N, M, and H refer to no alert, medium priority alert, and high priority alert, respectively. Similarly, C and A refer to the clients and the attacker, respectively.

**Remark 5.4** (Specification of the priors)**.** In practice, the prior over $\boldsymbol{\theta}_1$ for the instantiations described above can be defined based on domain knowledge or obtained through system measurements. Companies like google, meta, and ibm have documented procedures for estimating such distributions (Ford et al., 2010). Similarly, the prior over $\ell_{-k}$ can be obtained from opponent modeling (Shen and How, 2021).

### Evaluation results (Figs. 5.12–5.17, Table 5.3)

### Scenario 5.1 (Fig. 5.12)

Figures 5.12.a–b (on the next page) show the evolution of the defender's conjecture distribution $\mu_t$ (5.14b) and the discrepancy (5.15) of the conjecture $\overline{\ell}_{A,t}$ when $\mathcal{L} = \{1, 2\}$ and $\ell_A = 1$. We observe that $\mu_t$ converges and concentrates on the consistent conjecture (5.16) after 5 time steps, as predicted by Thm. 5.4.A. Figure 5.12.d shows the rate of convergence for different $|\mathcal{L}|$, indicating that a larger $|\mathcal{L}|$ leads to a slower convergence. This result is expected since a larger $|\mathcal{L}|$ means the defender has more uncertainty about $\ell_A$.

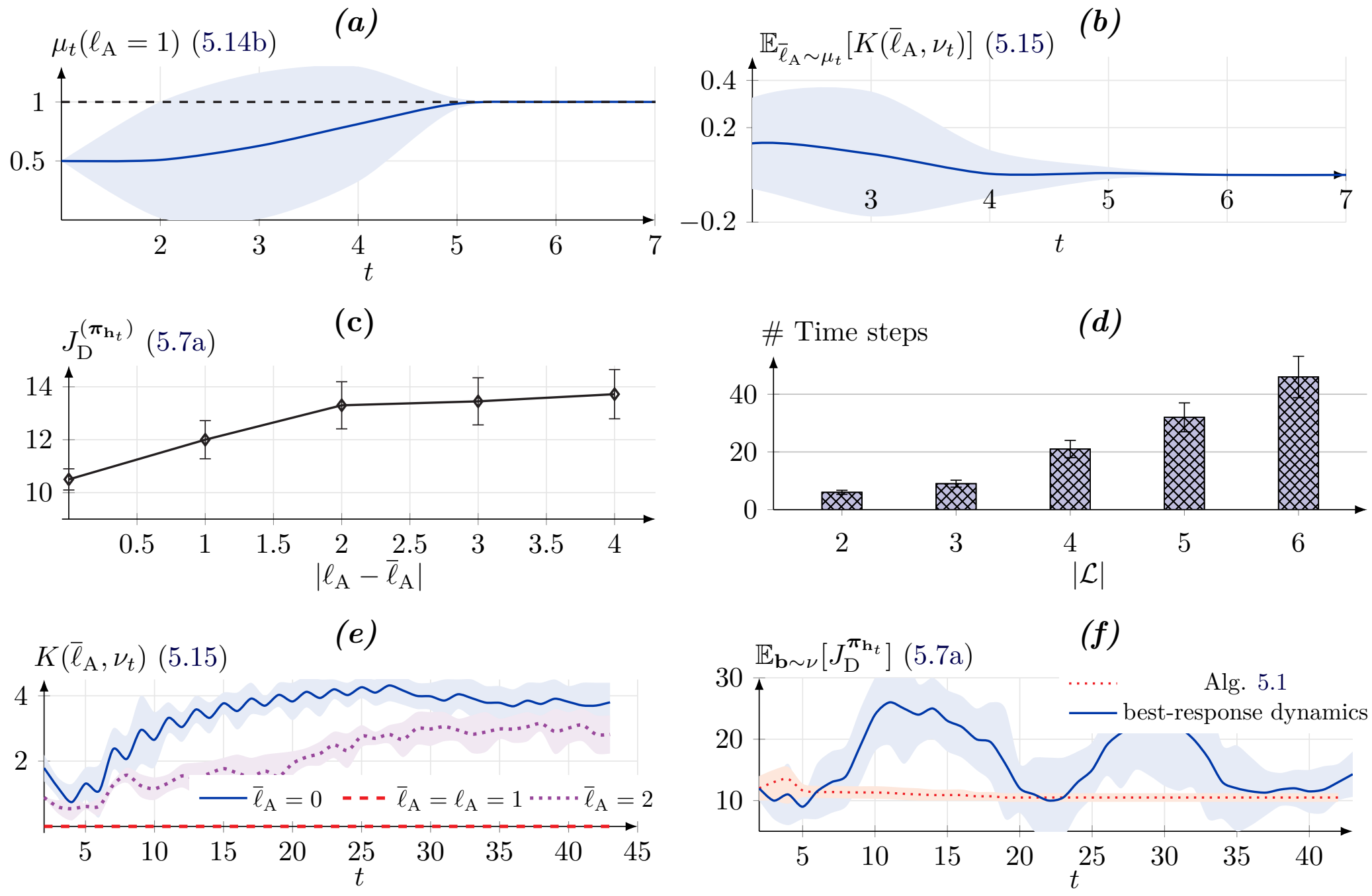

***Figure 5.12:*** *Evaluation results for Scenario 5.1; values indicate the mean; the shaded areas and the error bars indicate the 95% confidence interval based on* 20 *random seeds; hyperparameters are listed in Appendix F.*

Figure 5.12.c shows the expected cost of the defender as a function of $|\ell_A - \bar{\ell}_{A,t}|$, which quantifies the inaccuracy of the defender's conjecture $\bar{\ell}_{A,t}$. We observe that the defender's cost is increasing with the inaccuracy of its conjecture. Next, Fig. 5.12.e shows the evolution of the discrepancy (5.15) of different conjectures. We observe that the discrepancies of the incorrect conjectures increase over time and that the discrepancy of the correct conjecture is 0 (by definition).

Lastly, Fig. 5.12.f shows the expected cost of COL and the expected cost of best response reinforcement learning with CEM (Rubinstein, 1999) (i.e., *best response dynamics* (Nisan et al., 2007)). We note that the expected cost of reinforcement learning oscillates. Similar behavior of reinforcement learning has been observed in related work (Hernandez et al., 2019). The oscillation indicates that the players alternate between different best responses in a cycle. By contrast, the expected cost of COL is significantly more stable, and its behavior is consistent with convergence to a BNE. The strategy oscillations induced by reinforcement learning lead to unpredictability and computational inefficiency, making it an impractical solution for operational systems. In comparison, the BNE provides a robust and reliable strategy for the defender; see Fig. 5.12.f.

## Scenario 5.2 (Fig. 5.13)

Figures 5.13.a–b show the evolution of the defender's conjecture distribution $\rho_t^{(\mathrm{D})}$ (5.14a) and the discrepancy (5.15) of the conjecture $\overline{\boldsymbol{\theta}}_t^{(\mathrm{D})}$ when $\Theta_{\mathrm{D}} = \{12, 9\}$ and $\boldsymbol{\theta}_t = 12$ for all $t$. We observe that $\rho_t^{(\mathrm{D})}$ converges and concentrates on the consistent conjecture (5.16) after 18 time steps, as predicted by Thm. 5.4.B. Figure 5.13.d shows the rate of convergence for varying $|\Theta_{\mathrm{D}}|$. As expected, when the defender's uncertainty about $\boldsymbol{\theta}_t$ increases (i.e., when $|\Theta_{\mathrm{D}}|$ increases), the time it takes for the sequence of conjectures to converge increases.

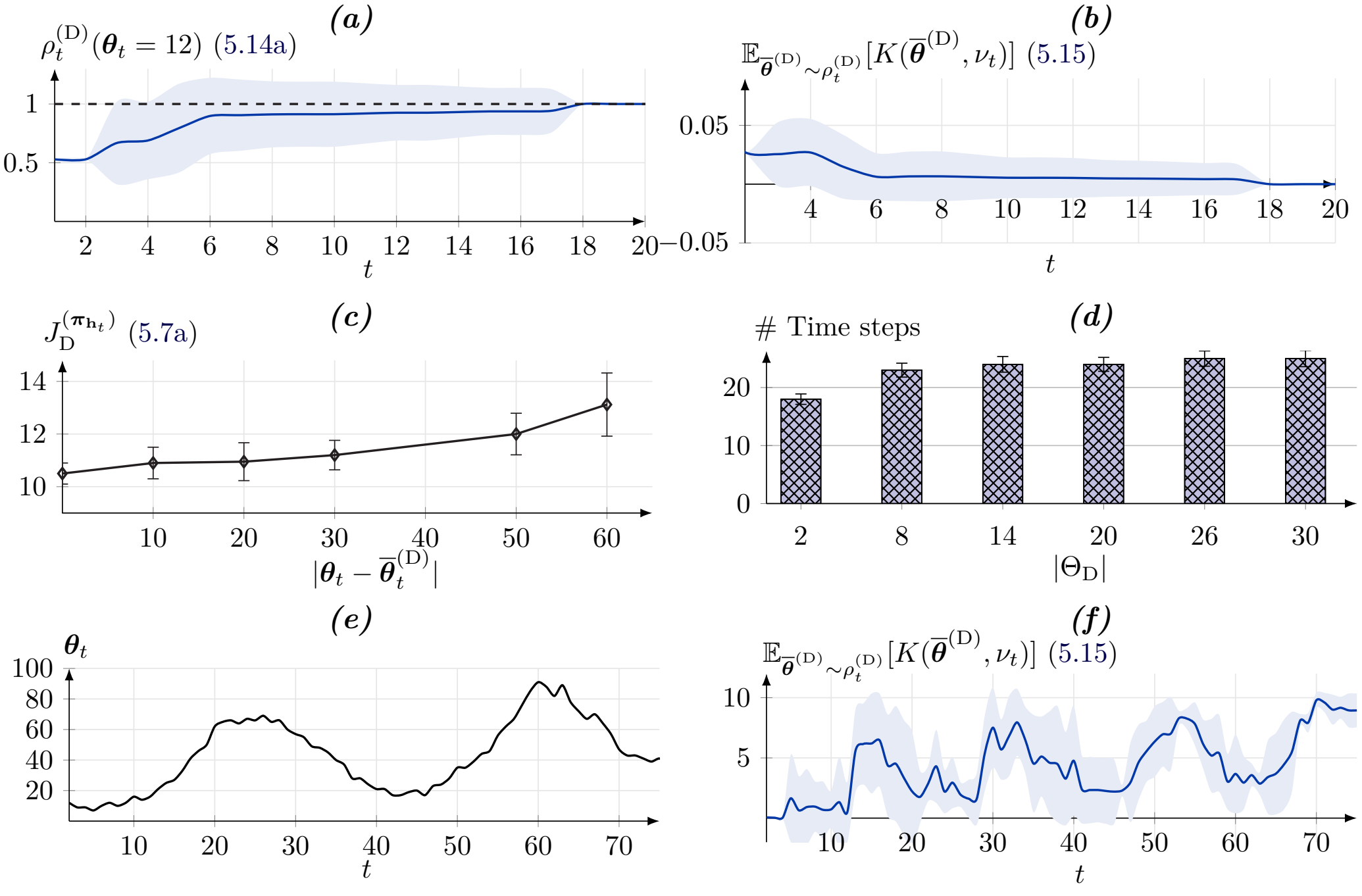

***Figure 5.13:*** *Evaluation results for Scenario 5.2; values indicate the mean; the shaded areas and the error bars indicate the 95% confidence interval based on* 20 *random seeds; hyperparameters are listed in Appendix F.*

Figure 5.13.c shows the expected cost of the defender as a function of $|\boldsymbol{\theta}_t - \overline{\boldsymbol{\theta}}_t^{(\mathrm{D})}|$, which quantifies the inaccuracy of the defender's conjecture $\overline{\boldsymbol{\theta}}_t^{(\mathrm{D})}$. We observe that the defender's cost is increasing with the inaccuracy of its conjecture. Lastly, Figs. 5.13.e–f show the expected discrepancy (5.15) of the posterior (5.14a) when $\boldsymbol{\theta}_t$ is changing at every time step, whereby $\rho_t^{(\mathrm{D})}$ does not converge. (Note that Thm. 5.4 only applies when $\boldsymbol{\theta}_t$ remains fixed.)

### Scenario 5.3 (Figs. 5.14–5.15)

Figures 5.14.a–b show the time for the defender's conjecture $\overline{\boldsymbol{\theta}}_t^{(\mathrm{D})}$ to converge for different sizes of $\Theta_{\mathrm{D}}$ when $\boldsymbol{\theta}_t$ is fixed. We observe that the time to converge increases with the size of $\Theta_{\mathrm{D}}$, which is expected since the size of $\Theta_{\mathrm{D}}$ represents the defender's degree of uncertainty about $\boldsymbol{\theta}_t$.

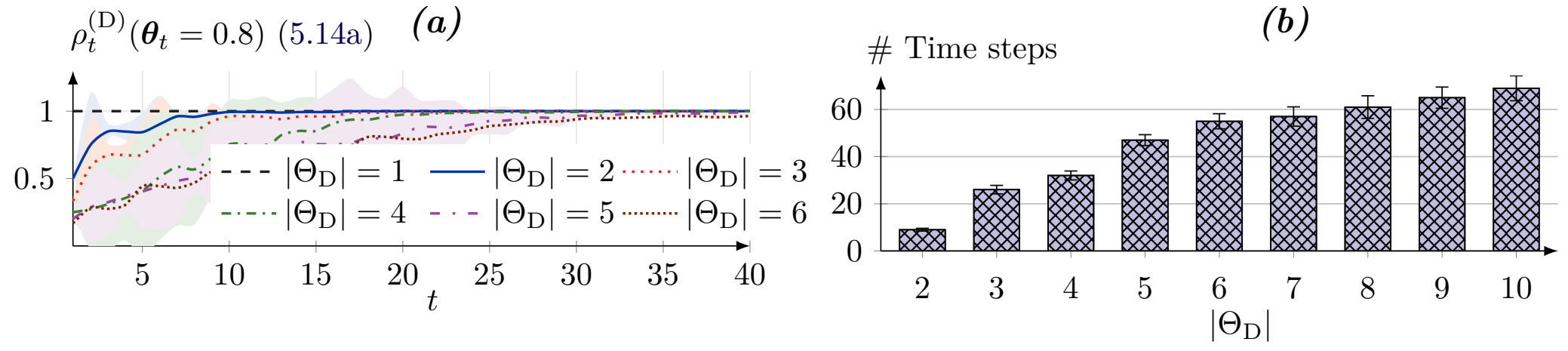

*Figure 5.14: Evaluation results for Scenario 5.3; values indicate the mean; the shaded areas and the error bars indicate the 95% confidence interval based on 20 random seeds; hyperparameters are listed in Appendix F.*

Figure 5.15 shows the evolution of $\rho_t^{(\mathrm{D})}$. We observe that $\rho_t^{(\mathrm{D})}$ starts from a uniform distribution over $\Theta_{\mathrm{D}}$ and as $t \to \infty$, it concentrates on the set of consistent conjectures $\Theta_{\mathrm{D}}^{\star}$ (5.16a).

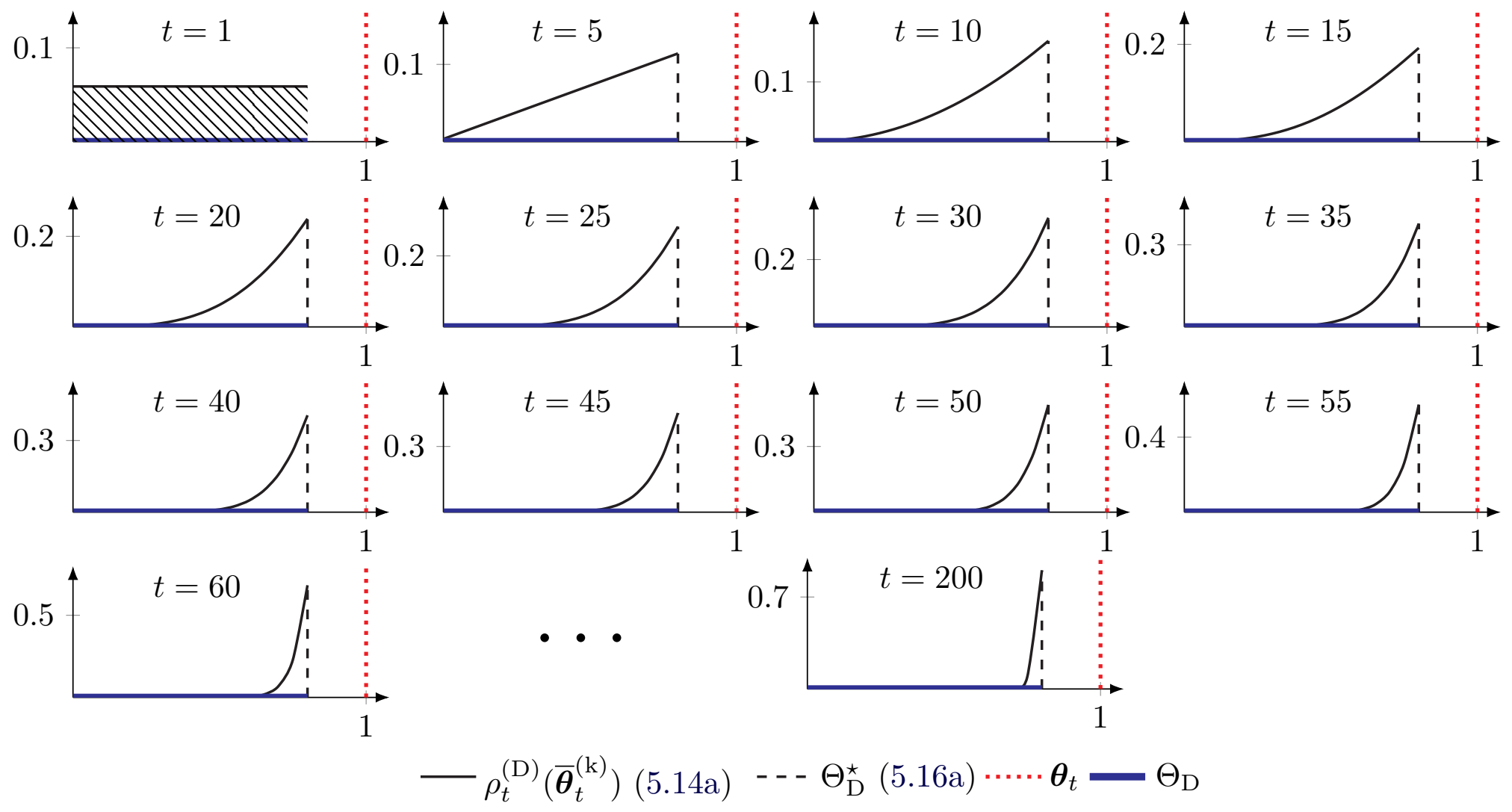

*Figure 5.15: Evolution of $\rho_t^{(\mathrm{D})}$ (5.14a) when $\Theta_{\mathrm{D}} = \{0.0, 0.05, \ldots, 0.8\}$ and $\boldsymbol{\theta}_t = 1$ for all $t$ (Scenario 5.3); hyperparameters are listed in Appendix F.*

## Scenario 5.4 (Fig. 5.16)

Figure 5.16 shows the evolution of the defender's conjecture distributions $\mu_t$ and $\rho_t^{(\mathrm{D})}$ (5.14). We observe that both distributions converge, which is consistent with Thm. 5.4. The convergence of $\mu_t$ is significantly faster than that of $\rho_t^{(\mathrm{D})}$. We believe this difference is because $|\mathcal{L}| < |\Theta_{\mathrm{D}}|$, which means that the defender is more uncertain about $\ell_{\mathrm{A}}$ than about $\boldsymbol{\theta}_t$.

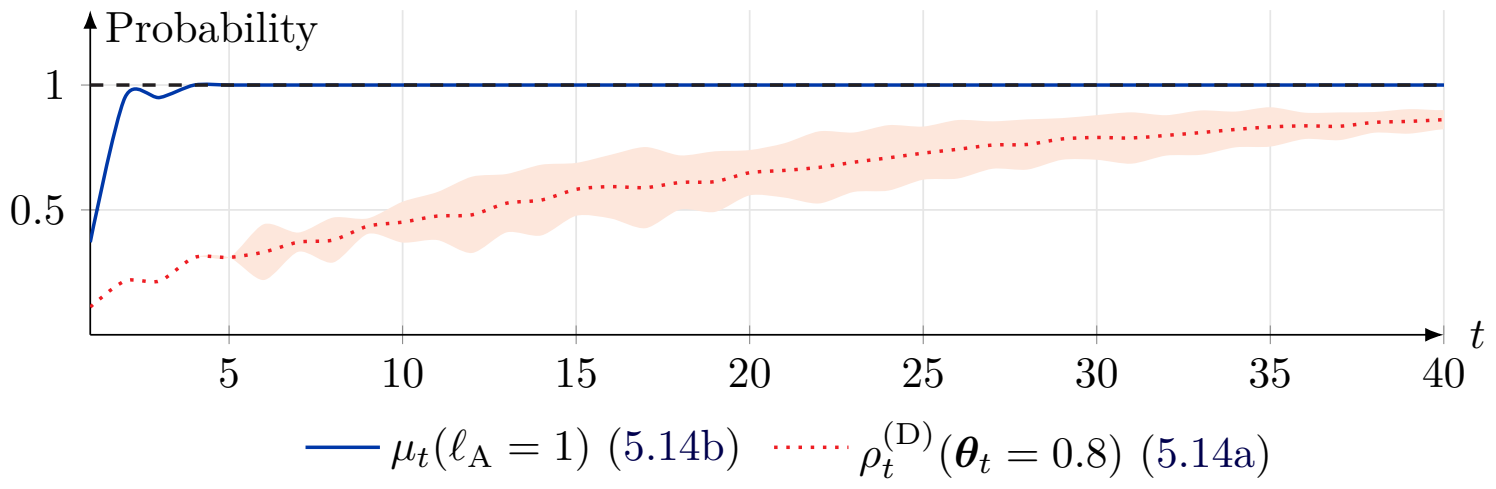

***Figure 5.16:*** *Evaluation results for Scenario 5.4; values indicate the mean; the shaded areas and the error bars indicate the 95% confidence interval based on* 20 *random seeds; hyperparameters are listed in Appendix F.*

## Scenario 5.5 (Fig. 5.17)

Figure 5.17 shows the percentage of blocked attacker and client traffic when running the SNORT IDPS (Roesch, 1999) and COL on the digital twin. We observe that both block some client traffic and fail to block some attacker traffic, which is expected considering the false IDS alarms generated by the clients. When comparing COL with SNORT we find that a) COL and IPS-HIGH blocks the least *client* traffic; and b) IPS-MEDIUM and COL blocks the most *attacker* traffic. This result suggests to us that COL balances the trade-off between blocking clients and the attacker based on the cost function (5.6). Further, COL adapts when $\boldsymbol{\theta}_t$ changes.

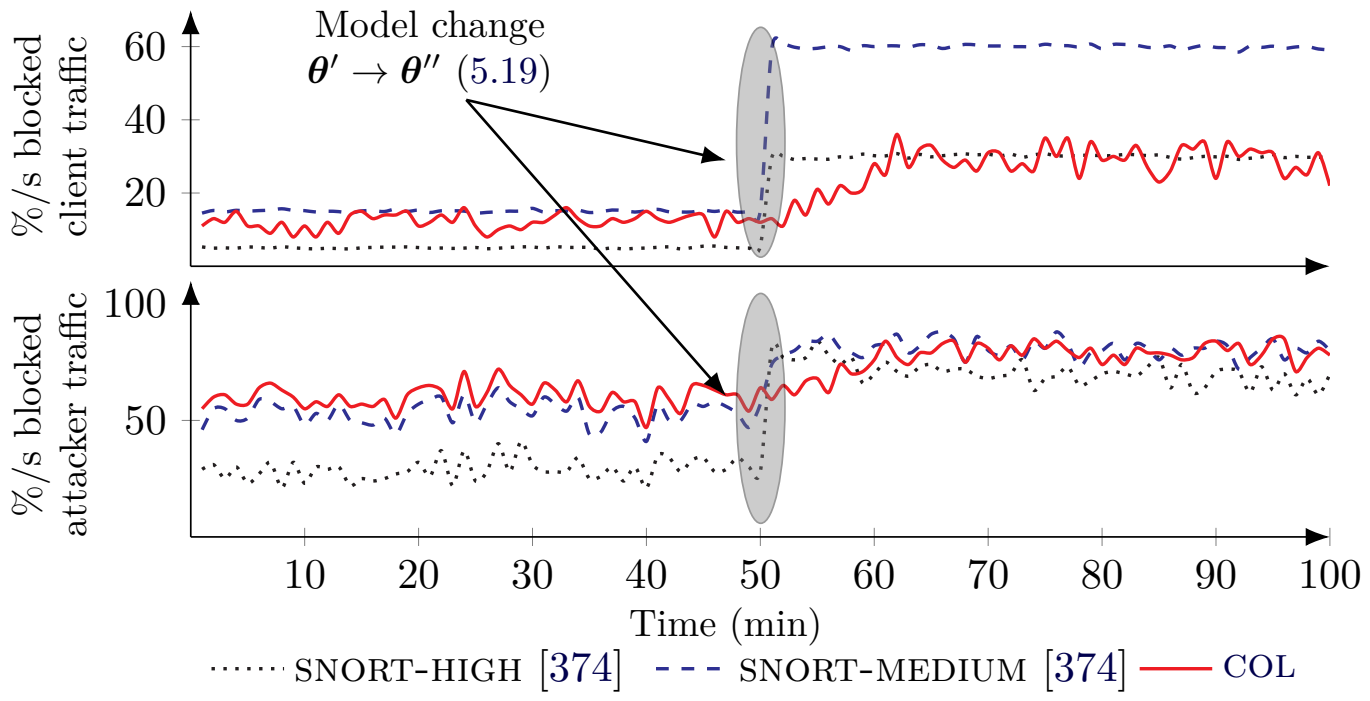

***Figure 5.17:*** *Evaluation results for Scenario 5.5; the curves show the percentage of blocked network traffic in the digital twin.*

### Comparison with state-of-the-art methods

The convergence times of COL (Alg. 5.1) and those used in prior work are listed in Table 5.3. While we observe that COL converges faster than the baselines, a direct comparison is not feasible for two reasons. First, the baselines do not consider the same solution concept as us, i.e., the BNE. Second, the baselines make different assumptions about the computational capacity and information available to the players. For example, fictitious play (Brown, 1951), which is a popular method among prior work, assumes that players a) have correctly specified models and b) have unlimited computational capacity. Similarly, popular reinforcement learning methods, such as PPO (Alg. 1, Schulman et al., 2017)[13] and NFSP (Alg. 9, Heinrich, 2017), are designed for offline rather than online learning.

| *Method* | *Fixed point* | *Time (min)* |
|---|---|---|
| COL, $\lvert\mathcal{L}\rvert = 2, \lvert\Theta_{\mathrm{k}}\rvert = 1$ | Berk-Nash equilibrium (Def. 5.1) | $6.7 \pm 0.7$. |
| COL, $\lvert\mathcal{L}\rvert = 3, \lvert\Theta_{\mathrm{k}}\rvert = 1$ | Berk-Nash equilibrium (Def. 5.1) | $11.4 \pm 0.9$. |
| COL, $\lvert\mathcal{L}\rvert = 4, \lvert\Theta_{\mathrm{k}}\rvert = 1$ | Berk-Nash equilibrium (Def. 5.1) | $39.1 \pm 1.3$. |
| COL, $\lvert\mathcal{L}\rvert = 8, \lvert\Theta_{\mathrm{k}}\rvert = 1$ | Berk-Nash equilibrium (Def. 5.1) | $137.3 \pm 2.8$. |
| COL, $\lvert\mathcal{L}\rvert = 1, \lvert\Theta_{\mathrm{k}}\rvert = 2$ | Berk-Nash equilibrium (Def. 5.1) | $12.9 \pm 0.9$. |
| COL, $\lvert\mathcal{L}\rvert = 1, \lvert\Theta_{\mathrm{k}}\rvert = 32$ | Berk-Nash equilibrium (Def. 5.1) | $17.9 \pm 1.0$. |
| COL, $\lvert\mathcal{L}\rvert = 1, \lvert\Theta_{\mathrm{k}}\rvert = 192$ | Berk-Nash equilibrium (Def. 5.1) | $29.9 \pm 1.2$. |
| COL, $\lvert\mathcal{L}\rvert = 4, \lvert\Theta_{\mathrm{k}}\rvert = 192$ | Berk-Nash equilibrium (Def. 5.1) | $194.6 \pm 3.7$. |
| Best-response dynamics (Fig. 5.12.f) | $\epsilon$-Nash equilibrium [322, Eq. 1] | DNC. |
| HSVI [201, Alg. 1] (Fig. 5.6) | $\epsilon$-Nash equilibrium [322, Eq. 1] | DNC. |
| NFSP [191, Alg. 9] | $\epsilon$-Nash equilibrium [322, Eq. 1] | $\approx 919$ min. |
| Fictitious play [76] | $\epsilon$-Nash equilibrium [322, Eq. 1] | $\approx 4800$ min. |
| PPO [396, Alg. 1], static $\pi_{\mathrm{A}}$ | Best response (5.8a) | $9.2 \pm 0.4$ min. |

***Table 5.3:*** *Comparison with baseline methods in terms of speed of convergence; DNC is short for "does not converge"; "≈" means that the algorithm nearly converges; numbers indicate the mean and the standard deviation from evaluations with 3 random seeds; hyperparameters are listed in Appendix F.*

### Discussion of the evaluation results

The key findings from the evaluation can be summarized as follows.

- The conjectures produced by COL converge to consistent conjectures once the model parameters $\boldsymbol{\theta}_t$ remain fixed (5.16) (Figs. 5.12–5.16, Thm. 5.4). The rate of convergence decreases as $\lvert\mathcal{L}\rvert$ and $\lvert\Theta_{\mathrm{k}}\rvert$ increase (Figs. 5.12–5.16).

- The defender's cost increases as the distance between its conjecture and the true model increases (Figs. 5.12–5.16).

[13]See Appendix D of Paper 3 for a derivation of the PPO algorithm.

- COL allows to configure the computational capacity and the uncertainty of each player k by tuning $\ell_k$ (Fig. 5.8) and $(\mathcal{L}, \Theta_k)$, respectively (Figs. 5.12–5.16).
- COL leads to effective strategies (Thm. 5.3) that are more stable than those obtained through reinforcement learning (Fig. 5.12.f, Table 5.3).
- COL outperforms the SNORT IDPS (Roesch, 1999) in several key metrics; see Fig. 5.17.
- Computation of perfect Bayesian equilibria and computation of exact rollout strategies is intractable for any non-trivial instantiation of $\Gamma$ (Fig. 5.6, Fig. 5.8.a).
- Approximate best response and rollout strategies of $\Gamma$ can be efficiently computed using stochastic approximation (Fig. 5.8.b, Fig. 5.4, Thm. 5.1).

The above findings suggest that COL can produce effective security strategies without relying on a correctly specified model of the environment. The practical implication of this result is that COL is suitable for dynamic IT infrastructures with short update cycles, which aligns with current trends of virtualization and zero-touch management.

## 5.8 Related Work

Since the early 2000s, researchers have studied automated security through modeling attacks and responses on an IT infrastructure as a game between an attacker and a defender; see textbooks (Alpcan and Basar, 2010), (Tambe, 2011), (Kamhoua et al., 2021) and surveys (Tan et al., 2023), (Nguyen and Reddi, 2023). The main difference between this paper and the prior work is that we propose a method for online learning in non-stationary security games in which players have misspecified game models. By contrast, prior work assumes that players have correctly specified models. While the study of learning with misspecified models has attracted long-standing interest in economics (Arrow and Green, 1973), engineering (Kagel and Levin, 1986), and psychology (Rabin, 2002), it remains unexplored in the security context. Related research in the security literature includes a) game-theoretic approaches based on imperfect and incomplete information; b) game-theoretic approaches based on bounded rationality; and c) model-free learning approaches. In the following subsections, we describe how these three approaches relate to this paper. The main differences are listed in Table 5.4 on page 230.

### Security games with imperfect and incomplete information

Security games with imperfect and incomplete information (i.e., partial observability and model uncertainty) include POSGs [471, 178, 183] and Bayesian games [33, 11, 312, 285]. These games capture scenarios where players have private knowledge represented by *types* or *observations.* Still, the players' perceptions about how the game works are identical (as defined by a common prior); the only thing distinguishing players is the information each has received (Harsanyi, 1967)(Horák et al., 2023). Conversely, our model allows to capture *misspecification*, where players have incorrect *conjectures* about the game's structure and the opponents' strategies. Such misspecification encapsulates players' subjective perception of how the game works and can include both game elements (Esponda and Pouzo, 2021) and other players' strategies (Esponda and Pouzo, 2016). Moreover, players can disagree on the very form of the game. For instance, one player may represent the game with a scalar, whereas another may represent it with a high-dimensional vector; see Prob. 5.1. Such disagreements are captured by the Berk-Nash equilibrium but not the perfect Bayesian equilibrium.

### Security games with bounded rationality

The concept of bounded rationality was introduced by Herbert Simon in the 1950s as a critique of the assumption of perfect rationality in classical game theory [417]. Research on security games with bounded rationality includes [90, 390, 151, 91, 483, 36, 1, 301, 450, 521]. These works differ from this paper in three main ways. First, they do not consider model misspecification as we do in this paper. Second, they study different types of games, i.e., one-stage games [90, 91]; network interdiction games [390]; differential games [301]; Stackelberg games [151]; behavioral games [301]; evolutionary games [450], and hypergames [483, 36]. Third, they study different types of equilibria, i.e., Nash equilibria [1]; Gestalt Nash equilibria [90, 91]; Stackelberg equilibria [390, 151]; evolutionary stable equilibria [450], and hyper Nash equilibria [483, 36]. By contrast, we study Berk-Nash equilibria [135, Def. 1] and a POSG where players have misspecified models. The benefit of our model and equilibrium concept is that they better capture the APT use case (§5.2).

### Learning in security games

Prior work that studies learning in security games includes [37, 280, 156, 279, 211, 527, 178, 179, 183, 185, 329, 506]. This paper differs from these works in two main ways. First, we design a novel way to update strategies using rollout with a conjectured model. This approach contrasts with all of the referenced works, which consider other types of strategy updates, e.g., offline learning [183, 185, 211], meta-learning [156], model-free learning [527, 329, 178, 179, 506], and learning with perfect rationality [37, 280]. The advantage of our approach is that it allows us to model players with limited computational capacity and varying degrees of misspecification. Second, we design a convergent Bayesian mechanism for online learning. In comparison, most of the prior work considers other types of learning mechanisms, e.g., fictitious play [183, 185], reinforcement learning [527, 329, 178, 179, 506], online gradient descent [3], and no-regret learning [37, 280].

## 5.9 Conclusion

This paper presents **C**onjectural **O**nline **L**earning (COL), a new game-theoretic method for online learning of security strategies that applies to dynamic IT environments where the attacker and defender are uncertain about the environment and the opponent's strategy. We formulate the interaction between an attacker and a defender as a non-stationary game where each player has a probabilistic conjecture about the game model, which may be misspecified in the sense that the true model has probability 0. Both players iteratively adapt their conjecture using Bayesian learning and update their strategy using rollout. We prove that the conjectures converge to best fits (Thm. 5.4), and we provide a bound on the performance improvement that rollout enables with a conjectured model (Thm. 5.3). To characterize the steady state of the game, we propose a novel equilibrium concept based on the Berk-Nash equilibrium, which represents a stable point where each player acts optimally given its conjecture (Def. 5.1). We present COL through an APT use case (§5.2). Evaluations on a testbed show that COL produces effective security strategies that adapt to a changing environment (Figs. 5.12–5.16). It also leads to faster convergence than current reinforcement learning techniques and outperforms the SNORT IDPS (Roesch, 1999) in several key metrics (Table 5.3 and Fig. 5.17).

In the context of the thesis, this paper addresses a drawback of papers 1–4, which assume that a perfect model of the underlying IT infrastructure can be obtained. In the next chapter of the thesis, we apply our methodology to a standard benchmark for automated security response, namely CAGE-2 (CAGE-2, 2022), which facilitates direct comparisons with the existing literature.

| *Paper* | *Use case* | *Method* | *Equilibrium* | *Evaluation* | *Game type* |
|---|---|---|---|---|---|
| [531] Zonouz, 2009 | Intrusion response | Dynamic programming | - | Testbed | Stationary POSG. |
| [527] Zhu, 2009 | Intrusion detection | Q-learning | Nash | Simulation | Stationary stochastic dynamic game. |
| [37] Balcan, 2015 | Resource allocation | No-regret learning | - | Analytical | Stationary repeated Stackelberg game. |
| [280] Lisý, 2016 | Resource allocation | No-regret learning | Nash | Analytical | Stationary NFGSS. |
| [90] Chen, 2019 | Risk management | Proximal optimization | Gestalt Nash | Simulation | Stationary one-stage game. |
| [390] Sanjab, 2020 | Drone operation | Prospect theory | Stackelberg | Simulation | Stationary network interdiction game. |
| [36] Bakker, 2020 | Intrusion response | Analytical | Hyper Nash | Simulation | Stationary repeated hypergame. |
| [1] Abdallah, 2020 | Power grid | Analytical | Nash | Simulation | Stationary behavioral game. |
| [211] Huang, 2020 | APT | Optimization | Perfect Bayes Nash | Simulation | Stationary POSG. |
| [522] Zhao, 2020 | Intrusion response | Analytical | Nash | Simulation | Non-stationary Markov game. |
| [33] Aydeger, 2021 | DDOS | Optimization | Perfect Bayes Nash | Testbed | Stationary signaling game. |
| [483] Wan, 2022 | APT | Analytical | Hyper Nash | Simulation | Stationary hypergame. |
| [151] Gabrys, 2023 | Cyber deception | Analytical | Stackelberg | Simulation | Stationary Stackelberg game. |
| [183] Hammar, 2023 | Intrusion response | Fictitious play | Nash | Testbed | Stationary POSG. |
| [301] Mavridis, 2023 | Cyber-physical security | Stochastic approximation | Nash | Simulation | Stationary differential game. |
| [156] Ge, 2023 | Zero-trust | Meta-learning | - | Simulation | Stationary POMDP. |
| **This paper, 2024** | APT | Bayesian learning & rollout | Berk-Nash | Testbed & simulation | Non-stationary and misspecified POSG. |

***Table 5.4:*** *Comparison between this paper and related work.*

# Acknowledgments

The authors would like to thank Branislav Bosanský for sharing the code of the HSVI algorithm.

# Appendix

## A Proof of Theorem 5.1

Given $(\pi_{\mathrm{A}}, \pi_{\mathrm{D}})$, the best response strategies $(\tilde{\pi}_{\mathrm{D}}, \tilde{\pi}_{\mathrm{A}})$ are optimal strategies in two POMDPs[14]: $\mathcal{M}^P_{\mathrm{D}}$ and $\mathcal{M}^P_{\mathrm{A}}$. Hence, it suffices to show that there exist optimal strategies $\pi^\star_{\mathrm{D}}$ and $\pi^\star_{\mathrm{A}}$ in $\mathcal{M}^P_{\mathrm{D}}$ and $\mathcal{M}^P_{\mathrm{A}}$ that satisfy (5.9a) and (5.9b), respectively. Towards this proof, we state the following three lemmas.

**Lemma 5.1.** *$\mathcal{M}^{\mathrm{P}}_{\mathrm{D}}$ can be formulated as a repeated optimal stopping problem.*

*Proof.* By definition, an optimal strategy $\pi^\star_{\mathrm{D}}$ in $\mathcal{M}^P_{\mathrm{D}}$ satisfies

$$
\begin{aligned}
\pi^\star_{\mathrm{D}} &\in \arg\min_{\pi_{\mathrm{D}} \in \Pi_{\mathrm{D}}} \mathbb{E}_{(\pi_{\mathrm{D}},\pi_{\mathrm{A}})}\left[\sum_{t=1}^{\infty} \gamma^{t-1} c(S_t, A^{(\mathrm{D})}_t) \mid S_1 = 0\right] \\
&\overset{(a)}{=} \arg\min_{\pi_{\mathrm{D}} \in \Pi_{\mathrm{D}}} \left[\mathbb{E}_{(\pi_{\mathrm{D}},\pi_{\mathrm{A}})}\left[\sum_{t=1}^{\tau_1} \gamma^{t-1} c(S_t, A^{(\mathrm{D})}_t) \mid S_1 = 0\right] + \right. \\
&\qquad \left. \mathbb{E}_{(\pi_{\mathrm{D}},\pi_{\mathrm{A}})}\left[\sum_{t=\tau_1+1}^{\tau_2} \gamma^{t-1} c(S_t, A^{(\mathrm{D})}_t) \mid S_{\tau_1} = 0\right] + \ldots\right] \\
&= \arg\min_{\pi_{\mathrm{D}} \in \Pi_{\mathrm{D}}} \left[\mathbb{E}_{(\pi_{\mathrm{D}},\pi_{\mathrm{A}})}\left[\gamma^{\tau_1} J_{\mathrm{D}}(\mathbf{e}_1) + \sum_{t=1}^{\tau_1} \gamma^{t-1} c(S_t, A^{(\mathrm{D})}_t) \mid S_1 = 0\right]\right],
\end{aligned}
\tag{5.20}
$$

where $\tau_1, \tau_2, \ldots$ are the stopping times; $J_{\mathrm{D}} : \mathcal{B} \to \mathbb{R}$ is the cost-to-go function in $\mathcal{M}^{\mathrm{P}}_{\mathrm{D}}$ (5.7); $\mathbf{e}_1 = (1, 0)$; and (a) follows by linearity of expectation. Since $\mathbb{E}_{(\pi_{\mathrm{D}},\pi_{\mathrm{A}})}[J_{\mathrm{D}}(\mathbf{e}_1)]$ can be seen as a fixed recovery cost, (5.20) defines an optimal stopping problem. □

**Lemma 5.2.** *If $s_t = N$, then $a^{(\mathrm{A})}_t = \mathsf{C}$ is a best response for any $\mathbf{b}_t \in \mathcal{B}$ and $\pi_{\mathrm{D}} \in \Pi_{\mathrm{D}}$.*

*Proof.* Let $J^\star_{\mathrm{A}}$ and $Q^\star_{\mathrm{A}}$ be the optimal cost-to-go function and $Q$-function in the attacker's best response POMDP $\mathcal{M}^P_{\mathrm{A}}$. Assume by contradiction that $a^{(\mathrm{A})}_t = \mathsf{C}$ is not a best response. Then the expected cost of $a^{(\mathrm{A})}_t = \mathsf{S}$ must be lower than that of $a^{(\mathrm{A})}_t = \mathsf{C}$, i.e.,

$$Q^\star_{\mathrm{A}}((\mathbf{b}_t, N), \mathsf{S}) < Q^\star_{\mathrm{A}}((\mathbf{b}_t, N), \mathsf{C})$$

[14]The components of a POMDP are defined the background chapter; see (15).

$$
\begin{aligned}
&\overset{(a)}{\implies} \mathbb{E}_{A_t^{(\mathrm{D})},\mathbf{B}_{t+1}}[-c(s_t, A_t^{(\mathrm{D})}) + \gamma J^{\star}_{\mathrm{A}}(\mathbf{B}_{t+1}) \mid s_t = N, a_t^{(\mathrm{A})} = \mathsf{S}] \\
&< \mathbb{E}_{A_t^{(\mathrm{D})},\mathbf{B}_{t+1}}[-c(s_t, A_t^{(\mathrm{D})}) + \gamma J^{\star}_{\mathrm{A}}(\mathbf{B}_{t+1}) \mid s_t = N, a_t^{(\mathrm{A})} = \mathsf{C}] \\
&\overset{(b)}{\implies} \mathbb{E}_{A_t^{(\mathrm{D})},S_{t+1}}\left[\sum_{o\in\mathcal{O}} z(o \mid S_{t+1}) J^{\star}_{\mathrm{A}}(\mathbb{B}(\mathbf{b}_t, A_t^{(\mathrm{D})}, o, \pi_{\mathrm{A}})) \mid s_t = N, a_t^{(\mathrm{A})} = \mathsf{S}\right] \\
&< \mathbb{E}_{A_t^{(\mathrm{D})},S_{t+1}}\left[\sum_{o\in\mathcal{O}} z(o \mid S_{t+1}) J^{\star}_{\mathrm{A}}(\mathbb{B}(\mathbf{b}_t, A_t^{(\mathrm{D})}, o, \pi_{\mathrm{A}})) \mid s_t = N, a_t^{(\mathrm{A})} = \mathsf{C}\right] \\
&\overset{(c)}{\implies} 0 < 0 \qquad \text{(contradiction)},
\end{aligned}
$$

where $\pi_{\mathrm{A}}$ is the attacker strategy assumed by $\pi_{\mathrm{D}}$, and $\mathbb{B}$ is defined in (22). Step (a) follows from Bellman's optimality equation; (b) follows because $c$ (5.6) is independent of $a_t^{(\mathrm{A})}$; and (c) follows because both $a_t^{(\mathrm{A})} = \mathsf{C}$ and $a_t^{(\mathrm{A})} = \mathsf{S}$ lead to the same state when $s_t = N$ (5.2). □

**Lemma 5.3.** *If $\pi_{\mathrm{D}}(\mathsf{S} \mid \mathbf{b}_t) = 1$ for some $\mathbf{b}_t \in \mathcal{B}$, then $a_t^{(\mathrm{A})} = \mathsf{C}$ is a best response.*

*Proof.* Let $J^{\star}_{\mathrm{A}}$ and $Q^{\star}_{\mathrm{A}}$ be the optimal cost-to-go function and $Q$-function in the attacker's best response POMDP $\mathcal{M}^{P}_{\mathrm{A}}$. Assume by contradiction that $a_t^{(\mathrm{A})} = \mathsf{C}$ is not a best response. Then the expected cost of $a_t^{(\mathrm{A})} = \mathsf{S}$ must be lower than that of $a_t^{(\mathrm{A})} = \mathsf{C}$, i.e.,

$$
\begin{aligned}
Q^{\star}_{\mathrm{A}}((\mathbf{b}_t, s_t), \mathsf{S}) < Q^{\star}_{\mathrm{A}}((\mathbf{b}_t, s_t), \mathsf{C}) &\overset{(a)}{\implies} J^{\star}_{\mathrm{A}}(\mathbf{e}_1) < J^{\star}_{\mathrm{A}}(\mathbf{e}_1) \\
&\overset{(b)}{\implies} 0 < 0 \qquad \text{(contradiction)},
\end{aligned}
$$

where (a) follows because $c$ (5.6) is independent of $a_t^{(\mathrm{A})}$ and because $\pi_{\mathrm{D}}(\mathsf{S} \mid \mathbf{b}_t) = 1 \implies \mathbf{b}_{t+1} = \mathbf{e}_1$ (5.2). □

## A.1 Proof of Theorem 5.1.A

Lemma 5.1, Lemma 1.1 of Paper 1, and the assumption that $N = 1$ means that $\mathscr{S} = [\alpha^{\star}, \kappa]$, where $0 \leq \alpha^{\star} \leq \kappa \leq 1$. Thus, it suffices to show that $\kappa = 1$. Bellman's optimality equation implies that

$$
\begin{aligned}
\pi^{\star}_{\mathrm{D}}(\mathbf{e}_2) &\in \operatorname*{arg\,min}_{a_t^{(\mathrm{D})}\in\{\mathsf{S},\mathsf{C}\}} \Bigg[\overbrace{c(1,\mathsf{S}) + \gamma J^{\star}_{\mathrm{D}}(\mathbf{e}_1)}^{a_t^{(\mathrm{D})}=\mathsf{S}}, \overbrace{c(1,\mathsf{C}) + \gamma J^{\star}_{\mathrm{D}}(\mathbf{e}_2)}^{a_t^{(\mathrm{D})}=\mathsf{C}}\Bigg] \\
&\overset{(a)}{=} \operatorname*{arg\,min}_{a_t^{(\mathrm{D})}\in\{\mathsf{S},\mathsf{C}\}} \left[c(1,\mathsf{S}) + \gamma J^{\star}_{\mathrm{D}}(\mathbf{e}_1), \gamma^{\tau-1} c(1,\mathsf{S}) + \gamma^{\tau} J^{\star}_{\mathrm{D}}(\mathbf{e}_1) + \sum_{t=1}^{\tau-1} \gamma^{t-1} c(1,\mathsf{C})\right]
\end{aligned}
$$

$$= \underset{a_t^{(\mathrm{D})} \in \{\mathsf{S},\mathsf{C}\}}{\arg\min} \left[ c(1,\mathsf{S}) + \gamma J_{\mathrm{D}}^{\star}(\mathbf{e}_1), \gamma^{\tau-1}c(1,\mathsf{S}) + \gamma^{\tau} J_{\mathrm{D}}^{\star}(\mathbf{e}_1) + \left(\frac{1-\gamma^{\tau-1}}{1-\gamma}\right) c(1,\mathsf{C}) \right]$$

$$= \underset{a_t^{(\mathrm{D})} \in \{\mathsf{S},\mathsf{C}\}}{\arg\min} \left[ q - r + \gamma J_{\mathrm{D}}^{\star}(\mathbf{e}_1), \gamma^{\tau-1}(q-r) + \gamma^{\tau} J_{\mathrm{D}}^{\star}(\mathbf{e}_1) + \left(\frac{1-\gamma^{\tau-1}}{1-\gamma}\right) 1^p \right]$$

$$= \underset{a_t^{(\mathrm{D})} \in \{\mathsf{S},\mathsf{C}\}}{\arg\min} \left[ q - r + \gamma J_{\mathrm{D}}^{\star}(\mathbf{e}_1), \gamma^{\tau-1}(q-r) + \gamma^{\tau} J_{\mathrm{D}}^{\star}(\mathbf{e}_1) + \left(\frac{1-\gamma^{\tau-1}}{1-\gamma}\right) \right] \overset{(b)}{=} \{\mathsf{S}\}$$

$$\implies \mathscr{S} = [\alpha^{\star}, 1],$$

where $\tau \geq 1$ denotes the stopping time; $\mathbf{e}_2 = (0,1)$; $\mathbf{e}_1 = (1,0)$; and $J_{\mathrm{D}}^{\star}$ is the optimal cost-to-go function. Step (a) follows because $s = 1$ is an absorbing state until the stop. Step (b) follows from (5.6) and the fact that $1 > q-r$, which implies that the cost per time-step is upper bounded by $c(N,\mathsf{C}) = N^p = 1^p = 1 > q-r$. □

## A.2 Proof of Theorem 5.1.B

Since $N = 1$, we have that $\mathcal{B} = [0,1]$ and $\mathbf{b}$ is uniquely determined by $\mathbf{b}(1)$. For ease of notation, we use $\mathbf{b}$ as a shorthand for $\mathbf{b}(1)$. Given Lemma 5.2, it suffices to consider the case when $s_t = 0$. From Lemma 5.3 and the assumption that $\pi_{\mathrm{D}}$ satisfies (5.9a), we know that $a_t^{(\mathrm{A})} = \mathsf{S}$ is a best response iff $\mathbf{b}_t \in [0, \alpha^{\star})$, where $\alpha^{\star} \leq 1$. It further follows from Lemma 5.3 that $\alpha^{\star} = 0 \implies \beta^{\star} = 0$. Thus, Thm. 5.1.B holds when $\alpha^{\star} = 0$. Next, consider the case when $\alpha^{\star} > 0$. We know from Thm. 2 in the background chapter and Lemma 1.1 of Paper 1 that the stopping set $\mathscr{S}_{\mathrm{A}} \subset \mathcal{B}$ for the attacker is a convex subset of $[0, \alpha^{\star})$. Since $0 \in \mathscr{S}_{\mathrm{A}}$ by assumption, it follows that $\mathscr{S}_{\mathrm{A}} = [0, \beta^{\star}]$ for some threshold $\beta^{\star}$. □

# B Proof of Theorem 5.3

As (5.12) implements one step of the policy iteration algorithm (Eqs. 6.4.1-22, Puterman, 1994), (5.13a) follows from standard results in dynamic programming theory, see e.g., (Prop. 1, Bhattacharya et al., 2020). To prove (5.13b) we adapt the proof in (Prop 5.1.1, Bertsekas, 2019). Let $\tilde{\boldsymbol{\pi}} \triangleq (\pi_{\mathrm{k},t}, \overline{\pi}_{-\mathrm{k},t})$ and $\boldsymbol{\pi} \triangleq (\pi_{\mathrm{k},1}, \overline{\pi}_{-\mathrm{k},t})$. Then define the Bellman operator[15]

$$(\mathscr{T}_{\mathrm{k},\boldsymbol{\pi}} J_{\mathrm{k}})(\mathbf{b}_t) \triangleq \mathbb{E}_{\boldsymbol{\pi}}[c(S_t, A_t^{(\mathrm{D})}) + \gamma J_{\mathrm{k}}(\mathbf{B}_{t+1}) \mid \boldsymbol{\pi}] \qquad \forall \mathbf{b} \in \mathcal{B}.$$

Since $J_{\mathrm{k}}^{(\boldsymbol{\pi})}$ is a fixed point of $\mathscr{T}_{\mathrm{k},\boldsymbol{\pi}}$, i.e., $\mathscr{T}_{\mathrm{k},\boldsymbol{\pi}} J_{\mathrm{k}}^{(\boldsymbol{\pi})} = J_{\mathrm{k}}^{(\boldsymbol{\pi})}$, and since $\mathscr{T}_{\mathrm{k},\boldsymbol{\pi}}^{k}$ is a contraction mapping (Prop. 6.2.4, Puterman, 1994), we have that $\lim_{j \to \infty} \mathscr{T}_{\mathrm{k},\boldsymbol{\pi}}^{j} J = J_{\mathrm{k}}^{(\boldsymbol{\pi})}$ for any $J$ (Thm. 6.4.6, Cor. 6.4.7, Puterman, 1994). As a consequence

$$\|\overline{J}_{\mathrm{k}}^{(\tilde{\boldsymbol{\pi}})} - J_{\mathrm{k}}^{\star}\| = \lim_{j \to \infty} \|\mathscr{T}_{\mathrm{k},\tilde{\boldsymbol{\pi}}}^{j} J_{\mathrm{k}}^{\star} - J_{\mathrm{k}}^{\star}\|. \tag{5.21}$$

[15] See the background chapter for details about Bellman operators.

By repeated application of the triangle inequality:

$$\|\mathscr{T}^{j}_{k,\tilde{\pi}}J^{\star}_{k} - J^{\star}_{k}\| \leq \|\mathscr{T}^{j}_{k,\tilde{\pi}}J^{\star}_{k} - \mathscr{T}^{j-1}_{k,\tilde{\pi}}J^{\star}_{k}\| + \|\mathscr{T}^{j-1}_{k,\tilde{\pi}}J^{\star}_{k} - J^{\star}_{k}\|$$
$$\leq \dots \leq \sum_{m=1}^{j}\|\mathscr{T}^{m}_{k,\tilde{\pi}}J^{\star}_{k} - \mathscr{T}^{m-1}_{k,\tilde{\pi}}J^{\star}_{k}\|.$$

Since $\mathscr{T}_{k,\tilde{\pi}}$ is a contraction mapping with modulus $\gamma < 1$,

$$\|\mathscr{T}_{k,\tilde{\pi}}(\mathscr{T}_{k,\tilde{\pi}}J^{\star}_{k}) - \mathscr{T}_{k,\tilde{\pi}}J^{\star}_{k}\| \leq \gamma\|\mathscr{T}_{k,\tilde{\pi}}J^{\star}_{k} - J^{\star}_{k}\|$$
$$\implies \|\mathscr{T}^{j}_{k,\tilde{\pi}}J^{\star}_{k} - \mathscr{T}^{j-1}_{k,\tilde{\pi}}J^{\star}_{k}\| \leq \gamma^{j-1}\|\mathscr{T}_{k,\tilde{\pi}}J^{\star}_{k} - J^{\star}_{k}\|.$$

The above inequality means that

$$\sum_{m=1}^{j}\|\mathscr{T}^{m}_{k,\tilde{\pi}}J^{\star}_{k} - \mathscr{T}^{m-1}_{k,\tilde{\pi}}J^{\star}_{k}\| \leq \sum_{m=1}^{j}\gamma^{m-1}\|\mathscr{T}_{k,\tilde{\pi}}J^{\star}_{k} - J^{\star}_{k}\|.$$

Since limits preserve non-strict inequalities,

$$\lim_{j\to\infty}\|\mathscr{T}^{j}_{k,\tilde{\pi}}J^{\star}_{k} - J^{\star}_{k}\| \leq \lim_{j\to\infty}\sum_{m=1}^{j}\gamma^{m-1}\|\mathscr{T}_{k,\tilde{\pi}}J^{\star}_{k} - J^{\star}_{k}\| = \frac{\|\mathscr{T}_{k,\tilde{\pi}}J^{\star}_{k} - J^{\star}_{k}\|}{1-\gamma}. \tag{5.22}$$

Now let $\widehat{J}^{(\tilde{\pi})}_{k} \triangleq \mathscr{T}^{\overline{\pi}_{-k,t}}_{k,\ell_k}\overline{J}^{(\tilde{\pi})}_{k}$, where $\mathscr{T}^{\overline{\pi}_{-k,t}}_{k,\ell_k}$ is defined in (5.12). Note that, since Thm. 5.3 assumes that the conjectures are correct, $\mathscr{T}_{k,\tilde{\pi}}\widehat{J}^{(\tilde{\pi})}_{k} = \mathscr{T}^{\overline{\pi}_{-k,t}}_{k,1}\widehat{J}^{(\tilde{\pi})}_{k}$ and $\mathscr{T}^{\overline{\pi}_{-k,t}}_{k,1}J^{\star}_{k} = J^{\star}_{k}$ by definition (Bertsekas, 2021).

Applying the triangle inequality to the numerator in (5.22),

$$\begin{aligned}
&\|\mathscr{T}_{k,\tilde{\pi}}J^{\star}_{k} - J^{\star}_{k}\| \\
&\leq \|\mathscr{T}_{k,\tilde{\pi}}J^{\star}_{k} - \mathscr{T}_{k,\tilde{\pi}}\widehat{J}^{(\tilde{\pi})}_{k}\| + \|\mathscr{T}_{k,\tilde{\pi}}\widehat{J}^{(\tilde{\pi})}_{k} - \mathscr{T}^{\overline{\pi}_{-k,t}}_{k,1}\widehat{J}^{(\tilde{\pi})}_{k}\| + \|\mathscr{T}^{\overline{\pi}_{-k,t}}_{k,1}\widehat{J}^{(\tilde{\pi})}_{k} - J^{\star}_{k}\| \\
&= \underbrace{\|\mathscr{T}_{k,\tilde{\pi}}J^{\star}_{k} - \mathscr{T}_{k,\tilde{\pi}}\widehat{J}^{(\tilde{\pi})}_{k}\|}_{\leq\gamma\|\widehat{J}^{(\tilde{\pi})}_{k} - J^{\star}_{k}\|} + \underbrace{\|\mathscr{T}^{\overline{\pi}_{-k,t}}_{k,1}\widehat{J}^{(\tilde{\pi})}_{k} - \mathscr{T}^{\overline{\pi}_{-k,t}}_{k,1}J^{\star}_{k}\|}_{\leq\gamma\|\widehat{J}^{(\tilde{\pi})}_{k} - J^{\star}_{k}\|} \\
&\leq 2\gamma\|\widehat{J}^{(\tilde{\pi})}_{k} - J^{\star}_{k}\| = 2\gamma\|\mathscr{T}^{\overline{\pi}_{-k,t}}_{k,\ell_k}\overline{J}^{(\tilde{\pi})}_{k} - \mathscr{T}^{\overline{\pi}_{-k,t}}_{k,\ell_k}J^{\star}_{k}\| \leq 2\gamma^{\ell_k}\|\overline{J}^{(\boldsymbol{\pi})}_{k} - J^{\star}_{k}\|.
\end{aligned} \tag{5.23}$$

Combining (5.21)–(5.23) gives (5.13b). □

## C Proof of Theorem 5.4.A

The idea behind the proof of Thm. 5.4.A. is to express $\mu_t$ (5.14b) in terms of log-likelihood ratios, which can be shown to converge using the martingale convergence theorem (Thm. 6.4.3, Ash, 1972). Towards this proof, we state and prove the following two lemmas.

**Lemma 5.4.** *Any sequence* $(\boldsymbol{\pi}_{\mathbf{h}_t})_{t\geq 1}$ *generated by* COL *induces a well-defined probability measure* $\mathbb{P}^{\mathscr{R}}$ *over the set of realizable histories* $\mathbf{h}_t \in \bigtimes_{t\geq 1}(\mathcal{H}_t^{(\mathrm{D})} \times \mathcal{H}_t^{(\mathrm{A})})$.

*Proof.* Since the sample space of the random vectors $(\mathbf{I}_t^{(\mathrm{D})}, \mathbf{I}_t^{(\mathrm{A})})$ (5.3) is finite (and measurable) for each $t$, the space of realizable histories $\mathbf{h}_t \in \mathcal{H}_t^{(\mathrm{D})} \times \mathcal{H}_t^{(\mathrm{A})}$ is countable. By the extension theorem of Ionescu Tulcea, it thus follows that a measure $\mathbb{P}^{\mathscr{R}}$ over $\mathcal{H}_t^{(\mathrm{D})} \times \mathcal{H}_t^{(\mathrm{A})}$ exists for all $t \geq 1$ (Ionescu Tulcea, 1949). □

**Lemma 5.5.** *For any* $\overline{\ell}_{\mathrm{A}} \in \mathcal{L}$ *and any sequence* $(\nu_t, \boldsymbol{\pi}_{\mathbf{h}_t})_{t\geq 1}$ *generated by* COL,

$$\lim_{t\to\infty}\left|\underbrace{t^{-1}\sum_{\tau=1}^{t}\ln\frac{\mathbb{P}[\mathbf{i}_{\tau+1}^{(\mathrm{D})} \mid \ell_{\mathrm{A}}, \mathbf{b}_\tau]}{\mathbb{P}[\mathbf{i}_{\tau+1}^{(\mathrm{D})} \mid \overline{\ell}_{\mathrm{A}}, \mathbf{b}_\tau]}}_{\triangleq Z_{t+1}(\overline{\ell}_{\mathrm{A}})} - K(\overline{\ell}_{\mathrm{A}}, \nu_t)\right| = 0 \ a.s. - \mathbb{P}^{\mathscr{R}}.$$

*Proof.* By definition of $Z_t$ and $\nu_t$,

$$\begin{aligned}
Z_{t+1}(\overline{\ell}_{\mathrm{A}}) &= t^{-1}\sum_{\tau=1}^{t}\ln\frac{\mathbb{P}[\mathbf{i}_{\tau+1}^{(\mathrm{D})} \mid \ell_{\mathrm{A}}, \mathbf{b}_\tau]}{\mathbb{P}[\mathbf{i}_{\tau+1}^{(\mathrm{D})} \mid \overline{\ell}_{\mathrm{A}}, \mathbf{b}_\tau]} \\
&\overset{(a)}{=} \sum_{\mathbf{b}\in\mathcal{B}} t^{-1}\sum_{\tau=1}^{t}\mathbb{1}_{\mathbf{b}=\mathbf{b}_\tau}\ln\frac{\mathbb{P}[\mathbf{i}_{\tau+1}^{(\mathrm{D})} \mid \ell_{\mathrm{A}}, \mathbf{b}_\tau]}{\mathbb{P}[\mathbf{i}_{\tau+1}^{(\mathrm{D})} \mid \overline{\ell}_{\mathrm{A}}, \mathbf{b}_\tau]} \\
&= \sum_{\mathbf{b}\in\mathcal{B}}\sum_{\tau=1}^{t} t^{-1}\mathbb{1}_{\mathbf{b}=\mathbf{b}_\tau}\ln\mathbb{P}[\mathbf{i}_{\tau+1}^{(\mathrm{D})} \mid \ell_{\mathrm{A}}, \mathbf{b}_\tau] - \sum_{\mathbf{b}\in\mathcal{B}}\sum_{\tau=1}^{t} t^{-1}\mathbb{1}_{\mathbf{b}=\mathbf{b}_\tau}\ln\mathbb{P}[\mathbf{i}_{\tau+1}^{(\mathrm{D})} \mid \overline{\ell}_{\mathrm{A}}, \mathbf{b}_\tau] \\
&= \mathbb{E}_{\mathbf{b}\sim\nu_t}\left[\sum_{\tau=1}^{t}\frac{\ln\mathbb{P}[\mathbf{i}_{\tau+1}^{(\mathrm{D})} \mid \ell_{\mathrm{A}}, \mathbf{b}]}{t} - \sum_{\tau=1}^{t}\frac{\ln\mathbb{P}[\mathbf{i}_{\tau+1}^{(\mathrm{D})} \mid \overline{\ell}_{\mathrm{A}}, \mathbf{b}]}{t}\right],
\end{aligned}$$

where we use $\sum_{\mathbf{b}\in\mathcal{B}}\mathbb{1}_{\mathbf{b}_\tau=\mathbf{b}} = 1$ in (a). This means that if the left sum above converges to $\mathbb{E}_{\mathbf{I}^{(\mathrm{D})}}[\ln\mathbb{P}[\mathbf{I}^{(\mathrm{D})} \mid \ell_{\mathrm{A}}, \mathbf{b}] \mid \ell_{\mathrm{A}}, \mathbf{b}]$ and the right sum converges to $\mathbb{E}_{\mathbf{I}^{(\mathrm{D})}}[\ln\mathbb{P}[\mathbf{I}^{(\mathrm{D})} \mid \overline{\ell}_{\mathrm{A}}, \mathbf{b}] \mid \ell_{\mathrm{A}}, \mathbf{b}]$, we obtain $Z_t \xrightarrow[t\to\infty]{} K(\overline{\ell}_{\mathrm{A}}, \nu_t)$ (5.15), which yields the desired result[16]. As these two proofs are almost identical, we only provide the first proof here.

Let $X_\tau \triangleq \ln\mathbb{P}[\mathbf{i}_{\tau+1}^{(\mathrm{D})} \mid \ell_{\mathrm{A}}, \mathbf{b}_\tau] - \mathbb{E}_{\mathbf{I}^{(\mathrm{D})}}\left[\ln\mathbb{P}[\mathbf{I}^{(\mathrm{D})} \mid \ell_{\mathrm{A}}, \mathbf{b}_\tau]\right]$. We will show that $(X_\tau)_{\tau\geq 1}$ is a martingale difference sequence (MDS). To show this, we need to prove that $(i)$ $\mathbb{E}[X_\tau \mid \mathbf{h}_{\tau-1}] = 0$; and $(ii)$ $\mathbb{E}[|X_\tau|] < \infty$. We start with $(i)$,

$$\begin{aligned}
\mathbb{E}[X_\tau \mid \mathbf{h}_{\tau-1}] &= \mathbb{E}_{\mathbf{I}^{(\mathrm{D})}}\left[\ln\mathbb{P}[\mathbf{I}^{(\mathrm{D})} \mid \ell_{\mathrm{A}}, \mathbf{b}_\tau] - \mathbb{E}_{\mathbf{I}^{(\mathrm{D})}}\left[\ln\mathbb{P}[\mathbf{I}^{(\mathrm{D})} \mid \ell_{\mathrm{A}}, \mathbf{b}_\tau]\right] \mid \mathbf{h}_{\tau-1}\right] \\
&\overset{(a)}{=} \mathbb{E}_{\mathbf{I}^{(\mathrm{D})}}\left[\ln\mathbb{P}[\mathbf{I}^{(\mathrm{D})} \mid \ell_{\mathrm{A}}, \mathbf{b}_\tau]\right] - \mathbb{E}_{\mathbf{I}^{(\mathrm{D})}}\left[\ln\mathbb{P}[\mathbf{I}^{(\mathrm{D})} \mid \ell_{\mathrm{A}}, \mathbf{b}_\tau]\right] = 0,
\end{aligned}$$

[16] Recall the definition $K(\overline{\ell}_{\mathrm{A}}, \nu) \triangleq \mathbb{E}_{\mathbf{b}\sim\nu}\mathbb{E}_{\mathbf{I}^{(\mathrm{k})}}\left[\ln\left(\frac{\mathbb{P}[\mathbf{I}^{(\mathrm{k})} \mid \ell_{\mathrm{A}}, \mathbf{b}]}{\mathbb{P}[\mathbf{I}^{(\mathrm{k})} \mid \overline{\ell}_{\mathrm{A}}, \mathbf{b}]}\right) \mid \ell_{\mathrm{A}}, \mathbf{b}\right]$.

where (a) follows because $\mathbf{I}^{(\mathrm{D})}$ is conditionally independent of $\mathbf{h}_{\tau-1}$ given $\mathbf{b}_\tau$ (5.5).

To prove (*ii*) we will write $X_\tau$ as an expression of the form $(\ln \mathbb{P}[\varphi])^2\mathbb{P}[\varphi]$, which is bounded by 1[17]. Towards this end, we start by applying Jensen's inequality to obtain

$$\mathbb{E}[|X_\tau|] = (\mathbb{E}[|X_\tau|]^2)^{1/2} \leq (\mathbb{E}[X_\tau^2])^{1/2}, \tag{5.24}$$

which means that it suffices to bound $\mathbb{E}[X_\tau^2]$.

Next, we use $\sum_{\mathbf{i}^{(\mathrm{D})}\in\mathcal{R}_{\mathbf{I}^{(\mathrm{D})}}} \mathbb{1}_{\mathbf{i}^{(\mathrm{D})}_{\tau+1}=\mathbf{i}^{(\mathrm{D})}} = 1$ to rewrite[18] $\ln \mathbb{P}[\mathbf{i}^{(\mathrm{D})}_{\tau+1} \mid \ell_{\mathrm{A}}, \mathbf{b}_\tau]$ as

$$\begin{aligned}
\ln \mathbb{P}[\mathbf{i}^{(\mathrm{D})}_{\tau+1} \mid \ell_{\mathrm{A}}, \mathbf{b}_\tau] &= \sum_{\mathbf{i}^{(\mathrm{D})}\in\mathcal{R}_{\mathbf{I}^{(\mathrm{D})}}} \mathbb{1}_{\mathbf{i}^{(\mathrm{D})}=\mathbf{i}^{(\mathrm{D})}_{\tau+1}} \ln \mathbb{P}[\mathbf{i}^{(\mathrm{D})}_{\tau+1} \mid \ell_{\mathrm{A}}, \mathbf{b}_\tau] \\
&= \sum_{\mathbf{i}^{(\mathrm{D})}\in\mathcal{R}_{\mathbf{I}^{(\mathrm{D})}}} \mathbb{1}_{\mathbf{i}^{(\mathrm{D})}=\mathbf{i}^{(\mathrm{D})}_{\tau+1}} \frac{\mathbb{P}[\mathbf{I}^{(\mathrm{D})} \mid \ell_{\mathrm{A}}, \mathbf{b}_\tau]}{\mathbb{P}[\mathbf{I}^{(\mathrm{D})} \mid \ell_{\mathrm{A}}, \mathbf{b}_\tau]} \ln \mathbb{P}[\mathbf{I}^{(\mathrm{D})} \mid \ell_{\mathrm{A}}, \mathbf{b}_\tau] \\
&= \frac{\mathbb{E}_{\mathbf{I}^{(\mathrm{D})}}\left[\mathbb{1}_{\mathbf{I}^{(\mathrm{D})}=\mathbf{i}^{(\mathrm{D})}_{\tau+1}} \ln \mathbb{P}[\mathbf{I}^{(\mathrm{D})} \mid \ell_{\mathrm{A}}, \mathbf{b}_\tau]\right]}{\mathbb{P}[\mathbf{i}^{(\mathrm{D})}_{\tau+1} \mid \ell_{\mathrm{A}}, \mathbf{b}_\tau]},
\end{aligned}$$

which means that we can write $X_\tau$ as

$$\begin{aligned}
X_\tau &= \ln \mathbb{P}[\mathbf{i}^{(\mathrm{D})}_{\tau+1} \mid \ell_{\mathrm{A}}, \mathbf{b}_\tau] - \mathbb{E}_{\mathbf{I}^{(\mathrm{D})}}\left[\ln \mathbb{P}[\mathbf{I}^{(\mathrm{D})} \mid \ell_{\mathrm{A}}, \mathbf{b}_\tau]\right] \\
&= \frac{\mathbb{E}_{\mathbf{I}^{(\mathrm{D})}}\left[\mathbb{1}_{\mathbf{I}^{(\mathrm{D})}=\mathbf{i}^{(\mathrm{D})}_{\tau+1}} \ln \mathbb{P}[\mathbf{I}^{(\mathrm{D})} \mid \ell_{\mathrm{A}}, \mathbf{b}_\tau]\right]}{\mathbb{P}[\mathbf{i}^{(\mathrm{D})}_{\tau+1} \mid \ell_{\mathrm{A}}, \mathbf{b}_\tau]} - \mathbb{E}_{\mathbf{I}^{(\mathrm{D})}}\left[\ln \mathbb{P}[\mathbf{I}^{(\mathrm{D})} \mid \ell_{\mathrm{A}}, \mathbf{b}_\tau]\right] \\
&= \frac{\mathbb{E}_{\mathbf{I}^{(\mathrm{D})}}\left[\mathbb{1}_{\mathbf{I}^{(\mathrm{D})}=\mathbf{i}^{(\mathrm{D})}_{\tau+1}} \ln \mathbb{P}[\mathbf{I}^{(\mathrm{D})} \mid \ell_{\mathrm{A}}, \mathbf{b}_\tau]\right] - \mathbb{P}[\mathbf{i}^{(\mathrm{D})}_{\tau+1} \mid \ell_{\mathrm{A}}, \mathbf{b}_\tau]\mathbb{E}_{\mathbf{I}^{(\mathrm{D})}}\left[\ln \mathbb{P}[\mathbf{I}^{(\mathrm{D})} \mid \ell_{\mathrm{A}}, \mathbf{b}_\tau]\right]}{\mathbb{P}[\mathbf{i}^{(\mathrm{D})}_{\tau+1} \mid \ell_{\mathrm{A}}, \mathbf{b}_\tau]} \\
&= \frac{\mathbb{E}_{\mathbf{I}^{(\mathrm{D})}}\left[\ln \mathbb{P}[\mathbf{I}^{(\mathrm{D})} \mid \ell_{\mathrm{A}}, \mathbf{b}_\tau]\left(\mathbb{1}_{\mathbf{I}^{(\mathrm{D})}=\mathbf{i}^{(\mathrm{D})}_{\tau+1}} - \mathbb{P}[\mathbf{i}^{(\mathrm{D})}_{\tau+1} \mid \ell_{\mathrm{A}}, \mathbf{b}_\tau]\right) \mid \ell_{\mathrm{A}}, \mathbf{b}_\tau\right]}{\mathbb{P}[\mathbf{i}^{(\mathrm{D})}_{\tau+1} \mid \ell_{\mathrm{A}}, \mathbf{b}_\tau]}.
\end{aligned} \tag{5.25}$$

Since we focus on realizable histories, we have that $\mathbb{P}[\mathbf{i}^{(\mathrm{D})}_{\tau+1} \mid \ell_{\mathrm{A}}, \mathbf{b}_\tau] \in (0, 1]$. Hence, it is safe to suppress the denominator in (5.25). Consequently, the square of (5.25) can be bounded by the Cauchy-Schwarz inequality as

$$\begin{aligned}
X_\tau^2 &\approx \mathbb{E}_{\mathbf{I}^{(\mathrm{D})}}\left[\ln \mathbb{P}[\mathbf{I}^{(\mathrm{D})} \mid \ell_{\mathrm{A}}, \mathbf{b}_\tau]\left(\mathbb{1}_{\mathbf{I}^{(\mathrm{D})}=\mathbf{i}^{(\mathrm{D})}_{\tau+1}} - \mathbb{P}[\mathbf{i}^{(\mathrm{D})}_{\tau+1} \mid \ell_{\mathrm{A}}, \mathbf{b}_\tau]\right) \mid \ell_{\mathrm{A}}, \mathbf{b}_\tau\right]^2 \\
&\leq \mathbb{E}_{\mathbf{I}^{(\mathrm{D})}}\left[\left(\underbrace{\ln \mathbb{P}[\mathbf{I}^{(\mathrm{D})} \mid \ell_{\mathrm{A}}, \mathbf{b}_\tau]}_{\triangleq \kappa}\right)^2\left(\underbrace{\mathbb{1}_{\{\mathbf{I}^{(\mathrm{D})}\}}(\mathbf{i}^{(\mathrm{D})}_\tau + 1) - \mathbb{P}[\mathbf{i}^{(\mathrm{D})}_{\tau+1} \mid \ell_{\mathrm{A}}, \mathbf{b}_\tau]}_{\triangleq \chi}\right)^2 \mid \ell_{\mathrm{A}}, \mathbf{b}_\tau\right].
\end{aligned}$$

[17] We use the standard convention that $(\ln 0)^2 0 = 0$.

[18] Recall that $\mathcal{R}_X$ denotes the range of the random variable $X$.

Since $\kappa \geq 0$ and $\chi \in [0,1]$, we obtain that

$$X_\tau^2 \leq \mathbb{E}_{\mathbf{I}^{(\mathrm{D})}}\left[\kappa^2\chi^2 \mid \ell_{\mathrm{A}}, \mathbf{b}_\tau\right] \leq \mathbb{E}_{\mathbf{I}^{(\mathrm{D})}}\left[\kappa^2 \mid \ell_{\mathrm{A}}, \mathbf{b}_\tau\right],$$

which means that

$$\mathbb{E}[X_\tau^2] \lesssim \mathbb{E}_{\mathbf{I}^{(\mathrm{D})}}\left[\mathbb{P}[\mathbf{I}^{(\mathrm{D})} \mid \ell_{\mathrm{A}}, \mathbf{b}_\tau]\left(\ln \mathbb{P}[\mathbf{I}^{(\mathrm{D})} \mid \ell_{\mathrm{A}}, \mathbf{b}_\tau]\right)^2 \mid \ell_{\mathrm{A}}, \mathbf{b}_\tau\right] \leq 1$$
$$\overset{(a)}{\Longrightarrow} \mathbb{E}[|X_\tau|] \lesssim 1 \implies \mathbb{E}[|X_\tau|] \leq \infty,$$

where (a) follows from (5.24). Therefore, $(X_\tau)_{\tau\geq 1}$ is an MDS. Consequently, the sequence $Y_t \triangleq \sum_{\tau=1}^{t} \frac{X_\tau}{\tau}$ is a martingale. By the martingale convergence theorem, $(Y_\tau)_{\tau\geq 1}$ converges to a finite and integrable random variable a.s.-$\mathbb{P}^{\mathscr{R}}$ (Thm. 6.4.3, Ash, 1972). This convergence means that we can invoke Kronecker's lemma, which states that $\lim_{t\to\infty} t^{-1}\sum_{\tau=1}^{t} X_\tau = 0$ a.s.-$\mathbb{P}^{\mathscr{R}}$ (p. 105, Pollard, 2001). As a consequence, the following holds a.s.-$\mathbb{P}^{\mathscr{R}}$ as $t \to \infty$

$$\lim_{t\to\infty}\sum_{\tau=1}^{t} \frac{\ln \mathbb{P}[\mathbf{i}_{\tau+1}^{(\mathrm{D})} \mid \ell_{\mathrm{A}}, \mathbf{b}_\tau] - \mathbb{E}_{\mathbf{I}^{(\mathrm{D})}}\left[\ln \mathbb{P}[\mathbf{I}^{(\mathrm{D})} \mid \ell_{\mathrm{A}}, \mathbf{b}_\tau]\right]}{t} = 0$$
$$\Longrightarrow \lim_{t\to\infty} \mathbb{E}_{\mathbf{b}\sim\nu_t}\left[\sum_{\tau=1}^{t} \frac{\ln \mathbb{P}[\mathbf{i}_{\tau+1}^{(\mathrm{D})} \mid \ell_{\mathrm{A}}, \mathbf{b}] - \mathbb{E}_{\mathbf{I}^{(\mathrm{D})}}\left[\ln \mathbb{P}[\mathbf{I}^{(\mathrm{D})} \mid \ell_{\mathrm{A}}, \mathbf{b}]\right]}{t}\right] = 0$$
$$\Longrightarrow \lim_{t\to\infty} \mathbb{E}_{\mathbf{b}\sim\nu_t}\left[\sum_{\tau=1}^{t} \frac{\ln \mathbb{P}[\mathbf{i}_{\tau+1}^{(\mathrm{D})} \mid \ell_{\mathrm{A}}, \mathbf{b}]}{t}\right] \overset{(a)}{=} \mathbb{E}_{\mathbf{b}\sim\nu_t}\left[\mathbb{E}_{\mathbf{I}^{(\mathrm{D})}}\left[\ln \mathbb{P}[\mathbf{I}^{(\mathrm{D})} \mid \ell_{\mathrm{A}}, \mathbf{b}]\right]\right],$$

where (a) follows because $\mathbb{E}_{\mathbf{I}^{(\mathrm{D})}}\left[\ln \mathbb{P}[\mathbf{I}^{(\mathrm{D})} \mid \ell_{\mathrm{A}}, \mathbf{b}]\right]$ is independent of $\tau$. □

## C.1 Proof of Theorem 5.4.A

To streamline analysis, we treat $\mu_t$ as a probability measure over $\mathcal{L}$ and use integral language. From Bayes rule and the Markov property of $\mathbb{P}[\mathbf{I}^{(\mathrm{D})} \mid \bar{\ell}_{\mathrm{A}}, \mathbf{b}_t]$, we have

$$\mu_{t+1}(\bar{\ell}_{\mathrm{A}}) = \frac{\mathbb{P}[\bar{\ell}_{\mathrm{A}}]\mathbb{P}[\mathbf{i}_2^{(\mathrm{D})}, \ldots, \mathbf{i}_{t+1}^{(\mathrm{D})} \mid \bar{\ell}_{\mathrm{A}}, \mathbf{h}_t^{(\mathrm{D})}]}{\mathbb{P}[\mathbf{i}_2^{(\mathrm{D})}, \ldots, \mathbf{i}_{t+1}^{(\mathrm{D})} \mid \mathbf{h}_t^{(\mathrm{D})}]} \overset{(a)}{=} \frac{\mu_1(\bar{\ell}_{\mathrm{A}})\prod_{\tau=1}^{t}\mathbb{P}[\mathbf{i}_{\tau+1}^{(\mathrm{D})} \mid \bar{\ell}_{\mathrm{A}}, \mathbf{b}_\tau]}{\int_{\mathcal{L}}\mu_1(\mathrm{d}\bar{\ell}_{\mathrm{A}})\prod_{\tau=1}^{t}\mathbb{P}[\mathbf{i}_{\tau+1}^{(\mathrm{D})} \mid \bar{\ell}_{\mathrm{A}}, \mathbf{b}_\tau]}$$
$$= \frac{\mu_1(\bar{\ell}_{\mathrm{A}})\exp\left(\ln\left(\prod_{\tau=1}^{t}\frac{\mathbb{P}[\mathbf{i}_{\tau+1}^{(\mathrm{D})}|\bar{\ell}_{\mathrm{A}},\mathbf{b}_\tau]}{\mathbb{P}[\mathbf{i}_{\tau+1}^{(\mathrm{D})}|\ell_{\mathrm{A}},\mathbf{b}_\tau]}\right)\right)}{\int_{\mathcal{L}}\mu_1(\mathrm{d}\bar{\ell}_{\mathrm{A}})\exp\left(\ln\left(\prod_{\tau=1}^{t}\frac{\mathbb{P}[\mathbf{i}_{\tau+1}^{(\mathrm{D})}|\bar{\ell}_{\mathrm{A}},\mathbf{b}_\tau]}{\mathbb{P}[\mathbf{i}_{\tau+1}^{(\mathrm{D})}|\ell_{\mathrm{A}},\mathbf{b}_\tau]}\right)\right)}$$
$$= \frac{\mu_1(\bar{\ell}_{\mathrm{A}})\exp\left(-tZ_{t+1}(\bar{\ell}_{\mathrm{A}})\right)}{\int_{\mathcal{L}}\mu_1(\mathrm{d}\bar{\ell}_{\mathrm{A}})\exp\left(-tZ_{t+1}(\bar{\ell}_{\mathrm{A}})\right)}, \tag{5.26}$$

where $Z_t$ is defined in Lemma 5.5. Step (a) above is well-defined by Assumption 5.1 and follows because $\mathbf{I}^{(\mathrm{D})}$ is conditionally independent of $\mathbf{h}^{(\mathrm{D})}_{t-1}$ given $\mathbf{b}_{t-1}$.

Using the expression in (5.26), we obtain

$$\mathbb{E}_{\overline{\ell}_{\mathrm{A}}\sim\mu_{t+1}}\Big[\overbrace{K(\overline{\ell}_{\mathrm{A}},\nu_t)-K^{\star}_{\mathcal{L}}(\nu_t)}^{\triangleq\Delta K(\overline{\ell}_{\mathrm{A}},\nu_t)}\Big] = \frac{\int_{\mathcal{L}}\Delta K(\overline{\ell}_{\mathrm{A}},\nu_t)\mu_1(\mathrm{d}\overline{\ell}_{\mathrm{A}})\exp\left(-tZ_{t+1}(\overline{\ell}_{\mathrm{A}})\right)}{\int_{\mathcal{L}}\mu_1(\mathrm{d}\overline{\ell}_{\mathrm{A}})\exp\left(-tZ_{t+1}(\overline{\ell}_{\mathrm{A}})\right)}$$
$$=\frac{\int_{\mathcal{L}}\Delta K(\overline{\ell}_{\mathrm{A}},\nu_t)\mu_1(\mathrm{d}\overline{\ell}_{\mathrm{A}})\exp\left(-tZ_{t+1}(\overline{\ell}_{\mathrm{A}})\right)}{\int_{\mathcal{L}}\mu_1(\mathrm{d}\overline{\ell}_{\mathrm{A}})\exp\left(-tZ_{t+1}(\overline{\ell}_{\mathrm{A}})\right)}\frac{\exp(K^{\star}_{\mathcal{L}}(\nu_t)t)}{\exp(K^{\star}_{\mathcal{L}}(\nu_t)t)}$$
$$=\frac{\int_{\mathcal{L}}\overbrace{\Delta K(\overline{\ell}_{\mathrm{A}},\nu_t)\mu_1(\mathrm{d}\overline{\ell}_{\mathrm{A}})\exp\left(-t\left(Z_{t+1}(\overline{\ell}_{\mathrm{A}})-K^{\star}_{\mathcal{L}}(\nu_{t+1})\right)\right)}^{\triangleq\sigma}}{\int_{\mathcal{L}}\mu_1(\mathrm{d}\overline{\ell}_{\mathrm{A}})\exp\left(-t(Z_{t+1}(\overline{\ell}_{\mathrm{A}})-K^{\star}_{\mathcal{L}}(\nu_{t+1}))\right)}. \tag{5.27}$$

By defining $\mathcal{L}_{\epsilon}\triangleq\{\overline{\ell}_A \mid \Delta K(\overline{\ell}_{\mathrm{A}},\nu_{t+1})\geq\epsilon\}$ we can write the numerator in (5.27) as

$$\int_{\mathcal{L}\setminus\mathcal{L}_{\epsilon}}\sigma+\int_{\mathcal{L}_{\epsilon}}\sigma\leq\epsilon+\int_{\mathcal{L}\setminus\mathcal{L}_{\epsilon}}\sigma. \tag{5.28}$$

Given this bound, it suffices to show that, for arbitrarily small $\epsilon>0$,

$$\lim_{t\to\infty}\frac{\int_{\mathcal{L}_{\epsilon}}\sigma}{\int_{\mathcal{L}}\mu_1(\mathrm{d}\overline{\ell}_{\mathrm{A}})\exp\left(-t(Z_{t+1}(\overline{\ell}_{\mathrm{A}})-K^{\star}_{\mathcal{L}}(\nu_{t+1}))\right)}=0.$$

Towards the proof of the limit above, we note that the exponent in $\sigma$ (5.27) can be written as

$$-t(Z_{t+1}(\overline{\ell}_{\mathrm{A}})-K^{\star}_{\mathcal{L}}(\nu_{t+1}))=-t(Z_{t+1}(\overline{\ell}_{\mathrm{A}})-K^{\star}_{\mathcal{L}}(\nu_{t+1}))+K(\overline{\ell}_A,\nu_{t+1})-K(\overline{\ell}_A,\nu_{t+1})$$
$$=-t(\Delta K(\overline{\ell}_{\mathrm{A}},\nu_{t+1})+Z_{t+1}(\overline{\ell}_{\mathrm{A}})-K(\overline{\ell}_A,\nu_{t+1})). \tag{5.29}$$

Therefore, we obtain

$$\frac{\int_{\mathcal{L}_{\epsilon}}\sigma}{\int_{\mathcal{L}}\mu_1(\mathrm{d}\overline{\ell}_{\mathrm{A}})\exp\left(-t(Z_{t+1}(\overline{\ell}_{\mathrm{A}})-K^{\star}_{\mathcal{L}}(\nu_{t+1}))\right)}$$
$$=\frac{\int_{\mathcal{L}_{\epsilon}}\Delta K(\overline{\ell}_{\mathrm{A}},\nu_{t+1})\mu_1(\mathrm{d}\overline{\ell}_{\mathrm{A}})\exp\left(-t(\Delta K(\overline{\ell}_{\mathrm{A}},\nu_{t+1})+Z_{t+1}(\overline{\ell}_{\mathrm{A}})-K(\overline{\ell}_A,\nu_{t+1}))\right)}{\int_{\mathcal{L}}\mu_1(\mathrm{d}\overline{\ell}_{\mathrm{A}})\exp\left(-t(\Delta K(\overline{\ell}_{\mathrm{A}},\nu_{t+1})+Z_{t+1}(\overline{\ell}_{\mathrm{A}})-K(\overline{\ell}_A,\nu_{t+1}))\right)}.$$

Next, we recall from Lemma 5.5 that for any $\epsilon>0$, there exists $\eta>0$ and $t_{\eta}\geq 1$ such that, for all $t\geq t_{\eta}$ and $\overline{\ell}_{\mathrm{A}}\in\mathcal{L}$, $|Z_t(\overline{\ell}_{\mathrm{A}}-K(\overline{\ell}_{\mathrm{A}},\nu_{t+1}))|<\eta$. ($t_{\eta}$ is uniform as $|\mathcal{L}|<\infty$, Assumption 5.1.) This convergence implies that

$$\frac{\int_{\mathcal{L}_{\epsilon}}\Delta K(\overline{\ell}_{\mathrm{A}},\nu_{t+1})\mu_1(\mathrm{d}\overline{\ell}_{\mathrm{A}})\exp\left(-t(\Delta K(\overline{\ell}_{\mathrm{A}},\nu_{t+1})+Z_{t+1}(\overline{\ell}_{\mathrm{A}})-K(\overline{\ell}_A,\nu_{t+1}))\right)}{\int_{\mathcal{L}}\mu_1(\mathrm{d}\overline{\ell}_{\mathrm{A}})\exp\left(-t(\Delta K(\overline{\ell}_{\mathrm{A}},\nu_{t+1})+Z_{t+1}(\overline{\ell}_{\mathrm{A}})-K(\overline{\ell}_A,\nu_{t+1}))\right)}$$

$$\leq \frac{\int_{\mathcal{L}_\epsilon} \Delta K(\overline{\ell}_{\mathrm{A}}, \nu_{t+1}) \mu_1(\mathrm{d}\overline{\ell}_{\mathrm{A}}) \exp\left(-t(\Delta K(\overline{\ell}_{\mathrm{A}}, \nu_{t+1}) - \eta)\right)}{\int_{\mathcal{L}} \mu_1(\mathrm{d}\overline{\ell}_{\mathrm{A}}) \exp\left(-t(\Delta K(\overline{\ell}_{\mathrm{A}}, \nu_{t+1}) + \eta)\right)} \qquad \forall t \geq t_\eta$$

$$= e^{2t\eta} \frac{\int_{\mathcal{L}_\epsilon} \Delta K(\overline{\ell}_{\mathrm{A}}, \nu_{t+1}) \mu_1(\mathrm{d}\overline{\ell}_{\mathrm{A}}) e^{-t\Delta K(\overline{\ell}_{\mathrm{A}}, \nu_{t+1})}}{\int_{\mathcal{L}} \mu_1(\mathrm{d}\overline{\ell}_{\mathrm{A}}) e^{-t\Delta K(\overline{\ell}_{\mathrm{A}}, \nu_{t+1})}}. \qquad (*)$$

Now, consider the numerator in $(*)$. Since $xe^{-tx}$ is decreasing for all $x > t^{-1}$ and since $\Delta K(\overline{\ell}_{\mathrm{A}}, \nu_{t+1}) \geq \epsilon$ for all $\overline{\ell}_{\mathrm{A}} \in \mathcal{L}_\epsilon$, we have that

$$\int_{\mathcal{L}_\epsilon} \Delta K(\overline{\ell}_{\mathrm{A}}, \nu_{t+1}) \mu_1(\mathrm{d}\overline{\ell}_{\mathrm{A}}) e^{-t\Delta K(\overline{\ell}_{\mathrm{A}}, \nu_{t+1})} \leq \epsilon e^{-t\epsilon} \qquad \forall t \geq \max\left[t_\eta, \frac{1}{\epsilon}\right].$$

Next, consider the denominator in $(*)$. By definition, $\exists \overline{\ell}_{\mathrm{A}} \in \mathcal{L}$ such that $K(\overline{\ell}_{\mathrm{A}}, \nu_t) = K^\star_{\mathcal{L}}(\nu_t) \forall t$. This fact, together with the assumption that $\mu_1$ has full support (Assumption 5.1), means that the denominator is a positive constant, which we denote by $k$. As a result, $(*) \leq e^{2t\eta} \epsilon e^{-t\epsilon} k^{-1}$. Let $\eta = \frac{\epsilon}{4}$. Then $e^{2t\eta}\epsilon e^{-t\epsilon} k^{-1} = e^{\frac{-t\epsilon}{2}} \epsilon k^{-1}$, which converges to 0 as $t \to \infty$. Consequently, $\lim_{t\to\infty} \mathbb{E}_{\overline{\ell}_{\mathrm{A}} \sim \mu_t}[\Delta K(\overline{\ell}_{\mathrm{A}}, \nu_t)] = 0$ a.s.-$\mathbb{P}^{\mathscr{R}}$.

## D Proof of Theorem 5.4.B

The proof of Thm. 5.4.B follows the same procedure as that of Thm. 5.4.A, with the difference that $\Theta_{\mathrm{k}}$ is allowed to be non-finite, whereas $\mathcal{L}$ in Thm. 5.4.A is finite (Assumption 5.1). Define $\Theta^+_{\mathrm{k},\epsilon} \triangleq \{\overline{\boldsymbol{\theta}} \mid \Delta K(\overline{\boldsymbol{\theta}}, \nu_t) \geq \epsilon\}$ and $\Theta^-_{\mathrm{k},\frac{\epsilon}{2}} \triangleq \{\overline{\boldsymbol{\theta}} \mid \Delta K(\overline{\boldsymbol{\theta}}, \nu_t) \leq \frac{\epsilon}{2}\}$. It then follows from (5.27)–(5.28) that

$$\int_{\Theta_{\mathrm{k}}} \Big(\overbrace{K(\overline{\boldsymbol{\theta}}, \nu_t) - K^\star_{\Theta_{\mathrm{k}}}(\nu_t)}^{\Delta K(\overline{\boldsymbol{\theta}}, \nu_t)}\Big) \rho^{(\mathrm{k})}_{t+1}(\mathrm{d}\overline{\boldsymbol{\theta}}) \leq \epsilon + \underbrace{\frac{\int_{\Theta^+_{\mathrm{k},\epsilon}} \Delta K(\overline{\boldsymbol{\theta}}, \nu_t) \exp\left(-t\left(Z_t(\overline{\boldsymbol{\theta}}) - K^\star_{\Theta_{\mathrm{k}}}(\nu_t)\right)\right) \rho^{(\mathrm{k})}_1(\mathrm{d}\overline{\boldsymbol{\theta}})}{\int_{\Theta^-_{\mathrm{k},\frac{\epsilon}{2}}} \exp\left(-t\left(Z_t(\overline{\boldsymbol{\theta}}) - K^\star_{\Theta_{\mathrm{k}}}(\nu_t)\right)\right) \rho^{(\mathrm{k})}_1(\mathrm{d}\overline{\boldsymbol{\theta}})}}_{\triangleq \star}, \tag{5.30}$$

where $\star$ is well-defined by Assumptions 5.1–5.2.

(5.30) implies that it suffices to prove that $\star \xrightarrow{t\to\infty} 0$ for arbitrarily small $\epsilon$. Applying Lemma 5.5 and $(*)$, we obtain

$$\star \overset{(a)}{\leq} e^{2t\eta} \frac{\int_{\Theta^+_{\mathrm{k},\epsilon}} \Delta K(\overline{\boldsymbol{\theta}}, \nu_t) e^{-t\Delta K(\overline{\boldsymbol{\theta}}, \nu_t)} \rho^{(\mathrm{k})}_1(\mathrm{d}\overline{\boldsymbol{\theta}})}{\int_{\Theta^-_{\mathrm{k},\frac{\epsilon}{2}}} e^{-t\Delta K(\overline{\boldsymbol{\theta}}, \nu_t)} \rho^{(\mathrm{k})}_1(\mathrm{d}\overline{\boldsymbol{\theta}})} \overset{(b)}{\leq} e^{2t\eta} \frac{\epsilon e^{-t\epsilon}}{\int_{\Theta^-_{\mathrm{k},\frac{\epsilon}{2}}} e^{-t\Delta K(\overline{\boldsymbol{\theta}}, \nu_t)} \rho^{(\mathrm{k})}_1(\mathrm{d}\overline{\boldsymbol{\theta}})}$$

$$\overset{(c)}{\leq} e^{2t\eta} \frac{\epsilon e^{-t\epsilon}}{e^{-t\frac{\epsilon}{2}} \int_{\Theta^-_{\mathrm{k},\frac{\epsilon}{2}}} \rho_1^{(\mathrm{k})}(\mathrm{d}\overline{\boldsymbol{\theta}})} = e^{2t\eta} \frac{\epsilon e^{-t\frac{\epsilon}{2}}}{\rho_1^{(\mathrm{k})}(\Theta^-_{\mathrm{k},\frac{\epsilon}{2}})} \qquad \forall t \geq \max\left[t_\eta, \epsilon^{-1}\right], \tag{5.31}$$

where (a) follows from (*); (b) follows because $xe^{-tx}$ is decreasing in $x$ for all $x > t^{-1}$; and (c) follows because $e^{-t\Delta K(\overline{\boldsymbol{\theta}},\nu_t)} \geq e^{-t\frac{\epsilon}{2}}$.

Let $\eta = \epsilon/8$. Then the numerator in the final expression above becomes $\epsilon e^{-t\frac{\epsilon}{4}}$, which converges to 0 as $t \to \infty$. Thus, what remains to show is that the denominator is positive in the limit, i.e., $\lim_{t\to\infty} \rho_1^{(\mathrm{k})}(\Theta^-_{\mathrm{k},\frac{\epsilon}{2}}) > 0$. We prove this statement by establishing uniform continuity of $\Delta K(\overline{\boldsymbol{\theta}}, \nu)$. Towards this end, we prove the following two lemmas.

**Lemma 5.6.** *$\mathcal{B}$ (5.5) is a compact subset of $\mathbb{R}^{|\mathcal{S}|}$ with the Euclidean metric $d$ and $\Delta(\mathcal{B})$ a compact metric space with the Wasserstein-$p$ distance $W_p$ ($p \geq 1$).*

*Proof.* Since $\mathcal{S}$ is finite, $\mathcal{B} = \Delta(\mathcal{S})$ is a compact subset of $\mathbb{R}^{|\mathcal{S}|}$ and $(\mathcal{B}, d)$ is a Polish space. To prove that $(\Delta(\mathcal{B}), W_p)$ is compact we will show that every sequence $(\nu_n)_{n=1}^{\infty} \subset \Delta(\mathcal{B})$ admits a subsequence converging to some limit point in $\Delta(\mathcal{B})$. Since $\mathcal{B}$ is compact and $\nu_n(\mathcal{B}) = 1$, this collection of measures is tight as $\exists \mathcal{C} \subseteq \mathcal{B}$ such that $\nu_n(\mathcal{C}) = 1 > 1 - \epsilon$ for any $\epsilon > 0$ and $\nu_n$. Therefore, $(\nu_n)_{n=1}^{\infty}$ admits a limit point $\nu^\star \in \Delta(\mathcal{B})$ w.r.t the topology of weak convergence (Prokhorov's theorem (Ch. 1, §5, Billingsley, 1999)). We will show that $\nu^\star$ is also a limit point under $W_p$. By Skorokhod's representation theorem (p. 70, Billingsley, 1999), there exists a sequence of $\mathcal{B}$-valued random variables $\{V_1, \ldots, V_n, \ldots, V^\star\}$ such that $V_n$ has the probability law $\nu_n$ and $V_n$ converges to $V^\star$ almost surely as $n \to \infty$. By the dominance convergence theorem and the facts that $\mathcal{B}$ is compact and $d$ is continuous, $\lim_{n\to\infty} \mathbb{E}[d(V_n, V^\star)^p] = 0$. Consequently, for any coupling $\xi$ between $\nu_n$ and $\nu^\star$,

$$\lim_{n\to\infty} \left( \int d(x,y)^p d\xi(x,y) \right)^{\frac{1}{p}} = 0.$$

Since $W_p(\nu_n, \nu^\star)$ is the infimum of the left-hand side above (by definition), $\lim_{n\to\infty} W_p(\nu_n, \nu^\star) = 0$. Hence, every sequence $(\nu_n)_{n=1}^{\infty} \subset \Delta(\mathcal{B})$ admits a subsequence converging to some limit point in $\Delta(\mathcal{B})$ under $W_p$. Thus, $(\Delta(\mathcal{B}), W_p)$ is compact. □

**Lemma 5.7.** $\Delta K(\overline{\boldsymbol{\theta}}, \nu) \triangleq K(\overline{\boldsymbol{\theta}}, \nu) - K^\star_{\Theta_\mathrm{k}}(\nu)$ *is a continuous map from $(\Theta_\mathrm{k}, d) \times (\Delta(\mathcal{B}), W_1)$ to $\mathbb{R}$, where $d$ and $W_1$ denote the Euclidean and the Wasserstein-1 distance, respectively.*

*Proof.* We start by showing that $K(\overline{\boldsymbol{\theta}}, \nu)$ is continuous by proving that for any convergent sequence $(\overline{\boldsymbol{\theta}}_n, \nu_n) \xrightarrow[n\to\infty]{} (\overline{\boldsymbol{\theta}}, \nu)$, the difference $|K(\overline{\boldsymbol{\theta}}_n, \nu_n) - K(\overline{\boldsymbol{\theta}}, \nu)|$ converges to 0. This difference can be bounded using the triangle inequality as

$$|K(\overline{\boldsymbol{\theta}}_n, \nu_n) - K(\overline{\boldsymbol{\theta}}, \nu)| \leq \underbrace{|K(\overline{\boldsymbol{\theta}}_n, \nu_n) - K(\overline{\boldsymbol{\theta}}_n, \nu)|}_{\triangleq ①} + \underbrace{|K(\overline{\boldsymbol{\theta}}_n, \nu) - K(\overline{\boldsymbol{\theta}}, \nu)|}_{\triangleq ②}.$$

Consider the left expression above (①). (5.15) implies that

$$① = \left| \int_{\mathcal{B}} \mathbb{E}_{\mathbf{I}^{(\mathrm{k})}} \left[ \ln \left( \frac{\mathbb{P}[\mathbf{I}^{(\mathrm{k})} \mid \boldsymbol{\theta}, \mathbf{b}]}{\mathbb{P}[\mathbf{I}^{(\mathrm{k})} \mid \overline{\boldsymbol{\theta}}_n, \mathbf{b}]} \right) \right] \nu_n(\mathrm{d}\mathbf{b}) - \int_{\mathcal{B}} \mathbb{E}_{\mathbf{I}^{(\mathrm{k})}} \left[ \ln \left( \frac{\mathbb{P}[\mathbf{I}^{(\mathrm{k})} \mid \boldsymbol{\theta}, \mathbf{b}]}{\mathbb{P}[\mathbf{I}^{(\mathrm{k})} \mid \overline{\boldsymbol{\theta}}_n, \mathbf{b}]} \right) \right] \nu(\mathrm{d}\mathbf{b}) \right|,$$

which is an integral probability metric (IPM) with the testing function $f(\mathbf{b}) \triangleq \mathbb{E}_{\mathbf{I}^{(\mathrm{k})}} \left[ \ln \left( \frac{\mathbb{P}[\mathbf{I}^{(\mathrm{k})} \mid \boldsymbol{\theta}, \mathbf{b}]}{\mathbb{P}[\mathbf{I}^{(\mathrm{k})} \mid \overline{\boldsymbol{\theta}}_n, \mathbf{b}]} \right) \right]$. This function is assumed to be Lipschitz continuous (Assumption 5.2.1). As the function can be rescaled, we can, without loss of generality, assume the Lipschitz constant to be 1. Since the Wasserstein distance is equivalent to the IPM w.r.t the class of 1-Lipschitz functions, ① is upper-bounded by $W_p(\nu_n, \nu)$. Hence, as $\nu_n$ converges to $\nu$ in $W_p$, ① converges to 0. Therefore, $\nu \mapsto K(\overline{\boldsymbol{\theta}}, \nu)$ is continuous. We now show that ② converges. Using Assumption 5.2.2 and the dominated convergence theorem, we obtain that

$$\lim_{n \to \infty} \mathbb{E}_{\mathbf{b} \sim \nu} \mathbb{E}_{\mathbf{I}^{(\mathrm{k})}} \left[ \ln \left( \frac{\mathbb{P}[\mathbf{I}^{(\mathrm{k})} \mid \boldsymbol{\theta}, \mathbf{b}]}{\mathbb{P}[\mathbf{I}^{(\mathrm{k})} \mid \overline{\boldsymbol{\theta}}_n, \mathbf{b}]} \right) \right] = \mathbb{E}_{\mathbf{b} \sim \nu} \mathbb{E}_{\mathbf{I}^{(\mathrm{k})}} \left[ \ln \left( \frac{\mathbb{P}[\mathbf{I}^{(\mathrm{k})} \mid \boldsymbol{\theta}, \mathbf{b}]}{\mathbb{P}[\mathbf{I}^{(\mathrm{k})} \mid \overline{\boldsymbol{\theta}}, \mathbf{b}]} \right) \right],$$

which implies that ② converges to 0 as $n \to \infty$. Consequently, $\overline{\boldsymbol{\theta}} \mapsto K(\overline{\boldsymbol{\theta}}, \nu)$ is continuous. Finally, since $\Theta_{\mathrm{k}}$ is compact (Assumption 5.1) and $K(\overline{\boldsymbol{\theta}}, \nu)$ is continuous in both $\overline{\boldsymbol{\theta}}$ and $\nu$, we can apply Berge's maximum theorem to $K(\overline{\boldsymbol{\theta}}, \nu)$ w.r.t. $\overline{\boldsymbol{\theta}}$. This theorem states that the mapping $\nu \mapsto K^{\star}_{\Theta_{\mathrm{k}}}(\nu)$ is continuous (5.16a). Since continuity is preserved under subtraction, it follows that $\Delta K$ also is continuous (Thm. 17.31, Aliprantis and Border, 2006). □

## D.1 Proof of Theorem 5.4.B

Lemmas 5.6–5.7 and the compactness of $\Theta_{\mathrm{k}}$ (Assumption 5.1) imply uniform continuity of $\Delta K(\overline{\boldsymbol{\theta}}, \nu)$ and that $|\Theta^{\star}_{\mathrm{k}}(\nu)| > 0$ (Thm. 17.31, Aliprantis and Border, 2006). As a consequence, for each $\overline{\boldsymbol{\theta}}_\nu \in \Theta^{\star}_{\mathrm{k}}(\nu)$, $\overline{\boldsymbol{\theta}}' \in \Theta_{\mathrm{k}}$, and $\nu', \nu \in \Delta(\mathcal{B})$, $\exists \delta_m$ such that $d(\overline{\boldsymbol{\theta}}_\nu, \overline{\boldsymbol{\theta}}') < \delta_m$, $W_1(\nu, \nu') < \delta_m$, and $d(\Delta K(\overline{\boldsymbol{\theta}}', \nu'), \Delta K(\overline{\boldsymbol{\theta}}_\nu, \nu)) \leq m \overset{(a)}{\Longrightarrow} \Delta K(\overline{\boldsymbol{\theta}}', \nu') < m$ for each $m > 0$, where (a) follows because $\overline{\boldsymbol{\theta}}_\nu \in \Theta^{\star}_{\mathrm{k}}(\nu) \implies \Delta K(\overline{\boldsymbol{\theta}}_\nu, \nu) = 0$. Define the ball $B(\nu, \delta_m) \triangleq \{\nu' \mid W_1(\nu, \nu') < \delta_m, \nu' \in \Delta(\mathcal{B})\}$. It follows that, for any $\nu \in \Delta(\mathcal{B})$ and $\nu' \in B(\nu, \delta_m)$,

$$\underbrace{\{\overline{\boldsymbol{\theta}}' \mid d(\overline{\boldsymbol{\theta}}', \overline{\boldsymbol{\theta}}_\nu) < \delta_m\}}_{\triangleq \Theta_\nu(\delta_m)} \subseteq \underbrace{\{\overline{\boldsymbol{\theta}}' \mid \Delta K(\overline{\boldsymbol{\theta}}', \nu') \leq m\}}_{\triangleq \Theta_{\nu'}(m)}.$$

Thus, for any $\nu$ and $\nu' \in B(\nu, \delta_m)$,

$$\rho_1^{(\mathrm{k})}(\Theta_{\nu'}(m)) \geq \rho_1^{(\mathrm{k})}(\Theta_\nu(\delta_m)) \overset{(a)}{>} 0,$$

where (a) follows because $\rho_1^{(\mathrm{k})}$ has full support (Assumption 5.1).

Since $\Delta(\mathcal{B})$ is compact (Lemma 5.6), the set $\{B(\nu, \delta_m)\}_{\nu \in \Delta(\mathcal{B})}$ forms an open cover for a compact space, which means that there exists a finite subcover $\{B(\nu_i, \delta_m)\}_{i=1}^{M}$. As a conseqeuence, each $\nu' \in \Delta(\mathcal{B})$ belongs to some Wasserstein ball $B(\nu_i, \delta_m)$. Let $r \triangleq \min_i \rho_1^{(\mathrm{k})}(\Theta_{\nu_i}(\delta_m)) > 0$. We then have that

$$\rho_1^{(\mathrm{k})}(\Theta_{\nu'}(m)) \geq \rho_1^{(\mathrm{k})}(\Theta_{\nu_i}(\delta_m)) \geq r.$$

Now recall the denominator $\rho_1^{(\mathrm{k})}(\Theta^-_{\mathrm{k},\frac{\epsilon}{2}})$ in (5.31). Let $m = \frac{\epsilon}{2}$. Then $\rho_1^{(\mathrm{k})}(\Theta^-_{\mathrm{k},\frac{\epsilon}{2}}) \geq r > 0$ for any $\epsilon > 0$. Hence, $\lim_{t\to\infty} \frac{\epsilon e^{-t\frac{\epsilon}{4}}}{\rho_1^{(\mathrm{k})}(\Theta^-_{\mathrm{k},\frac{\epsilon}{2}})} = 0$. □

## E Example Derivation of a Berk-Nash Equilibrium (BNE)

We use the following example to illustrate the steps required to find a BNE.

**Example.**

Consider Prob. 5.1 with $N \triangleq 1$, $p_{\mathrm{A}} \triangleq 1$, $\mathcal{O} \triangleq \{0,1\}$, $z_{\boldsymbol{\theta}_1}(\cdot \mid 0) \triangleq \mathrm{Ber}(p)$, $z_{\boldsymbol{\theta}_1}(1 \mid 1) \triangleq \mathrm{Ber}(q)$, $\mathbf{b}_1(1) = 0$, and $c$ (5.6) being defined as in Fig. 5.3. Let the rollout parameters be ($\ell_{\mathrm{A}} = 0$, $\ell_{\mathrm{D}} = 1$) and let $\boldsymbol{\pi}_1$ be threshold strategies with $\beta \geq 0$ and $\alpha \in (0,1]$ (Thm. 5.1). Finally, let $\mathcal{L} \triangleq \{\ell_{\mathrm{A}}\}$, $\Theta_{\mathrm{A}} \triangleq \{\boldsymbol{\theta}_1\}$, and $\Theta_{\mathrm{D}} \triangleq \{\overline{\boldsymbol{\theta}}_a, \overline{\boldsymbol{\theta}}_{\mathbf{b}(1)}\}$, where $z_{\overline{\boldsymbol{\theta}}_a}(0 \mid 0) \triangleq z_{\overline{\boldsymbol{\theta}}_a}(1 \mid 1) \triangleq z_{\overline{\boldsymbol{\theta}}_b}(1 \mid 0) \triangleq z_{\overline{\boldsymbol{\theta}}_b}(0 \mid 1) \triangleq 1$.

First note that the definition of $z_{\overline{\boldsymbol{\theta}}_a}$, $z_{\overline{\boldsymbol{\theta}}_b}$, and $\mathbf{b}_1$ imply that $\mathbf{b}_t(1) \in \{0,1\}$ for all $t$, which simplifies the following derivation. To derive a BNE, we start with condition $(i)$ in Def. 5.1. Since $\ell_{\mathrm{A}} = 0$, it suffices to consider $\pi_{\mathrm{D}}$. By the principle of optimality

$$\pi_{\mathrm{D}}(\mathbf{b}(1)) \in \operatorname*{arg\,min}_{a^{(\mathrm{D})} \in \mathcal{A}_{\mathrm{D}}} \mathbb{E}_{S,\mathbf{B}'(1)} \left[ c(S, a^{(\mathrm{D})}) + \gamma \overline{J}^{(\boldsymbol{\pi}_1)}_{\mathrm{D},\overline{\boldsymbol{\theta}}}(\mathbf{B}'(1)) \mid \mathbf{b}(1), \boldsymbol{\pi}_1 \right].$$

Let $\mathbf{P}_{\overline{\boldsymbol{\theta}},\boldsymbol{\pi}_1}$ and $\mathbf{c}_{\boldsymbol{\pi}_1}$ be the belief transition matrix and the vector of expected stage costs induced by $(\overline{\boldsymbol{\theta}}, \boldsymbol{\pi}_1)$, respectively. From the definition of $\Theta_{\mathrm{D}}$ we obtain that

$$\mathbf{P}_{\overline{\boldsymbol{\theta}}_a,\boldsymbol{\pi}_1} \overset{(a)}{=} \begin{bmatrix} 1-q & q \\ 1 & 0 \end{bmatrix}, \quad \mathbf{P}_{\overline{\boldsymbol{\theta}}_b,\boldsymbol{\pi}_1} \overset{(b)}{=} \begin{bmatrix} 1-p & p \\ 1 & 0 \end{bmatrix}, \quad \mathbf{c}_{\boldsymbol{\pi}_1} \overset{(c)}{=} \begin{bmatrix} 0 \\ -1 \end{bmatrix},$$

where (a)–(c) follow because $\alpha \in (0,1] \implies \pi_{\mathrm{D}}(1) = \mathsf{S}, \pi_{\mathrm{D}}(0) = \mathsf{C}$.

By definition, $\overline{J}^{(\boldsymbol{\pi}_1)}_{\mathrm{D},\overline{\boldsymbol{\theta}}} = (\mathbf{1}_2 - \gamma \mathbf{P}_{\overline{\boldsymbol{\theta}},\boldsymbol{\pi}_1})^{-1} \mathbf{c}_{\boldsymbol{\pi}_1}$, where $\mathbf{1}_2$ is the $2 \times 2$ identity matrix. Therefore,

$$\overline{J}^{(\boldsymbol{\pi}_1)}_{\mathrm{D},\overline{\boldsymbol{\theta}}_a} = (\mathbf{1}_2 - \gamma \mathbf{P}_{\overline{\boldsymbol{\theta}}_a,\boldsymbol{\pi}_1})^{-1} \mathbf{c}_{\boldsymbol{\pi}_1} = \left( \mathbf{1}_2 - \gamma \begin{bmatrix} 1-q & q \\ 1 & 0 \end{bmatrix} \right)^{-1} \begin{bmatrix} 0 \\ -1 \end{bmatrix}$$

$$= \begin{bmatrix} 1-\gamma+\gamma q & -\gamma q \\ -\gamma & 1 \end{bmatrix}^{-1} \begin{bmatrix} 0 \\ -1 \end{bmatrix} = \begin{bmatrix} \frac{-1}{(\gamma-1)(1+\gamma q)} & \frac{-\gamma q}{(\gamma-1)(1+\gamma q)} \\ \frac{-\gamma}{(\gamma-1)(1+\gamma q)} & \frac{\gamma-1-\gamma q}{(\gamma-1)(1+\gamma q)} \end{bmatrix} \begin{bmatrix} 0 \\ -1 \end{bmatrix}$$
$$= \frac{1}{(\gamma-1)(1+\gamma q)} \begin{bmatrix} \gamma q \\ 1+\gamma(q-1) \end{bmatrix}.$$

Similarly,

$$\overline{J}^{(\boldsymbol{\pi}_1)}_{\mathrm{D},\overline{\boldsymbol{\theta}}_b} = (\mathbf{1}_2 - \gamma \mathbf{P}_{\overline{\boldsymbol{\theta}}_b,\boldsymbol{\pi}_1})^{-1}\mathbf{c}_{\boldsymbol{\pi}_1} = \left(\mathbf{1}_2 - \gamma \begin{bmatrix} 1-p & p \\ 1 & 0 \end{bmatrix}\right)^{-1} \begin{bmatrix} 0 \\ -1 \end{bmatrix}$$
$$= \frac{1}{(\gamma-1)(1+\gamma p)} \begin{bmatrix} \gamma p \\ 1+\gamma(p-1) \end{bmatrix}.$$

Hence, to meet condition $(i)$, the defender's rollout strategy must satisfy $\pi_{\mathrm{D}}(0) = \mathsf{C}$ and $\pi_{\mathrm{D}}(1) = \mathsf{S}$, which is ensured by (5.12). As a consequence, $\boldsymbol{\pi} = \boldsymbol{\pi}_1$ in any BNE.

Now consider condition $(ii)$; (5.15) can be written as

$$K(\overline{\boldsymbol{\theta}}, \nu) = \mathbb{E}_{\mathbf{b}\sim\nu}\mathbb{E}_{\mathbf{I}^{(\mathrm{D})}}\left[\ln\left(\frac{\mathbb{P}[\mathbf{I}^{(\mathrm{D})} \mid \boldsymbol{\theta}_1, \mathbf{b}(1)]}{\mathbb{P}[\mathbf{I}^{(\mathrm{D})} \mid \overline{\boldsymbol{\theta}}, \mathbf{b}(1)]}\right) \mid \boldsymbol{\theta}_1, \mathbf{b}(1)\right]$$
$$= \sum_{\mathbf{b}(1)\in\{0,1\}} \nu(\mathbf{b}(1)) \sum_{o\in\{0,1\}} z_{\boldsymbol{\theta}_1}(o \mid \mathbf{b}(1)) \ln\left(\frac{z_{\boldsymbol{\theta}_1}(o \mid \mathbf{b}(1))}{z_{\overline{\boldsymbol{\theta}}}(o \mid \mathbf{b}(1))}\right)$$
$$= -\sum_{\mathbf{b}(1)\in\{0,1\}} \nu(\mathbf{b}(1)) \sum_{o\in\{0,1\}} z_{\boldsymbol{\theta}_1}(o \mid \mathbf{b}(1)) \ln z_{\overline{\boldsymbol{\theta}}}(o \mid \mathbf{b}(1)) + \text{const.}$$

Minimizing the above expression with respect to $\overline{\boldsymbol{\theta}}$ yields $\Theta^{\star}_{\mathrm{D}} = \{\overline{\boldsymbol{\theta}}_a\}$ if $(p = 0, q = 1)$. Conversely, $\Theta^{\star}_{\mathrm{D}} = \{\overline{\boldsymbol{\theta}}_b\}$ if $(p = 1, q = 0)$. Otherwise, $\Theta^{\star}_{\mathrm{D}}(\nu) = \{\overline{\boldsymbol{\theta}}_a, \overline{\boldsymbol{\theta}}_b\}$.

Lastly, condition $(iii)$ is satisfied iff $\mathbf{P}^T_{\overline{\boldsymbol{\theta}},\boldsymbol{\pi}_1}\nu = \nu$. Since $\mathbf{P}^T_{\overline{\boldsymbol{\theta}},\boldsymbol{\pi}_1} = \rho^{(\mathrm{D})}(\overline{\boldsymbol{\theta}}_a)\mathbf{P}^T_{\overline{\boldsymbol{\theta}}_a,\boldsymbol{\pi}_1} + (1-\rho^{(\mathrm{D})}(\overline{\boldsymbol{\theta}}_a))\mathbf{P}^T_{\overline{\boldsymbol{\theta}}_b,\boldsymbol{\pi}_1}$, solving this equation gives

$$\nu(0) = -\left(-1 - p + \rho^{(\mathrm{D})}(\overline{\boldsymbol{\theta}}_a)p - \rho^{(\mathrm{D})}(\overline{\boldsymbol{\theta}}_a)q\right)^{-1}, \tag{5.32}$$

which means the BNE is not unique and may not exist; see Fig. 5.18 on the next page. For example, if $p = 1$ and $q = 0$, then (5.32) requires that $\rho^{(\mathrm{D})}(\overline{\boldsymbol{\theta}}_a) = 1$, but this means that $\rho^{(\mathrm{D})} \notin \Delta(\Theta^{\star}(\nu)$, which violates condition $(ii)$.

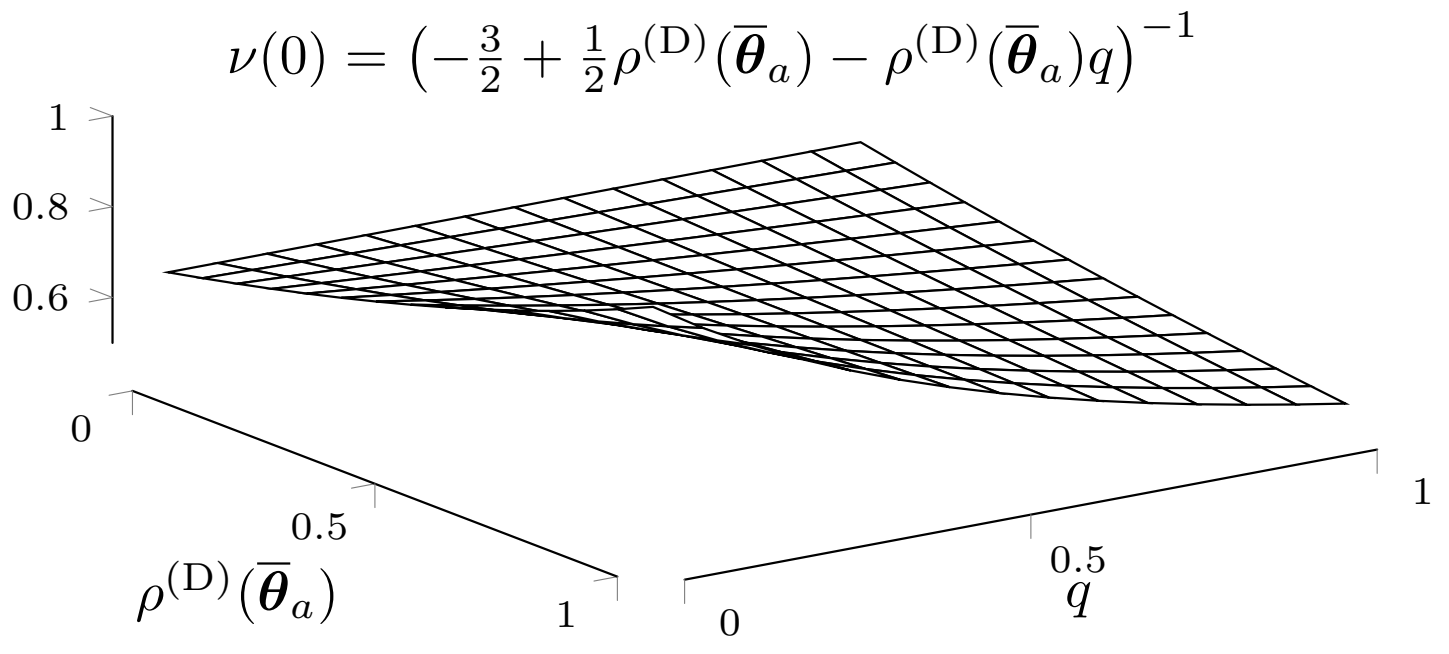

***Figure 5.18:*** *Berk-Nash equilibria of the example instantiation of* $\Gamma$ *when* $p = \frac{1}{2}$.

## F Hyperparameters

The hyperparameters used for the evaluation in this paper are listed in Table 5.5 on the next page and were obtained through grid search.

| *Figures and Tables* | *Values* |
|---|---|
| Figure 5.6 | $\mathcal{O} = \{0, \dots, 9\}$, $p_A = 0.1, \gamma = 0.99$ |
| | $z(\cdot \mid 0) = \text{BetaBin}(n = 10, \alpha = 0.7, \beta = 3)$ |
| | $z(\cdot \mid 1) = \text{BetaBin}(n = 10, \alpha = 1, \beta = 0.7$ |
| Figure 5.5 | $\mathcal{O} = \{0, \dots, 9\}$, $p_A = 0.1$, $N = 1$, $\gamma = 0.99$ |
| | $z(\cdot \mid 0) = \text{BetaBin}(n = 10, \alpha = 0.7, \beta = 3)$ |
| | $z(\cdot \mid 1) = \text{BetaBin}(n = 10, \alpha = 1, \beta = 0.7$ |
| Figure 5.8 | $N = 10$, $p_A = 0.1, \gamma = 0.99$ |
| | $\pi_{D,1}(\mathsf{S} \mid \mathbf{b}_t) = 1 \iff \mathbb{P}[S_t \geq 1 \mid \mathbf{b}_t] \geq 0.75$ |
| | $\pi_{A,1}(\mathsf{S} \mid \mathbf{b}_t, s_t) = 0.5$ |
| | Cost function of base strategy estimated |
| | using 100 MC samples w. horizon 50 |
| Figs. 5.12.a–e | $\ell_A = \ell_D = 1, \mathcal{L} = \{1, 2\}, p_A = 1$ |
| | $\pi_{D,1}(\mathsf{S} \mid \mathbf{b}_t) = 1 \iff \mathbb{P}[S_t \geq 1 \mid \mathbf{b}_t] \geq 0.75$ |
| | $\pi_{A,1}(\mathsf{S} \mid \mathbf{b}_t, s_t) = 0.05$ |
| | $N = 10$, $\mathcal{O}, z$ (Fig. 5.11) |
| | using 100 MC samples w. horizon 50 |
| Figure 5.4, Fig. 5.12.f | $\mathcal{O}, z$ (Fig. 5.11), $p_A = 0.1, \gamma = 0.99$ |
| | $\mathcal{L} = 0, 1, 2$, $N = 10$ |
| | $\pi_{D,1}(\mathsf{S} \mid \mathbf{b}_t) = 1 \iff \mathbb{P}[S_t \geq 1 \mid \mathbf{b}_t] \geq 0.75$ |
| | $\pi_{A,1}(\mathsf{S} \mid \mathbf{b}_t, s_t) = 0.05$ |
| | Best response computation: CEM [380] |
| | Best response parameterized following Thm. 5.1 |
| Figs. 5.13–5.16, 5.15 | $\ell_A = \ell_D = 1, p_A = 1, \gamma = 0.99$ |
| | $\pi_{D,1}(\mathsf{S} \mid \mathbf{b}_t) = 1 \iff \mathbb{P}[S_t \geq 1 \mid \mathbf{b}_t] \geq 0.75$ |
| | $\pi_{A,1}(\mathsf{S} \mid \mathbf{b}_t, s_t) = 0.05$ |
| | $N = 10$, $\mathcal{O}, z$ (Fig. 5.11) |
| | using 100 MC samples w. horizon 50 |
| Figure 5.13.e | $\Theta_D = \{0, \dots, 200\}$ |
| Table 5.3 | $\ell_A = \ell_D = 1, p_A = 1, \gamma = 0.99$ |
| | $\pi_{D,1}(\mathsf{S} \mid \mathbf{b}_t) = 1 \iff \mathbb{P}[S_t \geq 1 \mid \mathbf{b}_t] \geq 0.75$ |
| | $\pi_{A,1}(\mathsf{S} \mid \mathbf{b}_t, s_t) = 0.05$ |
| | $N = 10$, $\mathcal{O}, z$ (Fig. 5.11, $t = 10$) |
| | using 100 MC samples w. horizon 50 |
| Figure 5.10 | $\boldsymbol{\psi} = (\frac{1}{2}, 10^{-2}, -10^{5})$, $\boldsymbol{\chi} = (1.0593)$ |
| | $\boldsymbol{\phi} = (-0.5193)$, $\boldsymbol{\omega} = (0.054\pi)$ |
| | time step: 30s, service time: $\text{Exp}(\mu = 4)t$ |
| Figure 5.17 | $\mathcal{O}$ (Fig. 5.11), $\pi_{A,1}(\mathsf{S} \mid \cdot) = 1$, $\ell_D = 1$ |
| Confidence intervals | computed using the Student-t distribution |
| Base strategies $\pi_{D,1}\pi_{A,1}$ | Approximate threshold best responses |
| | against randomized opponents (Fig. 5.4) |
| Priors $\mu_1, \rho_1^{(D)}, \rho_1^{(A)}$ | uniform |
| Cost function (5.6), Fig. 5.3 | $p = 5/4$, $q = 1$, $r = 2$ |
| *HSVI Parameter* | |
| $\epsilon$ | 0.1 |
| *Cross-entropy method [380]* | |
| $\lambda$ (fraction of samples to keep) | 0.15, 100 |
| $K$ population size | 100 |
| $M$ number of samples for each evaluation | 50 |
| *PPO [396, Alg. 1] parameters* | |
| lr $\alpha$, batch, # layers, # neurons, clip $\epsilon$ | $10^{-5}$, $4 \cdot 10^{3}t$, 4, 64, 0.2, |
| GAE $\lambda$, ent-coef, activation | 0.95, $10^{-4}$, ReLU |
| *NFSP [191, Alg. 9] parameters* | |
| lr RL, lr SL, batch, # layers,# neurons, $\mathcal{M}_{RL}$ | $10^{-2}$, $5 \cdot 10^{-3}$, 64, 2,128, $2 \times 10^{5}$ |
| $\mathcal{M}_{SL}$,$\epsilon$, $\epsilon$-decay, $\eta$ | $2 \times 10^{6}$, 0.06, 0.001, 0.1 |

***Table 5.5:*** *Hyperparameters.*

# Paper 6[†]

# OPTIMAL DEFENDER STRATEGIES FOR CAGE-2 USING CAUSAL MODELING AND TREE SEARCH

Kim Hammar, Neil Dhir, and Rolf Stadler

**Abstract**

The CAGE-2 challenge is considered a standard benchmark to compare methods for automated security response. Current state-of-the-art methods evaluated against this benchmark are based on model-free (offline) reinforcement learning, which does not provide provably optimal defender strategies. We address this limitation and present a formal (causal) model of CAGE-2 together with a method that produces a provably optimal defender strategy, which we call **C**ausal-**P**artially **O**bservable **M**onte-**C**arlo **P**lanning (C-POMCP). It has two key properties. First, it incorporates the causal structure of the target system, i.e., the causal relationships among the system variables. This structure allows for a significant reduction of the search space of defender strategies. Second, it is an online method that uses tree search to update the defender strategy at each time step. Evaluations against the CAGE-2 benchmark show that C-POMCP achieves state-of-the-art performance with respect to effectiveness and is two orders of magnitude more efficient in computing time than the closest competitor method.

*All reasonings concerning matter of fact seem to be founded on the relation of cause and effect.*

— David Hume ***1748**, An enquiry concerning human understanding.*

## 6.1 Introduction

A Driving factor behind the research on automated security response is the development of evaluation benchmarks, which allow researchers to compare the performance of different methods. One such benchmark is csle, which is described in the methodology chapter and provides the basis for the experimental evaluations presented in Papers 1–5. Another popular benchmark is the **C**yber **A**utonomy **G**ym for **E**xperimentation **2** (cage-2) (cage-2, 2022), which involves defending a networked system against a simulated Advanced Persistent Threat (apt)[2]. At the time of writing, more than 30 methods have been evaluated against cage-2 (cage-2, 2022). Detailed descriptions of some methods can be found in [481, 492, 42, 140, 141, 190, 452, 241, 368, 357, 122, 22, 490, 495, 498, 93, 289]. While good results have been obtained, key aspects remain unexplored. For example, current methods are narrowly focused on *offline* reinforcement learning and require a lengthy training phase to obtain effective strategies. Further, these methods are *model-free* and do not provide provably optimal strategies. In addition, present methods provide limited ways to include domain expertise in the learning process, though attempts have been made with reward shaping [42].

In this paper, we address the above limitations and use the cage-2 scenario to illustrate our solution method. First, we develop a formal (causal) model of cage-2, which allows us to define and prove the existence of an optimal defender strategy. This model is based on the source code of cage-2 and is formalized as a Structural Causal Model (scm) (Def 7.1.1, Pearl, 2009). We prove that this scm is equivalent to a specific Partially Observed Markov Decision Process (pomdp)[3] (P.1, Åström, 1965). Compared to the pomdp, our scm offers a more expressive representation of the underlying causal structure, allowing us to understand the causal effects of defender strategies (Def. 3.2.1, Pearl, 2009).

Second, we design an online method that produces a *provably optimal* defender strategy, which we call ***C**ausal-**P**artially **O**bservable **M**onte-**C**arlo **P**lanning (c-pomcp)*. The method has two key properties: (1) it incorporates causal information of the target system in the form of a *causal graph* (Def. 2.2.1, Pearl, 2009), which allows us to prune the search space of defender strategies; and (2) it is an *online* method that updates the defender strategy at each time step via *tree search*.

Distribution Statement A (Approved for Public Release, Distribution Unlimited). This research is supported by the Defense Advanced Research Project Agency (darpa) through the castle program under Contract No. W912CG23C0029. The views, opinions, and/or findings expressed are those of the authors and should not be interpreted as representing the official views or policies of the Department of Defense or the U.S. Government.

[2]Unlike csle, which is based on virtual it infrastructures, cage-2 is a simulation benchmark.

[3]The components of a pomdp are defined the background chapter; see (15).

Our causal model represents one of many ways of formally modeling cage-2. A key question is the level of abstraction at which cage-2 is modeled. The more detailed we construct the model, the closer it can capture the cage-2 implementation. However, this comes at the expense of higher computational complexity and lower generalization ability. When balancing this trade-off, we follow the principle that a model should be detailed enough so that a theoretically optimal defender strategy exhibits state-of-the-art performance in a practical implementation (cage-2, 2022).

We evaluate c-pomcp against the cage-2 benchmark and show that it achieves state-of-the-art effectiveness while being two orders of magnitude more computationally efficient than the closest competitor method: cardiff-ppo (Vyas et al., 2023)[4]. The evaluation results also show that c-pomcp performs significantly better than its non-causal version: pomcp (Alg. 1, Silver and Veness, 2010). While prior work has focused on offline methods that require hours of training, c-pomcp produces equally effective defender strategies through 15 seconds of online search.

Our contributions can be summarized as follows:

- We present a causal model of the cage-2 scenario (M1). This model allows us to define and prove the existence of optimal defender strategies (Thm. 6.1).
- We design c-pomcp, an online method that leverages the causal structure of the target system to efficiently find an optimal defender strategy (Alg. 6.1). c-pomcp includes a novel approach to leverage causal information for tree search, which may be of independent interest.
- We prove that c-pomcp converges to an optimal strategy with increasing search time (Thm. 6.4).
- We evaluate c-pomcp against the cage-2 benchmark. The results show that c-pomcp outperforms the state-of-the-art methods in effectiveness and performs significantly better in computational efficiency (Vyas et al., 2023).

## 6.2 Related Work

To our knowledge, no prior work has provided a formal model of cage-2, nor considered tree search for finding effective defender strategies. Moreover, the only prior works that use causal inference for automated security response are (Andrew et al., 2022), (Highnam et al., 2023), (Shi et al., 2018), (Mueller et al., 2019), and (Maiti et al., 2023). This paper differs from them in two ways. First, the studies presented in (Highnam et al., 2023), (Shi et al., 2018), (Mueller et al., 2019), and (Maiti et al., 2023) use causality for analyzing the effects of attacks and countermeasures but do not present methods for finding defender strategies. Second, the method for finding defender strategies in (Andrew et al., 2022) uses Bayesian optimization and is *myopic*, i.e., it does not consider the future when selecting strategies. While this approach simplifies computations, the method is sub-optimal for most practical scenarios. By contrast, our method is non-myopic and produces optimal strategies (Thm. 6.4).

[4]See Appendix D of Paper 3 for a derivation of the ppo algorithm.

## 6.3 Causal Inference Preliminaries

This section covers notation and provides an overview of causal inference, which lays the foundation for the subsequent section, where we deduce a causal model of the CAGE-2 scenario.

### Structural causal models

A Structural Causal Model (SCM) is defined as

$$\mathscr{M} \triangleq \langle \mathbf{U}, \mathbf{V}, \mathbf{F}, \mathbb{P}[\mathbf{U}] \rangle, \qquad \text{(Def 7.1.1, Pearl, 2009)} \qquad (6.1)$$

where $\mathbf{U}$ is a set of *exogenous* random variables and $\mathbf{V}$ is a set of *endogenous* random variables[5]. Within $\mathbf{V} \cup \mathbf{U}$ we distinguish between five subsets that may overlap: the set of *manipulative* variables $\mathbf{X}$; *non-manipulative* $\mathbf{N}$; *observed* $\mathbf{O}$; *latent* $\mathbf{L}$; and *targets* $\mathbf{Y}$[6]. An SCM induces a *causal graph* $\mathcal{G}$ (Def. 2.2.1, Pearl, 2009), where nodes correspond to $\mathbf{V} \cup \mathbf{U}$ and edges represent (causal) functions $\mathbf{F} \triangleq \{f_i\}_{V_i \in \mathbf{V}}$. A function $f_i$ is a mapping from the ranges of a subset $\mathbf{K} \subseteq (\mathbf{U} \cup \text{pa}(V_i)_{\mathcal{G}})$ to the range of $V_i$, which is represented graphically by directed edges from the nodes in $\mathbf{K}$ to $V_i$; see Fig. 6.1. If each function is independent of time, the SCM is *stationary*.

Causal graphs with latent variables can be drawn in two ways; cf. Fig. 6.1.a and Fig. 6.1.b. One option is to include the latent variables in the graph (Fig. 6.1.a). Another option is to represent the latent variables with bidirected edges, where a bidirected edge between two observed variables means that they share an *unobserved confounder* (Def. 6.2.1, Pearl, 2009), i.e., a latent variable that influences both of them (Fig. 6.1.b).

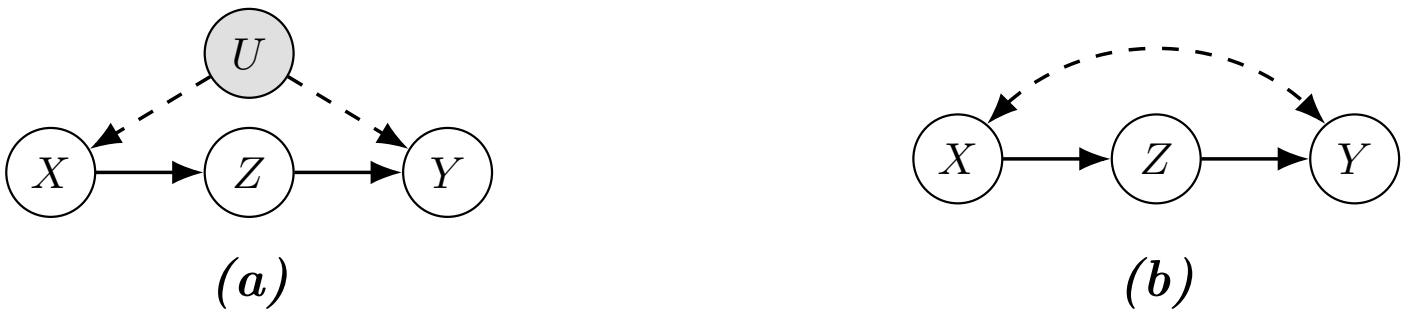

***Figure 6.1:*** *Causal graphs (Def. 2.2.1, Pearl, 2009); circles represent variables in an SCM (6.1); solid arrows represent causal relations, and dashed edges represent effects caused by latent variables; latent variables can either be represented with shaded circles or with bidirected dashed edges, i.e., the graphs in (a) and (b) represent the same causal structure.*

[5] With some abuse of notation, in this paper we use bold upper case letters to denote both sets of random variables and random vectors.

[6] SCMs with latent variables are called "partially observable" (Def. 1, Zhang et al., 2020).

We say that $\mathbb{P}[\mathbf{V}]$ is Markov relative to $\mathcal{G}$ if it admits the following factorization

$$\mathbb{P}[\mathbf{V}] = \prod_{i=1}^{|\mathbf{V}|} \mathbb{P}[\mathbf{V}_i \mid \text{pa}(\mathbf{V}_i)_{\mathcal{G}}]. \qquad \text{(Def. 1.2.2, Thm. 1.2.7, Pearl, 2009)} \tag{6.2}$$

Similarly, we say that an SCM is Markov if it induces a distribution over the observables $\mathbf{O}$ that satisfies (6.2) (Thm. 1.4.1, Pearl, 2009). If the SCM is not Markov and $\mathcal{G}$ is acyclic, we say that it is *semi*-Markov (Ch. 3, Pearl, 2009).

### Interventions

The operator $\text{do}(\mathbf{X} = \mathbf{x})$ represents an *atomic intervention* that fixes a set of endogenous variable(s) $\mathbf{X}$ to constant value(s) $\mathbf{x}$ irrespective of the functions $\mathbf{F}$ (Def. 3.2.1, Pearl, 2009). Similarly, $\text{do}(\mathbf{X} = \pi(\mathbf{O}))$ represents a *conditional intervention*, whereby the function(s) $\{f_i\}_{i \in \mathbf{X}}$ are replaced with a deterministic function $\pi$ of the observables. We call such a function an *intervention strategy*.

Interventions can be represented graphically by removing the incoming edges to the intervention set $\mathbf{X}$ (Def. 3.2.1, Pearl, 2009). We denote the resulting *mutilated graph* by $\mathcal{G}_{\overline{\mathbf{X}}}$. Examples of mutilated graphs are shown in Fig. 6.2.

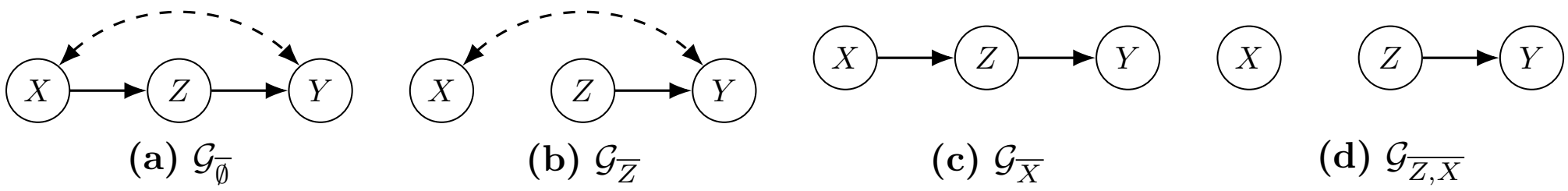

***Figure 6.2:*** *Mutilated causal graphs that represent post-intervention worlds where a specific intervention has been implemented, from left to right, these are* $\text{do}(\emptyset), \text{do}(Z), \text{do}(X)$ *and* $\text{do}(X, Z)$*; interventions are graphically represented with the incoming edges onto the intervened variables removed.*

### Causal inference

The standard way to estimate causal effects of interventions (Def. 3.2.1, Pearl, 2009) is through controlled experiments (Fisher, 1935). In practice, however, experimentation can be costly and is often not feasible in operational systems. This leads to the fundamental question of whether causal effects can be estimated only from observations. Such estimation can be performed using Pearl's *do-calculus*, which is an axiomatic system for replacing expressions containing the do-operator with conditional probabilities (Thm. 3.4.1, Pearl, 2009). It consists of three rules:

$$\mathbb{P}\left[Y \mid \text{do}(\mathbf{X} = \mathbf{x}), \mathbf{Z}, \mathbf{W}\right] = \mathbb{P}\left[Y \mid \text{do}(\mathbf{X} = \mathbf{x}), \mathbf{W}\right] \text{ if } (Y \perp\!\!\!\perp \mathbf{Z} \mid \mathbf{X}, \mathbf{W})_{\mathcal{G}_{\overline{\mathbf{X}}}} \tag{Rule 1}$$

$$\mathbb{P}\left[Y \mid \text{do}(\mathbf{X} = \mathbf{x}), \text{do}(\mathbf{Z} = \mathbf{z}), \mathbf{W}\right] = \mathbb{P}\left[Y \mid \text{do}(\mathbf{X} = \mathbf{x}), \mathbf{Z}, \mathbf{W}\right] \text{ if } (Y \perp\!\!\!\perp \mathbf{Z} \mid \mathbf{X}, \mathbf{W})_{\mathcal{G}_{\overline{\mathbf{X}}, \underline{\mathbf{Z}}}} \tag{Rule 2}$$

$$\mathbb{P}\left[Y \mid \mathrm{do}(\mathbf{X}=\mathbf{x}), \mathrm{do}(\mathbf{Z}=\mathbf{z}), \mathbf{W}\right] = \mathbb{P}\left[Y \mid \mathrm{do}(\mathbf{X}=\mathbf{x}), \mathbf{W}\right] \text{ if } (Y \perp\!\!\!\perp \mathbf{Z} \mid \mathbf{X}, \mathbf{W})_{\mathcal{G}_{\overline{X}, \overline{\mathbf{Z}(\mathbf{W})}}}, \tag{Rule 3}$$

where $\mathcal{G}_{\underline{\mathbf{Z}}}$ refers to the graph obtained by removing the outgoing edges from $\mathbf{Z}$; $(X \perp\!\!\!\perp Y \mid \mathbf{Z})_{\mathcal{G}}$ means that $X$ and $Y$ are conditionally independent given $\mathbf{Z}$ and $\mathcal{G}$; and $\mathbf{Z}(\mathbf{W})$ is the set of nodes in $\mathbf{Z}$ that are not ancestors of any node in $\mathbf{W}$.

### Causal effect identifiability

In case an SCM includes latent variables (Def. 2.3.2, Pearl, 2009), the question of *identifiability* arises:

**Definition 6.1** (Causal effect identifiability (Def. 3.2.4, Pearl, 2009))**.** *The causal effect (Def. 3.2.1, Pearl, 2009) of* $\mathrm{do}(\mathbf{X}=\pi(\mathbf{O}))$ *on* $Y$ *is identifiable from* $\mathcal{G}$ *if* $\mathbb{P}[Y \mid \mathrm{do}(\mathbf{X}=\pi(\mathbf{O})), \mathbf{O}]$ *is* uniquely *computable from* $\mathbb{P}[\mathbf{O}] > 0$ *in every* SCM *conforming to* $\mathcal{G}$*.*

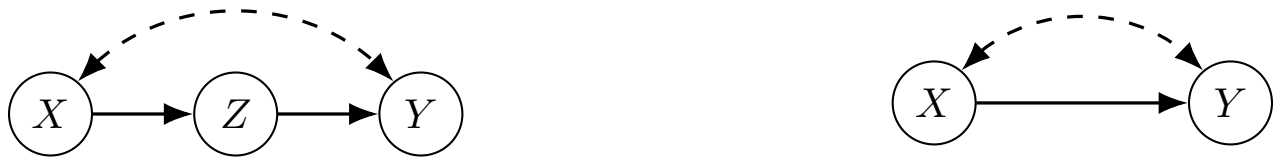

***(a)*** $\mathbb{P}[Y \mid \mathrm{do}(X=x)]$ *identifiable.* ***(b)*** $\mathbb{P}[Y \mid \mathrm{do}(X=x)]$ *unidentifiable.*

***Figure 6.3:*** *Determining causal effect identifiability (Def. 6.1) from causal graphs.*

Do-calculus is *complete* in that it allows us to derive all identifiable causal effects (Cor. 3.4.2, Pearl, 2009)(Thm. 23, Shpitser and Pearl, 2008). Consequently, one can prove identifiability by providing a do-calculus derivation that reduces the causal effect to an expression involving only $\mathbb{P}[\mathbf{O}]$. For example, consider the causal graph in Fig. 6.3.a and assume that $X, Y, Z$ are discrete random variables. The identifiability of $\mathbb{P}[Y \mid \mathrm{do}(X=x)]$ can be proven as follows.

$$\begin{aligned}
\mathbb{P}[Y \mid \mathrm{do}(X=x)] &\overset{(a)}{=} \sum_z \mathbb{P}[Y, Z=z \mid \mathrm{do}(X=x)] \\
&\overset{(b)}{=} \sum_z \mathbb{P}[Y \mid Z=z, \mathrm{do}(X=x)]\mathbb{P}[Z=z \mid \mathrm{do}(X=x)] \\
&\overset{(c)}{=} \sum_z \mathbb{P}[Y \mid Z=z, \mathrm{do}(X=x)]\mathbb{P}[Z=z \mid X=x] \\
&\overset{(d)}{=} \sum_z \mathbb{P}[Y \mid Z=z]\mathbb{P}[Z=z \mid X=x] \\
&\overset{(e)}{=} \sum_z \mathbb{P}[Z=z \mid X=x] \sum_{x'} \mathbb{P}[Y, X=x' \mid Z=z] \\
&\overset{(f)}{=} \sum_z \mathbb{P}[Z=z \mid X=x] \sum_{x'} \mathbb{P}[Y \mid X=x', Z=z]\mathbb{P}[X=x'],
\end{aligned}$$

where (a) follows from law of total probability; (b) uses C-component decomposition[7] (Tian and Pearl, 2002); (c) uses Rule 2 and the fact that $(X \perp\!\!\!\perp Z)_{\mathcal{G}_{\underline{X}}}$; (d) uses Rule 3 and the fact that $(X \perp\!\!\!\perp Y \mid Z)_{\mathcal{G}_{\overline{X,Z}}}$; (e) follows from the law of total probability; and (f) uses the chain rule of probability.

**Automatic intervention control**

The problem of finding a sequence of conditional interventions $\mathrm{do}(\mathbf{X}_1 = \pi(\mathbf{O})), \ldots, \mathrm{do}(\mathbf{X}_\mathcal{T} = \pi(\mathbf{O}))$ that maximizes a target variable $J$ can be formulated as a *feedback control problem* (Bertsekas, 2005), also known as a *dynamic treatment regime problem* (Murphy, 2003). We say that such a problem is identifiable if the effect on $J$ caused by every intervention strategy $\pi$ is identifiable:

**Definition 6.2** (Control problem identifiability (Ch. 4.4, Pearl, 2009))**.** *A control problem with target $J$ and time horizon $\mathcal{T}$ is identifiable from $\mathcal{G}$ if*

$$\mathbb{P}[J \mid \mathrm{do}(\mathbf{X}_1 = \pi(\mathbf{O})), \ldots, \mathrm{do}(\mathbf{X}_\mathcal{T} = \pi(\mathbf{O}))]$$

*is identifiable for each intervention strategy $\pi$ (Def. 6.1).*

Given a control problem and a causal graph, we can derive *possibly optimal minimal intervention sets* (pomiss):

**Definition 6.3** (pomis, adapted from (Def. 3, Lee and Bareinboim, 2019))**.** *Given a control problem with target $J$ and a causal graph $\mathcal{G}$, $\tilde{\mathbf{X}} \subseteq \mathbf{X}$ is a pomis if, for each scm conforming to $\mathcal{G}$, there is no $\mathbf{X}' \subset \tilde{\mathbf{X}}$ such that $\mathbb{E}_\pi[J \mid \mathrm{do}(\tilde{\mathbf{X}} = \tilde{\mathbf{x}})] = \mathbb{E}_\pi[J \mid \mathrm{do}(\mathbf{X}' = \mathbf{x}')]$ and there exists an scm such that*

$$\mathbb{E}_{\pi^\star}[J \mid \mathrm{do}(\tilde{\mathbf{X}} = \tilde{\mathbf{x}})] \geq \mathbb{E}_{\pi^\star}[J \mid \mathrm{do}(\mathbf{X}' = \mathbf{x}')] \qquad \textit{for all } \mathbf{X}' \textit{ and } \mathbf{x}', \tag{6.3}$$

*where $\pi^\star$ is an optimal intervention strategy satisfying $\mathbb{E}_{\pi^\star}[J] \geq \mathbb{E}_\pi[J]\ \forall \pi$.*

Let $\mathbf{P}^\star_\mathcal{G}$ denote the set of pomiss for a causal graph $\mathcal{G}$. $\mathbf{P}^\star_\mathcal{G}$ for two example graphs are shown in Fig. 6.4. As can be seen in Fig. 6.4.a, when $\mathcal{G}$ is Markovian, and all variables except the target are manipulative, the only pomis is the set of parents of the target (Prop. 2, Lee and Bareinboim, 2019). When there are unobserved confounders (Def. 6.2.1, Pearl, 2009), however, $\mathbf{P}^\star_\mathcal{G}$ generally includes many more sets, as shown in Fig. 6.4.b. An algorithm for computing $\mathbf{P}^\star_\mathcal{G}$ can be found in (Alg. 1, Lee and Bareinboim, 2019).

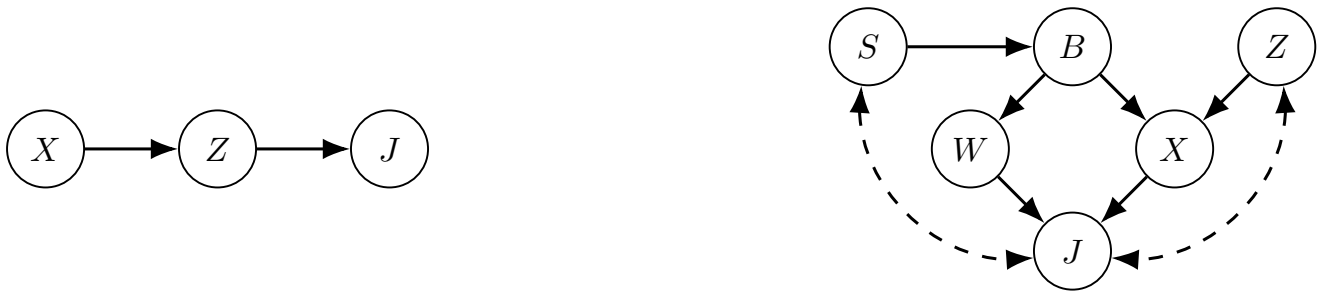

***(a)*** $\mathbf{P}^\star_\mathcal{G} = \{\{Z\}\}$. ***(b)*** $\mathbf{P}^\star_\mathcal{G} = \{\emptyset, \{X\}, \{W\}, \{Z\}, \{B, W\}, \{X, W\}, \{Z, W\}\}$.

***Figure 6.4:*** *Two causal graphs and the corresponding sets of pomiss; $J$ is the target variable, and all other variables are manipulative.*

[7] Since there is no bidirected edge between $Z$ and $Y$ in $\mathcal{G}_{\overline{X}}$, they belong to different C-components, which means that their joint probability factorizes.

## 6.4 The CAGE-2 Scenario

The CAGE-2 scenario involves defending a networked system against APTs (CAGE-2, 2022). The operator of the system, which we call the defender, takes measures to protect it against an attacker while providing services to a client population; see Fig. 6.5. The system is segmented into *zones* with *nodes* (servers and workstations) that run network services. Services are realized by *workflows* that clients access through a gateway, which is also open to the attacker. The detailed system configuration can be found in Appendix D.

The attacker aims to intrude on the system and disrupt service for clients. To achieve this goal, it can take five actions: scan the network to discover nodes; exploit a vulnerability to compromise a node; perform a brute force attack to obtain login credentials to a node; escalate privileges on a compromised node to gain root access; and disrupt the service on a compromised node. When selecting between these actions, the attacker follows a fixed strategy, i.e., the attacker is *static* and does not adapt its strategy to encountered defenses.

The defender monitors the system through log files and network statistics. It can make four types of *interventions* on a node to prevent, detect, and respond to attacks: analyze the node for a possible intrusion; start a decoy service on the node; remove malware from the node; and restore the node to a secure state, which temporarily disrupts its service. When deciding between these interventions, the defender balances two conflicting objectives: maximize service utility towards its clients and minimize the cost of attacks.

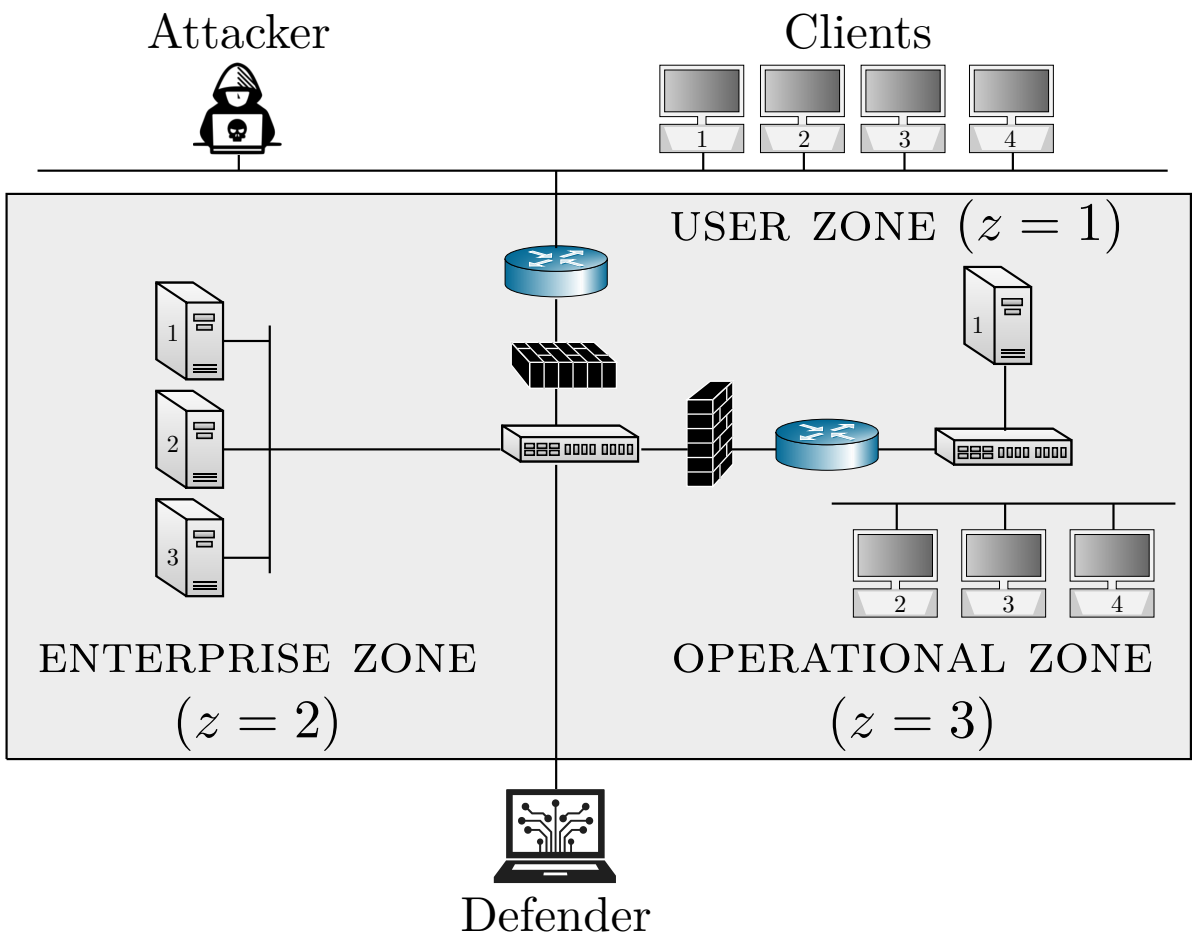

***Figure 6.5:*** *The CAGE-2 scenario (CAGE-2, 2022): a defender aims to protect a networked system against an Advanced Persistent Threat (APT) caused by an attacker while maintaining services for clients; the system configuration is listed in Appendix D.*

## 6.5 Causal Model of the cage-2 Scenario

We model the cage-2 scenario by constructing an scm and formulate the benchmark problem as the problem of finding an optimal intervention strategy for the defender. (A game-theoretic model is not needed since the attacker in cage-2 is static; i.e., it does not adapt its strategy to encountered defenses.) The requisite notation is listed in Table 6.1, and the scm components are defined below.

| *Notation(s)* | *Description* |
|---|---|
| $\mathcal{G}_S$, $\mathcal{G}_W$ | System graph and workflow graph (M1). |
| $\mathcal{V}$, $\mathcal{E}$, $\mathcal{Z}$ | Set of nodes and edges in $\mathcal{G}_S$, set of zones (M1). |
| $\mathbf{D}_t, \mathbf{I}_t, \mathbf{S}_t$ | Decoy states, intrusion states, and service states (M1). |
| $\mathsf{U}, \mathsf{K}, \mathsf{S}$ | The unknown, known, and scanned intrusion states (Fig. 6.6). |
| $\mathsf{C}, \mathsf{R}$ | The compromised and root intrusion states (M1). |
| $\mathbf{A}_t, \alpha_t$ | Attacker action and attacker action type (M1). |
| $V_t, P_t, T_t$ | Vulnerability, privileges, and target of attacker action (M1). |
| $\mathtt{S}, \mathtt{E}, \mathtt{P}$ | Scan, exploit, and privilege escalation attacker actions (M1). |
| $\mathtt{I}, \mathtt{D}$ | Impact and discover attacker actions (M1). |
| $f_I$, $f_S, f_C$ | Causal functions for $I_{i,t}$ (6.5), $S_{i,t}$ (6.6), and $C_t$ (6.8). |
| $Z_{i,t}$, $\mathbf{Z}_t$ | Observation of node $i$ and observations for all nodes (M1). |
| $f_{Z,i}$, $W_{i,t}$, $\mathbf{W}_t$ | Causal function for $Z_{i,t}$ (6.7), noise variable, noise variables. |
| $C_t$, $\mathscr{A}_t$, $\mathscr{D}_t$ | Number of clients, arrivals, and departures (6.8). |
| $\widehat{\mathbf{X}}_t, \widehat{\mathbf{x}}_t, \mathscr{T}$ | Intervention variables and values (M1), search operator (6.18). |
| $\text{do}(\widehat{\mathbf{X}}_t = \widehat{\mathbf{x}}_t)$ | Intervention at time $t$ (M1). |
| $\text{do}(\mathbf{X}^\star_t = \mathbf{x}^\star_t)$ | Optimal intervention at time $t$. |
| $R_t$, $J$, $\tilde{J}$ | Defender reward at time $t$, cumulative reward (6.9), objective function (6.13). |
| $f_R$, $f_J, f_D$ | Causal functions of $R_t$, $J$ (6.9), and $\mathbf{D}_{i,t}$. |
| $\mathbf{U}, \mathbf{V}$ | Exogenous and endogenous variables (M1). |
| $\mathbf{X}_t, \mathbf{N}_t$ | Manipulative, non-manipulative variables at time (M1). |
| $\mathbf{O}_t, \mathbf{L}_t, f_A$ | Observed and latent variables (M1) at time $t$, causal function of $\mathbf{A}_t$ (6.4). |
| $\mathbf{Y}_t$, $\mathcal{D}$ | Target variables (M1) at time $t$, set of decoys (Appendix D). |
| $\mathscr{M}, \mathbf{F}$ | scm and causal functions (M1). |
| $\mathbf{o}_t, \mathcal{T}$ | Observation (6.12) at time $t$ and time horizon (6.9). |
| $\mathbf{H}_t, \mathbf{h}_t, \gamma$ | History at time $t$ and its realization (6.11), discount factor. |
| $\pi_A, \pi_D \in \Pi$ | Attacker and defender strategies (M1). |
| $\pi^\star_D, \mathfrak{R}$ | Optimal defender strategy (Thm. 6.1), cumulative regret (6.20). |
| $\tilde{c}, \psi_{z_i}, \beta_{z_i}$ | Parameters in the defender objective (6.9) (6.13). |
| $\widehat{\mathbf{b}}_t, \mathbf{P}^\star_{\mathcal{G}}$ | Belief state (6.14), the set of pomiss (Def. 6.3). |
| $\mathbf{\Sigma}_t, \boldsymbol{\sigma}, \mathcal{G}$ | Markov state and its realization (Thm. 6.1), causal graph (Fig. 6.7). |
| $M, s_T, c$ | Number of particles, search time, exploration parameter (6.17). |

***Table 6.1:*** *Variables and symbols used in the model.*

### Target system (Fig. 6.5)

We represent the physical topology of the target system as a directed graph $\mathcal{G}_S \triangleq \langle \mathcal{V}, \mathcal{E} \rangle$, where nodes represent servers and workstations; edges represent network connectivity. Each node $i \in \mathcal{V}$ is (permanently) located in a zone $z_i \in \mathcal{Z}$ and has three state variables: an intrusion state $I_{i,t}$, a service state $S_{i,t}$, and a decoy state $\mathbf{D}_{i,t}$. $I_{i,t}$ takes five values: $\mathsf{U}$ if the node is unknown to the attacker, $\mathsf{K}$ if it is known, $\mathsf{S}$ if it has been scanned, $\mathsf{C}$ if it is compromised, and $\mathsf{R}$ if the attacker has root access; see Fig. 6.6. Similarly, $S_{i,t}$ takes two values: 1 if the service provided by node $i$ is accessible for clients, 0 otherwise. Lastly, the decoy state $\mathbf{D}_{i,t}$ is a vector $(\mathbf{D}_{i,t,1}, \dots, \mathbf{D}_{i,t,|\mathcal{D}|})$, where $\mathbf{D}_{i,t,j} = 1$ if decoy $j$ is active on node $i$, 0 otherwise. The set of decoys in CAGE-2 is available in Appendix D and is denoted by $\mathcal{D}$. The initial state of node $i$ is $(I_{i,1} = \mathsf{U}, S_{i,1} = 1, \mathbf{D}_{i,1} = \mathbf{0})$.

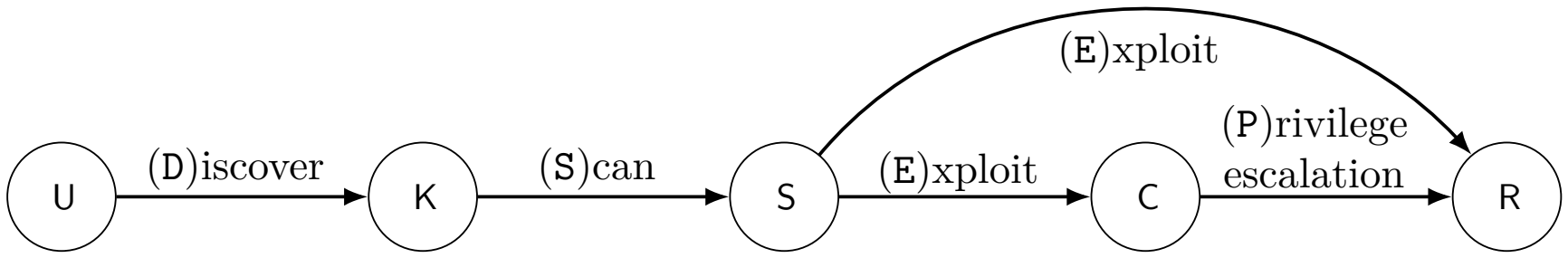

***Figure 6.6:*** *Transition diagram of the intrusion state $I_{i,t}$ (6.5); self-transitions are not shown; disks represent states; arrows represent state transitions; labels indicate conditions for state transition; the initial state is $I_{i,1} = \mathsf{U}$.*

Nodes of the target system provide services to clients (see Fig. 6.5). A *workflow graph* $\mathcal{G}_{\mathrm{W}}$ captures dependencies among these services. Specifically, a directed edge $i \rightarrow j$ in $\mathcal{G}_{\mathrm{W}}$ means that the service provided by node $i$ is used by node $j$. The configuration of $\mathcal{G}_{\mathrm{W}}$ for the target system is provided in Appendix D.

### Attacker

During each time step, the attacker performs an action $\mathbf{A}_t$, which targets a single node or all nodes in a zone (in case of a scan action). The action is determined by an attacker strategy $\pi_{\mathrm{A}}$. It consists of four components $\mathbf{A}_t \triangleq (\alpha_t, V_t, P_t, T_t)$: $\alpha_t$ is the action type, $V_t$ is the vulnerability, $P_t \in \{(\mathtt{U})\text{ser}, (\mathtt{R})\text{oot}\}$ is the privileges obtained by exploiting the vulnerability, and $T_t$ is the target, which can be either a single node $i \in \mathcal{V}$ or a zone $z \in \mathcal{Z}$.

There are five attack actions: (S)can, which scans the vulnerabilities of a node; (E)xploit, which attempts to exploit a vulnerability of a node; (P)rivilege escalation, which escalates privileges of a compromised node; (I)mpact, which stops the service on a compromised node; and (D)iscover, which discovers the nodes in a zone. These actions have the following causal effects on the intrusion state $I_{i,t}$ and the service state $S_{i,t}$ (Def. 3.2.1, Pearl, 2009).

$$\mathbf{A}_{t-1} = f_{\mathrm{A}}(\pi_{\mathrm{A}}, \{I_{i,t-1}\}_{i\in\mathcal{V}}) \tag{6.4}$$

$$I_{i,t} = f_{\mathrm{I}}(I_{i,t-1}, \mathbf{A}_{t-1}, \mathbf{D}_{i,t}, E_t) \triangleq \tag{6.5}$$

$$\begin{cases} \mathsf{K} \text{ if } T_{t-1} = z_i, \alpha_{t-1} = \mathtt{D} \\ \mathsf{K} \text{ if } T_{t-1} \in \mathrm{pa}(i)_{\mathcal{G}_\mathrm{W}}, \alpha_{t-1} = \mathtt{P} \\ \mathsf{S} \text{ if } T_{t-1} = i, \alpha_{t-1} = \mathtt{S} \\ \mathsf{C} \text{ if } T_{t-1} = i, \alpha_{t-1} = \mathtt{E}, P_{t-1} = \mathtt{U}, \mathbf{D}_{i,V_t,t} = 0, E_t = 1 \\ \mathsf{R} \text{ if } T_{t-1} = i, \alpha_{t-1} = \mathtt{E}, P_{t-1} = \mathtt{R}, \mathbf{D}_{i,V_t,t} = 0, E_t = 1 \\ \mathsf{R} \text{ if } T_{t-1} = i, \alpha_{t-1} = \mathtt{P} \\ I_{i,t-1} \text{ otherwise} \end{cases}$$

$$S_{i,t} = f_\mathrm{S}(\mathbf{A}_{t-1}, S_{i,t-1}) \triangleq \begin{cases} 0 & \text{if } T_{t-1} = i, \alpha_{t-1} = \mathtt{I} \\ S_{i,t-1} & \text{otherwise,} \end{cases} \tag{6.6}$$

where $E_t$ is a binary random variable and $\mathbb{P}[E_t = 1]$ is the probability that an exploit at time $t$ is successful; see Fig. 6.6.

The first two cases in (6.5) capture the transition $\mathtt{U} \to \mathtt{K}$, which occurs when the attacker discovers the zone of node $i$. The third case defines the transition $\mathtt{K} \to \mathtt{S}$, which happens when the attacker scans the node. The fourth case captures the transition $\mathtt{S} \to \mathtt{C}$, which occurs when the attacker compromises the node. The fifth and sixth cases define the transitions $\mathtt{S} \to \mathtt{R}$ and $\mathtt{C} \to \mathtt{R}$, which occur when the attacker obtains root privileges on the node. The final case captures the recurrent transition $I_{i,t} = I_{i,t-1}$. Lastly, (6.6) states that the (I)mpact action disrupts the service.

**Remark 6.1.** The dependence between (6.5) and $\mathcal{G}_\mathrm{W}$ is unintuitive but is warranted based on the source code of cage-2 (cage-2, 2022).

### Observations and clients

The defender knows the decoy state $\mathbf{D}_{i,t}$ and the service state $S_{i,t}$, but cannot observe the intrusion state $I_{i,t}$ nor the attacker action $\mathbf{A}_t$. Instead of $I_{i,t}$ and $\mathbf{A}_t$, the defender observes $Z_{i,t}$, which represents network activity at node $i$. Like the intrusion state $I_{i,t}$ (6.5), the activity $Z_{i,t}$ takes five values: (U)nknown, (K)nown, (S)canned, (C)ompromised, and (R)oot. The value of $Z_{i,t}$ is influenced both by attacker actions and by clients requesting service, which we express as

$$Z_{i,t} = f_{\mathrm{Z},i}(C_t, \mathbf{A}_{t-1}, W_{i,t}), \tag{6.7}$$

where $W_{i,t} \in \mathbb{N}$ is a noise variable and $C_t$ represents the number of clients requesting service at time $t$, which is determined as

$$C_t = f_\mathrm{C}(C_{t-1}, \mathscr{A}_t, \mathscr{D}_t) \triangleq \max\left[0, C_{t-1} + \mathscr{A}_t - \mathscr{D}_t\right], \tag{6.8}$$

where $\mathscr{A}_t$ and $\mathscr{D}_t$ are the number of clients that arrive and depart in the time interval $[t-1, t]$, respectively.

### Defender objective

The defender balances two objectives: maintain services to its clients and minimize the cost of intrusion. In CAGE-2, this bi-objective corresponds to maximizing

$$J = f_{\mathrm{J}}(\{R_t \mid 1 \leq t \leq \mathcal{T}\}) \triangleq \sum_{t=1}^{\mathcal{T}} \gamma^{t-1} R_t, \tag{6.9}$$

where $\gamma \in [0, 1]$ is a discount factor and $R_t$ is the reward at time $t$:

$$R_t = f_{\mathrm{R}}(\{I_{i,t}, S_{i,t}\}_{i \in \mathcal{V}}) \triangleq \overbrace{\sum_{i \in \mathcal{V}} \psi_{z_i}(S_{i,t} - 1)}^{\text{downtime cost}} \overbrace{-\beta_{z_i, I_{i,t}}}^{\text{intrusion cost}}.$$

Here $\psi_{z_i} \geq 0$ is the cost of service disruption in zone $z_i$; $\beta_{I_{i,t}, z_i} \geq 0$ is the cost of intrusion in zone $z_i$; and $\mathcal{T}$ is the time horizon. The configuration of $\psi_{z_i}$ and $\beta_{I_{i,t}, z_i}$ for the target system (Fig. 6.5) can be found in Appendix C.

### Defender interventions

During each time step, the defender performs an intervention that targets a single node. The defender can make four types of interventions: analyze the node for a possible intrusion, start a decoy service, remove malware, and restore the node to a secure state. We model these interventions as follows.

$$\begin{aligned}
&\mathrm{do}(Z_{i,t} = I_{i,t}) && \text{analyze;} && (6.10a)\\
&\mathrm{do}(\mathbf{D}_{i,j,t} = 1) && \text{decoy;} && (6.10b)\\
&\mathrm{do}(I_{i,t} = \mathsf{S}) \text{ if } I_{i,t-1} = \mathsf{C} && \text{remove;} && (6.10c)\\
&\mathrm{do}(\mathbf{D}_{i,t} = \mathbf{0}, I_{i,t} = \mathsf{S}) \text{ if } I_{i,t-1} \in \{\mathsf{C}, \mathsf{R}\} && \text{restore;} && (6.10d)\\
&\mathrm{do}(\emptyset) && \text{none.} && (6.10e)
\end{aligned}$$

Note that $\mathbf{D}_{i,t}$ remains constant if no interventions occur, i.e.,

$$\mathbf{D}_{i,t} = f_{\mathrm{D}}(\mathbf{D}_{i,t-1}) \triangleq \mathbf{D}_{i,t-1}.$$

When selecting interventions, the defender considers the *history*

$$\mathbf{H}_t \triangleq (\mathbf{V}_1, \mathrm{do}(\widehat{\mathbf{X}}_1), \mathbf{O}_2, \mathrm{do}(\widehat{\mathbf{X}}_2), \ldots, \mathrm{do}(\widehat{\mathbf{X}}_{t-1}), \mathbf{O}_t), \tag{6.11}$$

where the observables are defined as

$$\mathbf{O}_t \triangleq \{\mathbf{D}_{i,t}, S_{i,t}, Z_{i,t}, C_t \mid i \in \mathcal{V}\}. \tag{6.12}$$

Here, $\mathrm{do}(\widehat{\mathbf{X}}_t)$ is a shorthand for $\mathrm{do}(\widehat{\mathbf{X}}_t = \widehat{\mathbf{x}}_t)$ and $\mathbf{V}_1$ is the set of endogenous variables at time $t$ (defined below). The intervention at time $t$ can thus be expressed as $\mathrm{do}(\widehat{\mathbf{X}}_t = \pi_{\mathrm{D}}(\mathbf{h}_t))$, where $\pi_{\mathrm{D}}$ is a *defender strategy*.

**Remark 6.2** (Perfect recall)**.** The fact that the defender remembers the history $\mathbf{H}_t$ (6.11) means that it has *perfect recall* (Def. 7, Kuhn, 1953).

### A structural causal model

The variables and the causal functions (6.4)–(6.9) described above determine the SCM defined in (M1). Notable properties of (M1) are a) the causal graph is acyclic; b) the model is *stationary*; c) the model is *semi-Markov* (Thm. 1.4.1, Pearl, 2009); d) the exogenous variables are jointly independent; and e) $\mathbb{P}[\mathbf{V}_t]$ is Markov relative to $\mathcal{G}$ (6.2) (Thm. 1.2.7, Pearl, 2009).

---

Causal model of the CAGE-2 scenario (M1)

---

SCM: $\mathscr{M} \triangleq \langle \mathbf{U}, \mathbf{V}, \mathbf{F}, \mathbb{P}[\mathbf{U}] \rangle$. CAUSAL GRAPH: $\mathcal{G}$ (Fig. 6.7).

TARGET SYSTEM:

| | |
|---|---|
| $\mathcal{V}, \mathcal{T}$ | set of nodes and time horizon; see Appendix D |
| $\mathcal{G}_{\mathrm{S}} = \langle \mathcal{V}, \mathcal{E} \rangle$ | physical topology graph (Fig. 6.5) |
| $\mathcal{G}_{\mathrm{W}} = \langle \mathcal{V}_{\mathrm{W}} \subseteq \mathcal{V}, \mathcal{E}_{\mathrm{W}} \rangle$ | workflow graph; see Appendix D. |

RANDOM VARIABLES:

$$\mathbf{U} \triangleq \{E_t, \pi_{\mathrm{A}}, \mathscr{A}_t, \mathscr{D}_t, W_{i,t} \mid i \in \mathcal{V}, 2 \leq t \leq \mathcal{T}\}$$

$$\mathbf{V} \triangleq \{I_{i,t}, Z_{i,t}, S_{i,t}, \mathbf{D}_{i,t}, \mathbf{A}_t, C_t, R_t, J \mid i \in \mathcal{V}, 1 \leq t \leq \mathcal{T}\}$$

$$\mathbf{X} \triangleq \{\mathbf{D}_{i,t}, Z_{i,t}, I_{i,t} \mid i \in \mathcal{V}, 1 \leq t \leq \mathcal{T}\}$$

$$\mathbf{N} \triangleq (\mathbf{V} \cup \mathbf{U}) \setminus \mathbf{X}$$

$$\mathbf{O} \triangleq \{\mathbf{D}_{i,t}, S_{i,t}, Z_{i,t}, C_t, \mid i \in \mathcal{V}, 1 \leq t \leq \mathcal{T}\}$$

$$\mathbf{L} \triangleq (\mathbf{V} \cup \mathbf{U}) \setminus \mathbf{O}$$

$$\mathbf{Y} \triangleq \{R_t, J \mid 1 \leq t \leq \mathcal{T}\}.$$

INITIAL CONDITION: $I_{i,1} = Z_{i,1} = \mathsf{U}, S_{i,1} = 1, \mathbf{D}_{i,1} = \mathbf{0}, C_1 = R_1 = 0$ for all $i \in \mathcal{V}$.

CAUSAL FUNCTIONS: $\mathbf{F} \triangleq \{f_{\mathrm{I}}, f_{\mathrm{S}}, (f_{Z,i})_{i \in \mathcal{V}}, f_{\mathrm{C}}, f_{\mathrm{R}}, f_{\mathrm{J}}, f_{\mathrm{A}}, f_{\mathrm{D}}\}$.

OBSERVATIONAL DISTRIBUTIONS:

$$\mathbb{P}[\mathbf{U}] = \mathbb{P}[\pi_{\mathrm{A}}] \prod_{t=2}^{\mathcal{T}} \mathbb{P}[E_t]\mathbb{P}[\mathscr{A}_t]\mathbb{P}[\mathscr{D}_t]\mathbb{P}[(W_{i,t})_{i \in \mathcal{V}}]$$

$$\mathbb{P}[\mathbf{V}_t] = \prod_{i=1}^{|\mathbf{V}_t|} \mathbb{P}[\mathbf{V}_{i,t} \mid \mathrm{pa}(\mathbf{V}_{i,t})_{\mathcal{G}}].$$

INTERVENTIONS: $\mathrm{do}(\widehat{\mathbf{X}}_t = \pi_{\mathrm{D}}(\mathbf{H}_t))$ (6.10).

---

The size of the causal graph $\mathcal{G}$ associated with (M1) grows linearly with the time horizon $\mathcal{T}$ and with the number of nodes in the target system $|\mathcal{V}|$. A summary of $\mathcal{G}$ is shown in Fig. 6.7 on the next page.

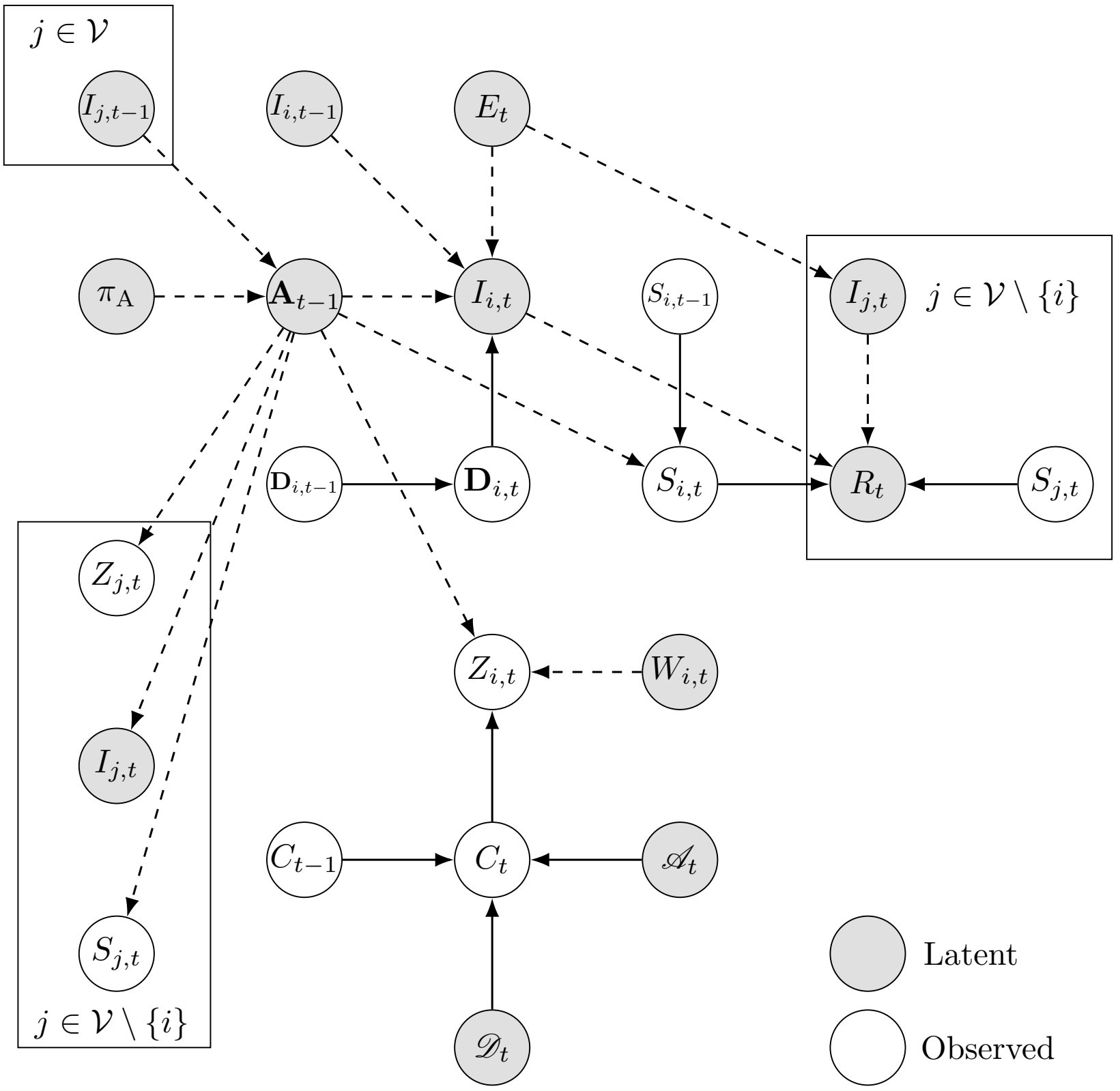

***Figure 6.7:*** *Causal (summary) graph of (M1) for node i (Ch. 10, Peters et al., 2017); plate notation is used to represent sets of variables (Buntine, 1994).*

### The defender problem in CAGE-2

Given (M1) and the defender objective $J$ (6.9), the problem for the defender can be stated as follows.

**Problem 6.1** (Optimal defender strategy in CAGE-2 (M1)).

$$\begin{aligned}
\underset{\pi_D}{\text{maximize}} \quad & \tilde{J}(\pi_D) \triangleq \mathbb{E}_{\pi_D}\left[J - \sum_{t=1}^{\mathcal{T}} \gamma^{t-1}\tilde{c}(\text{do}(\widehat{\mathbf{X}}_t = \pi_D(\mathbf{h}_t))) \mid \mathbf{V}_1 \setminus \{\mathbf{A}_1\}\right] && \text{(6.13a)}\\
\text{subject to} \quad & \text{do}(\widehat{\mathbf{X}}_t = \pi_D(\mathbf{h}_t)) \qquad \forall t \geq 1 && \text{(6.13b)}\\
& \pi_A \sim \mathbb{P}[\pi_A] && \text{(6.13c)}\\
& \mathscr{A}_t \sim \mathbb{P}[\mathscr{A}_t], \mathscr{D}_t \sim \mathbb{P}[\mathscr{D}_t] \qquad \forall t \geq 2 && \text{(6.13d)}\\
& E_t \sim \mathbb{P}[E_t], W_{i,t} \sim \mathbb{P}[W_{i,t}] \qquad \forall t \geq 2, i \in \mathcal{V} && \text{(6.13e)}\\
& \mathbf{A}_t = f_A(\pi_A, \{I_{i,t}\}_{i \in \mathcal{V}}) \qquad \forall t \geq 1 && \text{(6.13f)}
\end{aligned}$$

$$\mathbf{D}_{i,t} = f_{\mathrm{D}}(\mathbf{D}_{i,t-1}) \qquad \forall t \geq 2, i \in \mathcal{V} \quad (6.13\text{g})$$
$$I_{i,t} = f_{\mathrm{I}}(I_{i,t-1}, \mathbf{A}_{t-1}, \mathbf{D}_{i,t}, E_t) \qquad \forall t \geq 2, i \in \mathcal{V} \quad (6.13\text{h})$$
$$Z_{i,t} = f_{\mathrm{Z},i}(C_t, \mathbf{A}_{t-1}, W_{i,t}) \qquad \forall t \geq 2, i \in \mathcal{V} \quad (6.13\text{i})$$
$$C_t = f_{\mathrm{C}}(C_{t-1}, \mathscr{A}_t, \mathscr{D}_t) \qquad \forall t \geq 2 \quad (6.13\text{j})$$
$$S_{i,t} = f_{\mathrm{S}}(\mathbf{A}_{t-1}, S_{i,t-1}) \qquad \forall t \geq 2, i \in \mathcal{V} \quad (6.13\text{k})$$
$$R_t = f_{\mathrm{R}}(\{I_{i,t}, S_{i,t}\}_{i \in \mathcal{V}}) \qquad \forall t \geq 2 \quad (6.13\text{l})$$
$$J = f_{\mathrm{J}}(\{R_t\}_{t=1,\ldots,\mathcal{T}}), \quad (6.13\text{m})$$

where $t = 1, 2, \ldots, \mathcal{T}$; $\tilde{c}$ defines the cost of interventions (the configuration of $\tilde{c}$ can be found in Appendix C); (6.13b) defines the interventions; (6.13c)–(6.13e) capture the distribution of $\mathbf{U}$; and (6.13f)–(6.13m) define $\mathbf{F}$.

We say that a *defender strategy* $\pi^{\star}_{\mathrm{D}}$ is *optimal* if it solves Prob. 6.1. This problem is well-defined in the following sense.

**Theorem 6.1** (Existence of an optimal defender strategy for CAGE-2). *Assuming $C_t, \mathcal{A}, \mathcal{V}, \beta_{I_{i,t},z}, \psi_z$, are finite, $\tilde{c}$ is bounded, and $\mathcal{T}$ is finite or $\gamma < 1$, then there exists an optimal deterministic defender strategy $\pi^{\star}_{\mathrm{D}}$. If $\mathcal{T} = \infty$, then there exists a $\pi^{\star}_{\mathrm{D}}$ that is stationary.*

*Proof.* For notational convenience, let $\mathbf{S}_t \triangleq \{S_{i,t}\}_{i \in \mathcal{V}}$, $\mathbf{I}_t \triangleq \{I_{i,t}\}_{i \in \mathcal{V}}$, $\mathbf{D}_t \triangleq \{\mathbf{D}_{i,t}\}_{i \in \mathcal{V}}$, $\mathbf{W}_t \triangleq \{W_{i,t}\}_{i \in \mathcal{V}}$, and $\mathbf{Z}_t \triangleq \{Z_{i,t}\}_{i \in \mathcal{V}}$. We break down the proof into the following steps.

1) $\mathbf{\Sigma}_t \triangleq (\mathbf{I}_t, \mathbf{D}_t, C_t, \mathbf{A}_{t-1}, \pi_{\mathrm{A}}, \mathbf{S}_t)$ has the Markov property.
PROOF:

$$\mathbb{P}[\mathbf{\Sigma}_{t+1} \mid \mathbf{\Sigma}_1, \ldots, \mathbf{\Sigma}_t] = \mathbb{P}[\mathbf{A}_t \mid \pi_{\mathrm{A}}, \mathbf{I}_t]\mathbb{P}[\mathscr{A}_{t+1}]\mathbb{P}[\mathscr{D}_{t+1}] \times$$
$$\mathbb{P}[C_{t+1} \mid \mathscr{A}_{t+1}, \mathscr{D}_{t+1}, C_t]\mathbb{P}[E_{t+1}]\mathbb{P}[\mathbf{S}_{t+1} \mid \mathbf{S}_t, \mathbf{A}_t]\mathbb{P}[\mathbf{D}_{t+1} \mid \mathbf{D}_t] \times$$
$$\mathbb{P}[\mathbf{I}_{t+1} \mid \mathbf{I}_t, \mathbf{A}_t, \mathbf{D}_{t+1}, E_{t+1}] = \mathbb{P}[\mathbf{\Sigma}_{t+1} \mid \mathbf{\Sigma}_t].$$

2) $(\mathbf{O}_t, R_t \perp\!\!\!\perp \{\mathbf{V}_t\}_{t=1,\ldots,t} \mid \mathbf{\Sigma}_t)$.
PROOF:

$$\mathbb{P}[\mathbf{O}_t, R_t \mid \{\mathbf{V}_t\}_{t=1,\ldots,t}] = \mathbb{P}[\mathbf{W}_t]\mathbb{P}[\mathbf{Z}_t \mid C_t, \mathbf{A}_{t-1}, \mathbf{W}_t]\mathbb{P}[R_t \mid \mathbf{I}_t, \mathbf{S}_t] = \mathbb{P}[\mathbf{O}_t, R_t \mid \mathbf{\Sigma}_t].$$

3) $|\mathcal{R}_{\mathbf{O}_t}|$, $|\mathcal{R}_{\mathbf{\Sigma}_t}|$, and $R_t$ are finite. (Recall that $\mathcal{R}_X$ is the range of $X$.)
PROOF: Follows from the theorem assumptions.

4) Each defender strategy $\pi_{\mathrm{D}}$ induces a well-defined probability measure over the random sequence $(\mathbf{\Sigma}_t, \mathbf{O}_t)_{t \geq 1}$.
PROOF: 3) implies that the sample spaces of $(\mathbf{\Sigma}_t, \mathbf{O}_t)$ and $(\mathbf{\Sigma}_t, \mathbf{O}_t)_{t \geq 1}$ are measurable and countable, respectively. Further, the fact that $\mathcal{G}$ is acyclic implies that the interventional distributions induced by $(\mathrm{do}(\widehat{\mathbf{X}}_t = \widehat{\mathbf{x}}_t))_{t \geq 1}$ are well-defined (Ch. 3, Pearl, 2009). Consequently, the statement follows from the Ionescu-Tulcea extension theorem (Ionescu Tulcea, 1949).

5) $\mathbb{P}[\mathbf{\Sigma}_{t+1} \mid \mathbf{\Sigma}_t], \mathbb{P}[\mathbf{O}_t \mid \mathbf{\Sigma}_t]$, and $\mathbb{P}[R_t \mid \mathbf{\Sigma}_t]$ are stationary.
PROOF: Follows by stationarity of (M1).

Statements 1–5 imply that Prob. 6.1 defines a finite, stationary, and partially observed Markov decision process (POMDP) with bounded rewards, which satisfies assumptions 1–3 in the background chapter. The theorem thus follows from Thm. 2 in the background chapter. □

Theorem 6.1 states that an optimal defender strategy $\pi^{\star}_{\mathrm{D}}$ exists. Finding such a strategy requires estimating the causal effect

$$\mathbb{P}\left[J \mid \mathrm{do}(\widehat{\mathbf{X}}_1 = \pi_{\mathrm{D}}(\mathbf{H}_1)), \ldots, \mathrm{do}(\widehat{\mathbf{X}}_{\mathcal{T}} = \pi_{\mathrm{D}}(\mathbf{H}_{\mathcal{T}}))\right] \quad \text{(Def. 3.2.1, Pearl, 2009)}$$

for different strategies $\pi_{\mathrm{D}}$. A key question is thus whether the effect is identifiable, i.e., whether it can be estimated from the observables. The following theorem states that the answer is negative.

**Theorem 6.2.** *Problem 6.1 is not identifiable (Def. 6.2).*

*Proof.* To prove non-identifiability, it suffices to present two sets of causal functions $\mathbf{F}', \mathbf{F}''$ that induce identical distributions over the observables $\mathbf{O}$ but have different causal effects (Def. 6.2) (Lem. 1, Shpitser and Pearl, 2006). For simplicity, consider $\mathcal{T} = 2$ and $|\mathcal{V}| = 1$. In this case $J = R$ (6.9). Let

$$\mathbf{F}' \triangleq \{f_{\mathrm{R}}(S, I) \triangleq \mathbb{1}_{I=\mathsf{R}}, f_{\mathrm{I}}(\mathbf{A}, E, \mathbf{D}) = \mathsf{R}, f_{\mathrm{S}}(S, \mathbf{A}) = 1, f_{\mathrm{Z}}, f_{\mathrm{C}}, f_{\mathrm{J}}, f_{\mathrm{A}}, f_{\mathrm{D}}\},$$

where $\{f_{\mathrm{Z}}, f_{\mathrm{C}}, f_{\mathrm{J}}, f_{\mathrm{A}}, f_{\mathrm{D}}\}$ are defined arbitrarily. Define $\mathbf{F}''$ as $\mathbf{F}'$ except for $f_{\mathrm{R}}(S, I) \triangleq S$. Thus, $\mathbb{P}[\mathbf{O}]$ is the same with both $\mathbf{F}'$ and $\mathbf{F}''$. However, $\mathbb{P}[J = 0 \mid \mathrm{do}(I = \mathsf{S})] = 1$ with $\mathbf{F}'$ but $\mathbb{P}[J = S \mid \mathrm{do}(I = \mathsf{S})] = 1$ with $\mathbf{F}''$. □

Theorem 6.2 states that causal effects of defender interventions (6.10) *cannot* be identified from observations. This statement is obvious in hindsight but has important ramifications. It implies that to evaluate a defender strategy $\pi_{\mathrm{D}}$, the defender must either know (M1) or perform controlled experiments to measure the effects of the interventions prescribed by $\pi_{\mathrm{D}}$.

While it is likely that the defender is aware of certain components of (M1), it is unrealistic that it knows the entire model. A more reasonable assumption is that the defender knows the causal graph $\mathcal{G}$ (Fig. 6.7), which does not capture all nuances of the causal mechanisms but provides structural information. Leveraging this structure, we next present a method for finding an optimal strategy $\pi^{\star}_{\mathrm{D}}$ which only requires access to the causal graph and a simulator of (M1).

**Remark 6.3** (Simulation-based optimization)**.** Access to a simulator is assumed by virtually all existing methods for automated security response [481, 492, 42, 140, 141, 190, 452, 241, 368, 357, 122, 22, 490, 495, 498, 93].

## 6.6 Causal-Partially Observable Monte-Carlo Planning (c-pomcp)

In this section, we present c-pomcp, an online method for obtaining an optimal defender strategy $\pi^{\star}_{\mathrm{D}}$ for Prob. 6.1. The method involves three consecutive actions at each time step $t$; see Fig. 6.8. The first action uses the observation $\mathbf{o}_t$ and a *particle filter* to compute the defender's *belief* $\widehat{\mathbf{b}}_t$ in the form of a probability distribution over the latent variables $\mathbf{L}$ in (M1). The second action constructs a *search tree* of possible future histories $\mathbf{h}_k$ (6.11), which is initialized with a root node that represents the current history $\mathbf{h}_t$. Each edge extends this history by either an observation or an intervention: if $\mathbf{h}_{k+1}$ is a child node of $\mathbf{h}_k$, then either $\mathbf{h}_{k+1} = (\mathbf{h}_k, \mathbf{o}_{k+1})$ or $\mathbf{h}_{k+1} = (\mathbf{h}_k, \mathrm{do}(\widehat{\mathbf{X}}_k = \widehat{\mathbf{x}}_k))$. c-pomcp then *prunes* the tree by *excluding* histories that contain interventions that do not belong to a pomis. The third action uses the belief $\widehat{\mathbf{b}}_t$ and the pruned tree to perform *Monte-Carlo tree search*, which involves estimating $\tilde{J}(\pi_{\mathrm{D}})$ (6.13) through simulations. Once the search has been completed, the intervention from the root node that leads to the highest value of $\tilde{J}$ (6.13) is returned. The pseudocode of c-pomcp is listed in Alg. 6.1 on page 267, and the main components of the method are described below.

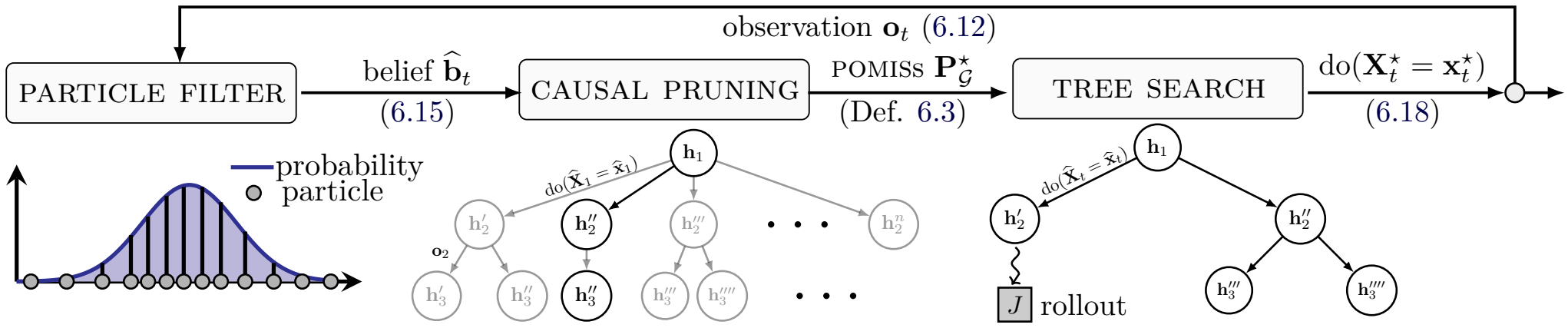

*Figure 6.8:* *Causal-Partially Observed Monte-Carlo Planning (c-pomcp, Alg. 6.1); the figure illustrates one time step during which (i) a particle filter is used to compute an approximate belief state* $\widehat{\mathbf{b}}_t$ *(6.14); (ii) a causal graph (Def. 2.2.1, Pearl, 2009) (see Fig. 6.7) is used to prune the search tree of possible histories* $\mathbf{h}_k$ *(6.11) by only considering histories with interventions in pomiss; and (iii) tree search is used to find an optimal intervention* $\mathrm{do}(\mathbf{X}^{\star}_t = \mathbf{x}^{\star}_t)$ *(6.18).*

### Particle filtering to estimate latent variables

The particle filter is a method for state estimation in partially observed dynamical systems (Thrun et al., 2005). Since (M1) can be formulated as such a system (Thm. 6.1), we use the particle filter to estimate the values of the latent variables $\mathbf{L}$ in (M1), e.g., the intrusion state $I_{i,t}$ (6.5).

We define the defender's *belief state* as

$$\mathbf{b}_t(\boldsymbol{\sigma}_t) \triangleq \mathbb{P}[\boldsymbol{\Sigma}_t = \boldsymbol{\sigma}_t \mid \mathbf{h}_t] \overset{\text{(Bayes)}}{=} \eta \mathbb{P}[\mathbf{o}_t \mid \boldsymbol{\sigma}_t, \mathbf{h}_{t-1}] \mathbb{P}[\boldsymbol{\sigma}_t \mid \mathrm{do}(\widehat{\mathbf{X}}_t = \widehat{\mathbf{x}}_t), \mathbf{h}_{t-1}] \quad (6.14)$$
$$\overset{\text{(Markov)}}{=} \eta \mathbb{P}[\mathbf{o}_t \mid \boldsymbol{\sigma}_t] \mathbb{P}[\boldsymbol{\sigma}_t \mid \mathrm{do}(\widehat{\mathbf{X}}_t = \widehat{\mathbf{x}}_t), \mathbf{h}_{t-1}]$$

$$\overset{\text{(Markov)}}{=} \eta \mathbb{P}[\mathbf{o}_t \mid \boldsymbol{\sigma}_t] \sum_{\boldsymbol{\sigma}_{t-1}} \mathbb{P}[\boldsymbol{\sigma}_t \mid \boldsymbol{\sigma}_{t-1}, \text{do}(\widehat{\mathbf{X}}_t = \widehat{\mathbf{x}}_t)] \mathbf{b}_{t-1}(\boldsymbol{\sigma}_{t-1}),$$

where $\mathbf{h}_t$ is the history (6.11), $\boldsymbol{\sigma}_t$ is a realization of $\boldsymbol{\Sigma}_t$ (see Thm. 6.1), and $\eta$ is a normalizing constant (as defined in (17) in the background chapter). The sum in (6.14) is over all possible realizations of $\boldsymbol{\Sigma}_{t-1} \in \mathcal{R}_{\boldsymbol{\Sigma}_{t-1}}$.

The computational complexity of (6.14) is $O(|\mathcal{R}_{\boldsymbol{\Sigma}}|^2)$, which grows quadratically with the size of the state space and exponentially with the number of state variables. For this reason, the particle filter approximates (6.14) by representing $\mathbf{b}_t$ by a set of $M$ sample states (particles) $\mathcal{P}_t = \{\widehat{\boldsymbol{\sigma}}_t^{(1)}, \ldots, \widehat{\boldsymbol{\sigma}}_t^{(M)}\}$ (Crisan and Doucet, 2002). These particles are sampled recursively as

$$\overline{\mathcal{P}}_t \triangleq \bigcup_{i=1}^{M} \left\{ \widehat{\boldsymbol{\sigma}}_t^{(i)} \sim \mathbb{P}\left[\cdot \mid \widehat{\boldsymbol{\sigma}}_{t-1}^{(i)}, \text{do}(\widehat{\mathbf{X}}_{t-1} = \widehat{\mathbf{x}}_{t-1})\right] \right\} \tag{6.15a}$$

$$\mathcal{P}_t \triangleq \bigcup_{i=1}^{M} \left\{ \widehat{\boldsymbol{\sigma}}_t^{(i)} \overset{\propto \mathbb{P}[\mathbf{o}_t \mid \widehat{\boldsymbol{\sigma}}_t^{(i)}]}{\sim} \overline{\mathcal{P}}_t \right\}, \tag{6.15b}$$

where $\widehat{\boldsymbol{\sigma}}_{t-1}^{(i)} \in \mathcal{P}_{t-1}$ and $x \overset{\propto\varphi}{\sim}$ means that $x$ is sampled with probability proportional to $\varphi$. (6.15a)–(6.15b) focus the particle set to regions of the state space with a high probability of generating the latest observation $\mathbf{o}_t$ (Thrun et al., 2005). This sampling ensures that the belief state induced by the particles converges to (6.14) when $M \to \infty$, as stated below.

**Theorem 6.3.** *Let* $\widehat{\mathbf{b}}_t(\boldsymbol{\sigma}_t) = \frac{1}{M}\sum_{i=1}^{M} \mathbb{1}_{\boldsymbol{\sigma}_t = \widehat{\boldsymbol{\sigma}}_t^{(i)}}$, *then* $\lim_{M\to\infty} \widehat{\mathbf{b}}_t \to \mathbf{b}_t$ *a.s.* $\forall t$.

This is a standard result in particle filtering. Proof is given in Appendix A.

### Causal pruning of the search tree

We use the causal graph $\mathcal{G}$ (Fig. 6.7) to prune the search tree by excluding histories $\mathbf{h}_k$ (6.11) that contain interventions that do not belong to a POMIS $\mathbf{P}^{\star}_{\mathcal{G}}$. For example, when $t = \mathcal{T} - 1$, then $Z_{i,t}$ (6.7) and $J$ (6.9) are *d-separated* in $\mathcal{G}$ (Def. 1.2.3, Pearl, 2009). This separation means that the intervention $\text{do}(Z_{i,t} = I_{i,t})$ (6.10a) has no causal effect on $J$ and thus $Z_{i,t} \notin \mathbf{P}^{\star}_{\mathcal{G}}$. By restricting the possible interventions at time $t$ to the set of POMISs $\mathbf{P}^{\star}_{\mathcal{G}}$, the number of interventions in the search tree is reduced by a factor of

$$\prod_{t=1}^{\mathcal{T}} \frac{\sum_{\tilde{\mathbf{X}} \in \mathbf{P}^{\star}_{\mathcal{G}}} |\mathcal{R}_{\tilde{\mathbf{X}}}|}{\sum_{\widehat{\mathbf{X}} \in 2^{\mathbf{X}_t}} |\mathcal{R}_{\widehat{\mathbf{X}}}|}, \tag{6.16}$$

where $\mathbf{X}_t$ is the set of manipulative variables at time $t$, $\mathcal{R}_{\mathbf{X}_t}$ is the combined range of the random variables in $\mathbf{X}_t$, and $2^{\mathbf{X}_t}$ is the power set. Hence, even if only a small subset of interventions does not belong to an POMIS, a significant reduction in the search tree size can be expected; see Fig. 6.9 on the next page.

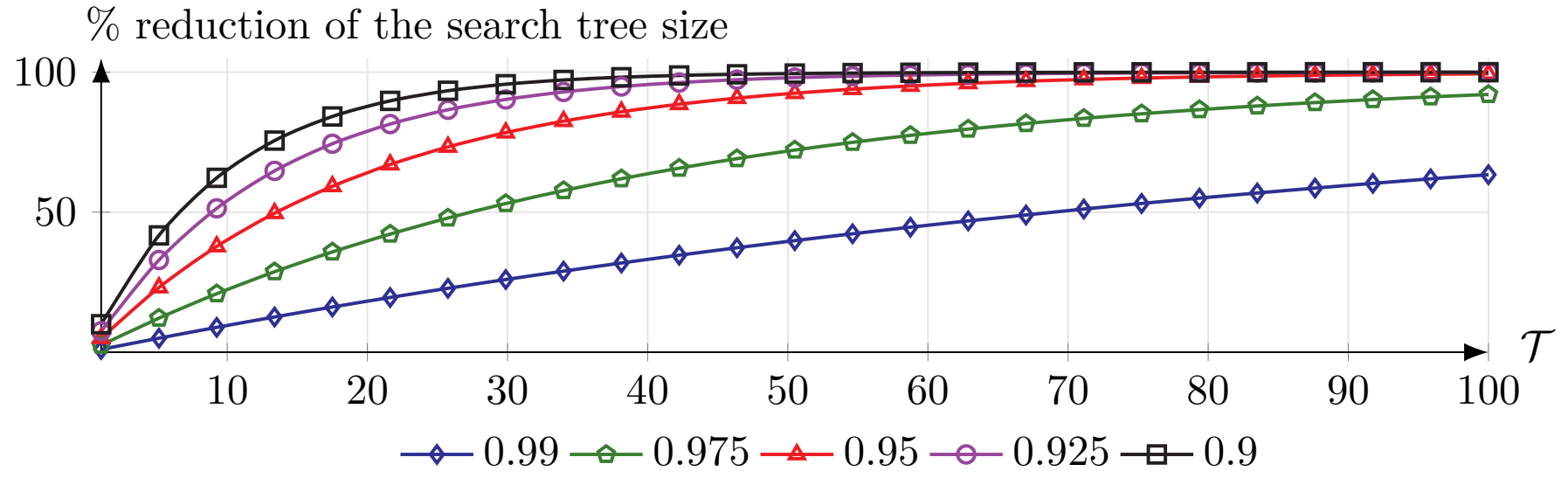

***Figure 6.9:*** *Reduction of the size of the search tree by pruning the intervention space $|2^{\mathbf{X}_t}|$ to the set of* POMISs $\mathbf{P}^{\star}_{\mathcal{G}}$*; the x-axis indicates the tree depth $\mathcal{T}$; curves relate to the factor in (6.16).*

Unfortunately, computing $\mathbf{P}^{\star}_{\mathcal{G}}$ is generally intractable, as stated below.

**Proposition 6.1.** *Computing $\mathbf{P}^{\star}_{\mathcal{G}}$ (Def. 6.3) is* PSPACE*-hard.*

*Proof.* We prove the PSPACE-hardness by reduction to the problem of solving a POMDP, which is PSPACE-hard (Thm. 6, Papadimitriou and Tsitsiklis, 1987). Let $x$ be an instance of the problem of computing $\mathbf{P}^{\star}_{\mathcal{G}}$ (Def. 6.3). Finding a solution to $x$ involves checking (6.3) for each $\tilde{\mathbf{X}}_t \in \mathbf{P}^{\star}_{\mathcal{G}}$. This means a solution to $x$ allows constructing an optimal solution to Prob. 6.1. By Thm 6.1, such a solution also provides a solution to a POMDP. □

Given the impracticality of computing $\mathbf{P}^{\star}_{\mathcal{G}}$ (Prop. 6.1), we approximate $\mathbf{P}^{\star}_{\mathcal{G}}$ as follows. First, we reduce the causal graph to a *subgraph* $\mathcal{G}[\mathbf{U}_{t-1} \cup \mathbf{U}_t \cup \mathbf{V}_{t-1} \cup \mathbf{V}_t]$ (Def. 7.1.2, Pearl, 2009). We then remove all variables in the subgraph whose values are uniquely determined by $\widehat{\mathbf{b}}$ (6.15). Subsequently, we add a node to the subgraph that represents the target $J$ (6.9), whose causal parents (Def. 1.2.1, Pearl, 2009) are determined using a *base strategy* $\widehat{\pi}$, which can be chosen freely. It can, for example, be based on heuristics or be designed by a domain expert. Finally, we compute a POMIS for the reduced graph using (Alg. 1, Lee and Bareinboim, 2019).

**Remark 6.4.** Since (Alg. 1, Lee and Bareinboim, 2019) is sound and complete (Thm. 9, Lee and Bareinboim, 2019), the approximation described above is exact when the base strategy $\widehat{\pi}$ is optimal.

When applying the above procedure to the CAGE-2 scenario, we identify the following types of defender interventions that are never included in a POMIS: (*i*) interventions that start decoys that are already running; (*ii*) defensive interventions on nodes that are not compromised according to $\widehat{\mathbf{b}}$ (6.15); and (*iii*) forensic and deceptive interventions on nodes that are compromised according to $\widehat{\mathbf{b}}$ (6.15).

**Remark 6.5.** The pruning of the search tree based on the POMISs occurs *during* the construction of the tree. The complete search tree is generally too large to construct.

### Monte-carlo tree search

Given the particle filter (6.14) and the pomiss, c-pomcp searches for optimal interventions using the tree search algorithm described in (Alg. 1, Silver and Veness, 2010). This algorithm constructs a search tree iteratively by repeating five steps (see Fig. 6.10): (*i*) it selects a path from the root to a leaf node using the *tree policy* described below; (*ii*) it expands the tree by adding children to the leaf, each of which corresponds to an intervention (6.10) on a pomis; (*iii*) it executes a rollout simulation from the leaf; (*iv*) it adds a child to the leaf that corresponds to the first observation (6.12) in the simulation; and (*v*) it records the value of $\tilde{J}$ (6.13) and backpropagates the value up the tree.

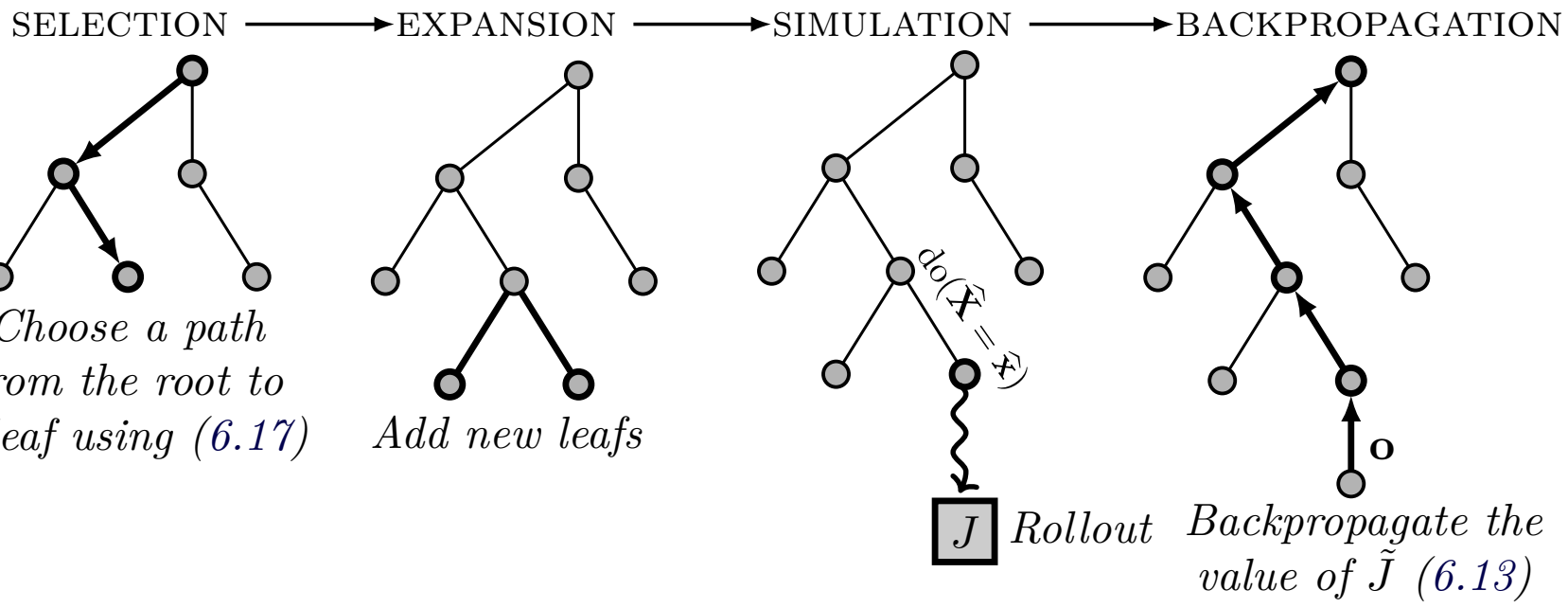

***Figure 6.10:*** *Tree search in c-pomcp; a search tree is constructed iteratively where each iteration consists of the four phases above.*

***Tree policy*** A node at depth $k$ of the tree is associated with a history $\mathbf{h}_k$ (6.11) and stores two variables: the average objective value $\widehat{J}(\mathbf{h}_k)$ (6.13) of simulations on the subtree emanating from the node, and the visit count $N(\mathbf{h}_k) \geq 1$, which is incremented whenever the node is visited during the search. Using these variables, we implement the tree policy by selecting nodes that maximize the upper confidence bound

$$\widehat{J}(\mathbf{h}_k) + c\sqrt{\frac{\ln N(\mathbf{h}_{k-1})}{N(\mathbf{h}_k)}}, \tag{6.17}$$

where $c > 0$ controls the exploration-exploitation trade-off.

***Rollout*** The initial state of a rollout simulation is sampled from the belief state $\widehat{\mathbf{b}}_k$ (6.15), and the intervention at each time step is selected using the *base strategy* $\widehat{\pi}$. The simulation executes for a depth of $\delta_{\mathrm{R}}$, after which the objective value for the remainder of the simulation is estimated using a *base value function* $J_{\widehat{\pi}}$. Like the base strategy, this function can be chosen freely. It can, for example, be obtained through offline reinforcement learning. After the simulation has completed, the discounted sum of the rewards $R_1, R_t, \ldots R_{\delta_{\mathrm{R}}}$ (6.9) and $J_{\widehat{\pi}}$ is used to update $\widehat{J}(\mathbf{h}_k)$.

***Convergence*** The process of running simulations and extending the search tree continues for a *search time* $s_{\mathrm{T}}$, after which the intervention that leads to the largest value of $\widehat{J}$ is returned, i.e.,

$$\mathrm{do}(\tilde{\mathbf{X}}_t = \tilde{\mathbf{x}}_t) \in \operatorname*{arg\,max}_{\mathrm{do}(\widehat{\mathbf{X}}_t = \widehat{\mathbf{x}}_t)} \widehat{J}((\mathbf{h}_t, \mathrm{do}(\widehat{\mathbf{X}}_t = \widehat{\mathbf{x}}_t))).$$

We can express this search procedure as

$$\mathrm{do}(\tilde{\mathbf{X}}_t = \tilde{\mathbf{x}}_t) \leftarrow \mathscr{T}(\mathbf{h}_t, \widehat{\mathbf{b}}_t, \widehat{\pi}, s_{\mathrm{T}}, \mathscr{S}, \mathcal{G}, \mathbf{P}^{\star}_{\mathcal{G}}), \tag{6.18}$$

where $\mathscr{T}$ is a tree search operator.

**Theorem 6.4** (Convergence of C-POMCP)**.** *Under the assumptions made in Thm. 6.1 and further assuming that the* POMIS *computation is exact,* $M \to \infty$*,* $s_{\mathrm{T}} \to \infty$*,* $\mathcal{T} < \infty$*, and* $c$ *is chosen such that*

$$\mathbb{P}\left[\widehat{J}(\mathbf{h}_k) \leq \mathbb{E}[\widehat{J}(\mathbf{h}_k)] \pm c\sqrt{\frac{\ln N(\mathbf{h}_{k-1})}{N(\mathbf{h}_k)}}\right] \leq k^{-4} \qquad \forall k \geq 1, \tag{6.19}$$

*Then the intervention prescribed by* C-POMCP *(6.18) for any* $\mathbf{h}_t$ *converges in probability to an optimal intervention* $\mathrm{do}(\mathbf{X}^{\star}_t = \mathbf{x}^{\star}_t)$ *as the search time is increased, i.e.,* $\lim_{s_{\mathrm{T}} \to \infty} \mathbb{P}[\mathscr{T}(\mathbf{h}_t, \widehat{\mathbf{b}}_t, \widehat{\pi}, s_{\mathrm{T}}, \mathscr{S}, \mathcal{G}, \mathbf{P}^{\star}_{\mathcal{G}}) \neq \mathrm{do}(\mathbf{X}^{\star}_t = \mathbf{x}^{\star}_t)] = 0.$

The proof of Thm. 6.4 relies on mapping an execution of C-POMCP to an execution of the UCT algorithm (Fig. 1, Kocsis and Szepesvári, 2006), which is known to converge as $s_{\mathrm{T}} \to \infty$ (Thm. 7, Kocsis and Szepesvári, 2006). We provide the proof in Appendix B. Note that (6.19) can always be satisfied by choosing a large $c$.

**Remark 6.6.** Theorem 6.4 is not confined to CAGE-2 (M1). Rather, the theorem is general and applies to any control problem based on an SCM with interventions that can be formulated as a finite and stationary POMDP (Thm. 6.1).

***Algorithm 6.1:*** *C-POMCP: **C**ausal-**P**artially **O**bservable **M**onte-**C**arlo **P**lanning.*

**Input:** Simulator $\mathscr{S}$ of (M1), causal graph $\mathcal{G}$ (Fig. 6.7), search operator $\mathscr{T}$, search time $s_{\mathrm{T}}$, horizon $\mathcal{T}$, number of particles $M$, base strategy $\widehat{\pi}$.
**Output:** Interventions $\mathrm{do}(\tilde{\mathbf{X}}_1 = \tilde{\mathbf{x}}_1), \ldots, \mathrm{do}(\tilde{\mathbf{X}}_{\mathcal{T}} = \tilde{\mathbf{x}}_{\mathcal{T}})$.

**procedure** C-POMCP($\mathscr{S}$, $\mathcal{G}$, $\mathscr{T}$, $s_{\mathrm{T}}$, $\mathcal{T}$, $M$, $\widehat{\pi}$)
  $\mathbf{h}_1 = (\mathbf{V}_1)$.
  **for** $t = 1, 2, \ldots, \mathcal{T}$ **do**
    Compute $\widehat{\mathbf{b}}_t$ using (6.15) with $M$ particles.
    Compute $\mathbf{P}^{\star}_{\mathcal{G}}$ (Def. 6.3).
    $\mathrm{do}(\tilde{\mathbf{X}}_t = \tilde{\mathbf{x}}_t) \leftarrow \mathscr{T}(\mathbf{h}_t, \widehat{\mathbf{b}}_t, \widehat{\pi}, s_{\mathrm{T}}, \mathscr{S}, \mathcal{G}, \mathbf{P}^{\star}_{\mathcal{G}})$ (6.18).
    Perform intervention $\mathrm{do}(\tilde{\mathbf{X}}_t = \tilde{\mathbf{x}}_t)$ (6.10).
    Observe $\mathbf{o}_{t+1}$.
    Update history $\mathbf{h}_{t+1} = (\mathbf{h}_t, \mathrm{do}(\tilde{\mathbf{X}}_t = \tilde{\mathbf{x}}_t), \mathbf{o}_{t+1})$ (6.11).

### Comparison with other methods

C-POMCP (Alg. 6.1) distinguishes itself from existing methods evaluated against the CAGE-2 benchmark ([481, 492, 42, 140, 141, 190, 452, 241, 368, 22, 490, 495, 93, 498]) in four key aspects. First, it incorporates the causal structure of the target system. Second, it guarantees an optimal solution (Thm. 6.4). No such guarantees are available for the existing methods. Third, while the above-referenced methods ignore the latent variables, C-POMCP explicitly models the uncertainty of the latent variables and how this uncertainty changes in light of new observations (6.14). Fourth, in contrast to the existing *offline* methods, C-POMCP is an *online* method that updates the defender strategy at each time step.

## 6.7 Evaluating C-POMCP Against CAGE-2

We implement C-POMCP (Alg. 6.1) in Python and run it to learn defender strategies for the CAGE-2 scenario (CAGE-2, 2022). The system configuration is listed in Appendix D; the hyperparameters are listed in Appendix C; and the computing environment is an M2-ultra processor.

### Baselines

We compare the performance of C-POMCP with that of two baselines: CARDIFF-PPO, a current state-of-the-art method for CAGE-2 (Vyas et al., 2023), and POMCP (Alg. 1, Silver and Veness, 2010), a non-causal version C-POMCP. Note that, while we only compare against CARDIFF-PPO from the CAGE-2 leaderboard, it represents all methods on the leaderboard since it achieves better performance than the other methods (CAGE-2, 2022).

### Evaluation metrics

We use two evaluation metrics: $\tilde{J}(\pi_{\mathrm{D}})$ (6.13) and the cumulative regret

$$\mathfrak{R}_n \triangleq n\tilde{J}(\pi^\star_{\mathrm{D}}) - \sum_{l=1}^{n} \tilde{J}(\pi_{l,\mathrm{D}}), \qquad \text{(Lattimore and Szepesvári, 2020)} \qquad (6.20)$$

where $n$ is the total computational time in minutes and $\pi_{l,\mathrm{D}}$ is the strategy after $l$ minutes (e.g., $l$ minutes of tree search). Since computing $\pi^\star_{\mathrm{D}}$ is PSPACE-hard (Thm. 6, Papadimitriou and Tsitsiklis, 1987), we estimate $\tilde{J}(\pi^\star_{\mathrm{D}})$ using the current state-of-the-art value when computing (6.20).

### CAGE-2 Scenarios

CAGE-2 can be instantiated with different attacker strategies $\pi_{\mathrm{A}}$ (6.4) as well as different topologies within each zone; see Fig. 6.5 (CAGE-2, 2022). Based on these parameters, we define the following evaluation scenarios.

**Scenario 6.1** (b-line attacker). In this scenario, the attacker strategy $\pi_A$ represents the b-line attacker from cage-2 (cage-2, 2022), which attempts to move directly to the operational zone. The topology is shown in Fig. 6.5.

**Scenario 6.2** (meander attacker). In this scenario, $\pi_A$ represents the meander attacker from cage-2 (cage-2, 2022). meander explores the network one zone at a time, seeking to gain privileged access to all hosts in a zone before moving on to the next one, eventually arriving at the operational zone. The topology is shown in Fig. 6.5.

**Scenario 6.3** (random attacker). In this scenario, $\pi_A$ is b-line with probability 0.5 and meander with probability 0.5. The topology is shown in Fig. 6.5.

**Scenario 6.4** (random topology). This scenario is the same as Scenario 6.1 except that the topologies of the enterprise and user zones are randomized at the start of each evaluation episode.

### CAGE-2 benchmark results

The evaluation results are summarized in Figs. 6.11–6.12 and Tables 6.2–6.3 on the subsequent pages. The results show that c-pomcp achieves the highest objective value (6.13) and the lowest regret (6.20) across all evaluation scenarios and time horizons $\mathcal{T}$. (The results are not statistically significant in all cases, though.)

The green curves in Fig. 6.11 relate to c-pomcp. The blue and red curves relate to the baselines. The leftmost column in Fig. 6.11 shows the regret (6.20). Notably, the regret of c-pomcp is two orders of magnitude lower than the regret of cardiff-ppo and one order of magnitude lower than the regret of pomcp.

The three rightmost columns in Fig. 6.11 show the objective value (6.13) obtained by c-pomcp and pomcp in function of the search time $s_T$. We observe that c-pomcp achieves a significantly higher value than pomcp, although the difference diminishes with increasing $s_T$, which is expected (Thm. 6.4). We explain the improvement of c-pomcp compared to pomcp by the pruned search tree, which is obtained by leveraging the causal structure (Def. 6.3). The reduction in search tree size achieved by the pruning is shown in Fig. 6.12. We see that the pruning reduces the size of the search tree by around 90–95%.

Lastly, Table 6.3 contains the results for Scenario 6.4. We find that c-pomcp and pomcp are agnostic to changes in the topology within each zone. By contrast, the performance of cardiff-ppo reduces drastically when the topology changes, indicating that its strategy is overfitted to the training environment (Vyas et al., 2023). cardiff-ppo has shown similar behavior in (Wolk et al., 2022).

| Method | Training / search (minutes) (seconds) | $\mathcal{T} = 30$ SCENARIO 6.1 | SCENARIO 6.2 | SCENARIO 6.3 | $\mathcal{T} = 50$ SCENARIO 6.1 | SCENARIO 6.2 | SCENARIO 6.3 | $\mathcal{T} = 100$ SCENARIO 6.1 | SCENARIO 6.2 | SCENARIO 6.3 |
|---|---|---|---|---|---|---|---|---|---|---|
| CARDIFF | 2000 / $10^{-4}$ | −3.57 ± 0.06 | −5.69 ± 1.68 | −4.76 ± 1.90 | −6.44 ± 0.16 | −9.23 ± 2.87 | −7.64 ± 2.78 | −13.69 ± 0.533 | −17.16 ± 4.41 | −15.28 ± 4.18 |
| C-POMCP | 0 / 0.05 | −4.64 ± 0.5 | −5.73 ± 0.08 | −5.18 ± 0.13 | −9.20 ± 0.38 | −9.35 ± 0.16 | −9.27 ± 0.67 | −25.05 ± 3.02 | −18.29 ± 0.13 | −21.67 ± 3.19 |
| C-POMCP | 0 / 0.1 | −3.89 ± 0.25 | −5.62 ± 0.14 | −4.75 ± 0.34 | −8.46 ± 0.27 | −8.92 ± 0.23 | −8.69 ± 0.47 | −21.28 ± 0.72 | −17.38 ± 0.20 | −19.33 ± 1.03 |
| C-POMCP | 0 / 0.5 | −4.00 ± 0.14 | −5.61 ± 0.02 | −4.81 ± 0.24 | −7.38 ± 0.19 | −8.62 ± 0.18 | −8.00 ± 0.44 | −18.08 ± 1.32 | −16.81 ± 0.14 | −17.45 ± 1.14 |
| C-POMCP | 0 / 1 | −3.64 ± 0.13 | **−5.52 ± 0.16** | −4.58 ± 0.27 | −6.60 ± 0.32 | −8.55 ± 0.08 | −7.58 ± 0.29 | −17.42 ± 1.08 | −16.34 ± 0.44 | −16.88 ± 1.29 |
| C-POMCP | 0 / 5 | −3.50 ± 0.11 | −5.65 ± 0.11 | −4.57 ± 0.23 | −6.52 ± 0.34 | −8.46 ± 0.11 | −7.49 ± 0.46 | −13.23 ± 0.43 | −16.46 ± 0.30 | −14.85 ± 0.79 |
| C-POMCP | 0 / 15 | **−3.37 ± 0.08** | −5.66 ± 0.11 | **−4.52 ± 0.19** | −6.57 ± 0.38 | −8.57 ± 0.13 | −7.57 ± 0.52 | **−12.98 ± 1.55** | **−15.87 ± 0.67** | **−14.43 ± 1.99** |
| C-POMCP | 0 / 30 | −3.42 ± 0.09 | −5.70 ± 0.09 | −4.56 ± 0.14 | **−6.34 ± 0.28** | **−8.52 ± 0.18** | **−7.43 ± 0.52** | −13.32 ± 0.18 | −16.05 ± 0.96 | −14.68 ± 1.02 |
| POMCP | 0 / 0.05 | −6.87 ± 0.21 | −9.50 ± 0.19 | −8.19 ± 0.37 | −13.90 ± 0.24 | −22.26 ± 0.44 | −18.08 ± 0.72 | −38.71 ± 1.99 | −50.24 ± 2.67 | −44.48 ± 3.11 |
| POMCP | 0 / 0.1 | −6.31 ± 0.12 | −8.70 ± 0.07 | −7.51 ± 0.19 | −13.71 ± 0.22 | −20.20 ± 0.47 | −16.96 ± 0.76 | −38.02 ± 0.53 | −46.40 ± 0.64 | −42.21 ± 0.79 |
| POMCP | 0 / 0.5 | −5.32 ± 0.24 | −8.28 ± 0.13 | −6.80 ± 0.33 | −12.89 ± 0.20 | −19.16 ± 0.09 | −16.03 ± 0.33 | −34.92 ± 0.96 | −47.29 ± 0.39 | −41.11 ± 1.23 |
| POMCP | 0 / 1 | −5.27 ± 0.65 | −7.68 ± 0.10 | −6.48 ± 0.75 | −12.57 ± 0.41 | −18.38 ± 0.50 | −15.48 ± 0.94 | −34.50 ± 0.65 | −47.02 ± 1.75 | −40.76 ± 2.34 |
| POMCP | 0 / 5 | −5.11 ± 0.32 | −7.58 ± 0.05 | −6.35 ± 0.36 | −12.03 ± 0.93 | −18.22 ± 0.19 | −15.13 ± 1.18 | −33.06 ± 0.21 | −45.15 ± 0.54 | −39.13 ± 0.66 |
| POMCP | 0 / 15 | −5.18 ± 0.66 | −7.30 ± 0.38 | −6.24 ± 1.22 | −11.32 ± 0.84 | −17.68 ± 0.58 | −14.50 ± 1.37 | −30.88 ± 1.41 | −45.19 ± 0.35 | −38.04 ± 1.57 |
| POMCP | 0 / 30 | −4.50 ± 0.17 | −6.92 ± 0.59 | −5.71 ± 0.77 | −9.88 ± 1.63 | −17.55 ± 0.44 | −13.72 ± 2.09 | −29.51 ± 2.00 | −44.27 ± 1.13 | −36.89 ± 2.49 |

***Table 6.2:*** *Comparing C-POMCP with baselines: CARDIFF-PPO (Vyas et al., 2023) and POMCP (Silver and Veness, 2010); columns indicate the time horizon $\mathcal{T}$ (6.9); subcolumns indicate the evaluation scenario; numbers indicate the mean and the standard deviation of the objective $\tilde{J}(\pi_D)$ (6.13) from evaluations with 3 random seeds.*

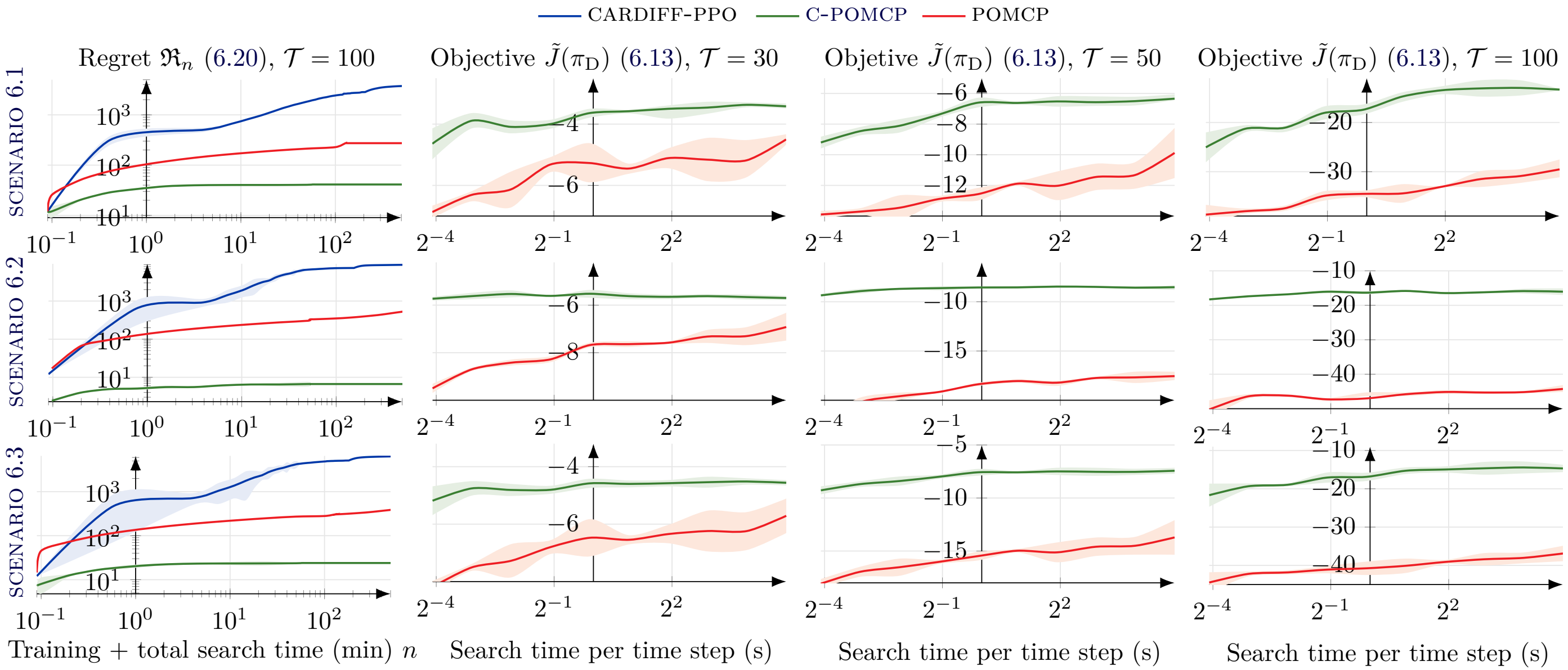

***Figure 6.11:*** *Comparing* C-POMCP *(green curves) with baselines:* CARDIFF-PPO *(Vyas et al., 2023) (blue curves) and* POMCP *(Silver and Veness, 2010) (red curves); rows indicate the evaluation scenario; the curves show the mean value from evaluations with* 3 *random seeds; shaded areas indicate the standard deviation; the number of data points on the $x$-axis is* 2000 *for the left-most column and* 10 *for the other columns.*

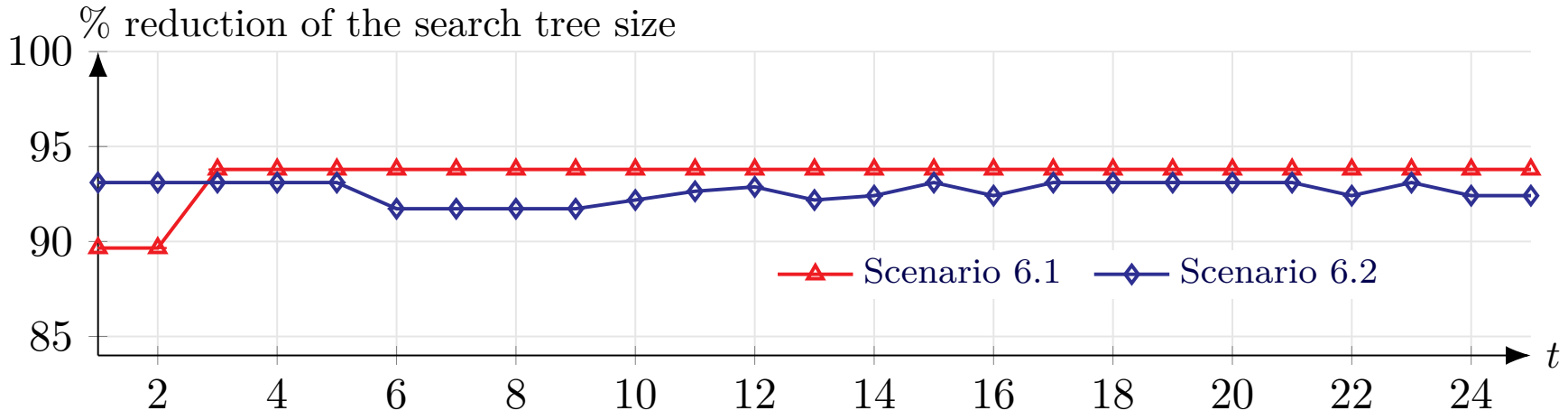

***Figure 6.12:*** *Effect of the pruning of the search tree in* C-POMCP.

| *Method* | *Training (min)* | *Search (s)* | *Objective* $\tilde{J}(\pi_{\mathrm{D}})$ *(6.13)* |
|---|---|---|---|
| CARDIFF-PPO[481] | 2000 | $10^{-4}$ | $-429 \pm 167$ |
| C-POMCP | 0 | 30 | $\mathbf{-13.32 \pm 0.18}$ |
| POMCP [416] | 0 | 30 | $-29.51 \pm 2.00$ |

***Table 6.3:*** *Evaluation results for Scenario 6.4 with* $\mathcal{T} = 100$.

### Discussion of the CAGE-2 benchmark results

The key findings from the CAGE-2 benchmark results are:

- Leveraging the causal structure of the target system, C-POMCP achieves state-of-the-art performance (Figs. 6.11–6.12, Table 6.2).
- The interventions prescribed by C-POMCP are guaranteed to converge to optimal interventions as $s_{\mathrm{T}} \to \infty$ (Thm. 6.4), which is consistent with the evaluation results.
- C-POMCP is two orders of magnitude more efficient in computing time than the state-of-the-art method CARDIFF-PPO (Fig. 6.11).
- C-POMCP is an online method and can adapt to changes in the topology of the target system (Table 6.3).

Surprisingly, the results demonstrate that C-POMCP requires only $5 - 15$ seconds of search to achieve competitive performance on the CAGE-2 benchmark. The fact that C-POMCP performs significantly better than its non-causal version POMCP (Alg. 1, Silver and Veness, 2010) indicates that the main enabler of the efficiency is the causal structure, which we exploit for pruning the search space. This observation suggests limitations of existing methods that narrowly focus on *model-free* reinforcement learning and do not consider the causal structure of the underlying system. While the results demonstrate clear benefits of C-POMCP compared to the existing methods, C-POMCP has two drawbacks. First, execution of C-POMCP is slower than that of pre-trained methods; see Table 6.2, typically $10^{-4}$s vs 10s. Second, the performance of C-POMCP depends on the causal structure of the target system (Def. 2.2.1, Pearl, 2009). If no causal structure is known, the performance of C-POMCP drops; cf. the performance of C-POMCP and POMCP in Table 6.2.

## 6.8 Conclusion

This paper presents a formal (causal) model of CAGE-2 (M1), which is considered a standard benchmark to evaluate methods for automated security response (CAGE-2, 2022). Based on this model, we prove the existence of optimal defender strategies (Thm. 6.1) and design an iterative method that converges to such a strategy (Thm. 6.4). The method, which we call **C**ausal-**P**artially **O**bservable **M**onte-**C**arlo **P**lanning (C-POMCP), leverages causal structure to prune, construct and traverse a search tree (Alg. 6.1). C-POMCP has four advantages over the state-of-the-art methods that have been proposed in the context of CAGE-2: (*i*) it is two orders of magnitude more computationally efficient (Fig. 6.11); (*ii*) it achieves better performance (Table. 6.2); (*iii*) it is an online method which adapts to topology changes in the target system (Table 6.3); and (*iv*), it produces provably optimal defender strategies (Thm. 6.4).

In the context of this thesis, this paper demonstrates two new things: (1) how our methodology for automated security response can be integrated with causal modeling[8]; and (2) how the defender strategies produced by our methodology can be evaluated against the CAGE-2 benchmark.

## ■ Acknowledgments

The authors are grateful to DARPA and program manager Tejas Patel for supporting this research. The authors would also like to thank the Siemens research team for discussions and feedback: Enrico Lovat, Jagannadh Vempati, Anton Kocheturov, Arun Ramamurthy, Arif Haque, and Abhishek Ramchandran. Finally, the authors thank KTH researchers Forough Shahab Samani, Xiaoxuan Wang, and Duc Huy Le for constructive comments.

## ■ Appendix

### A Proof of Theorem 6.3

1. CASE: $t = 1$.
   PROOF: $\mathbf{b}_1$ is given, hence $\widehat{\mathbf{b}}_1 = \mathbf{b}_1$.

2. CASE: $t > 1$.
   PROOF: Assume $\widehat{\mathbf{b}}_{t-1} = \mathbf{b}_{t-1}$. Let

$$g(\overline{\boldsymbol{\sigma}}_t) \triangleq \sum_{\boldsymbol{\sigma}_{t-1}} \widehat{\mathbf{b}}_{t-1}(\boldsymbol{\sigma}_{t-1})\mathbb{P}[\overline{\boldsymbol{\sigma}} \mid \boldsymbol{\sigma}_{t-1}, \mathrm{do}(\widehat{\mathbf{X}}_{t-1} = \widehat{\mathbf{x}}_{t-1})].$$

[8] See the methodology chapter for details about our experimental methodology.

We then have that

$$\begin{aligned}\mathbf{b}_t(\boldsymbol{\sigma}) &= \mathbb{E}_{\overline{\boldsymbol{\sigma}}\sim\mathbf{b}_t}[\mathbb{1}_{\boldsymbol{\sigma}=\overline{\boldsymbol{\sigma}}}] = \sum_{\overline{\boldsymbol{\sigma}}}\mathbf{b}_t(\overline{\boldsymbol{\sigma}})\mathbb{1}_{\boldsymbol{\sigma}=\overline{\boldsymbol{\sigma}}} \overset{(a)}{=} \sum_{\overline{\boldsymbol{\sigma}}}\frac{g(\overline{\boldsymbol{\sigma}})}{g(\overline{\boldsymbol{\sigma}})}\mathbf{b}_t(\overline{\boldsymbol{\sigma}})\mathbb{1}_{\boldsymbol{\sigma}=\overline{\boldsymbol{\sigma}}}\\
&= \mathbb{E}_{\overline{\boldsymbol{\sigma}}\sim g}\left[\frac{\mathbf{b}_t(\overline{\boldsymbol{\sigma}})}{g(\overline{\boldsymbol{\sigma}})}\mathbb{1}_{\boldsymbol{\sigma}=\overline{\boldsymbol{\sigma}}}\right] \overset{(b)}{=} \mathbb{E}_{\overline{\boldsymbol{\sigma}}\sim g}\left[\frac{\eta\mathbb{P}[\mathbf{o}_t \mid \boldsymbol{\sigma}]g(\overline{\boldsymbol{\sigma}})}{g(\overline{\boldsymbol{\sigma}})}\mathbb{1}_{\boldsymbol{\sigma}=\overline{\boldsymbol{\sigma}}}\right]\\
&= \mathbb{E}_{\overline{\boldsymbol{\sigma}}\sim g}\left[\eta\mathbb{P}[\mathbf{o}_t \mid \boldsymbol{\sigma}_t]\mathbb{1}_{\boldsymbol{\sigma}=\overline{\boldsymbol{\sigma}}}\right],\end{aligned} \tag{6.21}$$

where (b) follows from the definition of $\mathbf{b}$ (6.14).

As $\widehat{\mathbf{b}}_{t-1} = \mathbf{b}_{t-1}$, the denominator in (a) is non-zero for all $\overline{\boldsymbol{\sigma}}$ where $\mathbf{b}_t(\overline{\boldsymbol{\sigma}}) > 0$. Since the particles are distributed according to $\eta'\eta\mathbb{P}[\mathbf{o}_t \mid \boldsymbol{\sigma}_t]g(\widehat{\boldsymbol{\sigma}}_t)$ ($\eta'$ is a normalizing factor), it follows from the strong law of large numbers (Thm. 6.2, Çınlar, 2011) that $\widehat{\mathbf{b}}(\boldsymbol{\sigma}_t) = \frac{1}{M}\sum_{i=1}^{M}\mathbb{1}_{\boldsymbol{\sigma}_t=\widehat{\boldsymbol{\sigma}}_t^{(i)}}$ converges almost surely to (6.21) as $M \to \infty$. (Remark: the probability measure $\mathbb{P}$ in (6.21) exists since $|\mathcal{R}_{\boldsymbol{\Sigma}_t} \cup \mathcal{R}_{\mathbf{O}_t}| < \infty$ (Thm. 6.1, (Thm. 2.2.1, Rosenthal, 2006)).) □

## B Proof of Theorem 6.4

It follows from Thm. 6.3 and (Lem. 1–2, Silver and Veness, 2010) that C-POMCP corresponds to the UCT algorithm (Fig. 1, Kocsis and Szepesvári, 2006) when $M \to \infty$. Hence, we can base the proof of convergence of C-POMCP on the proof of convergence for UCT, which was originally published in (Kocsis and Szepesvári, 2006). The key insight behind the proof is that the decision problem at each node in the search tree corresponds to a non-stationary multi-armed bandit (MAB), which becomes stationary if the prescribed actions at the child nodes converge. Further, the tree policy corresponds to the UCB1 algorithm (Fig. 1, Auer et al., 2002). Therefore, it suffices to prove that UCB1 converges at each node in the search tree. Towards this proof, we state and prove the following six lemmas.

### Notation

$K = O(|2^{\mathbf{X}_t}|)$ is the number of arms in the MAB at each node in the search tree; $t$ indexes the MAB rounds; $R_{i,t}$ is the reward of arm $i$ at round $t$; $\overline{R}_{i,n} = \frac{1}{n}\sum_{k=1}^{n} R_{i,n}$ is the mean reward of arm $i$ based on $n$ samples; $\mu_{i,n}$ is the mean of $\overline{R}_{i,n}$; $\mu_i \triangleq \lim_{n\to\infty}\mu_{i,n}$; $\mu_i \triangleq \mu_{i,n} - \delta_{i,n}$; $T_i(t)$ is the number of times arm $i$ has been pulled at round $t$; $\Delta_i \triangleq \mu^\star - \mu_i$; $I_t$ is the arm picked at round $t$; and $c_{t,n}$ is the exploration term in UCB1 for an arm that has been pulled $n$ times at round $t$. Quantities related to the optimal arm are superscripted by $\star$, i.e., $\mu^\star, T^\star(t)$, etc.

**Assumption 6.1** (Bounded rewards and asymptotic stationarity).

1. $R_{i,n} \in [0,1]$ *for all* $i$ *and* $n$.
2. *The limit* $\mu_i = \lim_{t\to\infty}\mu_{i,t}$ *exists for each arm* $i$.

3. *There exists a constant $C_p$ and an integer $N_p$ such that for $n \geq N_p$ and any $\delta \geq 0$, the following bounds hold.*

$$\mathbb{P}\left[n\overline{R}_{i,n} \geq n\mu_{i,n} + C_p\sqrt{n\ln(1/\delta)}\right] \leq \delta$$
$$\mathbb{P}\left[n\overline{R}_{i,n} \geq n\mu_{i,n} - C_p\sqrt{n\ln(1/\delta)}\right] \leq \delta.$$

**Lemma 6.1.** *Given Assumption 6.1, if $c_{t,n} = 2C_p\sqrt{\frac{\ln t}{n}}$, then*

$$\mathbb{P}\left[\overline{R}_{i,n} \geq \mu_{i,n} + c_{t,n}\right] \leq t^{-4} \qquad n \geq N_p$$
$$\mathbb{P}\left[\overline{R}_{i,n} \geq \mu_{i,n} - c_{t,n}\right] \leq t^{-4} \qquad n \geq N_p.$$

*Proof.*

$$\mathbb{P}\left[n\overline{R}_{i,n} \geq n\mu_{i,n} + C_p\sqrt{n\ln(1/\delta)}\right] \overset{\text{(Assumption 6.1)}}{\leq} \delta$$
$$\implies \mathbb{P}\left[\overline{R}_{i,n} \geq \mu_{i,n} + C_p\sqrt{\frac{\ln(1/\delta)}{n}}\right] \leq \delta \overset{(\delta=t^{-4})}{\Longrightarrow} \mathbb{P}\left[\overline{R}_{i,n} \geq \mu_{i,n} + 2C_p\sqrt{\frac{\ln(t)}{n}}\right] \leq t^{-4}.$$

□

**Lemma 6.2.** *If Assumption 6.1 holds, then there exists an integer $N_0(\epsilon)$ such that $t \geq N_0(\epsilon) \implies |\delta_{i,t}| \leq \frac{\epsilon\Delta_i}{2}$ and $|\delta_{j^\star,t}| \leq \min_i \frac{\epsilon\Delta_i}{2}$ for all $\epsilon > 0$.*

*Proof.* Follows by the definition of Assumption 6.1. □

**Lemma 6.3.** *Given Assumption 6.1, if the exploration term used by* UCB1 *(Fig. 1, Auer et al., 2002) is $c_{t,s} = 2C_p\sqrt{\frac{\ln t}{s}}$, then*

$$\mathbb{E}[T_i(t)] \leq \frac{16C_p^2 \ln t}{(1-\epsilon)^2\Delta_i^2} + N_0(\epsilon) + N_p + 1 + \frac{\pi^2}{3} \tag{6.22}$$

*for all $\epsilon > 0$ and each sub-optimal arm $i$.*

*Proof.* Let

$$A_0(t,\epsilon) \triangleq \min\{s \mid c_{t,s} \leq (1-\epsilon)\Delta_i/2\}$$
$$A(t,\epsilon) \triangleq \max[A_0(t,\epsilon), N_0(\epsilon), N_p].$$

Next, note that

$$c_{t,s} \leq (1-\epsilon)\Delta_i/2 \implies 2C_p\sqrt{\frac{\ln t}{s}} \leq (1-\epsilon)\Delta_i/2$$

$$\implies 16C_p\frac{\ln t}{s} \leq (1-\epsilon)^2\Delta_i^2 \implies s \geq \frac{16C_p\ln t}{(1-\epsilon)^2\Delta_i^2} \implies A_0(t,\epsilon) = \left\lceil \frac{16C_p\ln t}{(1-\epsilon)^2\Delta_i^2}\right\rceil.$$

Now consider $T_i(n)$. By definition,

$$\begin{aligned}
T_i(n) &= 1 + \sum_{t=K+1}^{n} \mathbb{1}\{I_t = i\} \qquad (6.23)\\
&= 1 + \sum_{t=K+1}^{n} \mathbb{1}\{I_t = i, T_i(t-1) \geq A(n,\epsilon)\} + \sum_{t=K+1}^{n} \mathbb{1}\{I_t = i, T_i(t-1) < A(n,\epsilon)\}\\
&\overset{(a)}{\leq} A(n,\epsilon) + \sum_{t=K+1}^{n} \mathbb{1}\{I_t = i, T_i(t-1) \geq A(n,\epsilon)\}\\
&\overset{(b)}{\leq} A(n,\epsilon) + \sum_{t=K+1}^{n} \mathbb{1}\Big\{\overline{R}^{\star}_{T^{\star}(t-1)} + c_{t-1,T^{\star}(t-1)} \leq\\
&\qquad \overline{R}_{i,T_i(t-1)} + c_{t-1,T_i(t-1)}, T_i(t-1) \geq A(n,\epsilon)\Big\}\\
&\leq A(n,\epsilon) + \sum_{t=K+1}^{n} \mathbb{1}\Big\{\min_{0<s<t} \overline{R}^{\star}_{s} + c_{t-1,s} \leq \max_{A(n,\epsilon)<s_i<t} \overline{R}_{i,s_i} + c_{t-1,s_i}\Big\}\\
&\leq A(n,\epsilon) + \sum_{t=K+1}^{n}\sum_{s=1}^{t-1}\sum_{s_i=A(n,\epsilon)}^{t-1} \mathbb{1}\Big\{\overline{R}^{\star}_{s} + c_{t-1,s} \leq \overline{R}_{i,s_i} + c_{t-1,s_i}\Big\},
\end{aligned}$$

where (a) follows because the second sum is upper bounded by $n-K$ and $n < A(n,\epsilon) \implies T_i(n) < A(n,\epsilon)$. (b) follows from the arm-selection rule in UCB1 (Fig. 1, Auer et al., 2002).

Next note that for $t \geq A(n,\epsilon) \geq N_0(\epsilon)$, we have $\mu^{\star}_t \geq \mu_{i,t} + 2c_{t,s_i}$. This inequality holds because

$$\begin{aligned}
\mu^{\star}_t \geq \mu_{i,t} + 2c_{t,s_i} &\iff \mu^{\star}_t - \mu_{i,t} - 4C_p\sqrt{\frac{\ln t}{s_i}} \geq 0\\
&\overset{(t\geq A_0(n,\epsilon))}{\iff} \mu^{\star}_t - \mu_{i,t} - (1-\epsilon)\Delta_i \geq 0\\
&\iff \mu^{\star}_t + \delta^{\star} - \mu_{i,t} - \delta_{i,t} - (1-\epsilon)\Delta_i \geq 0\\
&\overset{(\text{Lemma 6.2})}{\iff} \mu^{\star}_t - \epsilon\Delta_i - \mu_{i,t} - (1-\epsilon)\Delta_i \geq 0\\
&\iff \mu^{\star}_t - \mu_{i,t} - \Delta_i \geq 0 \iff \Delta_i - \Delta_i \geq 0. \qquad (6.24)
\end{aligned}$$

Using the above inequality, we deduce that

$$\mathbb{P}[\overline{R}^{\star}_{s} + c_{t-1,s} \leq \overline{R}_{i,s_i} + c_{t-1,s_i}] \leq \mathbb{P}[\overline{R}^{\star}_{s} \leq \mu^{\star}_t + c_{t,s}] + \mathbb{P}[\overline{X}_{i,s_i} \geq \mu_{i,t} + c_{t,s_i}]. \qquad (6.25)$$

This follows because if the left inequality above holds and the right inequalities do not hold, we obtain

$$\overline{R}^{\star}_{s} + c_{t-1,s} \leq \overline{R}_{i,s_i} + c_{t-1,s_i} \implies \mu^{\star}_t - c_{t,s} + c_{t-1,s} < \mu_{i,t} + c_{t,s_i} + c_{t-1,s_i}$$

$$\implies \mu_t^\star < \mu_{i,t} + 2c_{t,s_i},$$

which by (6.24) is false for $t \geq A(n,\epsilon) \geq N_0(\epsilon)$. Now we take expectations of both sides of the inequality in (6.23) and plug in (6.25), which gives

$$\mathbb{E}[T_i(n)] \leq A(n,\epsilon) + \sum_{t=K+1}^{n} \sum_{s=1}^{t-1} \sum_{s_i=A(n,\epsilon)}^{t-1} \mathbb{P}[\overline{R}_s^\star \leq \mu_t^\star + c_{t,s}] + \mathbb{P}[\overline{X}_{i,s_i} \geq \mu_{i,t} + c_{t,s_i}]$$

$$\overset{\text{(Lemma 6.1)}}{\leq} A(n,\epsilon) + \sum_{t=K+1}^{n} \sum_{s=1}^{t-1} \sum_{s_i=A(n,\epsilon)}^{t-1} 2t^{-4} \leq A(n,\epsilon) + \sum_{t=1}^{\infty} \sum_{s=1}^{t} \sum_{s_i=1}^{t} 2t^{-4}$$

$$= A(n,\epsilon) + \sum_{t=1}^{\infty} t^{-2} + t^{-3} \leq A(n,\epsilon) + \sum_{t=1}^{\infty} t^{-2} + \sum_{t=1}^{\infty} t^{-2}$$

$$\overset{(a)}{=} A(n,\epsilon) + \frac{\pi^2}{3} \leq \frac{16C_p \ln t}{(1-\epsilon)^2 \Delta_i^2} + 1 + N_0(\epsilon) + N_p + \frac{\pi^2}{3},$$

where (a) follows from the Riemann zeta function $\zeta(2) = \sum_{t=1}^{\infty} t^{-2} = \frac{\pi^2}{6}$. □

**Lemma 6.4** (Lower bound)**.** *Given Assumption 6.1, there exists a positive constant $\rho$ such that for all $i$ and $t$, $T_i(t) \geq \lceil \rho \log t \rceil$.*

*Proof.* Since $R_{i,t} \in [0,1]$ and $T_i(t-1) \geq 1$ for all $t \geq K$, there exists a constant $M$ such that

$$\mu_{i,t} + 2C\sqrt{\frac{\ln t}{T_i(t-1)}} \leq M \implies T_i(t-1) \geq \frac{4C^2 \ln t}{(M - \mu_i - \delta_{i,t})^2}$$

for all $i$ and $K \leq t < \infty$. Assumption 6.1 implies that $\lim_{t\to\infty} \delta_{i,t} = 0$, which means that there exists a constant $\rho \geq \frac{4C^2}{(M-\mu_i-\delta_{i,t})^2}$. Hence, $T_i(t) \geq \lceil \rho \log t \rceil$. □

**Lemma 6.5.** *Let $\overline{R}_n = \sum_{i=1}^{K} \frac{T_i(n)}{n} \overline{X}_{i,T_i(n)}$ and $N_0 = N_0(\epsilon = \frac{1}{2})$. Then, the following holds under Assumption 6.1.*

$$|\mathbb{E}[\overline{R}_n] - \mu^\star| \leq |\delta_n^\star| + O\left(\frac{K(C_p^2 \ln n + N_0)}{n}\right).$$

*Proof.* By the triangle inequality,

$$|\mathbb{E}[\overline{R}_n] - \mu^\star| \leq |\mu^\star - \mu_n^\star| + |\mu_n^\star - \mathbb{E}[\overline{R}_n]| = |\delta_n^\star| + |\mu^\star - \mathbb{E}[\overline{X}_n]|.$$

Hence, it only remains to bound $|\mu^\star - \mathbb{E}[\overline{X}_n]|$. By definition:

$$|\mu^\star - \mathbb{E}[\overline{X}_n]| = \left|\mu^\star - \mathbb{E}\left[\sum_{i=1}^{K} \frac{T_i(n)\overline{R}_{i,T_i(n)}}{n}\right]\right| \tag{6.26}$$

$$\implies n|\mu^\star - \mathbb{E}[\overline{X}_n]| = \left|\sum_{t-1}^{n}\mathbb{E}[R_t^\star] - \mathbb{E}\left[\sum_{i=1}^{K} T_i(n)\overline{R}_{i,T_i(n)}\right]\right|$$
$$\overset{(a)}{=} \left|\sum_{t-1}^{n}\mathbb{E}[R_t^\star] - \mathbb{E}\left[T^\star(n)\overline{R}^\star_{T^\star(n)}\right]\right| - \mathbb{E}\left[\sum_{i\neq i^\star} T_i(n)\overline{R}_{i,T_i(n)}\right],$$

where (a) follows because $\overline{R}_{i,t} \in [0,1]$ for all $i$ and $t$ (Assumption 6.1). We start by bounding the second term in (6.26):

$$\mathbb{E}\left[\sum_{i\neq i^\star} T_i(n)\overline{R}_{i,T_i(n)}\right] \leq \mathbb{E}\left[\sum_{i\neq i^\star} T_i(n)\right]$$
$$\overset{\text{(Lemma 6.3)}}{\leq} K\left(\frac{16C_p^2 \ln t}{(1-\epsilon)^2\Delta_i^2} + N_0(\epsilon) + N_p + 1 + \frac{\pi^2}{3}\right) = O\left(K\left(C_p^2 \ln n + N_0(\epsilon)\right)\right).$$

Now we consider the first term in (6.26). Note that $T^\star(n)\overline{R}^\star_{T^{(\star)}(n)} = \frac{T^\star(n)}{T^\star(n)}\sum_{t=1}^{T^\star(n)} \overline{R}_t^\star = \sum_{t=1}^{T^\star(n)} \overline{R}_t^\star$. Using this expression we obtain:

$$\left|\sum_{t-1}^{n}\mathbb{E}[R_t^\star] - \mathbb{E}\left[T^\star(n)\overline{R}^\star_{T^\star(n)}\right]\right| = \left|\mathbb{E}\left[\sum_{t-1}^{n} R_t^\star - \sum_{t-1}^{T^\star(n)} R_t^\star\right]\right|$$
$$\overset{(a)}{=} \sum_{t=T^\star(n)+1}^{n} \mathbb{E}\left[R_t^\star\right] \leq \mathbb{E}[n - T^\star(n)] = \sum_{i\neq i^\star}\mathbb{E}[T_i(n)]$$
$$\overset{\text{(Lemma 6.3)}}{\leq} K\left(\frac{16C_p^2 \ln t}{(1-\epsilon)^2\Delta_i^2} + N_0(\epsilon) + N_p + 1 + \frac{\pi^2}{3}\right) = O\left(K\left(C_p^2 \ln n + N_0(\epsilon)\right)\right),$$

where (a) follows from the fact that $\overline{R}_{i,t} \in [0,1]$ for all $i$ and $t$ (Assumption 6.1). □

**Lemma 6.6.** *Let $n_0$ be such that $\sqrt{n_0} \geq O(K(C_p^2 \ln n_0 + N_0(\frac{1}{2})))$. Given Assumption 6.1, the following holds for any $n \geq n_0$ and $\delta > 0$:*

$$\mathbb{P}\left[n\overline{X}_n \geq n\mathbb{E}[\overline{X}_n] + 9\sqrt{2\ln(2/\delta)}\right] \leq \delta$$
$$\mathbb{P}\left[n\overline{X}_n \geq n\mathbb{E}[\overline{X}_n] - 9\sqrt{2\ln(2/\delta)}\right] \leq \delta.$$

The above lemma was originally proved by Kocsis & Szepesvári (Thm. 5, Kocsis and Szepesvári, 2006). A more accessible version of the proof can be found in (Thm. 5, Dam et al., 2021). We omit the full proof for brevity. For the sake of completeness, we briefly outline the main ideas behind the proof here. The proof involves defining a counting process that represents the number of times a sub-optimal arm is pulled and then bounding the deviation of this process. Key to this argument is the Hoeffding-Azuma inequality (Lem. 8–10, Kocsis and Szepesvári, 2006). By leveraging this inequality and martingale theory, it is possible to conclude that the desired inequalities must hold.

**Lemma 6.7.** *Given Assumption 6.1,* $\lim_{t\to\infty} \mathbb{P}[I_t \neq i^\star] = 0$.

*Proof.* Let $p_{i,t} = \mathbb{P}[\overline{R}_{i,T_i(t)} \geq R^\star_{T^\star(t)}]$. Clearly, $\mathbb{P}[I_t \neq i^\star] \leq \sum_{i\neq i^\star} p_{it}$. Hence, it suffices to show that $p_{i,t} \leq \frac{\epsilon}{K}$ for all $i$ and any $\epsilon > 0$. Towards this proof, note that if $\overline{R}_{i,T_i(t)} < \mu_i + \frac{\Delta_i}{2}$ and $\overline{R}^\star_{T^\star(t)} > \mu^\star - \frac{\Delta}{2}$, then

$$\overline{R}_{i,T_i(t)} < \mu_i + \frac{\Delta_i}{2} = \mu^\star - \frac{\Delta_i}{2} < \overline{R}^\star_{T^\star(t)}.$$

As a consequence,

$$p_{i,t} \leq \mathbb{P}\left[\overline{R}_{i,T_i(t)} < \mu_i + \frac{\Delta_i}{2}\right] + \mathbb{P}\left[\overline{R}^\star_{T^\star(t)} > \mu^\star - \frac{\Delta}{2}\right].$$

Since $T_i(t)$ grows slower than $T^\star(t)$, it suffices to bound the first of the two terms above. By definition:

$$\mathbb{P}\left[\overline{R}_{i,T_i(t)} < \mu_i + \frac{\Delta_i}{2}\right] = \mathbb{P}\left[\overline{R}_{i,T_i(t)} < \mu_{i,T_i(t)} - |\delta_{i,T_i(t)}| + \frac{\Delta_i}{2}\right].$$

Next note that $|\delta_{i,T_i(t)}|$ converges to 0 by Assumption 6.1. Hence, we can assume that $|\delta_{i,T_i(t)}|$ is decreasing in $t$ without loss of generality. It then follows from Lemma 6.4 that $|\delta_{i,T_i(t)}| \leq |\delta_{i,\lceil\rho\log t\rceil}|$. Now consider $t \geq \lceil\rho\log t\rceil \geq N_0(\frac{\Delta}{4})$, then $|\delta_{i,T_i(t)}| \leq |\delta_{i,\lceil\rho\log t\rceil}| \leq \frac{\Delta_i}{4}$ by Lemma 6.2. Therefore,

$$\mathbb{P}\left[\overline{R}_{i,T_i(t)} < \mu_{i,T_i(t)} - |\delta_{i,T_i(t)}| + \frac{\Delta_i}{2}\right] \leq \mathbb{P}\left[\overline{R}_{i,T_i(t)} < \mu_{i,T_i(t)} + \frac{\Delta_i}{4}\right] \tag{6.27}$$
$$\leq \mathbb{P}\left[\overline{R}_{i,T_i(t)} < \mu_{i,T_i(t)} + \frac{\Delta_i}{4}, T_i(t) \geq a\right] + \mathbb{P}\left[T_i(t) \leq a\right].$$

Since $\lim_{t\to\infty} T_i(t) = \infty$ (Lemma 6.4), we have that $\lim_{t\to\infty} \mathbb{P}[T_i(t) < a] = 0$ for any $a$. Hence, it suffices to bound the first term in (6.27), which can be done as follows.

$$\mathbb{P}\left[\overline{R}_{i,T_i(t)} < \mu_{i,T_i(t)} + \frac{\Delta_i}{4}, T_i(t) \geq a\right] \leq$$
$$\mathbb{P}\left[\overline{R}_{i,T_i(t)} < \mu_{i,T_i(t)} + \frac{\Delta_i}{4} \mid T_i(t) \in [a,b]\right] \mathbb{P}\left[T_i(t) \in [a,b]\right] + \mathbb{P}\left[T_i(t) \notin [a,b]\right].$$

Next note that

$$\mathbb{P}\left[\overline{R}_{i,T_i(t)} < \mu_{i,T_i(t)} + \frac{\Delta_i}{4} \mid T_i(t) \in [a,b]\right] \leq \sum_{k=a}^{b} \mathbb{P}\left[\overline{R}_{i,k} < \mu_{i,k} + \frac{\Delta_i}{4}\right]$$

$$\leq (b-a+1) \max_{a \leq k \leq b} \mathbb{P}\left[\overline{R}_{i,k} < \mu_{i,k} + \frac{\Delta_i}{4}\right].$$

Now we use Assumption 6.1, which implies that the value of $a$ in the expression above can be chosen such that for all $t \geq a$, $(t+1)\mathbb{P}[\overline{X}_{i,t} \geq \mu_{i,t} + \frac{\Delta_i}{4}] < \frac{\epsilon}{2K}$, which means that

$$(b-a+1) \max_{a \leq k \leq b} \mathbb{P}\left[\overline{R}_{i,k} < \mu_{i,k} + \frac{\Delta_i}{4}\right] =^{b=2a} (a+1) \max_{a \leq k \leq b} \mathbb{P}\left[\overline{R}_{i,k} < \mu_{i,k} + \frac{\Delta_i}{4}\right]$$

$$\leq (a+1)\mathbb{P}\left[\overline{R}_{i,a} < \mu_{i,a} + \frac{\Delta_i}{4}\right] \leq \frac{\epsilon}{2K}.$$

Putting the bounds above together, we get:

$$\mathbb{P}[I_t \neq i^\star] \leq \sum_{i \neq i^\star} p_{it} \leq K\frac{\epsilon}{K} = \epsilon.$$

□

Now we use Lemmas 6.1–6.7 to prove Thm. 6.4.

## B.1 Proof of Theorem 6.4

The proof uses mathematical induction on the time horizon $\mathcal{T}$, which corresponds to the search tree's depth. Without loss of generality, we assume that the target variables are normalized to the interval $[0,1]$. For the inductive base case where $\mathcal{T} = 2$, c-pomcp corresponds to a stationary mab problem, which means that Assumption 6.1.1–2 hold. Further, Assumption 6.1.3 follows from Hoeffding's inequality:

$$\mathbb{P}\left[\overline{R}_{i,n} \leq \mu_{i,n} \pm \frac{1}{n}\sqrt{\frac{\ln t}{n}}\right] \leq e^{-2\frac{\sqrt{\frac{2\ln t}{n}}}{n}} = e^{-4\ln t} = t^{-4}.$$

As a consequence, we can invoke Lemma 6.5 and Lemma 6.7, which asserts that the theorem statement holds when $\mathcal{T} = 2$. Assume by induction that the theorem statement holds for time horizons $3, 4, \ldots, \mathcal{T}-1$ and consider $\mathcal{T}$. By Thm. 6.3, the problem of finding an optimal intervention from the root node when the horizon is $\mathcal{T}$ corresponds to a non-stationary mab with correlated rewards. This mab satisfies Assumption 6.1.1 if the rewards are divided by $\mathcal{T}$. Further, it follows from the induction hypothesis that the reward distributions in each subtree converge, which means that Assumption 6.1.2 holds. Moreover, the induction hypothesis and Lemma 6.6 imply that Assumption 6.1.3 is satisfied. Consequently, we can apply Lemma 6.7, which ensures that the intervention prescribed at the root converges to an optimal intervention. Similarly, we can apply Lemma 6.5, which states that

$$\left|\mathbb{E}\left[\overline{R}_n - \mu_n^\star\right]\right| \leq |\delta^\star| + O\left(\frac{KC_p^2 \ln n + N_0}{n}\right),$$

where $\overline{R}_n$ is the average reward at the root. By the induction hypothesis,

$$|\delta^\star| = O\left(\frac{K(\mathcal{T}-1)\log n + K^{\mathcal{T}-1}}{n}\right).$$

Hence,

$$\left|\mathbb{E}[\overline{R}_n - \mu_n^\star]\right| \leq O\left(\frac{K(\mathcal{T}-1)\log n + K^{\mathcal{T}-1}}{n}\right) + O\left(\frac{KC_p^2 \ln n + N_0}{n}\right).$$

Thus, it only remains to bound $N_0$. It follows from Lemma 6.2 that $N_0$ is upper bounded by the smallest value of $n$ for which the following inequality holds

$$\frac{K(\mathcal{T}-1)\log n + K^{\mathcal{T}-1}}{n} \geq \frac{\Delta_i}{2} \implies \frac{2\left(K^{\mathcal{T}-1} + K(\mathcal{T}-1)\log n\right)}{\Delta_i} \geq n$$
$$\implies N_0 = O(K^{\mathcal{T}-1}).$$

Consequently,

$$\left|\mathbb{E}[\overline{R}_n - \mu_n^\star]\right| \leq O\left(\frac{K(\mathcal{T}-1)\log n + K^{\mathcal{T}-1}}{n}\right)$$
$$+ O\left(\frac{KC_p^2 \ln n + K^{\mathcal{T}}}{n}\right) = O\left(\frac{(K\mathcal{T}\log n + K^{\mathcal{T}})}{n}\right).$$

Thus, by induction, the theorem holds for all $\mathcal{T}$. □

## C Hyperparameters

The hyperparameters used for the evaluation are listed in Table 6.4 and were obtained through random search.

| *Parameters* | *Values* |
|---|---|
| $\beta_{\mathsf{R},1}$, $\beta_{\mathsf{R},2}$, $\beta_{\mathsf{R},3}$ | 0.1, 1, 1 |
| $\beta_{I_{i,t},z}$ | 0 if $I_{i,t} \neq \mathsf{R}$ |
| $\tilde{c}(\text{do}(\widehat{\mathbf{X}}_t = \pi_{\text{D}}(\mathbf{h}_t)))$ | 1 for restore interventions, 0 otherwise |
| $\psi_{z_i}$ | 10 if $z_i = 3$, 0 otherwise |
| Cyborg version [440] | commit 9421c8e |
| c-pomcp and pomcp [416, Alg. 1] | |
| search time, default node value | 0.05s–30s, 0 |
| $M$, $\gamma$, $c$ | 1000, 0.99, 0.5 |
| rollout depth, maximum search depth | 4, 50 |
| intervention space | intervention space described in [481] |
| base strategy, base value | $\widehat{\pi}_{\text{D}}(\text{do}(\emptyset)) = 1$, $J_{\widehat{\pi}}(\cdot) = 0$ |
| cardiff-ppo [396, Alg. 1] [481] | |
| learning rate, # hidden layers, | $5148 \cdot 10^{-5}$, 1, |
| # neurons per layer, # steps between updates, | 64, 2048, |
| batch size, discount factor $\gamma$ | 16, 0.99 |
| gae $\lambda$, clip range, entropy coefficient | 0.95, 0.2, $2 \cdot 10^{-4}$ |
| value coefficient, max gradient norm | 0.102, 0.5 |
| feature representation | the original cyborg features [440] & one-hot encoded scan-state & decoy-state for each node |

***Table 6.4:*** *Hyperparameters.*

## D Configuration of the Target System (Figure 6.5)

The configuration of the target system in cage-2 (Fig. 6.5) is available in Table 6.5. The attacker actions are listed in Table 6.6, and the defender interventions are listed in Table 6.7. The decoy services are listed in Table 6.8, and the workflow graph is shown in Fig. 6.13.

| *Node* ID*, hostname* | *Processes* | *Ports* | *Users* | *Vulnerabilities* |
|---|---|---|---|---|
| 1, CLIENT-1 | SSHD.EXE | 22 | SSHD_SERVER | CWE-251. |
| | FEMITTER.EXE | 21 | SYSTEM | CVE-2020-26299. |
| 2, CLIENT-2 | SMSS.EXE | 445,139 | SYSTEM | -. |
| | SVCHOST.EXE | 135 | SYSTEM | -. |
| | SVCHOST.EXE | 3389 | NETWORK | CVE-2019-0708. |
| 3, CLIENT-3 | MYSQL | 3389 | ROOT | CVE-2019-0708. |
| | APACHE2 | 80,443 | WWW-DATA | CWE-89, HTTP-(S)RFI. |
| | SMTP | 25 | ROOT | CVE-2016-1000282. |
| 4, CLIENT-4 | SSHD | 22 | ROOT | CWE-251. |
| | MYSQL | 3390 | ROOT | CWE-89. |
| | APACHE2 | 80, 443 | WWW-DATA | CWE-89, HTTP-(S)RFI. |
| | SMTP | 25 | ROOT | CVE-2016-1000282. |
| 5, ENTERPRISE-1 | SSHD.EXE | 22 | ROOT | CWE-251. |
| 6, ENTERPRISE-2 | SSHD.EXE | 22 | SSHD_SERVER | CWE-251. |
| | SVCHOST.EXE | 135 | SYSTEM | -. |
| | SVCHOST.EXE | 3389 | SYSTEM | CVE-2019-0708. |
| | SMSS.EXE | 445,139 | SYSTEM | CVE-2017-0144. |
| | TOMCAT8.EXE | 80,443 | NETWORK | CWE-89,HTTP-(S)RFI. |
| 7, ENTERPRISE-3 | SSHD.EXE | 22 | SSHD_SERVER | CWE-251. |
| | SVCHOST.EXE | 135 | SYSTEM | -. |
| | SVCHOST.EXE | 3389 | SYSTEM | CVE-2019-0708. |
| | SMSS.EXE | 445,139 | SYSTEM | CVE-2017-0144. |
| | TOMCAT8.EXE | 80,443 | NETWORK | CWE-89,HTTP-(S)RFI. |
| 8, OPERATIONAL-1 | SSHD | 22 | ROOT | CWE-251. |
| 9, OPERATIONAL-2 | SSHD | 22 | ROOT | CWE-251. |
| 10, OPERATIONAL-3 | SSHD | 22 | ROOT | CWE-251. |
| 11, OPERATIONAL-4 | SSHD | 22 | ROOT | CWE-251. |
| 12, DEFENDER | SSHD | 22 | ROOT | CWE-251. |
| | SYSTEMD | 53,78 | SYSTEMD+ | |
| 13, ATTACKER | SSHD.EXE | 22 | SSHD_SERVER | CWE-251. |
| | FEMITTER.EXE | 21 | SYSTEM | CVE-2020-26299. |

***Table 6.5:*** *Configuration of the target system in* CAGE-2 *(Fig. 6.5); vulnerabilities are identified by their identifiers in the Common Vulnerabilities and Exposures (*CVE*) database (The MITRE Corporation, 2022) and the Common Weakness Enumeration (*CWE*) list (The MITRE Corporation, 2023). (Note that each node in Fig. 6.5 is labeled with an identifier.)*

| *Type* | *Actions* | mitre att&ck technique |
|---|---|---|
| Reconnaissance | Subnet scan for nodes | t1018 system discovery. |
| | Port scan on a specific node | t1046 service scanning. |
| Exploits | cve-2017-0144, http-srfi | t1210 service exploitation. |
| | sql injection (cwe-89) | t1210 service exploitation. |
| | cve-2016-1000282 | t1210 service exploitation. |
| | cve-2020-26299 http-rfi | t1210 service exploitation. |
| Brute-force | ssh | t1110 brute force. |
| Escalate | Escalate privileges of user to root | t1068 privilege escalation. |
| Impact | Stop services running on node | t1489 service stop. |

***Table 6.6:*** *Attacker actions in cage-2 (cage-2, 2022); the actions are linked to the corresponding attack techniques in the mitre att&ck taxonomy (Strom et al., 2018).*

| *Type* | *Interventions* | mitre d3fend technique |
|---|---|---|
| Monitor | Network monitoring | d3-nta network analysis. |
| | Forensic analysis | d3-fa file analysis. |
| Start decoys | apache, femitter | d3-de decoy environment. |
| | haraka, smss | d3-de decoy environment. |
| | sshd, svchost tomcat | d3-de decoy environment. |
| Restore | Restore node to a checkpoint | d3-ra restore access. |
| | Attempt to remove attacker | d3-fev file eviction. |

***Table 6.7:*** *Defender interventions in cage-2 (cage-2, 2022); the interventions are linked to the corresponding defense techniques in the mitre d3fend taxonomy (Kaloroumakis and Smith, 2021).*

| id | *Name* | *Description* |
|---|---|---|
| 1 | decoy-apache | Starts a vulnerable apache http server decoy. |
| 2 | decoy-femitter | Starts a vulnerable femitter ftp server decoy. |
| 3 | decoy-smtp | Starts a vulnerable haraka smtp server decoy. |
| 4 | decoy-smss | Starts a vulnerable smss server decoy. |
| 5 | decoy-sshd | Starts an ssh server decoy with a weak password. |
| 6 | decoy-svchost | Starts a vulnerable svchost.exe process decoy. |
| 7 | decoy-tomcat | Starts a vulnerable tomcat http server decoy. |
| 8 | decoy-vsftpd | Starts a vulnerable vsftpd ftp server decoy. |

***Table 6.8:*** *Decoy services in cage-2 (cage-2, 2022).*

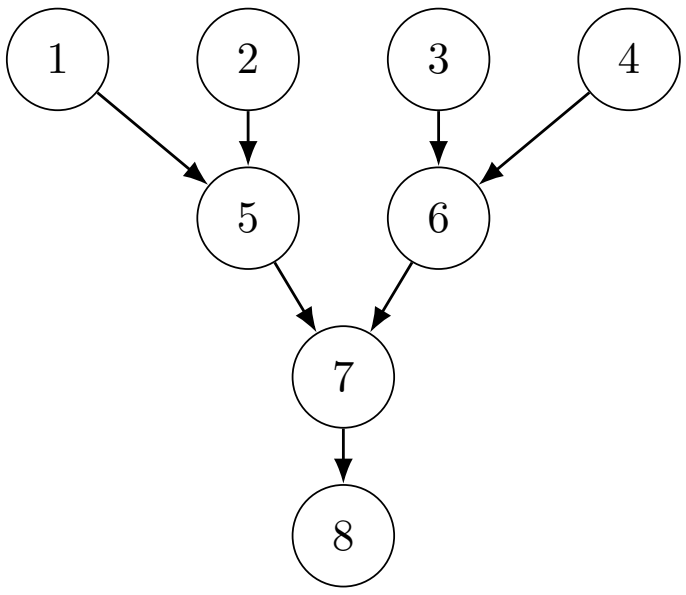

***Figure 6.13:*** *Workflow graph $\mathcal{G}_{\mathrm{W}}$ in* CAGE-2 *(*CAGE-2*, 2022); circles represent nodes of the target system (Fig. 6.5) and edges represent service dependencies. (Note that each node in Fig. 6.5 is labeled with an identifier.)*

# CONCLUSION

*We know the past but cannot control it.*
*We control the future but cannot know it.*

— Claude Shannon ***1959***, *Coding theorems for a discrete source.*

OUR reliance on online services and IT systems has never been more apparent; they provide critical societal functions, enable global communication, and drive economic growth. The extent of this dependence was clearly demonstrated during the COVID-19 pandemic, when remote work, online education, and digital healthcare became necessary. While this digitalization provides many benefits, it simultaneously introduces vulnerabilities that adversaries can exploit. As these vulnerabilities are of growing concern to society, *automated security response* has been recognized as an essential countermeasure (Montesino and Fenz, 2011).

By *response*, we understand the coordinated actions taken to contain, mitigate, and recover from cyber attacks. Historically, such actions have been automated (to some extent) using rules designed by human experts (Denning, 1987). Though this approach has been accepted practice, the rapidly growing number of attacks renders it impractical today. Framing security response as an optimal control problem[†] offers a promising solution to this limitation by enabling the automatic optimization of responses based on system measurements. Such framing facilitates a rigorous treatment of security response where trade-offs between different security objectives can be systematically studied through mathematical models. Prior art demonstrates the advantages of this approach for simulation-based environments (Nguyen and Reddi, 2023). However, due to the gap between a simulation and an IT infrastructure, the feasibility of such approaches for operational use remains elusive. Motivated by this limitation, this thesis studies the following question.

> *What methodology can be used to develop an automated security response system that guarantees scalability and optimality, and how can the system be rigorously validated on a testbed?*

We answer this question by presenting a practical methodology for *optimal security response* in IT infrastructures, which constitutes our main contribution. It includes two systems; see Fig. 12 in the introduction chapter. First, it includes an *emulation system* that replicates key components of the target infrastructure. We use this system to gather measurements and logs, based on which we identify a game-theoretic model. Second, it includes a *simulation system* where response strategies are optimized through stochastic approximation to meet a given objective, e.g., mitigate attacks while maintaining service availability. These strategies are then evaluated and refined in the emulation system to bridge the gap between theoretical and operational performance. We argue that this methodology provides a foundation for the next generation of response systems. The response strategies computed through our methodology can be used in these systems to decide when an automated response should be triggered or an operator should be alerted. They can also suggest, in real-time, which responses to execute. Unlike the simulation-based solutions proposed in prior research, our response strategies are experimentally validated and useful in practice: they can be integrated with existing response systems, they satisfy safety constraints, and they are computationally efficient.

[†]Here "*control problem*" encompasses both control- and game-theoretic problem formulations.

We present CSLE, a novel open-source platform that implements our methodology (Hammar, 2023). This platform allows us to prove the methodology's practical applicability and superiority to current solutions by validating it experimentally on several instances of the security response problem, including intrusion prevention (Paper 1), intrusion response (Paper 2 and Paper 3), intrusion tolerance (Paper 4), and defense against advanced persistent threats (Paper 5 and Paper 6). Besides the experimental validation, in these papers we also develop fundamental mathematical tools for security response and prove structural properties of optimal response strategies, such as decomposability (Thm. 3.2) and threshold structure (Thms. 1.1, 2.1, 4.3, 4.5, and 5.1). Leveraging these structural results, we develop scalable computational techniques (Alg. 1.1–6.1), which have optimality guarantees (Thms. 5.3, 5.4, and 6.4). Our dual focus on practical applicability and computational efficiency enables us to narrow the gap between theoretical and operational performance; see the results in Fig. 1.10, Fig. 2.7, Fig. 3.9, Fig. 4.15, and Fig. 5.17.

The main experimental finding of this thesis is that the most important factor for scalable and optimal security response is to *exploit structure*, both structure in theoretical models (e.g., optimal substructure) and structure of the IT environment (e.g., network topology). The former enables efficient computation of optimal response strategies and the latter is key to managing the high-dimensional nature of IT infrastructures. Another notable finding is that stochastic approximation is a suitable method for such computation, striking a good balance between satisfactory performance and computational expenditure.

**Future research**

Optimal security response is a vast field that spans numerous areas of expertise, including cybersecurity, optimization theory, game theory, control theory, causality, stochastic approximation, reinforcement learning, and systems engineering. As such, there is an endless road of possible generalizations to the results of this thesis. Some extensions were deliberately excluded to maintain coherence (e.g., extensions to different use cases), while other limitations were imposed for practical reasons (e.g., evaluation in a production environment). The next page highlights the research directions that we consider the most relevant for further investigation.

1. **Adaptive digital twin**
   Online execution of our methodology requires updating the digital twin to reflect changes in the physical infrastructure; see Fig. 19 in the methodology chapter. While we have performed a preliminary study of this problem (Hammar and Stadler, 2022), we have not studied it in depth. Therefore, this problem requires additional examination.

2. **Response strategies for diverse infrastructures**
   In each case study presented in this thesis, we focus on the problem of finding an optimal response strategy for a specific target infrastructure. We have not considered the problem of finding a general strategy for a class of infrastructures; this problem presents an opportunity for future research.

3. **Resilience against adversarial perturbations**
   Control strategies optimized through stochastic approximation are vulnerable to adversarial attacks introducing small perturbations to the strategy input (Russo and Proutiere, 2021). The problem of developing robust response strategies against such perturbations requires further analysis.

4. **Integration with human-in-the-loop architectures**
   This thesis focuses on fully automated security response systems. However, not all responses can be fully automated due to the complexity, ambiguity, and ethical considerations of some security incidents. Future work should explore how to effectively interface automated systems with human operators.

5. **Extensions to additional use cases**
   In each response scenario studied in this thesis, we make assumptions about the underlying IT infrastructure and the attacker's capabilities. Applying our methodology to new scenarios with different assumptions will likely uncover additional challenges that must be dealt with. For example, many of our techniques for IT security response can be extended to cyber-physical systems.

## Conclusion

The rising frequency of cyber attacks and their widespread impact on society necessitates automated security response. Achieving such automation is one of the greatest challenges in cybersecurity. In this thesis, we have demonstrated optimal security response in the most realistic attack scenarios considered to date. We achieved our results using a novel methodology that combines a digital twin with system identification and simulation-based optimization. While technical challenges remain, our general outlook on automated security response is positive. We foresee future IT infrastructures where virtualization technologies combined with increasing data rates will enable dynamic and automated security response through our data-driven methodology. Such automation has the potential to significantly reduce the impact of cyber attacks, decrease operational costs, and shorten recovery times, thereby contributing to more sustainable and reliable digital infrastructures.